LBNL-190005

# LUX-ZEPLIN(LZ)
# Conceptual Design Report

March 2015

# The LUX-ZEPLIN (LZ) Collaboration

A. Harris, R. Rosero, M. Yeh
*Brookhaven National Laboratory (BNL) P.O. Box 5000, Upton, NY 11973-5000, USA*

C. Chan, S. Fiorucci, R.J. Gaitskell, D.Q. Huang, D.C. Malling,[1] M Pangilinan,[2] C.Rhyne, W. Taylor, J.R. Verbus
*Brown University, Department of Physics, 182 Hope Street, Providence, RI 02912-9037, USA*

W.H. Lippincott, D.J. Markley, T.J. Martin, M. Sarychev
*Fermi National Accelerator Laboratory (FNAL), Batavia, IL 60510-0500, USA*

H.M. Araújo, A.J. Bailey, A. Currie, T.J. Sumner, A. Tomás
*Imperial College London, High Energy Physics, Blackett Laboratory, London, SW7 2AZ, London, UK*

J.P. da Cunha, L. de Viveiros,[3] A. Lindote, M.I. Lopes, F. Neves, J.P. Rodrigues, C. Silva, V.N. Solovov
*Laboratório de Instrumentação e Física Experimental de Partículas (LIP), Department of Physics, University of Coimbra, Rua Larga, 3004-516, Coimbra, Portugal*

M.J. Barry, A. Dobi, W.R. Edwards, C.H. Faham,[2] N.J. Gantos, V.M. Gehman, M.G.D. Gilchriese, M.D. Hoff, K. Kamdin,[4] K.T. Lesko, K.C. Oliver-Mallory,[4] S.J. Patton, J.S. Saba, P. Sorensen, K.J. Thomas,[5] C.E. Tull, W.L. Waldron
*Lawrence Berkeley National Laboratory (LBNL), 1 Cyclotron Road, Berkeley, CA 94720-8156 USA*

A. Bernstein, K. Kazkaz
*Lawrence Livermore National Laboratory (LLNL), 7000 East Avenue, Livermore, CA 94550-9698, USA*

D. Yu. Akimov, A.I. Bolozdynya, A.V. Khromov, A.M. Konovalov, A.V. Kumpan, V.V. Sosnovtsev
*National Research Nuclear University (NRNU MEPhI), Kashirkoe sh. 31, Moscow, 115409, Russia*

C.E. Dahl[6]
*Northwestern University, Department of Physics & Astronomy, 2145 Sheridan Road, Evanston, IL 60208-3112, USA*

C. Fu, X. Ji,[7] Y. Ju, J. Liu
*Shanghai Jiao Tong University, INPAC at the Department of Physics and Astronomy, 800 Dongchuan Road, Shanghai, 200240, China*

D.S. Akerib,[ba] T.P. Biesiadzinski,[ba] R. Bramante,[ba] W.W. Craddock,[b] C.M. Ignarra,[ba] W. Ji,[ba] H.J. Krebs,[b] C. Lee,[ba,8] S. Luitz,[b] M.E. Monzani,[ba] F.G. O'Neill,[b] K.J. Palladino,[ba] B.N. Ratcliff,[b] T.A. Shutt,[ba] K. Skarpaas,[b] W.H. To,[ba] J. Va'vra,[b] T.J. Whitis,[ba] W.J. Wisniewski[b]
*[a]Kavli Institute for Particle Astrophysics and Cosmology, M/S 29; [b]SLAC National Accelerator Laboratory, 2575 Sand Hill Road, Menlo Park CA 94205-7015, USA*

[1]Now at: Lincoln Laboratory, 244 Wood Street, Massachusetts Institute of Technology, Lexington, MA 02421-6426, USA
[2]Now at: Insight Data Science, 260 Sheridan Avenue Suite 310, Palo Alto, CA 94306-2010, USA
[3]Now at: University of California (UC), Santa Barbara, Department of Physics, Broida Hall, Santa Barbara, CA 93106-9530, USA
[4]Also at: University of California (UC), Berkeley, Department of Physics, 366 LeConte Hall MC 7300, Berkeley, CA 94720-7300, USA
[5]Also at: University of California (UC), Berkeley, Department of Nuclear Engineering, 4155 Etcheverry Hall MC 1730, CA 94720-1730, USA
[6]Also at: Fermi National Accelerator Laboratory (FNAL), Batavia, IL 60510-0500, USA
[7]Also at: University of Maryland, Department of Physics, College Park, MD 20742-4111, USA
[8]Now at: Institute for Basic Science, 70, Yuseong-daero 1689-gil, Yuseong-gu, Daejeon, Korea, 305-811

X. Bai, R. Bunker, E.H. Miller, J. Reichenbacher, R.W. Schnee, M.R. Stark, K. Sundarnath, D.R. Tiedt
*South Dakota School of Mines and Technology, 501 East Saint Joseph Street, Rapid City, SD 57701-3901, USA*

P. Bauer, B. Carlson, M. Johnson, D.J. Taylor
*South Dakota Science and Technology Authority (SDSTA), Sanford Underground Research Facility, 630 East Summit Street, Lead, SD 57754-1700, USA*

S. Balashov, V. Francis, P. Ford, E. Holtom, A. Khazov, P. Majewski, J.A. Nikkel, J. O'Dell, R.M. Preece, M.G.D. van der Grinten, S.D. Worm
*STFC Rutherford Appleton Laboratory (RAL), Didcot, OX11 0QX, UK*

R.L. Mannino, T.M. Stiegler, P.A. Terman, R.C. Webb
*Texas A&M University, Department of Physics and Astronomy, 4242 TAMU, College Station, TX 77843-4242, USA*

J. Mock, M. Szydagis, S.K. Young
*University at Albany (SUNY), Department of Physics, 1400 Washington Avenue, Albany, NY 12222-1000, USA*

M. Cascella, J.E.Y. Dobson, C. Ghag, X. Liu, L. Manenti, L. Reichhart,[9] S. Shaw
*University College London, Department of Physics and Astronomy, Gower Street, London WC1E 6BT, UK*

J.K. Busenitz, M. Elnimr,[10] Y. Meng, A. Piepke, I. Stancu
*University of Alabama, Department of Physics & Astronomy, 206 Gallalee Hall, 514 University Boulevard, Tuscaloosa, AL 34587-0324, USA*

S.A. Hertel,[11] R.G. Jacobsen, D.N. McKinsey,[11] K. O'Sullivan,[11]
*University of California (UC), Berkeley, Department of Physics, 366 LeConte Hall MC 7300, Berkeley, CA 94720-7300, USA*

J.E. Cutter, R.M. Gerhard, B. Holbrook, B.G. Lenardo, A.G. Manalaysay, J.A. Morad, S. Stephenson, J.A. Thomson, M. Tripathi, S. Uvarov, E. Woods,[12]
*University of California (UC), Davis, Department of Physics, One Shields Avenue, Davis, CA 95616-5270, US*

M.C. Carmona-Benitez, S.J. Haselschwardt, S. Kyre, H.N. Nelson, D.T. White, M.S. Witherell
*University of California (UC), Santa Barbara, Department of Physics, Broida Hall, Santa Barbara, CA 93106-9530, USA*

P. Beltrame, T.J.R. Davison, M.F. Marzioni, A.St.J. Murphy
*University of Edinburgh, SUPA, School of Physics and Astronomy, Edinburgh EH9 3FD, UK*

S. Burdin, S. Powell, H.J. Rose
*University of Liverpool, Department of Physics, Liverpool L69 7ZE, UK*

J. Balajthy, T.K. Edberg, C.R. Hall
*University of Maryland, Department of Physics, College Park, MD 20742-4111, USA*

[9]Now at: IMS Nanofabrication AG, Wolfholzgasse 20-22, 2345 Brunn am Gebirge, Austria
[10]Now at: University of California (UC), Irvine, Department of Physics & Astronomy, 4129 Frederick Reines Hall, Irvine, CA 92697-4575, USA
[11]Also at: Yale University, Department of Physics, 217 Prospect Street, New Haven, CT 06511-8499, USA
[12]Now at: 729 Willow Street, Oakland, CA 94607-1335, USA

C.W. Akerlof, W. Lorenzon, K. Pushkin, M.S.G. Schubnell
*University of Michigan, Randall Laboratory of Physics, 450 Church Street, Ann Arbor, MI 48109-1040, USA*

K.E. Boast, C. Carels, H. Kraus, F.-T. Liao, J. Lin, P.R. Scovell
*University of Oxford, Department of Physics, Oxford OX1 3RH, UK*

E. Druszkiewicz, D. Khaitan, W. Skulski, F.L.H. Wolfs, J. Yin
*University of Rochester, Department of Physics and Astronomy, 206 Bausch & Lomb Hall, P.O. Box 270171 Rochester, NY 14627-0171, USA*

E.V. Korolkova, V.A. Kudryavtsev, D. Woodward
*University of Sheffield, Department of Physics and Astronomy, Sheffield S3 7RH, UK*

A.A. Chiller, C. Chiller, D.-M. Mei, L. Wang, W.-Z. Wei, M. While, C. Zhang
*University of South Dakota, Department of Physics, 414 East Clark Street, Vermillion, SD 57069-2307, USA*

S.K. Alsum, D.L. Carlsmith, J.J. Cherwinka, S. Dasu, B. Gomber, C.O. Vuosalo
*University of Wisconsin-Madison, Department of Physics, 1150 University Avenue Room 2320, Chamberlin Hall, Madison, WI 53706-1390, USA*

J.H. Buckley, V.V. Bugaev, M.A. Olevitch
*Washington University in St. Louis, Department of Physics, One Brookings Drive, St. Louis, MO 63130-4862, USA*

E.P. Bernard, E.M. Boulton, B.N. Edwards, W.T. Emmet, T.W. Hurteau, N.A. Larsen, E.K. Pease, B.P. Tennyson, L. Tvrznikova
*Yale University, Department of Physics, 217 Prospect Street, New Haven, CT 06511-8499, USA*

## ABSTRACT

The design and performance of the LUX-ZEPLIN (LZ) detector is described as of March 2015 in this Conceptual Design Report. LZ is a second-generation dark-matter detector with the potential for unprecedented sensitivity to weakly interacting massive particles (WIMPs) of masses from a few $GeV/c^2$ to hundreds of $TeV/c^2$. With total liquid xenon mass of about 10 tonnes, LZ will be the most sensitive experiment for WIMPs in this mass region by the end of the decade. This report describes in detail the design of the LZ technical systems. Expected backgrounds are quantified and the performance of the experiment is presented. The LZ detector will be located at the Sanford Underground Research Facility in South Dakota. The organization of the LZ Project and a summary of the expected cost and current schedule are given.

# TABLE OF CONTENTS

## ACKNOWLEDGMENTS

The work was partially supported by the U.S. Department of Energy (DOE) under award numbers DE-SC0012704, DE-SC0010010, DE-AC02-05CH11231, DE-SC0012161, DE-SC0014223, DE-FG02-13ER42020, DE-FG02-91ER40674, DE-NA0000979, DE-SC0011702, DESC0006572, DESC0012034, DE-SC0006605, and DE-FG02-10ER46709; by the U.S. National Science Foundation (NSF) under award numbers NSF PHY-110447, NSF PHY-1506068, NSF PHY-1312561, and NSF PHY-1406943; by the U.K. Science & Technology Facilities Council under award numbers ST/K006428/1, ST/M003655/1, ST/M003981/1, ST/M003744/1, ST/M003639/1, ST/M003604/1, and ST/M003469/1; and by the Portuguese Foundation for Science and Technology (FCT) under award numbers CERN/FP/123610/2011 and PTDC/FIS-NUC/1525/2014.

We acknowledge additional support from Brown University for the Center for Computation and Visualization, from the South Dakota Governor's Research Center for the Center for Ultra-Low Background Experiments (CUBED), from the Boulby Underground Laboratory in the U.K., and from the University of Wisconsin for grant UW PRJ82AJ.

We acknowledge many types of support provided to us by the South Dakota Science and Technology Authority, which developed the Sanford Underground Research Facility (SURF) with an important philanthropic donation from T. Denny Sanford as well as support from the State of South Dakota. SURF is operated by Lawrence Berkeley National Laboratory for the DOE, Office of High Energy Physics.

The technical editor of this document was Ms. Christine Steers, and we acknowledge here dedicated and tireless efforts over many months.

# 1 Direct Detection of Dark Matter

In the past two decades, a standard cosmological picture of the universe (the Lambda Cold Dark Matter or LCDM model) has emerged, which includes a detailed breakdown of the main constituents of the energy density of the universe. This theoretical framework is now on a firm empirical footing, given the remarkable agreement of a diverse set of astrophysical data [1,2]. Recent results by Planck largely confirm the earlier Wilkinson Microwave Anisotropy Probe (WMAP) conclusions and show that the universe is spatially flat, with an acceleration in the rate of expansion and an energy budget comprising ~5% baryonic matter, ~26% cold dark matter (CDM), and roughly ~69% dark energy [3,4]. With the generation-2 (G2) dark-matter experiments, we are now are in a position to identify this dark matter through sensitive terrestrial *direct detection* experiments. Failing to detect a signal in the next (or subsequent generation [G3] of experiments) would rule out most of the natural parameter space that describes weakly interacting massive particles (WIMPs), forcing us to reassess the WIMP paradigm and look for new detection techniques. In the following sections, we introduce the cosmological and particle physics evidence pointing to the hypothesis that the dark matter is composed of WIMPs, detectable through nuclear recoil (NR) interactions in low-background experiments. We then give the motivation for a very massive liquid xenon (LXe) detector as the logical next step in the direct detection of dark matter.

## 1.1 Cosmology and Complementarity

While the Large Hadron Collider (LHC) experiments continue to verify the Standard Model of particle physics to ever-greater precision, the nature of the particles and fields that constitute dark energy and dark matter remains elusive. The gravitational effects of dark matter are evident throughout the cosmos, dominating gravitational interactions of objects as small as dwarf satellites of the Milky Way, up to galaxy clusters and superclusters. Simple application of Kepler's laws leads to the inescapable conclusion that our own galaxy (and all others) are held together by the gravitational pull of a dark halo that outweighs the combined mass of stars and gas by an order of magnitude, and appears to form an extended halo beyond the distribution of luminous matter.

At the same time, very weakly interacting CDM (particles or compact objects that were moving non-relativistically at the time of decoupling) appears to be an essential ingredient in the evolution of structure in the universe. N-body simulations of CDM can explain much of the structure, ranging from objects made of tens of thousands of stars to galaxy clusters. In the past few years, more realistic simulations, including both baryonic matter (gas and stars) and dark matter, are beginning to reveal how galaxy-like objects can arise from the primordial perturbations in the early universe [5,6]. While we know much about the impact of dark matter on a variety of astrophysical phenomena, we know very little about its nature.

An attractive conjecture is that dark-matter particles were in equilibrium with ordinary matter in the hot early universe. We note, however, that there are viable dark-matter candidates, including axions, where the conjecture of thermal equilibrium is not made [7]. Thermal equilibrium describes the balance between annihilation of dark matter into ordinary particle-antiparticle pairs, and vice versa. As the universe expanded and cooled, the reaction rates (the product of number density, cross section, and relative velocity) eventually fell below the level required for thermal equilibrium, leaving behind a relic abundance of dark matter. The lower the annihilation cross section of dark matter into ordinary matter, the higher the relic abundance of dark matter. An annihilation cross-section characteristic of the weak interaction results in a dark-matter energy consistent with that observed by cosmological measurements [8].

Remarkably, models of supersymmetry (SUSY) predict the neutralino, a new particle that has properties appropriate to be a WIMP. SUSY posits a fermion-like partner for every Standard Model boson, and a boson-like partner for every Standard Model fermion. A principal feature of SUSY is its natural means for achieving cancellations in quantum field theory amplitudes that could cause the Higgs mass to be much higher than the 125 GeV/$c^2$ recently observed [9,10]. The neutralino is a coherent quantum state formed from the SUSY partners of the photon, the $Z^0$, and Higgs boson, and is a "Majorana" particle, meaning it is its own antiparticle.

Astrophysical measurements show that dark matter behaves like a particle and not like a modification of gravity. Gravitational lensing of distant galaxies by foreground galactic clusters can provide a map of the total gravitational mass, showing that this mass far exceeds that of ordinary baryonic matter. By combining the distribution of the total gravitational mass (from lensing) with the distribution of the dominant component of baryonic matter (evident in the X-ray-emitting cluster gas) one can see whether the dark mass follows the distribution of baryonic matter. For some galaxy clusters, in particular the Bullet cluster [11], the total gravitational mass (dominated by dark matter) follows the distribution of other non-interacting test particles (stars) rather than the dominant component of baryonic matter in the cluster gas. Combining this evidence with other observations of stellar distributions and velocity measurements for galaxies with a wide range of mass-to-light ratios, it appears that the total gravitational mass does not follow the distribution of baryonic matter as one would expect for modified gravity, but behaves like a second, dark component of relatively weakly interacting particles.

There are three complementary signals of WIMP dark matter. The dark matter of the Milky Way can interact with atomic nuclei, resulting in NRs that are the basis of direct detection (DD) experiments like LZ. At the LHC, the dark matter will appear as a stable, non-interacting particle that causes missing energy and momentum. Out in the cosmos, dark matter collects at the centers of galaxies and in the sun, where pairs of dark-matter particles will annihilate with one another, if the dark matter is a Majorana particle, as expected in SUSY theories. The annihilations will produce secondary particles, including positrons, antiprotons, neutrinos, and gamma rays, providing the basis for "indirect" detection (ID) by gamma ray, cosmic ray, and neutrino telescopes.

In Figure 1.1.1, we show the results of a recent analysis of the complementarity of the three signals from WIMP dark matter. The LHC has already provided constraints on the simplest SUSY parameter space, and the Higgs mass is in some tension with the most constrained versions of SUSY, requiring theorists to relax simplifying assumptions. One slightly less restrictive choice of parameters is the so-called phenomenological minimal supersymmetric standard model (pMSSM) model [12,13]. In Figure 1.1.1, each point represents a choice of pMSSM parameters that satisfies all known physics and astrophysics constraints [14]. The color of the points show which experiments have adequate sensitivity to test whether that point is valid. The three experimental tests are: DD with the proposed LZ experiment, current LHC data, and ID from a proposed ID experiment (an enhanced version of the Cherenkov Telescope Array [15] with twice the number of telescopes compared with the current baseline). Each of the three experimental techniques tends to be most sensitive to one region in Figure 1.1.1, although there are regions of overlapping sensitivity.

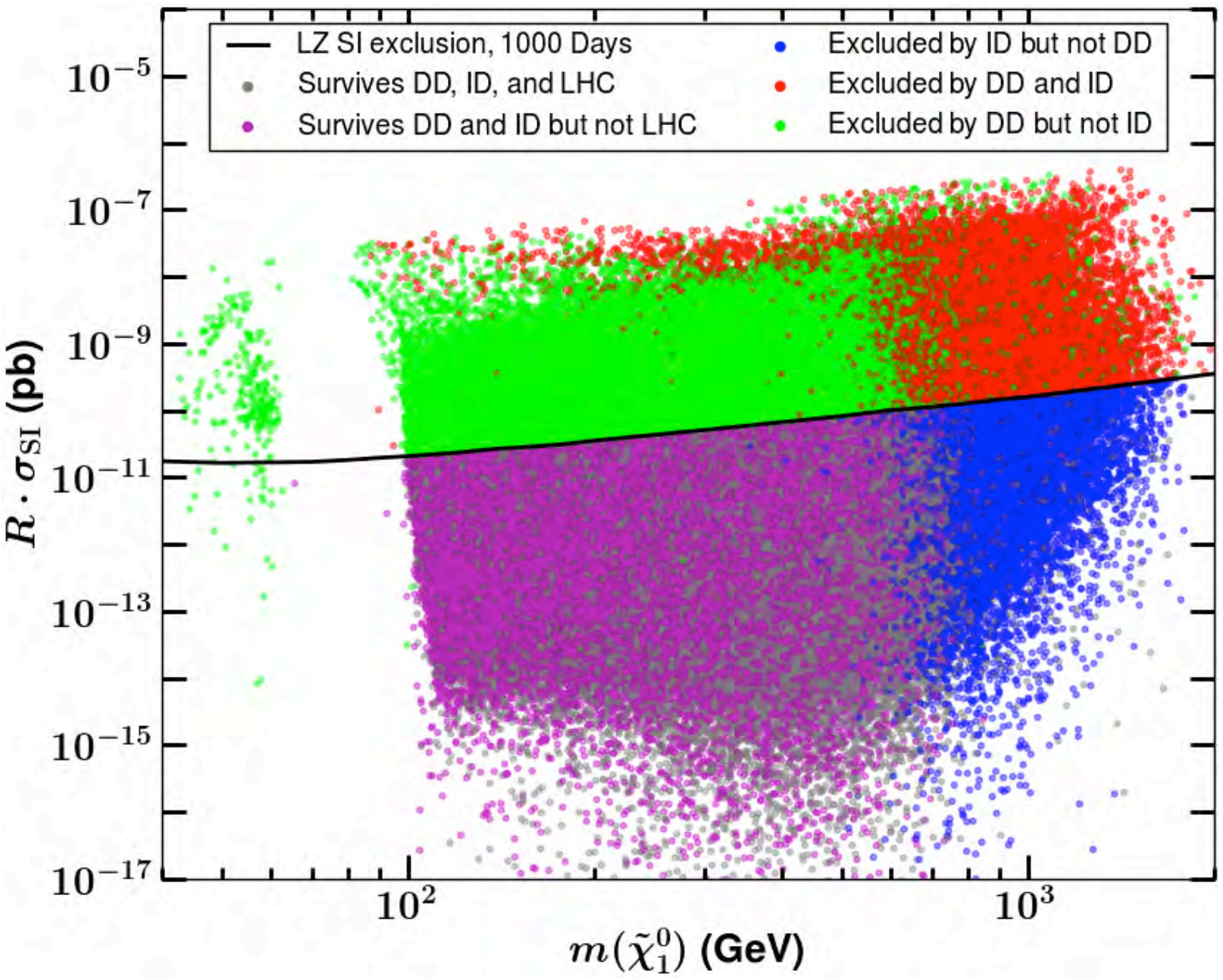

**Figure 1.1.1. Scan of pMSSM parameter space and complementarity. Each point in this space of the cross section, scaled by abundance, of WIMPs with nucleons versus mass of the WIMP represents a SUSY model. The colors show models that can be tested by the three techniques of detection: direct detection (DD), LHC, and indirect detection (ID), and their combinations. The expected LZ sensitivity is shown as the black line [14].**

The three experimental techniques are subject to very different systematic limitations. Direct-detection experiments like LZ depend on the extrapolation of interaction cross sections from Big Bang conditions to the current astrophysical situation, but are relatively free from uncertainty due to imperfect knowledge of the local mass density of dark matter. The LHC production of dark matter also depends on specific implementation of the processes that mediate dark-matter creation, but can have sensitivity down to the zero WIMP mass. Indirect detection signals depend sensitively on the assumed density of dark matter in astrophysical sources, but the annihilation cross section is closely related to the decoupling cross section.

The most important goal for the G2 program is to produce the best constraints over the natural mass range for dark matter. The LZ experiment accomplishes this goal. A broader goal of dark-matter research is to use the three complementary approaches to establish the validity of any signal, and to identify the properties of the dark-matter particle. For example, if LZ first sees a signal, ID could follow up with measurements of the gamma-ray spectrum and angular distribution that aid in determining the mass, annihilation channels, and galactic dark-matter halo density, and the LHC or future colliders could fully explore the nature of the dark-matter particle through experiments that actually create it in the laboratory.

## 1.2 Direct Detection Experiments

The direct detection of dark matter in earthbound experiments depends on the local properties of the Milky Way's dark matter and on the properties of the dark-matter particles themselves. The local properties of the Milky Way's dark halo are determined by astrophysical studies, and include the local dark-matter mass density as well as the distribution of the velocities of dark-matter particles. The conjecture that the dark-matter particles are WIMPs implies that its scattering with the nucleus is non-relativistic two-body scattering; in LZ, we seek to observe the xenon nuclei that recoil after having been struck by an incoming WIMP. The mass assumed for the WIMP determines the kinematics of the scattering, and the rate of WIMP-nucleus scatters seen in a WIMP detector depends additionally on the exposure defined as the product of the target mass and the live time, the WIMP-nucleus cross section, and the energy threshold for detection of the NR.

The dark-matter halo of the Milky Way is harder to quantify than that of other galaxies. The density of all matter, including dark matter, in galaxies is quantified with rotation curves, which describe the average circular velocity of matter in orbit about the galactic center as a function of radius from the galactic center. The rotation curve of the Milky Way for radii larger than 8.0 kpc, that of our sun, is a challenge to quantify. Estimates of the local dark-matter density range from 0.235 ± 0.030 GeV/cm$^3$ to 0.389 ± 0.025 GeV/cm$^3$ [16-18]. Most DD experiments adopt, for ease of inter-comparison, a standard value for the local dark-matter mass density of $\rho_0$=0.3 GeV/cm$^3$. The experiments also adopt a standard distribution function for the velocity of dark-matter particles, characterized by a Maxwell-Boltzmann distribution with solar circular velocity $v_0$=220 km/s, which is cut off at the galactic escape velocity of $v_{esc}$=544 km/s, and with proper accounting for the sun's peculiar velocity and the periodic annual motion of the Earth [19,20].

The minimum WIMP mass detectable with a particular DD experiment depends on the maximum velocity in the galactic WIMP spectrum, the atomic mass $A$ of the target nucleus, and the energy threshold $E_{min}$ for NR detection in that experiment. Straightforward kinematics gives the minimum detectable WIMP mass as $M_{WIMP}$(GeV) $\approx (1/4)\sqrt{E_{min}\ (keV) A}$, for a maximum velocity of $v_{esc}$=544 km/s. This minimum $M_{WIMP}$ is 2 GeV for the recent CDMSlite Ge-target result [21], and 6 GeV for the recent LUX result [22] using LXe.

These results are shown in Figure 1.2.1, from Ref. [23], along with the current experimental situation for the spin-independent (SI) WIMP-nucleon cross section. We discuss the details of spin independence and other types of WIMP-nucleon interactions in Chapter 4. The SI cross section is the standard benchmark, and would result from a WIMP coupling to the Standard Model Higgs.

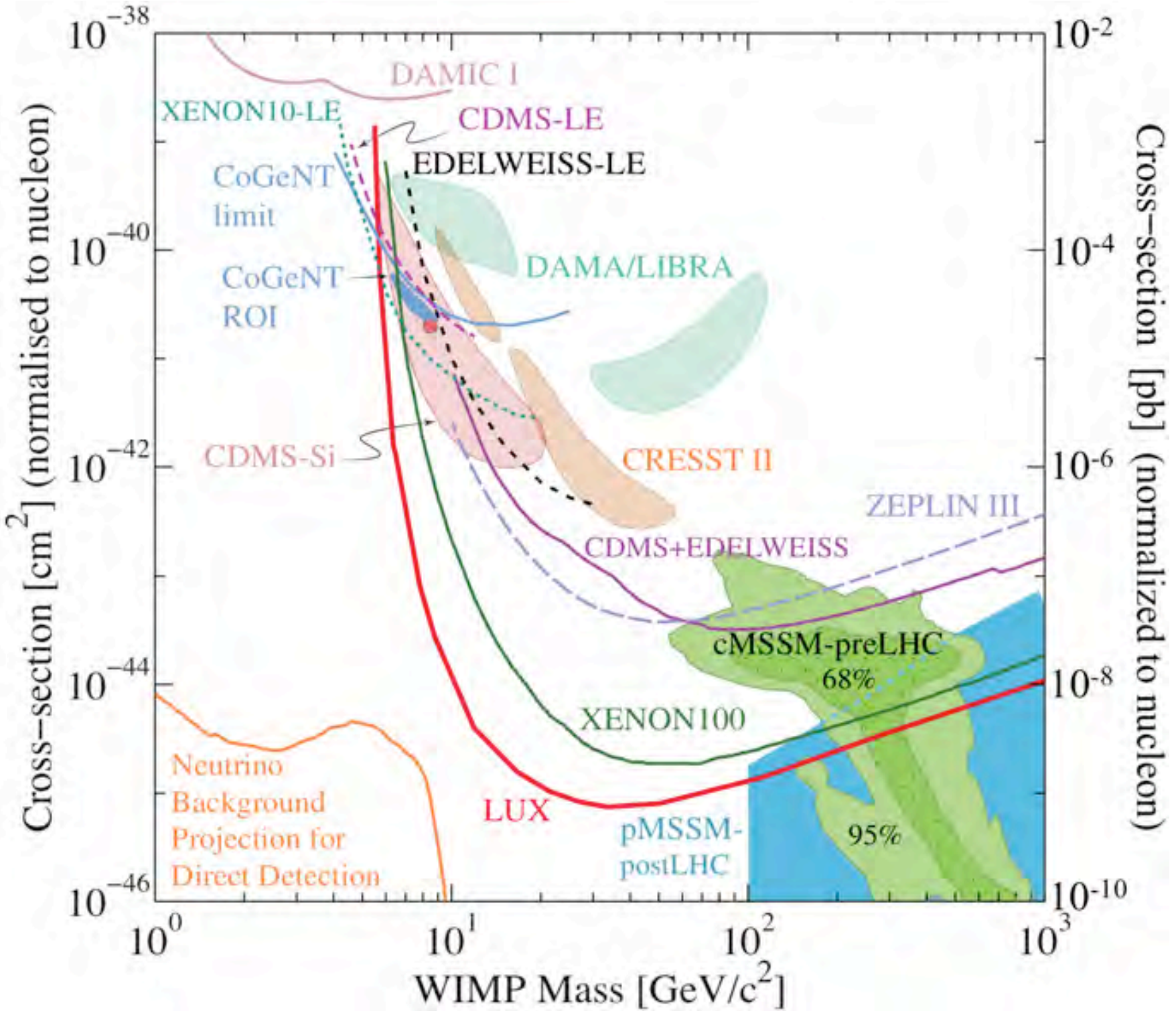

**Figure 1.2.1. A compilation of current WIMP-nucleon SI cross-section upper limits at 90% confidence level (solid and dashed curves, labeled by experiment names), and hints for WIMP signals (pale shaded closed contours, labeled by experiment names). Regions favored by supersymmetric models are the bold closed curves at the lower right. From Ref. [23].**

As the presumed $M_{\mathrm{WIMP}}$ rises above this value, the portion of the WIMP velocity distribution function that permits NR above the detectable threshold rises rapidly. This rapid rise drives the improvement in sensitivities as $M_{\mathrm{WIMP}}$ rises above 5 GeV, as shown in Figure 1.2.1. As $M_{\mathrm{WIMP}}$ approaches the mass of the atom used in the target, the kinematics of energy transfer to the target nuclei becomes most efficient, and experiments reach their maximum sensitivity.

This region, roughly $M_{\mathrm{WIMP}}$>10 GeV, is the region most likely for a WIMP, as other weak-interaction particles, including the $Z^0$, the $W$, and the Higgs particles, have masses in that region. As $M_{\mathrm{WIMP}}$ grows yet larger, the number density of WIMPs implied by the astrophysical mass density, 0.3/$M_{\mathrm{WIMP}}$(GeV) 1/cm$^3$, falls, reducing the sensitivity of direct-detection experiments, as shown in Figure 1.2.1. The maximum WIMP mass consistent with thermal equilibrium in the early universe is 340 TeV, above which unitarity can no longer be satisfied [24].

Figure 1.2.1 also shows a number of closed "regions of interest" from a variety of experiments, which indicate potential signals. All of the regions of interest result from excesses of events just above the thresholds of the respective experiments, where backgrounds are in general the highest and most challenging to understand thoroughly. The most robust regions of interest are those from the DAMA/LIBRA collaboration, which employed most recently 250 kg of sodium iodide crystals for a total exposure of 1.17 tonne-years [25].

The DAMA/LIBRA experiment sought an effect caused by the Earth's orbit about the sun. In June of each year, the Earth's orbital velocity adds to that of the sun's circular velocity around the center of the Milky Way, and in December of each year the Earth's orbital velocity cancels a small portion of the sun's circular velocity. The number of events just above threshold seen by DAMA increases in June, and was diminished in December; the strength of DAMA/LIBRA's signal is 8.9 standard deviations. Detailed criticisms of the DAMA/LIBRA result are presented in Ref. [23]. As can be seen from Figure 1.2.1, results from experiments that use Ge targets (CDMS and EDELWEISS) or Xe targets (ZEPLIN-III,

XENON100, and LUX) have attained sensitivities that exceed those of DAMA/LIBRA by many orders of magnitude, under the most simple SI interpretation.

The history of and future projections for WIMP sensitivity are shown in Figure 1.2.2. There are three distinct eras: (1) 1986-1996; (2) 2000-2010; and (3) post-2010. In the first era, low-background Ge and NaI crystals dominated. In these experiments, discrimination between the dominant background of electron recoils (ERs) from gamma rays and NRs was not available. The cross-section sensitivity per nucleon achieved, $10^{-41}$ cm$^2$ ($10^{-5}$ pb), was sufficient to rule out the most straightforward WIMP implementation: that the WIMP is a heavy Dirac neutrino, which was the original suggestion of Ref. [8]. In the second era, cryogenic Ge detectors with the ability to distinguish NR from the ER background dominated, and achieved a cross-section per nucleon sensitivity of $5 \times 10^{-44}$ $cm^2$($5 \times 10^{-8}$ pb). This sensitivity began to probe dark matter that is a Majorana fermion, which couples to nucleons via the Higgs particle. In the third era, still under way, LXe time projection chambers (TPCs) have been most sensitive, and with the LUX experiment have achieved a sensitivity of $7.6 \times 10^{-46}$ $cm^2$($7.6 \times 10^{-10}$ pb). The LXe TPC has the ability to discriminate ER and NR, and can be expanded to large, homogenous volumes. It is possible to make accurate predictions for the backgrounds in the central LXe TPC region, particularly with the added power of outer detectors that can characterize the radiation field in the TPC vicinity.

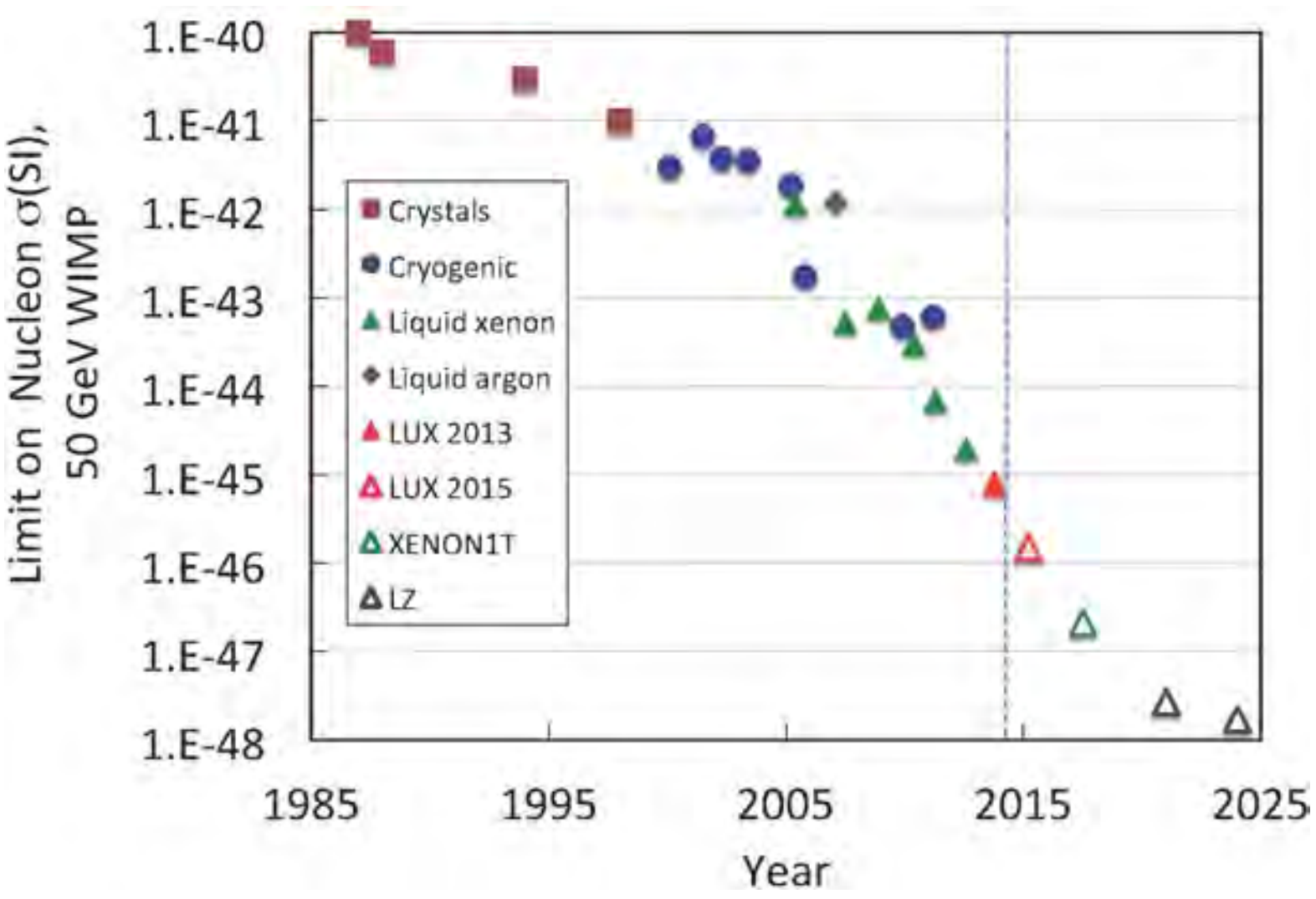

**Figure 1.2.2. The evolution of cross-section limit for 50 GeV WIMPs as a function of time. Past points are published results. Future points are from the Snowmass meetings [26].**

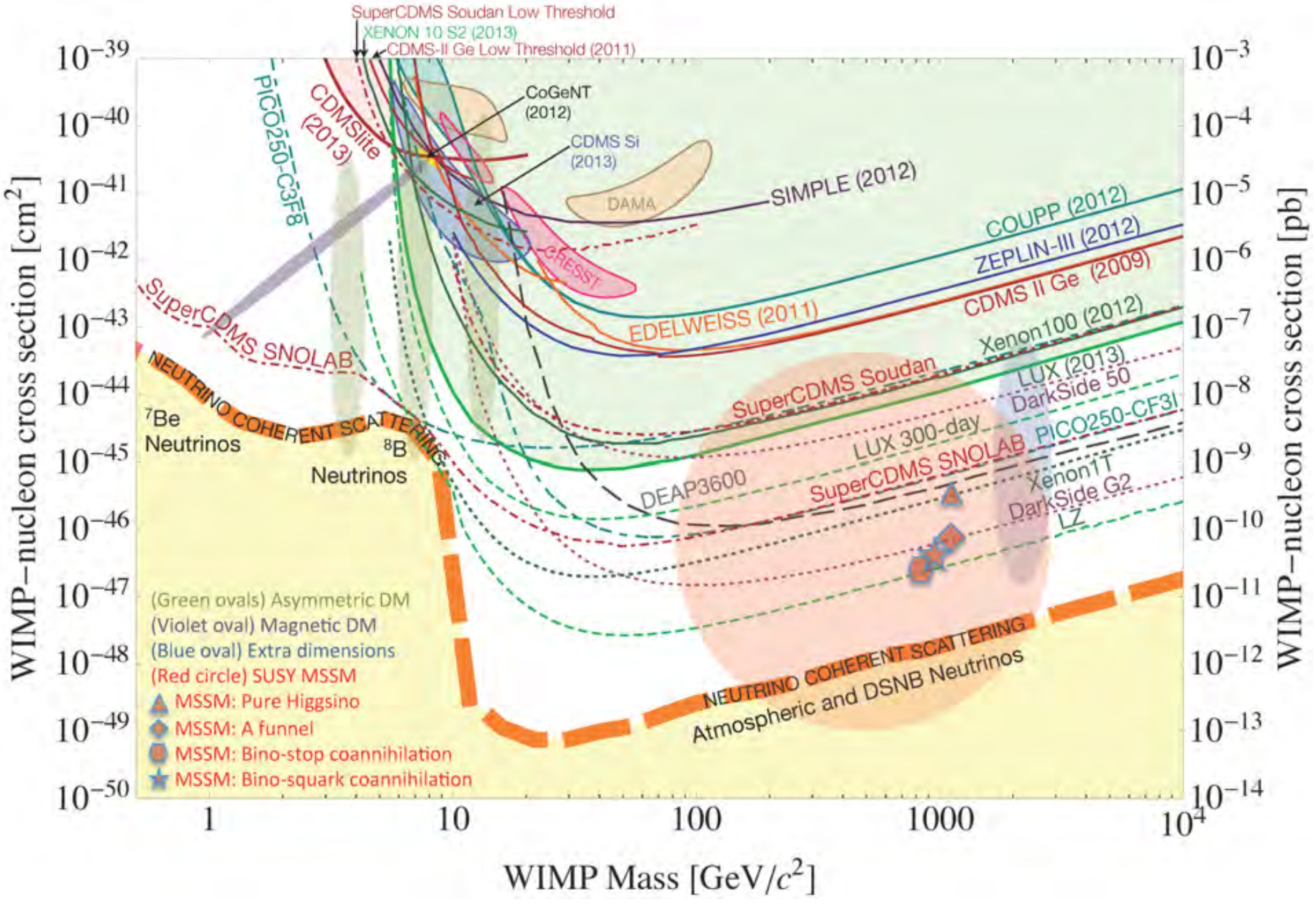

**Figure 1.2.3. A compilation of WIMP-nucleon SI cross-section sensitivity (solid curves), hints for WIMP signals (shaded closed contours), and projections (dot and dot-dash curves) for DD experiments of the past and projected into the future. Also shown is an approximate band where the coherent nuclear scattering of $^8$B solar neutrinos, atmospheric neutrinos, and diffuse supernova neutrinos will limit the sensitivity of DD experiments to WIMPs. Finally, a suite of theoretical model predictions is indicated by the shaded regions, with model references included [26].**

A more detailed portrayal of the variety of experiments, both from the past and projected into the future, is given in Figure 1.2.3 [26]. In the region $M_{\mathrm{WIMP}}$>10 GeV, most likely for a WIMP due to proximity to the masses of the weak bosons, LZ will be the most sensitive experiment. Figure 1.2.3 also shows the "neutrino floor," where NRs from coherent neutrino scattering, a process that has not yet been observed, will greatly influence progress in sensitivity to WIMP interactions. A 1,000-day run of the LZ experiment will just begin to touch this background. Searches for direct interactions of dark matter with exposure substantially greater than LZ will see many candidate events from NRs in response to, primarily, atmospheric neutrinos.

## Chapter 1 References

# 2 Instrument Overview

The core of the LZ experiment is a two-phase xenon (Xe) time projection chamber (TPC) containing about 7 fully active tonnes of liquid Xe (LXe). Scattering events in LXe create both a prompt scintillation signal (S1) and free electrons. Various electric fields are employed to drift the electrons to the liquid surface, extract them into the gas phase above, and accelerate them to create a proportional scintillation signal (S2). Both signals are measured by arrays of photomultiplier tubes (PMTs) above and below the central region. The difference in time of arrival between the signals measures the position of the event in *z*, while the *x,y* position is determined from the pattern of S2 light in the top PMT array. Events with an S2 signal but no S1 are also recorded. A 3-D model of the LZ detector located in a water tank is shown in Figure 2.1. The water tank is located at the 4,850-foot level (4850L) of the Sanford Underground Research Facility (SURF). The heart of the LZ detector (including the inner titanium [Ti] cryostat) will be assembled on the surface at SURF, lowered in the Yates shaft to the 4850L of SURF, and deployed in the existing water tank in the Davis Cavern (where LUX is currently located). The principal parameters of the LZ experiment are given in Table 2.1, along with the proposed Work Breakdown Structure (WBS) for the LZ Project.

The LZ design is enhanced by several added capabilities beyond the successfully demonstrated LUX and ZEPLIN designs. The most important addition is a hermetic liquid organic scintillator (gadolinium-loaded linear alkyl benzene [LAB]) outer detector, which surrounds the central cryostat vessels and TPC. The outer detector and the active Xe "skin" layer operate as an integrated veto system, which has several benefits. The first is rejecting gammas and neutrons generated internally (e.g., in the PMTs) that scatter a single time in the fully active region and would otherwise escape without detection; this could mimic a weakly interacting massive particle (WIMP) signal. As these internally generated backgrounds interact primarily at the outer regions of the detector, the veto thus allows an increase in the fiducial volume.

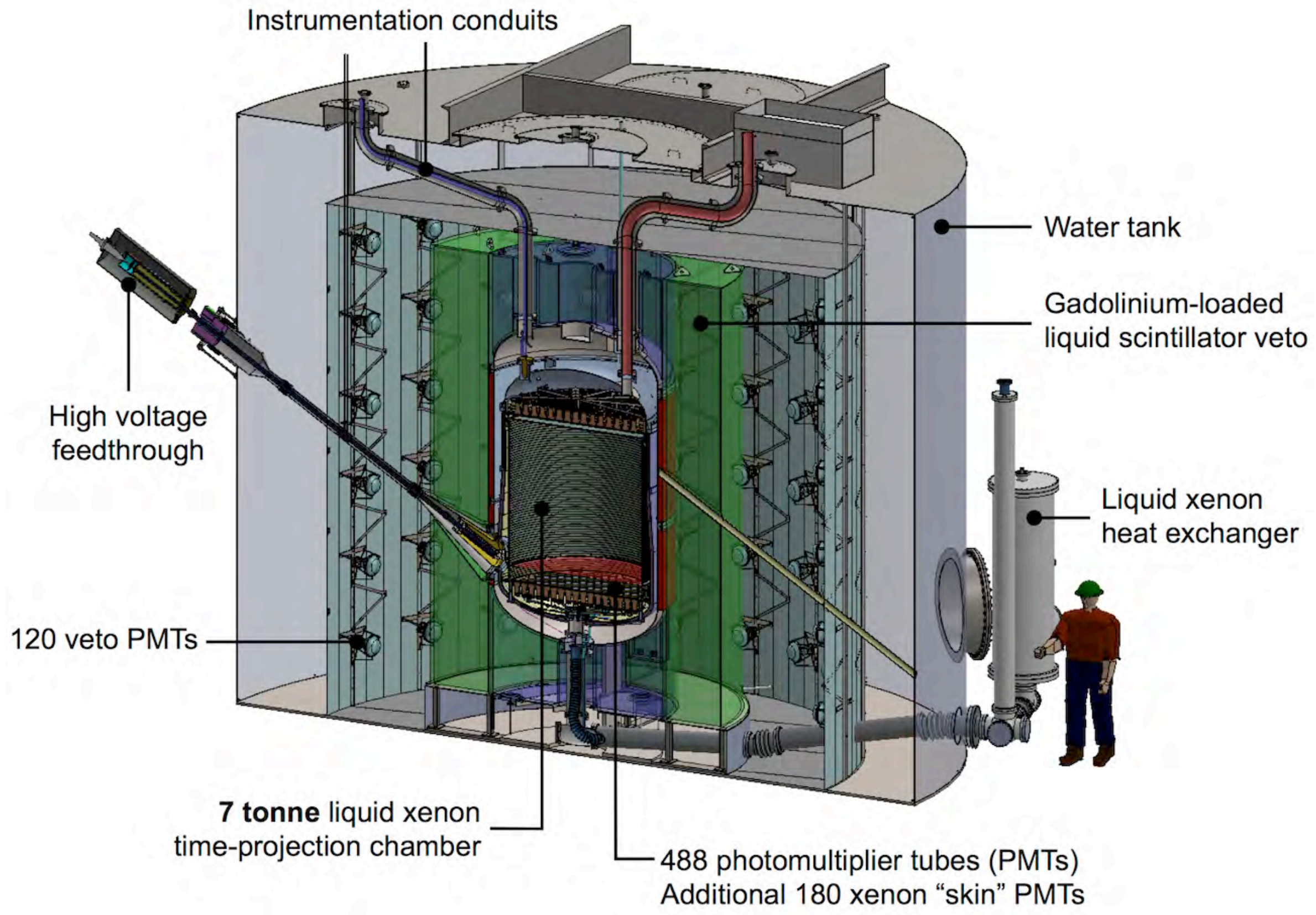

**Figure 2.1. LZ detector concept.**

**Table 2.1 Work Breakdown Structure (WBS) and principal parameters of the LZ detector.**

| WBS | Description | Quantity |
|---|---|---|
| **1.1** | **Xenon Procurement** | |
| | Approximate total mass | 10 tonnes |
| | Mass inside TPC active region | 7 tonnes |
| | Approximate fiducial mass | 5.6 tonnes |
| **1.2** | **Cryostat** | |
| | Inner cryostat — inside diameter (tapered), height, wall thickness | 1.58–1.66 m, 2.52 m, 8 mm |
| | Outer cryostat — inside diameter, height, wall thickness | 1.83 m, 3.09 m, 8 mm |
| | Approximate cryostat weights — inner, outer | 0.74 tonne, 1.09 tonne |
| **1.3** | **Cryogenic System** | |
| | Cooling power | 1 kW @ 80 K |
| | Input electrical power | 11 kW |
| **1.4** | **Xenon Purification** | |
| | Krypton content | <0.015 ppt (g/g) |
| | $^{222}$Rn content in active Xe | <0.67 mBq |
| | Recirculation rate | 500 slpm |
| | Electron lifetime | >0.75 ms |
| | Charge attenuation length | >1.5 m |
| **1.5** | **Xenon Detector** | |
| | Top (Bottom) TPC 3-inch PMT array | 247 (241), 488 total tubes |
| | "Side skin readout" and "bottom" 1-inch PMT | 60 top skin, 60 bottom skin, 60 bottom |
| | Nominal (Design) cathode operating voltage | 100 (200) kV |
| | Reverse field region (cathode to bottom tube shield) | 0.146 m |
| | TPC height (cathode to gate grid) | 1.456 m |
| | TPC effective diameter | 1.456 m |
| **1.6** | **Outer Detector System** | |
| | Weight of Gd-loaded LAB scintillator | 20.8 tonnes |
| | Number of acrylic vessels, total acrylic mass | 9 vessels, 4 tonnes |
| | Number of 8-inch PMTs | 120 |
| | Minimum thickness of scintillator | 0.61 m |
| | Diameter of water tank | 7.62 m |
| | Height of water tank | 5.92 m |
| | Approximate weight of water | 228 tonnes |
| **1.7** | **Calibration System** | |
| | Number of source ($\gamma$, n) calibration tubes | 3 |
| | Other calibration tools | $^{83m}$Kr, n-generator, tritiated $CH_4$, $^{37}$Ar |
| **1.8** | **Electronics, DAQ, Controls, and Computing** | |
| | Trigger rate (all energies, 0-40 keV) | 40 Hz, 0.4 Hz |
| | Average event size (noncalibration, uncompressed) | 0.2–1.0 MB |
| | Data volume per year | 350–850 TB |
| **1.9** | **Integration and Installation** | |
| **1.10** | **Cleanliness and Screening** | |
| **1.11** | **Offline Computing** | |
| **1.12** | **Project Management** | |

Additionally, this direct vetoing is an important means of risk mitigation against one of the detector components (e.g., the cryostat materials or the PMTs) having a higher-than-expected background. A second benefit of the outer detector is that the combination of outer detector and segmented Xe detector will form a nearly hermetic detection system for all internal radioactivity. This will not only directly measure the internal backgrounds with an unprecedented level of detail and completeness, but also provide a similarly dramatic improvement in our understanding of the detector's response to those backgrounds.

The design of this scintillator veto is informed by proven designs of other experiments. As shown in Figure 2.1, the liquid scintillator volume is confined in a segmented, clear acrylic housing encapsulating the Ti cryostat of the central LZ detector. PMTs mounted on ladders in the outer water shield simultaneously view the light from both the scintillator and inner water volumes.

An important enhancement beyond LUX is the treatment of the "skin" layer of LXe located between the PTFE (polytetrafluoroethylene, or Teflon®) structure that surrounds the fully active region and the cryostat wall, as well as the region beneath the bottom PMT array. A skin of some (~few cm) thickness is difficult to avoid given the TPC geometry, the need for high-voltage standoff, and the strong mismatch in thermal expansion between the PTFE panels and Ti vessels. The skin readout alone has a limited veto efficiency, but has sufficient gamma-stopping power to augment the scintillator veto. The combination of skin readout and outer detector creates a highly efficient integrated veto system. Scintillation light from the skin region is observed by a sparse PMT array dedicated to this region. Thin (6-mm) PTFE panels are attached to the inner cryostat wall to enhance light collection in this region.

For detector calibration, neutron and gamma-ray sources will be brought next to the wall of the inner cryostat via an array of three source tubes that penetrate the water and organic scintillator. The principal calibrations, metastable krypton and tritiated methane, will be introduced directly into the LXe via the Xe gas-handling system to allow in situ calibration, which has been demonstrated in LUX. An external neutron generator will also be employed, as has also been the case for LUX.

Another key determinant of the sensitivity of the experiment is the level of discrimination of electron recoil (ER) backgrounds from nuclear recoils (NRs). This depends on the electric field established in the TPC. The nominal operating voltage (cathode-to-anode) is 100 kV but all components will be designed to a voltage of 200 kV to have sufficient operating margin.

The LZ TPC detector will employ Hamamatsu R11410-20 3-inch-diameter PMTs with a demonstrated low level of radioactive contamination and high quantum efficiency. Care will be taken in the design of the highly reflective PTFE TPC structure to isolate light produced in the central volume from the skin region, and vice-versa. Additional 1-inch PMTs will be utilized in the skin region.

As in LUX, LZ will employ an array of liquid nitrogen (LN)-cooled thermosyphons to control the detector temperature and minimize thermal gradients. The baseline purification system is a scaled-up version of the LUX room-temperature gas-phase purification system, and will exploit the liquid/gas heat-exchanger technology developed for LUX to minimize LN consumption. Krypton will be removed from the gaseous Xe using scaled-up techniques already demonstrated successfully for LUX.

The electronics front-end, trigger and data acquisition, and slow controls are based on the LUX experience but will be significantly expanded and improved. Software and computing systems for LZ are based on the successful operation of LUX and analysis of LUX data but will be substantially augmented to accommodate the larger LZ data volume. LZ data will be buffered locally at SURF and then transmitted to primary data storage that will be mirrored in the United States and the UK. The simulation of the LZ detector and its response is already well advanced, and the results are given in other chapters of this *Conceptual Design Report*.

Components of LZ will be manufactured and brought to SURF. Assembly of the Xe detector and the inner cryostat will occur in a dedicated clean room at SURF that includes air handling to remove radon.

The inner cryostat with the Xe detector will be lowered as a unit down the Yates shaft and transported to the Davis Cavern water tank. All other components will be staged at SURF, lowered via the Yates shaft, and similarly transported. This includes the segmented acrylic vessels for the outer detector system and other large components.

# 3 Design Drivers for WIMP Identification

Having established the motivation to perform direct searches for WIMP dark matter, we introduced in the previous section the proposed configuration of LZ. Searching for events that are rare ($\lesssim$0.1 per day per tonne of target mass) and that involve very small energy transfers ($\lesssim$100 keV) is extremely challenging. This section focuses on the more salient features of the experiment and the detection medium, and how these will contribute to the identification of a galactic WIMP signal with low systematic uncertainty. The detailed design and its technical implementation are described in later sections; here, we address the key requirements that drive the conceptual design and how we propose to address remaining technical challenges.

## 3.1 Overview of the Experimental Strategy

Xenon has long been recognized as a very attractive WIMP target material [1-3]. Its high atomic mass provides a good kinematic match to intermediate WIMP masses of O(100 GeV/$c^2$) and the largest spin-independent scattering cross section among the available detector technologies, as illustrated in Figure 3.1.1. Sensitivity to lighter WIMPs, with masses of O(10 GeV/$c^2$), can be also be achieved, given the excellent low-energy scintillation and ionization yields in the liquid phase [4]. Xenon contains neither long-lived radioactive isotopes with troublesome decays nor activation products that remain significant after the first few months of underground deployment. It is also sensitive to spin-dependent interactions via the odd-neutron isotopes $^{129}$Xe and $^{131}$Xe, which account for approximately half of the natural isotopic abundance. If a WIMP discovery were made, the properties of the new particle could be studied by altering the isotopic composition of the target. This broad WIMP sensitivity confers maximum discovery potential to LZ.

The liquid phase is preferred over the gas phase due to its high density (3 g/cm$^3$) and high scintillation yield, and because its charge quenching of NRs provides a powerful particle ID mechanism. Early experiments such as ZEPLIN-I [5] exploited simple pulse shape discrimination (PSD) of the scintillation signal to reject electronic backgrounds; however, this achieved modest rejection efficiencies and only at relatively high recoil energies. When the first double-phase Xe detectors were deployed for dark-matter searches, in the ZEPLIN-II/III [9,10] and XENON10 [11] experiments, the increase in engineering complexity soon paid off in sensitivity, and this technique has been at the forefront of the field ever since. Comprehensive reviews on the application of the noble liquids to rare-event searches can be found in the literature [12,13].

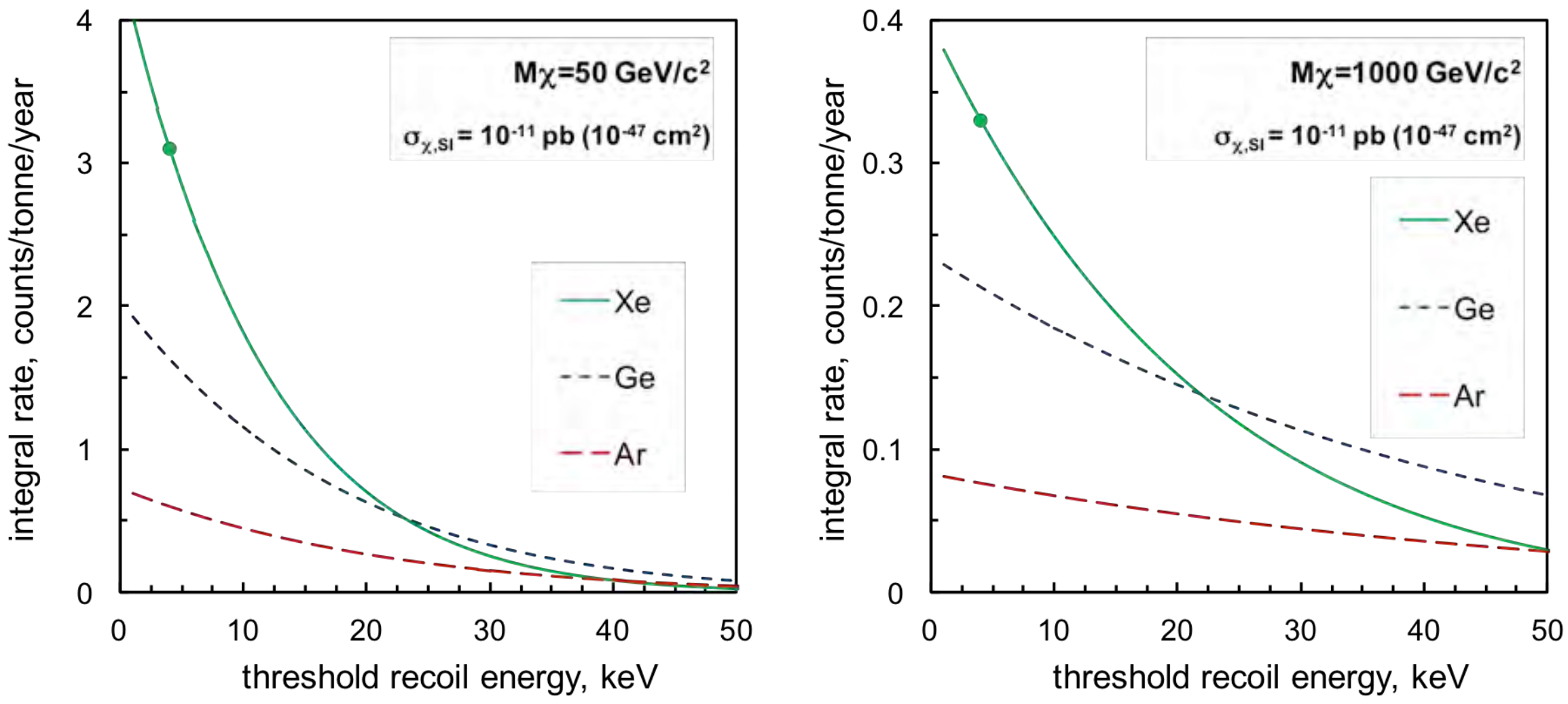

**Figure 3.1.1. Integrated rate above threshold per tonne·year of exposure for WIMP elastic scattering on Xe, Ge, and Ar targets for 50 GeV/$c^2$ and 1 TeV/$c^2$ WIMP masses and $10^{-47}$ cm$^2$ interaction cross section per nucleon. The green marker indicates the 4.3 keV WIMP-search threshold in LUX with nominal ER/NR discrimination [4]. CDMS II searched above 10 keV in their Ge target; selected SuperCDMS detectors allowed a 1.6-keV threshold with lower discrimination [6]. In LAr, the WARP (WIMP Argon Programme) 2.3-liter chamber achieved 55 keV [7], and the DarkSide-50 experiment has recently conducted a WIMP search above 38 keV [8].**

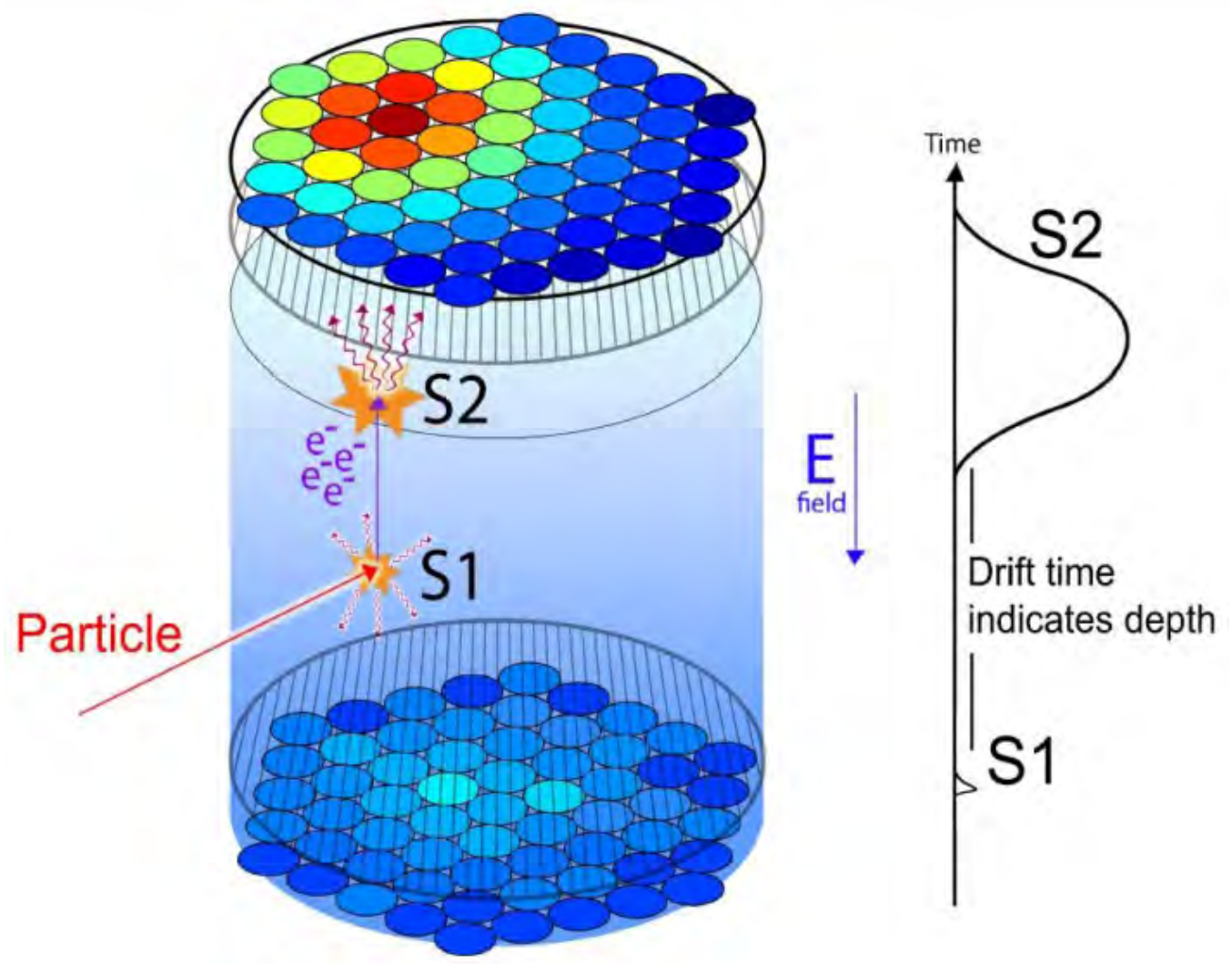

**Figure 3.1.2. Operating principle of the double-phase Xe TPC. Each particle interaction in the LXe (the WIMP target) produces two signatures: one from prompt scintillation (S1) and a second, delayed one from ionization, via electroluminescence in the vapor phase (S2). This allows precise vertex location in three dimensions and discrimination between nuclear and electron recoils.**

The TPC configuration at the core of double-phase detectors, illustrated in Figure 3.1.2, has several notable advantages for WIMP searches, in that two signatures are detected for every interaction: a prompt scintillation signal (S1) and the delayed ionization response, detected via electroluminescence in a thin gaseous phase above the liquid (S2). These permit precise event localization in three dimensions (to within a few mm [14]) and discrimination between electron and nuclear recoil events (potentially reaching 99.99% rejection [15]).

Both channels are sensitive to very low NR energies. The S2 response enables detection of single ionization electrons extracted from the liquid surface due to the high photon yield that can be achieved with proportional scintillation in the gas [16-18]. In LUX we have demonstrated sufficient S1 light collection to achieve a NR energy threshold below 5 keV [4].

The combination of accurate 3-D imaging capability within a monolithic volume of a readily purifiable, highly self-shielding liquid is nearly an ideal architecture for minimizing backgrounds. It allows optimal exploitation of the powerful attenuation of external gamma rays and neutrons into LXe, distinguishes multiply-scattered backgrounds from single-site signals, and precisely tags events on the surrounding surfaces. This latter feature is important, given the difficulty of achieving contamination-free surfaces. The low surface-to-volume ratio of the large, homogeneous TPC lowers surface backgrounds in comparison to signal, and stands in stark contrast to the high surface-to-volume ratio of segmented detectors.

These concepts are illustrated in Figure 3.1.3, which shows neutron interactions occurring just a few millimeters apart in the ZEPLIN-III detector. The S1 signals are essentially time-coincident, but the S2 pulses have different time delays corresponding to different vertical coordinates, making the rejection of such multiple scatters extremely efficient. The figure shows also a pulse observed in delayed coincidence in the surrounding veto detector, indicating radiative capture of this neutron on the gadolinium-loaded plastic installed around the WIMP target. LZ will utilize a similar anticoincidence detection technique to characterize the radiation environment around the Xe detector and to further reduce backgrounds.

Nevertheless, when the first tonne-scale Xe experiments were proposed just over a decade ago, it was unclear whether LXe technology could be monolithically scaled as now proposed for LZ, or if it would be necessary to replicate smaller devices with target masses of a few hundred kilograms each. The latter option, while conceptually simple, fails to fully exploit the power of self-shielding. Since then, several aspects of the double-phase TPC technique have been further developed to make LZ technologically feasible. First, the ionization and scintillation yields of LXe and their dependence on energy, electric field, and particle type have now been established down to a few keV by a comprehensive development program carried out around the globe, including substantial work by members of the LZ Collaboration. Second, good acceptance for the primary scintillation light must be maintained as the detector becomes larger, and the remarkably high reflectance (>95%) of PTFE at the 178 nm LXe scintillation wavelength has made this practical. Third, considerable control over electronegative impurities is required to drift charge over a distance of a meter or more, and commercial purification technology and new screening and detection methods developed by LZ scientists have made this routinely achievable. Fourth, the extraordinary self-shielding of an LZ-class instrument requires the use of internal calibration sources, and these have now been developed and deployed within LUX by LZ groups. Fifth, radioactive impurities such as Kr must be reliably removed from the Xe, and other sources of internal radioactivity such as radon must be tightly controlled. Finally, a large detector requires a substantial cathode voltage, or else fluctuations in the charge recombination near the interaction site will degrade the recoil discrimination. Very recently, LUX has demonstrated a rejection efficiency of 99.6% (for 50% NR acceptance) even at a modest field of 180 V/cm [4] — matching already the baseline assumption for LZ. From our present understanding of the physics of recombination, we expect further gains in discrimination at higher fields, and this challenge is a major focus of our current R&D effort.

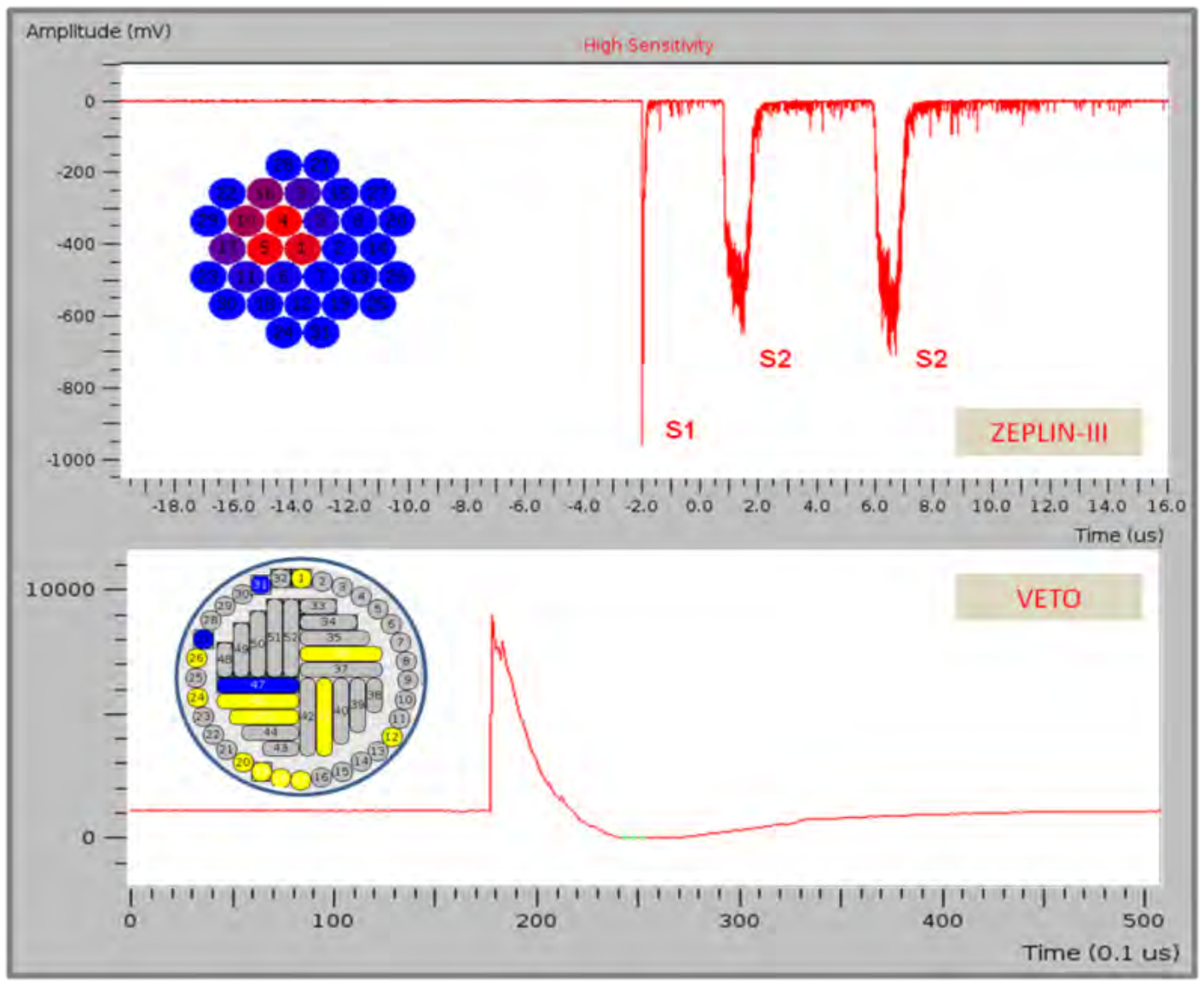

**Figure 3.1.3. A double-scatter neutron event recorded in ZEPLIN-III. The upper panel shows two elastic vertices clearly resolved in drift time (two S2 pulses, representing different vertical coordinates), although both have similar horizontal positions. The lower panel shows the summed waveform from the 52-module veto detector which surrounded the main instrument, indicating radiative capture of this neutron some 17 μs after interacting in the LXe target. Recording additional particle scatters (either in the WIMP target or in an ancillary veto detector) provides a powerful rejection of backgrounds.**

On the whole, the progressive nature of our program has contributed to an increase in the readiness level of this technology: LZ entails a twentyfold scale-up from LUX, the latter being also an order of magnitude or so larger than the ZEPLIN and XENON10 targets. Besides having the favorable properties of the WIMP target material and the proven sensitivity of the technology to small energy deposits, a successful experiment must achieve very low background rates over a significant fraction of its active medium. Indeed, it is worth noting that LZ will be a factor of $10^4$ times more sensitive than current limits from the EDELWEISS and Cryogenic Dark Matter Search (CDMS) experiments, which led in sensitivity only one decade ago [19,20]. This implies a corresponding reduction in the background rate. This is achieved to first order by the power of self-shielding of local radioactivity, in combination with an outer layer of instrumented LXe and a hermetic gadolinium-loaded scintillator "veto" shield capable of tagging neutrons and gamma rays with high efficiency. The construction of a veto instrument at the required scale builds on two decades of development work in the field of reactor neutrino physics, and its development within LZ is led by scientists with considerable expertise in this area. Three other important developments, again pioneered by LZ groups, have also made this possible: the development, in collaboration with Hamamatsu, of very-low-background PMTs compatible with LXe [21]; the identification via the LUX program of radio-clean titanium for cryostat fabrication [22]; and the development of krypton-removal and -screening technology capable of delivering sub-ppt concentrations [23,24].

This strategy leads to a WIMP-search background of order 1 event in 1,000 days of live exposure for a 5.6-tonne fiducial mass. Remarkably, the remaining component will be due to astrophysical neutrinos, dominated by solar pp neutrino scattering from electrons, with a small fraction of these events mimicking NRs due to the finite S2/S1 discrimination power. Coherent scattering of atmospheric neutrinos from Xe nuclei (CNS) will constitute an even smaller, but irreducible, background. These rates are well understood and background expectations are calculable with small systematic uncertainty (e.g., these events are spatially uniform and their energy spectra are well known). With its pioneering capability, LZ will be sensitive to these ultrarare processes.

## 3.2 Self-shielding in Liquid Xenon

At the core of any WIMP search experiment is a substantial screening and materials-selection program that controls the trace radioactivity of the detector components. In the case of LZ, however, backgrounds from detector radioactivity will also be rejected to unprecedented levels by the combination of self-shielding of external particles and operation in anticoincidence with outer veto detectors. This will render external gamma rays and neutrons less problematic than in other experiments.

The self-shielding strategy, in particular, relies on the combination of a large, dense, high-*Z* and continuous detection medium with the ability to resolve interaction sites in three dimensions with high precision. An outer layer of the target can therefore be defined (in data analysis) that shields a *fiducial* region with extremely low background at the center of the active medium. The *nonfiducial* layer will be only a few centimeters thick. Because the size of the LZ detector is much larger than the interaction lengths for MeV gamma rays and neutrons, as shown in Figure 3.2.1, when these particles penetrate more than a few cm they will scatter multiple times and be rejected (Figure 3.2.2). X-rays, with energies similar to WIMP events, penetrate only a few mm into the LXe.

Double-phase Xe detectors implement this strategy very successfully, and this is reflected in their present dominance in WIMP sensitivity — with LUX being a prime example of this concept. In LZ, a fiducial mass of nearly 6 tonnes will be practically free of external gamma-ray or neutron backgrounds, which represents 85% of active mass within the TPC, compared to about 50% for LUX [4] and XENON100 [27]. In addition to the required high density and high-*Z* of the detection medium, we highlight the importance of the precise spatial resolution that can be achieved in these detectors, which is of the order of 1 cm or better at threshold. The fiducial fraction can be much smaller in single-phase, scintillation-only detectors, which typically achieve a vertex location of about 10 cm. For example, only (roughly) 5% of the LXe mass was utilized in the recent search for inelastic WIMP scattering in XMASS [28].

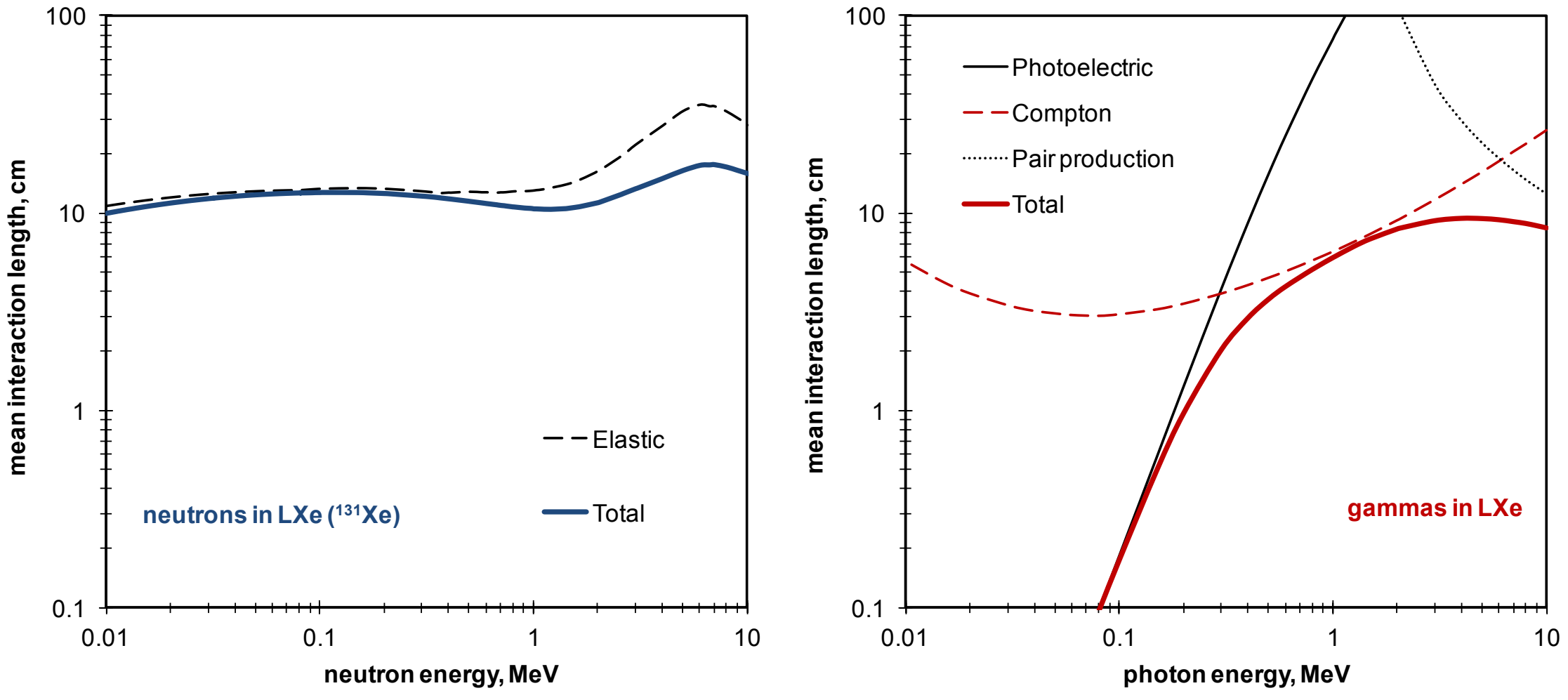

**Figure 3.2.1. Mean interaction lengths for neutrons [25] and gamma rays [26] in LXe.**

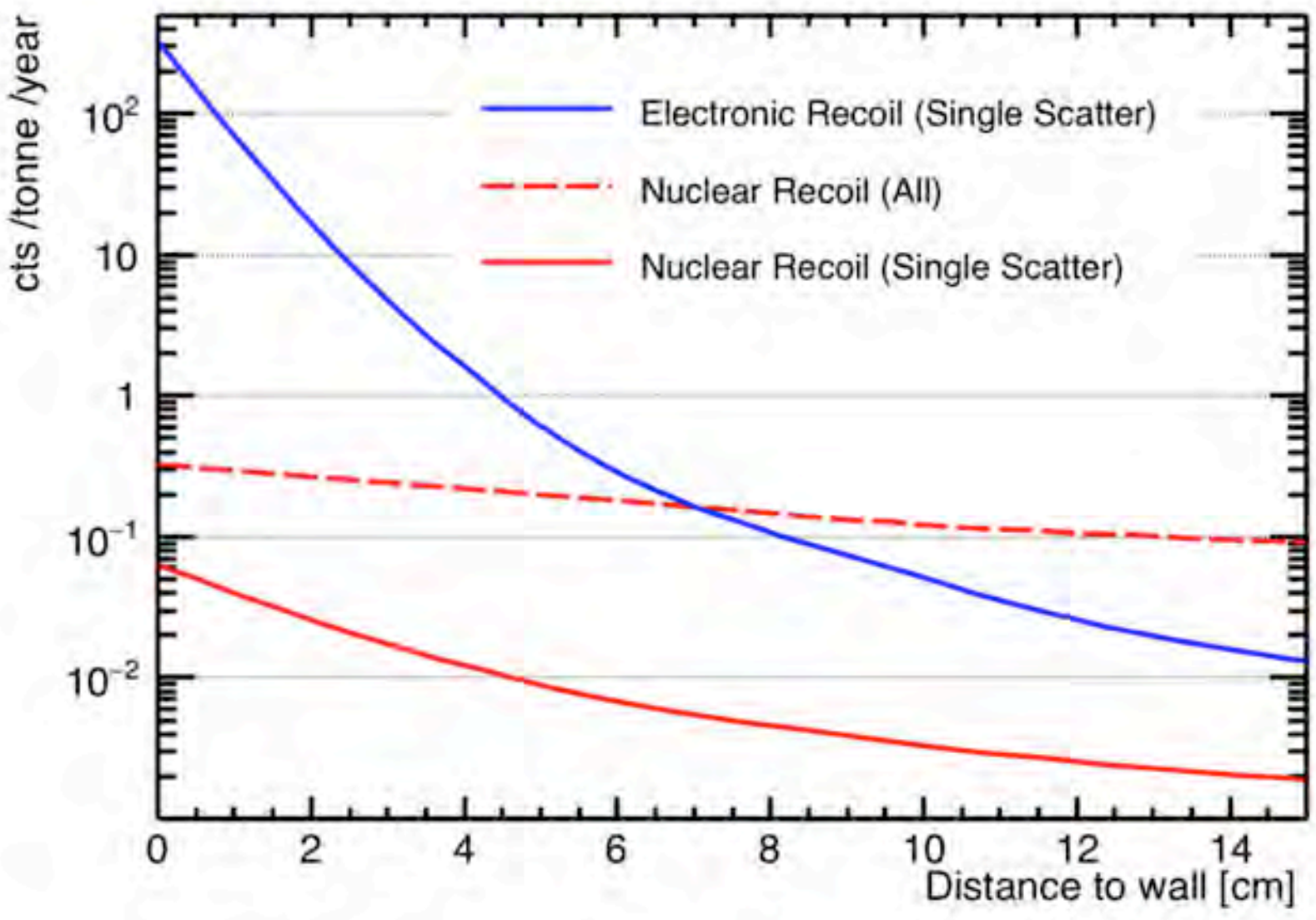

**Figure 3.2.2. Self-shielding of external neutrons and gamma rays in LXe. The red lines indicate the number of elastic neutron scatters creating 6–30 keV NRs as a function of distance to the lateral TPC wall; the continuous line shows single scatters only, while the dashed line includes all multiplicities adding up to the same total energy; the input spectrum is that from (α,n) neutron production in PTFE, an important background near the TPC walls. The blue line represents single-site ER interactions from U/Th gamma rays from PTFE with energy 1.5–6.5 $keV_{ee}$. A tenfold decrease is achieved at ~2 cm and ~6 cm from the wall for gamma rays and neutrons, respectively.**

## 3.3 Low-energy Particle Detection in Liquid Xenon

The potential of this medium for particle detection was recognized in the mid-20th century, when the combination of good scintillation and ionization properties was first noted (see [12] and references therein). In the 1970s, the first double-phase detectors were demonstrated, originally using argon [29]. Initially, our understanding of the mechanisms involved in generating the scintillation and ionization responses in the noble liquids progressed slowly, especially regarding the response to low-energy nuclear and electron recoils. However, great steps have been taken in the past decade, with LZ collaborators taking a central role. This effort continues around the world.

In this section, we summarize those LXe properties that affect the detection of low-energy nuclear and electronic recoils via scintillation and ionization; the next section discusses how to discriminate between them. The response of LXe to electron and nuclear recoils is now well understood over the energy range of interest for "standard WIMP" searches ($\gtrsim$3 keV). Significant progress has equally been made in modeling its behavior as a function of incident particle species, energy, and electric field, in order to optimize detector design and the physics analyses. Naturally, the increasing WIMP scattering rates with decreasing recoil energy and the need to probe lighter dark-matter candidates mean that pushing further down in threshold is a perennial concern for any detection technology.

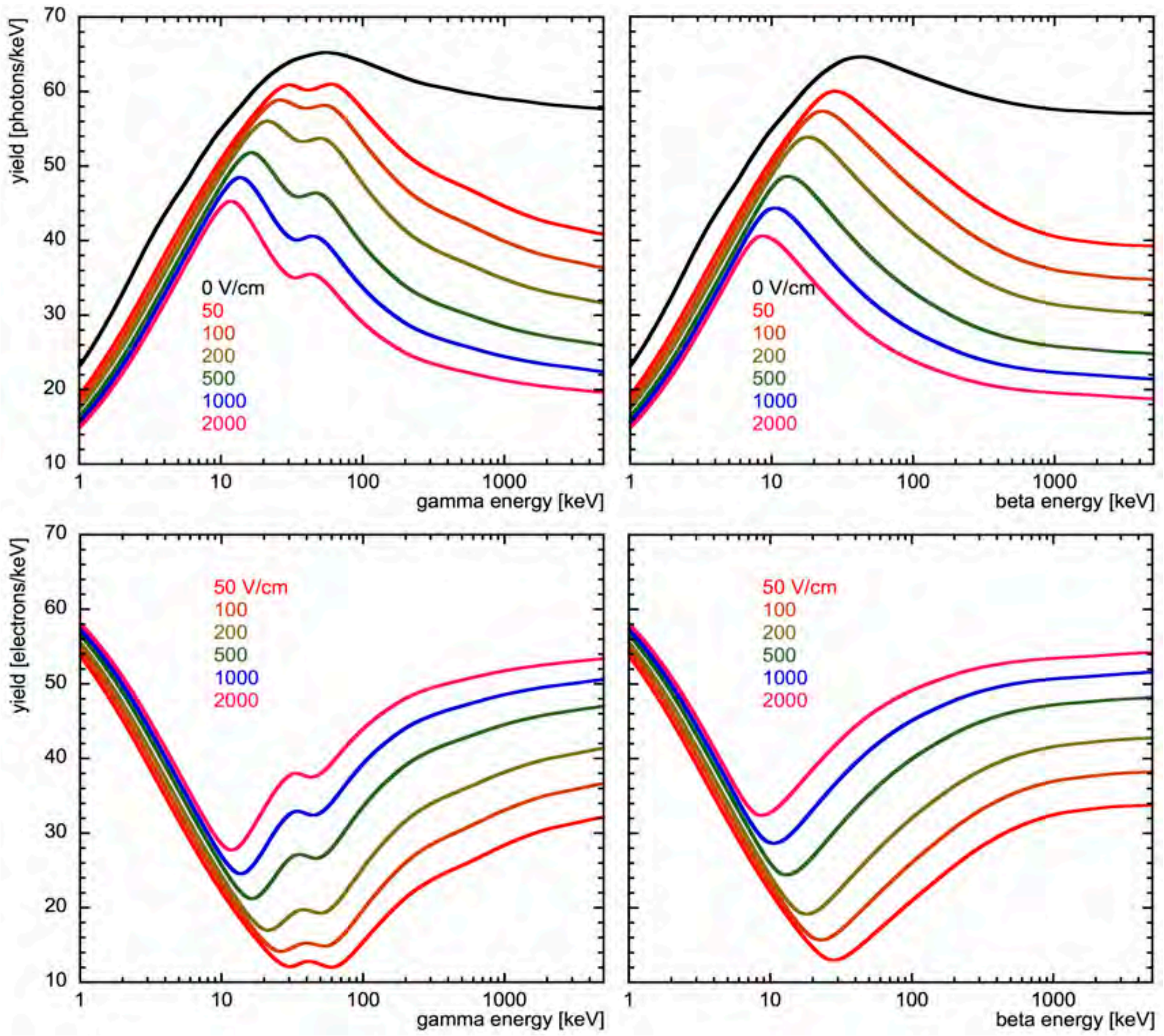

**Figure 3.3.1. NEST predictions of light (top) and ionization yields for ERs in LXe, for incident gamma rays (left), and primary electrons (β-particles, δ-rays) (right) [30]. Increasing the electric-field strength reduces recombination, raising the charge yield at the expense of light. The dip in the gamma-ray curves is due to the Xe K-shell X-ray that creates a second interaction site, displaced from the initial energy deposition.**

Scintillation and ionization yields for ERs in LXe are shown in Figure 3.3.1, as predicted by the Noble Element Simulation Technique (NEST) model (see [30-32] and the brief description in the next section); data for Compton electrons now reach down to 1.5 keV [31,33], and NEST shows good agreement with these results [34]. LXe compares favorably to the best scintillators and is also a good ionization medium. For example, the maximum photon yield at a few tens of keV is some 40% higher than that of liquid argon. This is important for a number of reasons: It reduces the variance of the ER response, which is important for particle discrimination; it permits effective detector calibration; and it is directly relevant to some leptophilic dark-matter searches.

As Figure 3.3.1 suggests, the scintillation yield is suppressed with increasing electric-field strength, while the ionization yield improves by the same amount. This behavior is also observed for individual events: A fraction of the photon yield comes from recombination luminescence, whereby VUV photons are generated from electron-ion recombination occurring near the interaction site, and therefore some electrons contribute either to scintillation (S1) or to ionization (S2), but not to both. This event-by-event anticorrelation of the two signatures can be exploited very effectively at higher energies in double-phase detectors to obtain good energy resolution for the spectroscopy of ERs, with application to gamma-ray background studies and searches for 0νββ decay.

For NRs, both data and modeling have progressed markedly in recent years. The picture here is more complex than for electron interactions. Most of the energy deposited by electrons is shared between ionization and excitation of the medium, making much of it observable. In NR interactions, a larger fraction is spent in atomic collisions, which is dissipated as heat and not detected. However, data obtained by scattering experiments at neutron beams ex situ, e.g. [35,36], are now in good agreement with those from indirect in situ techniques [37], down to ~3 keV.

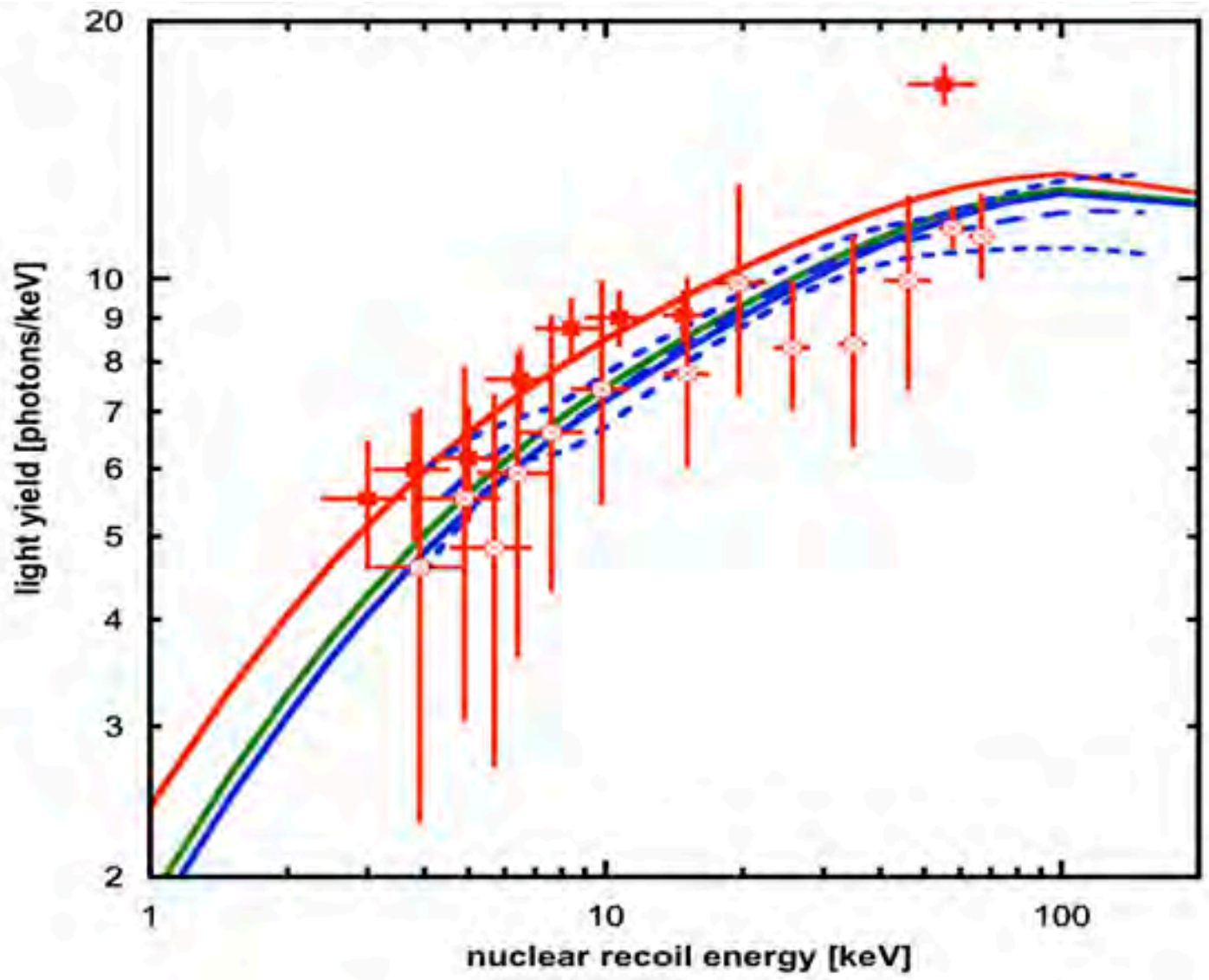

**Figure 3.3.2. Absolute NR scintillation yield in LXe. Hollow red markers are from neutron-beam measurements at Yale [35] and filled markers from [36] — both at zero field. Blue dashed lines are the combined mean and 1-σ curves from two in situ measurements with Am-Be neutron sources via fitting to MC simulation from ZEPLIN-III [37] (3,650 V/cm). The NEST model [33] is shown in red, green, and blue for zero field, 700 V/cm (LZ baseline), and 3,650 V/cm, respectively. The green curve conservatively zeroed below 3 keV is used for LZ sensitivity calculations.**

The ability of our models to reproduce these results has also improved, as shown in Figure 3.3.2, which summarizes the scintillation yield from Xe recoils in LXe. As has been the case with ERs, new NR calibration techniques will continue to decrease the uncertainties still affecting the lowest energies; in fact, LUX is pursuing such calibration work. The impact of this remaining uncertainty in the LZ sensitivity predictions is discussed in Chapter 4.

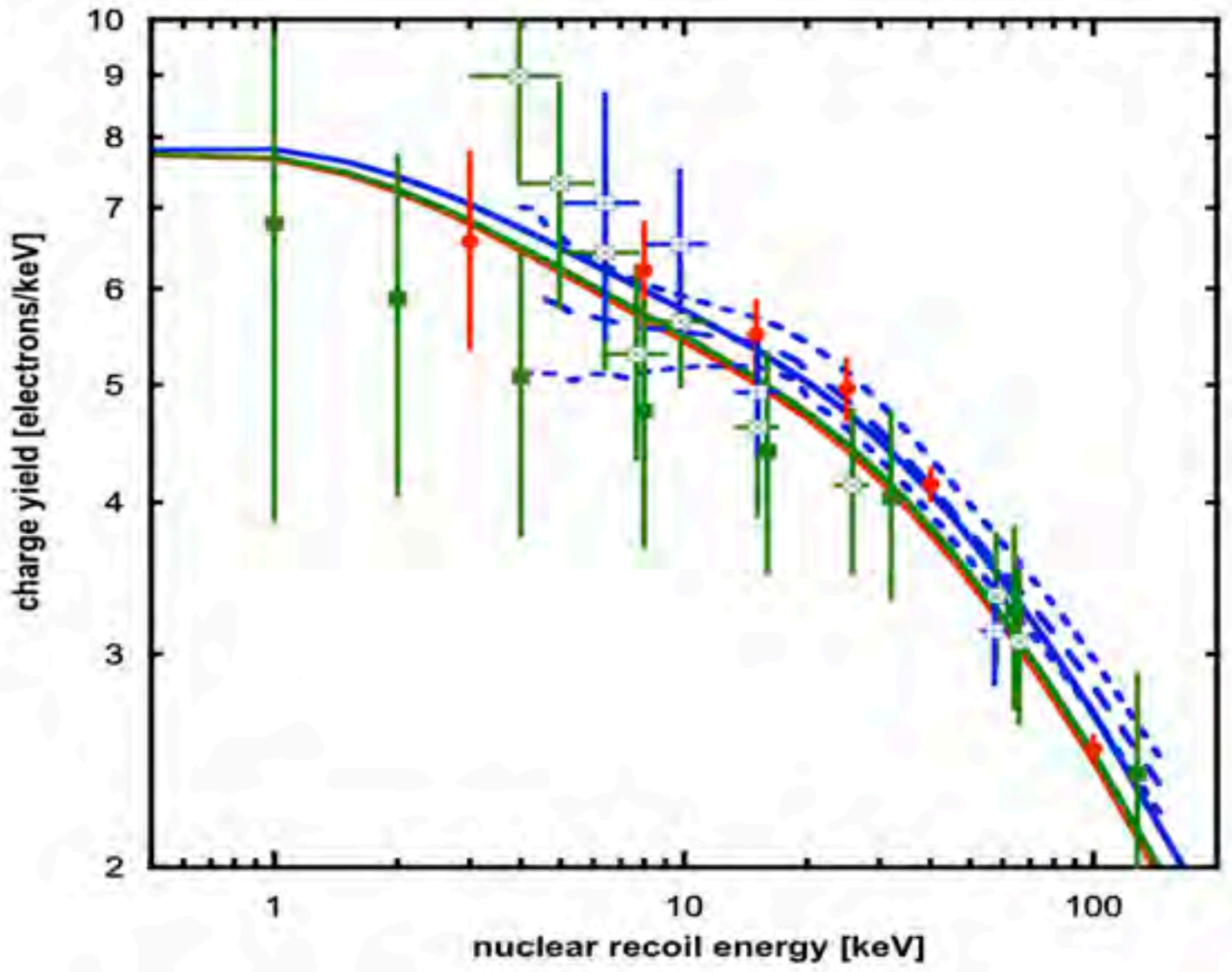

**Figure 3.3.3. NR ionization yield in LXe. Data are as follows: blue and green hollow squares from neutron beam data from Yale at 4 kV/cm and 1 kV/cm, respectively [35]; dashed blue curves from Monte Carlo matching from ZEPLIN-III [37] at 3,650 V/cm; solid green squares from XENON10 at 730 V/cm [38]; red markers from XENON100 [39] at 530 V/cm. The NEST prediction [33] is shown in red, green and blue for 530 V/cm, 700 V/cm (LZ baseline) and 3,650 V/cm, respectively. The green curve is used for LZ sensitivity calculations.**

As the NR scintillation yield declines gently at least down to 3 keV, the ionization yield increases accordingly, as shown in Figure 3.3.3. We note that the experimental data suggest a remarkably high yield even for 1-keV recoils, with several electrons being released per interaction. As with the scintillation yield, the electric field dependence is modest.

The high ionization yield allied to the ability to detect single electrons with high efficiency is a very attractive feature of this technology: Not only does it provide a low-threshold channel for light WIMP searches but, from a practical standpoint, it allows very high triggering efficiency (on S2) for the lowest-energy events, which are associated with very small S1 pulses.

### 3.3.1 Low Energy and Low Mass Sensitivity

Double-phase Xe detectors achieve the best NR energy threshold among the leading WIMP-search technologies — while maintaining discrimination and good vertex location. Of all such detectors operated so far, LUX can claim the lowest NR threshold of approximately 4 keV. WIMP masses down to ~10 GeV are directly accessible to an instrument such as LZ operating in the "normal" TPC mode, requiring one S1 pulse and one S2 pulse. A low-energy ER interaction (~1.5 $keV_{ee}$) is shown in Figure 3.3.1.1 — 3-D position resolution and discrimination are fully effective even at these energies.

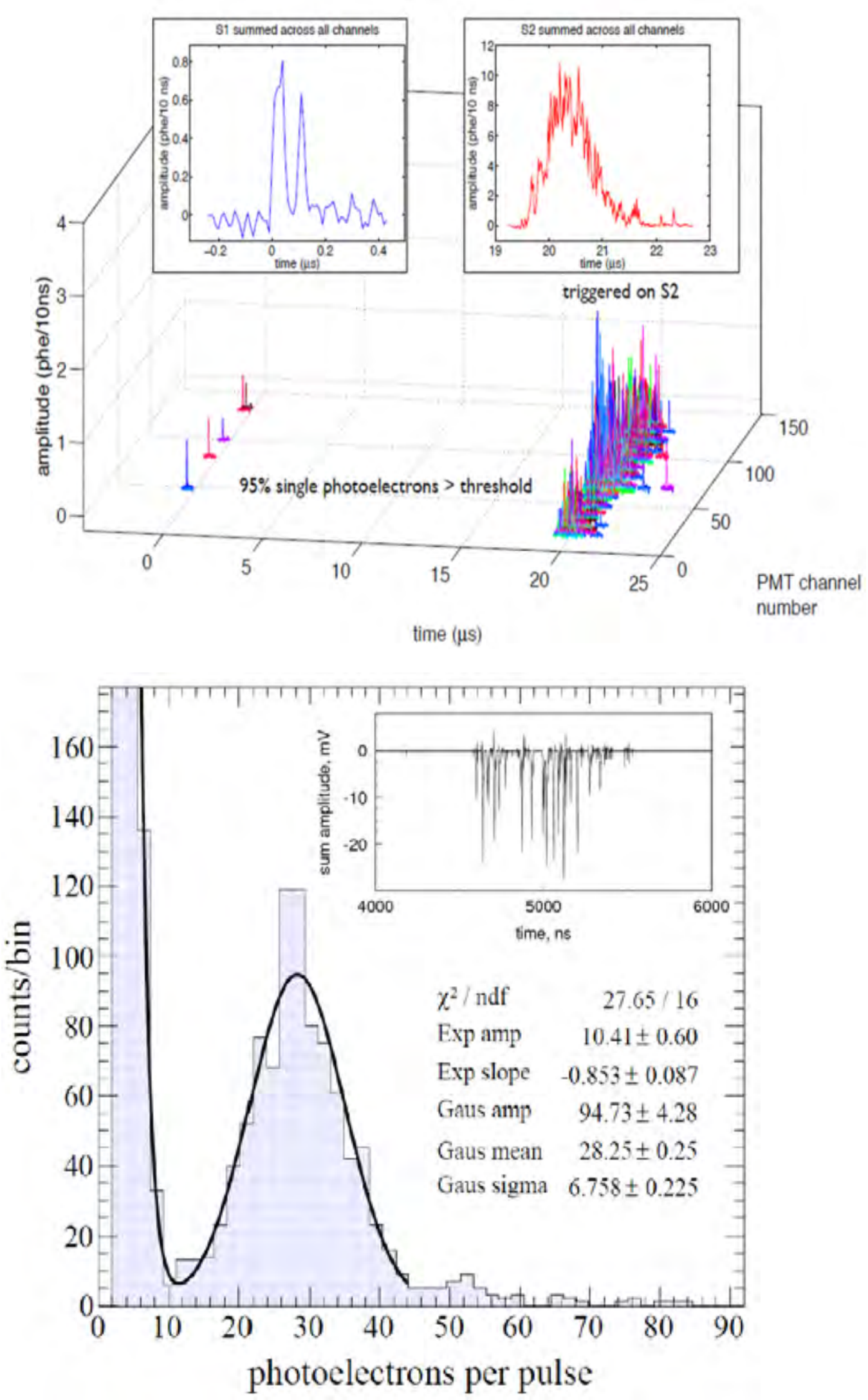

**Figure 3.3.1.1. Low-energy performance of double-phase Xe detectors. Top: A 1.5-$keV_{ee}$ electron interaction in LUX [41], showing a fivefold coincidence for S1 and the corresponding (much larger) S2 delayed by 20 µs. Bottom: Pulse size distribution of single electrons measured by electroluminescence in ZEPLIN-III, showing a mean of 28 photoelectrons per emitted electron (one such waveform is shown in inset) [18].**

At the smallest NR energies (≲4 keV), it is clear that the S1 signal is often absent but S2 is still easily detectable, so that LZ can recover sensitivity in this regime by performing an "S2-only" analysis [40]. Discrimination based on S2/S1 ratio is not possible in this instance, but the detector retains the ability to reject edge events in $(x,y)$. A more limited but still useful degree of $z$ position reconstruction is possible based on the broadening of the S2 pulse due to longitudinal diffusion of electrons as they drift in the liquid. As a result, an S2-only search can still exploit the extremely radio-quiet inner region of the WIMP target, and place upper limits on the dark-matter scattering cross section. Naturally, a thorough understanding of backgrounds is required for this type of analysis; several background mechanisms create single S2 electrons, while the two-electron random coincidence rate might still be significant. This technique is particularly applicable to particle masses lower than about 10 GeV.

One class of NR event that inevitably will be visible below the (3-phe) S1 threshold is due to coherent elastic scattering of $^{8}B$ solar neutrinos off Xe nuclei. The electron counting technique (S2-only) was in fact suggested a decade ago to allow a first observation of this process [42]. Due to energy resolution broadening of the scintillation signal, some events will register both S1 and S2 — from this and other neutrino fluxes. Despite significant interest in this signal per se, coherent neutrino-nucleus scattering is also a fundamental background for dark-matter searches, which is quantified in Chapter 4.

## 3.4 Electron/Nuclear Recoil Discrimination

Discrimination of ERs is key to the positive identification of a WIMP signal, both by directly reducing the effect of the dominant electronic backgrounds in the detector, and by confirming a NR origin. The physical basis for discrimination is the difference in the ratio of ionization electrons to scintillation photons that emerge from the interaction site and subsequently create the measured S2 and S1 signals, respectively. In a plot of the logarithm of S2/S1 as a function of S1, as in Figure 3.4.1, electron and nuclear recoils each form a distinct band, with NRs having a lower average charge/light ratio.

Discrimination is commonly quantified by the ER leakage past the median of the NR population (i.e., retaining a flat 50% NR acceptance). Previous values are between 99.5% in XENON10 [11] and 99.99% in ZEPLIN-III [15]. For the purpose of sensitivity calculations, we assume a baseline discrimination value of 99.5%, a conservative assumption given the performance already obtained in LUX, as discussed below.

Electron/nuclear recoil discrimination is determined by the separation of the bands as well as their widths, and in particular the "low tail" in $\log_{10}(S2/S1)$ of the ER band. Remarkably, the bands are mostly Gaussian when binned in slices of S1. Some skewness was observed in the electron band in ZEPLIN-III, although this was measured with external gamma rays rather than internally dispersed sources, and at very high field [15].

The physics determining both the position of the bands and their widths has been studied and we are increasingly able to model it successfully [43]. The overall separation of the bands is mostly due to NRs producing less initial ionization and more direct excitation (leading to scintillation) than do ERs. In turn, the bandwidths depend strongly on the physics of electron-ion recombination at the interaction site. A recombination episode generates an excited Xe atom that de-excites through scintillation (via the $Xe_2^*$ state). Therefore, initial ionization is either measured as charge (via S2) or light (via S1), and event-by-event fluctuations in the amount of recombination are one of the primary sources of band broadening. These fluctuations increase with recoil energy over the range of interest.

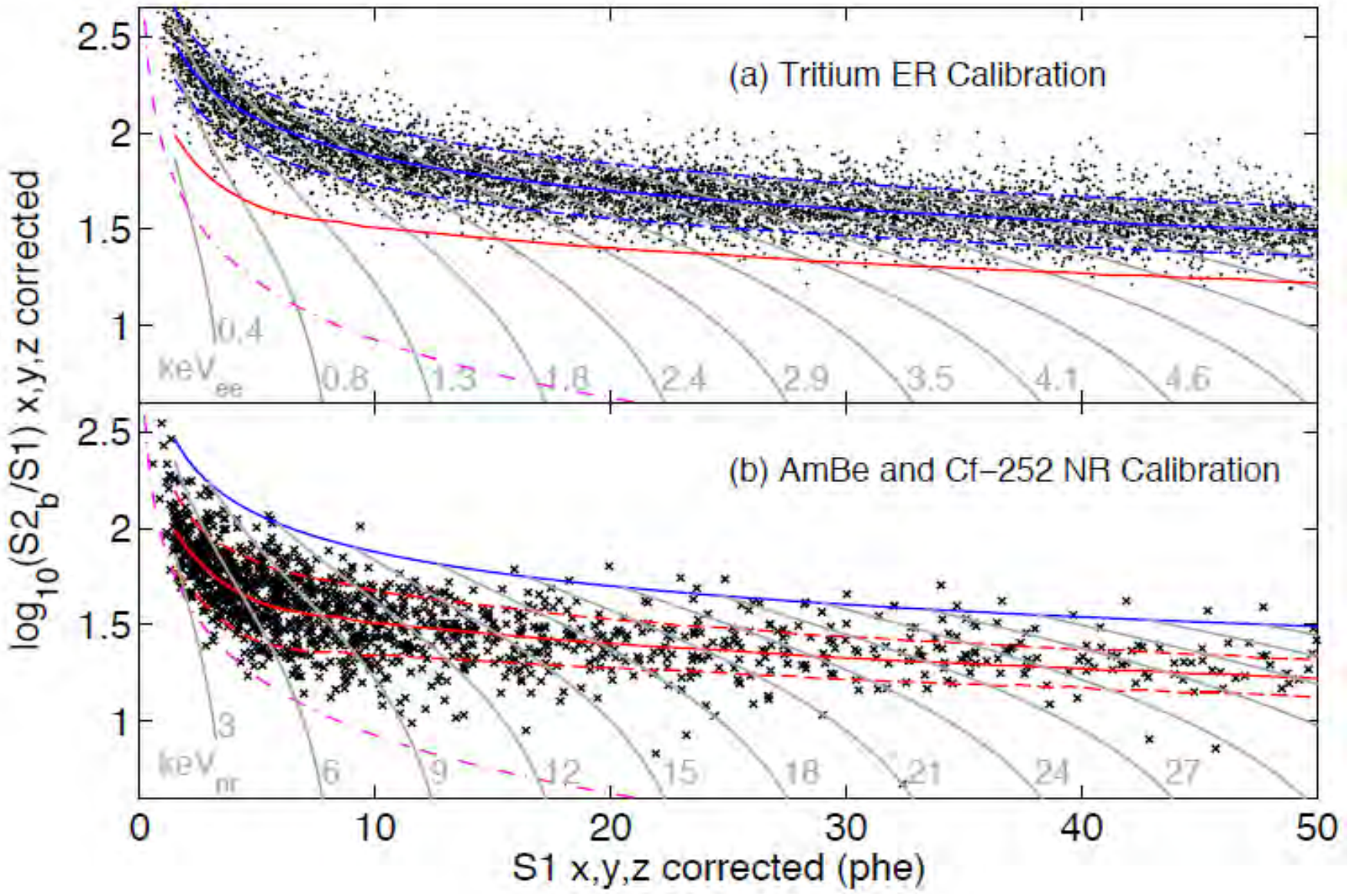

**Figure 3.4.1. Discrimination parameter $\log_{10}(S2/S1)$ as a function of S1 signal obtained with LUX calibration [4]. (a) ER band calibrated with beta decays from a dispersed $^{3}H$ source; the median is shown in blue, with 80% population contours indicated by the dashed blue lines. (b) NR band populated by elastic neutron scattering from AmBe and $^{252}Cf$ neutron sources; the median and 80% bandwidth are indicated in red, but in this instance they are defined via simulation to account for systematic effects present in neutron-calibration data (but not expected in a WIMP signal). The mostly vertical gray lines are contours of constant energy deposition. For more information, see Chapter 4.**

At the lowest energies, however, the distributions "flare up" due to statistical fluctuations in the S1 signal. This broadening is therefore reduced in chambers with higher light yield, improving discrimination. In fact, ER rejection in this technology is better just above threshold (~5–15 keV), where WIMP-induced recoil rates are highest, than at intermediate energies (~15–40 keV), and then improves dramatically beyond ~40 keV. The excellent ER discrimination at low energies is due in part to a decrease of recombination fluctuations, but is also caused by the curvature of both bands at low energy, as shown in Figure 3.4.1: The bands are largely *parallel* to the direction of S1 fluctuations, sharply reducing their impact on discrimination.

In addition to light collection, the drift field is also expected to affect discrimination. Although largely determined by the amount of initial ionization, the positions of the band medians have a residual dependence on the different amounts of field-dependent recombination for the two recoil species, and their separation increases at higher fields [43]. This may explain the world-best discrimination observed in ZEPLIN-III, which operated with close to 4 kV/cm drift field, and is an important driver of the LZ design. The bandwidth should, in principle, also have some field dependence, though this has not been well measured. The model developed in [43] has been incorporated in the NEST Monte Carlo package [33,30], which, as described below, has informed the sensitivity projections for this report. However, caution must be exercised when comparing values from different experiments, since instrumental effects other than electric field and light yield can impact discrimination very severely. A case in point is the degraded discrimination measured in the second science run (SSR) of ZEPLIN-III [44] relative to the first science run (FSR) [15], benchmarked essentially in the same detector and with the same software, but following upgrade of the TPC with underperforming photomultipliers.

The question of discrimination is of great important in LZ, as its dominant background is ERs from solar neutrinos. To predict the LZ sensitivity, the *electric field strength* and the *light-collection efficiency* in the WIMP target are the main ingredients required. These are key performance parameters and we motivate their choice in two separate sections below; we also describe the steps we are taking to achieve the required performance. We use these parameters in conjunction with a full Geant4 Monte Carlo simulation [45] based on the LUXSim package [46] and incorporating NEST, which we also describe briefly below.

The adopted values — a drift field of 700 V/cm and an S1 photon detection efficiency (PDE) of 7.5% — motivate an average nominal discrimination of 99.5% for a flat ER spectrum such as that from solar pp neutrinos (an ER leakage past the NR median of 1:200). This is supported by both NEST-based simulations and by XENON10 [11], which achieved that level of discrimination at similar field and light collection as proposed here. PANDA-X recorded 99.7% at 667 V/cm, for a PDE of 10.5% [47]. Significantly, LUX initially reported 99.6% discrimination at only 180 V/cm in Run 3 [4], increasing to 99.8% in a subsequent reanalysis with improved algorithms. Therefore, we are confident of reaching the 99.5% value assumed in this report.

Figure 3.4.2 shows how the discrimination generally improves for smaller S1 signals, as measured in LUX. In addition, a light-WIMP signal is not distributed symmetrically around the NR median: Upward fluctuations in S1 that cause low-energy events to be considered above a given analysis threshold also dictate that the distribution of $\log_{10}(S2/S1)$ is systematically lower than the NR median, and so the acceptance for light WIMPs is higher than the 50% generally mentioned here. We will assess how this higher acceptance translates to light-WIMP sensitivity in Chapter 4.

A potential limitation to ER discrimination comes from multiply-scattered gamma rays, in which a well-measured, low-energy scatter in the active TPC is accompanied by a further interaction in a region that yields a light signal but no charge (e.g., the liquid below the cathode). If the pathology of these so-called

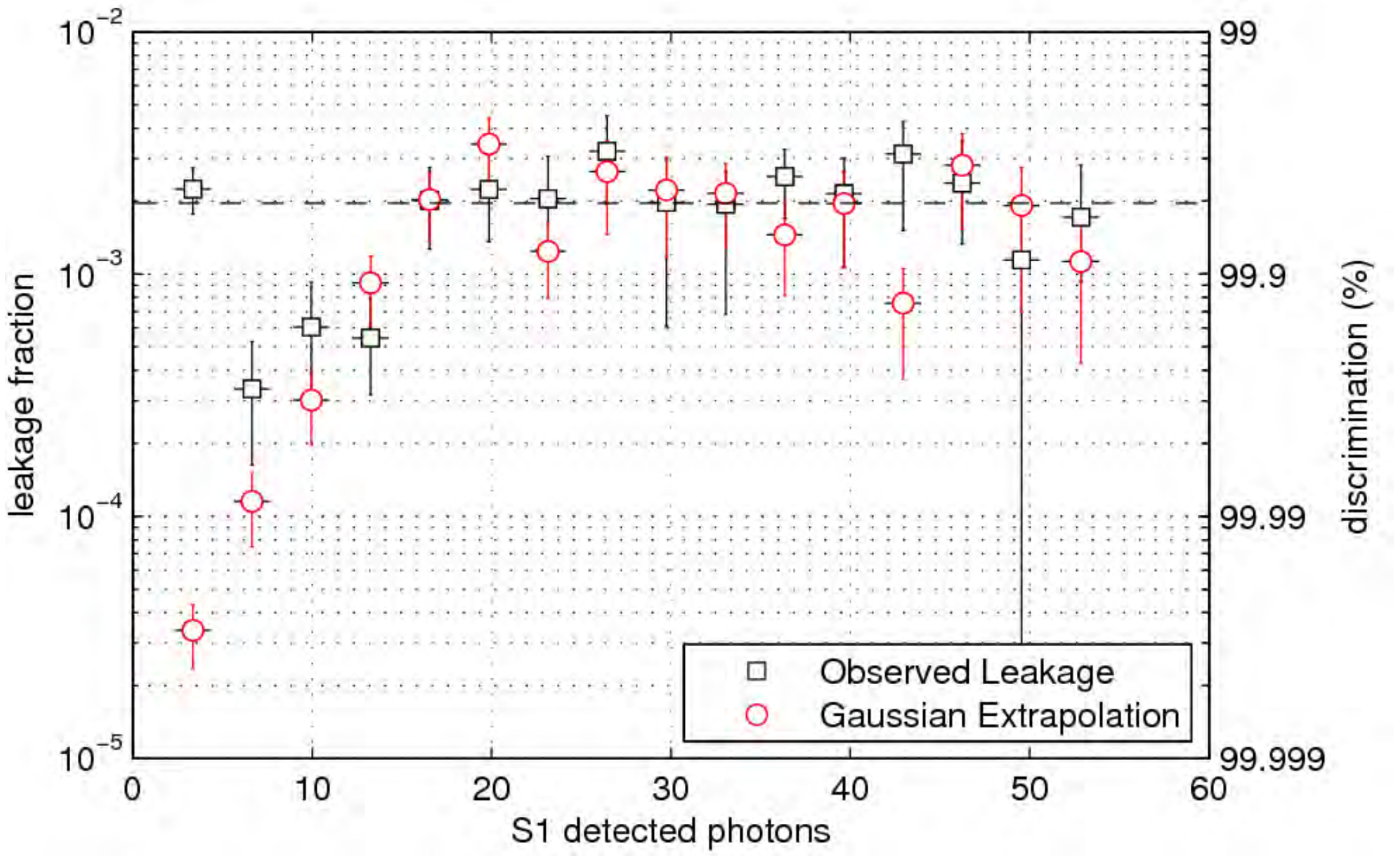

**Figure 3.4.2. ER leakage fraction past the NR median line measured with tritium data in LUX. Black squares are from event counting and the red circles are from integrating the tails of Gaussian fits to the ER population. The dashed line indicates the average leakage of 0.002 (99.8% discrimination) in the S1 range 2–50 phe. The general improvement of discrimination at low energies can be clearly seen, with the exception of the very lowest S1 data point where the ER band starts to flare up due to photoelectron statistics.**

"gamma-X" events is not recognized (by, for instance, the light pattern in the bottom PMT array), the result is a suppressed S2/S1 ratio, potentially mimicking a NR. Such events are subdominant in an instrument as large as LZ, where gamma-ray interactions are very rare in the fiducial volume. In addition, LZ minimizes inactive "skin" volumes by instrumenting the LXe outside of the TPC, effectively turning these potentially problematic regions into an additional veto detector. In defining a preliminary fiducial volume in Section 3.8.5, used for the sensitivity calculations presented in Chapter 4, we have assumed, rather conservatively, that events with one vertex in these regions (e.g., sub-cathode) are not vetoed. In reality, most will exhibit S1 light patterns that are strongly peaked in a single bottom-array PMT, which makes them atypical when compared with the deepest fiducial interactions and provides therefore a basis for their removal.

### 3.4.1 Discrimination Modeling with NEST

NEST (Noble Element Simulation Technique) provides a model for both the scintillation light and ionization charge yields of nuclear and electron recoils as a function of electric field and energy or $dE/dx$ [33,30,32]. "NEST" refers both to a collection of microscopic models for energy deposition in noble elements and to the Monte Carlo simulation code that implements these models. NEST provides mean yields and intrinsic fluctuations due to the physics of excitation, ionization, and recombination, including both Gaussian and non-Gaussian components of the energy resolution. To properly model the discrimination of nuclear versus electron recoils, the mean S1 and S2 yields must be known, but also their variances, which are made up of the intrinsic fluctuations referred to above as well as of instrumental fluctuations and data-analysis effects.

Since discrimination is a function of electric-field strength, recoil energy, and light-collection efficiency, all of these must be modeled together to predict the baseline LZ sensitivity. To validate this methodology,

NEST was initially trained on data from the small double-phase detector from Case Western Reserve University (Xed), which yielded comprehensive data sets in terms of energy range and field sweep [43].

### 3.4.2 High Voltage: Results, Design, and Program

The cathode HV in LZ is an important performance parameter that will directly affect the science reach of the instrument because of its impact on ER rejection. Introduction of HV into the Xe space is challenging due to possible charge buildup and sparking, and also because high-field regions can produce unwanted electroluminescence and electron emission that blind the detector to the flashes of scintillation light produced by WIMP interactions.

The LZ *operational* and *design* voltages were determined by considering the trade-offs between dark-matter sensitivity, project cost, and risk. Between December 2012 and April 2013, a dedicated LZ Task Force of 10 engineers and scientists examined the various design proposals and critically evaluated their technical feasibility, with the scope covering the electrode grids, portions of the field cage, internal connections, and the cathode feedthrough. The Task Force recommended an operational cathode HV of 100 kV, so as to generate a 700 V/cm drift field. This should ensure an ER rejection efficiency of 99.5%, as demonstrated in XENON10 and supported by NEST (and now surpassed in LUX). The design goal for cathode HV was set at 200 kV; all subsystems will be designed to withstand this higher value to help ensure that the 100 kV operational voltage can be met with high probability.

No double-phase Xe detector developed for rare-event searches has yet been able to operate at nominal-design electric fields “as built.” In the past, maximum design fields were considered that were a relatively modest factor below the onset of electroluminescence in the liquid (~400 kV/cm), as reported in the literature [48-50]. However, breakdown or some form of “electroluminescence” has invariably been observed at fields at least an order of magnitude lower than expected. The choice of a reasonable “allowable field” — to be adhered to everywhere within the LXe space — is therefore a useful concept. We surveyed several double-phase Xe chambers to determine the average fields observed at the surface of cathode wires, typically where the highest voltage gradients are located, which could be applied stably and for which this stability could be established down to single quanta of light and charge. This level of stability is essential, for example, for the cathode and gate wire grids within the TPC, although it may well be a conservative assumption for external metal surfaces.

We do not yet have a complete understanding of the breakdown mechanisms at play in previous detectors. Several hypotheses have been considered, namely ion-related, UV-related, or particulate/dust effects, as well as enhanced field emission from asperities, Xe ice layers, and oxide layers, to mention a few. This remains the focus of R&D at several LZ institutes (see Section 6.7) and was the subject of a workshop organized in part by LZ scientists in November 2014 [51]. There is considerable evidence that it is indeed possible to achieve up to 400 kV/cm in LXe, at least on some cathodic surfaces. The surface gradient on cathode wires may be a special case, as it will involve ion processes directly and with much higher density than any other electrode in the system.

We identified several examples of HV gradients achieved stably at the cathodes of several LXe experiments. In all cases, the detector operated just below the onset of instability. For example, ZEPLIN-III sustained 40–60 kV/cm in two long runs [15,44], using 100-micron stainless steel wires, and so did its prototype chambers at Imperial [52] and ITEP-Moscow [53,54]. Significantly, the Xed prototype reached a value of 220 kV/cm [55], but with different material wires (Cu-Be, 40 micron diameter). This is the only case we identified in which a significantly higher field was sustained for long periods within the liquid — but it is also the case that the total span of wire used in this chamber was small: 0.5 m, compared with 117 m for the ZEPLIN-III cathode. The XENON100 cathode voltage was limited to −16 kV for stable operation, resulting in a drift field of 0.53 kV/cm across the TPC. Higher cathode voltages resulted in additional light pulses, likely caused by field-emitted electrons and subsequent scintillation in the strong electric field near sharp features of the cathode mesh [56]. In the LUX experiment, the cathode wire grid was built from 206-micron stainless steel wire, with 5 mm wire pitch.

**Table 3.4.2.1. High field regions in LZ at the operating cathode HV of 100 kV.**

| Electrode | Medium | Voltage Gradient | Safety Factor |
|---|---|---|---|
| Cathode wire surface | LXe | 37 kV/cm | 1.4 |
| Gate wire surface | LXe | 69 kV/cm | 0.7 |
| Field cage† | LXe | 23 kV/cm | 2.2 |
| Cathode ring OD | LXe | 34 kV/cm | 1.5 |
| Reverse field region | LXe | 25 kV/cm | 2.0 |
| HV umbilical | LXe | 41 kV/cm | 1.2 |
| HV cable | Polyethylene | 194 kV/cm | 3.0* |
| HV feedthrough | Epoxy | 30 kV/cm | 4.9‡ |

†Near start of vessel taper; *580 kV/cm maximum rating; ‡150 kV/cm maximum rating

During underground commissioning, light production was seen when the field at the cathode wire surface exceeded ~20 kV/cm, assuming a cylindrical wire without defects. LUX is still in operation, and it has not been determined why light was generated at such low nominal values. However, it is suspected that metal or plastic debris attached to the wire creates higher fields than would otherwise be present. This light production limited LUX to a cathode voltage of −10 kV, which resulted in a drift field of 0.18 kV/cm in its 48-cm drift region.

At the moment, we are adopting a value of 50 kV/cm as the maximum allowable field for stable operation of immersed metal surfaces throughout the LZ detector. Table 3.4.2.1 summarizes high field regions in the LZ design; our goal is to achieve safety factors of at least 2 so that we are compliant with the maximum allowable field at the full design voltage. Additional design work is needed and will be informed by our R&D results, which may lead us to make minor design modifications to some components (e.g., wire grids) or it may motivate an increase in allowable field from the 50 kV/cm considered presently — at least for some materials or surfaces. In any case, we note that LUX has already matched the discrimination assumed in LZ sensitivity calculations at much lower drift field.

Three approaches to HV delivery are being developed, all depicted in Figure 3.4.2.1. The first approach (and also the LUX baseline design) is a warm feedthrough, placed at room temperature outside the water tank. The second approach is a cold, low-radioactivity feedthrough, located at LXe temperature at the base of the cryostat [57,58]. The third approach is a Cockcroft-Walton generator, located in LXe-filled conduit within the water tank.

The warm feedthrough approach places the cathode HV insertion at the end of a long, vacuum-insulated, Xe-filled umbilical, outside the water shield at room temperature. With the dominant cable material being polyethylene, radon emanation is minimized. The outgassing properties of the polyethylene to be used in LZ have been measured and found to be acceptable from the perspective of Xe purity (see Chapter 9). With the feedthrough at room temperature and far away from the active LXe, there are no concerns about thermal contraction compromising a leak-tight seal to the Xe space, and no major constraints from feedthrough radioactivity. A feedthrough at the warm end of the umbilical allows a commercial polyethylene-insulated cable to pass from a commercial power supply, through an epoxy plug, and into the gaseous Xe. The cable then passes through the center of the umbilical and routes the HV through LXe and to a field-graded connection to the cathode. A smaller version of this feedthrough is installed in LUX, and was successfully tested up to 100 kV. A warm feedthrough prototype has already been successfully tested at Yale up to 200 kV, with the HV cable terminated in transformer oil. A more detailed description of the warm feedthrough approach may be found in Chapter 6.

The benefit of a cold feedthrough is that right up until the port on the cryostat, the HV can be delivered through a vacuum space rather than Xe. This alleviates concerns regarding radon permeation and outgassing, impurities contaminating the Xe, and complications with Xe recirculation through such conduits. It also allows commercial HV cable to be used all the way to the feedthrough, without the additional heat flux from such cable to the liquid, and avoiding risk from gas pockets and virtual leaks. Two insulating materials are being considered: quartz and polyethylene. Challenges to this approach include reliable sealing of the feedthrough at cryogenic temperatures, field distortions that might be produced after thermal contraction, and feedthrough radioactivity.

Internal HV generation using a rectifier circuit (Cockcroft-Walton) is being investigated. This AC-DC conversion would be done in the LXe, using a stack of capacitors and diodes to shuttle charge up to a high potential. This approach avoids the difficulties of passing high voltages through Xe gas, or passing extremely high voltages through a leak-tight feedthrough. The Cockcroft-Walton approach is commonly used for reaching high voltages in Van de Graaff accelerators but, in the case of LZ, the generator must not produce excessive radioactive background, nor generate RF radiation that could compromise the PMT arrays and instrumentation electronics [59]. By spatially separating the Cockcroft-Walton generator from the TPC, both radioactive background and RF noise may be greatly mitigated. The HV must then be transported through LXe to the cathode, as in the warm or cold feedthrough approaches.

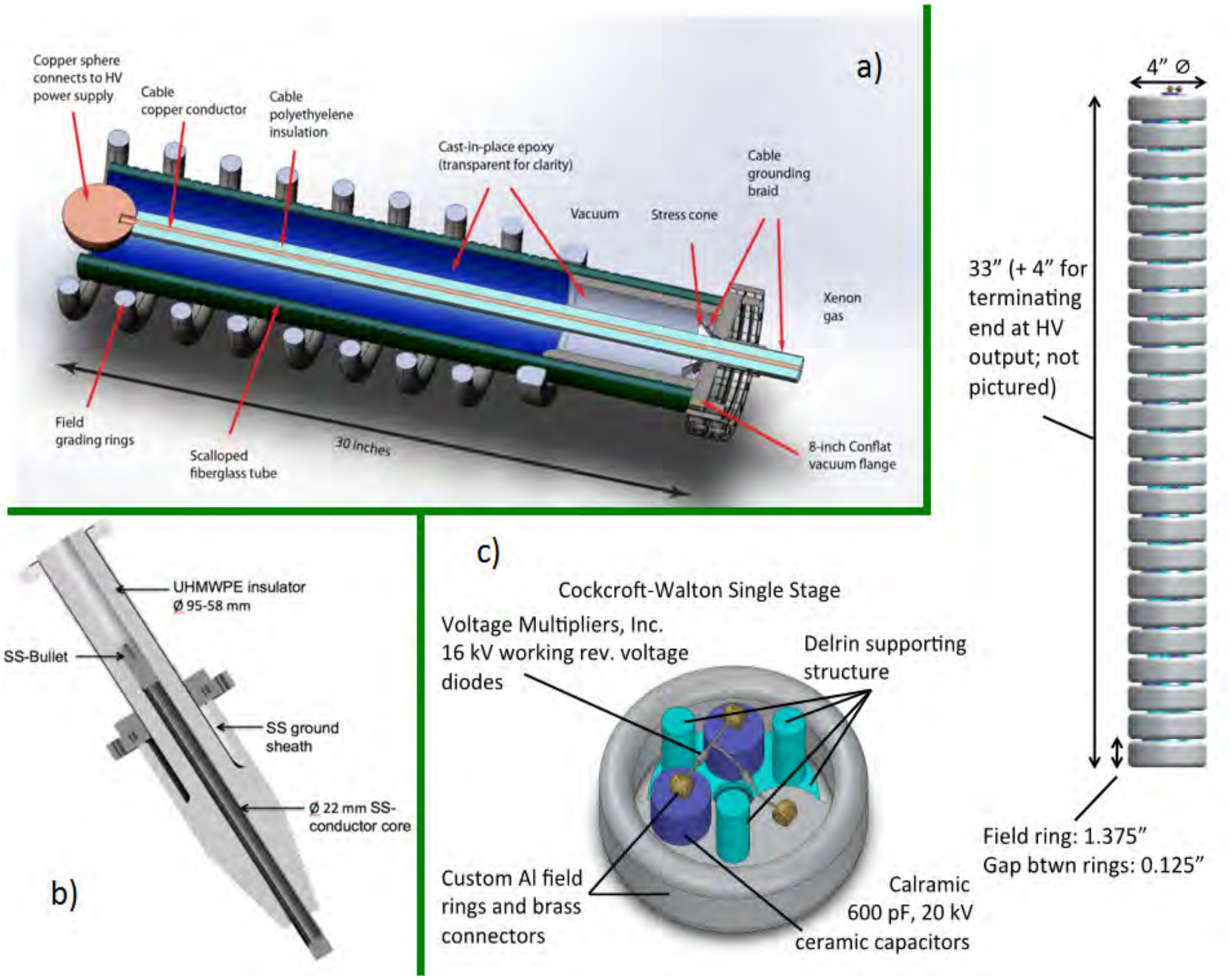

**Figure 3.4.2.1. HV delivery options in LZ. (a): The warm feedthrough approach (baseline design); (b) the cold feedthrough approach; (c) the Cockcroft-Walton generator approach.**

### 3.4.3 Light Collection: Results, Design, and Program

The energy thresholds associated with S1 and S2 are determined in each case by the microscopic light and charge yields of LXe and the light-collection efficiency of the chamber. For a given energy acceptance window, the magnitude of S1 also impacts discrimination, as highlighted above (e.g., Figure 3.4.2).

Maximizing the sensitivity of this response channel is therefore an experimental priority. Scintillation yields for electron and nuclear recoils, described in Section 3.3, depend also on energy and electric field. For reference, at the nominal LZ field we expect 470 scintillation photons to be emitted from a 10-keV electron track (47 ph/keV$_{ee}$), and 75 photons from a 10-keV NR (7.5 ph/keV). The experimental challenge is to maximize how many are recorded as photoelectrons in the PMT arrays.

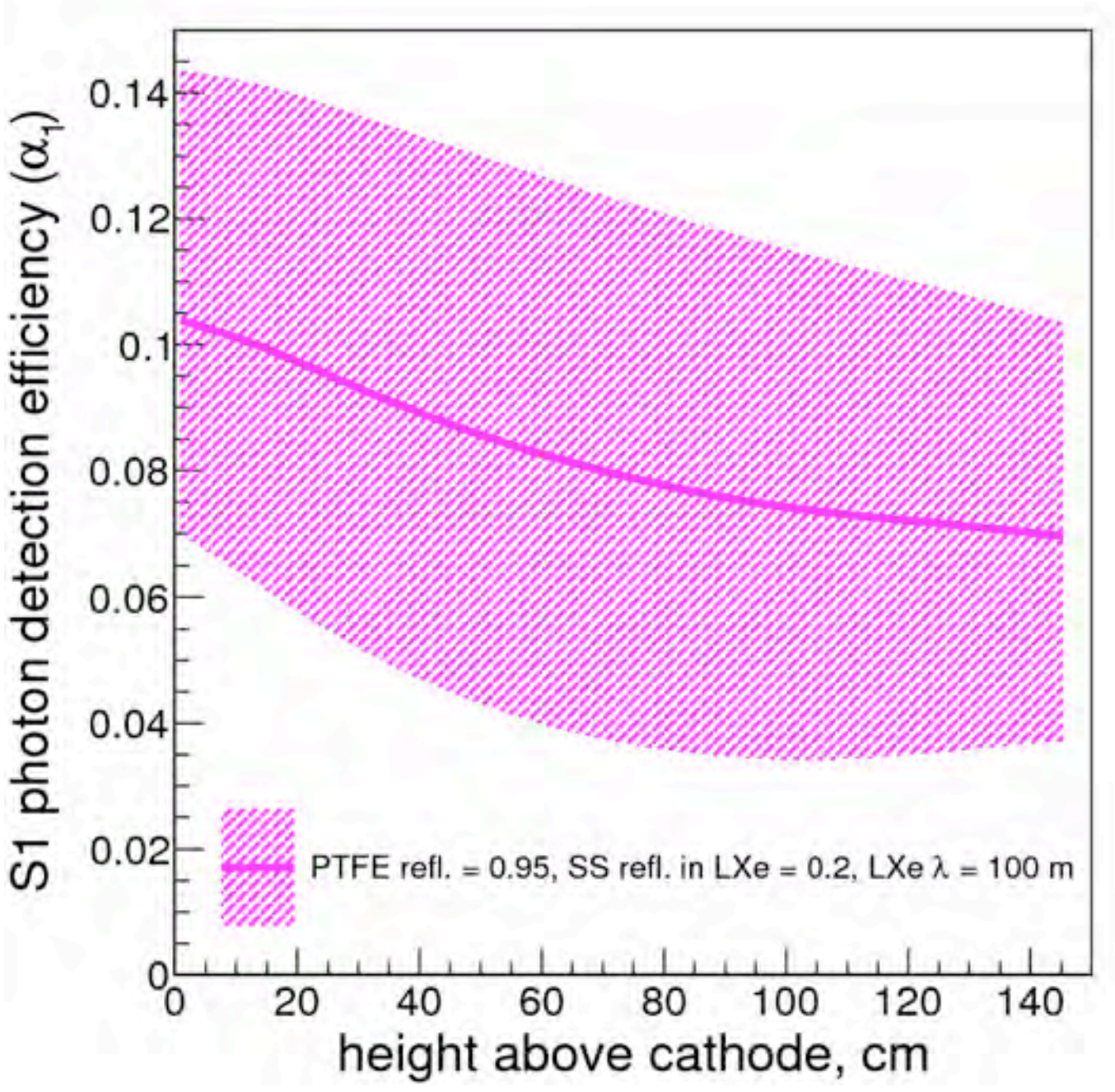

**Figure 3.4.3.1. Simulated S1 photon-detection efficiency as a function of distance from the cathode for three light-collection scenarios, averaging 4.4% (lower bound), 8.3% (line), and 12.5% (upper bound). Varied parameters are the PTFE and grid reflectivities and the photon absorption length; the number of PMTs (488) and their QE at 178 nm (25%) is the same in all cases. The baseline assumed for sensitivity calculations is $\alpha_1$=7.5%, with 5.0% and 10% also assessed to represent pessimistic and optimistic scenarios.**

In comparison, the electroluminescence yield in Xe vapor is high (typically ~1000 photons/cm per emitted electron [60], as discussed in Chapter 6); additionally, S2 light is recorded with high efficiency by the upper PMT array. This allows sub-keV$_{ee}$ detection thresholds to be easily achieved, and the sensitivity of the S2 channel is usually not a concern — in fact, single electrons emitted from the liquid can be detected straightforwardly, as mentioned previously.

The LXe scintillation emission is centered at 178 nm, with FWHM = 14 nm [61]. Light from electroluminescence in Xe vapor has a similar spectrum, but not quite identical [12]. Wavelength shifting is not required since the Xe luminescence spectrum is compatible with quartz-windowed photomultipliers. The basic optical properties of LXe are established: The refractive index for scintillation light is $n$ = 1.67 [62], which is well matched to that of quartz ($n$ = 1.57). This allows good optical coupling to the PMTs immersed in the liquid phase. The Rayleigh scattering length is 30–50 cm (see [12] and references therein), which must be considered in optical simulations.

The key issue is then to maximize the S1 photon-detection efficiency, $\alpha_1$, which measures the fraction of emitted scintillation photons that generate detected photoelectrons. The main factors affecting $\alpha_1$ in LZ are: (1) the VUV reflectivity of internal surfaces made from PTFE; (2) the photon absorption length in the liquid bulk; (3) the geometric transparency and reflectivity of all grids; (4) the PMT photocathode coverage fraction; and (5) the PMT optical performance.

Below we discuss briefly these parameters and how they translate into the three light-collection scenarios assumed for LZ in this report; these are illustrated in Figure 3.4.3.1. Optical simulations for the baseline justify a volume-averaged $\alpha_1$ = 7.5%, and we adopt also pessimistic and optimistic combinations of parameters that support $\alpha_1$ =5% and 10%, respectively.

The phototube specification and array layout are addressed in Chapter 6; here we assume tightly packed top and bottom arrays of 3-inch tubes (top and bottom arrays with 247 and 241 units, respectively), with average QE = 25%. This conservative QE matches the manufacturer specification and allows for double-photoelectron emission observed in some photocathodes at these wavelengths (and not captured by the QE specification). This is an important issue that we are presently investigating in the LUX and LZ PMTs [63]. We use the lower value in all three optical simulation scenarios, although typical QEs reported by the manufacturer for LXe scintillation are more like 30%.

The photon absorption length in the bulk LXe depends on the purity of the liquid with respect to trace amounts of contaminants with absorption bands overlapping the LXe scintillation spectrum, mostly $H_2O$ and $O_2$. For our tight purity requirements for those electronegative species (0.1 ppb, see Chapter 9), a value $l \sim 100$ m is reasonable (see Figure 15 in [12]), but we vary this parameter from 10 m to 1,000 m for the other scenarios.

The five grids in the LZ TPC all affect light collection through obscuration and wire reflectivity and, in a chamber where other sources of optical extinction have been minimized, these grids can have a significant effect in the light yield, especially those immersed in the liquid. Our baseline assumption is 20%, varied to fully absorptive or to 50% reflective for the other two scenarios.

Finally, the last and perhaps most important optical parameter is the reflector used in constructing the TPC. Based on a decade of experience, PTFE is indeed the best reflector for LXe scintillation, and also for properties other than optical: The manufacturing process yields very radiopure material (~ppt in U/Th); it has good mechanical properties (despite about 1.2% thermal contraction to LXe temperatures); and outgassing rates are relatively low (see Chapter 9). Optically, reflectivities of at least 90–95% have been reported in LXe chambers [64,65,41], and evidence from LUX suggests even higher values. Unlike metallic coatings such as aluminum, also a good VUV reflector [66], the optical properties of PTFE will not degrade during the lifetime of the experiment.

The VUV reflectivity of PTFE is a critical parameter in these detectors and for this reason we initiated a study of its optical properties at LIP-Coimbra. In vacuum, the reflectivity depends strongly on the manufacturing process and surface finish; best results were obtained with molded PTFE after polishing [67]. It is also known that optical properties differ for low- and high-density versions of the fluoropolymer [68]. However, we found also that the properties of the PTFE-LXe interface are not well predicted by straightforward modification of the optical models derived from vacuum measurements with Xe scintillation: An average hemispherical reflectance of 90% is predicted [67,69], which is consistently lower than measured in the liquid. A dedicated test chamber was constructed for this purpose, which involves a small LXe cell with lateral walls made from the PTFE under test, photomultiplier readout from the top, and an α-particle source at the bottom. The chamber is filled progressively and the resulting response is fitted by a five-parameter optical model that returns the reflectivity of the liquid and gas interfaces, among others. Preliminary results for the same PTFE as used in LUX are confirming 97% hemispherical reflectance when immersed in the liquid; this is shown in Figure 3.4.3.2. This result is in good agreement with that derived by comparing LUX data with optical simulations from LUXSim. Candidate materials (different types, finishes, and manufacturers) can be tested in this way before procurement for construction. We adopt a somewhat conservative reflectivity of 95% in our optical simulations for the baseline, with 90% for the pessimistic and 99% for the optimistic scenarios.

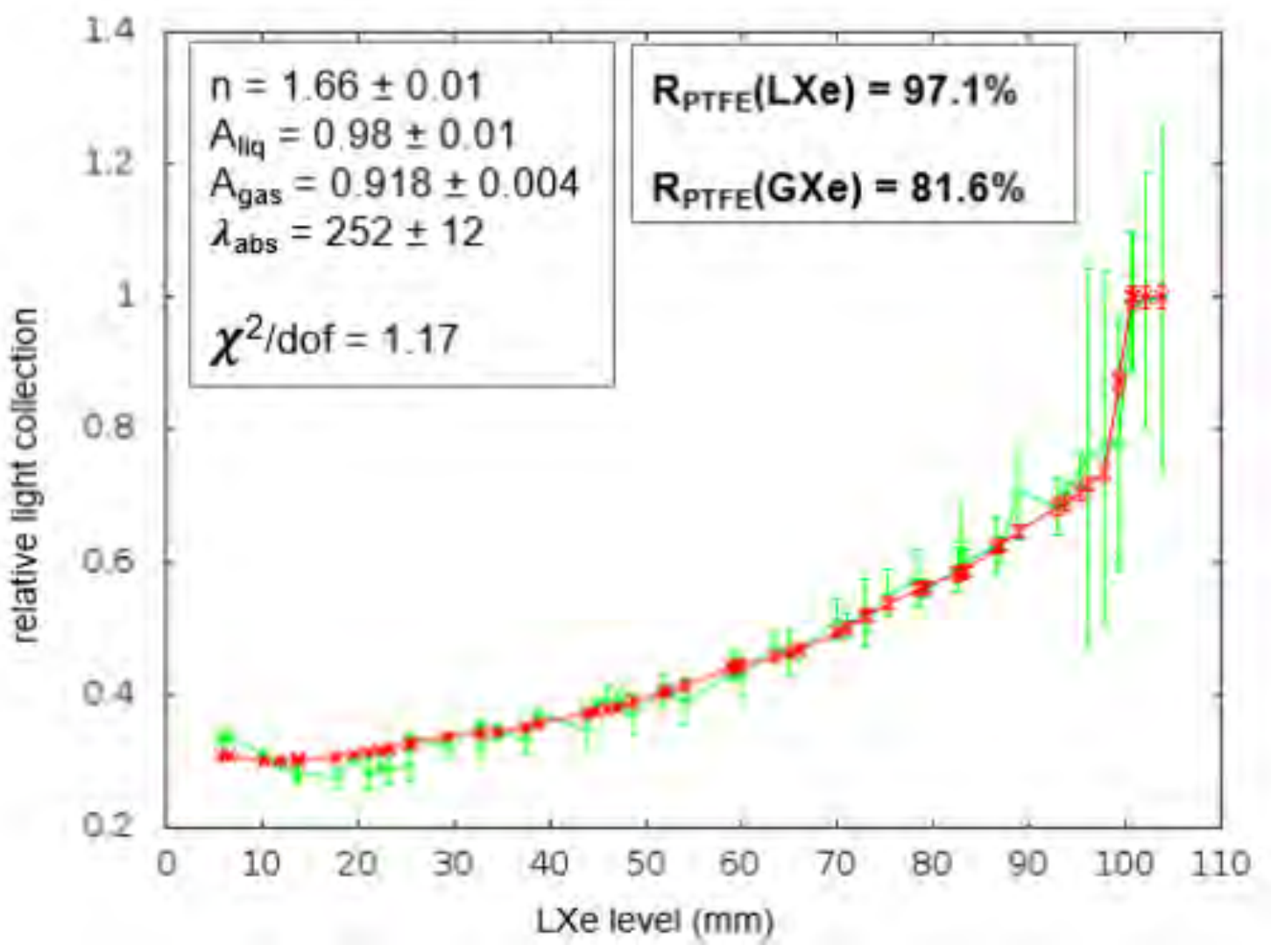

**Figure 3.4.3.2. Measurement (green) and simulation-derived fit (red) of the total hemispherical reflectance of the PTFE/LXe interface in a LXe scintillation cell. The PTFE sample under study is the material used in LUX, and it makes up the lateral walls of the test chamber.**

To reduce the dead volume around the active LXe, as well as outgassing and potential backgrounds, it is desirable to minimize the thickness of the PTFE walls of the TPC and skin detectors. A lower limit is established by the transmittance of PTFE to Xe scintillation light, and the need for optical isolation between the TPC and skin regions as well as between these and any dead regions containing LXe. The transmittance of the PTFE used in LUX was measured as a function of thickness and for different wavelengths: Xe gas scintillation (178 nm) as well as 255, 340, and 470 nm. Results for Xe scintillation light show a transmittance <0.1% for 1.5 mm PTFE thickness, but rising significantly to as much as 10% for 5 mm in the case of blue light (470 nm).

The baseline $\alpha_1$ of 7.5% translates to an S1 response of 4.7 phe/keV at zero field for $^{57}$Co gamma rays (a traditional measure of light yield in LXe chambers) — cf. 8.8 phe/keV in LUX [4], 6.6 in XENON10 [11,17], 5.0 in ZEPLIN-III [15], 4.3 in XENON100 [56], and 1.1 in ZEPLIN-II [9]. According to NEST, the corresponding NR energy threshold is approximately 6 keV for a 3-phe coincidence requirement, which we adopt for our sensitivity estimates. We note that a lower, twofold coincidence may be possible (as in LUX Run 3) that would lower the 6 keV threshold.

## 3.5 Outer Detector Systems

A WIMP scatter would deposit a few keV in the central volume of the LXe TPC, with no simultaneous energy deposit in surrounding materials. Radioactivity neutrons, which can fake WIMP interactions when they scatter elastically, are likely to interact again either within the TPC or nearby, and so it is broadly desirable to replace as much of the neighboring material as possible with additional radiation detectors. It also helps to minimize intervening material between the active LXe in the TPC and any such ancillary detectors, namely by decreasing the thickness of the field-cage and of the cryostat vessels. Active material surrounding the central LXe volume also permits assessment of the local radioactivity environment, and thus to infer additional information on the backgrounds in the WIMP search region. A persuasive WIMP discovery will require excellent understanding of all background sources, which is best done through the characterization of those sources in situ.

The LZ apparatus will feature two distinct regions where active material surrounds the LZ TPC. The first is a “skin” of LXe, formed by liquid between the field cage and the inner vessel of the cryostat. The second is surrounding detectors of liquid scintillator (LS), which is doped with a small amount of gadolinium, to enhance its capability for neutron detection. We envisage a threshold of 100–200 $keV_{ee}$ for both systems, with the gadolinium-doped LS detecting neutrons more effectively and the skin detector performing best for internal gamma rays.

### 3.5.1 Xenon Skin Veto

The lateral skin region consists of an unavoidable 4–8 cm of LXe between the outer boundary of the field cage and the inner boundary of the cryostat. An even thicker LXe region exists below the bottom PMT array. It is highly desirable to read out scintillation light generated in these regions for two main reasons: Besides constituting anti-coincidence detectors in their own right, they will also tag external LXe interactions where VUV photons can leak into the TPC and fake S1 light there. Our approach has been to maximize the optical isolation between active and inactive LXe volumes as far as practicable, and to instrument as much passive LXe as possible. The sensitivity to deposited energy in the skin will be far less than that of the central Xe TPC and, naturally, no ionization can be detected.

To get reasonable utility for vetoing gamma rays, the threshold should be a small fraction of the typical energy of an environmental gamma. A threshold of about 100 $keV_{ee}$ is adequate to detect Compton recoils from MeV gamma rays from radiogenic backgrounds. The skin detector will have low sensitivity for neutron detection via elastic scattering due to the mismatch in mass between neutrons and Xe nuclei, but inelastic neutron interactions can be still detected.

The LXe skin veto has some advantages over the outer detector for gamma detection in that some gamma rays do not penetrate the various vessels all the way to the LS. Our design studies indicate that a threshold of 100 $keV_{ee}$ can be reached if the walls of the skin volumes are sufficiently reflective, and as long as the photocathode coverage is sufficient. Approximately 120 1-inch Hamamatsu R8520 PMTs, split into top and bottom arrays, will monitor the cylindrical shell between the sides of the TPC and the cryostat wall, with a threshold of three photoelectrons providing a 100 $keV_{ee}$ energy threshold. In the bottom region below the TPC, a further 60 1-inch PMTs provide a similar sensitivity.

Should the threshold be worse than 100 $keV_{ee}$, there will be additional ER background in the LZ fiducial mass. For 300 $keV_{ee}$, the rate of ERs in the 5.6-tonne LZ fiducial mass roughly doubles, but would still be an order of magnitude below the ER rate from solar neutrinos. The skin detector is further described in Section 6.7.

### 3.5.2 Scintillator Outer Detector

The goal of the outer detector is to surround the LZ cryostat with a near-hermetic gamma-ray and neutron anticoincidence system. Much of the challenge is of a practical nature: Deep underground, the safety of combustible LS solvents is a prominent concern. Fabricating the large (~5-m-tall) tanks for underground deployment successfully and economically are other important considerations.

LZ will employ linear alkyl benzene (LAB), an LS solvent developed by the reactor neutrino community in the past decade [70]. The flash point of LAB is >120$^{o}$C, exceeding that of diesel fuel (commonly used underground for backup generators). Small quantities of a standard fluor and wavelength shifter will be added to the solvent to provide the scintillation signal. A PMT system located in the water space outside of the clear acrylic tanks containing the LS will view this scintillation light.

To enhance neutron detection, 0.1% by weight of natural gadolinium will be dissolved in the LAB with a chelating agent to form Gd-loaded LS, or GdLS. Two isotopes of Gd, $^{155}$Gd and $^{157}$Gd, have neutron-capture cross sections that are 61 and 254 kilobarn, respectively. Each of these isotopes constitutes about 15% of natural Gd and, at 0.1% concentration by weight, capture on Gd is about 1 order of magnitude more probable than is capture on hydrogen.

A neutron that can cause a Xe NR in the same energy range as the recoil from a WIMP will have an energy between about 0.5 and 5 MeV. The source of most of these neutrons will be from the (α,n) process from material around the edges of the Xe, and their energy spectrum will be toward the low end of this interval. A dangerous neutron will enter the LXe TPC and then scatter back out after one interaction. Many of those neutrons will traverse the intervening material and then thermalize and capture in the Gd in the GdLS. The length scale for thermalization and capture is a few centimeters and the typical capture time is ~30 μs, which is small compared with the 670 μs maximum drift time of the TPC.

After capture on Gd, a total energy of about 8 MeV is emitted as a burst of several gamma rays, which then interact in the LS (or the skin, or the TPC). This large energy release separates neutron captures from the gamma rays from natural radioactivity, which die out above 3 MeV. The thickness of the GdLS is determined by the gamma scattering length, which is ~25 cm. The thickness of the GdLS layer is 75 cm, or three scattering lengths. For a fraction of gamma rays that deposit energy in the central LXe TPC, the outer detector system also functions as a gamma-ray veto, for those that propagate through the LXe skin and cryostat; these share the same detection requirements as the capture gammas. To achieve good efficiency as a gamma-ray veto, we need a threshold to Compton electrons near 100 $keV_{ee}$.

In reactor neutrino experiments, typically a single acrylic cylinder contains the Gd-loaded scintillator. Transport logistics preclude this solution for LZ, and instead segmentation of the volume into nine smaller tanks that can be transported through the shafts and drifts that lead to the Davis Cavern is the adopted solution. To enhance light collection, the outer surface of the LZ cryostat will be affixed with a diffuse white reflective layer of Tyvek© to reflect light into the 8-inch Hamamatsu R5912 PMTs that will surround the tanks. The 600-μm multilayer Tyvek© we plan to use has a reflectivity in excess of 95%.

The PMT collecting power must be sufficient to achieve a 100 $keV_{ee}$ threshold. Our preliminary studies indicate that, by covering the cryostat in Tyvek© and employing another reflective cylinder outside of the PMT system, the collecting power of 120 8-inch PMTs is sufficient to achieve the required threshold.

The GdLS tanks will be surrounded by ultrapure water, and the distance to the water-space PMT system that detects the LS scintillation light must be sufficient to attenuate gamma rays from PMT radioactivity (these are not low-background models). A distance of 80 cm from the scintillator tanks to the PMTs reduces the rate from this source to less than 5 Hz.

The tightest specifications on radioactive impurities in the GdLS arise from considerations of deadtime caused by the outer detector system. Asking that the false veto probability not exceed 1% over 4 neutron capture times results in requirements of <1.7 ppt U, <3.2 ppt Th, <0.6 ppt $^{40}$K, and <2 × $10^{-18}$ g/g in $^{14}$C. While LAB itself generally exceeds this requirement, the additives must be purified somewhat more completely than has been achieved in the reactor neutrino experiments. The purity achieved by the Borexino Collaboration greatly exceeds the LZ requirements for U/Th/K. Borexino demonstrated a $^{14}$C impurity slightly above that needed for LZ [71]. Chapter 7 details the implementation and performance of the outer detector more fully.

## 3.6 Internal Calibration with Dispersed Sources

The physics of self-shielding allows LZ to achieve its unprecedented sensitivity by reducing the rate of gamma-ray scatters in the energy range of interest to a level of secondary importance. Arguably, this is the central feature of the LZ detector design. Conversely, the same effect presents a challenge for a calibration program based solely on external gamma sources such as $^{137}$Cs. Such a calibration campaign may be useful for monitoring the edges of the detector and calibrating at higher energies, but will be of limited utility for probing the detector response to low energy interactions in the fiducial volume.

To take full advantage of a monolithic dark-matter detector such as LZ, it is necessary to implement internal calibration methods that can temporarily defeat the self-shielding effect. The development of such sources has been accomplished through a global R&D effort with major contributions from LZ scientists (e.g., [72]), and LUX is the first working experiment to rely primarily on internal calibration sources as the workhorses of its calibration campaign.

Two internal sources have been successfully deployed within LUX: $^{83m}$Kr, a source of 9.4 keV and 32.1 keV conversion electrons, separated in time by an average of 154 ns; and tritium ($^{3}$H), a $\beta^-$ emitter with a $Q$-value of 18.6 keV.

The radioisotope $^{83m}$Kr has a half-life of 1.8 hours, and is ideal for calibrating the spatial response of LUX to S1 and S2, for monitoring the free-electron lifetime, and for setting the energy scale in both response channels. It has also been useful for imaging the fluid flow in the LUX detector. (However, note that the scintillation yield of the second [9.4 keV] electron is affected by the ionization left behind by the first transition [31].) LUX deployed $^{83m}$Kr on average about twice per week throughout the physics running period, and these calibration data sets have been the primary method for monitoring the detector response during its first WIMP search run. LZ will capitalize on this success.

While the $^{83m}$Kr conversion electrons are ideal sources for routine monitoring of S1 and S2, they are less useful for measuring the ER discrimination factor in the energy range of interest (<6.5 $keV_{ee}$). Tritium, however, emits a beta particle with a maximum energy of 18.6 keV and a most probable energy of only 3 keV, allowing the discrimination efficiency to be determined over the full WIMP-search energy range, including probing the detector threshold at low energy. Tritium has been deployed with good results in LUX in the chemical form of tritiated methane ($CH_3T$). The primary challenge presented by tritium is its long half-life (12.3 years), which necessitates that the source must be removed by active purification. The effectiveness of the purification was carefully studied by LUX in bench-top tests, and those good results culminated in the decision to proceed with a $CH_3T$ injection in the summer of 2013. The result of this injection was shown previously in Figure 3.4.1: This confirms that the ER band is uniformly populated down to the lowest energies, with a source of ERs that allows calibration of the intrinsic bandwidth — as opposed to gamma rays, which interact resonantly with bound atomic electrons. The experience gained through the LUX calibration program is driving the design and implementation of the LZ effort.

## 3.7 Xenon Purity for Detector Performance

The fluid nature of LXe provides an opportunity to manipulate the purity of the LZ target material. The previous section describes how this allows internal calibration sources to be temporarily introduced into the LXe. This section considers how the purity can be maximized for low-background physics running. We consider two classes of impurities: the radioactive noble gases $^{85}$Kr and $^{222}$Rn — although the latter is addressed fully only in Chapter 12 — and electronegative contaminants such as oxygen and water.

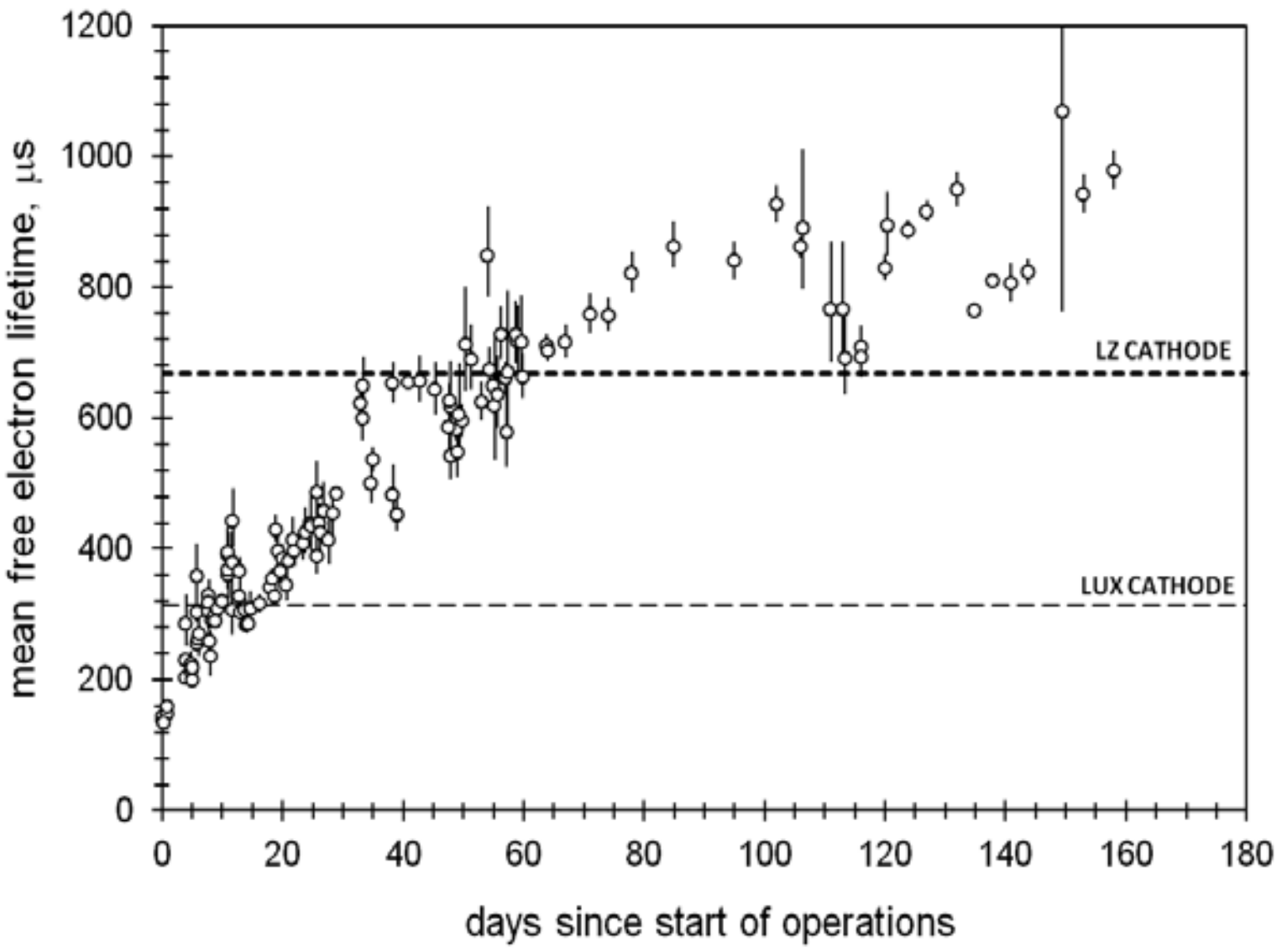

**Figure 3.7.1. Evolution of the mean free electron lifetime in LUX from the start of underground operations. The electronegative purity required to observe signals from the LZ cathode (~670 µs for the 147-cm long TPC) has already been achieved in LUX.**

Electronegative impurities are introduced during operations by the outgassing of detector materials, and they must be continuously suppressed to the level of ~0.1 ppb by the purification system to ensure good charge and photon transport. Previous detectors such as ZEPLIN-III achieved this through clean construction techniques with low-outgassing materials (i.e., no plastics), which allowed these systems to maintain purity without recirculation [73]. In larger detectors such as LUX, the need to utilize large volumes of PTFE for the reasons outlined above demands active recirculation in the gas phase. The efficient removal of electronegative contaminants is made possible by purification technology developed for the semiconductor industry. The central elements of this technology are a heated zirconium getter for the removal of non-noble species, UHV-compatible plumbing and instrumentation, and a gas-circulation pump.

Despite the availability of this technology, until recently the achievement of good electronegative purity in a LXe TPC was considered a significant technical challenge. What was lacking was an economical and sensitive monitoring technique to allow the purification technology to be fully exploited. While free electron lifetime monitors for LXe were developed over 20 years ago, these devices cannot identify individual impurity species, are insensitive to noble gas impurities, and provide little guidance on the origin of any impurities.

In the past few years, however, LZ scientists have developed a mass spectrometry method that allows most electronegative and noble gas impurities to be individually monitored in real time [24]. Most crucially, the method provided, for the first time, information on the impurity source. For example, the presence of an air leak introduces $N_2$, $O_2$, and Ar in a characteristic ratio, while a large excess of $N_2$ is a signature of a saturated purifier. Outgassing, on the other hand, leads to a uniform rate of increase of all common impurity species.

LZ scientists have gained experience with the mass spectrometry method by applying it not only to LUX, but also to the EXO-200 $0\nu\beta\beta$-decay experiment [74]. Both experiments achieved their purity goals with relative ease due in part to the effectiveness of this program, and this valuable experience can be brought to bear on LZ. We have learned, for example, that vendor-supplied Xe is often relatively pure of electronegatives; that zirconium getters are effective for Xe purification but that their performance degrades at higher gas flow rates; and that purifier performance improves at elevated temperature. This experience is of direct relevance for the design of the LZ purification system. This experience is bearing fruit in LUX, where electron lifetimes of the order of 1 ms have been demonstrated, as shown in Figure 3.7.1. This is already sufficient to drift charge from interactions near the LZ cathode.

An additional benefit is that the method is sensitive not only to electronegatives but also to trace quantities of noble gas impurities, which is of critical importance for control of krypton. Krypton is a particularly dangerous impurity for LZ because of the presence of the beta emitter $^{85}$Kr. This isotope, whose abundance at present is $\sim 2 \times 10^{-11}$ ($^{85}$Kr/$^{nat}$Kr), presents the leading purification challenge for LZ because its noble nature makes it impervious to the zirconium getter technology. Vendor-supplied Xe typically contains about 100 ppb of $^{nat}$Kr, which, if left untreated, would give rise to an $^{85}$Kr beta decay rate of 29 mBq per kilogram of Xe.

Previous experiments had difficulty achieving both their free electron lifetime and krypton-removal goals. These tasks are made more difficult by a lack of tools to monitor the performance of krypton-removal systems and an inability to identify and isolate leaks during detector operation. On the other hand, by applying advanced mass spectrometry methods, LUX was able to identify, isolate, and correct small problems with its krypton-removal system; confirm that its average krypton concentration was suitable prior to physics running; and has continuously monitored $O_2$, $N_2$, Ar, and Kr during physics operations to ensure that no leaks are present. This led directly to a successful WIMP search run for LUX.

LZ will capitalize on the success of EXO-200 and LUX by integrating sensitive monitoring into its Xe handling program from the time of Xe procurement until the conclusion of the experiment. In addition, because the origin of impurities is now understood to be primarily due to outgassing, a comprehensive purification plan can be developed and implemented to ensure that LZ achieves its performance goals. In fact, we have already carried out an extensive materials outgassing screening program to aid in developing such a plan. This allows us to develop an outgassing budget for the experiment and to design an appropriate purification system to mitigate it.

While sensitive impurity monitoring will lay the groundwork for the LZ Xe purification program, the heart of the program will be krypton removal using the chromatographic technique developed at Case Western by LZ scientists [23]. This program has been successfully applied to LUX and will be scaled up in mass-throughput by a factor of 10 for LZ. The krypton concentration goal is 0.02 ppt (g/g), a factor of 200 below the LUX goal. In parallel, the krypton detection limit of the mass spectrometry method will be improved more than tenfold from the currently demonstrated value of ~0.3 ppt (g/g). The ultimate source of krypton impurities in LZ will be the outgassing of detector materials, which we can control through careful materials selection and through our outgassing plan.

Similarly, $^{222}$Rn must be controlled by limiting the emanation sources within the detector and the gas system via a careful screening program. In fact, the need to limit radon and krypton in the LZ Xe will be a driving consideration in the choice of key Xe system components such as the gas-recirculation pump. More generally, the materials-screening program must work hand-in-hand with the purification effort, as the detector materials are the ultimate irreducible source of all classes of impurities.

## 3.8 Dominant Backgrounds

LZ has a clear background-control strategy with optimal exploitation of self-shielding to pursue an unprecedented science reach. These are the most salient features of this strategy: (1) underground operation within an instrumented water tank to mitigate cosmogenic backgrounds; (2) deployment of a target mass large enough to self-shield external radioactivity backgrounds, working in combination with outer, anticoincidence detectors; (3) construction from low-activity materials and purification of the target medium to render intrinsic backgrounds subdominant; and (4) rejection of the remaining ER backgrounds by S2/S1 discrimination.

The dominant radioactive backgrounds come from the Xe-space PMTs and their bases, field cage structures including the PTFE reflectors, and the cryostat vessels. Our goal is to self-shield those sources over the first few centimeters of active liquid and thereby be sensitive to a population of ERs caused by elastic scattering of solar pp neutrinos — which can be further discriminated by their S2/S1 ratio. An irreducible but very small background of NRs will also arise from coherent neutrino-nucleus scattering. To achieve this, we must render negligible any intrinsic ER backgrounds contained within the Xe itself, with $^{85}$Kr, $^{39}$Ar, and Rn progeny being a particular challenge. Electron recoils caused by the $2\nu\beta\beta$ decay of $^{136}$Xe, now confirmed by the EXO-200 [75] and KamLAND-Zen [76] experiments, are subdominant below about 20 $keV_{ee}$. Cosmogenic (muon-induced) backgrounds are not significant during operation due to the tagging capability of the instrumented water tank and the scintillator veto, but cosmogenic activation of detector materials prior to deployment (including the Xe) must be addressed.

We describe briefly some of these background categories here (radioactivity external to the TPC, intrinsic contamination of the LXe, and cosmogenic backgrounds), as these are important LZ sensitivity drivers, but we postpone a full discussion on their mitigation to Chapter 12. Neutrino backgrounds, which determine the LZ sensitivity to first order, are explored further in Chapter 4. The solar pp neutrino scattering rate of $0.8 \times 10^{-5}$ events/kg/day/$keV_{ee}$ is the benchmark against which other rates are assessed.

### 3.8.1 Backgrounds from Material Radioactivity

Radioactivity backgrounds have limited nearly all dark-matter experiments so far and, in spite of the power of self-shielding, we are not complacent in addressing them in LZ. They impact the thickness of the sacrificial layer of LXe that shields the fiducial mass, and they may cause rare-event topologies that may be of consequence (from random coincidences, atypical surface interactions, or Cherenkov emission in PMT glasses, for example). It is important, therefore, to minimize the rate of $\alpha$, $\beta$, and $\gamma$ activity around the active volume, as well as neutron production from spontaneous fission of $^{238}$U and from $(\alpha,n)$ reactions. In addition, the rate and spatial distribution of such backgrounds must be well characterized to build an accurate background model for the experiment. The LZ background model is derived from a high-fidelity simulation of the experiment in the LUXSim framework mentioned previously, which was successfully used for the LUX background model [77].

All materials to be used in LZ will be subject to stringent constraints as part of the comprehensive screening campaign described in Chapter 12, with 10% of the solar pp neutrino scattering rate and a maximum of ~0.2 NRs at 50% signal acceptance being the target for the total contribution from material radioactivity within the fiducial volume. The dominant rates come from the various PMT systems and the LZ cryostat, based on the large masses and close proximity to the active region of the detector. Table 3.8.1.1 summarizes contamination and count rates from neutron and gamma-ray emission expected from detector materials and other backgrounds, which are discussed in greater detail in Chapter12.

The PMTs chosen for the LZ TPC are Hamamatsu R11410s, which have achieved very low radioactivity values; LZ scientists have been involved in a long campaign to establish their performance for dark-matter experiments, working actively with the manufacturer to enable this [21]. The PTFE required to fabricate the TPC field cage, skin reflectors, and other internal components may also be an important source of neutron emission from the bulk material. We use upper limits on the contamination measured by EXO-200 in calculating its impact [78]. The cryostat is another dominant component, mostly owing to its large mass. For the titanium baseline design (2,140 kg), the total neutron emission rate is estimated at 0.6 n/day based on recent titanium samples procured for LZ. As a result of a 2-year material search campaign, we were able to find titanium with U/Th contamination, which is a factor of 2 lower than that used in LUX [22] as explained in Chapter 8.

**Table 3.8.1.1. Summary of backgrounds in LZ, showing radioactivity levels for some dominant components, their neutron emission rates, and the number of counts expected in 1,000 live days in an indicative 5.6-tonne fiducial mass between 1.5–6.5 keVee (ER) and 6–30 keV (NR). A comprehensive list can be found in Table 12.2.2.**

| Item | Mass kg | U mBq/kg | Th mBq/kg | $^{60}$Co mBq/kg | $^{40}$K mBq/kg | n/yr | ER cts | NR cts |
|---|---|---|---|---|---|---|---|---|
| R11410 PMTs | 93.7 | 2.7 | 2.0 | 3.9 | 62.1 | 373 | 1.24 | 0.20 |
| R11410 bases | 2.7 | 74.6 | 29.1 | 3.6 | 109.2 | 77 | 0.17 | 0.03 |
| Cryostat vessels | 2,140 | 0.09 | 0.23 | ≈0 | 0.54 | 213 | 0.86 | 0.02 |
| OD PMTs | 122 | 1,507 | 1,065 | ≈0 | 3,900 | 20,850 | 0.08 | 0.02 |
| Other components | - | - | - | - | - | 602 | 9.5 | 0.05 |
| **Total components** | | | | | | | 11.9 | 0.32 |
| **Dispersed radionuclides (Rn, Kr, Ar)** | | | | | | | 54.8 | - |
| **$^{136}$Xe 2νββ** | | | | | | | 53.8 | - |
| **Neutrinos (ν-e, ν-A)** | | | | | | | 271 | 0.5 |
| **Total events** | | | | | | | 391.5 | 0.82 |
| **WIMP background events (99.5% ER discrimination, 50% NR acceptance)** | | | | | | | 1.96 | 0.41 |
| **Total ER+NR background events** | | | | | | | 2.37 | |

### 3.8.2 Surface Plating of Radon Progeny

The noble gas radon consists solely of radioactive isotopes, of which four are found in nature: $^{222}$Rn and $^{218}$Rn produced in the $^{238}$U decay chain, $^{220}$Rn from the $^{232}$Th decay chain, and $^{219}$Rn from the $^{228}$Ac series. As a result of its chemical inertness, radon exhibits long diffusion lengths in solids. $^{222}$Rn is the most stable isotope ($T_{1/2}$ = 3.82 days), and is present in air at levels of about ten to hundreds of Bq/m$^3$. Charged radon progeny — especially metallic species such as $^{218}$Po — plate out onto macroscopic surfaces that are exposed to radon-laden air. A fraction will deposit and even implant into material surfaces during detector construction or installation [79].

Backgrounds from surface beta and gamma radioactivity, as well as recoiling nuclei (e.g., $^{206}$Pb from the alpha decay of $^{210}$Po), are largely mitigated by short half-lives and the self-shielding of LXe. However, α-particles released in the decay chain, particularly from $^{210}$Po — a granddaughter of the long-lived $^{210}$Pb ($T_{1/2}$ = 22.3 years) — result in neutron emission following (α,n) reactions. This is problematic for TPC materials with large (α,n) yields such as PTFE ($10^{-5}$ n/α, due to the presence of fluorine). Additionally, because PTFE is produced in granular form before being sintered in molds, plate-out poses further risk because surface contamination of the granular form becomes bulk contamination when the granules are poured into molds.

A second concern relates to our ability to correctly reconstruct surface interaction at the TPC inner walls, since the imperfect reconstruction of these events leads to a background population leaking radially toward the fiducial volume [4,9]. This concern drives the design of the top PMT array and places tight requirements on the plate-out of radon progeny on the TPC walls (see Section 6.5.3).

Controls to mitigate background from radon plate-out will include limiting the exposure of detector parts to radon-rich air; monitoring from point of production through transport and storage in Rn-proof materials; and employing surface cleaning techniques, such that neutron emission is negligible relative to material radioactivity from bulk uranium and thorium contamination.

### 3.8.3 Intrinsic Backgrounds

We are confident that our challenging goals for intrinsic radioactive contamination from $^{85}$Kr and $^{222}$Rn can be met with the Xe-purification techniques described in Chapter 9, coupled with the radon emanation screening of Xe-wetted materials described in Chapter 12. We note that most of these backgrounds can be estimated with low systematic uncertainty. In addition to direct sampling, the $^{85}$Kr $\beta^-$ decay spectrum is well understood and the decay rate can be measured during operation with delayed $\beta$–$\gamma$ coincidences.

Other delayed coincidence techniques as well as $\alpha$ spectroscopy allow precise estimation of radon-induced backgrounds. In fact, it is possible to follow dynamically the spatial distribution of these decays throughout the detector, which was done very successfully in LUX [80]. The two main concerns in this instance are a "weak" naked beta decay from $^{214}$Pb in the bulk of the TPC ($E_{max}$ = 1,019 keV, BR = 11%), and the possibility of gamma-ray escape for peripheral events from the dominant $^{214}$Pb decay modes.

Our goal is to control each of these two backgrounds to <10% of the solar pp neutrino rate, limiting the $^{222}$Rn activity to 0.67 mBq and the krypton concentration to 0.02 ppt (g/g). In a more conservative scenario, we allow the sum of these two components to match the ER background from pp neutrinos.

Trace quantities of argon are also a concern due to $\beta$-emitting $^{39}$Ar, with a 269-year half-life and 565 keV endpoint energy. This background is constrained to be less than 10% of $^{85}$Kr, resulting in a specification of $4.5 \times 10^{-10}$ (g/g) or 2.6 μBq. The Kr-removal system, which also removes Ar, should easily achieve this.

### 3.8.4 Cosmogenic Backgrounds

A rock overburden of 4,300 mwe above the Davis Cavern at SURF reduces the muon flux by a factor of $3.7 \times 10^6$ relative to the surface [81]. Muons crossing the water tank are readily detected via Cherenkov emission, and any coincident energy deposition in LZ is similarly easily identified. However, muon-induced neutron production through spallation, secondary spallation, or photonuclear interactions by photons from muon-induced EM showers in high-Z materials may generate background [82,83]. The total muon-induced neutron flux at SURF from the surrounding rock is calculated to be $0.54 \times 10^{-9}$ n/cm$^2$/s, with approximately half of this flux coming from neutrons above 10 MeV, and some 10% from energies above 100 MeV [84]. This neutron rate is considerably lower than that from internal component radioactivity.

Muon-induced neutrons generated in the water shield produce a similarly low rate, despite the several hundred tonnes of target mass, due to the low atomic number of water and consequent low neutron yield (~$2.5 \times 10^{-4}$ n/μ/(g/cm$^2$), translating to a production rate of order $10^{-9}$ n/kg/s). The water itself attenuates the flux on the cryostat to ~0.2 n/day, a small contribution compared with that from the cryostat itself.

Cosmogenic activation — radioisotopes production within materials, largely through spallation reactions of fast nucleons from cosmic rays while on the Earth's surface — can present electromagnetic background in LZ. $^{46}$Sc produced in the titanium cryostat decays through emission of 889 keV and 1120 keV gamma rays (with $T_{1/2}$ = 84 days). Cosmogenic $^{60}$Co ($T_{1/2}$ = 5.3 years) in copper components will produce gamma rays of 1,173 keV and 1,332 keV.

Activation of the Xe itself during storage or transport generates several radionuclides, some of which are important, especially in the first few months of operation. Tritium ($T_{1/2}$ = 12.3 years and production rate of ~15/kg/day at the Earth's surface [85]) was previously a concern; however, this is effectively removed through purification during operation. Production of Xe radioisotopes, such as $^{127}$Xe ($T_{1/2}$ = 36.4 days), $^{129m}$Xe ($T_{1/2}$ = 8.9 days), and $^{131m}$Xe ($T_{1/2}$ = 11.9 days), are more problematic, as they cannot be mitigated through self-shielding or purification. In particular, atomic de-excitation of the 2s and 3s shells in $^{127}Xe$ generates 5.2 and ≤1.2 keV energy deposits, respectively, which are an important background in the WIMP search energy region for certain event topologies [4]. Since $^{127}Xe$ is produced efficiently through neutron resonance capture, shielding against thermal neutrons is desirable even when stored underground. Furthermore, these backgrounds soon reach negligible levels once underground operation starts.

### 3.8.5 Fiducialization

To assess the impact of all sources of background in LZ, it is sensible to define a fiducial volume, even if only approximately at this stage. A rigorous definition will involve sophisticated statistical analyses that take into account their measured spatial distribution. However, for the purpose at hand as well as for reasons of design and engineering, it is sensible to know the approximate location of such a volume.

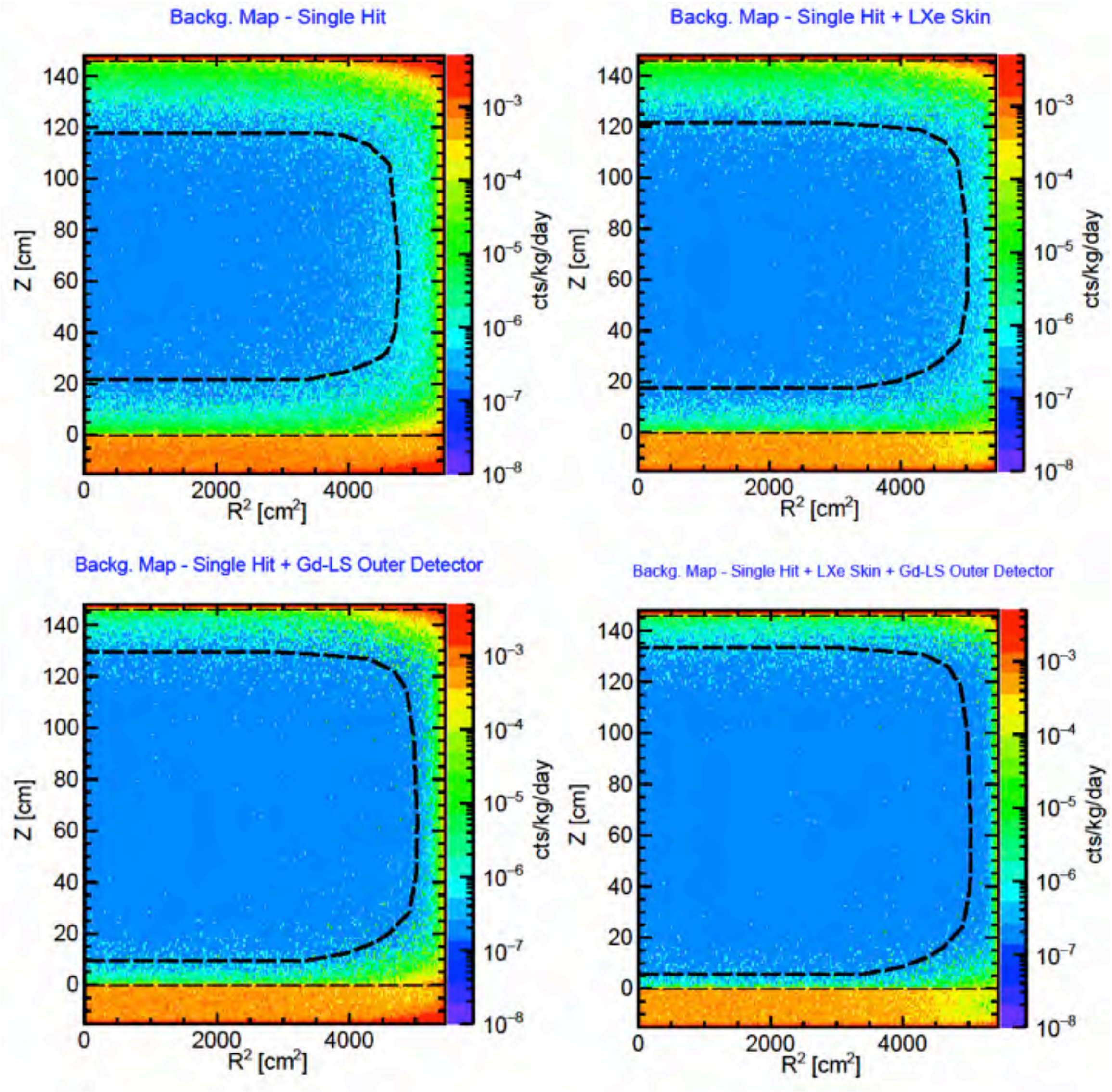

**Figure 3.8.5.1. Total NR background plus ER leakage from sources external to the LXe in the TPC, counted over a 6–30 keV acceptance region; a discrimination efficiency of 99.5% is applied to ERs from gamma rays and solar pp neutrinos. Top left: Single scatters only, no vetoing by the anti-coincidence systems. Top right: Adding LXe skin only. Bottom left: Adding GdLS outer detector only. Bottom right: Adding the combination of both vetoes. In each panel, approximate fiducial contours are shown in the black dashed line and weigh 3.3, 4.2, 5.1, and 5.6 tonnes, respectively.**

Two main factors determine how far the fiducial boundary should lie from the lateral TPC walls. The prime consideration is to ensure a sufficiently thick layer of LXe to self-shield against the external radioactivity backgrounds; this is related to the mean attenuation length for those particles: Figure 3.2.2 confirmed that ~2 cm of liquid decreases the gamma-ray background tenfold, and as much as ~6 cm is needed to mitigate neutrons by the same factor. However, the outer detector is very efficient for neutron tagging, which brings these two requirements closer together.

Secondly, it is essential that the reconstructed (*x*,*y*) positions of low-energy interactions occurring near the TPC walls do not "leak" into the fiducial volume. As mentioned above (and discussed in Chapter 6) interactions from radon progeny plating the lateral PTFE are of particular concern: These can generate events with very small S2 signals due to trapping of charge drifting too close to the PTFE. If allied with poor position resolution, this can constitute a very challenging background [9]. In the vertical direction, only the former consideration arises. We point out that the reverse field region below the cathode will provide much of the required self-shielding (>14 cm), whereas at the top, the small thickness of liquid above the gate (0.5 cm) will have a limited impact.

Figure 3.8.5.1 shows the simulated background rate from material radioactivity in the WIMP region of interest (6–30 keV) as a function of radius squared and height above the cathode grid. Nuclear and electron recoil backgrounds were combined, with 50% acceptance applied to the former and 99.5% discrimination applied to the latter. The neutrino contributions listed in Table 3.8.1.1 were included. TPC events with additional sub-cathode vertices (gamma-X) are not rejected, making this a conservative assessment (see brief discussion in Section 3.4). In each panel, we show the number of background counts per kg per day, with the dashed black line indicating the position of an approximate fiducial volume. This is drawn so as to constrain the ER background from radioactivity to one-tenth of the solar pp neutrino rate, and separately to constrain the NR background from radioactivity to ~0.2 events in 1,000 days. The four panels are for single hit interactions with no further vetoes, then adding either the skin or the outer detector, then combining the two anti-coincidence systems. In the first case (top left), with no additional vetoes applied, the fiducial cut is 15 cm away from the wall, retaining a fiducial mass of 3.3 tonnes.

The final panel illustrates what can be gained by our dual "veto" strategy. Using the combination of skin and outer detectors (both with 100 $keV_{ee}$ threshold) improves the estimated fiducial mass to 5.6 tonnes. In this instance, the lateral cut is 4 cm away from the wall, which is also a reasonable distance to allow good position reconstruction for wall events and to avoid electric field non-uniformity near the field-cage surface. The total estimated background count in this fiducial volume in a 1,000-day data set is ~2.4 events, as indicated in Table 3.8.1.1.

## Chapter 3 References

# 4 LZ Sensitivity

The LZ detector system described in previous and subsequent chapters is highly sensitive to a variety of physics signals. The principal signal we seek is that of NRs distributed uniformly throughout the LXe TPC volume, in response to an impinging flux of nonrelativistic galactic WIMPs. In the first sections of this chapter, we describe the sensitivity to various WIMP-particle cross sections.

The first step in selecting the sample of WIMP candidates is to define a suitable search region in the two variables: S1 (the prompt scintillation light) and S2 (the delayed electroluminescence light, a measure of primary ionization). The use of both variables enables the separation of NRs from the much more numerous ERs, resulting in a search region that is nearly or completely background-free for a multitonne Xe fiducial mass. The second step is to remove from this sample all events associated with time-coincident energy deposit in the Xe skin or the outer detector.

We first describe our sensitivity and discovery potential for the spin-independent (SI) interpretation of the WIMP-nucleon interaction. We then discuss interpretations involving more general forms of the WIMP-nucleon interaction.

The S2 signal is sensitive to smaller energy depositions than is S1 and, as described in Chapter 3, NR thresholds <1 keV can be sensed in the ionization channel due to its sensitivity to individual electrons emitted from the liquid. Use of the S2 signal alone, when the NR is too feeble to cause a detectable S1, can provide enhanced sensitivity to WIMPs of the lowest mass, which cause the softest NR spectra. Use of the S2 signal alone removes the capability to distinguish NRs from ERs, and consequently the ERs provide a substantial and irreducible background to the "S2-only" analysis. We note, however, that this is true of other WIMP-search technologies (e.g., p-type point-contact germanium detectors) that are also searching for light WIMPs. We estimate the limiting sensitivity of an "S2-only" analysis of LZ data.

Should LZ see a WIMP signal, the distribution of that signal in NR energy will allow constraints on the WIMP-Xe scattering cross section, the WIMP-Xe reduced mass, and on the velocity distribution of galactic WIMPs [1].

A variety of other physics processes can be probed by selective detection of NRs and ERs as defined with S1 and S2. The central fiducial region of the LZ detector will be an extraordinarily quiet laboratory for processes that deposit energy. Among the physics processes that can be probed are:

1. Interactions of WIMPs with atomic electrons.
2. Solar and certain dark-matter axion-like particles (ALPs), which interact via the axioelectric effect.
3. Solar neutrinos emitted by the pp fusion process in the sun.
4. Neutrinos emitted by a nearby supernova and detected by coherent neutrino-nucleus scattering.
5. Neutrinoless double-beta decay of $^{136}$Xe.
6. Neutrino oscillations with parameters motivated by the current "reactor/source anomalies," and a neutrino magnetic moment.

We also present a summary of top-level requirements at the end of this chapter. We have a well-developed process to capture and flow down science requirements. The dependencies of key requirements on critical performance characteristics are also summarized.

## 4.1 WIMP Sensitivity and Discovery Potential

The principal physics analyses of the LZ experiment will be searches for the recoils of Xe atoms caused by the interaction of WIMPs with the Xe nucleus. As discussed above, two types of signal are formed in the LXe response to the recoils: S1 and S2. In the principal LZ search, the energy of the recoil is reconstructed from a combination of S1 and S2, and the ratio S2/S1 provides discrimination of NRs from the background of ERs. The value of the reconstructed energy depends on whether the event is an NR or an ER.

An auxiliary LZ search for NRs exploits the S2 signal alone (the S2-only analysis), which is more sensitive to energy deposits than is the S1 signal. The S2-only analysis provides additional sensitivity to small energy deposits from low-mass WIMPs, at the cost of the ability to discriminate against the ER background.

### 4.1.1 S1+S2 Analysis

The S1+S2 analysis in LZ will follow the general framework of the published first LUX search for NRs in response to WIMPs [2]. The experimental details that influence the analysis were discussed in Chapter 3. We define a search region in the plane of log(S2/S1) versus S1, shown in Figure 4.1.1.1. The definition of the LZ baseline search region in this plane is described in Table 4.1.1.1.

The baseline detector performance assumed for LZ is in most cases more conservative than that achieved by LUX. The most prominent exception in Table 4.1.1.1 is the liquid/gas emission probability, where we presume that the limitations of the LUX electric field will be removed in the LZ experiment. In practice, the NR acceptance exceeds 50% for the lowest-energy ERs, but we make a conservative assumption that the acceptance is 50%.

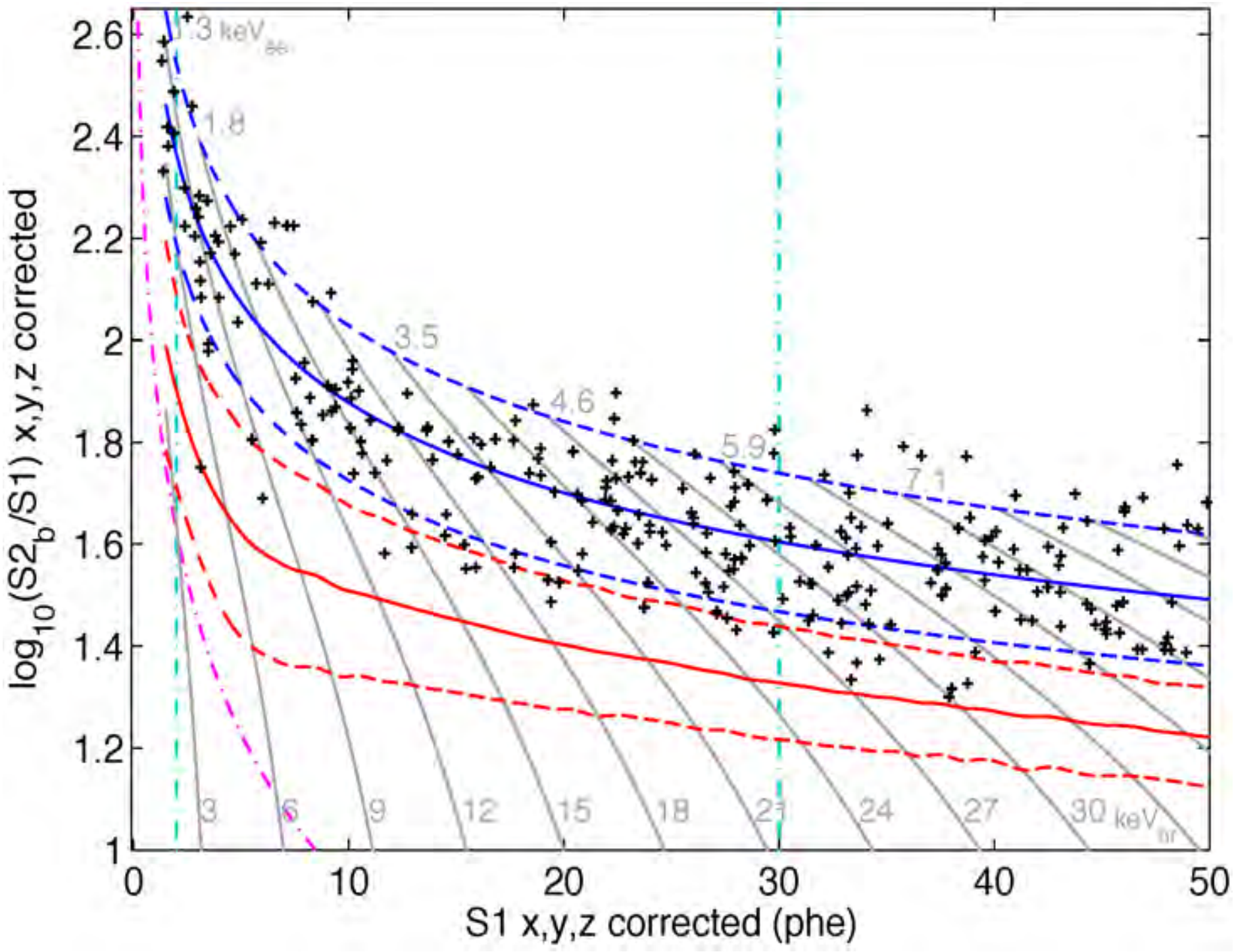

**Figure 4.1.1.1. The LUX WIMP search data [2]. The logarithm of the ratio S2/S1 is plotted versus S1, after spatial corrections. The centroid (solid) and search region boundaries (dotted) are red for the signal (NR) region or "band," and corresponding lines in blue describe the primary background (ER) band. The dotted lines are $\pm 1.28\sigma$ around the centroid. Contours of equal recoil energy for NR ($keV_{nr}$) and ER ($keV_{ee}$) interpretations are shown in grey. The LUX data, consistent with a background of ERs, is shown, and the LUX NR search region is between the vertical light-blue dot-dash lines and the solid red and dashed red lines.**

**Table 4.1.1.1. Comparison of the key performance assumptions for LZ compared to published values for LUX.**

| Quantity | Units | LZ Assumption | LUX [2] |
|---|---|---|---|
| **Recoil threshold, 50% efficiency** | $keV_{nr}$ | 6 | 4.3 |
| **Maximum recoil energy** | $keV_{nr}$ | 30 | n/a |
| **S1 range** | Detected photoelectrons | 3-30 | 2-30 |
| **S2 range** | Detected photoelectrons | >250 | >200 |
| **S1 light-collection efficiency** | Absolute | 7.5% | 14% |
| **Photocathode efficiency** | Absolute | 25% | 30% |
| **Liquid/gas emission probability** | Absolute | >95% | 65% |
| **ER discrimination** | Absolute | 99.5% | 99.6% |
| **NR acceptance assumed for sensitivity estimation** | Absolute | 50% | 50% |

The benchmark process we will use to interpret NRs will be the interaction of WIMPs via an SI process, such as exchange of a Higgs particle [3], with the gluons in the nucleons in the Xe nucleus [4]. This process produces a WIMP-nucleus scattering rate that is independent of the identity, neutron or proton, of the nucleon in the nucleus. For the low-momentum transfers of typical WIMP interactions, the scattering amplitude is proportional to A, the number of nucleons in the nucleus. The scattering cross section includes the density of states, which also favors larger A, while the threshold for energy detection favors smaller A. The nuclear form factor is employed to account for quantum-mechanical interference attributable to the non-zero nuclear size [6], and the standard halo model (SHM) of the distribution of WIMP velocities in the Milky Way is used [7].

The backgrounds expected for LZ are described in detail in Chapter 12 and summarized in Table 3.8.1.1. A background arises from solar-neutrino-induced ERs that leak into the NR region. In the ER band and energy range of 1.5-6.5 $keV_{ee}$, we expect ERs from neutrinos originating predominantly in the pp fusion process in the sun, and scattering in LZ off of atomic electrons. The flux of pp neutrinos is predicted by solar models to better than 1%, assuming the solar luminosity constraint, but there is uncertainty that arises from atomic binding effects of electrons in the Xe atom. The ultimate level of ERs from solar neutrinos in the LXe TPC will be well constrained by studies of the "sideband" in ER energy extending from 6.5–20 $keV_{ee}$. For energies above 20 $keV_{ee}$, ERs from the 2ν double-beta decay from $^{136}Xe$ dominate the spectrum of ER events.

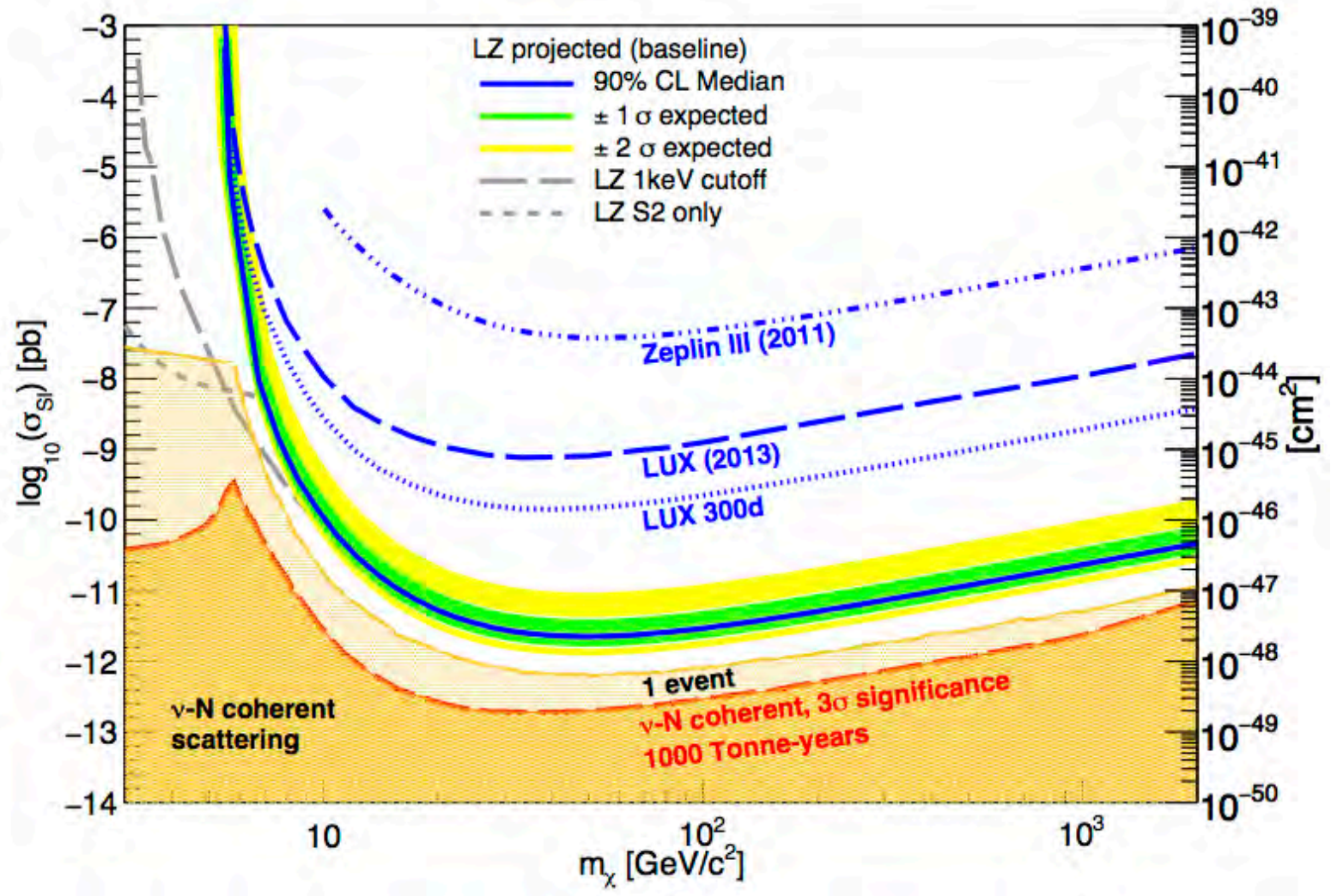

**Figure 4.1.1.2. Projected 90% confidence level (CL) sensitivity for the SI WIMP-nucleon cross sections for LZ (solid blue) along with the current world's-best limits from LUX (dashed blue), the LUX 300-day projection (dotted blue), and the final ZEPLIN result (dot-dashed blue). Regions above the curves are excluded. The green and yellow bands display the 68% (1σ) and 95% (2σ) ranges of the expected LZ 90% CL limit. The grey small-dashed line is an estimate of the 90% CL for the S2-only technique. The grey long-dashed line indicates the potential improved low-mass reach if the lower energy threshold is lowered. The regions where background NRs from cosmic neutrinos emerge, and an ultimate neutrino floor [5], are shown.**

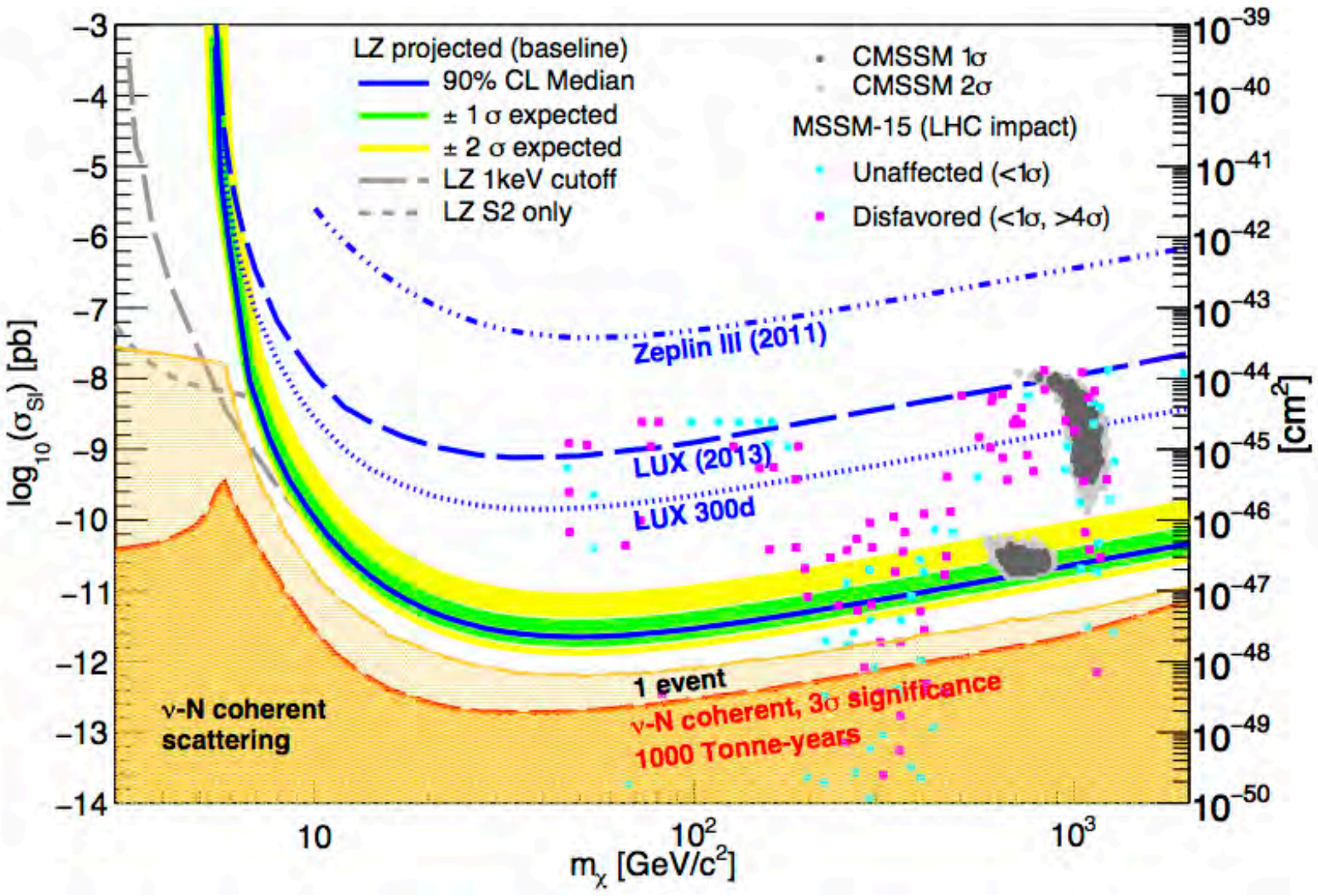

**Figure 4.1.1.3. The same, in part, information as Figure 4.1.1.2, very recent SUSY theoretical expectations included. The grey-colored regions are favored by recent scans of the five-parameter CMSSM, which include the most current constraints from LHC results [8]. The purple and blue points are pMSSM models, where 15 parameters are scanned [9]. The number of standard deviations (σ) that quantify consistency are higher for models that are more inconsistent with very recent LHC data.**

Another source of background is the as-yet-unobserved coherent nuclear scattering from atmospheric neutrinos and neutrinos from the diffuse supernova neutrino background (DSNB), which contribute events in the NR band. Nuclear recoils from the coherent scattering of solar neutrinos fall below the LZ energy threshold for the standard S1+S2 analysis.

An additional source of background is beta-decay electrons, generating ERs, emitted from $^{85}Kr$ and the $^{222}Rn$ chain. The ultimate level of radon in the TPC will be very well constrained by the measurement of alpha-particles in the radon decay chain.

We use the backgrounds described in Chapter 12 and summarized in Table 3.8.1.1 to derive the projected sensitivity of LZ to WIMPs. The resulting sensitivity plot is shown in Figure 4.1.1.2, along with the current LUX limit and the projected LUX sensitivity. The best (lowest) sensitivity shown, at a mass of approximately 50 GeV/$c^2$, is about 2 $\times 10^{-48}$ $cm^2$, which is 2 $\times 10^{-12}$ pb.

Figure 4.1.1.3 redisplays the information in Figure 4.1.1.2, but includes regions and points in the same parameter space that are consistent with very recent evaluations made with SUSY models, which have included the most recent constraints on those models by LHC experiments. A tightly constrained SUSY model with only five parameters, known as the constrained minimal supersymmetric model (CMSSM), favors WIMP masses and WIMP-nucleon scattering cross sections that are largely within the sensitivity of LZ [8]. These CMSSM models are generally out of reach of future LHC runs, but future gamma-ray telescopes could make a detection complementary to one made by LZ. Figure 4.1.1.3 also shows points that arise from an analysis of a much-less-restricted ensemble of SUSY models, the 15-parameter pMSSM [9] (there is also a 19-parameter phenomenological MSSM [pMSSM] discussed in Chapter 1). For the 15-parameter pMSSM, the recent LHC results have been largely inconsistent with many models that predict WIMP-nucleon cross sections well below the LZ sensitivity.

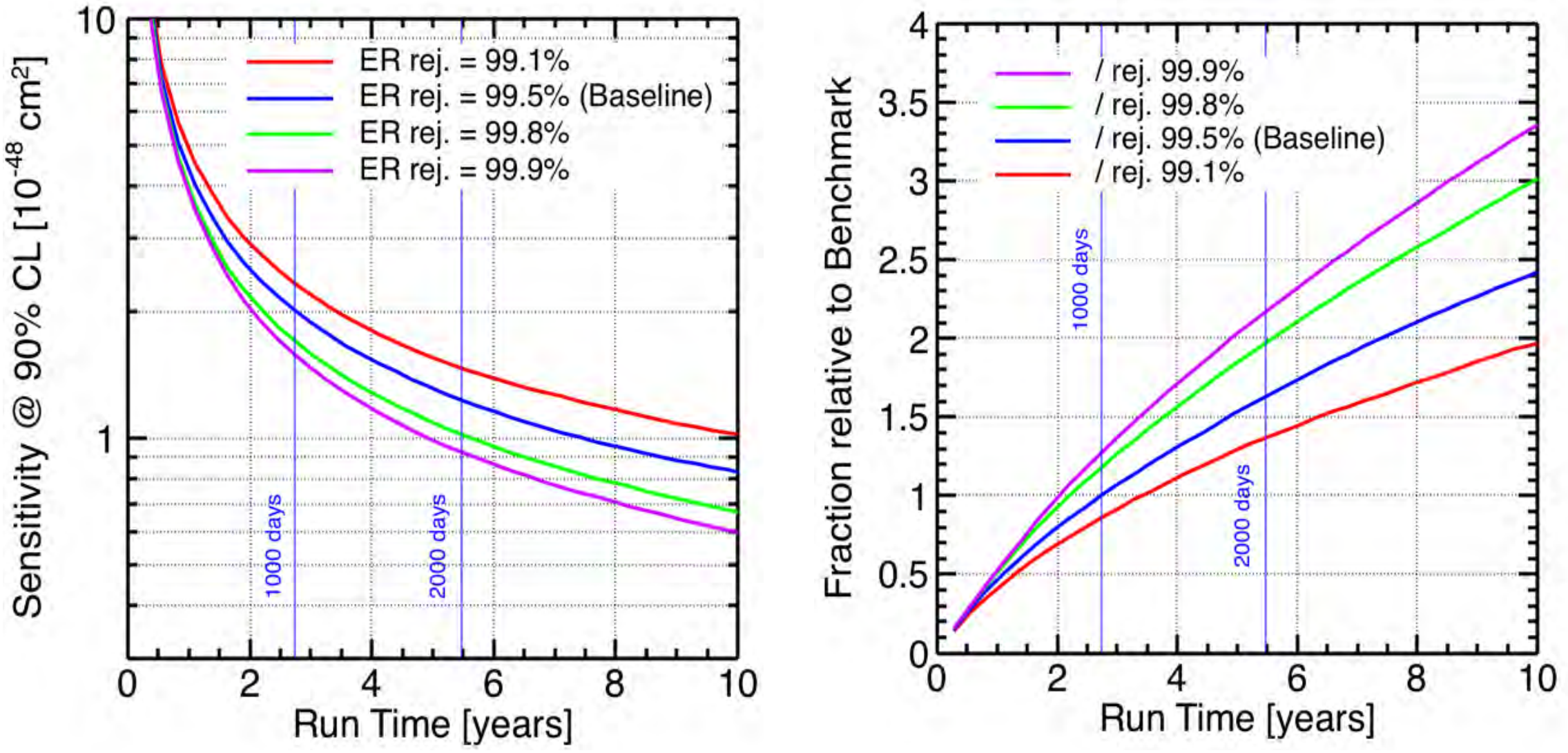

**Figure 4.1.1.4. Sensitivity as a function of running time. The left panel shows the 90% CL upper limit on the SI WIMP-nucleon cross section that can be achieved by LZ as a function of exposure in years. The baseline case is blue. The limiting background is misidentification of ERs, which originate principally from solar neutrinos. Improvement of the ER rejection permits improvement between 1,000 and 2,000 live days that is close to the inverse of the time. The right panel shows the improvement relative to the baseline case, which is normalized to 1 unit at 1,000 days exposure.**

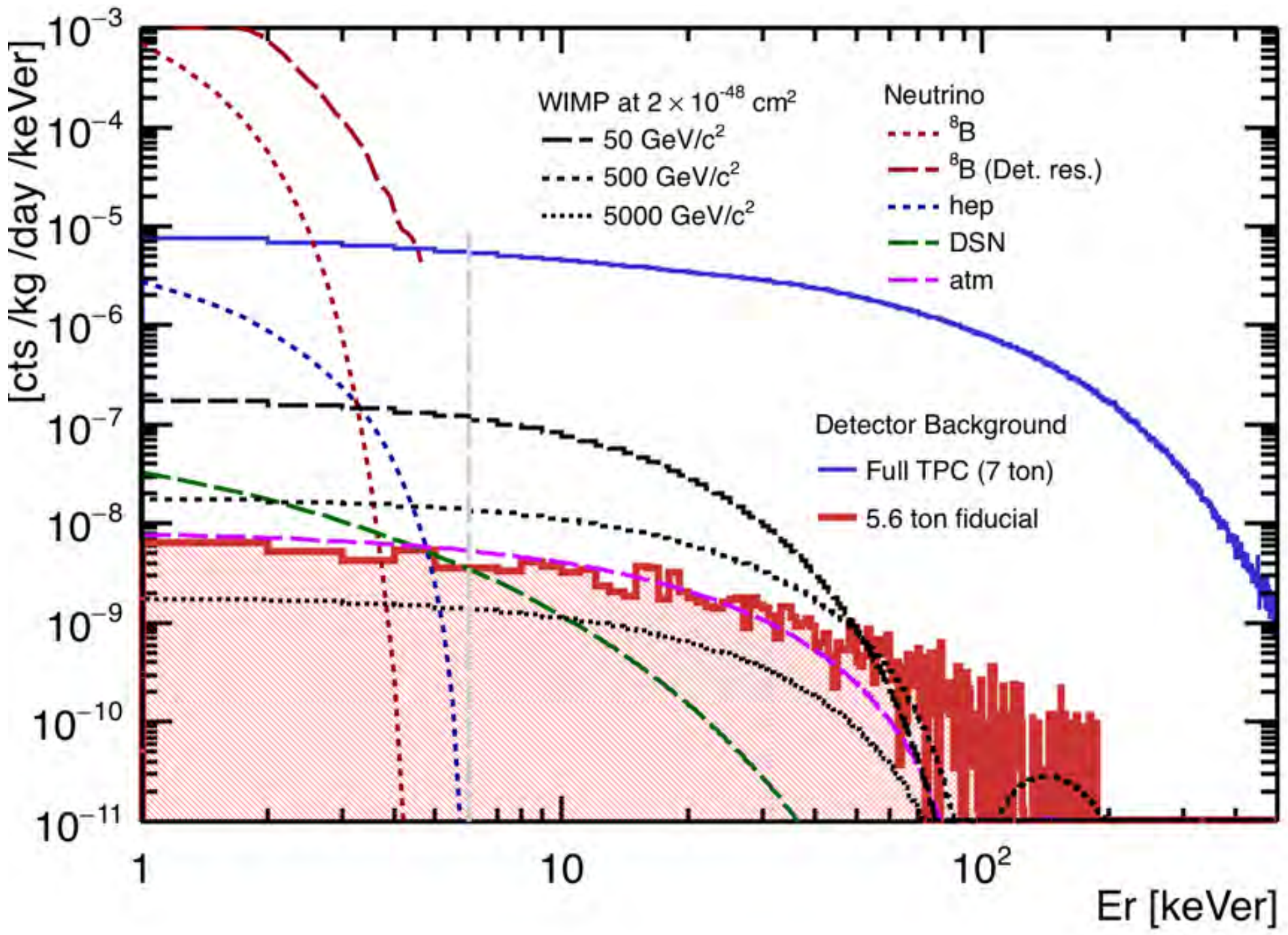

**Figure 4.1.1.5. Energy spectra of signal and backgrounds, in the NR band. The expected counts per keV$_{nr}$ per tonne per 1,000 days is shown versus the NR energy. The expected signals from WIMPs of three masses and cross sections are plotted in black. The signal expected from the coherent scattering of solar $^{8}$B neutrinos is shown in dotted brown; the dashed brown is after convolution with the expected energy resolution. The LZ threshold is shown in grey dash. Expected signals from coherent nuclear scattering of the diffuse supernova neutrino background (green) and from atmospheric neutrinos (magenta) are shown. The background from external radioactivity in the complete LXe volume is shown in blue, and the portion that survives to the 5.6-tonne fiducial volume is shown in red. The ER rejection is assumed to be the LZ baseline of 99.5%. Contributions from leakage of ERs from pp neutrinos, double-beta decay, $^{222}$Rn, and $^{85}$Kr are not shown.**

The sensitivity of LZ to SI WIMP-nucleon scattering as a function of running time of the experiment is shown in Figure 4.1.1.4, for the three different cases of ER leakage probability into the NR signal region presented in Table 4.1.1.2, and for the ER leakage probability achieved by the ZEPLIN-III experiment. The main physical difference between the LZ baseline, shown in blue, and the case of the ZEPLIN-III ER leakage probability are mainly events from ER leakage.

With twice the live time, the sensitivity at 50 GeV/c$^2$ improves by about 40%. The scaling of the sensitivity with exposure is less than linear, ultimately limited by backgrounds. If better ER discrimination than 99.5% can be achieved, backgrounds would be reduced and the sensitivity improved. With the longer exposure and the highest discrimination, the 90% CL sensitivity could become about $1 \times 10^{-48}$ cm$^2$. The energy spectra from potential signals and many of the expected backgrounds are shown in Figure 4.1.1.5 for baseline assumptions.

### 4.1.2 S2-only Analysis

The center of the LZ LXe volume is very well shielded from gamma rays and neutrons that originate from radioactive impurities outside. The rate of events in the center, measured with either S1 or S2, provides an interesting measurement of impinging particles that do not originate locally. The energy threshold for a detected S2 signal is somewhat lower than that for S1, because electroluminescence provides an amplification mechanism such that a signal electron is well above threshold. For this reason, an analysis using only S2 signals can probe lower energy deposits than one using both S1 and S2.

Because there is no discrimination between NR and ER events when the S2 signal alone is used, the background level in an S2-only search is higher than in the standard S1+S2 analysis. Nevertheless, at the smallest WIMP masses, the S2-only search can provide better sensitivity than the S1+S2 analysis, as depicted in Figure 4.1.1.2.

The S2-only analysis for Figure 4.1.1.2 uses a smaller fiducial mass of 1 tonne, and a threshold of 2.5 ionization electrons extracted from the LXe, which corresponds to 100 phe detected in the Xe PMT system. The reduction of fiducial mass is achieved by requiring the S2 pulse to fall within a radius of 40 cm.

The z-coordinate of the S2 pulse cannot be reconstructed from drift time because of the absence of the S1 pulse from which the start time is determined. However, the electrons in the S2 pulse diffuse as they drift to the cathode, and so the width of the pulse can be used to deduce the z-coordinate of the event, with a resolution of about 10 cm. The reconstruction of the z-coordinate is most effective for events with the shortest drift distance, at the top of the LXe TPC. We assume that the 40 cm closest to the top of the TPC and the 40 cm at the bottom of the TPC will be rejected, resulting in a column height of 70 cm for the fiducial volume.

The principal limitations of the S2-only analysis are instrumental backgrounds that are difficult to extrapolate from the ZEPLIN program and from LUX to LZ. Bursts of ionization electrons take place in LXe TPCs [10]. The origin of these bursts is not yet fully understood, but is likely to be related to electrons trapped on the liquid-vapor interface, vacuum ultraviolet (VUV) light emitted from electroluminescence in the vapor, phosphorescence in materials local to the TPC, and field emission from the cathode. As part of the LZ program, we plan to characterize the contributions of these phenomena. The experience in LUX has been that these bursts can be identified and removed with a loss of live time of approximately 15%.

For the estimate of the S2-only sensitivity in Figure 4.1.1.2, we have neglected instrumental backgrounds. Our experience in LUX indicates that at a threshold of 2.5 ionization electrons, these backgrounds should not be dominant.

### 4.1.3 General WIMP-Nucleon Couplings

Although the SI coupling provides an important standard for interpreting experimental results, it is one of many possible WIMP-nucleon couplings. Two long-standing themes have guided alternate interpretations: first, that the WIMP-neutron and WIMP-proton couplings, SI or otherwise, might differ; second, that the couplings might involve the spins of the nucleons. For two decades, experimental interpretations of NR experiments have employed the limit where the nucleons are taken to be static, and the WIMP is nonrelativistic. In this "static-nucleon" limit, two classes of terms survive: (1) SI terms, with contributions from scalar, vector, and tensor interactions; and (2) spin-dependent (SD) terms, with a contribution from the axial vector interaction [11].

Recently, the fact that the nucleon velocity is near-relativistic has been applied to the WIMP-nucleus interaction and has led to consideration of couplings that involve the orbital angular momentum of the nucleon [12].

No single target material possesses sensitivity to the complete set of generalized WIMP-nucleon couplings. Xe is sensitive to a wide variety of the general couplings, and targets such as fluorine, sodium, and iodine complement Xe by providing sensitivity to interactions that couple exclusively to proton spin and angular momentum. Should a WIMP signal be seen in LZ, it would be possible to exchange the target of natural Xe with one of isotopically enriched or depleted Xe, to deduce whether the WIMP-nucleon coupling is SI, SD, or something more complex.

Within the context of SI interactions, the coupling coefficient ($f_p$) to protons may be different from that ($f_n$) to neutrons. For example, if WIMPs interact via the vector current that results from exchange of the $Z^0$, $f_p/f_n = -(1-4\sin^2\theta_w) \simeq -0.04$, and the coupling is SI [13]; this possibility tends to be neglected, because the most-favored models specify that the WIMP is a Majorana fermion, for which the vector current vanishes. A variety of extensions to the Standard Model, including most implementations of SUSY, do result in $f_p \simeq f_n$ for scalar interactions of Majorana fermions, but there are alternates that violate that near-equality [14,15].

The experimental consequences of $f_p \neq f_n$ have recently been examined in depth [16]. Natural Xe has an advantage due to its variety of stable isotopes: Each of seven isotopes constitutes more than 1% of natural Xe. If a signal is seen with any other target, the variety of isotopes in natural Xe makes it impossible to completely suppress its SI interaction cross section by adjusting $f_p/f_n$ to achieve destructive interference. Complete suppression is only possible for elements that consist of a single isotope.

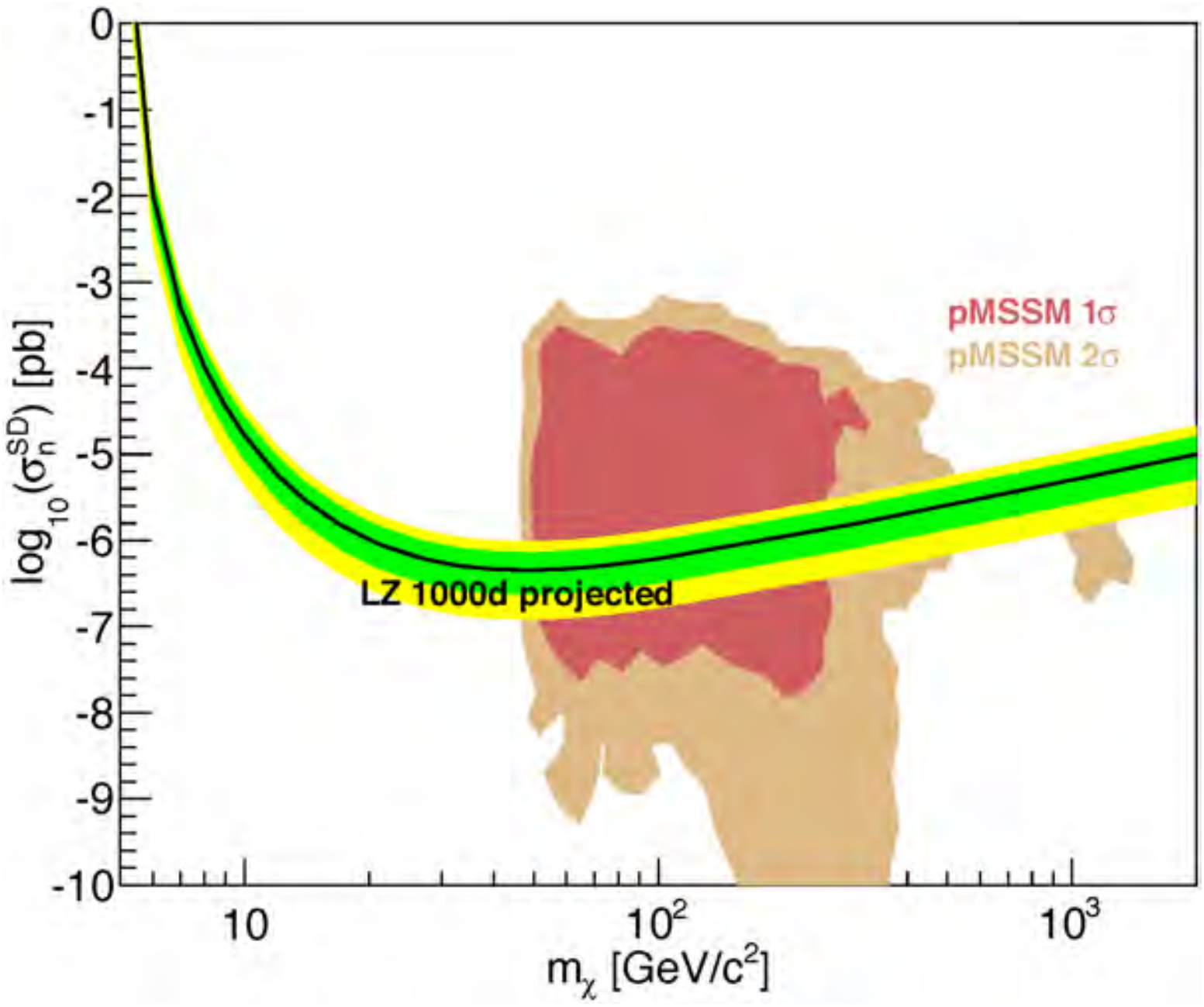

**Figure 4.1.3.1. The LZ projected sensitivity to an SD WIMP-neutron interaction. The median LZ 90% CL sensitivity is in black, and the green and yellow bands display the range of 68% (1σ) and 95% (2σ) of the expected 90% CL limits. A fiducial mass of 5.6 tonnes and a live time of 1,000 days is assumed. Expectations from the 15-parameter pMSSM are shown in rose and beige, prior to consideration of the latest LHC constraints [9]. 1 picobarn is $10^{-36}$ cm$^2$.**

Targets such as argon, iodine, fluorine, and neon do consist dominantly of a signal isotope. It is possible to completely suppress, through a devious choice of $f_p/f_n$, WIMP interaction rates in these targets, should a signal be seen in another target. Conversely, should a signal be seen in argon, iodine, fluorine, or neon, a sensitivity in Xe that exceeds the isospin-conserving SI interpretation by factors of 40, 2, 90, or 170 would conclusively test the isospin-violating SI interpretation. The large fiducial mass of LZ enables the achievement of these sensitivities relative to experiments with other targets [17].

An SD interaction arises if the WIMP is a Majorana fermion, as expected in most implementations of SUSY, and if the dominant WIMP-quark interaction proceeds through the $Z^0$. The coupling coefficient ($a_p$) to proton spin and that ($a_n$) to neutron spin would be in proportion $a_p/a_n \simeq -1.14$ [15]. Implementations of SUSY can result in a very wide range of values for $a_p/a_n$ [18]. Because most of the neutrons and protons in a nucleus form spin-coupled pairs, the dominant interaction, when possible, arises with an unpaired (odd) neutron or proton. The SD interaction thus fails to benefit from the quantum coherence over nucleons, which greatly enhances the WIMP-nucleus SI cross-section. Nevertheless, the SI and SD WIMP-nucleus cross sections can be, in some cases, of similar magnitude [14].

In the static-nucleon limit, the sensitivity of a Xe target to SD interactions arises primarily from the two isotopes $^{129}$Xe and $^{131}$Xe, which have unpaired neutrons. These isotopes make up nearly half of natural abundance. There is also sensitivity to SD coupling to the spin of the proton, but this sensitivity is suppressed because the protons in Xe are all paired.

For a given experimental run using natural Xe, the limit for the SD WIMP-neutron cross section is about $10^5$ times weaker than the corresponding limit for the isospin-conserving SI cross section. The coherence in the SI case provides the additional sensitivity. To evaluate the LZ sensitivity, we employee the SD form factors of Ref. [19]. There is an uncertainty due to form-factor variation of a factor of 2 documented in the literature [20], but Ref. [19] uses a recent large-scale nuclear structure calculation to achieve an error at the 10% level. A 1,000-day LZ run would provide a maximum sensitivity to the WIMP-neutron SD cross section of $4 \times 10^{-43}$ cm$^2$ for a WIMP mass near 50 GeV/c$^2$. We portray the expected LZ mass-dependent WIMP-neutron cross-section sensitivity in Figure 4.1.3.2. Additional important sensitivity to a WIMP-neutron SD cross section arises from inelastic scattering with the $^{129}$Xe and $^{131}$Xe isotopes [21].

When the WIMP-neutron and WIMP-proton cross sections are nearly equal [15], the LZ sensitivity to SD interactions will exceed the current sensitivity of the IceCube and Super-Kamiokande detectors to the annihilation of WIMPs in our sun by 2 orders of magnitude [22,23].

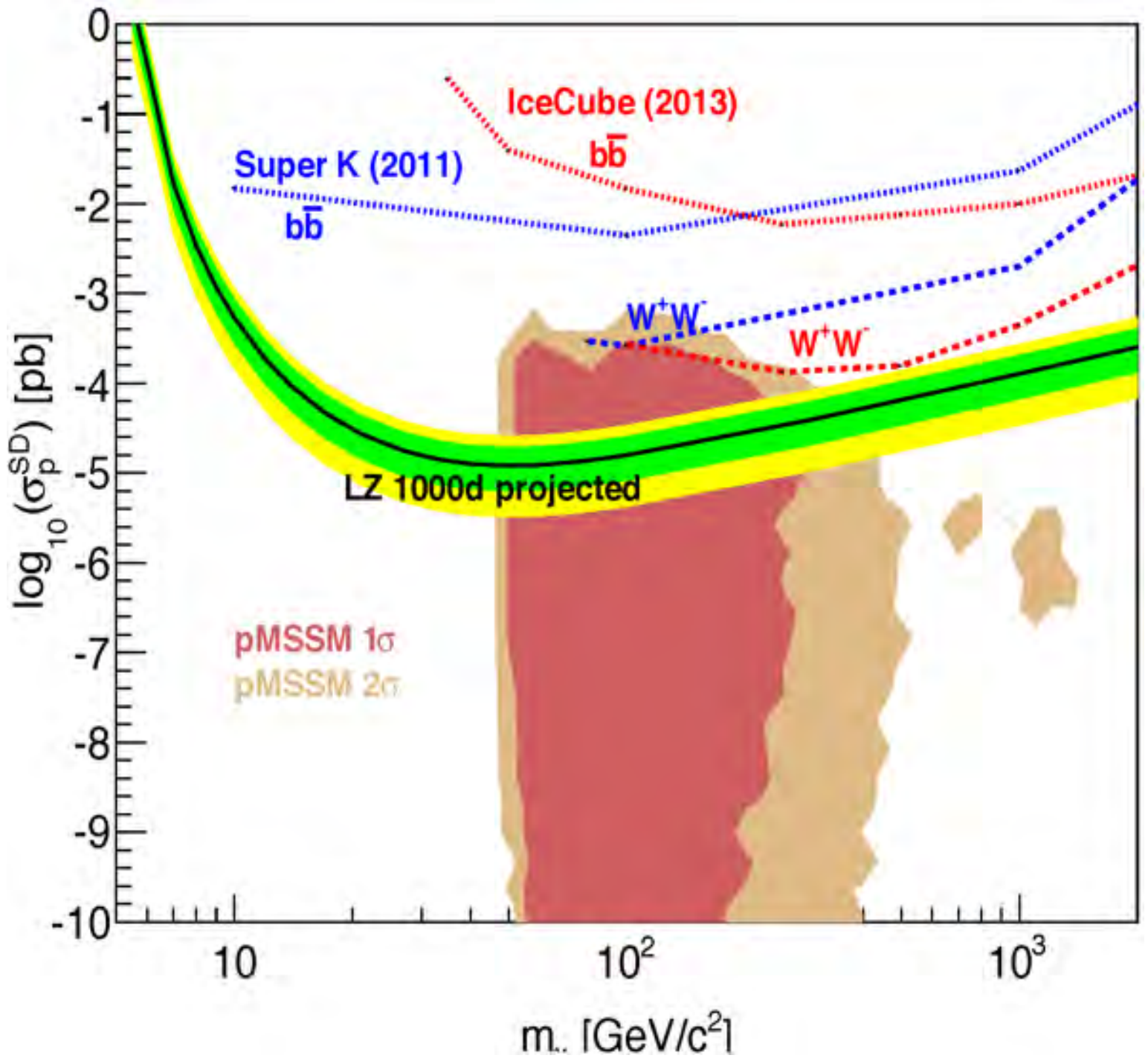

**Figure 4.1.3.2. The LZ projected sensitivity to an SD WIMP-proton interaction. The median LZ 90% CL sensitivity is in black, and the green and yellow bands display the range of 68% (1σ) and 95% (2σ) of the expected 90% CL limits. A fiducial mass of 5.6 tonnes and a running time of 1,000 days is assumed. Current indirect detection results from Super-Kamiokande [22] and IceCube [23] are shown. Expectations from the 15-parameter pMSSM are shown in rose and beige, prior to consideration of the latest LHC constraints [9]. 1 picobarn is $10^{-36}$ cm$^2$.**

If WIMP-Xe scattering is driven by the SD WIMP-proton interaction, and because there are no unpaired proton spins in Xe, the net scattering amplitude can nearly vanish. To evaluate the LZ sensitivity, we employee the SD form factors of Ref. [19], and note that an uncertainty of a factor of $10^3$ is documented in the literature [20]. However, Ref. [19] uses a large-scale nuclear structure calculation to achieve an error at the 50% level. The greatest sensitivity of a 1,000-day LZ run, at a WIMP mass near 50 GeV/c$^2$, would be about $1 \times 10^{-41}$ cm$^2$, somewhat better than the current sensitivity of the IceCube or Super-Kamiokande detectors to the annihilation of WIMPs in our sun [22,23].

Recently, the validity of the static-nucleon limit has been examined and found to be substantially incomplete [12]. Ref. [12] employs effective field theory to analyze all the Lorentz structures that can contribute to WIMP-nucleus scattering. In addition to the long-considered SI interactions, the relativistic motion of the nucleons induces three additional terms:

1. An orbital angular-momentum (L) dependent (LD) term.
2. A combined angular-momentum (L) and spin (S) dependent (LSD) term. This term is particularly interesting because its contribution builds coherently among nucleons, like the SI term.
3. The SD term breaks into two independent terms, one transverse and one longitudinal to the momentum transfer.

The five terms can be distinct for neutrons and for protons, resulting in a total of 10 coefficients to completely specify the WIMP-nucleus interaction. Among the targets considered in [12], Xe provides sensitivities to the broadest range of the 10 WIMP-nucleus parameters. For five of those parameters, Xe gives the best sensitivity per kilogram, and sensitivity is meager for only three of the parameters. Targets that complement a measurement in Xe include fluorine, sodium, and iodine, all of which have an unpaired proton. Analyses of existing experimental constraints based on effective field theory are now under way [24,25].

## 4.2 Beyond Nuclear Recoils from WIMPs

### 4.2.1 Electrophilic WIMPs

One type of WIMP-matter coupling that does not cause NRs, at least at tree-level, is the coupling of a WIMP to a charged lepton. A WIMP-charged lepton vector coupling induces a WIMP-nucleon interaction at one loop in perturbation theory, where the charged lepton loop interacts with the nucleon via photon exchanges [26]. This interaction is surprisingly sensitive. The WIMP-nucleon SI cross-section sensitivity of $2 \times 10^{-48}$ $cm^2$ achievable by LZ at a WIMP mass of 50 $GeV/c^2$ corresponds, when converted via a one-loop calculation, to a WIMP-electron cross section of $1 \times 10^{-50}$ $cm^2$. Should the interaction be exclusively WIMP-muon, the LZ sensitivity at 50 $GeV/c^2$ corresponds to a vector-mediated WIMP-muon cross section of $4 \times 10^{-50}$ $cm^2$; for a tau, the corresponding WIMP-tau cross section is $3 \times 10^{-49}$ $cm^2$.

If the WIMP is a Majorana particle, all its vector couplings vanish, but an SD axial-vector coupling is still possible. The axial-vector coupling does not induce an interaction at higher order in perturbation theory with the nucleus; the only observable consequence in LZ of an axial-vector coupling of a WIMP to an electron is WIMP-electron scattering.

The physical situation for WIMP-electron scattering resembles that for WIMP-nucleon scattering, described at the end of Section 4.1.5, where the nucleon motion is important. The electron motion in the atom is crucial, and it is the very highest momentum tails of the electron wavefunction that determine the cross section for an impinging WIMP to ionize a Xe atom. The resulting events are ERs, and their energy spectrum rises very quickly as the energy deposition falls. Limits on axial-vector WIMP-electron scattering depend critically on the low energy threshold [26].

Interpretations of the DAMA [27] event excess as axial-vector WIMP-electron scattering imply a WIMP-electron cross section of $2 \times 10^{-32}$ $cm^2$ at a WIMP mass 50 $GeV/c^2$. The LZ experiment will observe an ER background, primarily from pp neutrinos, about 5 orders of magnitude lower than DAMA backgrounds, so LZ should achieve a limit, assuming background subtraction, of approximately $2 \times 10^{38}$ $cm^2$. This sensitivity is comparable to the indirect astrophysical limits on the SD WIMP-electron scattering cross sections deduced from Super-Kamiokande data [22].

### 4.2.2 Axions and Axion-like Particles

The axion was introduced to describe the absence of CP-violation in the strong interaction. These particles, known as QCD axions, have a specific relationship between their mass and their coupling to fermions [28-30]. A particle with properties similar to the axion, but without the relationship between mass and fermion coupling, is known as an axion-like particle (ALP) [31].

The LZ experiment will be sensitive to axions and ALPs via the axioelectric effect, where an axion is absorbed and an atomic electron is ejected [32]. In contrast to the photoelectric effect, the mass of the axion or ALP is available for transfer to the atomic electron.

Two sources of axions or ALPs contribute to a possible signal in LZ [33]:

1. Nonrelativistic ALPs that might constitute the dark matter of our galaxy could cause signals in LZ, if their masses are sufficient to provide enough energy to ionize a Xe atom.
2. Axions or ALPs with a mass less than about 15 keV emitted by bremsstrahlung, Compton scattering, or other atomic processes in the sun also can ionize the Xe atoms in LZ [34].

Events caused by axions or ALPs in LZ would be ERs with energy up to a few tens of $keV_{ee}$. The neutrinos emitted by pp fusion in the sun will be the dominant background. The signal identification relies on the distinct shape of the energy spectrum of the axion or ALP signal.

The signal for a galactic dark-matter ALP would be a peak in ERs with energy at the mass of the particle. Our studies indicate that the LZ sensitivity to the coupling between electrons and galactic dark-matter ALPs ranges from a coupling constant $g_{Ae}$ of $10^{-14}$ to one of $10^{-13}$, for masses between 1 keV/c$^2$ and 20 keV/c$^2$, as shown in Figure 4.2.2.1.

The signal for solar ALPs is a broad thermal spectrum caused principally by bremsstrahlung and the Compton effect in the sun convolved with the axioelectric cross section. Our studies indicate that LZ is sensitive to a coupling constant $g_{Ae}$ between solar ALPs and the electron of about $1.3 \times 10^{-12}$ for masses between 0 keV/c$^2$ and approximately 1 keV/c$^2$, as shown in Figure 4.2.2.2.

### 4.2.3 Neutrino Physics

The LZ detector is sufficiently large and sensitive that neutrinos cause interesting signals that are uniform throughout the LXe volume, and which cannot be shielded. We have studied possible LZ observations of astrophysical, reactor, and geophysical neutrinos. Solar and atmospheric neutrinos have been studied as both signal and background to a WIMP search, and the prospective neutrino signal from a nearby supernova has been evaluated. LZ can make the first real-time observations of the neutrinos from pp fusion via elastic $\nu e \rightarrow \nu e$ scattering, and would be sensitive to the neutrino burst from a nearby supernova via the as-yet-unobserved process of coherent nuclear scattering. We have also estimated the potential of LZ to observe neutrinoless double-beta decay ($0\nu\beta\beta$) from $^{136}$Xe, and considered the impact on the reactor/source neutrino anomaly and on searches for a neutrino magnetic moment of a prolonged exposure of LZ to a nearby $^{51}$Cr neutrino source.

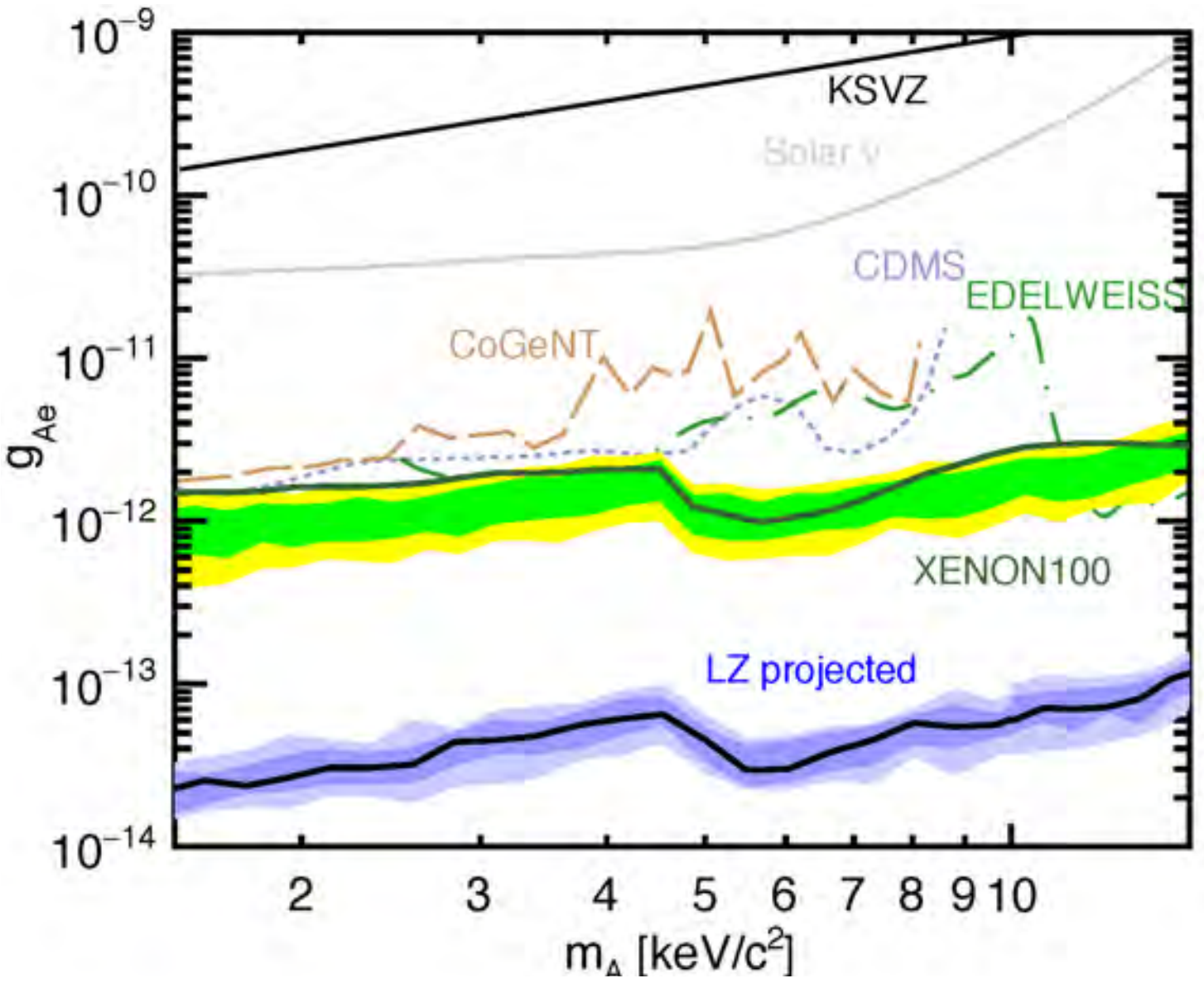

**Figure 4.2.2.1. Dark-matter axion-like particle sensitivity. The LZ projected sensitivity for ALPs at 90% CL is shown by the dark/light blue bands, which show the 68%(1σ) and 95%(2σ) bands for that sensitivity. The line that defines KSVZ axions [35,36], an astrophysical upper limit from solar neutrinos [37], is shown. Upper limits by the experiments CDMS [38], EDELWEISS [39], CoGeNT [40], and XENON100 [41] are also shown.**

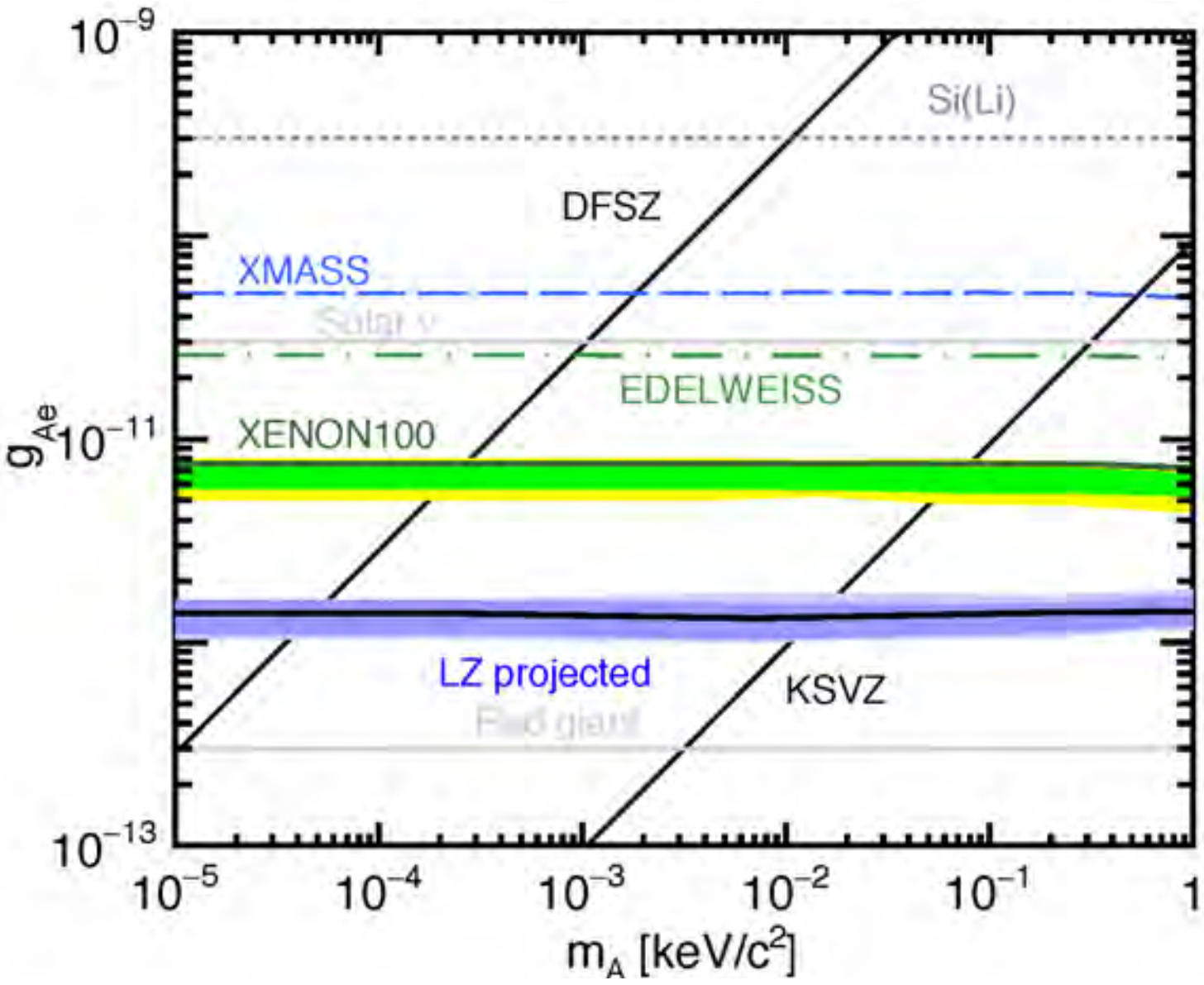

**Figure 4.2.2.2. Solar axion-like particle sensitivity. Horizontal lines all extend down to $m_A$=0. The LZ projected sensitivity for ALPs at 90% CL is shown by the dark/light blue bands, which show the 68%(1σ) and 95%(2σ) bands for that sensitivity. The lines that define DFSZ axions [43,44] and KSVZ axions [35,36], astrophysical upper limits from solar neutrinos [37], and from red giants [45], are shown. Upper limits by the experiments XMASS [46], EDELWEISS [39], and XENON100 [41] are also shown.**

Most events in LZ from neutrino-related processes are ERs from solar neutrinos originating from the pp burning reaction in the sun [47,48], and also from the decay electrons from two-neutrino double-beta decay ($2\nu\beta\beta$) of $^{136}$Xe [49]. For ER energies between 1.5 and roughly 20 $keV_{ee}$, ERs from pp solar neutrinos dominate [50], and contribute 850 observable events through $\nu e \rightarrow \nu e$ scattering in the LZ fiducial mass of 5.6 tonnes and a run of 1,000 days. We have neglected atomic effects that suppress the rate by of order 10% [51]. For ER energies above 20 $keV_{ee}$, ($2\nu\beta\beta$) events from $^{136}$Xe dominate.

The LZ experiment alone compares very favorably with the existing world experimental data on pp solar neutrinos. The SAGE experiment, consisting of approximately 50 tonnes of gallium, observed 854 events attributed to pp solar neutrinos in 18 years of operations [52]. The SAGE experiment detected solar neutrinos via inverse beta decay, while LZ will detect solar neutrinos via $\nu e \rightarrow \nu e$ scattering. The threshold neutrino energy for LZ is 20 keV, while that of a gallium experiment is 233 keV, giving LZ sensitivity to a different portion of the pp fusion neutrino spectrum than was measured with SAGE. The LZ experiment will identify the time of pp solar neutrino events to a few nanoseconds, in contrast to the multiday time delay of the radiochemical process in SAGE.

Although the LZ experiment will open up new experimental territory in the study of pp solar neutrinos, the current consensus in the solar neutrino community is that the accuracy of pp solar neutrino measurement must be better than 1% to improve understanding of solar neutrinos [48]. To achieve 1% accuracy, LZ would need to observe several tens of thousands of pp neutrino-induced ER events, and also control systematics at a sub-1% level. Elimination of the $^{136}$Xe isotope and a live time of 2,000 to 4,000 days would allow the accuracy of an LZ measurement of pp solar neutrinos to approach 1%.

The ERs from solar pp neutrinos, after the rejection in the S1+S2 analysis, are the largest source of background in the NR search region of LZ. However, there are also NR backgrounds originating from the as-yet-unobserved process of neutrino scattering that is coherent across nucleons in the nucleus [53,54]. For a given energy of the incident neutrino, the energy of the NR from coherent neutrino is typically suppressed by $\approx m_e/m_N$ relative to the energy of the analogous ER. Nuclei that recoil from solar pp neutrinos, and indeed from the entire spectrum of solar neutrinos, fall below the LZ S1+S2 analysis threshold of 6 $keV_{nr}$ [55,56]. The S2-only analysis should be sensitive to solar neutrinos from $^{8}$B.

There are other sources of neutrinos (from the diffuse supernova neutrino background (DSNB) and from atmospheric neutrinos) with energies above the 19 MeV necessary to cause a Xe recoil above the LZ threshold, and below 50 MeV where scattering off of nucleons in the nucleus becomes incoherent. We estimate an irreducible background in the NR search region of 0.05 (DSNB) and 0.25 (atmospheric neutrinos) for an LZ fiducial mass of 5.6 tonnes and a run duration of 1,000 days.

The nearest power reactors are about 800 km away, in Fort Calhoun, NE (0.5 GWe), and Cooper, NE (0.8 GWe). The power/distance$^2$ distribution shows a broad peak for reactors in Illinois and Wisconsin. The net flux is small enough, however, that we expect negligible detected events from power-reactor neutrinos in LZ.

Geophysical neutrinos from $^{238}$U and $^{232}$Th decays have been seen by the KamLAND [57-59] and Borexino [60] detectors. Those detectors have an energy threshold for neutrinos of about 1.8 MeV. They are unable to detect neutrinos from the decay of $^{40}$K, which have an energy just below 1.5 MeV. Using the Reference Earth Model and neutrino flux calculations from the KamLAND work, we estimate 1.5 ER events/year from $^{40}$K decay, 0.3 ER events/year from $^{238}$U decay, and 0.2 ER events/year from $^{232}$Th decay. With LZ's ER/NR rejection ratio, these provide negligible backgrounds for the dark-matter search.

Should a supernova occur in our galaxy during LZ operation, neutrinos emitted from the supernova would be detected via coherent neutrino-nucleus scattering, which is blind with respect to neutrino flavor. The energy spectrum of neutrinos emitted from a typical supernova peaks near 10 MeV, and has a tail that extends above 50 MeV, which causes NRs above the LZ threshold [61]. Coherent neutrino-nucleus scattering is mediated by the weak neutral current, and thus provides important information on the flux and spectrum of muon and tau neutrinos from supernovae, complementary to the signals that would be seen in other detectors. From a supernova in our own galaxy at 10 kpc, LZ would see ~50 NR events of energy greater than 6 $keV_{nr}$ in a rapid 10-sec burst [62,63]. The NR recoil spectrum increases as the recoil energy decreases; a threshold of 3 $keV_{nr}$ would allow detection of ~100 supernova neutrino-induced NR events. The current world sample of 19 supernova neutrino-induced events were detected from supernova 1987a, 50 kpc from Earth, by detectors with total mass 1,200 times greater than LZ. A supernova 10 kpc from Earth would cause about 7,000 neutrino-induced events in the 32,000 tonnes of water in the Super-Kamiokande detector [61].

The sensitivity of LZ to neutrinoless double-beta decay of $^{136}Xe$ (Q-value 2,458 keV) depends strongly on the radioactivity levels achieved in detector materials and on the energy resolution of the combined S1+S2 signal (we note that LZ is not optimized for such large dynamic range). We have performed Monte Carlo simulations of various backgrounds for neutrinoless double-beta decay, and find that the most significant contributions are the 2,448-keV gamma line and the 2,614-keV gamma line, from $^{214}Bi$ and $^{208}Tl$ decay, respectively, which can penetrate deeply into the active region. The $^{60}Co$ sum peak (1,173 keV + 1,333 keV =2,506 keV) is also an important background, arising especially in stainless steel components. These gamma-ray backgrounds can be accurately measured using the full detector active mass, and can be substantially reduced through self-shielding and multiple scattering cuts. Solar neutrino and 2-neutrino double-beta decay backgrounds are found to be small in comparison. Energies are determined through an optimal linear combination of S1 and S2 signals, with a predicted 1-sigma energy resolution of 0.8% at the 2,458-keV Q-value. With a natural Xe target and 1,000 live days, LZ should be sensitive to $^{136}Xe$ half-lives from $2 \times 10^{25}$ years to $2 \times 10^{26}$ years, depending on achieved background, spatial, and energy resolution. The shorter value corresponds to an increase of 10 times over baseline radiopurity, an energy resolution of 2%, and a spatial resolution of 6 mm. Improvements in spatial and energy resolution, background reductions, and enriching the Xe target would improve these limits to perhaps $2 \times 10^{27}$ years. For comparison, the July 2012 half-life limit from EXO-200 [65] was 1.6 × $10^{25}$ years at 90% confidence limit, and KamLAND-Zen has placed a limit of 1.9 × $10^{25}$ years [66].

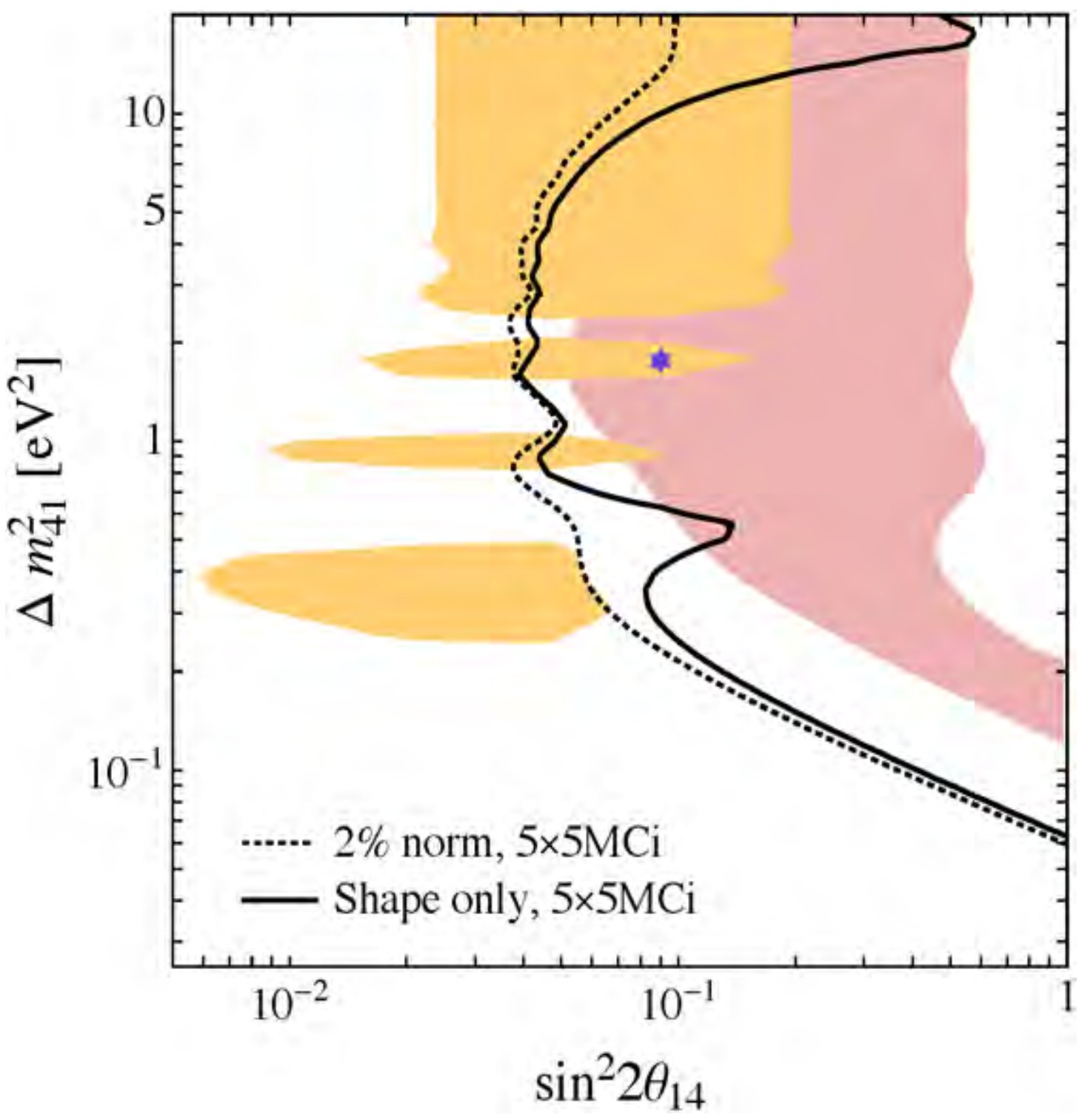

**Figure 4.2.3.1. Sensitivity to sterile neutrino oscillations as a function of mass-difference and mixing angle. The parameter space to the right of each line would be excluded at 95% CL. The shaded areas show the 95% CL allowed regions for source (pink) and reactor (yellow) anomalies. The blue star is the joint best fit. The black solid line shows the expected contours for five 100-day deployments of a 5 MCi $^{51}Cr$ source next to LZ, without use of the source normalization. The dotted line shows the contour if a 2% normalization of the source is available. From Ref. [64].**

There are long-standing anomalies arising from the detailed study of antineutrinos from reactors, and from source-calibration of solar neutrino experiments [67]. A recent study has evaluated the capabilities of deployment of a 5 MCi $^{51}Cr$ electron neutrino source near to the LZ detector [64]. The excellent spatial resolution of the LXe TPC allows the spatial pattern of electron neutrino oscillation into a sterile neutrino to be detected. A neutrino source experiment with LZ would not be part of the principal LZ science goal, which is the WIMP search, and would constitute a distinct follow-on experiment after the WIMP search had achieved significant results.

The sensitivity achievable by five source deployments of a 5 MCi $^{51}Cr$ source near LZ is shown in Figure 4.2.3.1. Numerous proposals are under way to probe the origin of the reactor/source anomalies [68], but the potential LZ advantage is a diminished need to control the source normalization due to LZ's excellent spatial resolution. In addition, a source deployment near LZ will bring sensitivity to an electron neutrino magnetic moment that is close to the limits deduced from astrophysical considerations [64].

## 4.3 Key Requirements

In this section, we summarize the key high-level requirements and their dependence on some of the critical detector performance assumptions. The LZ collaboration has established a small number of such requirements to guide and evaluate the design and later fabrication of the detector systems. The top-level scientific requirement is the sensitivity to WIMPs. Subsidiary high-level science requirements and the flow-down from the overall sensitivity are shown in Figure 4.3.1. The high-level requirements, including the key infrastructure requirements, are summarized in Table 4.3.1. These requirements flow down to the detector subsystems and are captured in a concise form available to the collaboration. There are two practical high-level requirements. First, all equipment and subassemblies must be transported via the Yates shaft (see Chapter 13), which imposes dimensional and weight limits. Second, the existing water tank now housing the LUX detector must be reused (rather than be made anew).

The collaboration has also captured the requirements for detector subsystems at WBS level 2. There is a well-identified process for requirements flow-down and verification that will be used first in the design phase and then in the fabrication (or installation) phase.

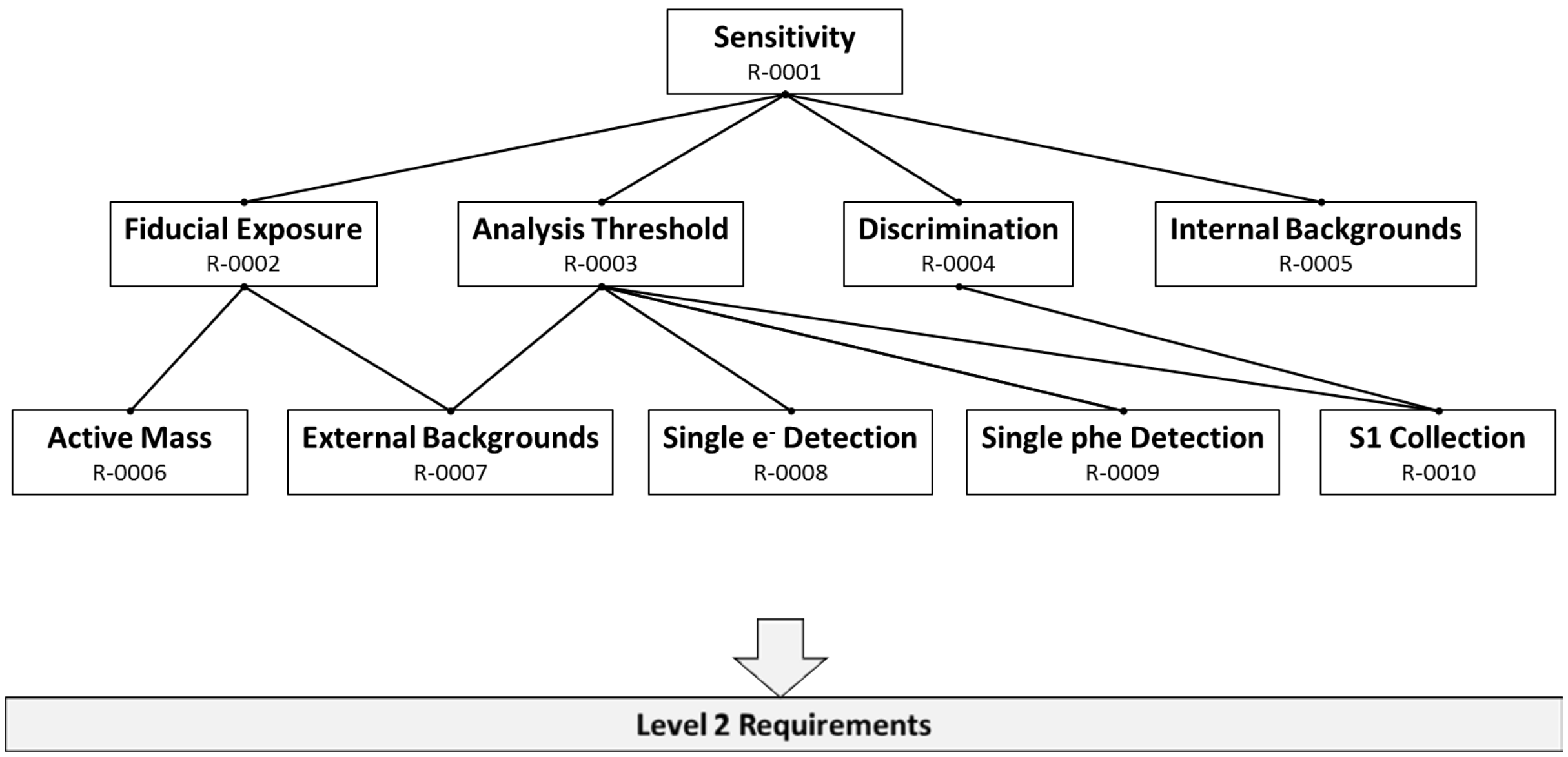

**Figure 4.3.1. High-level science requirements, leading to the overall sensitivity to WIMPs.**

**Table 4.3.1. Summary of high-level requirements established by LZ to guide the design and fabrication of the experiment.**

| Requirement Number | Type | Name | Value | Description |
|---|---|---|---|---|
| **Primary** | | | | |
| R-0001 | Science | WIMP Sensitivity | Sensitivity to 50 GeV/$c^2$ WIMPs is 2 x $10^{-48}$ $cm^2$ or better. | Probe limit of LXe technology set by solar neutrino background. Approach sensitivity to atmospheric neutrinos. Test prominent supersymmetric and extra-dimension models of dark matter. |
| **Secondary** | | | | |
| R-0002 | Science | Fiducial Exposure | 5,600 tonne-days | Minimum fiducial mass of 5.6 tonnes and assumed running period of 1,000 live days. |
| R-0003 | Science | Analysis Threshold | 50% efficiency at 6 keVnr | Probe WIMP mass range down to 5 GeV with non-negligible sensitivity. |
| R-0004 | Science | ER Discrimination | 99.5% | Limit background from ERs so as to reach WIMP sensitivity requirement. NR acceptance 50%. |
| R-0005 | Science | Internal Backgrounds | ER events from Kr+Rn <20% of pp solar neutrino ER rate | Limit ERs from internal backgrounds to be significantly less than ERs from solar neutrinos. |
| **Tertiary** | | | | |
| R-0006 | Science | Active Mass | 7.0 tonnes | Required to reach fiducial exposure |
| R-0007 | Science | External Backgrounds | Backgrounds from radioactivity of the detector components (not including internal backgrounds, R-0005). ER counts before discrimination <21. NR counts ≤0.1 | ER counts constrained to be <10% of ERs from solar neutrinos, including uncertainty in this rate. NR events constrained to be small in comparison to total background. We rely on veto efficiency to reduce the NR rate contribution. This rate, and to a lesser extent external ER contributions, define the fiducial mass. Analysis threshold also depends on size of these backgrounds. |
| R-0008 | Science | Single Electron Detection | 50 photoelectrons detected per emitted electron | Sufficiently large S2 signal for accurate reconstruction of peripheral interactions, such as those arising from contamination on the TPC walls. |
| R-0009 | Science | Single Photoelectron Detection | Single S1 photoelectron detection with >90% efficiency, so as to reach >70% efficiency for 3 phe | Main determinant of analysis threshold |
| R-0010 | Science | S1 Light Collection | Volume-averaged S1 photon-detection efficiency (geometric light-collection times effective PMT quantum efficiency) of ≥7.5% | Good discrimination and low-energy threshold, equal to or better than past Xe experiments. Exponentially falling (in recoil energy) WIMP spectrum means more recoils at lower energies, and low-energy recoils produce less S1 (both total and per-unit-energy) driving the S1 light collection efficiency requirement. |

| Requirement Number | Type | Name | Value | Description |
|---|---|---|---|---|
| **Infrastructure** | | | | |
| R-0100 | General | All parts fit down Yates shaft | All detector elements must be sized so that they can be lowered via the Yates shaft. | Yates shaft is primary access to the Davis Campus. |
| R-0110 | General | Reuse Davis water tank | Existing Davis water tank is reused. Include minor modifications and refurbishment. | Not practical or cost-effective to replace water tank. Insufficient underground space to make larger tank. |

Requirements validation is a key element of internal reviews of LZ detector systems and will be an important aspect of configuration control.

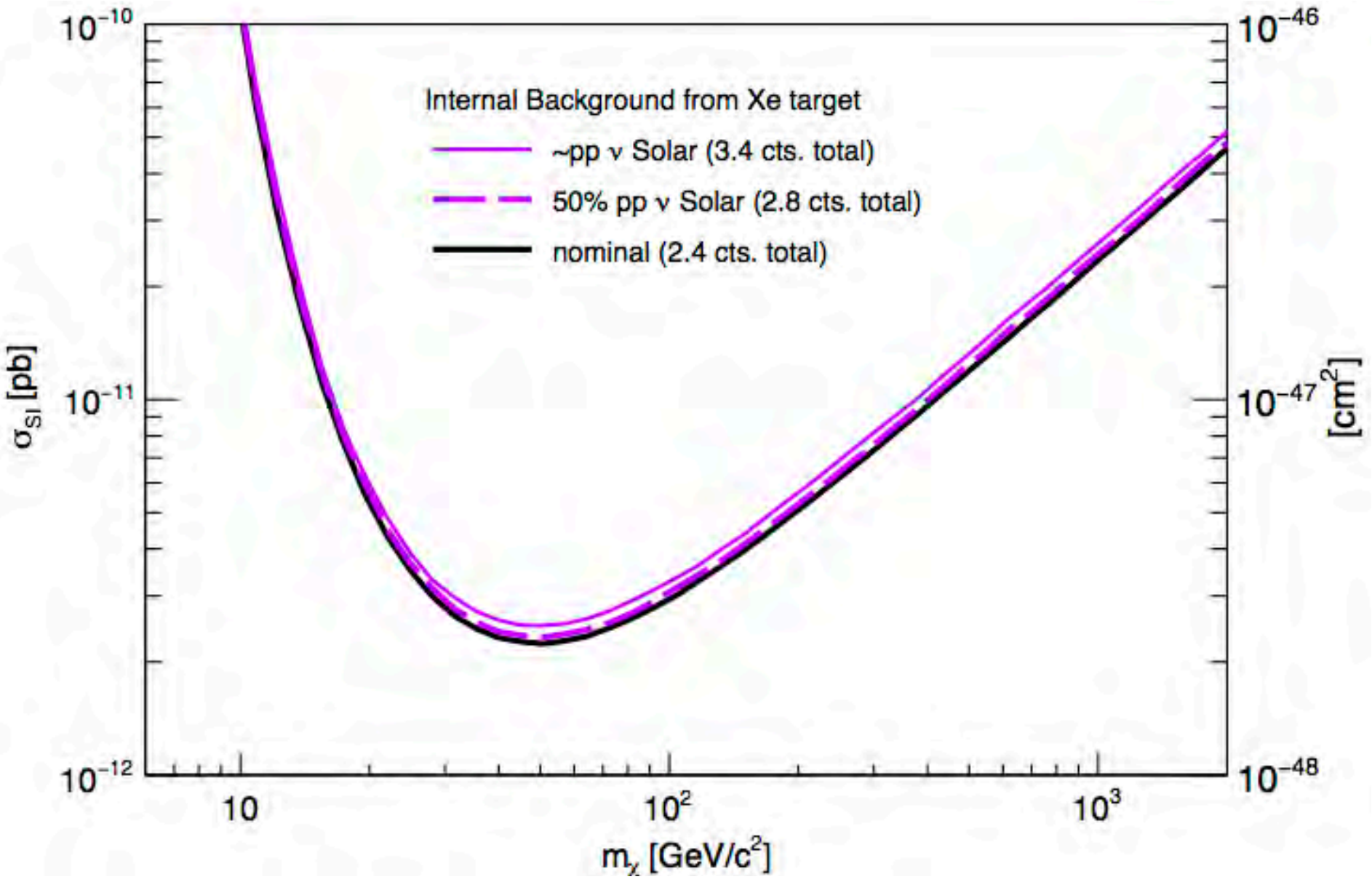

**Figure 4.3.2. Sensitivity if internal backgrounds (Kr and Rn) increase beyond baseline assumptions.**

We have examined the dependency of the LZ sensitivity on some critical performance assumptions and present key parts of these studies of below. The projected sensitivity for the baseline fiducial exposure of 5,600 tonne-days was shown in Figure 4.1.1.2, and the dependence of the sensitivity on less and more fiducial exposure is given in Figure 4.1.1.4. Even if the ER discrimination is poorer than the baseline assumption (99.5%), we can likely achieve the sensitivity requirement by additional running time.

The dependence of the background (with baseline background assumptions) in case the ER discrimination is better than 99.5% or lower at 99.1% is given in Table 4.1.1.2. Our assumption of 99.5% ER discrimination is conservative, and better discrimination would somewhat reduce the overall background levels. Conversely, ER discrimination somewhat poorer than the baseline would have a modest degrading effect and require more running time, as noted above.

Our baseline assumptions of internal Kr and Rn backgrounds are shown in Table 3.8.1.1. The dependence of the sensitivity in case these internal backgrounds increase is shown in Figure 4.3.2. These backgrounds, particularly Rn, are among the most difficult to control but we are not near a critical point with our baseline assumption, which is 10% of the pp solar neutrino rate.

Our baseline assumptions for key external backgrounds are also given in Table 3.8.1.1. The dependence of the sensitivity in case the external backgrounds increase is given in Figure 4.3.3. Note that the baseline backgrounds correspond to about 10% of the pp neutrino solar background. We have prudent headroom in case the external backgrounds are larger than our baseline assumptions.

The approximate dependence of the sensitivity on active mass is given in Figure 4.3.4. This is calculated from a simple right-cylinder model of obtaining the fiducial volume from the baseline 7-tonne active volume, and then simple scaling for smaller volumes. We note that at least the full 7-tonne active volume and three years of operation are required to start to be sensitive to the neutrino background at larger WIMP masses.

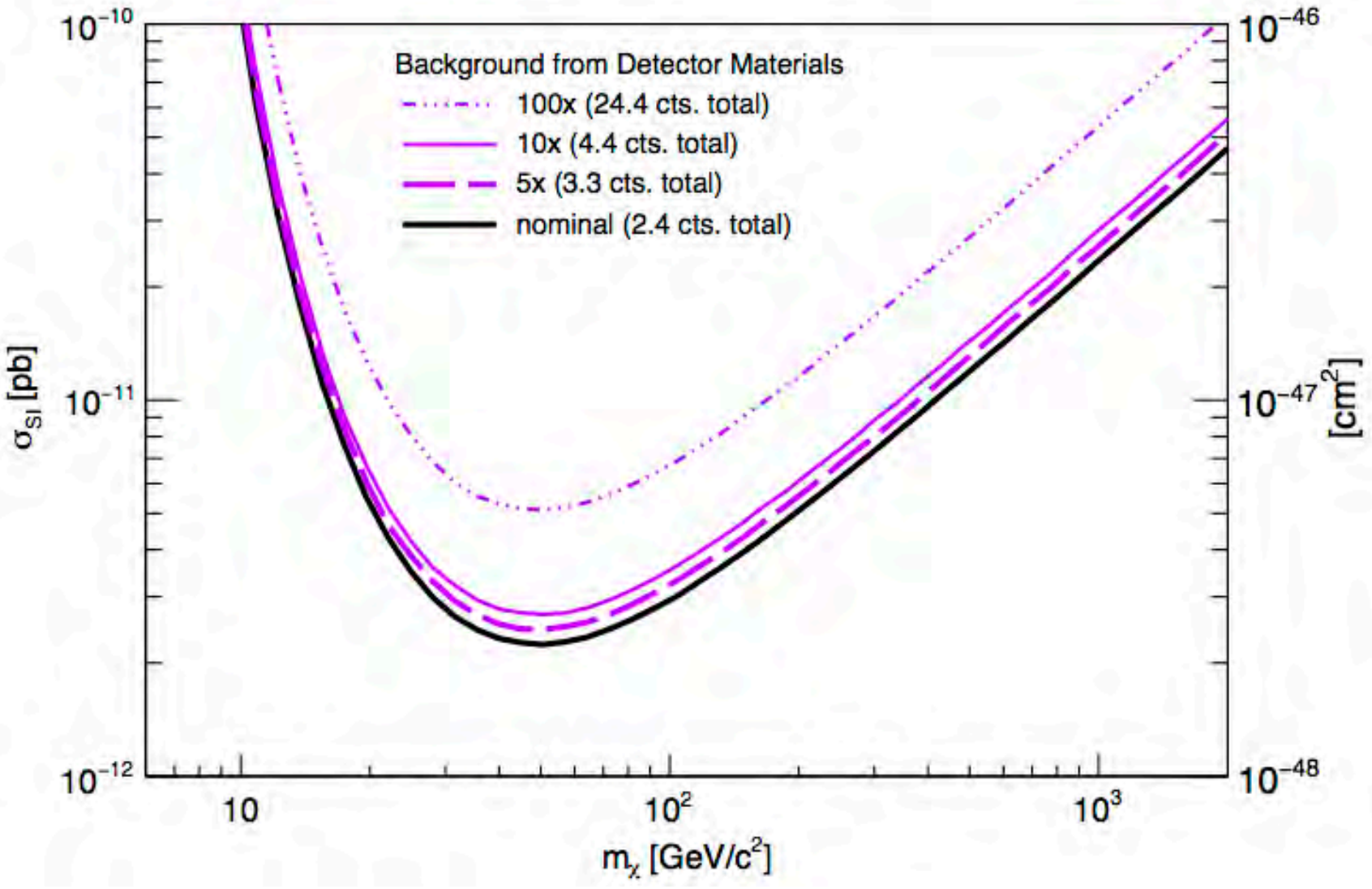

**Figure 4.3.3. Sensitivity if external backgrounds increase beyond baseline assumptions.**

Our baseline assumption for the average S1 light collection is 7.5%. The light-collection efficiency affects the ER discrimination, which is discussed in more detail in Chapter 6. The dependence of the sensitivity on average light collection is shown in Figure 4.3.5. For example, a reduction in light-collection efficiency to 4% would yield an ER discrimination of 99.1%, and the effect of this has been described previously. The effects of changes to the single-electron and single-photoelectron detection efficiency are also discussed in more detail in Chapter 6. The dependence of the sensitivity on the single-photoelectron detection efficiency is shown in Figure 4.3.6.

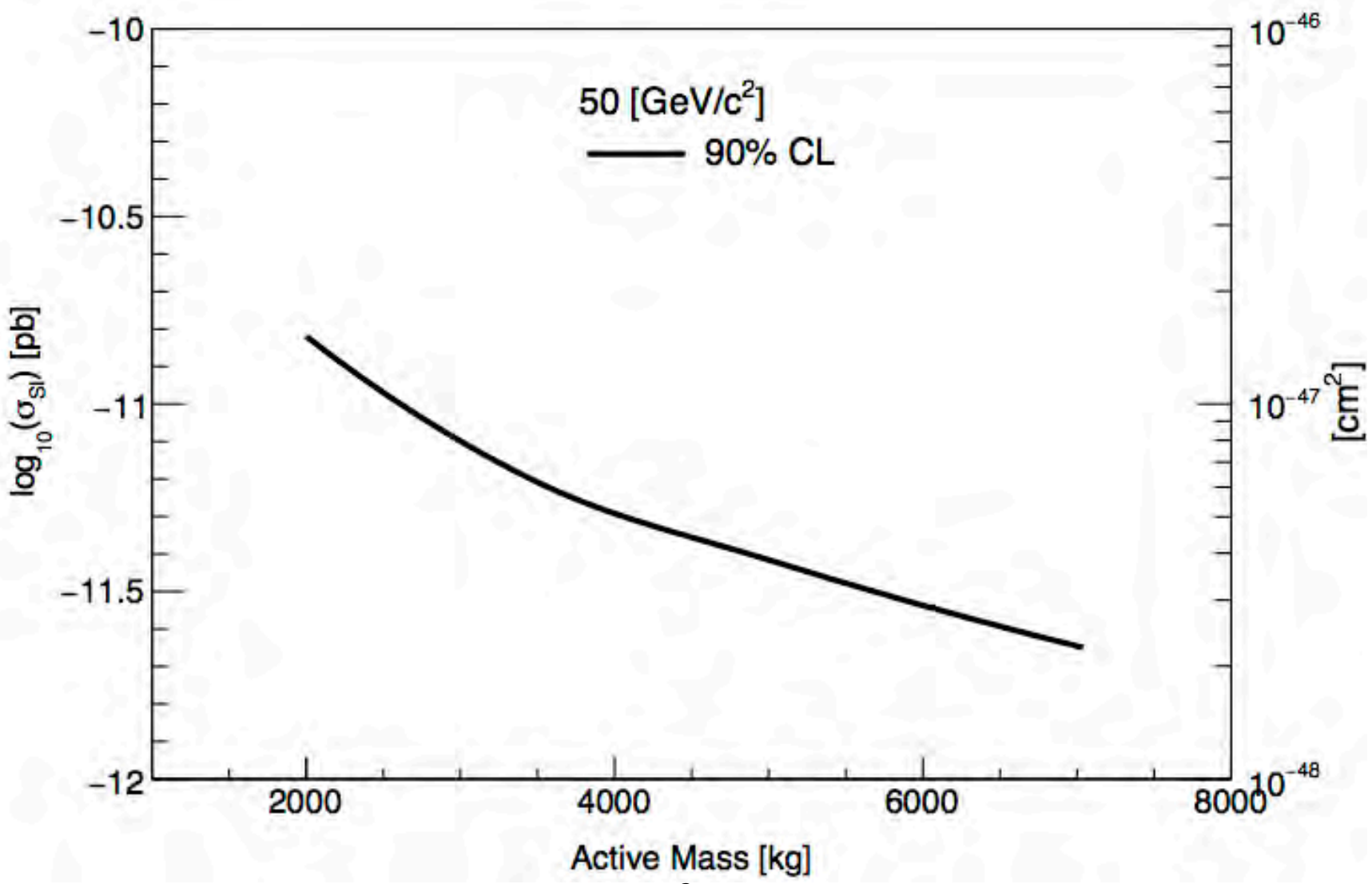

**Figure 4.3.4. Sensitivity at 50 GeV/c$^2$ as a function of the total active Xe mass.**

The sensitivity also could be affected by the purity of the xenon. The dependence of the sensitivity on the characteristic drift times is shown in Figure 4.3.7. There is significant margin, unless the drift time becomes less than one-half the nominal value. We note that drift times in excess of 500 microseconds have routinely been obtained in LUX.

Finally, we show in Figure 4.3.8 both the nominal sensitivity and a curve representing a 3σ discovery.

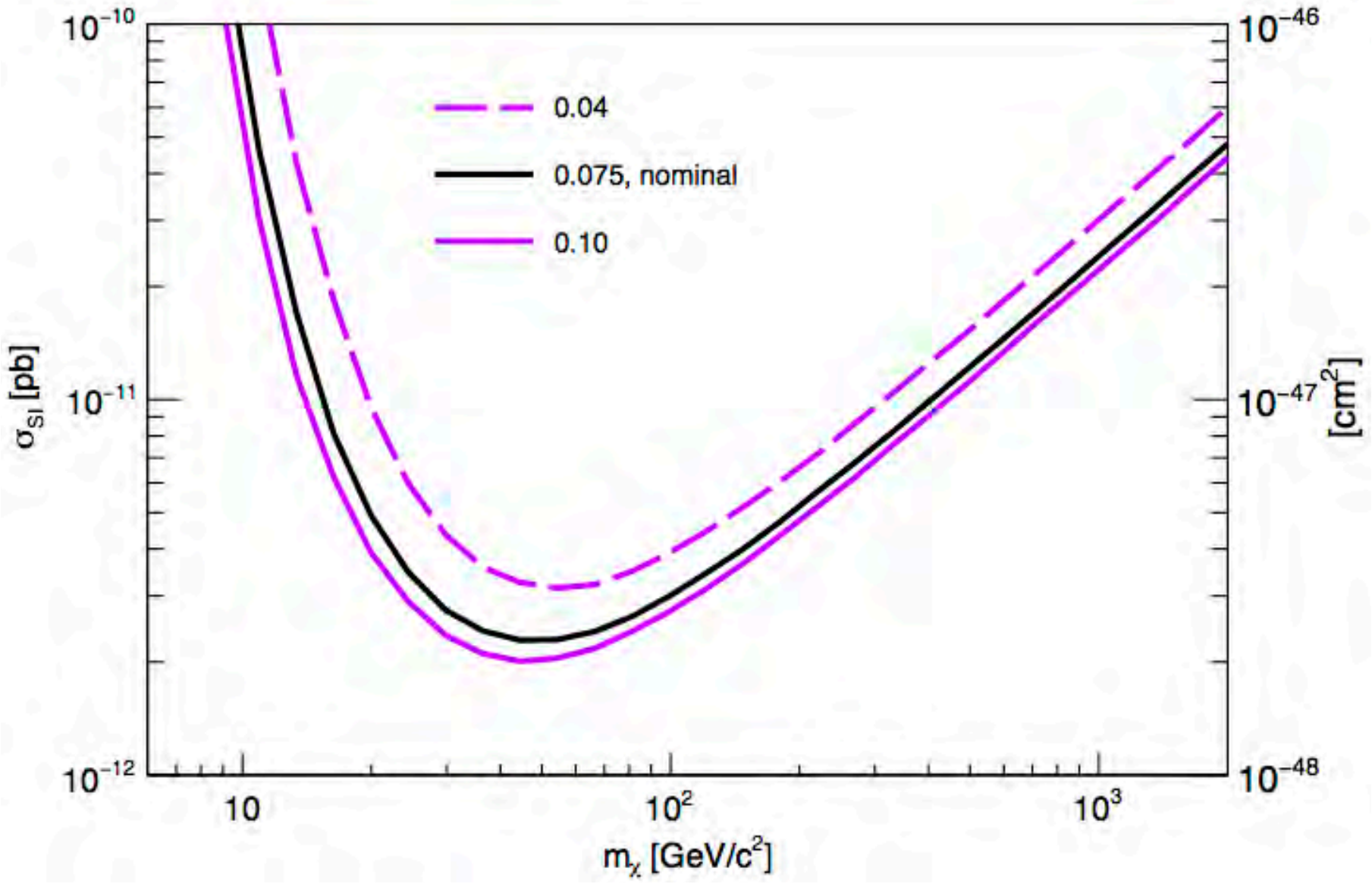

**Figure 4.3.5. Dependence of the sensitivity on the average light collection as it varies from 4% to 10%. The baseline value is 7.5%.**

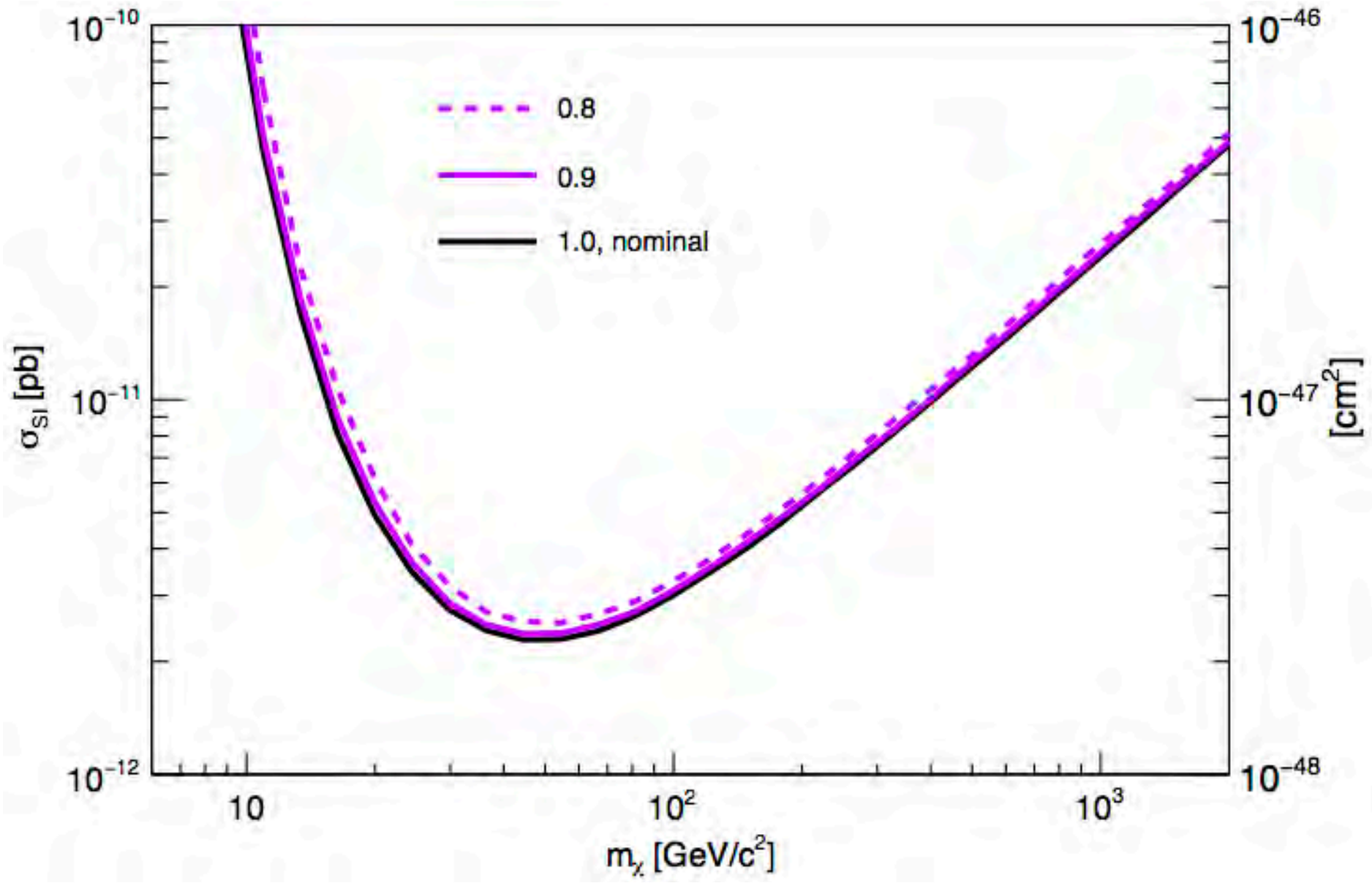

**Figure 4.3.6. Dependence of the sensitivity on the single photoelectron detection efficiency.**

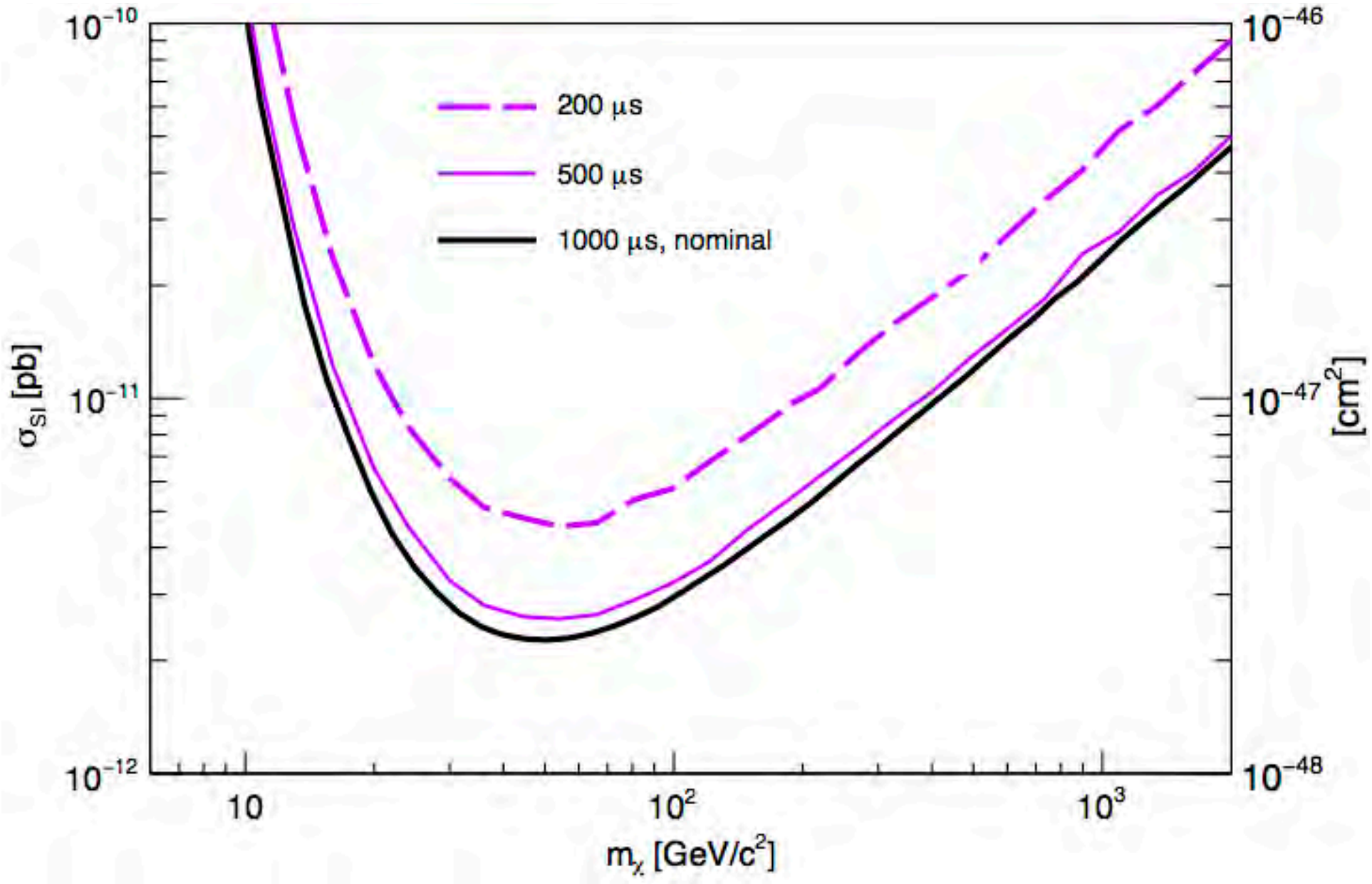

**Figure 4.3.7. Dependence of the sensitivity on the characteristic drift time.**

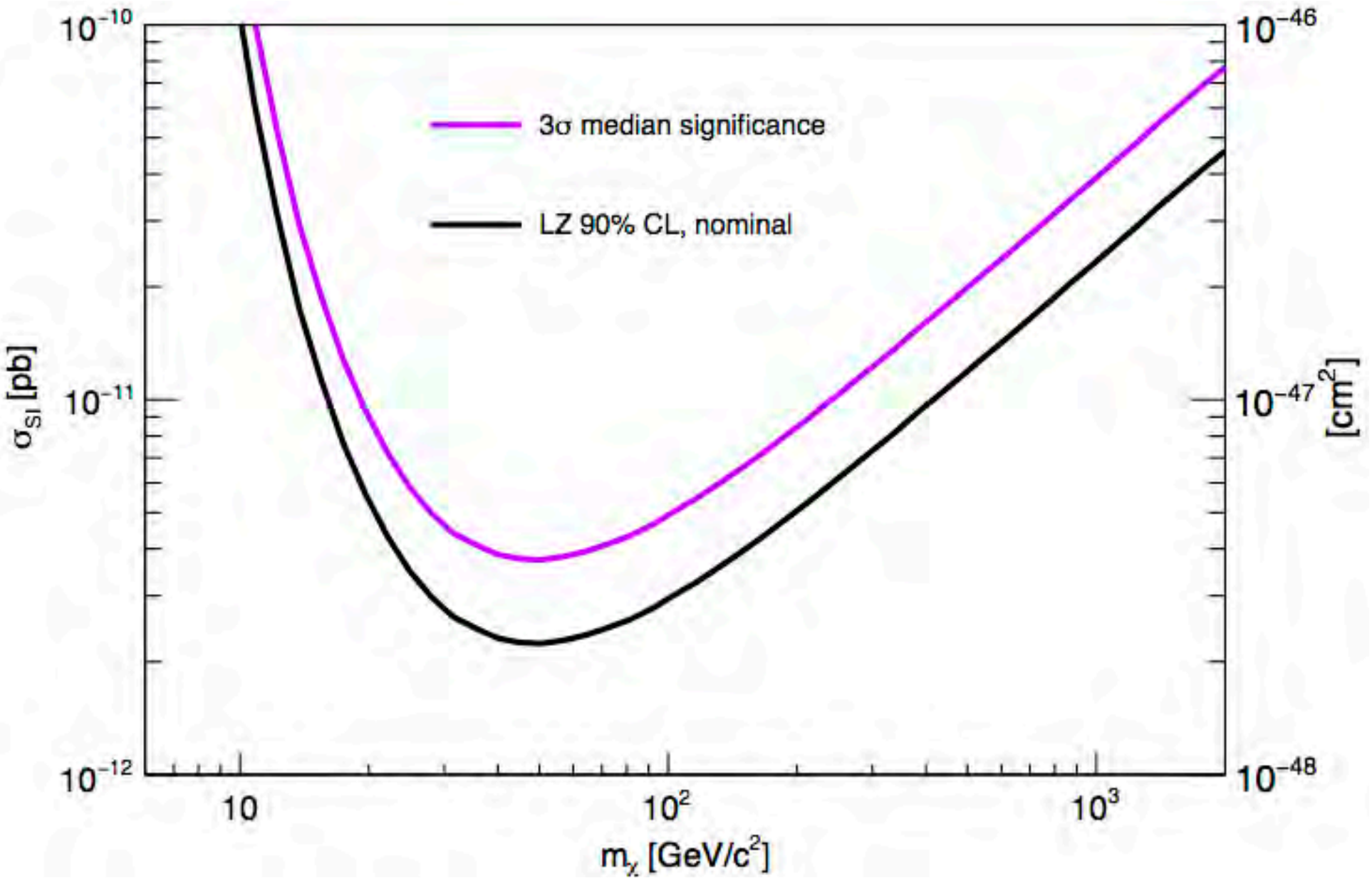

**Figure 4.3.8. The nominal sensitivity 90% confidence level limit and a 3σ significance discovery potential.**

## Chapter 4 References

# 5 ZEPLIN and LUX

Behind the potential of LZ lie a highly developed and well-understood technology and a team with a long track record in dark-matter searches with LXe. This section describes the ZEPLIN and LUX programs, with emphasis on the technical developments relevant to LZ. The confluence of these two programs led directly to LZ — although we draw also from significant experience from XENON10, CDMS, CRESST, Edelweiss, EXO-200, Daya Bay, and other rare-event searches, which have engaged many LZ collaborators in the past.

## 5.1 The ZEPLIN Program at Boulby

The Xe-based ZEPLIN program at the Boulby mine (UK) dates back to the 1990s [1-3]. It coalesced around the UK Dark Matter Collaboration (UKDMC), which had been exploring the viability of various WIMP search technologies for a few years. The first dark-matter results were published from a sodium iodide crystal [4], leading subsequently to the NAIAD (Nal Advanced Detector) experiment, which operated until 2003 [5,6]. The ZEPLIN and DRIFT (Directional Recoil Identification From Tracks) programs followed, the latter developing gaseous TPC detectors to measure recoil directionality [7,8].

ZEPLIN-I exploited pulse-shape discrimination (PSD) in an LXe scintillation detector, publishing final results in 2005 [9]. It featured at its core a PTFE-lined chamber containing a 5 kg LXe target viewed by three 3-inch photomultipliers (PMTs) coupled to quartz windows, as shown in Figure 5.1.1 (left). This was followed by the first double-phase Xe TPCs, ZEPLIN-II and ZEPLIN-III, in which the ionization response was also detected via electroluminescence developed in a thin layer of vapor above the liquid [10]. Besides affording much better discrimination than the simple PSD technique exploited in ZEPLIN-I, this second signature allows precise 3-D reconstruction of the interaction site and a very low NR energy threshold, being sensitive down to individual electrons emitted from the liquid [11-13]. ZEPLIN-II, shown in Figure 5.1.1 (center and right), became the first double-phase system to operate underground, completing in 2007 [14,15]. It featured a deep, high-reflectance PTFE chamber containing 31 kg of LXe with readout from seven PMTs in the gas phase. ZEPLIN-III [16] concluded the Boulby program, with science runs in 2008 [17,18] and, following an upgrade phase, in 2010-11 [19]; it utilized 31 PMTs immersed in the liquid, viewing a thin disc geometry of 12.5 kg of LXe at high electric field.

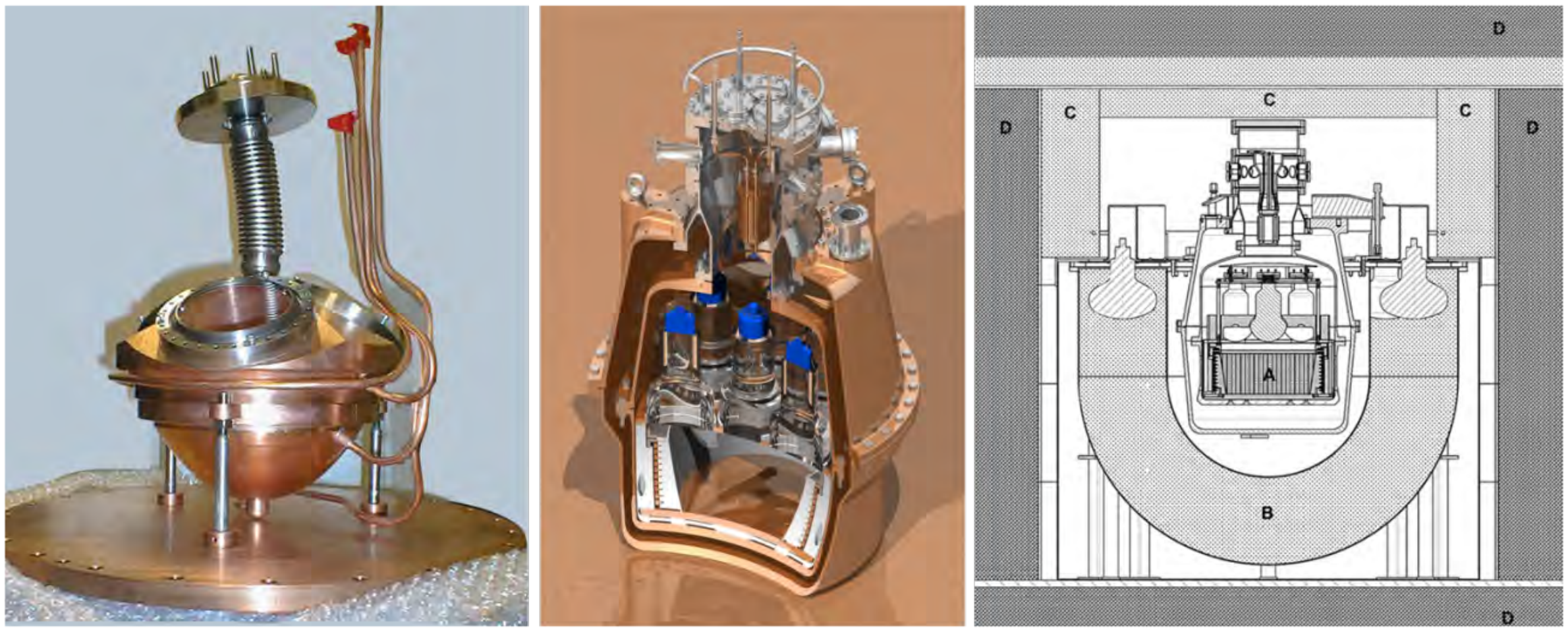

**Figure 5.1.1. Left: Liquid xenon chamber of the ZEPLIN-I detector as built; three quartz windows permitted viewing of the 5 kg WIMP target by photomultipliers operating warm. Center: Schematic representation of the ZEPLIN-II detector, where the PTFE-lined chamber is viewed by seven PMTs in the gas phase. Right: Both systems were operated within a liquid scintillator veto detector (B), shielded by Gd-loaded polyethylene (C) and lead (D).**

A main aim of the UK program was to evaluate the distinct technical solutions adopted in both detectors, with a view to building a tonne-scale experiment, ZEPLIN-MAX. However, with the timely development of LUX, a merger between the ZEPLIN-III and LUX teams became the sensible continuation of the UK program and a memorandum of understanding (MOU) was signed in 2008, leading to LZ.

### 5.1.1 ZEPLIN-III

The ZEPLIN-III experiment achieved the best WIMP sensitivity of the Boulby program and demonstrated important features that now inform the design and exploitation of double-phase Xe experiments. The instrument construction is described in [16]; the main components of the experiment are illustrated in Figure 5.1.1.1. Its most innovative features were the thin disc geometry, to permit application of a strong electric field to the target, and the immersion of the PMTs directly in the cold liquid phase, for improved light collection. Most elements were built from high-purity copper to minimize background. The outer cryostat vessel enclosed two chambers; the lower one contained the $LN_2$ coolant, which boiled off through a heat exchanger attached to the Xe vessel above it. The latter housed a 12.5-kg LXe WIMP target, with the immersed PMTs viewing upward to maximize detection efficiency for the primary scintillation. The active volume was formed by an anode disc 39.2 cm in diameter and a cathode wire grid located 4 cm below it, and a few mm above the PMT array.

Contrary to ZEPLIN-II, where a wire-grid just below the liquid surface helped with cross-phase emission, in ZEPLIN-III the planar geometry allowed application of a strong field to the whole liquid phase with only two electrodes, thus enhancing the efficiency for charge extraction from the particle tracks. Typical operating fields were 3–4 kV/cm in the liquid and approximately twice as strong in the gas [17,19]. A second wire-grid just above the PMTs isolated their input optics from the external field. Only Xe-friendly, low-outgassing materials were used within this chamber, in particular avoiding any plastics, in order to maintain sufficient electron lifetime in the liquid without continuous purification. This was indeed achieved, with the lifetime even improving steadily over one year of operation in the closed system [22].

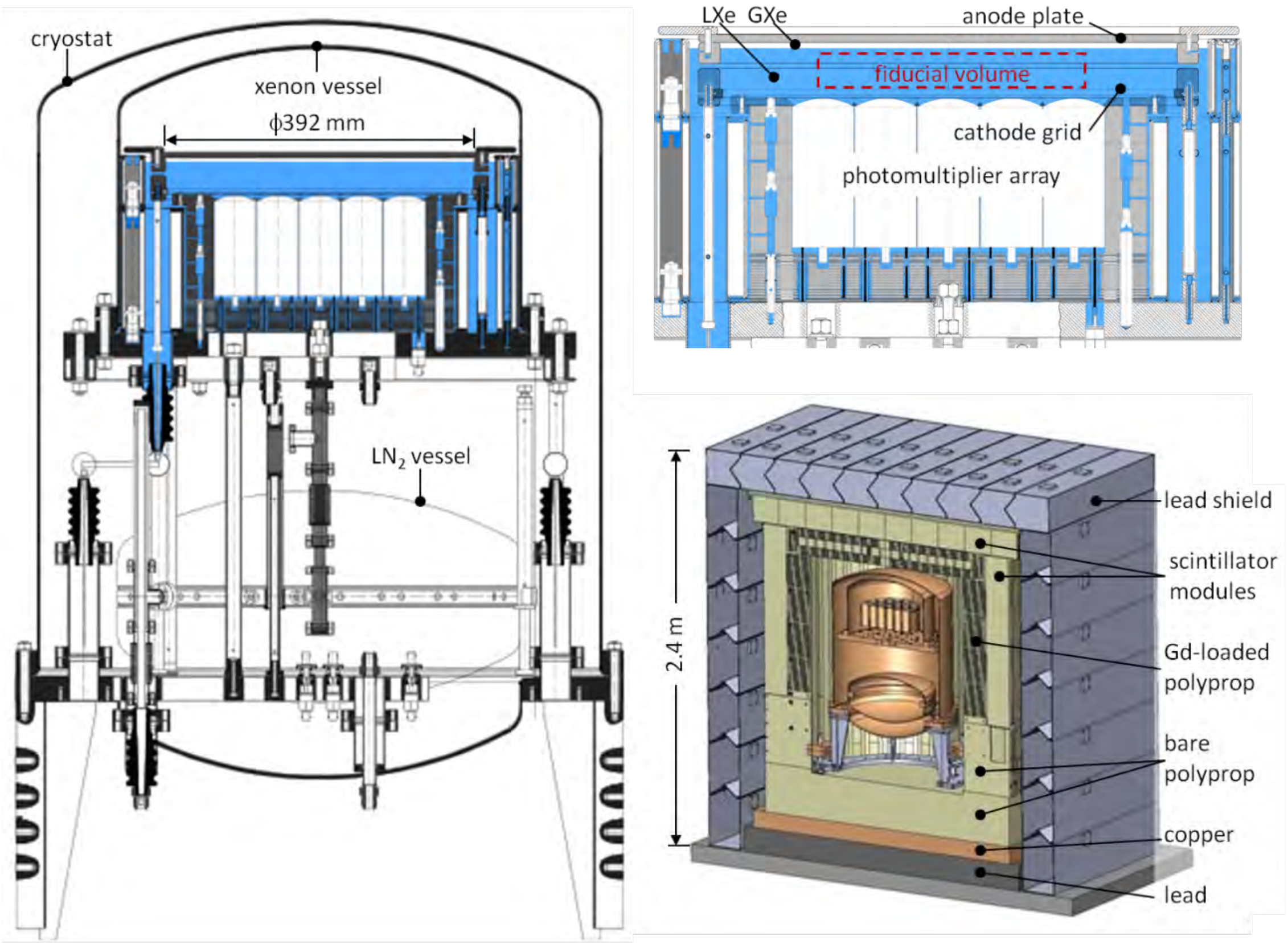

**Figure 5.1.1.1. Schematic drawings of the ZEPLIN-III experiment. Left: The WIMP target, with LXe in blue. Top right: The double-phase chamber, with an approximate fiducial volume indicated in dashed red. Lower right: The fully shielded configuration at Boulby (including a plastic veto instrument surrounding the WIMP target) [20,16,21].**

In the first science run, custom-made PMTs (ETEL D730Q/9829QA) were used; these had bialkali photocathodes with metal fingers deposited on quartz windows under the photocathode for low-temperature operation. The average (cold) quantum efficiency for Xe light was 30% [23]. For the second science run, those PMTs were replaced with a pin-by-pin compatible model with 40 times lower radioactivity (35 mBq per unit in gamma activity), lowering the overall electromagnetic background of the experiment to 750 mdru at low energy [24]. Unfortunately, their optical performance was much poorer, with only 26% mean quantum efficiency and very large gain dispersion [22]. For this reason, ETEL PMTs are not considered as a viable option for LZ.

Between the two science runs, an anticoincidence "veto" instrument was fitted around the WIMP target (shown in Figure 5.1.1.1), replacing some of the hydrocarbon shielding. This veto detector counted 52 plastic scintillator modules with independent PMT readout, arranged into barrel and roof sections, surrounding a Gd-loaded polypropylene structure tailored for neutron moderation and efficient radiative capture (vetoing ~60% of neutrons) [21,25,19]. Events tagged promptly during science running — exclusively gamma rays, vetoed with 28% efficiency [19] — provided access to a low-energy data set that could be used without compromising a blind analysis. The veto system also allowed the independent measurement of muon-induced neutron production from the lead shield around the experiment [26]. Accurate position reconstruction of particle interactions in three dimensions allows a fiducial volume to be defined very precisely, well away from any surfaces and avoiding outer regions with non-uniform electric field and light collection. A typical gamma-ray event is shown in Figure 5.1.1.2 (left). The depth coordinate was obtained with precision of a few tens of μm from the drift time of the ionization charge.

The horizontal coordinates were reconstructed from S2 signals from all PMTs; a spatial resolution of 1.6 mm (FWHM) was achieved for 122 keV gamma rays [27], using the novel Mercury algorithm now applied also to LUX. Other significant analysis algorithms were developed by the project, namely

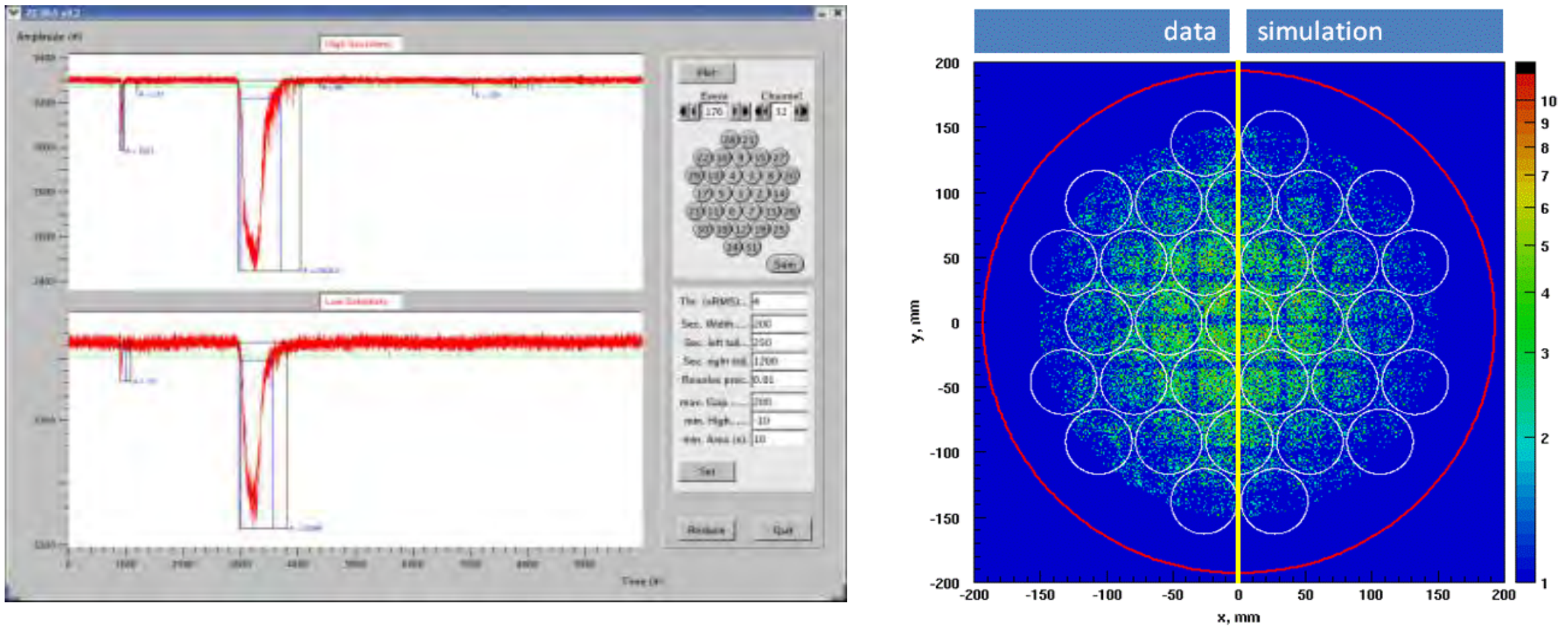

**Figure 5.1.1.2. Left: Gamma ray interaction in ZEPLIN-III, showing a fast scintillation signal (S1) followed by a large electroluminescence pulse (S2); a high-sensitivity channel is displayed in the upper panel, and a lower gain channel in the lower one. Right: Calibration and Monte Carlo data for $^{57}$Co gamma rays incident from above the detector. The grid-like structure arises from a copper absorber placed directly on top of the solid anode plate; the simulation assumes perfect position resolution; the data are reconstructed with the Mercury algorithm [27].**

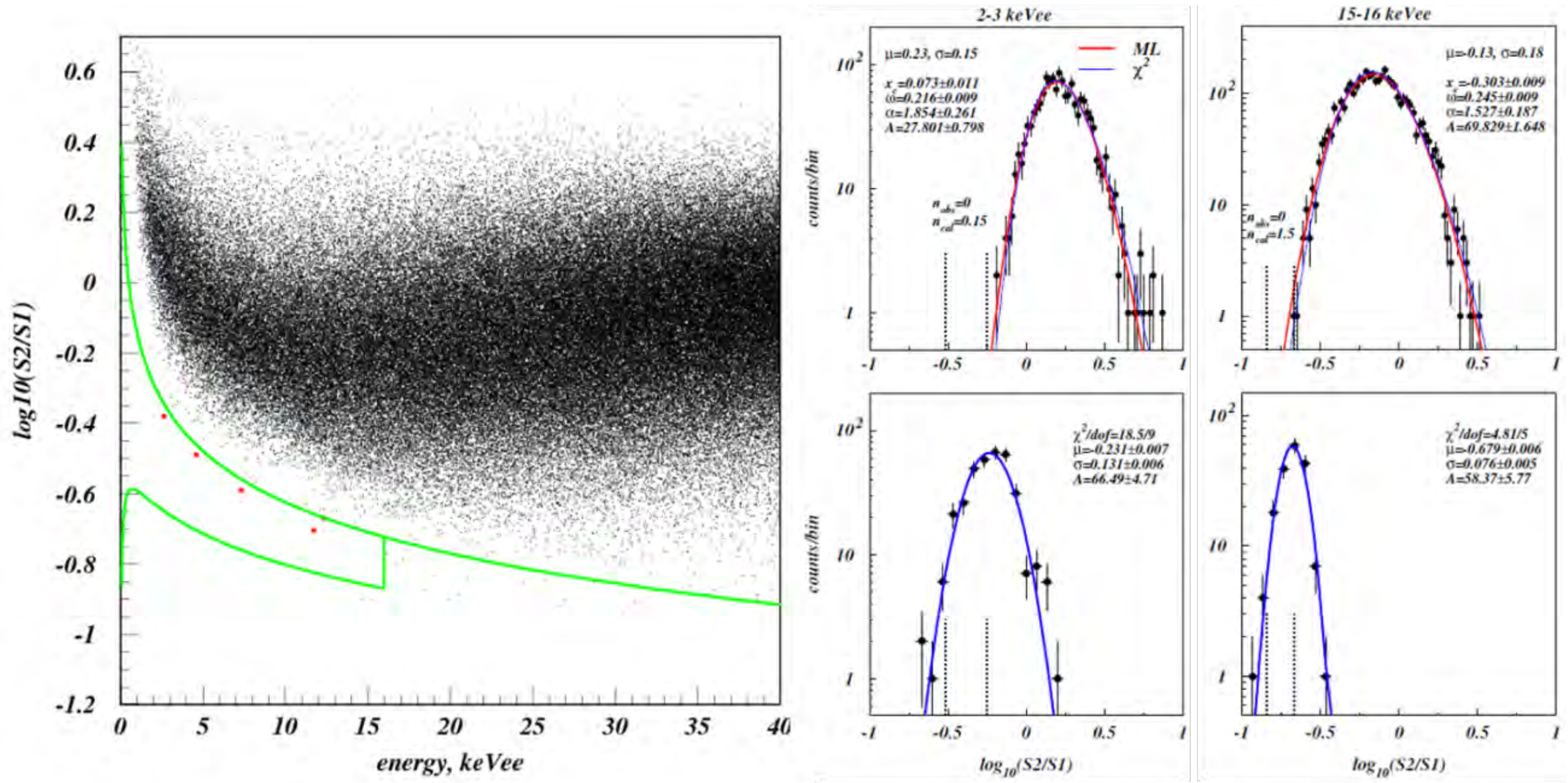

**Figure 5.1.1.3. Left: First (83-day) WIMP-search run of ZEPLIN-III; the average electron/nuclear recoil discrimination in the 2–16 keVee acceptance region was 99.99% for 50% NR acceptance. Right: Fitting of ER band for lowest and highest 1-keVee wide bins to a skew-Gaussian function (upper panel) and of the NR band, obtained with an Am-Be neutron source, with a Gaussian function [17].**

ZE3RA, a full data-reduction and display software tool [28], and a new technique to calibrate photomultiplier arrays under exact data conditions [29].

A fiducial volume containing 6.5 kg of LXe was defined for the 83-day first run of ZEPLIN-III [17], decreasing to 5.1 kg for the 319-day second run, owing to the poorer PMT performance. The NR threshold for WIMP searches was ~7 keV in both runs [30], determined by the scintillation yield of the chamber (5.0 phe/keV for $^{57}$Co gamma rays at zero electric field); the ionization threshold was five S2 electrons. A rejection efficiency of 99.99% (for ER leakage past the NR median) was achieved at WIMP-search energies in the first run, which remains the best reported for double-phase Xe. This is shown in Figure 5.1.1.3, where the S2/S1 discrimination parameter is plotted for first WIMP search; histograms of this parameter for the nuclear and electron recoil populations are also shown.

In both runs, a handful of events were observed within the signal-acceptance region, consistent with background expectations in both cases. The combined result excluded a WIMP-nucleon scalar cross section above $3.9\times10^{-44}$ cm$^2$ at 90% CL for a 50 GeV/c$^2$ WIMP mass [19].

The early parallel development of ZEPLIN-II and ZEPLIN-III contributed to the success of double-phase Xe, pursued subsequently by the XENON program and now by LUX. Different approaches were deliberately explored for most subsystems. At the core of the detectors, different designs were implemented for light collection (PMTs in the gas or in the liquid, high-reflectance PTFE chamber or shallow, disc-like target, respectively), electric field in the drift region (1 kV/cm and 4 kV/cm), design of the electroluminescence region (3-electrode and 2-electrode chambers), readout granularity, and position resolution (seven 3-inch or 31 2-inch PMTs). Other subsystems were likewise dissimilar: liquefaction method and thermal control (LXe "raining" from a "cryocooler" cold-head above the target or internal $LN_2$ heat exchanger at the bottom plate), the approach to LXe purity (external recirculation or clean chamber construction), the powering of the PMTs (internal voltage divider bases versus common "dynode plates" fed externally), general construction materials (faster construction using cast metal or machined ultrapure copper). This invaluable experience propagated to the design of other systems around the world.

## 5.2 LUX: The Large Underground Xenon Experiment

The LUX experiment [31] is the most recent two-phase Xe dark-matter detector to begin operations, and in its first result [32] it has achieved world-leading WIMP sensitivity for WIMP masses above 6 GeV/$c^2$. The basic technology follows on more than a decade of effort from the ZEPLIN and XENON programs.

LUX has introduced a number of important innovations that will be important for LZ, including a low-radioactivity titanium cryostat, nitrogen thermosyphons, high-flow Xe purification, two-phase Xe heat exchangers, internal calibration with gaseous sources of $^{83m}$Kr and $^{3}$H, and nuclear recoil calibration using multiple scatters of monoenergetic neutrons. A schematic of LUX is shown in Figure 5.2.1.

The LUX cryostat vessels are fabricated from Ti with very low levels of radioactivity [33], rivaling even copper, which is highly radiopure but has inferior mechanical properties and is much denser. The high strength-to-weight ratio and low Z of Ti also give it low stopping power for gamma rays, which will enhance the efficiency of the outer detector system in LZ.

The two-phase TPC technique requires precise control of the thermodynamic environment, and this was achieved in LUX through the development of an innovative system of nitrogen thermosyphons. These feature precise, automated control of cooling up to hundreds of kW, plus simple and reliable remote operation. Among other things, this system has allowed highly controlled initial cooling of the system [34], which is necessary to avoid warping of the large plastic structures of the TPC — also a concern in LZ. A related development is the fluid-circulation system that allows rapid circulation of LXe through an external gas-phase getter at flows exceeding 25 standard liters per minute, while maintaining a stable liquid surface through the use of a weir. Using an innovative two-phase heat exchanger system [35], it obtained very efficient heat transfer between evaporating and condensing Xe streams, and thus negligible overall heat load on the detector. Xe purification was greatly aided by the use of an innovative gas trapping and mass spectrometry system [36,37] that is sensitive to impurities at the sub-ppb concentration needed for good electron transport and light collection. This qualitatively new level of diagnostic capability allowed us to monitor various portions of the gas system and detector and efficiently track down any sources of contaminant. These systems were demonstrated during LUX's first science run, where cooldown was achieved in only nine days, and sufficient purity in terms of electron drift lifetime to begin science operations was achieved only ~1 month after filling with LXe (see Figure 5.2.2). Afterward, we achieved stable operation of the detector with mostly unattended operation over a several-month period, during which the pressure and liquid level had sufficient stability (1% and <500 µm, respectively) to introduce no measurable variations in the S2 or S1 signals.

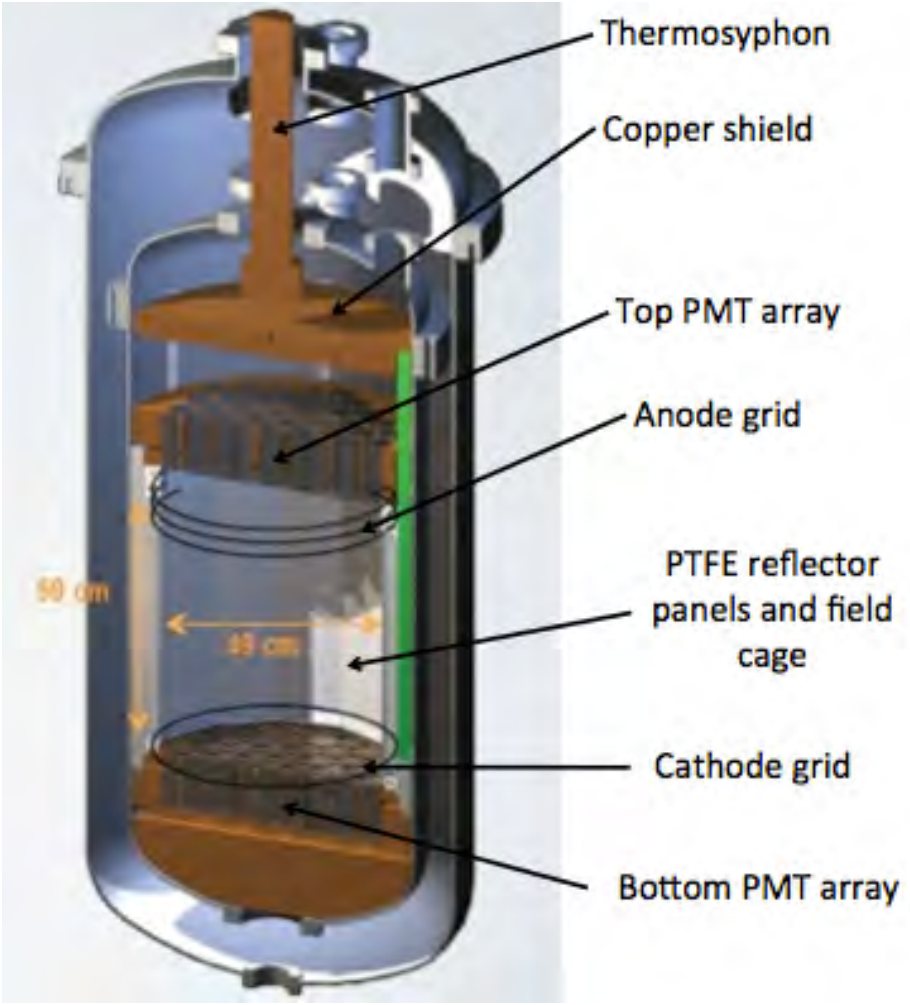

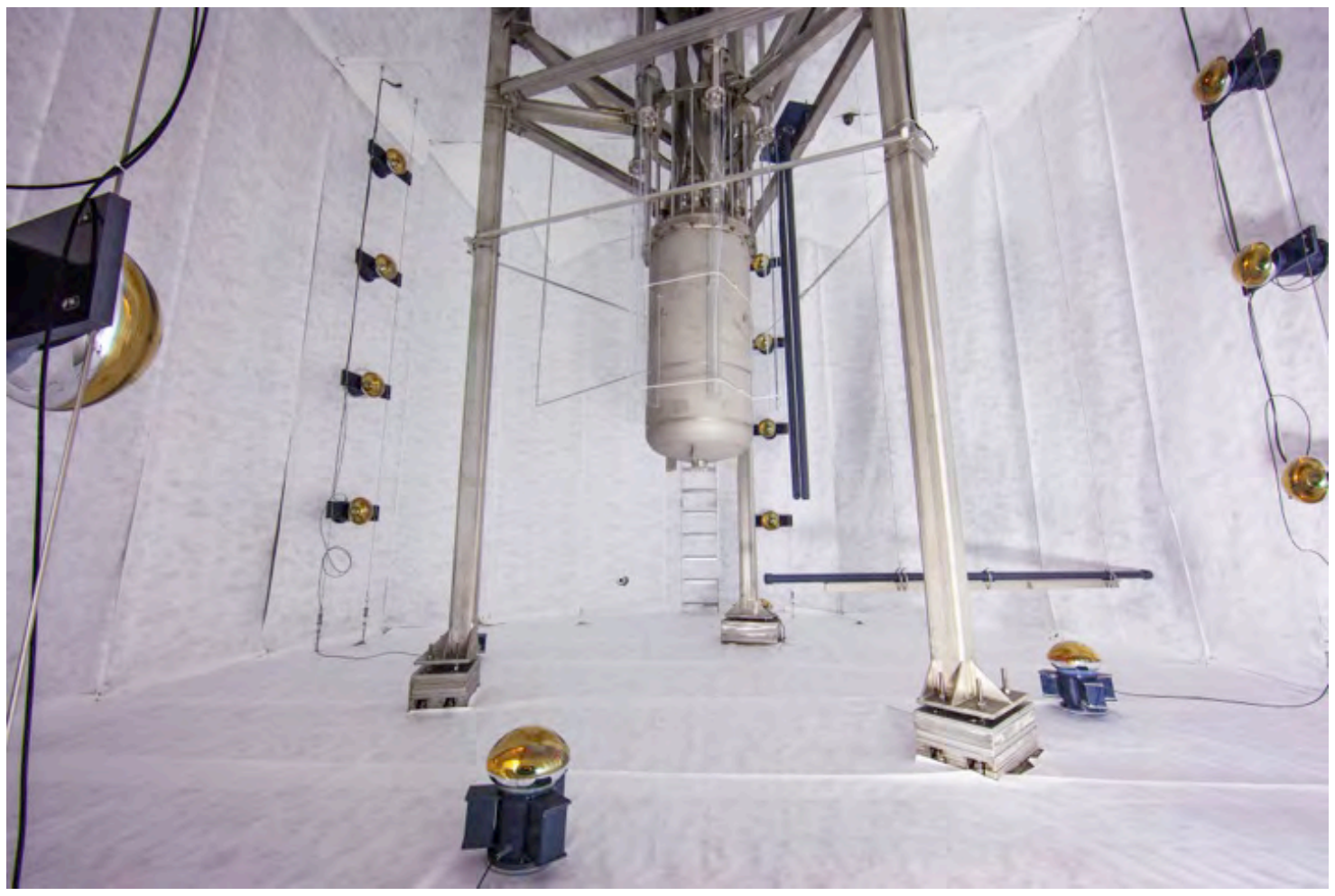

**Figure 5.2.1. Left: LUX features a ~50-cm-tall, 50-cm-diameter TPC containing 250 kg of active LXe mass with PMTs at top and bottom. Right: LUX as installed in the Davis Cavern water shield.**

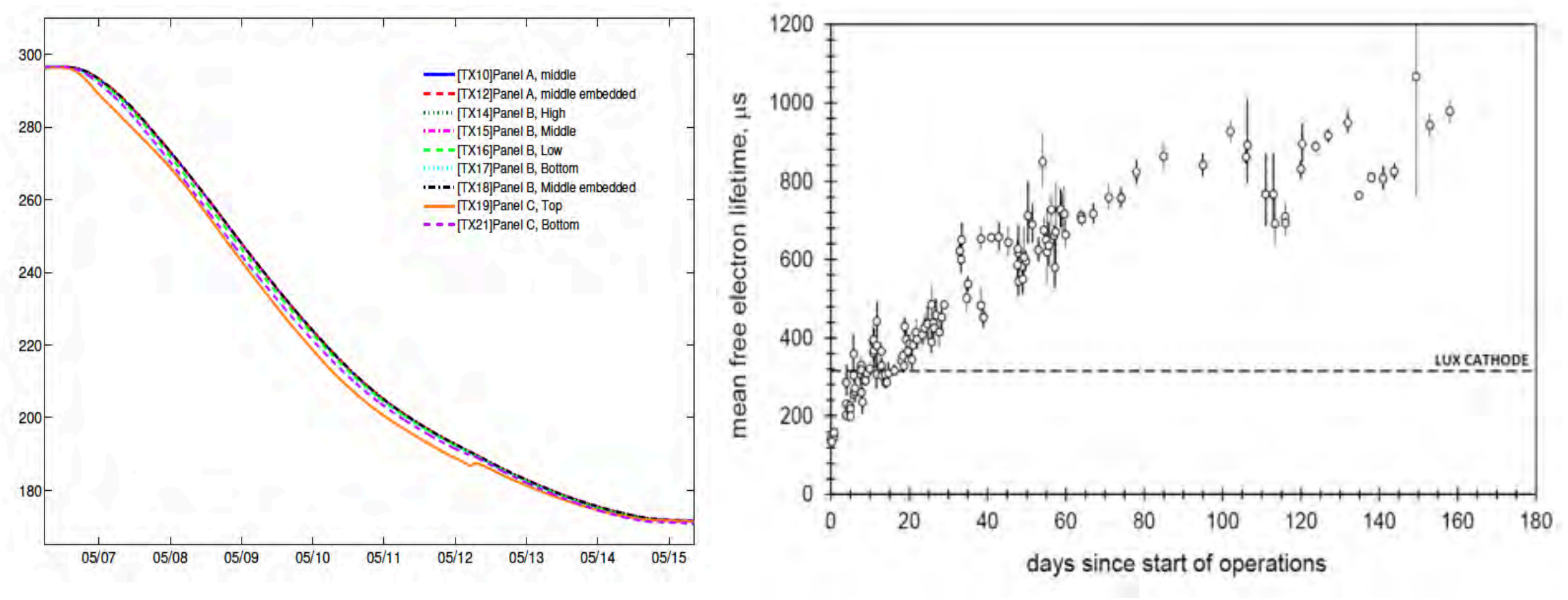

**Figure 5.2.2. Left: Measured temperatures by thermometry in the polyethylene panels surrounding the active region during LUX cooldown, which was completed over a period of nine days. Right: Mean free electron lifetime, measured from the start of underground operations.**

The 2-inch-diameter Hamamatsu R8778 PMTs used in LUX have high quantum efficiency and a low radioactive background of roughly 10 mBq/unit in terms of gamma-ray emission [38]. The low-noise amplifiers and electronics system used to read out the PMTs have resulted in ≥95% efficiency for single phe detection. The data acquisition (DAQ) system [39] features pulse-only digitization that significantly reduces the data-set size; especially important given the long drift time in LUX. The system also utilizes a sophisticated digital trigger system that will form the basis for the LZ DAQ system.

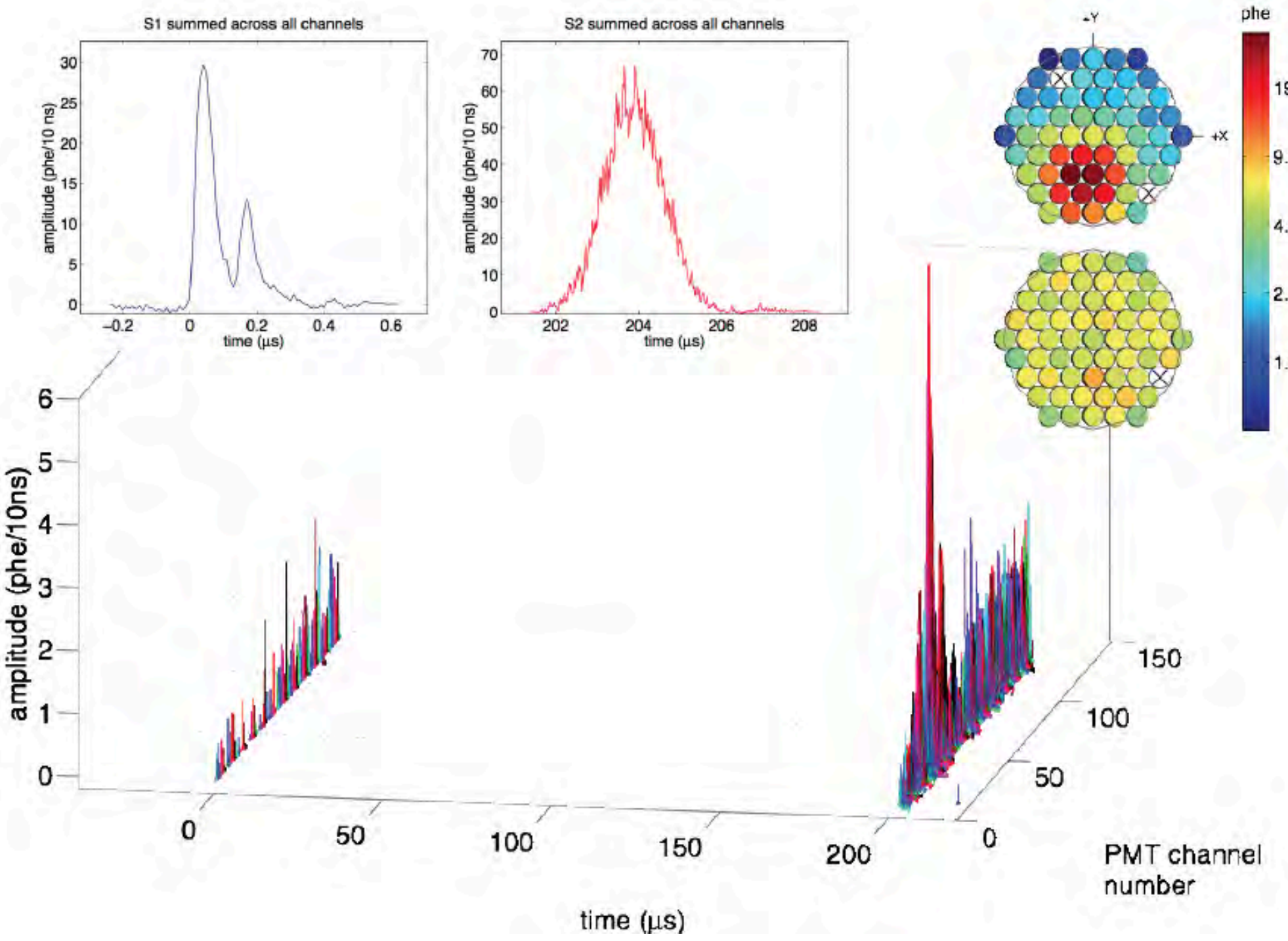

**Figure 5.2.3. An example calibration event from LUX, produced from $^{83m}$Kr decay.**

Calibration of a large two-phase Xe TPC is difficult because of the strong self-shielding of the LXe, and the near impossibility of introducing a sealed source into the TPC interior without perturbing the fields. LUX therefore developed two novel gaseous internal calibration sources, both of which will be used for LZ. The first uses $^{83m}$Kr [40], which has a ~41 keV decay and 1.8-hour half-life to stable $^{83}$Kr. This was used in LUX for the first time in an extensive and ongoing way to map the position response of both S1 and S2, and the stability of these signals and other aspects of the detector, including fluid flow, electric field distribution, and electron drift length. A sample $^{83}$Kr event is shown in Figure 5.2.3. The second provides a calibration of the response to the dominant background of ERs in the WIMP energy range. Here we used the beta decay of $^{3}$H (18.6 keV endpoint, 12.3-year half-life), dispersed in the form of tritiated methane. The key issue is removal of this long-lived species. This was accomplished using the getter purification system, which reduced the $^{3}$H concentration with a ~7-hour exponential time constant, down by at least a factor of ~$10^4$, at which point the $^{3}$H could not be seen above other backgrounds. Finally, we have completed a novel in situ neutron calibration using mono-energetic neutrons from a DD generator deployed outside the water shield and an evacuated tube in the shield serving as a collimator.

The final backgrounds in the central fiducial region in LUX agree well with expectations based on the screening of components prior to construction, and within the WIMP region of interest was measured to be $3.6 \times 10^{-3}$ events/keV/kg/day, which is the lowest ER background ever recorded below 100 keV. Background distributions within the fiducial volume are shown in Figure 5.2.4. The gamma portion is dominated by PMTs and cosmogenic activation of Cu, while the internal activity is currently dominated by $^{127}$Xe (36.4-day half-life), but also has small contributions from a $^{214}$Pb, a Rn daughter, and $^{85}$Kr. $^{85}$Kr is an especially problematic background as it is present in commercial Xe at unacceptable levels. LUX developed a chromatographic system [41] that processed 50 kg of Xe per week to an average level of 4 ppt (the initial goal was 5 ppt), with one batch doubly processed to better than a measurement limit of 0.2 ppt. Measurements at this level were made using LUX's custom mass spectrometry plus trapping technique [37].

The ER discrimination has been expected to improve with both light collection and drift field, as described in Chapter 3. Light collection was optimized in the design by maximizing surface coverage with PTFE and using high-transparency grids. The measured light-collection efficiency (phe per initial photon) of $\alpha_1$=0.14 (8.4 phe/keV at 662 keV gamma rays at zero field) is the highest value yet obtained in a large LXe TPC. This high light collection has resulted in a very low S1 threshold of 4.3 keV for a 2-phe coincidence level. The drift field in LUX was limited to ~180 V/cm for stable operation in 2013. Despite this, and possibly in part due to the excellent light collection, discrimination is measured to be >99.6% using $^{3}$H calibration (see Figure 5.2.5). This exceeds the original LUX goal — and in fact exceeds the LZ baseline assumption for a higher electric field. All these factors have combined to give LUX the world's best WIMP sensitivity in an initial limited duration run with a maximum sensitivity at 33 GeV/c$^2$ of $7.6 \times 10^{-46}$ cm$^2$, and a powerful reach at low WIMP masses (see Chapter 1). A second, longer LUX run with improved reach started in October 2014 and is anticipated to continue into 2016.

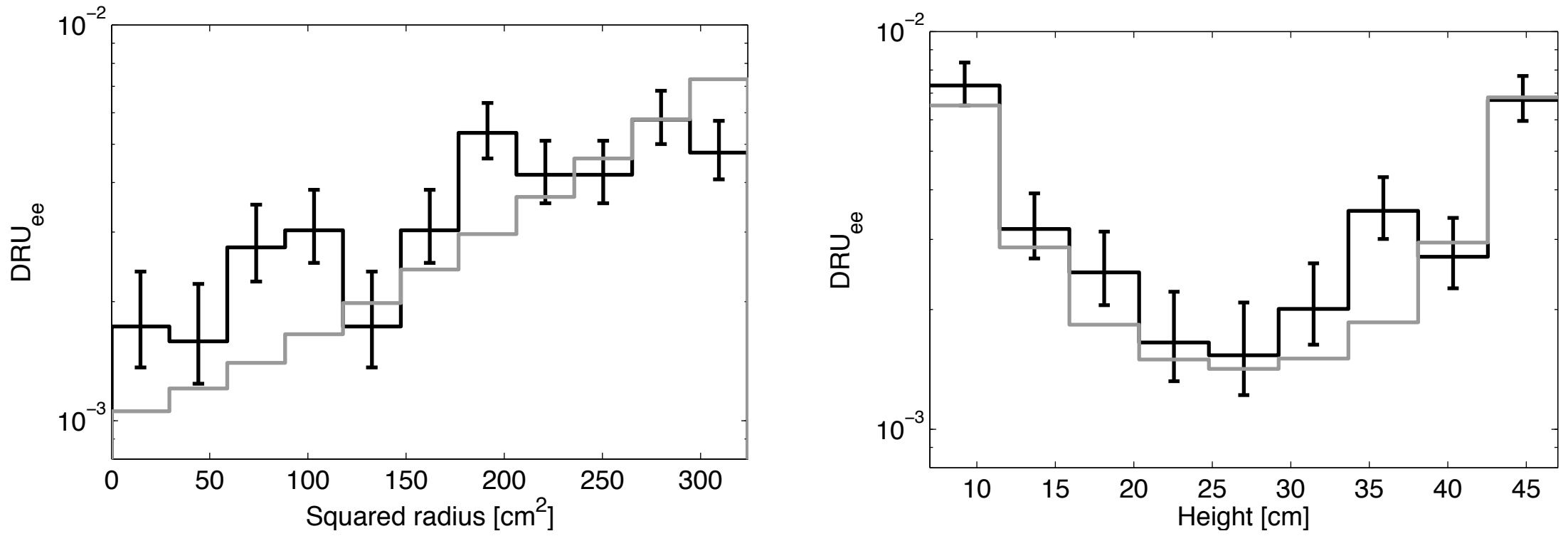

**Figure 5.2.4. Low-energy measured distributions in (left) squared radius, and (right) height, within the LUX 118 kg fiducial mass, measured over the full 85.3-day WIMP search run [32]. Measured data are indicated by the black histogram with error bars. Simulation data are shown as the gray histogram.**

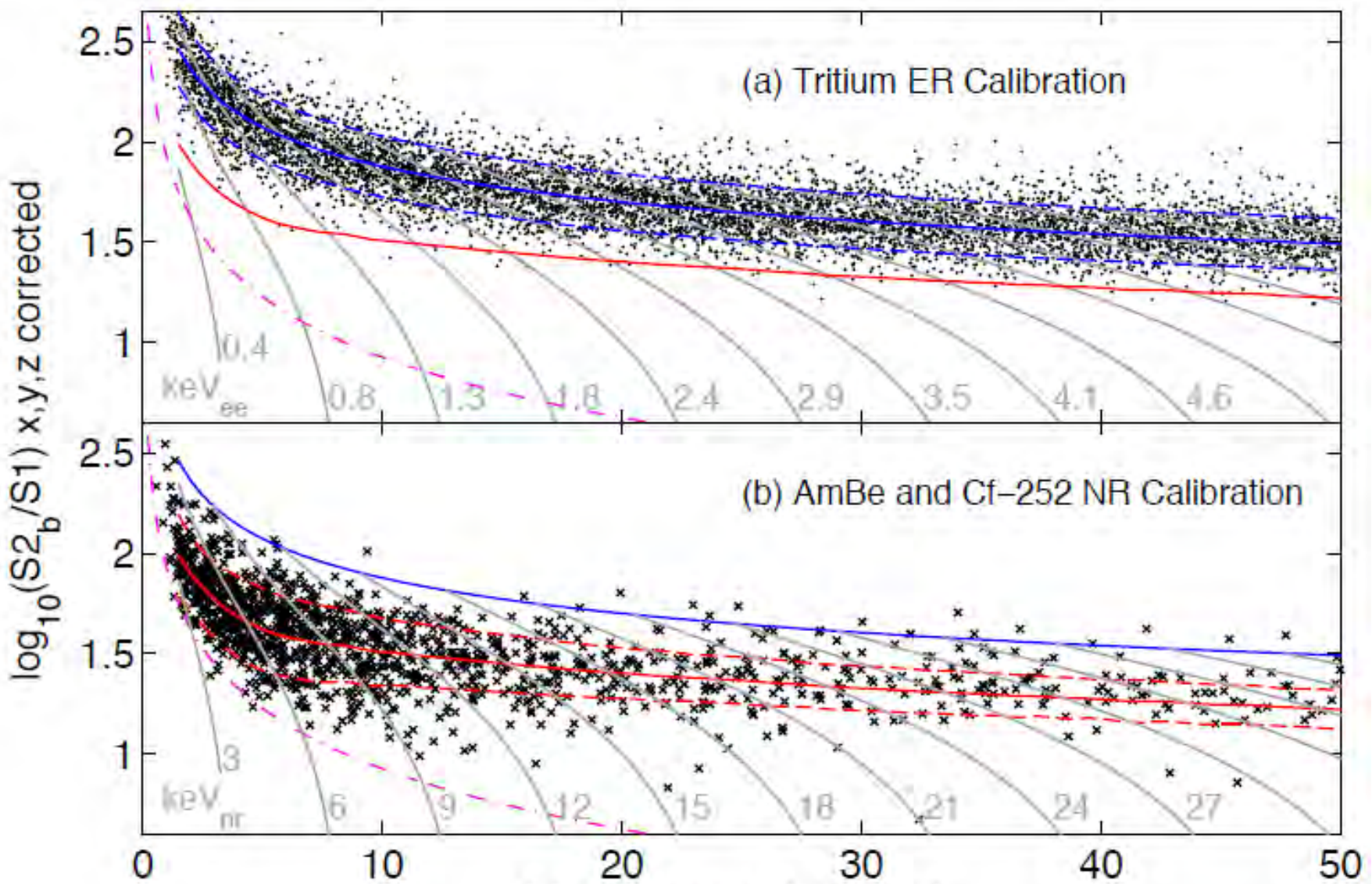

**Figure 5.2.5. Discrimination demonstrated in LUX [3] with (a) ER events from beta decay of tritiated methane (discussed below) and (b) NR events from neutrons from AmBe and $^{252}$Cf sources. Overlaid curves are the means and 80% inclusive bands for the ER (blue) data and NR (red) distributions. The discrimination defined by the leakage of ERs to below the NR band mean in the 2-30 pe S1 range is 99.6±0.1%. Energy contours are shown, along with (purple) the approximate location of a 200 phe cut on raw S2 size.**

## 5.3 Beyond LUX and ZEPLIN

For all the experience accumulated through the LUX and ZEPLIN programs and others, it is necessarily the case that LZ must push many aspects of this technology into new regimes in order to reach its unprecedented sensitivity. The scale-up of some subsystems is more challenging than for others, and we conclude this section with a table highlighting the key differences between LZ and LUX (Run 3) or other LXe detectors — which we indicate in brackets if not LUX. We also assign a "technical difficulty" (TD) indicator to relevant parameters, representing how significant a challenge we deem this scaling to be with respect to what has been achieved previously. We note, however, that this cannot be simply interpreted as a risk factor since we will normally have in place additional mitigation for the more challenging issues. This mitigation (e.g., R&D) is also indicated. The TD indicator ranges from "o" for minimal, to "ooo" for significant.

**Table 5.3.1. Key differences between the LZ baseline design and LUX, ZEPLIN, EXO, and other experiments.**

| WBS Item | LZ | LUX et al. | Comments/Mitigation | TD |
|---|---|---|---|---|
| **1: XENON** | | | | |
| **Total mass** | 10,000 kg | 370 kg | 27x | ooo |
| **Active mass** | 7,000 kg | 250 kg | 28x | |
| **Fiducial mass** | 5,600 kg | 118 kg | 47x | |
| **2: CRYOSTAT** | | | | |
| **Material** | Titanium (CP-1) | Titanium (CP-1) | Stainless steel design backup | o |
| **U/Th activity** | <0.75/0.75 mBq/kg | 0.25/0.2 mBq/kg | 2-year search campaign | oo |
| **IV/OV wall** | 6-8/8 mm | 6/6 mm | | o |
| **Max Op Pressure** | 4.0 bar(a) | 4.0 bar(a) | | o |
| **Height** | 3.2 m | 2 m | Excluding supports | o |
| **Weight** | 1,800 kg | 325 kg | Excluding supports | o |
| **IV support** | Suspended | Suspended | Fine leveling by suspension rods | oo |
| **OV support** | Bottom legs | Suspended | | o |
| **Conveyance** | OV in 3 parts | Whole | Detector delivered with IV | o |
| **3: CRYOGENICS** | | | | |
| **Detector cooling** | External delivery of subcooled LXe | Internal thermosyphons | Reduces detector complexity<br>Allows separate system test | oo |
| **Cooling power** | 1 kW | 0.4 kW | Inc. Purification Tower | o |
| **Liquid nitrogen** | Cryocooler | Delivery u/g | Delivery u/g is backup | o |
| **4: PURIFICATION** | | | | |
| **Electron lifetime** | 3 ms goal<br>0.75 ms req. | 1 ms (also Z2)<br>(EXO >4 ms) | Emanation budget<br>Improved gas sampling | o |
| **Recirculation** | Compressors | PTFE-diaph. pump | Higher flow, more reliable | o |
| **Recirculation rate** | 500 slpm<br>(4,300 kg/day) | 25 slpm<br>(230 kg/day) | | oo |
| **Xenon recovery** | Compressor-based recovery | Cold recovery vessel + bladder | Passive capture not viable | ooo |
| **Kr requirement** | 0.015 ppt | 5 ppt | <0.2 ppt demonstrated | ooo |
| **Kr sampling sensitivity** | 0.015 ppt | 0.2 ppt | LZ sampling fully automated | oo |
| **Radon** | 0.67 mBq | ~10 mBq<br>0.4 mBq in EXO | Screening program (WBS 1.10) | ooo |
| **5: XENON DETECTOR** | | | | |
| **TPC length/width** | 1.46/1.46 m | 0.5/0.5 m | | o |
| **TPC construction** | Segmented rings | Vertical panels | Better electrostatics/mechanics | o |
| **Therm. contraction** | ~2 cm | ~0.7 cm | TPC anchored at bottom | oo |
| **S1 Photon Det. Eff.** | 7% | 11% | Control sources of optical extinction | ooo |
| **Skin dielectric** | 4-8 cm LXe | 5 cm UHMWPE | Instrumented (enhances veto) | oo |

| WBS Item | LZ | LUX et al. | Comments/Mitigation | TD |
|---|---|---|---|---|
| **Cathode voltage** | –100 kV (–200 kV design) | –10 kV (Z3: –10 kV) | LUX field matched at – 30 kV Extensive R&D & System Test | ooo |
| **Cathode delivery** | Own conduit | Thru skin | Risk mitigation | ooo |
| **Rev. field region** | Integral design, voltage grading | LXe gap | Better electrostatics/mechanics, Optical isolation from skin region | ooo |
| **Field in gas** | 10.6 kV/cm goal | 6.6 kV/cm (Z2: 8.4 kV/cm) | R&D & System Test | oo |
| **Emission prob.** | >95% | 65% | Decrease delayed e-emission | oo |
| **S2 gain** | 50 phe/e | 14 phe/e (Z3: 30 phe/e) | Top array only | oo |
| **PMT numbers** | 247/241/180 | 61/61/0 | Top/Bottom/Skin | o |
| **PMT model (TPC)** | 3-inch R11410 | 2-inch R8778 | | o |
| **PMT activity (TPC)** | 3/3/30/3 mBq | 10/3/66/3 | $^{238}U/^{232}Th/^{40}K/^{60}Co$ (requirements) | oo |
| **PMT mounts** | ¼-inch Ti plate + trusses | Solid Cu block with cutouts | Lightweight construction Higher OD/Skin veto efficiency | o |
| **Fluid circulation** | Full sweep Min. dead space | Single in/outlet | | o |
| **TPC monitoring** | Many sensors | LXe level and thermometry | HV diagnostic, protection | o |
| **6: OUTER DETECTOR** | | | | |
| **Vetoing media** | Liquid scintillator Water Cherenkov | (Z1/2/3 vetos) Water Cherenkov | Acrylic tanks in water, common PMTs | ooo |
| **Scintillator type** | Gd-loaded LAB | (Z2 LS, Z3 plastic) | cf. Borexino, Daya Bay, SNO+ | oo |
| **Scintillator mass** | 27 tonnes | (Z1/2/3 ~1 tonne) | | o |
| **Scintillator purity** | 1.2/4.7 ppt U/Th | n/a | U at 10% Daya Bay; Th is similar | ooo |
| **Transparency** | >10 m at 430 nm | n/a | | oo |
| **Light yield** | 9000 ph/MeV | n/a | | o |
| **PMT channels** | 120 | (Z2 10, Z3 52) | Using Xenon Detector DAQ | o |
| **PMT calibration** | Fibres | LEDs | Experience: SNO+, Z3 | o |
| **7: CALIBRATION** | | | | |
| **NR calibration** | AmBe, $^{252}Cf$, YBe D-D generator | AmBe, $^{252}Cf$ D-D generator | Special YBe port and shielding | o |
| **ER calibration (int)** | $^{83m}Kr$, $CH_3T$, $^{37}Ar$ | $^{83m}Kr$, $CH_3T$ | | o |
| **Gamma sources** | Pipes in OV/IV | Pipes in water | | o |
| **8: ELECTRONICS** | | | | |
| **Channels** | 1,276 | 122 | Inc. TPCx2+Skin+OD | oo |
| **Trigger rate** | 40 c/s | 10 c/s | All energies | o |
| **Pre/post-amplifiers** | 100 | 16/16 | Pre- and post-amps are integrated | o |
| **Front-end BW** | 30 MHz | 30 MHz | At digitizer input | o |
| **Noise** | 0.2 $mV_{rms}$ | 0.2 $mV_{rms}$ | At digitizer input | o |

| WBS Item | LZ | LUX et al. | Comments/Mitigation | TD |
|---|---|---|---|---|
| **Digitizers** | DDC32 | Struck SIS3301 | | o |
| **Operating mode** | POD | POD | POD: Pulse Only Digitization | o |
| **Resolution** | 14 bits | 14 bits | Dual range | o |
| **Sampling** | 100 MS/s | 100 MS/s | | o |
| **Memory depth** | >163k samples for 32 channels | 128k samples | ~1 ms maximum drift time; For Spartan 6 FPGA; significantly larger for Kintex FPGA | o |
| **Trigger** | DDC-32 | DDC-8 DSPs, Trigger Builder | LZ trigger and DAQ firmware run in parallel on the same FPGA | o |
| **Data volume** | 750 TB/yr | 40 TB/yr | Size of event files | o |
| **Slow Control** | MySQL custom sys + PLC system | MySQL custom sys | | oo |
| **Power** | 110 kW | 18 kW | Maximum, full load | oo |
| **9: ASSEMBLY & INSTALLATION** | | | | |
| **Detector assembly** | Rn-scrubbed CR, Static precautions | Surface CR, No static prec. | SURF surface building | oo |
| **U/G deployment** | Detector in IV, conv. horizontally | Detector in IV/OV + conduits, whole | | ooo |
| **10: SCREENING** | | | | |
| **Cleanliness control** | Full QA/QC | Limited QC | Control of PTFE debris | oo |
| **HPGe screening** | 100 ppt, in-house | 100 ppt, in-house | U/Th; already demonstrated | o |
| **ICP-MS screening** | 10 ppt, in-house | Commercial | U/Th; already demonstrated | o |
| **GD-MS screening** | 10 ppt, commercial | None | U/Th; already demonstrated | o |
| **Neutron activation** | 20 ppt, in-house | Limited | U/Th demonstrated for PTFE | o |
| **Rn screening** | 0.03 mBq, in-house | None | 0.1 mBq already demonstrated | oo |
| **Radon plate-out: wall events, ($\alpha$,n)** | Full QA/QC, Rn-reduced air CR, | Fiducial reduction limited Z2 | QA/QC, Rn-scrubbed CR Emanation budget | ooo |

## Chapter 5 References

# 6 Xenon Detector System

## 6.1 Overview

The direct observation of WIMP dark matter scattering within a detector of any kind is a significant experimental challenge. Searches for these elusive particles require an extremely sensitive, low-background detector able to separate NR events at the few-keV energy from a dominant background of ER interactions, some created by particles external to the WIMP target and others arising within it from radioactive contaminants. The LZ experiment addresses part of the background issue by operating deep underground, and surrounding the instrument with a set of concentric water and Gd-LS veto shields that are described in detail in Chapter 7. However, observing these small energy depositions in space and time requires a highly instrumented LXe TPC assembled from high-performance, low-radio-background components operating in the cold liquid that have been developed explicitly for this purpose. The design of these elements has been influenced extensively by our collaboration's experience with operating both the LUX and ZEPLIN experiments as described in Chapter 5.

While LXe is inherently a very radio-quiet detector material with enough density and Z to very effectively self-shield from external backgrounds, the design of this new detector nevertheless requires that attention be paid to the radiopurity of a number of significant detector elements, such as the PMTs, support structures, and reflecting surfaces. This imposes serious constraints on material composition and their location, adding significant complication to the instrument's design. The details of these developments will be described in the related sections below.

The Xe detector system includes the TPC and ancillary systems required for its readout, control, and monitoring (cables and conduits, monitoring sensors, etc.). An additional anticoincidence detector is formed by a layer of LXe enveloping the TPC, which we term the "skin" detector. The main components of these two instruments are described in this chapter: the TPC, including HV delivery, PMT systems, and internal liquid flow and monitoring instrumentation; and the skin detector and its readout.

The TPC itself has a three-electrode configuration: a cathode grid at the bottom, a gate grid just below the liquid surface, and an anode grid just above the liquid surface. It features two arrays of PMTs, one immersed in the LXe viewing up, and the other in the gas phase viewing down. The WIMP target contains some 7 tonnes of active LXe, located vertically between the cathode and gate grids and enclosed laterally by a cylindrical arrangement of PTFE reflector panels. Interactions in this region generate prompt VUV scintillation light detected by the PMTs (S1 pulse). The applied electric field sweeps the ionization charge liberated at the interaction site and drifts it upward past the gate electrode; these electrons are extracted into the vapor phase, where they generate electroluminescence — which is again detected by the same two PMT arrays (S2 pulse). This double-phase (liquid/gas) technique, which generates two pulses per interaction, resolves the energy deposition sites with great spatial accuracy down to very low energies, allowing identification of multiple scatter events and, as described previously, providing discrimination between ER and NR interactions.

Table 6.1.1 lists the key design parameters of the Xe detector system and performance specifications needed to meet the scientific goals described previously.

An important enhancement beyond LUX is the treatment of the skin layer of LXe located between the stack of PTFE reflector panels that surround the active region and the cryostat wall, as well as the region beneath the bottom PMT array. A very-high-quality dielectric standoff is needed between the very-high electric field portions of the field cage and the grounded metallic vessel wall. A few-cm-thick layer of LXe is excellent for this role, with the added advantage of allowing measurement of any energy deposited in this layer, from which we read out the scintillation light. Operated as a stand-alone veto, this layer is insufficiently thick to have high efficiency. However, the combination of this skin detector and the outer LS detector forms a highly efficient tag of internal and external backgrounds. The efficiency is further enhanced by the overall minimization of inert materials that can absorb gammas and neutrons: The TPC is constructed of the minimum needed mass of PTFE and field-shaping rings, and the vessels and PMT support structures are made of Ti. Both PTFE and Ti are low density and low Z, and thus highly transparent to gamma rays. Important design drivers for the skin are its optical decoupling from the TPC, and compatibility between the skin readout and the TPC HV design.

**Table 6.1.1. Major parameters of the Xe detector system.**

| Item | Parameter |
|---|---|
| Liquid Xenon | Total mass = 9.6 tonnes<br>Active mass= 7.0 tonnes |
| **Vertical dimensions** | |
| Drift region (cathode-anode) | 1.46 m |
| Extraction region (gate-anode) | 1.0 cm (0.5 cm liquid, 0.5 cm gas) |
| Reverse field region (sub-cathode) | 14.0 cm |
| **Lateral dimensions** | |
| TPC Diameter | 1.46 m |
| Field cage wall thickness | 2.0 cm |
| Skin thickness — wall region | Min (max) = 4.0 (8.0) cm |
| **Grid transparencies at normal incidence** | |
| Bottom shield | 96 % |
| Cathode | 92 % |
| Gate | 98 % |
| Anode | 76 % |
| Top shield | 99 % |
| **Operating conditions** | |
| Cathode voltage | –100 kV |
| Gate voltage | –4 kV |
| Anode voltage | +4 kV |
| Gas region field | 10.6 kV/cm |
| Drift region field | 0.7 kV/cm |
| Design target highest surface field (in LXe) | 50.0 kV/cm |
| Operating pressure | 1.6 bar |
| **Photomultipliers** | |
| TPC 3” Ø phototube count | Top (Bottom) = 247 (241) tubes |
| Xenon skin 1”-square phototube count | Sides (Bottom) = 120 (60) |

Another area of major difference between the device proposed here and the previous LUX and ZEPLIN detectors is the side-entry method of bringing in the very-high-voltage connection to the cathode, and the short “reverse-field” region between the cathode and the lower PMT array. This reverse-field region is especially challenging in LZ because of the very-high electric field there, which results from having the highest possible voltage on the cathode while simultaneously minimizing the mass of LXe between the cathode and the bottom PMT array. Our approach to these issues is described below in separate sections on the reverse-field region and cathode HV delivery system. An overview of the Xe detector system is shown in Figure 6.1.1.

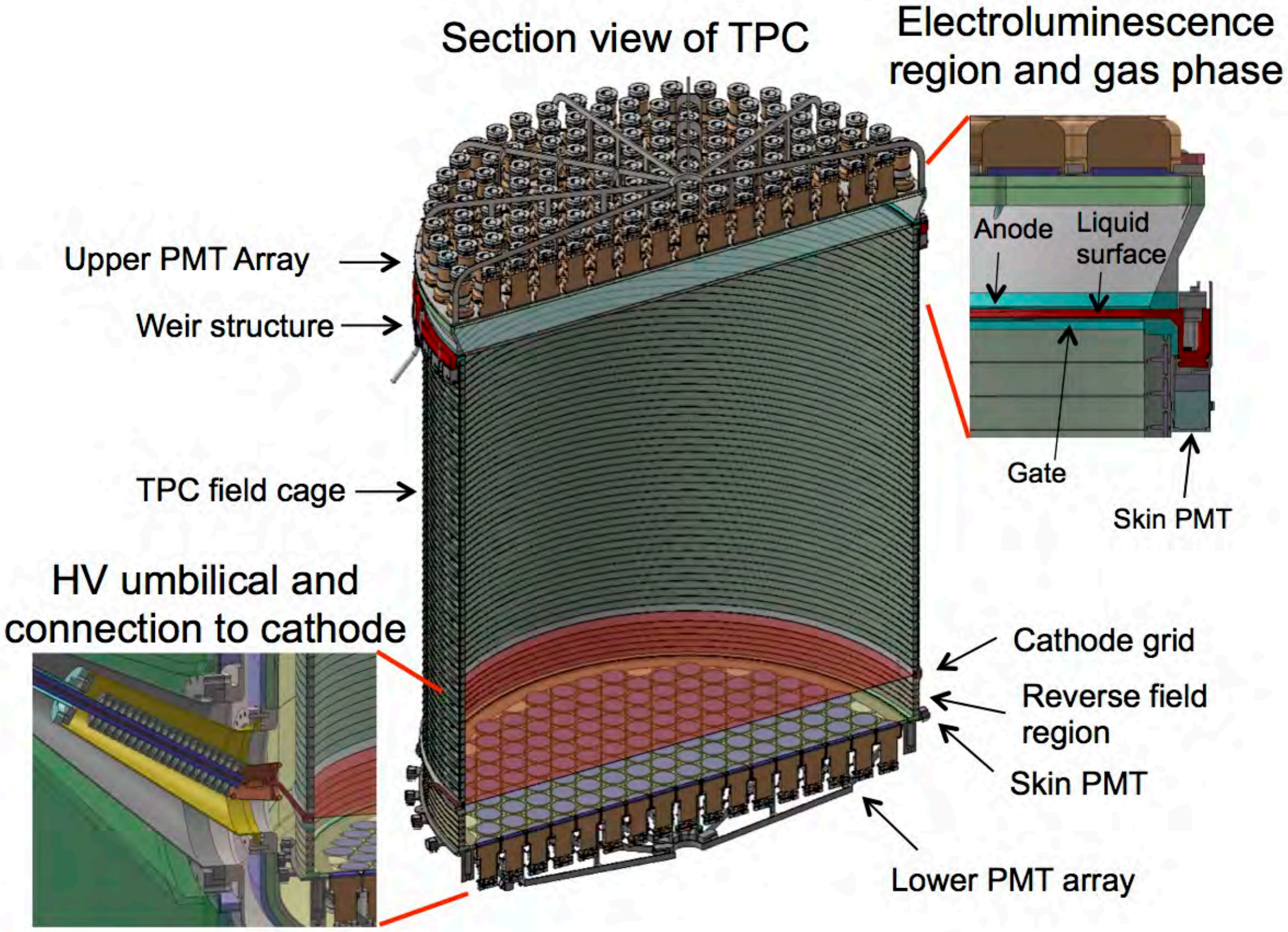

**Figure 6.1.1. Schematic views of the Xe detector. The 7-tonne active region is contained in the TPC field cage between the cathode and gate electrodes, viewed by PMT arrays in the vapor and liquid phases. S2 signal generation occurs between the liquid surface and the anode (right inset). The HV connection to the cathode (left inset) uses a dedicated conduit leading from outside of the water tank. Below the TPC, the reverse-field region grades the cathode potential to low voltage at the bottom PMT grid. The lateral skin PMT readout is shown outside of the TPC field cage.**

By design, the structures surrounding the central Xe volume are as lightweight as possible for transparency to gammas and neutrons, and this also helps keep their total radioactivity low. The most challenging requirements on the intrinsic radioactivity (i.e., radioactivity per mass or area) are in the largest or most massive components — the PTFE walls and field-shaping rings, and the PMTs with their bases and cables. This section discusses the approach to obtaining the needed radioactivity levels for a number of these major items. However, the absolute level of radioactivity of everything in the detector system must be held at acceptable levels, so all components must be carefully selected and screened. We discuss the screening program that ensures this in Chapter 12.

## 6.2 Central TPC: Field Cage, PTFE Reflectors, and Grids

At the heart of the TPC are the field cage embedded in the reflective PTFE panels, and the grids. The grids and field cage create the set of electric fields that drift the electrons to create the S2 signal, and the highly reflective PTFE panels are essential to efficient measurement of the initial S1 scintillation signal.

## 6.2.1 Electric Field Design

The electric field configuration inside the TPC volume is made up of three distinct regions, described in detail in this section: (1) the drift region, (2) the extraction and electroluminescence region, and (3) the reverse-field region.

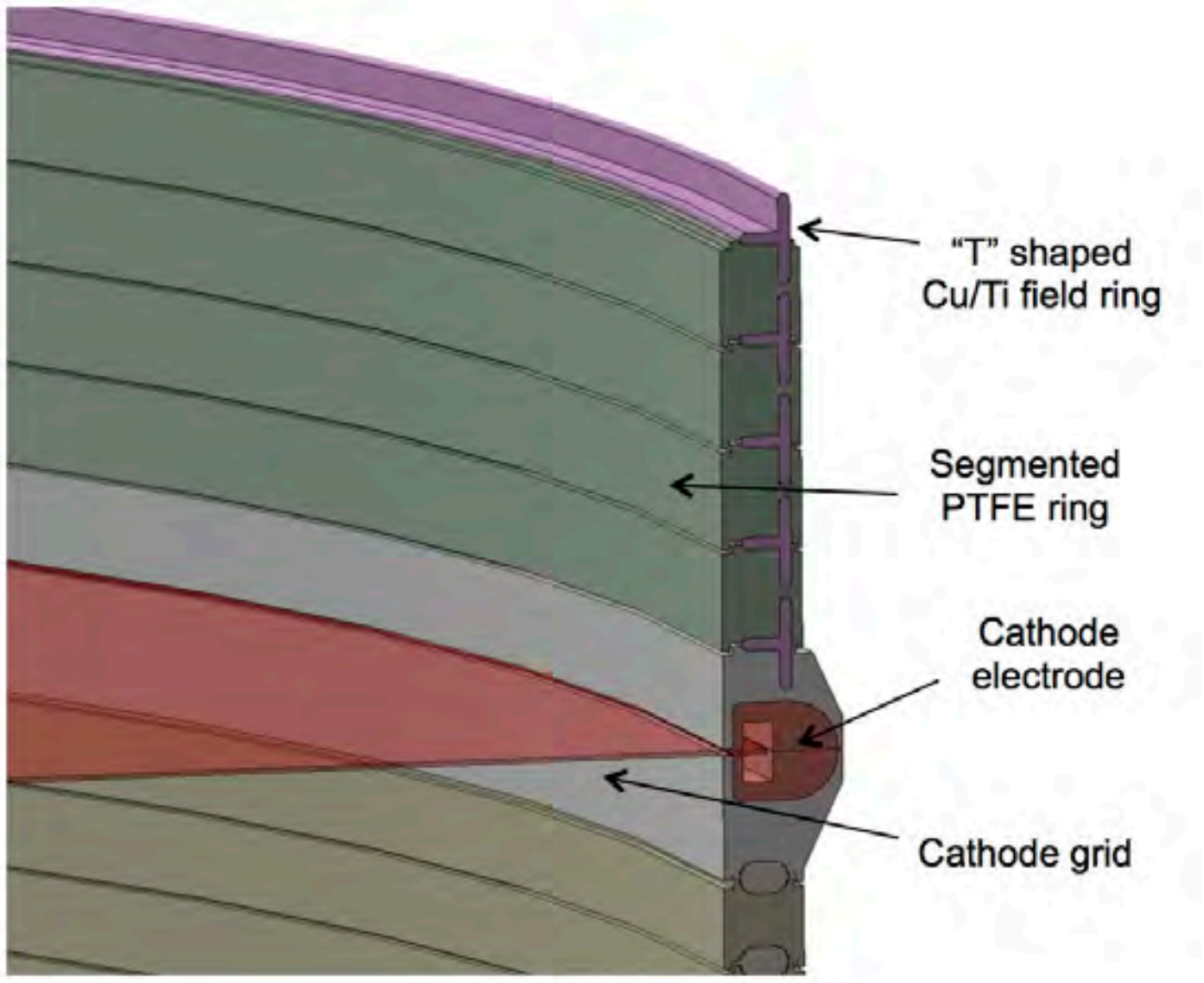

**Figure 6.2.1.1. Cross section of the TPC walls in the drift region, with field-shaping structures embedded in the PTFE walls.**

### Cathode and Drift Region

The region between the cathode and gate contains the fiducial volume and is therefore one of the most important regions of the detector. This is where electrons are drifted up to the extraction region; hence, the electric field uniformity in this region has a major impact on the ability to fiducialize events in the detector. It is important that the electric field in this region be vertical and that the field lines are parallel to the surfaces of the cylindrical PTFE reflectors that set the outer boundaries of this drift region.

To produce a uniform electric field between the cathode and the gate electrodes, we use a set of 57 equally spaced field rings embedded in PTFE and connected by pairs of 1 GΩ HV resistors. The rings will be made from either C101 OFHC copper, or titanium from the same source as that used for the cryostat. The details of this design are shown in Fig. 6.2.1.1. The rings are T-shaped to help maintain the uniform field pattern needed within the TPC region by keeping the equipotential surfaces nearly normal to the inner surface of the PTFE rings. Figure 6.2.1.2 shows the calculated fields produced by this structure. The field-shaping rings are embedded in vertically and laterally segmented rings of PTFE that have been precision machined and then assembled in a stack to produce the completed field cage. The sharp difference in thermal contraction between PTFE and the metal field-shaping rings is accommodated by having these segmented pieces of PTFE slide laterally along the conducting rings when the detector is cooled. This approach is discussed in Section 6.2.3. The field cage structure will be mounted to the lower reverse-field region and lower PMT support, which in turn is supported from the bottom of the cryostat.

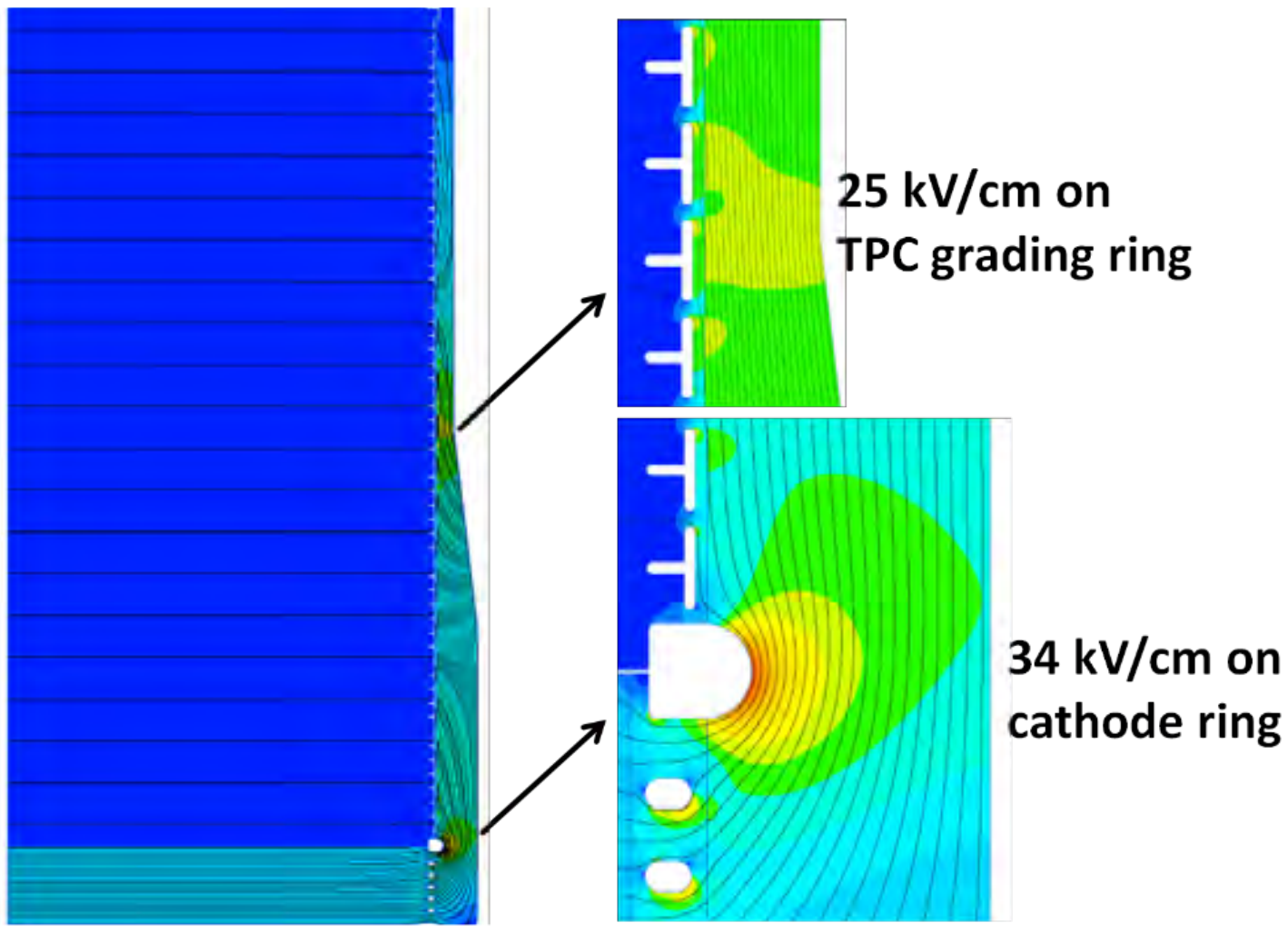

**Figure 6.2.1.2. Field uniformity calculations for the bottom of the drift region and the reverse-field region of the TPC for LZ. High-field regions are yellow and orange while lower-field regions are green and blue. The right edge is the location of the grounded wall of the cryostat. The close-up views on the right show the maximum fields in the skin region on the cathode grid and field-shaping rings.**

The cathode grid will be constructed using a large circular 316 stainless steel (SS) frame that will hold two wire planes, each with 200-μm diameter ultrafinish SS wire planes oriented at 90° to one another, with the wires spaced every 1 cm in each plane. A large electrostatic attractive force exists between the cathode and bottom grids because of the high field in the reverse region; the tension on the wires in these grids must be large enough to limit distortions of the plane from this force. Setting a limit of ~2 mm deflection for 200 kV on the cathode, a 1 kg load is required in 200 µm Ø wires, which is well within the yield strength of available SS wires, but nonetheless represents an important mechanical requirement on the assembly. Note that the tension needed to minimize this overall deflection is larger than the minimum tension needed to prevent the well-known wire-to-wire "sawtooth" instability encountered in a single plane of wires in a wire chamber. This crossed set of wires will act as a large-opening wire mesh, with a 92% transparency (at normal incidence), providing the necessary field termination point at the cathode as well as allowing the lower PMTs a less obstructed view of the light produced in the TPC above. Two construction techniques for the grids are under consideration. One, based roughly on the method used by LUX, features a wire-crimping system with individually tensioned wires. This method has the advantage of all-metal construction, which minimizes outgassing or other contamination and provides good stability through temperature cycles. The other, which is optimal for a woven-mesh grid, is to capture the wires between two rings that are glued together. This method has the advantage of mechanical simplicity and having a smaller footprint than the crimp system. Details of the crimp method are shown in Figure 6.2.1.3. Small-diameter prototypes will be constructed and tested (see Section 6.10).

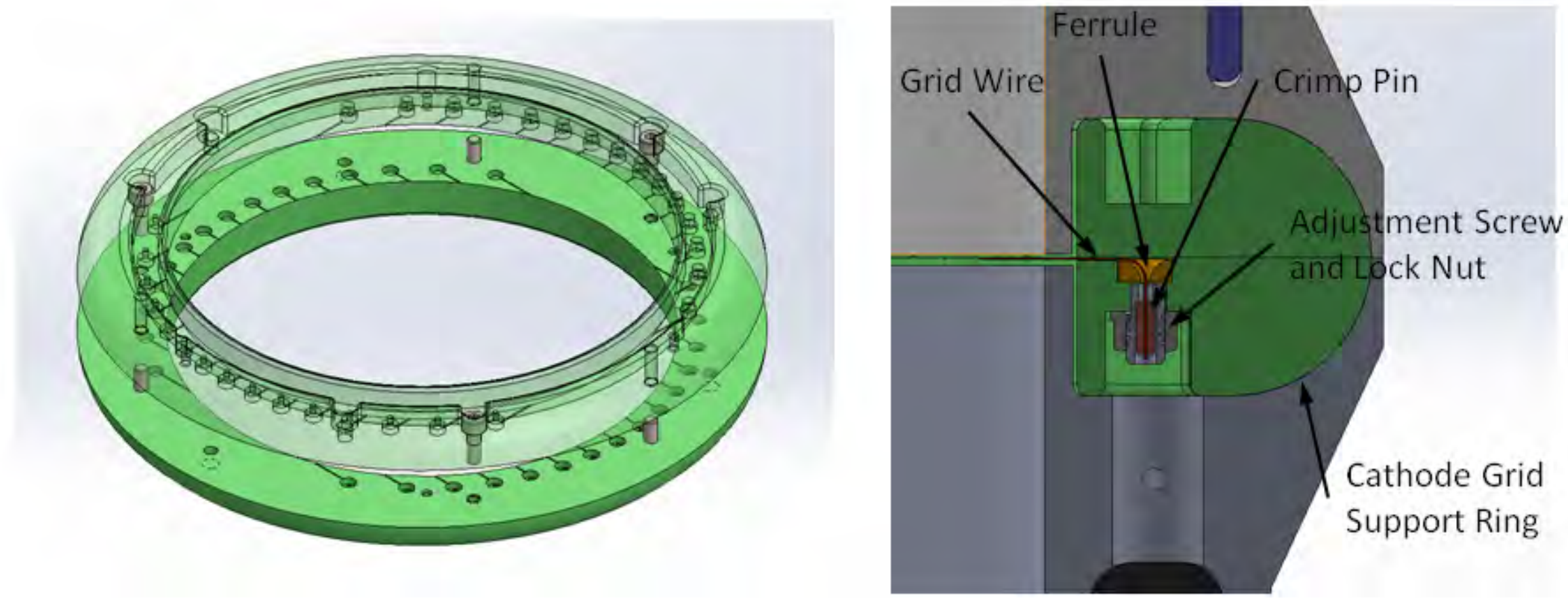

**Figure 6.2.1.3. Left: View through the cathode grid ring for the first prototype assembly for Phase I system testing (see Section 6.10). The full LZ design will be essentially the same in all dimensions, apart from the overall diameter. The grid is composed of two planes of wire oriented at 90° to each other. In the figure, the top plane of the wire frame is transparent and the bottom plane is shown as solid. Right: The design of the wire-fixing mechanism that will be used to mechanically secure the wires in these grid frames.**

### Reverse-Field Region

The reverse-field region between the cathode grid and the bottom PMT shield grid is one of the biggest challenges of constructing the LZ TPC because of the very high field involved. We must grade the cathode voltage from −100 kV to ground while keeping all surfaces in the fields in this region below the 50 kV/cm target described in Chapter 3. At the same time, we must try to keep this space as small as possible, both to reduce the amount of Xe in this region, and to reduce the rate of events that scatter in both the reverse-field region and the central TPC. Such events are a class of background that can mimic WIMP signals, but, as discussed in Chapter 3, have an acceptably low rate for the baseline design presented here. In the LUX detector, due to the much lower cathode voltages and the shorter drift region in the TPC, this was handled with a 4-cm spacing and no field grading between the cathode and PMT shield grids, along with a near-zero field region of 2 cm between the shield and the PMT front surfaces. For the LZ configuration, we have chosen a voltage-grading structure similar to that in the drift region.

This better defines the fields, and is a more robust approach to the more challenging LZ voltage requirements.

The current design, shown in detail in Figure 6.2.1.4, is composed of a stack of six PTFE "rings," each ~2.5 cm high and embedded with a copper or titanium field-shaping ring. These conducting rings are placed near the outside wall of the PTFE rings to keep the field inside the TPC volume as uniform as possible above the PMTs while keeping the fields between the TPC region and the grounded outer cryostat below the required 50 kV/cm. The most recent calculated fields in this region can be seen in Figure 6.2.1.2. The smooth shape of these rings, compared with the T-shape in the drift field region, creates much lower surface fields on the rings, but results in a less-uniform field in the central LXe region. This is allowed because there is not a strong uniformity requirement in the reverse-field region. The voltages between each of the field rings are graded down from the cathode potential using a set of series resistors, similar to those used in the drift region, but we need 4 times the number of resistors between each ring to accomplish this stronger field grading.

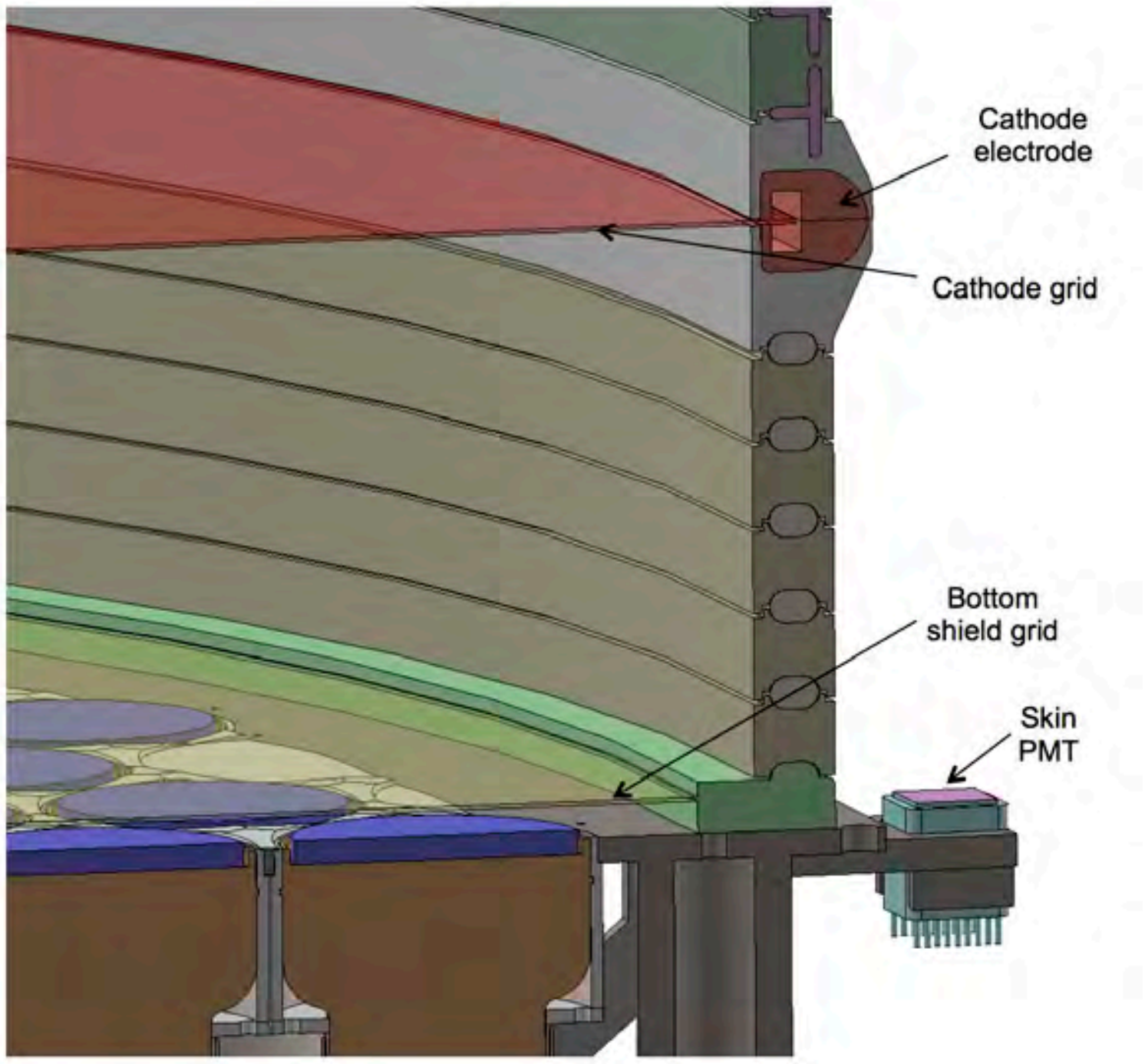

**Figure 6.2.1.4. The reverse-field region, with the cathode (red) and bottom shield (green) grids visible, and the oval field-shaping rings used in this region.**

The resistors in the reverse-field region are more challenging for radioactivity than those in the drift region because they are larger. The main radioactive challenge in electronic components is ceramic, which in all standard (non-"synthetic") forms is very high in radioactivity. Our baseline plan, following LUX, is to use standard surface-mount resistor components that have the smallest ceramic mass for the required voltage rating. We are also considering custom fabrication of a film resistor on a base material made from synthetic quartz or sapphire, an approach that was successfully used by EXO [1].

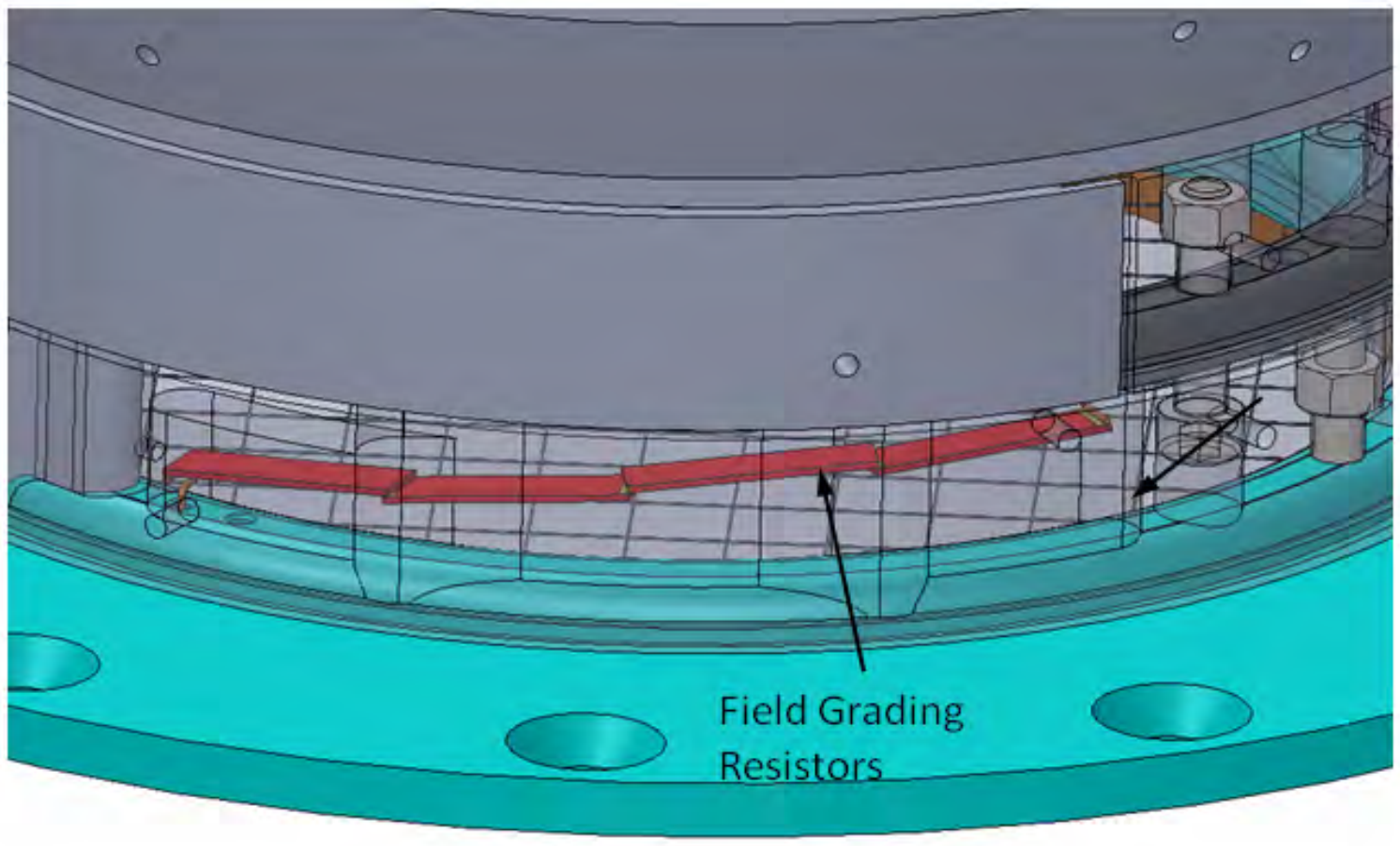

**Figure 6.2.1.5. Placement of the four field-grading resistors in the reverse-field region, shown for the first system test prototype. The resistors are embedded inside the PTFE ring structure and attached to successive field-shaping rings.**

Figure 6.2.1.5 shows the current design and location of these grading resistors inside the PTFE spacers. The lowest PTFE "ring" will be attached to the top of the lower PMT shield grid and this grid will be spaced another 2 cm above the PMT surfaces, also using a PTFE spacer ring. This entire assembly will in turn be attached to the lower PMT support structure, which will then be fixed to the cryostat for final mechanical support. While extensive electrostatic and mechanical studies of this region have been carried out, we consider this design to be preliminary, in particular until it has been tested as one of the first elements of the system test program described later.

### Electroluminescence Region

The design of region above the gate grid, where the electrons are extracted from the liquid and create the S2 signals before they are collected on the anode grid, presents several challenges. The fields are high; the optimization of the grids to create the S2 signal requires care, as is discussed in detail in Section 6.6; the mechanics of having gate and anode grids with very low deformation from the large electric fields is challenging; both the weir structure (Section 6.8) and skin PMTs (Section 6.7) must be accommodated in a tight space; and the overall structure must maintain a very low level of distortion in the rings supporting the anode and gate grids, so that a parallel arrangement of grids and liquid surface can be obtained (tip-tilt adjustment of the detector to assure parallelism of grids and liquid surface is discussed in Chapter 8). A close-up view of this region is shown in Figure 6.2.1.6.

The field in the liquid above the gate must be significantly stronger than the field in the drift region, because a ~5-kV/cm "extraction" field is needed in order to give the electrons sufficient kinetic energy to overcome an energy barrier at the liquid surface and be extracted into the gas phase with near-unity probability. Once electrons enter the gas phase, where the field is approximately twice as strong due to the lower dielectric constant there ($\varepsilon_{r.liq}$=1.96), they are accelerated and produce electroluminescence photons in the 5-mm drift distance until they are collected on the anode grid. The photon yield is ~550 photons per emitted electron at 1.6-bar operating pressure, with 10 kV/cm in the gas. For these operating conditions, the electron transit time to the anode is ~0.7 μs, which, along with diffusion while the electrons drift in the liquid, determines the width of the S2 pulse.

The gate electrode decouples the field applied to the drift region — which tends to be limited to ~1 kV/cm or lower due to the length of the chamber — from the ~5-kV/cm extraction field above it. The gate grid is assembled onto a circular SS frame, with a single wire plane stretched and fixed in a similar fashion to the cathode grid described previously. It will employ 100-μm ultrafinish SS wires wound with a spacing of 5.0 mm, with the wires fixed on this frame after stretching using the staking system described for use on the cathode. This choice of wire and spacing results in an optical obscuration of only 2.5%, while the transparency for drifting electrons is approximately 100%. The upper element of this electrode assembly is the anode, located nominally 5 mm above the liquid at the top of the TPC.

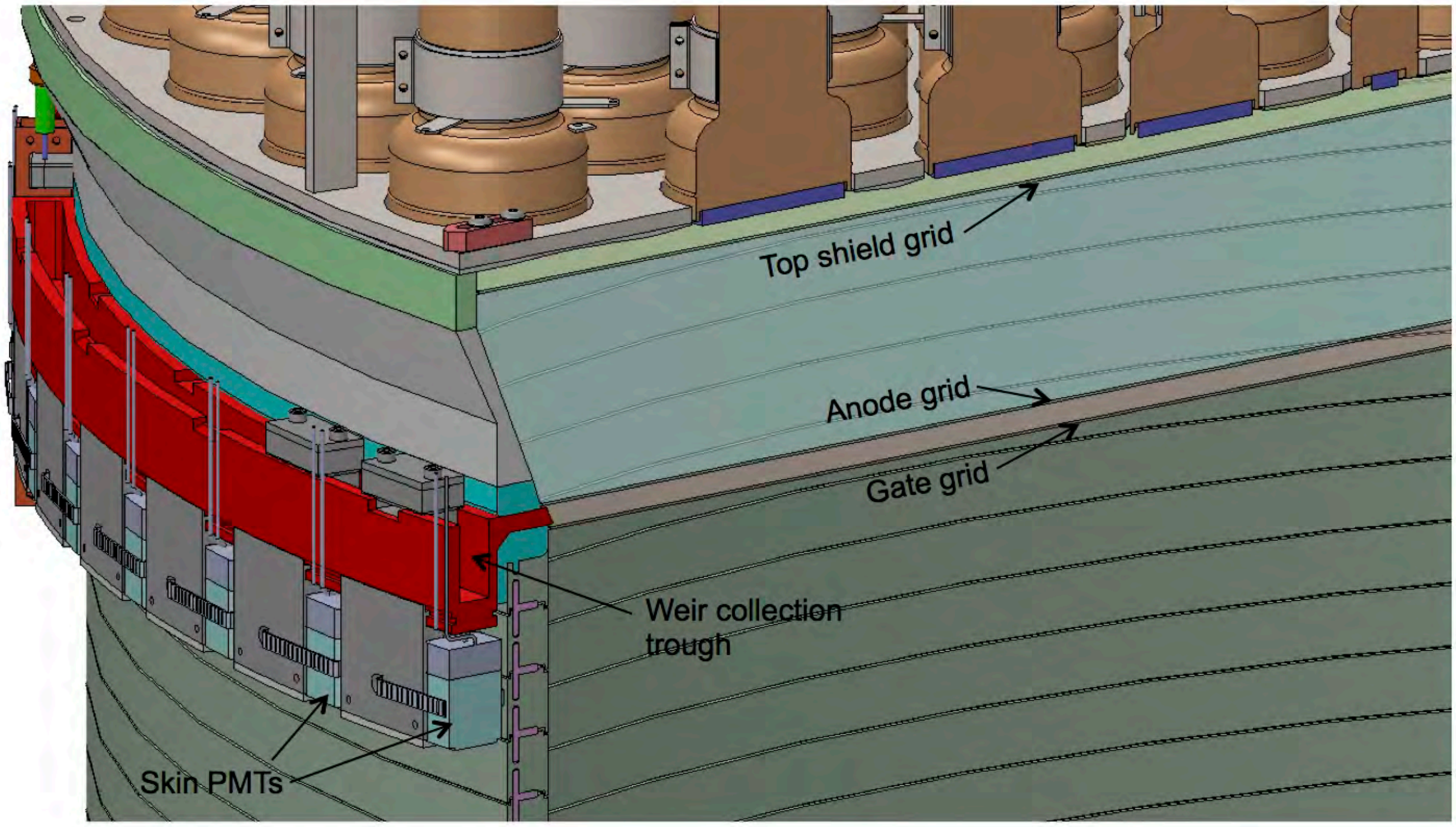

**Figure 6.2.1.6. The electroluminescence region, with the gate, anode, and top shield grids shown, along with the weir, and top skin PMTs.**

Because the S2 signal develops as the electrons drift from the liquid surface to the anode electrode, it is essential to minimize the variance of S2 photon production for different electron emission points, as this relates directly to the energy resolution achieved in the S2 channel (and hence to discrimination). Firstly, the tension on the both the gate and anode wires must be significant in order to minimize sagging as well as electrostatic deflection, ensuring a relatively uniform electroluminescence response across the entire surface of the detector. A small deflection can be calibrated to first order by mapping the width of S2 pulses in (x,y) using calibration data, since this is proportional to the transit time in the gas, but our goal is to keep deflection at the center of both grids to <1 mm. This is feasible in terms of wire strength, but will require the entire gate and anode grid assembly to have high mechanical integrity. That stability will be aided by robust coupling of those grids to the upper PMT array, as indicated in Figure 6.2.1.6. Secondly, to minimize S2 variability on the scale of individual wires, these anodes tend to be constructed from densely packed fine wires, chemically etched meshes, or woven meshes. Our baseline design, based on LUX, uses a woven mesh of 30-μm wires on a 250-μm pitch. Figure 6.2.1.7 shows a prototype anode using a stretched mesh employed in LUX construction. The optimization of the design for S2 signal production is further discussed in Section 6.6.

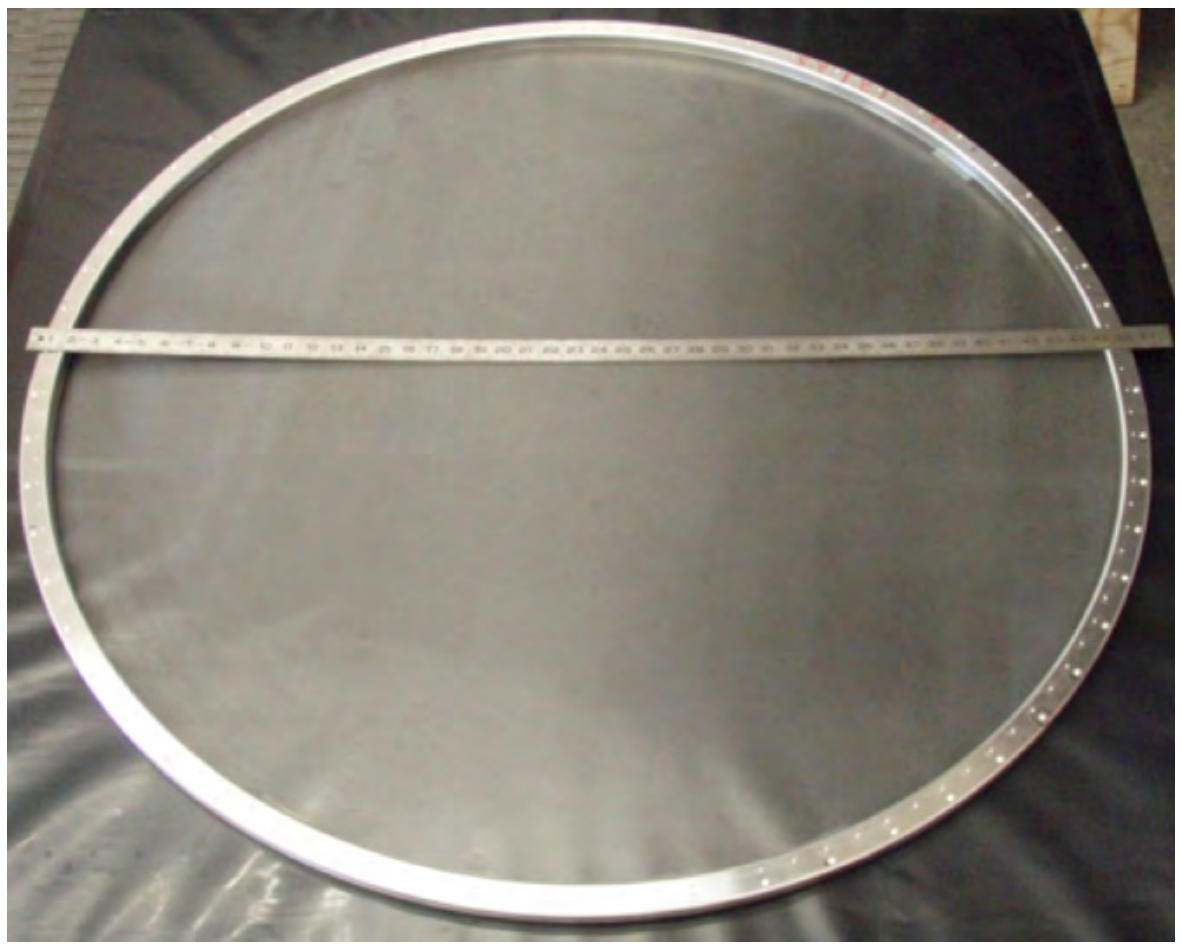

**Figure 6.2.1.7. A 1.16-m-diameter stretched prototype of the LZ anode grid.**

A final element in this region is the upper PMT shield grid. The function of this electrode is to match the potentials from the TPC to those of the photocathode of the phototubes. This grid will be located about 4 cm above the anode and about 1 cm below the PMT windows. It will be constructed once again using a circular SS frame that we will stretch 100-μm ultrafinish SS wires at a pitch of 1 cm, similar to the other individual wire grids used in the TPC.

### 6.2.2 VUV Reflectors

Based on the experience of the LUX experiment in optimizing light collection within the TPC volume through the use of high-UV-reflectivity PTFE panels, we intend to use the same material, albeit in a slightly different configuration to maintain this same high reflectivity within the LZ TPC and skin regions. Our plan is to use machined "rings" of high-purity PTFE approximately 2 cm thick and 2.5 cm tall to form the inner reflecting surface in the TPC region as well as the outer reflecting surface between the TPC and the cryostat wall, which itself will have a few-mm-thick segmented lining of PTFE. As mentioned earlier, the reverse-field region will be composed of six such rings stacked on one another, and the drift field region will be formed from a stack of 57 such "rings" where the field-shaping electrodes are embedded inside the PTFE. This configuration provides the support for necessary electrode structure needed to produce a uniform drift field.

The radioactivity of PTFE must be held low both because of direct gamma production and, more importantly, neutron production from (alpha,n) reactions on F from alpha decays in the U and Th chains. The raw precursor material for PTFE structural material is a powder form produced by DuPont and a few other suppliers, and is expected to be extremely low in radioactivity because of the gas-phase process used to produce it. A large number of smaller manufacturers produce structural shapes from these powders, and the final material can be very low in radioactivity if there is sufficient care in controlling contamination (e.g., from dust) in this second manufacturing step. We assume numbers equal to the limits achieved by EXO [2] working with a company with which we are in contact (see Chapter 12 for further discussion). These are 0.01, 0.002, and 0.06 mBq/kg in U, Th, and K, for which the gamma background is negligible and neutron backgrounds are somewhat subdominant to the assumptions for the PMTs and much below conservative Ti estimates.

### 6.2.3 Thermal Considerations

Given that the inner detector region is composed of PTFE, SS, and copper/titanium pieces, attention must be given to the issue of differential thermal contraction as the detector is cooled to LXe temperatures. The PTFE that makes up the majority of the surface area of the TPC is expected to shrink by ~1.7% linearly, or about 7.6 cm in circumference and 1 cm in TPC radius when cooled from room temperature to ~170 K. Stainless steel, by contrast, contracts only ~0.2% over the same temperature range, and titanium even less. We have chosen to cope with these differences by constructing the metallic field cage rings as solid assemblies, while the PTFE rings are segmented both horizontally (i.e., into rings) and vertically, so that each ring is itself composed of several segments. These latter segments contract and slide circumferentially along the solid metal field cage rings. In this way, the overall diameter of the TPC is determined by the metal field cage rings, and thus undergoes a relatively small thermal contraction. As the PTFE contracts, the seams between the segments open, but the design has overhangs such that there will continue to be a reflecting surface in the exposed gaps. In the vertical direction, the dimension of the field cage is determined by the PTFE panels, and there is an overall height (top PMT array to bottom PMT array) contraction of about ~2.6 cm. To minimize the movement in the critical region where the HV connection to the cathode is made (discussed in Section 6.3), we have chosen to support the entire TPC assembly from the bottom PMT array, which will be connected to the cryostat vessel. This means that the top PMT array will contract downward, increasing the Xe gas-filled region in the top dome.

### 6.2.4 Field Uniformity

In LUX it was observed that field lines at the side edges of the TPC, particularly near the top and bottom, are not fully parallel to the PTFE surfaces. We have come to understand this as being intrinsic to its design: The overall fields resulting from the grids and field cage structure were designed using 2-D electrostatics calculations that treated the grids as continuous conducting sheets. It is well known [3,4] that the 3-D stretched-wire grids have a "transparency" such that the bulk electric fields are somewhat (O(10%)) different than the values calculated assuming the grids are conducting planes, and this effect was taken into account in establishing the operating fields. But a subtler additional effect happens at the top and bottom of the TPC cylinder, where the transparency of the grids causes some bleed-through of the concentrated fields that terminate on the vessel and other grounded structures just outside the main part of the TPC. A more complete calculation using transparent grids reproduces the observed pattern in LUX. Such an effect was in fact previously observed in XENON100 and understood as described above [5].

In LUX, this effect caused electrons at the bottom edge of the detector to deflect ~2-3 cm inward as they followed distorted field lines. This did not pose a fundamental problem for the science data, since the effect could be readily corrected for in analysis. Nonetheless, we will seek to better control the fields in LZ. Based on preliminary electrostatic calculations, we believe we can mitigate this effect by adjusting the values of the last few resistors at the top and bottom of the field cage, and possibly modify the geometry of the electrodes in this area. Another design change over LUX is the vertically segmented design of the PTFE field cage walls. The essentially uninterrupted PTFE surfaces of the field cage are necessary for good light collection, but not ideal from the point of view of good high-voltage design practice, because insulating surfaces can at least in principle accumulate charge that distorts fields. LUX was constructed of vertically continuous slabs of PTFE, whereas the 2.5-cm-tall segments in LZ provide much shorter paths to the conducting field rings from any location on the PTFE walls.

## 6.3 Cathode HV Delivery System

### 6.3.1 Cathode HV Requirements

The cathode HV for LZ is a critical performance parameter that will directly affect the science reach of the instrument because of its impact on ER rejection. Introduction of HV into the Xe space is challenging because of possible charge buildup and sparking, and also because high-field regions can produce electroluminescence that blinds the detector to the flashes of light produced by WIMP interactions.

The LZ operational and design voltages were determined through a combination of task-force activity, evaluation of dark-matter sensitivity, and project cost and risk. Between December 2012 and April 2013, a dedicated LZ task force of 10 engineers and scientists examined the various design ideas and critically evaluated their technical feasibility, with the scope covering the grids, portions of field cage, internal connections, and the cathode feedthrough. The task force culminated in a 46-page report [6]. The operational cathode HV for LZ will be -100 kV, so as to generate a ~700 V/cm drift field. At this drift field, an ER rejection efficiency of 99.5% is expected at 50% NR acceptance, as demonstrated in previous two-phase Xe detectors and modeled through the Noble Element Simulation Technique (NEST) simulation package. The LZ design cathode HV goal is -200 kV; all subsystems in LZ will be designed to withstand a -200 kV cathode voltage to help ensure that a -100 kV operational voltage can be met.

### 6.3.2 Cathode HV System Overview

An overview schematic of the cathode HV system is shown in Figure 6.3.2.1. The baseline LZ design places the cathode HV feedthrough (from air into Xe space) outside the shield at room temperature, at the end of a long, vacuum-insulated, Xe-filled umbilical. With the dominant cable material being polyethylene, Rn emanation is minimized. Polyethylene is known to be a safe material in LXe, mitigating concerns about emanation of electronegative contaminants. With the feedthrough at room temperature and far away from the active LXe, there are no concerns of thermal contraction compromising a leak-tight seal to the Xe space, and no concerns about feedthrough radioactivity. A feedthrough at the warm end of the umbilical allows a commercial polyethylene-insulated cable to pass from a commercial power supply, through an HV epoxy plug, and into the gaseous Xe. The cable then travels through the center of the umbilical and routes the HV through LXe and to a field-graded connection to the cathode. A smaller version of this feedthrough is installed in LUX, and was successfully tested up to 100 kV in gaseous argon before installation. A prototype warm feedthrough has been successfully tested at Yale up to 200 kV, with the cable terminated in transformer oil.

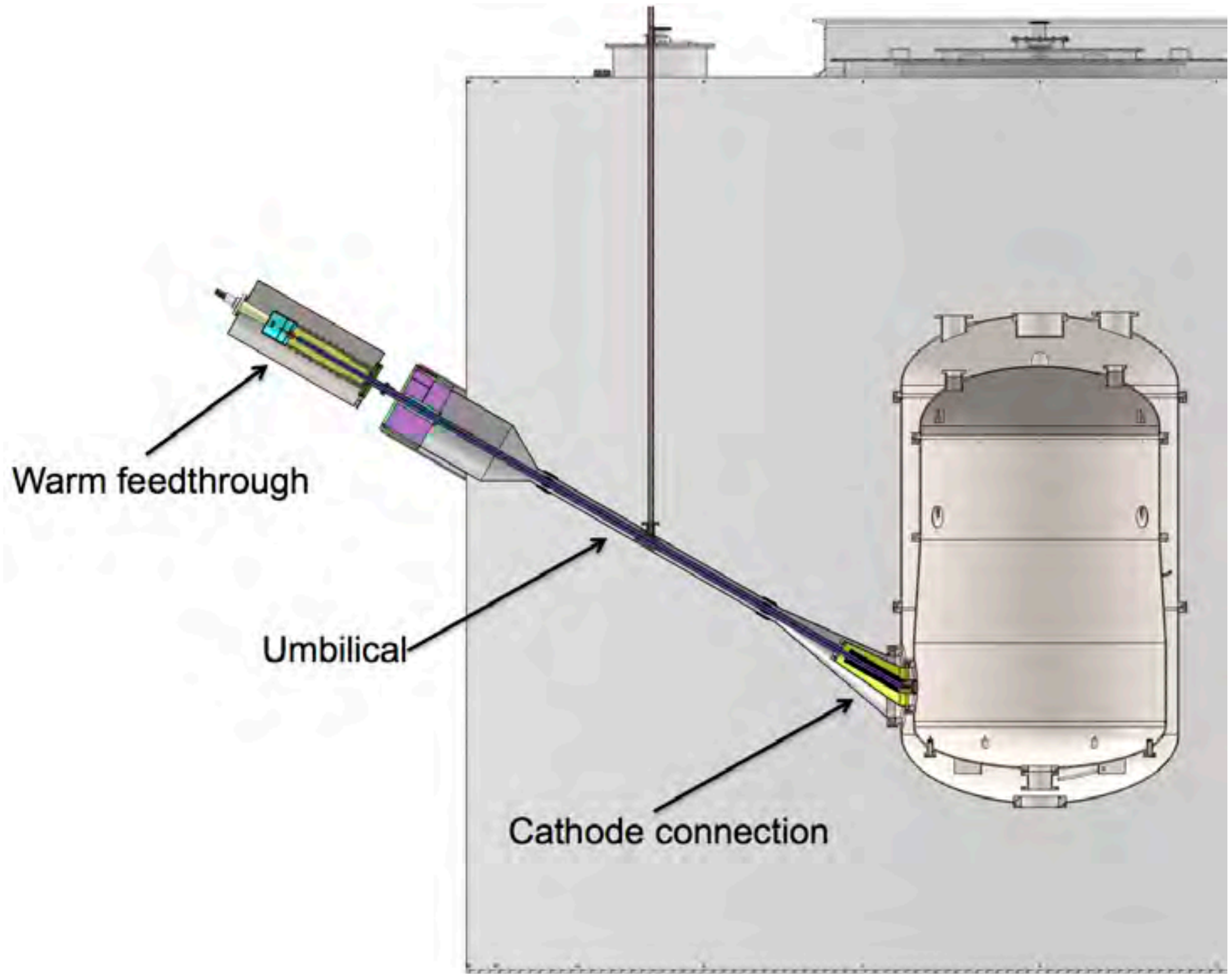

**Figure 6.3.2.1. Overview of the cathode HV system.**

### 6.3.3 Cathode HV Supply and Cable Connection

The maximum design voltage for the cathode grid is -200 kV. The cathode grid power supply is an XRV series power supply rated at -225 kV from Spellman High Voltage and is a standard model for medical and industrial X-ray applications. Several modifications to the standard model limit the maximum current to <1 mA, enhance the resolution of the internal current monitor for accurate measurement of the load current, and reduce the stored energy in the internal capacitance to limit available fault energy. The power supply output connector is a standard R28 connector from Essex X-ray and is rated for 225 kV. The output cable is type Q HV cable from Parker Medical that has an internal resistance of 75 ohms/ft to further limit fault current and energy in the event of an HV breakdown at the load. This cable is terminated into a commercial vacuum feedthrough from Parker Medical (H1827P03) that is rated at 220 kV. By leveraging these commercially available components designed for medical and industrial X-ray applications, the design of the HV generation and delivery system into the warm feedthrough is safe, reliable, cost-effective, and readily available.

### 6.3.4 Cathode HV Feedthrough

The warm cathode HV feedthrough, shown in Figure 6.3.4.1, supplies negative HV to the cathode of the LZ detector. The feedthrough is a specialized termination of an HV polyethylene cable, Dielectric Sciences model 2077, which is rated for 300-kV DC operation. The far warm end of this cable is symmetrically encased in epoxy plastic, which forms a vacuum seal to the cable. This epoxy also forms a vacuum seal to fiberglass tubing that is in turn sealed to a standard 8-inch conflat vacuum flange. This combination forms a helium-leak-tight seal between the conflat and the cable, while confining all strong electric fields within the epoxy plastic. The cable emerges from the conflat-flanged end of the feedthrough, within the Xe space of the detector, while at the opposite end of the termination a metal sphere embedded in the epoxy acts as a terminal for the HV connection. The other side of the sphere is housed in vacuum, where a connection is made to a commercial feedthrough leading to a commercial HV power supply. An electric field simulation of the warm feedthrough and photograph of a prototype warm feedthrough are shown in Figure 6.3.4.2.

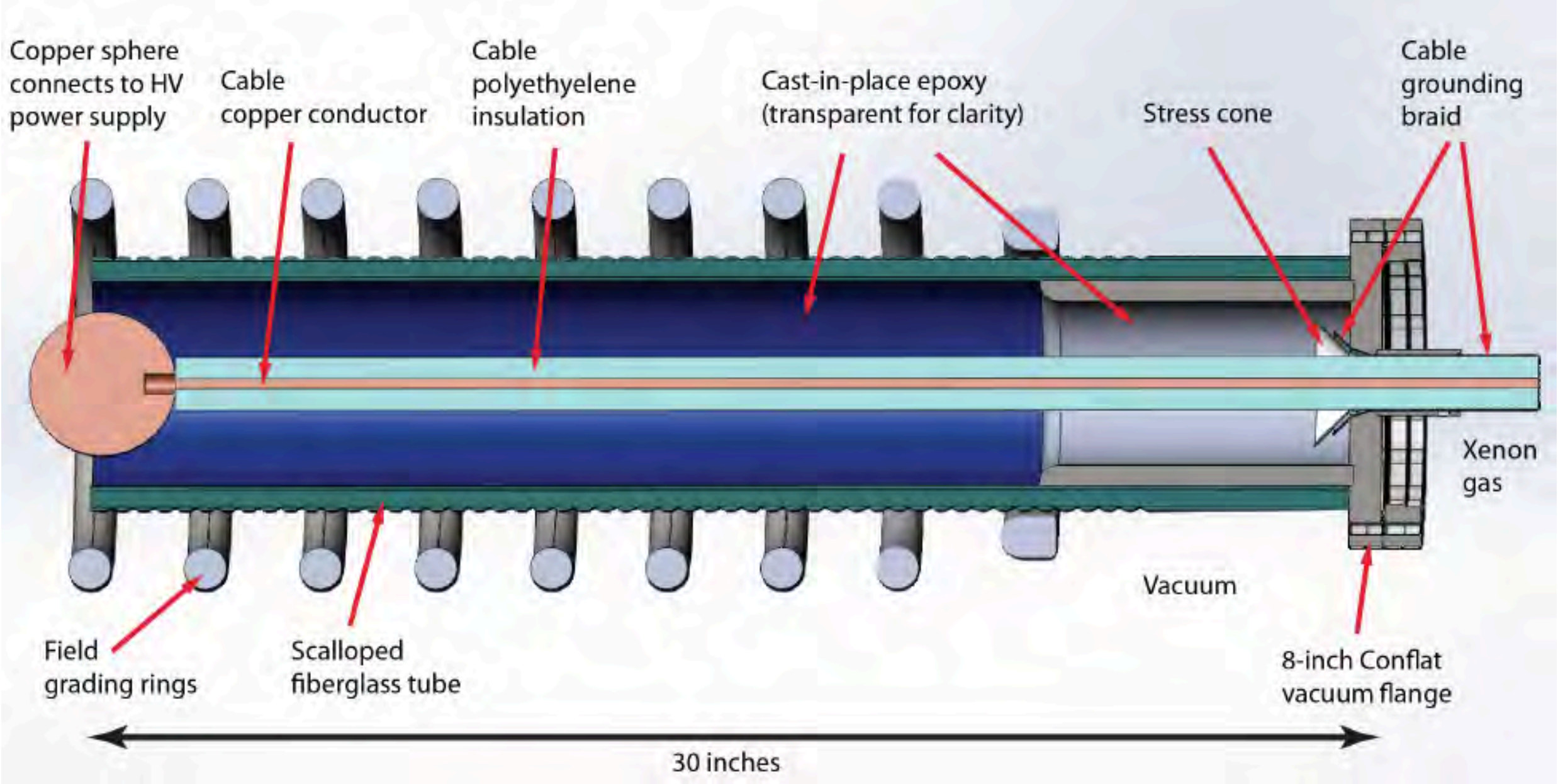

**Figure 6.3.4.1. Warm feedthrough detail for the HV connection to the cathode.**

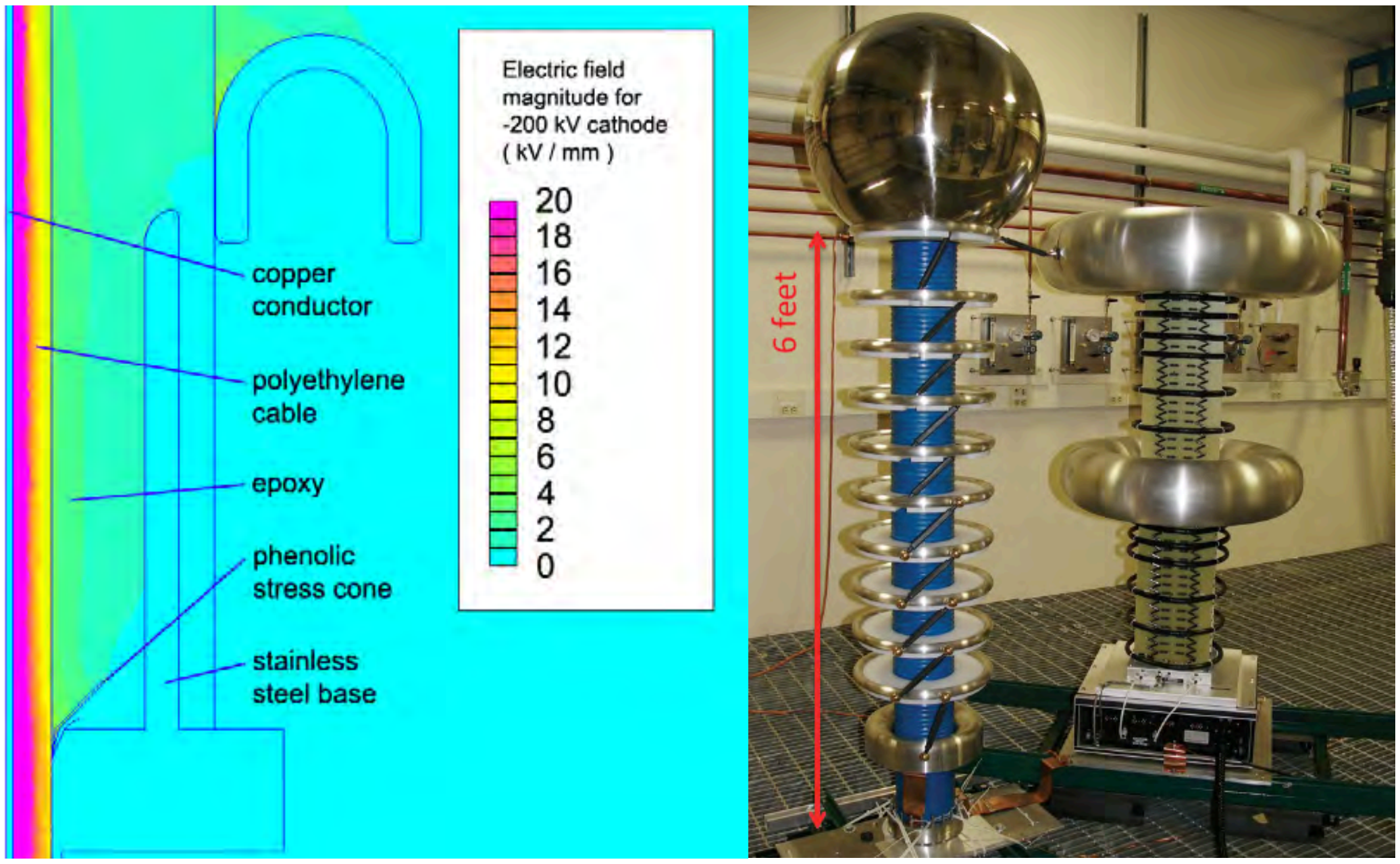

**Figure 6.3.4.2. Left: Electric field simulation of warm feedthrough, at the critical region where the polyethylene cable passes through the flange dividing vacuum space from Xe space (this is on the right end of the assembly in Figure 6.3.4.1). Right: Photograph of warm feedthrough (blue structure on right), under test at Yale.**

### 6.3.5 Cathode HV Umbilical

The cathode HV umbilical, shown in Figure 6.3.5.1, is designed to carry the Dielectric Sciences HV cable from the warm feedthrough to the cathode of the detector. It is formed of a nested pair of tubes that protrude from the side of the detector at about the height of the cathode. These rise upward in at an angle of 30° from the horizontal and penetrate the water-tank side wall at approximately 3/4 of the height of the tank. The inner tube of the umbilical is connected to the Xe space and is joined to a protrusion from the inner vessel of the detector by a short bellows. The outer tube of the umbilical contains vacuum and is similarly connected to the outer vessel of the detector. The outside of the outer tube is immersed in the water of the tank. The evacuated space between the tubes contains super-insulation reflective wrap and acts to thermally isolate the Xe space from the water. This allows

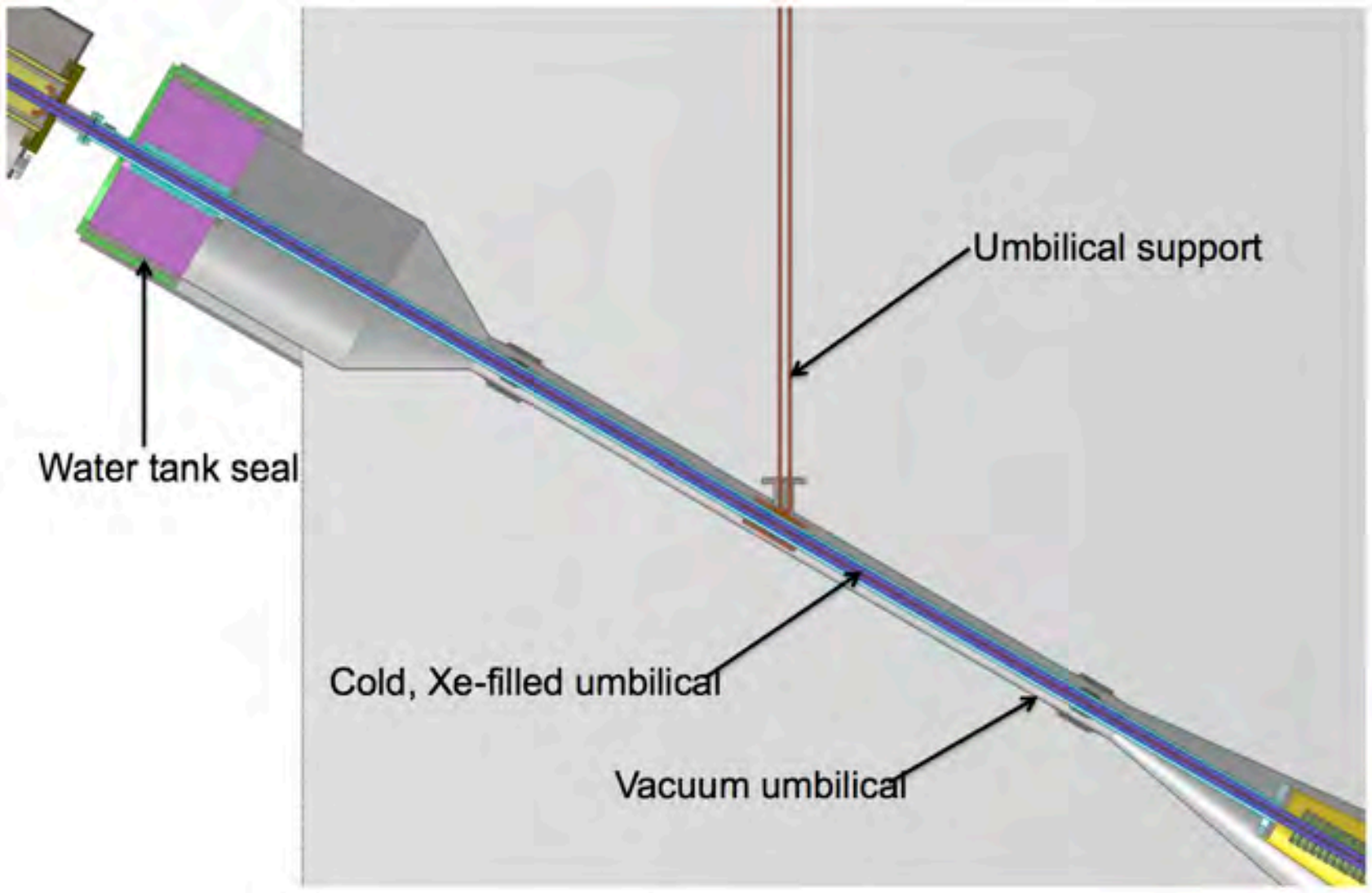

**Figure 6.3.5.1. Detail of the HV umbilical that spans between the HV cable at room temperature on the left, and the connection to the cold cathode grid on the right.**

LXe to fill the inner tube of the umbilical until it reaches a height equal to the level of the Xe surface inside the detector. Thus, the lower part of the umbilical is filled with LXe, while the upper part contains Xe gas. The long length of LXe is necessary to accommodate the field-grading region of the HV cable. A port near the high end of the umbilical connects to the Xe circulation system to allow control over the flow of Xe through the umbilical. Finally, the high end of the umbilical connects to the warm HV feedthrough. The umbilical is heavy and is supported from a structure standing on the floor of the water tank.

### 6.3.6 Spark and Discharge Mitigation

The field-grading structure at the cold end of the HV cable, shown in Figure 6.3.6.1, allows for the ground braid of the cable to terminate while the polyethylene insulation and conductive center of the cable continue. This structure is long in order to minimize the electric field parallel to the surface of the cable. The cable is surrounded by 20 field rings made of conductive plastic. These rings enclose coil springs that grip the cable circumferentially and provide electrical contact to its surface. The field rings are connected in series by small resistors to establish a uniform voltage grading between them. The highest potential ring (lower right of figure) is connected to the center conductor of the cable, while the lowest potential ring (upper left of figure) is connected to the cable ground braid. The surfaces of the rings are heavily rounded, and the resistors are nested between them. This minimizes the field within the LXe that surrounds the grading structure and separates it from the grounded wall of the inner tube of the umbilical. The grading ring structure is supported entirely by the HV cable, so there is no need for a "stand-off" to the grounded wall of the umbilical. The entire grading structure is immersed within the LXe; all sections of the cable within Xe gas have an intact ground shield. An alternative design being considered has a more gradual departure of the cable ground braid from the cable surface. This further reduces the field within the LXe near the cable surface.

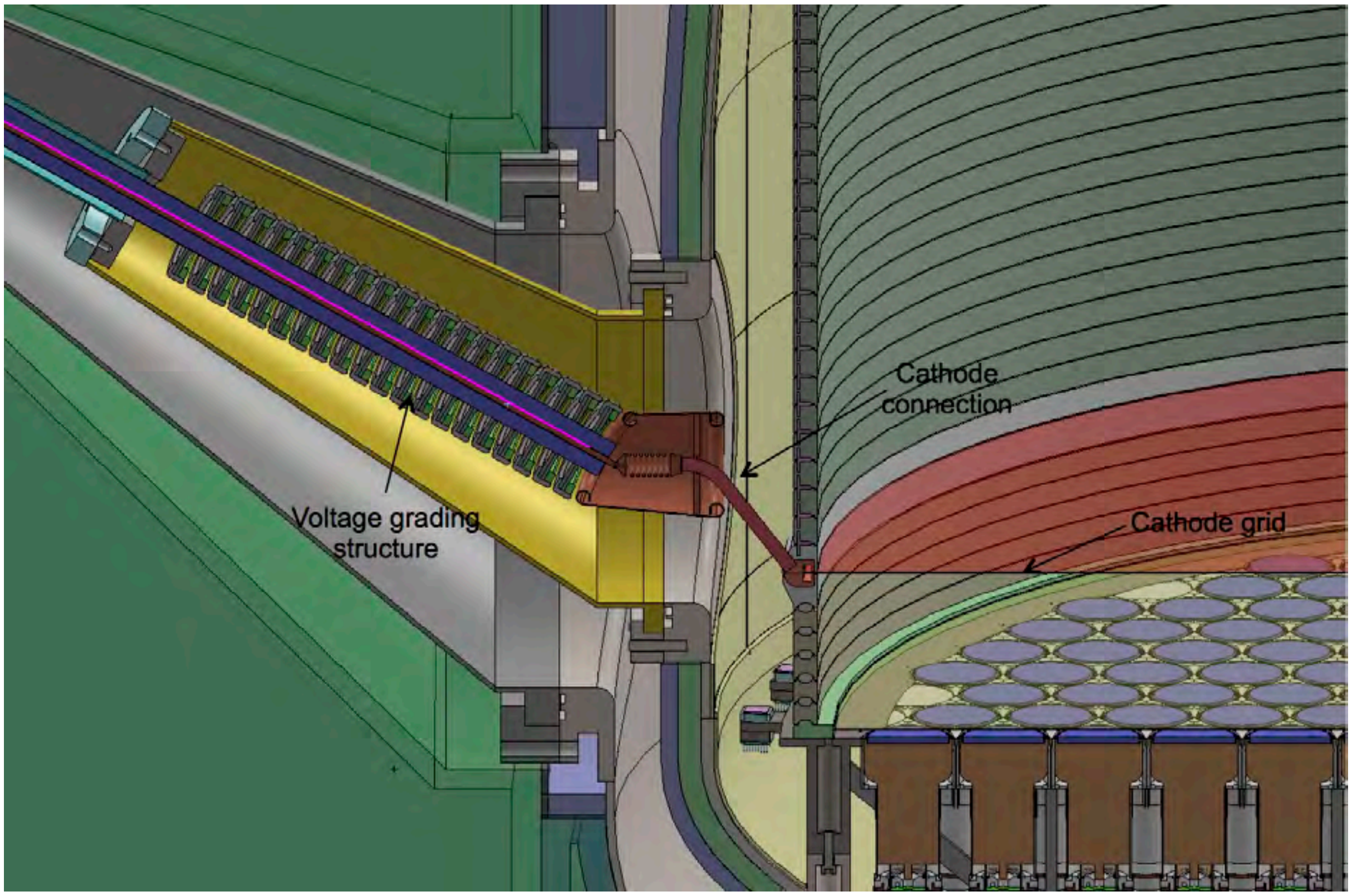

**Figure 6.3.6.1. Schematic of the flexible HV connection to the cathode grid, showing details of the field-grading structures on the incoming HV cable required to keep the fields in the LXe below 50 kV/cm.**

### 6.3.7 HV Connection to the Cathode

A schematic of the HV connection to the cathode is also shown in Figure 6.3.6.1. Because the TPC (including the cathode grid) is supported from the bottom of the vessel, the cathode grid moves down approximately 2 mm as the PTFE TPC components contract when the system is brought from room temperature to operating temperature (~172 K). To account for this movement, a compliant spring connection between the end of the HV cable and the hardware is fixed to the cathode grid ring. The hardware that extends radially from the cathode grid ring is designed to be stiff with minimal weight, as it is a cantilevered load, and to provide sufficient electric field shaping to shield the field enhancements of the small connection components.

### 6.3.8 HV Safety Issues

The combined stored energy from the cathode power supply output capacitance, output cable capacitance, warm feedthrough and umbilical capacitance, and TPC capacitance is approximately 8 J at the 100-kV operating voltage, and is classified as yellow 3.2C [7]. This hazard class indicates that injury or death could occur by contact (shock). To mitigate this shock hazard, engineering controls are required for operation and administrative controls are required for electrical work. Specific lockout/tagout and grounding procedures will be implemented for various operations such as unplugging the output cable and accessing the internals of the warm feedthrough, umbilical, and the TPC. Each worker who is authorized to perform these tasks will have energized work training and specific HV, high-current, and high-power safety training.

## 6.4 Photomultiplier Tubes

To reach the performance specifications described previously, the Xe detector is equipped with top and bottom arrays of 3-inch-diameter PMTs (Hamamatsu R11410-22) to view the active region of the TPC, and top and bottom rings of smaller, 1-inch-square PMTs (R8520) to view the scintillation light emitted in the Xe skin — the region outside the TPC and inside the cryostat inner vessel. Both types of PMT have been developed to meet important performance requirements, including good spectral response in the VUV, good single-photoelectron definition, low dark noise, and the ability to operate at LXe temperature, in addition to having ultralow levels of radioactivity of ~mBq/unit in U/Th. This section describes in detail the properties and deployment of the PMT for the TPC and skin. Subsequent sections discuss the design and optimization of S1, S2, and skin light signals.

The LZ Collaboration has been pursuing the development of ultralow-background PMTs tailored specifically for use in LXe with a radioactivity goal for U/Th of 1/1 mBq per unit and QE >30% at 178 nm wavelength [8]. The LZ experiment configuration requires ~500 3-inch PMTs, and double that number if 2-inch-diameter tubes were used instead. Because of its outstanding radioactivity performance, the 3-inch Hamamatsu R11410-22 model has been adopted; this tube contains ~1,000 times less radioactivity than a standard off-the-shelf item and is the result of our coordinated development with the manufacturer and a very rigorous screening campaign of subcomponents before the items are even manufactured.

The dynode optics in the R11410 are electrically identical to those used in LUX (2-inch R8778), exhibiting similar gain and single photoelectron response. The distribution of QE at 178 nm is also compatible with that of the previous model (26% typical). The photocathode diameter is 64 mm. This tube has 12 dynodes and provides a gain of $5 \times 10^6$ at 1,500 V bias voltage. The PMTs are assembled to passive voltage divider bases and will be negatively biased so that the signal can be collected by directly coupling the amplifier electronics at near-ground potential. Very high peak-to-valley ratios >2 are obtained for the single photoelectron response, which is a key parameter to ensure high detection efficiency for the smallest S1 signals that are composed of single photoelectrons appearing in multiple PMTs.

Besides good VUV sensitivity, these quartz-windowed PMTs are designed to be operated at LXe temperature featuring a special low-temperature bialkali photocathode with low surface resistivity. This obviates the need for metallic underlayers or conductive fingers [9]. In any case, we will confirm correct operation for every unit through a comprehensive low-temperature test program to confirm optical and electrical performance during and after thermal cycling.

### 6.4.1 PMT Radioactivity Specifications and Radioassay Program

Due to their complexity, total mass (~100 kg), and proximity to the active volume, the Xe-space PMTs are a significant source of radioactivity background in LZ. For this reason, they will be subject to a thorough screening campaign using HPGe detectors (see Chapter 12). Screening of fabrication materials and subcomponents will take place prior to PMT manufacture, and every assembled PMT will again be screened after delivery from Hamamatsu.

The 3-inch R11410 has been delivered in part through a 4-year NSF S4 development program by Brown University with Hamamatsu, which achieved unprecedented radioactivity performance compared with previous generation tubes [10]. Further analysis of results from the screening of 25 of the R11410-20 first-generation model of this series is shown in Table 6.4.1.1. The $^{60}$Co levels have been reduced in R11410 production that followed these earlier prototypes. Measurement sensitivity on early-chain $^{238}$U activity is being further improved using new detectors as discussed in Chapter 12. That same model PMT has also been advanced by other collaborations, notably XENON1T with variant R11410-21. Comprehensive radioactive screening results for 216 of these PMTs are publicly available [11] as shown in Table 6.4.1.1. Results from the screening programs are in broad agreement. LZ intends to procure the most recent R11410-22 version of these tubes.

We use the values in Table 6.4.1.1 to establish our background levels in simulations, as reported in Chapter 12. The 3-inch PMTs account for 1.2 ER events in a 5.6T fiducial volume in 1,000 days, before discrimination, and 0.2 NR events. They are a significant source of NR events among internal detector components, providing neutrons both through spontaneous fission (majority component) and (alpha,n) reactions in the PMT materials. As discussed in Chapter 12, spontaneous fission neutrons can be vetoed even more effectively than those from (alpha,n) reactions; however this has not yet been taken into account in the current NR estimate (see the discussion in Table 12.2.2). In addition, as also discussed in Chapter 12, new detectors are improving screening sensitivities to $^{238}$U, so we expect to further improve the errors/upper limits on the presence of the dominant spontaneous fission emitter.

The 1-inch R8520 PMTs used in the skin detector have radioactivity levels that are well understood thanks to their wide use in past detectors. Nevertheless, we will adopt for them the same screening procedures as for the larger R11410 as described above. The contribution of the 180 skin PMTs to the background of the instrument is subdominant (see Table 12.2.2), given their comparable specific activity but more peripheral location and smaller number.

**Table 6.4.1.1. Radioactivity summary per unit for LZ skin detector PMTs (R8520) from [12], for LUX PMTs (R8778) [8], and for LZ TPC PMTs (R11410) based on 25 early-production R11410-20 LZ PMTs. Values for the R11410-21 model studied extensively in [8] are also shown. Average activities per PMT are quoted per parent decay. Errors are 1σ and upper limits are 90% CL.**

| PMT | $^{238}$U (early) mBq | $^{226}$Ra (late) mBq | $^{232}$Th mBq | $^{40}$K mBq | $^{60}$Co mBq |
|---|---|---|---|---|---|
| **R8520 (1”)** | < 1.39 | 0.12 ± 0.01 | 0.11 ± 0.01 | 7.6 ± 0.9 | 0.55 ± 0.04 |
| **R8778 (2”)** | < 3.0 | 9.5 ± 0.6 | 2.7 ± 0.3 | 66 ± 2 | 2.6 ± 0.1 |
| **R11410-20 (3”)** | < 26 | 1.1 ± 0.4 | 1.5 ± 0.5 | 25 ± 4 | 2.1 ± 0.2 |
| **R11410-21 (3”)** | < 12.9 | 0.52 ± 0.1 | 0.39 ± 0.1 | 11.9 ± 0.2 | 0.74 ± 0.1 |

### 6.4.2 PMT Bases and Cabling

Individual passive voltage-divider bases and two coaxial cables (for HV bias and signal) are attached to each PMT. Given their locations, these components are also under detailed scrutiny and selection as part of the radioactivity and radon-emanation screening programs. The latter is a critical consideration for the ~1,400 cables that terminate in the feedthroughs at the warm end. Sleeved and sleeveless candidate cables are presently being assayed for this purpose.

The voltage-divider bases are made from thick polyimide PCB (Cirlex) with surface-mounted passive components. Cirlex is an excellent material for this purpose, having very high dielectric strength, a low thermal expansion coefficient, high tensile strength, and low internal stress. The voltage-divider circuits are as recommended by Hamamatsu. High-resistance chains are used for low power dissipation in the LXe (24 mW/unit), which is essential to prevent bubbling and to have minimal impact on the thermal design of the detector. Charge supply capacitors are added to the last few dynodes to improve linearity.

The radioactivity performance of the PMT bases is of special concern, both from the point of view of neutron/gamma emission and from radon emanation. Although Cirlex is intrinsically a radio-clean material, the discrete components make more significant U/Th contributions, in spite of the small masses employed. Of particular concern are the ceramic (barium titanate) capacitors, the high concentration of $^{210}$Pb commonly found in resistors, and the spring materials used within pin receptacles to connect to the PMTs and the cables. The current program for gamma-ray screening of the passive components for the bases has already identified a design that delivers lower gamma-ray activity than the PMTs. Our target for the bases is to keep these components at one-third of the PMT radioactivity or less. We will continue to survey potential components to further reduce this contribution prior to finalizing component choice as production nears.

The PMT signals and HV supplies are carried separately between the PMT bases and the warm breakout interface by low-radioactivity coax cables. The baseline design is to use Gore 3007 Coax with no outer jacket, the same cable that was used in this role for LUX. The 50-ohm characteristic impedance cable uses an AWG 30 silver-plated, Cu-clad steel, surrounded by an AWG 40 SS braid.

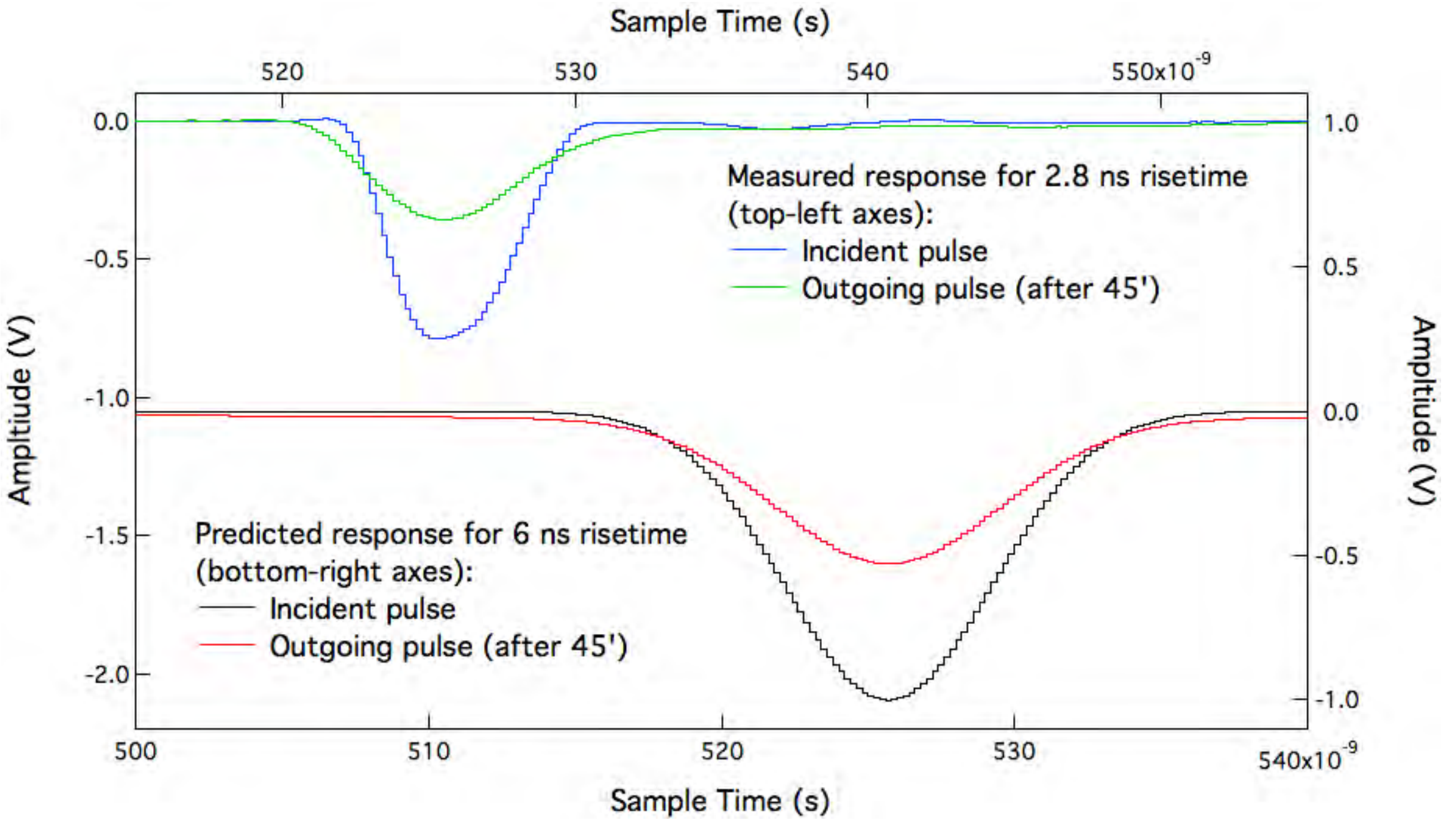

**Figure 6.4.2.1. Measured signal attenuation from 13.7 m (45 ft) of Gore 3007 Coax in response to 2.8 ns rise-time pulses, and calculated response for 6-ns rise-time signals, which is approximately the anode pulse rise time for the R11410 PMTs.**

The cables from the PMTs associated with the upper and lower parts of the TPC are housed in separate conduits, so that no cabling is routed through the side skin region. This could interfere with the ability to hold a high voltage on the cathode and it would degrade the light collection efficiency from the skin region.

The lengths of typical top and bottom cables are 12.8 m and 11.6 m, respectively. The upper routing consists of 672 cables for 247 TPC and 60 skin-veto PMTs, plus internal monitoring sensors. The lower routing consists of 744 cables for 241 TPC and 60 (side) and 60 (dome) skin veto PMTs, plus sensors. Twelve cables included in this count will be run down both paths as dummies for electrical troubleshooting. The total heat load calculated from the cables is less than 6 W, which is a subdominant contribution to the thermal model. A screening program has been initiated to measure the Rn emanation from the baseline cabling, as well as possible alternatives, to ensure that the finally selected cable will meet the overall Rn requirements discussed in Chapter 12. The Rn emanation from the cable is expected to be dominated by the warm region, which is 8 m in length for the upper routing, and 1 m for the lower routing. Emanation from the feedthroughs (which were previously used in the LUX experiment) will also be measured.

The Gore 3007 cable has been tested and shown to support 2 kV, comfortably meeting the HV requirements of all PMTs. The signal characteristics for a 13.7-m length are shown in Figure 6.4.2.1. For a 6-ns rise-time pulse, expected for the R11410 PMTs, we predict an amplitude reduction of 47% and a pulse area loss of 20%.

### 6.4.3 Assembly and Integration with TPC

The 247/241 PMTs per top/bottom array will be assembled onto titanium support frames. The PMTs will be held in position using Kovar rings fabricated by Hamamatsu from the same material used for PMT body production. Three PTFE columns will then be used to hold the collar to the PMT mounting plates. This mounting system is specified to hold the PMTs in place in the Ti mounting plate both when the plate is in the vertical and horizontal orientations (e.g., during assembly and transport). The arrays are shown in Figure 6.4.3.1, and details of the mounting of the PMTs to the arrays are shown in Figure 6.4.3.2.

The support frames for the PMTs consist of a Ti flat plate with a supporting truss-work. The loads on this structure are substantial, particularly in the case of the lower array. For the submerged PMTs, the buoyancy force on the lower array far exceeds the gravitational force. The net upward load in the lower array is approximately 8 N per PMT, and collectively the total load for the array is approximately 2200 N. Many configurations of the support frame were considered and simulated using finite element analysis. Starting with a bare plate (no truss-work), a single 6-7-mm thick Ti plate deflected upward approximately 19 mm. Other options include successively thicker plates, double plates, curved plates, honeycomb reinforcement, and truss reinforcement. The truss reinforcement had the best overall performance when trying to limit deflection, minimize mass (and therefore background radiation), and provide a relatively open volume for scintillation light in the bottom skin to find its way to skin veto PMTs. The baseline lower PMT support frame is expected to deflect approximately 1 mm upward in operation. The upper PMT support frame will have a similar design, but since the net force is dominated by the weight (the top PMTs reside in gas phase Xe), the expected deflection is approximately 0.3 mm downward in operation.

The Ti surfaces surrounding the front faces of the PMTs, in both the top and bottom arrays, will be covered by PTFE pieces designed to increase the recycling of photons, and so increase photon detection in the main chamber, as discussed in Section 6.5. The pieces are designed to provide >95% coverage of the Ti structural elements, while accommodating the differential thermal contraction coefficients of the PTFE and the Ti mount.

The lower LXe region, below the bottom PMTs and mounting frame, forms part of the Xe skin veto in which the goal is to maintain >95% detection efficiency for ER events above 100 keV. The rear of the bottom PMTs, which project into this volume, are also sleeved in PTFE in order to increase photon recycling in the LXe below the array — this includes both a PTFE cylinder for the PMT body, and end-caps to cover the PMT bases. The PTFE base covers also prevent stray light leaking into the PMT envelope, and avoid any pin short-circuits.

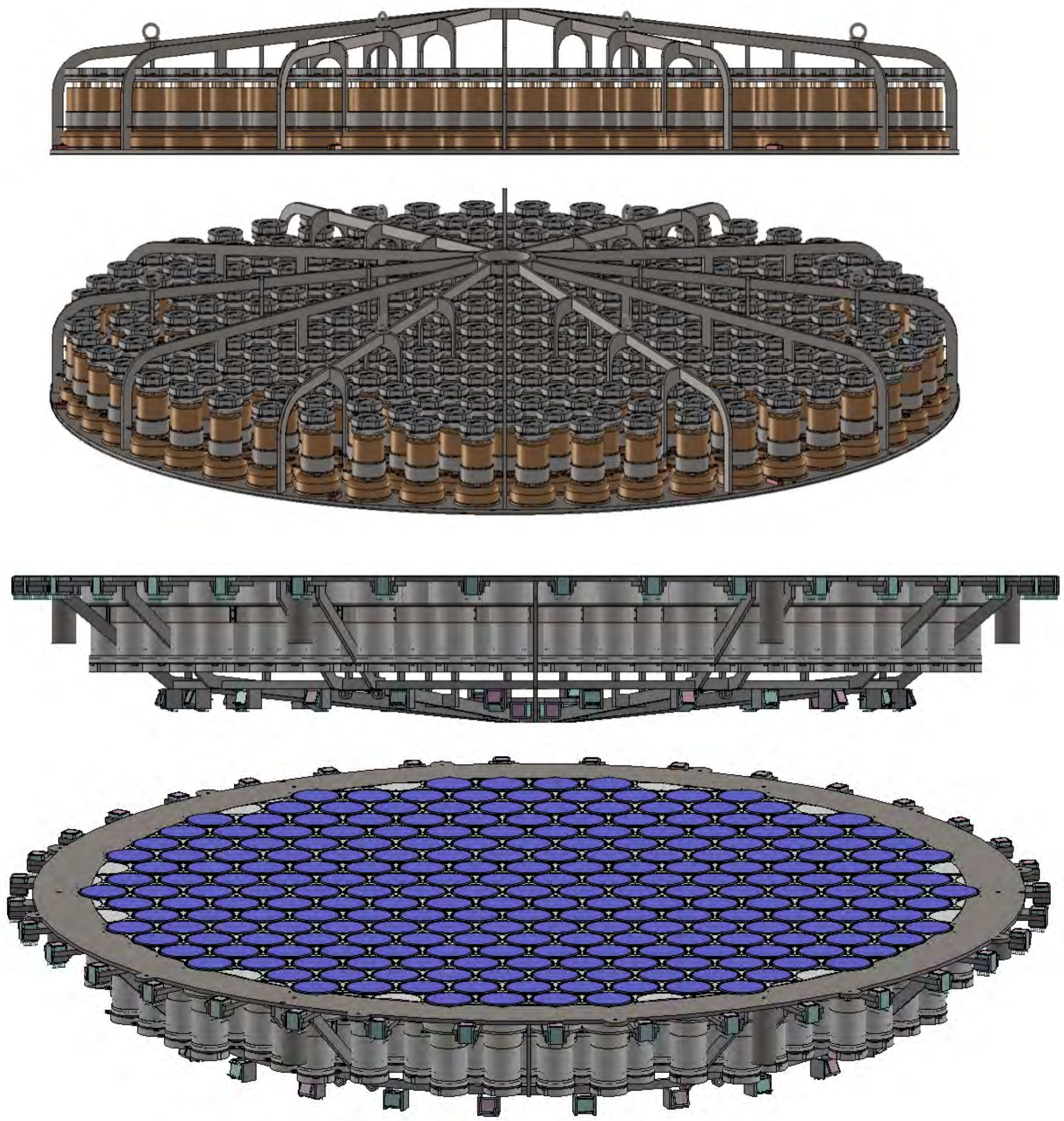

**Figure 6.4.3.1. Top (upper two figures) and bottom (lower two figures) arrays of 247 and 241 PMTs. Ti mounting plates and trusses also shown. Details of PTFE reflector system and PMT mounts are shown in Figure 6.4.3.2.**

The underside of the PMT mounting structure and braces will also be covered in PTFE reflectors where required, to increase the overall photon detection efficiency in the skin region.

The PTFE components will be fabricated from material that has been prescreened to achieve the intrinsic activity budget with respect to both gamma and neutron emission, as discussed in Section 6.2 and Chapter 12. During machining of the components and the assembly of the top and bottom PMT arrays, the PTFE components will be maintained in purge boxes to reduce the plating of alpha emitters associated with airborne Rn, and ensure that the additional ($\alpha$,n) neutron generation is significantly below the intrinsic neutron emission goals.

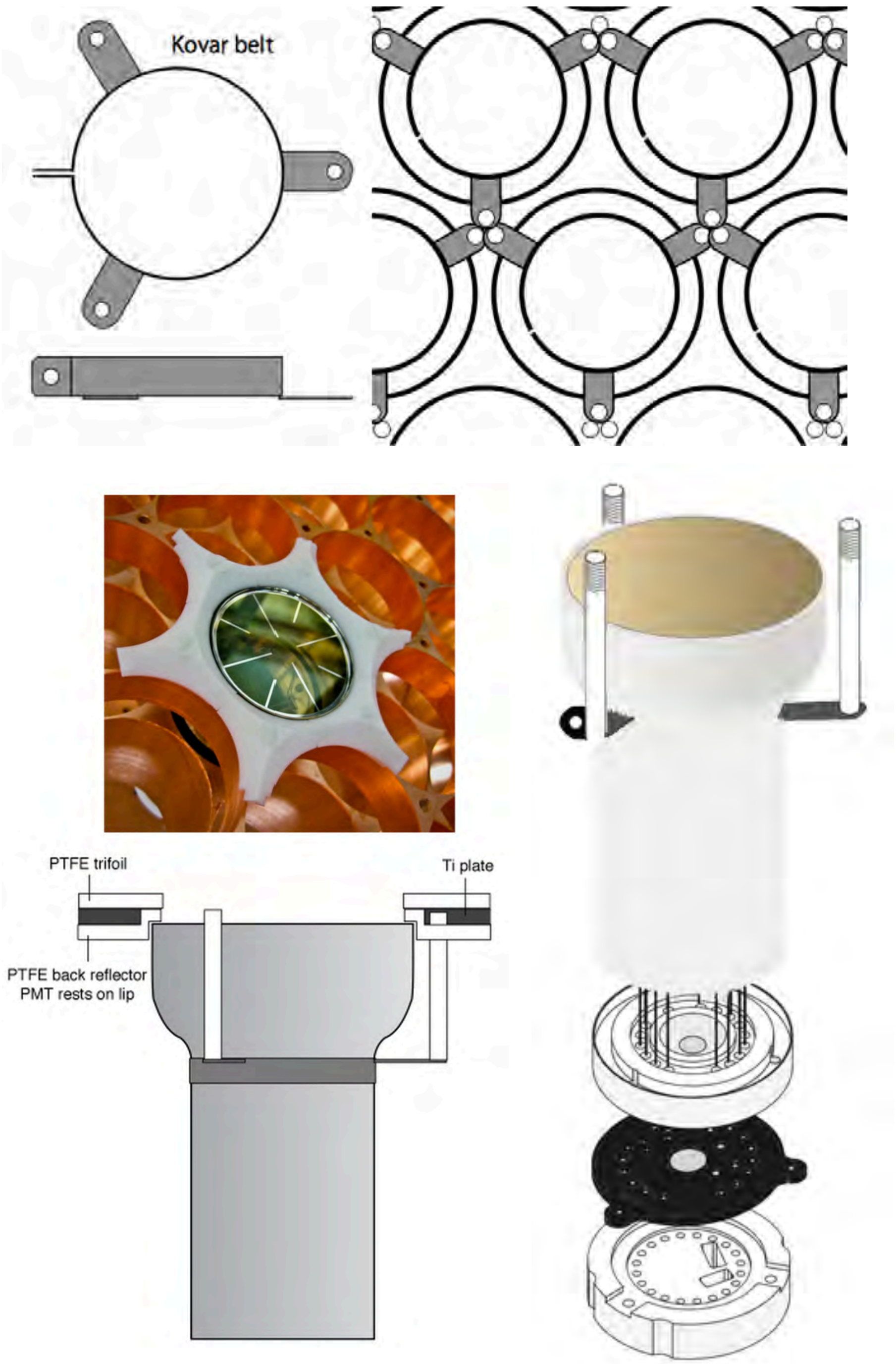

**Figure 6.4.3.2. Top: Arrangement of PMT holders and Kovar mounting belt in hexagonal region of PMT array. Photo: Example of front-facing trifoil PTFE reflector used in LUX and baseline design for LZ. Lower left: PTFE reflector arrangement on either side of Ti mounting plate used on bottom array. Top array uses only trifoil front reflectors. Lower right: Assembly of the R11410-20 tube with PTFE sleeve (only used in bottom array) and PTFE end caps (used in both arrays). The latter prevent strain on the PMT stem, protect against electrical shorts in the base, and help relieve strain on the cables. The dark structure represents the Cirlex voltage-divider base.**

Calibration LEDs, driven by signals sent down cables that are fed from pulsers in the room-temperature electronics racks, will be mounted in some locations on the face of each array to shine on the PMTs opposite. The LZ calibration system is closely modeled on the system developed for LUX, which was shown to satisfactorily meet all performance requirements.

## 6.5 Optimization of Light Collection

### 6.5.1 Overview of Design and Optical Performance of the TPC

The S1 energy threshold for NR detection in LZ is determined by several factors: (1) the primary scintillation yield of LXe, which is particle-, energy-, and field-dependent; (2) the VUV reflectivity of all PTFE surfaces; (3) the photon absorption length in the liquid, which is determined to be >10 m from operational data in LUX; (4) the geometric transparency and reflectivity of the grids, especially those in the liquid; and (5) the optical performance of the PMTs. The magnitude of S1 also has a strong impact on discrimination, and maximizing the sensitivity of this response channel is an experimental priority. Based on optical simulations and expected PMT performance, we predict a volume-averaged photon detection efficiency $\alpha_1 \approx 7.5\%$ for S1 light in LZ, which will allow us to reach the target threshold and discrimination efficiencies (see Chapter 3).

The S2 channel can easily provide as much gain as necessary, but the layout of the upper PMT array is important to achieve a position resolution <10 mm (rms) at threshold — especially for interactions near the edge of the TPC such as those arising from any contaminants plated out on the field cage walls. The layout of this region has been optimized for this purpose, and the result of these studies is presented in Section 6.5.3 and in Table 6.5.1.1. The adopted baseline is a bottom array of 241 units with close-packed hexagonal layout (third entry in Table 6.5.1.1), and a top array with a hybrid hexagonal/circular arrangement containing 247 tubes (last entry in Table 6.5.1.1). The photocathode coverage represents ≈40% of the TPC cross section for each array.

### 6.5.2 TPC Optical Performance as a Function of Main Parameters

A full optical Monte Carlo based on Geant4 has been used to obtain the baseline photon detection efficiency for S1 light, $\alpha_1$, defined as the number of photoelectrons per initial photon generated at the event site. This is discussed in Chapter 3, and summarized in Section 3.4.3. We have adopted a baseline $\alpha_1$ of 7.5%, along with a range that varies from 5% to 10% or more that reflects a conservatively broad assessment of the possible range of optical properties of the detector materials. Here we describe results from a streamlined Monte Carlo code that complements the full Geant4-based simulation and was used to scan the leading parameters that determine optical performance: the number of PMTs, the reflectivity of the PTFE-LXe interface, absorption length in the liquid, and both the mechanical transparency and reflectivity of the electrode grid wires. The light-collection values from both Monte Carlos agree in most cases to between ~5–20%. In Figure 6.5.2.1, we first show a map of $\alpha_1$ as a function of event location for the baseline parameters (described also in Chapter 3): 25% QE in the PMTs, 95% fully diffusive reflectivity for PTFE in liquid, 20% reflectivity for all grids, 75% reflectivity for PTFE in the gas, and100 m absorption length in the LXe, as well as the baseline number of PMTs and geometric grid transparencies. The value of $\alpha_1$ for these parameters in this Monte Carlo is quite close to our adopted 7.5% baseline value. The set of pessimistic and optimistic optical parameters scanned over is indicated in the figures that follow. A key feature evident in Figure 6.5.2.1 is the effect of total internal reflection at the liquid surface, and the optical mismatch between Xe gas and the quartz windows of the PMTs. Photon collection is higher in the bottom array (the index of refraction of LXe and quartz are fairly well matched) than the top array, except just below the liquid surface. Another notable feature is that the average S1 photon path length is ~6 m, and the average number of scatters on surfaces that could result in absorption is ~5: To a large extent, the detector is a "mirrored box" in which the value of light collection is the result of a competition between a photon being detected at a photocathode, and absorption that occurs with low probability but a high number of chances as the photon scatters around the detector.

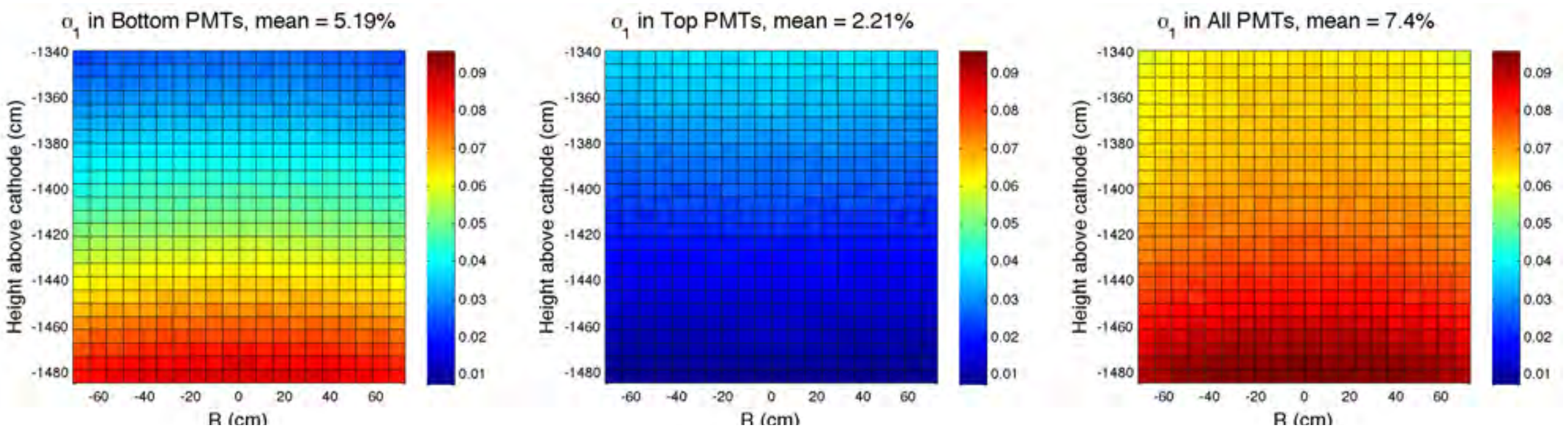

**Figure 6.5.2.1. The photon-detection efficiency $\alpha_1$ as a function of event location in the fully active part of the TPC between the cathode and gate grids, with the baseline values of all optical parameters. The right panel shows the full S1 signal in all PMTs, while in the left and middle panels show those fractions of the signal measured in the top and bottom arrays, respectively.**

**Table 6.5.1.1. Summary of light collection and position reconstruction performance for candidate top array layouts. The lower half of the table presents spatial resolution *rms* values away from the detector edge with nominal optical parameters; L28 quantifies the fractional leakage of wall events reconstructed more than 28 mm into the TPC.**

| | | | | | | | | | | |
|---|---|---|---|---|---|---|---|---|---|---|
| | 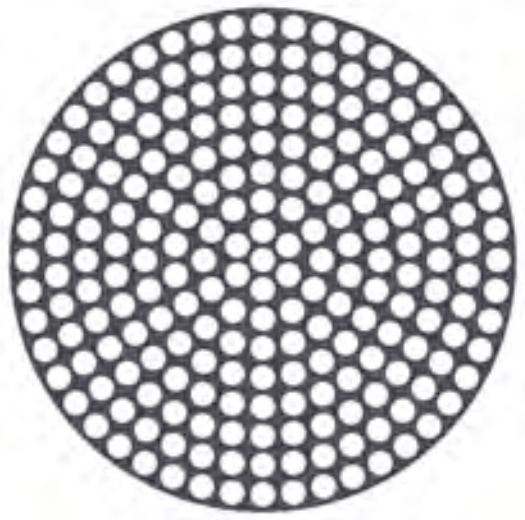 | | 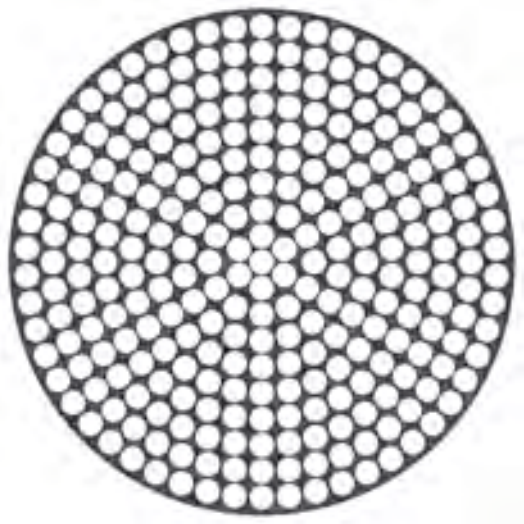 | | 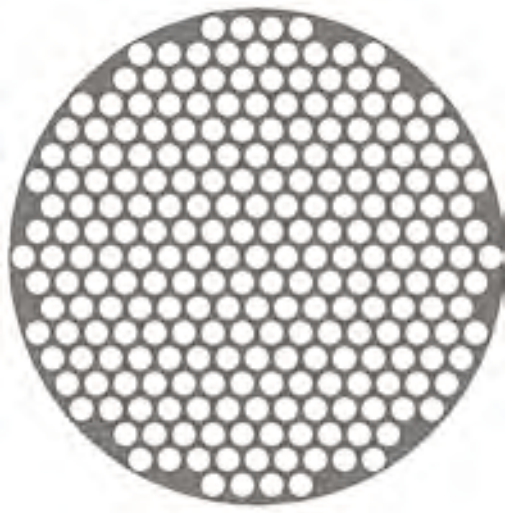 | | 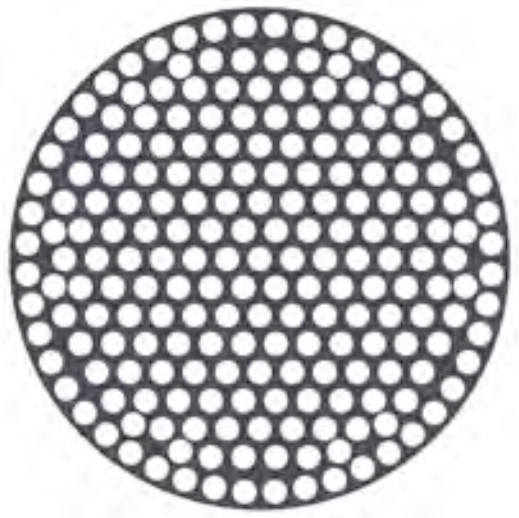 | | 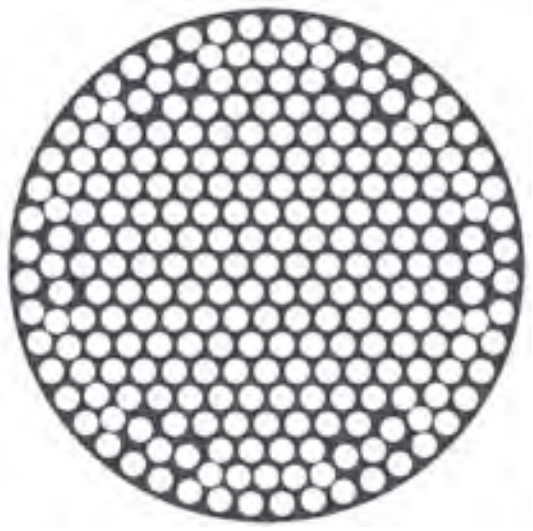 | |
| **Number** | **217** | | **271** | | **241** | | **211** | | **247** | |
| **Configuration** | **Circular 1** | | **Circular 2** | | **Hexagonal dilated** | | **Hybrid 1** | | **Hybrid 2** | |
| **Spacing** | 91.0 mm shell radius | | 80.90 mm shell radius | | 92.38 mm (all) | | 83.51 mm (closest) | | 80.11 mm (closest) | |
| **Notes** | | | Not viable mechanically | | | | 1 row circular | | 2 rows circular, adjusted | |
| **PhC coverage** | 37.3% | | 47.1% | | 43.7% | | 36.1% | | 43.1% | |
| **S1 PDE[1]** | 7.8% | | 8.2% | | 8.1% | | 7.7% | | 8.0% | |
| **S2 pulse: 3,000 emitted photons (270 pe in top array)** | | | | | | | | | | |
| **PMT-anode** | ***rms*, mm** | **L28** | ***rms*, mm** | **L28** | ***rms*, mm** | **L28** | ***rms*, mm** | **L28** | ***rms*, mm** | **L28** |
| **30 mm** | 5.3 | 0.001% | – | – | 4.5 | 2.0% | 5.2 | 0.008% | 4.6 | 0.0001% |
| **50 mm** | 6.2 | 0.045% | – | – | 5.6 | 2.6% | 6.4 | 0.075% | 5.7 | 0.028% |
| **S2 pulse: 1,000 emitted photons (90 pe in top array)** | | | | | | | | | | |
| **PMT-anode** | ***rms*, mm** | **L28** | ***rms*, mm** | **L28** | ***rms*, mm** | **L28** | ***rms*, mm** | **L28** | ***rms*, mm** | **L28** |
| **30 mm** | 9.3 | 1.3% | – | – | 7.8 | 8.5% | 9.3 | 1.1% | 8.1 | 0.2% |
| **50 mm** | 11.0 | 5.5% | – | – | 9.9 | 13.2% | 11.3 | 4.7% | 10.0 | 2.0% |

[1]From streamlined MC, with the baseline optical parameters discussed in the text, and averaged over events at $r=0$, for which the yield is ~7% higher than a full volume average.

With fixed PMT count and grid opacity, Figure 6.5.2.2 shows some of the main results of the parameter scan. In the left panel, we vary the reflectivity of PTFE in the liquid while holding the other parameters fixed at three plausible sets of values. A range of measurements and modeling of light collection in previous detectors strongly indicate that the PTFE will be diffusively reflective over no less than the range of values shown. The net effect of varying the PTFE alone is as much as ~50%, while the range from varying all parameters is more than a factor of 3 in the most extreme cases. In the right panel, we vary the photon absorption length *($l_{abs}$)* over the range of 10 m to 1 km. Values below ~30 m have significantly reduced light collection. We believe such strong absorption is unlikely based on the VUV absorption cross sections of dominant contaminants such as $O_2$ and $H_2O$ and their required concentrations to obtain the mean electron lifetimes specified for LZ.

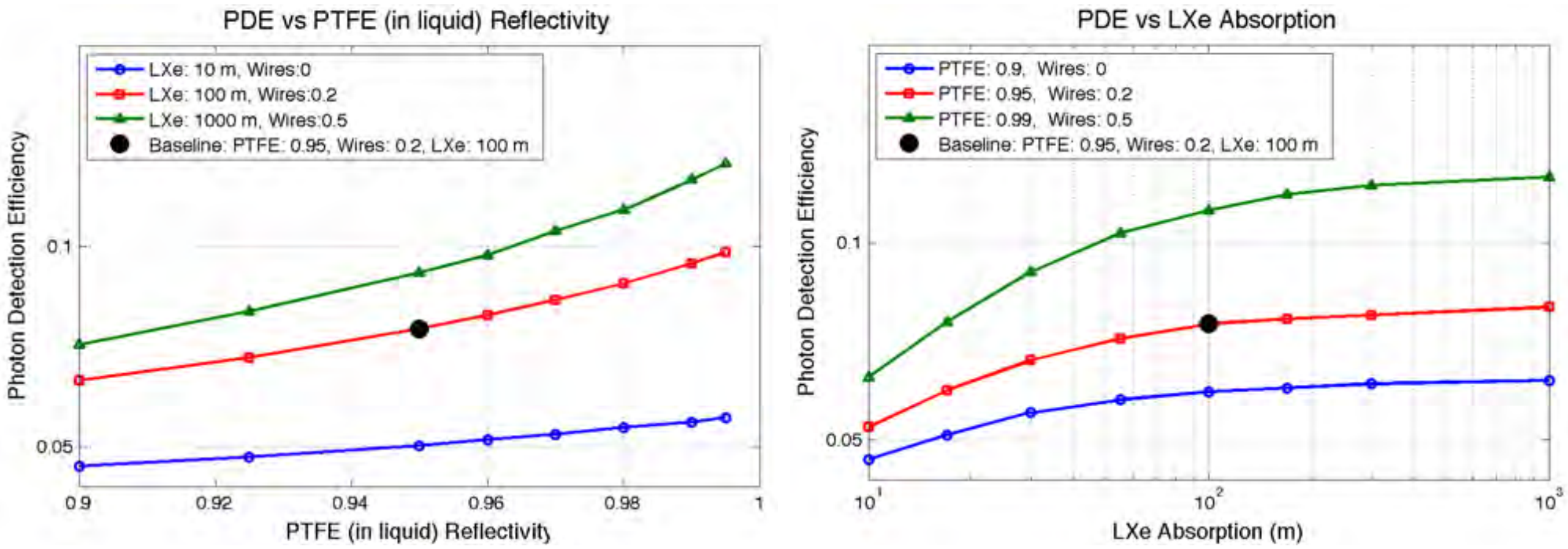

**Figure 6.5.2.2. Varying the main optical TPC parameters affects the photon-detection efficiency, $\alpha_1$, of the LZ TPC. Left: Dependence of the S1 photon-detection efficiency on PTFE-LXe reflectivity for three scenarios of photon-absorption length and SS grid reflectivity given in the legend. Right: Varying the photon-absorption length in LXe.**

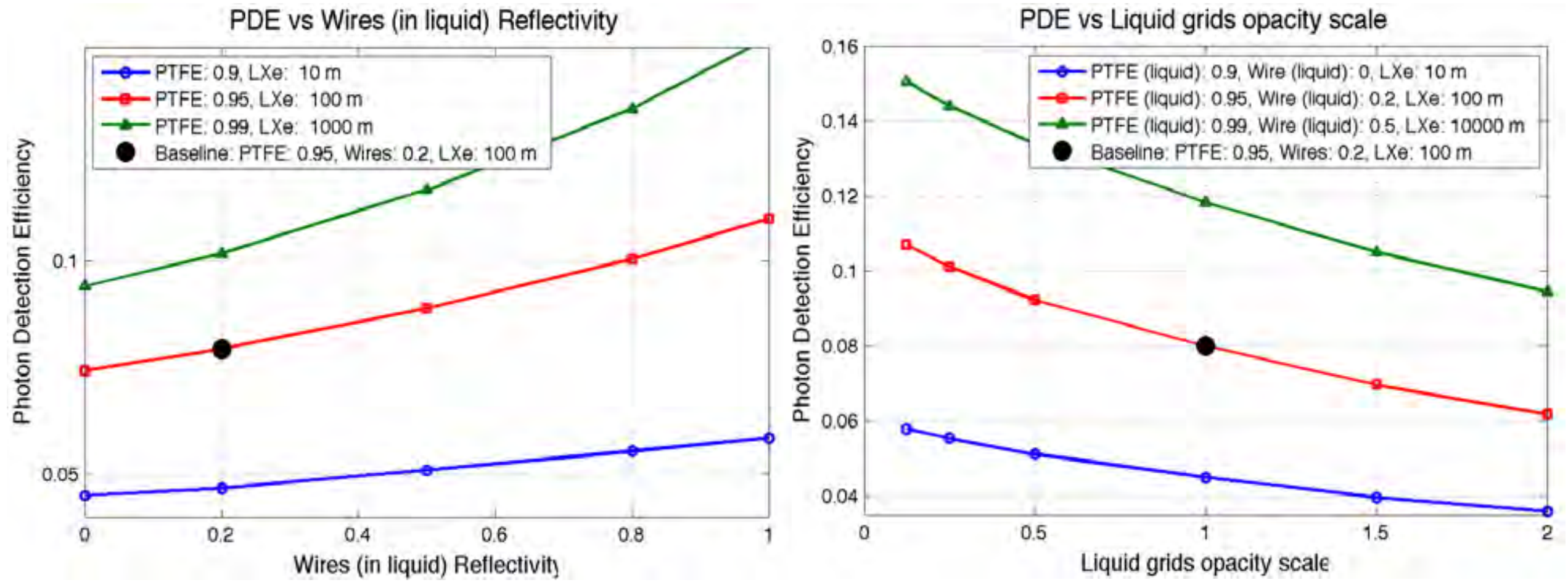

**Figure 6.5.2.3. The effect on photon-detection efficiency, $\alpha_1$, from varying parameters of the grids. In the left panel, the reflectivity of SS grids is varied, while in the right panel, the opacities (i.e., wire diameter/wire spacing) of all of the grids under the liquid surface are varied together relative to the baseline values (listed in Table 6.1.1), and where transparency = 1 – opacity.**

The baseline design of LZ has grids that are relatively opaque in order to reduce electric fields on the wire surfaces. The grids thus have an important effect on light collection, which is explored in Figure 6.5.2.3. The left panel considers a variation in the grid reflectivity, which, if possible, would be achieved by changing the grid material or coating. The limited literature on reflectivity for metals at 178 nm indicates that very high values are unlikely. However, the ongoing HV tests of grids (Section 6.10) may show that more transparent grids are possible. This is explored in the right panel of Figure 6.5.2.3, where all the grids under the liquid surface (bottom shield, cathode, and gate) have their opacity scaled together over a plausible range of values. A 30% improvement or a further 20% loss from the grid opacity alone is possible. A separate scan of the opacity of the grids in the gas (anode and top shield) shows a much weaker (± ~10%) effect. This is fortunate because, as discussed later, achieving uniform S2 signal pushes the anode grid to be more opaque than any other grid in the detector.

In general, when varied over plausible ranges, PTFE reflectivity, grid reflectivity, and liquid absorption length have roughly comparable effects. Somewhat modest gains (or losses) can be achieved by maximizing (or doing less well in) any one of these parameters, while the combined effect of improving (or doing worse in) all three could result in perhaps doubling (or near halving) of the overall light collection. We believe the baseline parameters we have adopted are somewhat conservative, and hope to be able to select materials so that we achieve $\alpha_1$ above 7.5%.

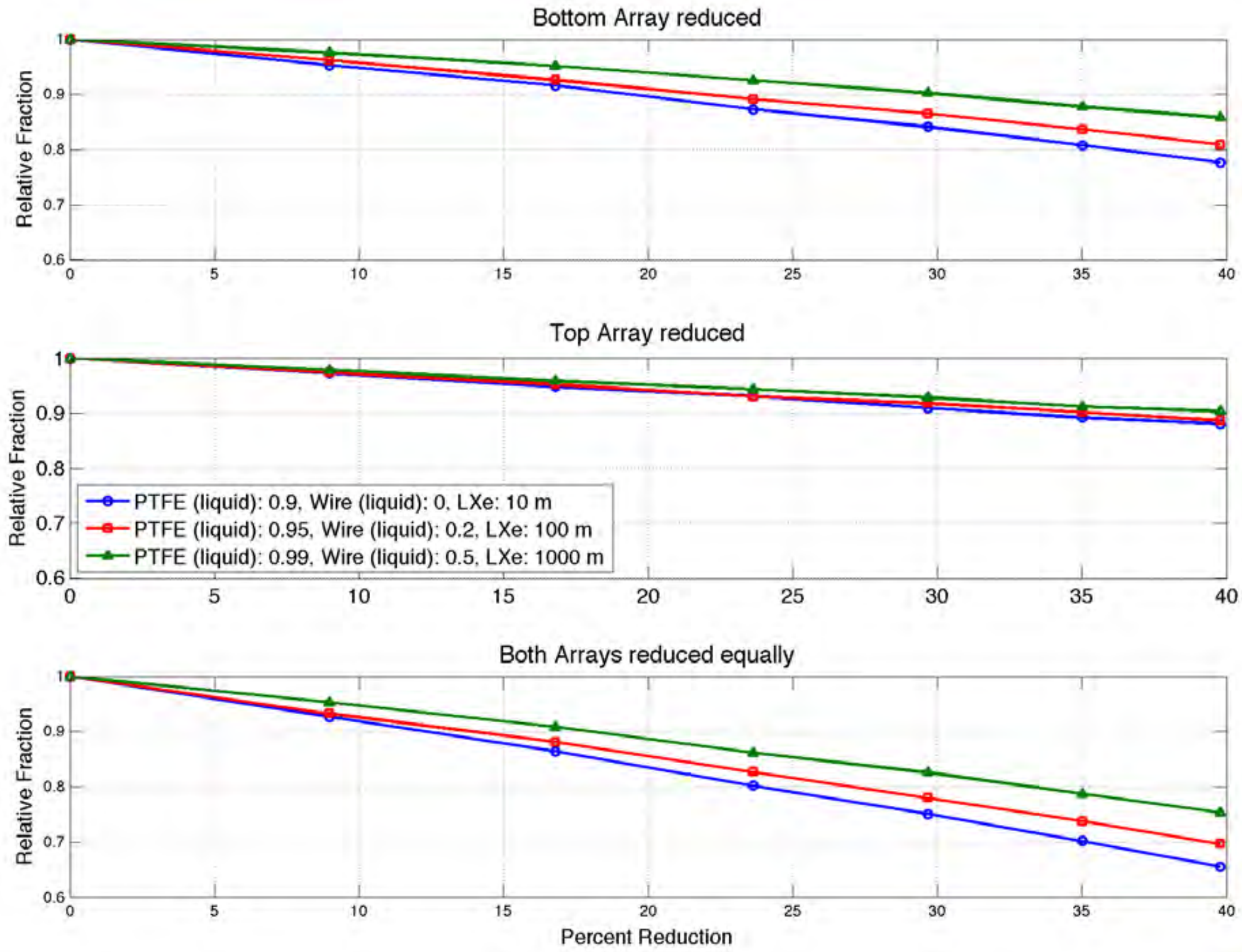

**Figure 6.5.2.4. Effect of reducing the number of PMTs in both arrays on the relative photon-detection efficiency for S1 light for the three scenarios indicated in the legend.**

Concerning the PMT count, the best performance for S1 light is naturally obtained with the maximum packing fraction on the bottom array, which is 241 units. The top array configuration is driven by other considerations, which are described in detail later. The same streamlined simulation was used to assess how the light yield is worsened as the number of PMTs in either or both arrays is reduced. This is shown in Figure 6.5.2.4. As expected, descoping the bottom array is more damaging than descoping the top if the remaining optical parameters are not all extremely good: With more extinction in the chamber, the refractive index mismatches at the liquid/gas and gas/quartz interfaces lower the fraction of S1 light collected by the top array. The bottom array avoids both effects since quartz and LXe are well matched optically.

### 6.5.3 Optimizing the TPC Array Configuration

The design drivers for the two PMT arrays that read out the TPC are not identical, and this motivates the different layouts adopted. The bottom array provides most of the detection efficiency for S1 photons (≈70%), which determines the NR energy threshold and discrimination efficiency. This is mostly due to the total internal reflection at the liquid surface above a critical angle of 36° and the good match in the VUV of refractive indices between the quartz in the PMT windows and LXe. Maximum photocathode coverage of the TPC cross section is therefore the main requirement. In contrast, the principal function of the top array is to reconstruct S2 events and provide spatial resolution in the horizontal plane. Especially critical is the accuracy of reconstructing the x-y position of "wall events" that result from interactions near the vertical cylindrical surface of the PTFE that defines the TPC. The placement of the outer few PMT rows is critical: Misreconstruction of peripheral interactions further into the TPC volume can lead to a significant reduction in fiducial mass.

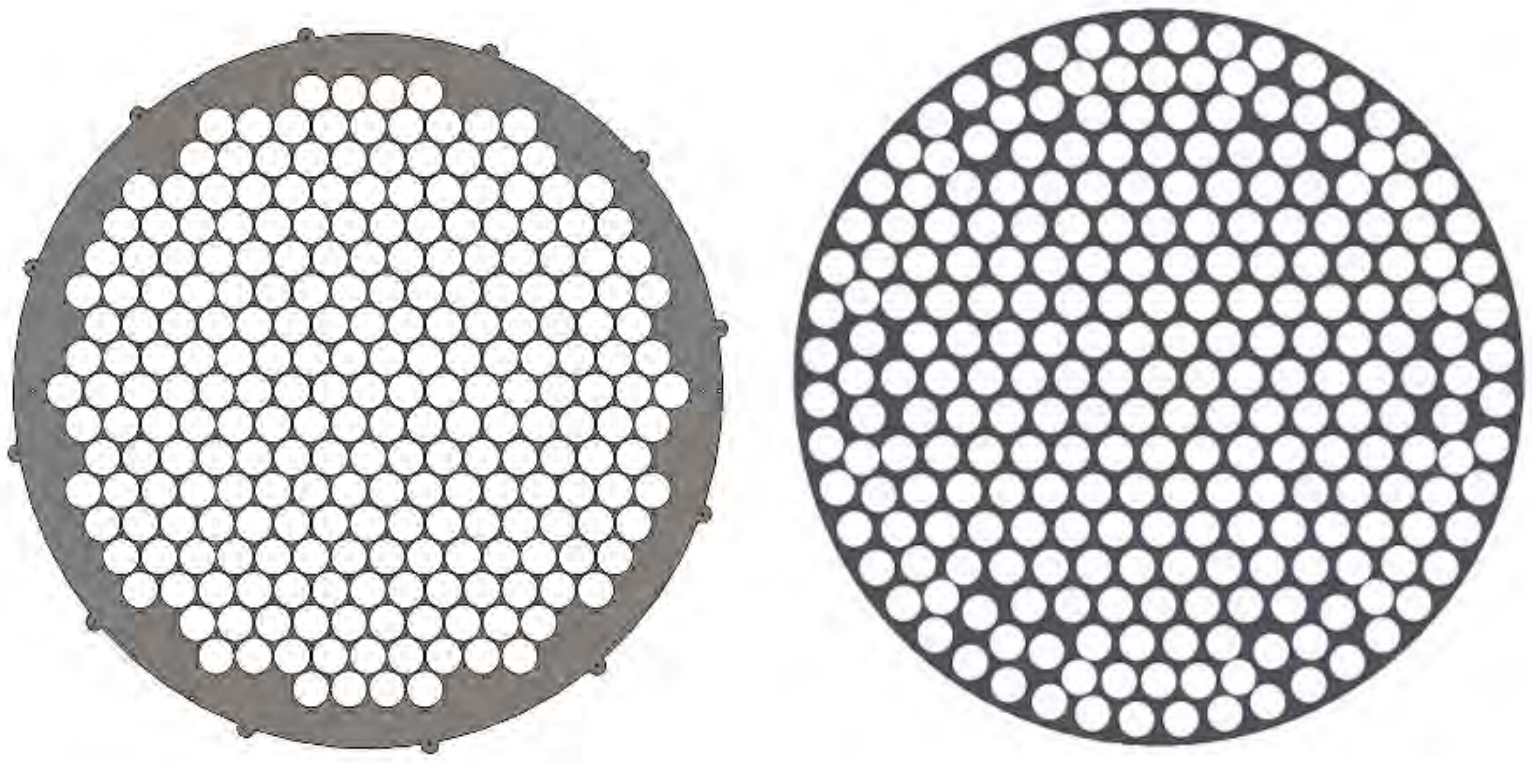

**Figure 6.5.3.1. Left: Bottom array configuration with 241 tubes in close-packed hexagonal configuration. Right: Top array layout with 247 units, a hybrid configuration consisting of a hexagonal matrix with two (nearly) circular outer rows.**

The optimization of the bottom array was straightforward, involving mechanical considerations and Monte Carlo simulation of S1 light-collection efficiency. A close-packed hexagonal layout of 241 tubes fully contained within the TPC diameter (Figure 6.5.3.1, left) was selected as the baseline. The axis-to-axis separation of the PMTs is 91 mm, for a cut-out diameter of 80 mm. Reducing PMT numbers in the bottom array lowers the overall S1 photon-detection efficiency, especially if the photon extinction provided by surface and bulk absorption is more severe than anticipated.

The top array layout is driven to a significant extent by the need to correctly reconstruct low-energy background events from the TPC walls. In particular, radon progeny plated out on the PTFE can lead to a significant population of events, including NRs from $\alpha$ decay (where the $\alpha$-particle goes into the wall) and low-energy ERs, which can also be dangerous due to loss of charge at the wall. In contrast to the bottom array design, which is fully contained within the TPC diameter, the top array must overhang the edge of the TPC or all the reconstruction bias will point inward. Ideally, at least a full row of tubes would be located beyond the inner radius of the chamber. This is not possible due to the proximity of the inner cryostat vessel, and instead we locate the outermost circle of tubes at the largest-possible radius, which aligns the PMT centers above the TPC wall.

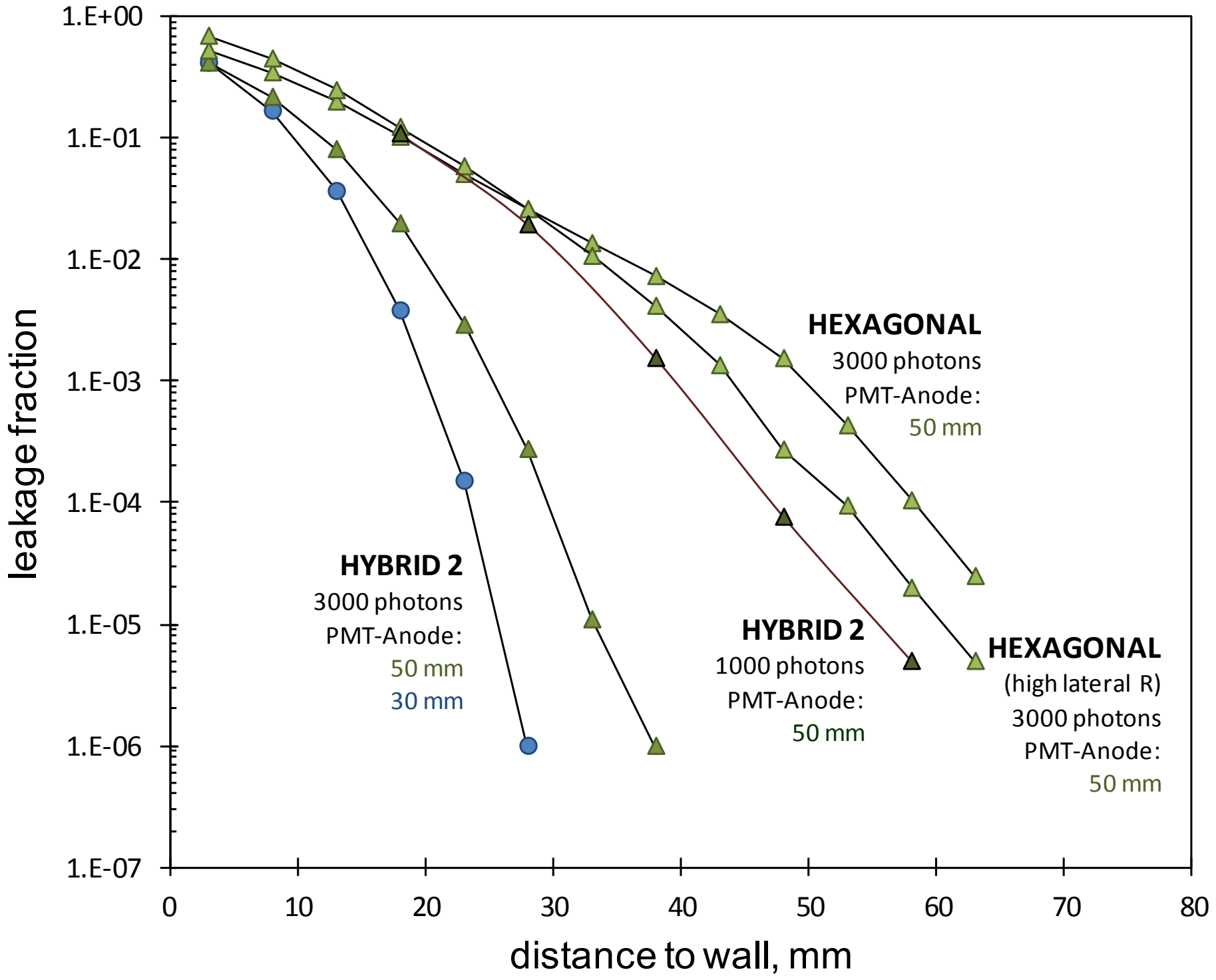

**Figure 6.5.3.2. How the leakage fraction varies with the array configuration for the best and poorest layouts studied, as well as lateral reflectivity in the gas phase (60% for the high-reflectivity scenario and 0% for the others), and the distance between the PMT windows and the anode electrode. The fiducial volume with 5.6-tonne mass starts 39 mm from the wall.**

Five layouts were considered; these are depicted in Table 6.5.1.1, which also summarizes key parameters and the results from the optimization exercise. We include two circular arrays of 217 and 271 units (the latter was subsequently found not to be mechanically viable), a hexagonal array of 241 units slightly dilated with respected to the bottom array, and two circular/hexagonal hybrid arrays containing 211 and 247 PMTs.

The methodology employed to determine which array configuration is best suited to minimize wall leakage involved extensive optical Monte Carlo coupled to the Mercury vertex reconstruction algorithm used in ZEPLIN-III and LUX Run 3 [13]. This provided a realistic assessment of the position resolution of the chamber for very small S2 signals and, in particular, the fraction of peripheral events that is misreconstructed into the TPC volume. This “leakage” fraction was the main design criterion used to select the best array configuration.

Two other design parameters influence the peripheral position resolution and were therefore included in this study. These are the distance between the anode grid and the PMT windows, and the reflectivity of the lateral wall in the gas. Regarding the latter issue, a low-reflectance material is desirable so as not to distort the response of the outer PMTs. Titanium has ≈16% reflectance at 178 nm, but its oxides can be significantly more reflective in the VUV [14]. We studied a high-reflectance scenario (PTFE) as well as values in the range of 0–30% that could be achieved by anodizing titanium or employing a thin layer of a polyimide such as Kapton, which is essentially black in the VUV [15].

For each array configuration, in an initial study the Mercury algorithm was trained to obtain axially symmetric light response functions (LRFs) for each PMT using simulated S2 light. Then, very small S2 signals were simulated and randomly distributed in (*x,y*) or at the TPC walls with 1,000 photons (~90 phe collected in the top array, coming from an initial ~4 emitted electrons) and 3,000 photons (270 phe in the top array, and ~12 electrons). The position was obtained by fitting to all channels simultaneously. The reconstruction *rms* for events near the middle of the detector (more than 130 mm from the wall), and the leakage fraction, are summarized in Table 6.5.1.1. The latter is defined here as the fraction of events located at the wall, which are reconstructed to more than 28 mm from the wall. The hexagonal array performs noticeably worse due to the poor coverage in some positions. The hybrid with 247 PMTs (Figure 6.5.3.1, right) is the best, and this is our baseline. With the exception of the hexagonal array, the other layouts do not perform significantly worse from this point of view. We also considered the distances from anode to PMT array, and the reflectivity of the wall in the gas (at and above where the S2 light is generated). The best performance is obtained with the anode as close as possible to the PMTs.

The reflectance of the wall in the gas region has a mixed effect, sometimes improving and sometimes decreasing performance, an issue that is under further study. Our baseline design is based on a distance of 48 mm between anode and top PMTs, of which 38 mm are between the anode and top shield grid, and 10 mm between this grid and the top PMTs.

Finally, in Figure 6.5.3.2 we show the results from an improved and higher-statistics study of the leakage past the 39-mm distance of the nominal 5.6-tonne fiducial volume. This was done for the baseline hybrid array, and the 241 PMT hexagonal array. The strong signal size dependence of the leakage is apparent, as well as the improvement of the hybrid array over the hexagonal array. The leakage into the fiducial volume is small except at the lowest-possible values of S2.

## 6.6 Optimization of the Electroluminescent (S2) Signal Production

The extraction/electroluminescence region of the TPC is located at the top of the field cage, with the gate and anode electrodes (nominally 10 mm apart) straddling the liquid surface. The liquid level is controlled by a weir system at the edge of the TPC, which is detailed below. This region generates a light signal proportional to the number of electrons drifted away from the interaction site via proportional scintillation in the gas phase, readily providing sensitivity to single electrons emitted from the liquid, as discussed in Chapter 3.

Three main parameters characterize the S2 response: the photon yield, which depends on both the electroluminescence production and the surface extraction probability for electrons; the pulse width, which is proportional to the electron transit time in the gas phase to first order; and the resolution of the S2 signal, which depends on several parameters discussed below. These characteristics depend on operating parameters such as vapor pressure, length of the gas gap, and the voltage applied between the anode and gate electrodes. The S2 performance is also affected by other electrostatic considerations (e.g., maximum fields that can be sustained at the wire surfaces) and mechanical feasibility (e.g., limitations on the manufacture of large wire grids, wire sagging, etc.). In the following sections, we highlight baseline and “maximum” design scenarios, and describe how the S2 response depends on the operating conditions. Key parameters are presented in Table 6.6.1. The size of the S2 signal also depends on the efficiency of light collection. We here assume 9% based on a preliminary Monte Carlo, which uses a crude treatment of the grids, though we anticipate that this value may decrease as the grids are treated more accurately.

#### Baseline Design

For the smallest S2 signals, generated by 1 to a few ionization electrons, the main S2 requirements are: (1) definition of the single-electron response with a high S/N ratio, to allow absolute calibration of the ionization channel, and to enable physics searches down to S2 signals as small as a few electrons; and (2) sufficiently large S2 signal for accurate reconstruction of the x-y location of peripheral interactions, such

**Table 6.6.1. Summary of electroluminescence region design parameters.**

| | Baseline | Maximum | |
|---|---|---|---|
| **Gate, kV** | −4 | −7 | Grid: 5 mm / 100 µm* |
| **Anode, kV** | +4 | +7 | Grid: Woven wire mesh |
| **Top grid, kV** | −1.0 | 0 | Grid: 5 mm / 50 µm* |
| **Extraction field, kV/cm** | 10.6 | 11.1 | In gas |
| **Distance from liquid surface to anode, mm** | 5 | 10 | |
| **S2 yield** | | | |
| **Photons/e-** | 550 | 1,200 | (Parallel field approx.) |
| **phe/e-** | 50 | 110 | (PDE=9% top array) |
| **Fields (kV/cm)** | | | |
| **On surface of gate wires** | 75 | 79 | In liquid, with cathode at 100 kV |
| **Above anode** | 1.3 | 1.8 | In gas |
| **On surface of top shield wires** | 25 | 10 | In gas |

* Wire pitch / diameter

as those arising from contamination on the TPC walls. This motivates a photon yield of at least ~50 photoelectrons detected in the top array per emitted electron (cf. ≈31 in the first run of ZEPLIN-III [16] and ≈25 in LUX Run 3 for both arrays [17]). More details on how the S2 pulse size affects the reconstruction of wall events are given in Section 6.5.3.

Considering an S2 photon-detection efficiency of ~9% for the top array, predicted by simulation, the above photoelectron yield implies 550 photons generated per emitted electron. For a gate-anode distance of 10 mm with the liquid level halfway between them — our nominal design — this is achieved with a gate-anode voltage of 8 kV at the operating pressure of 1.6 bar, as shown in Figure 6.6.1. We plan to apply +4 kV to the anode and −4 kV to the gate, leaving the liquid surface near −1.3 kV.

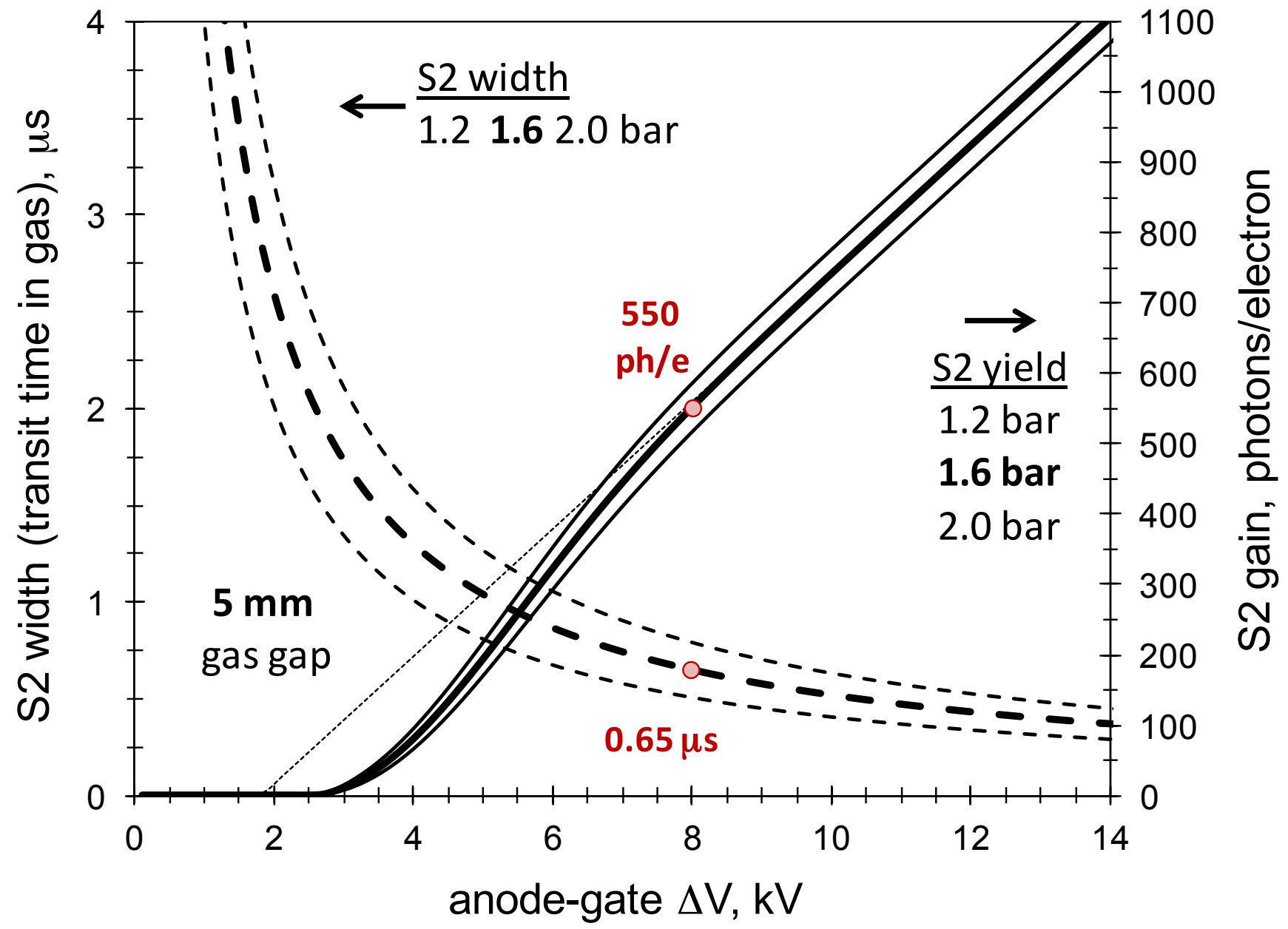

**Figure 6.6.1. Dependence of the S2 photon yield and S2 pulse width (for emitted electrons, i.e., ignoring longitudinal diffusion in the liquid) on the voltage between anode and gate electrodes. At the nominal ΔV=8 kV, the photon yield [18], including the electron emission probability [19] and the electron transit time in the gas phase (S2 pulse width) [20] (for operating pressures around the 1.6 bar nominal and a gas gap of 5 mm) are shown.**

The required yield can be achieved with other combinations of anode-gate separation ($L$), length of the gas gap ($L_g$), overall applied gate-anode voltage ($\Delta V$), and vapor pressure (P). All of these parameters are intimately connected to S2 light production: Both the electroluminescence yield and the electron drift velocity in the gas are determined by the reduced electric field in that region, $E/P$; in addition to the applied voltages, the electric field depends on both $L$ and $L_g$. Therefore, these parameters must be studied together and their optimization is subtle. We describe below some of the main arguments that motivated our baseline design for the electroluminescence region, with reference to Figures 6.6.1 and 6.6.2. We postpone the discussion of how the actual electrodes are implemented until the end of this section, focusing here on mean yield values only.

The electron emission probability at the liquid surface decreases rapidly when the field in the gas drops below 10 kV/cm [19]. In Figure 6.6.1, the S2 yield assuming full extraction efficiency is represented by the dotted line, while the continuous lines include the field-dependent extraction probability. For our nominal parameters, that probability is close to unity. Increasing the S2 yield by increasing the length of the gas phase may appear desirable, but it may lead to low extraction if nominal voltages fail to be achieved, or if they need to be reduced to preserve linearity for larger signals.

Longer electron transit times in the gas also hide the effect of electron diffusion in the liquid, which encodes interaction-depth information on the S2 pulse shape. This information allows some coarse fiducialization, which is important for the S2-only analysis. On the other hand, too short an S2 signal may be adverse for robust pulse identification, causing confusion with S1 pulses and other topologies; we wish to maintain this parameter to be $\gtrsim$0.2 μs. Very short gas gaps are also problematic for practical reasons: the need to level the detector with extremely high precision, to cope with the inevitable sagging and electrostatic deformation of the anode and gate grids, and to achieve homogeneous fields above the gate, which has a wire pitch of order millimeters. A nominal anode-gate distance of 10 mm with a gas thickness of 5 mm is a reasonable compromise, leading to a transit time of ~0.65 μs for ΔV=8 kV.

### High-yield Design Maximum

Our understanding of the reconstruction of "wall events" is still evolving, informed by LUX data and optical simulations such as those described in Section 6.5.3. Although the S2 gain cannot be arbitrarily large due to the limited dynamic range of the optical readout, we will continue to consider in parallel a more aggressive S2 design with a larger gas gap of 10 mm (for an electrode separation of 15 mm) and higher applied voltages, $\Delta V$=14 kV. These changes require only modest hardware modifications. Although this design involves higher fields near the electrodes and may compromise the diffusion information due to longer transit times in the gas, it is also less aggressive on mechanical requirements involving detector leveling and grid sagging — besides doubling the S2 gain. The mean S2 yield for this scenario is ~1,200 photons per emitted electron (110 phe/e) and the mean transit time in the gas is 1.3 μs. This design involves application of higher voltages to electrodes in the gas phase, which may create regions with electric field above the electroluminescence threshold of ~2 kV/cm; we avoid this in the region between the anode and the top grid (which protects the top PMT array) by bringing that electrode to ground in this case. Another critical parameter is the maximum field at the surface of the anode wires, which is addressed below.

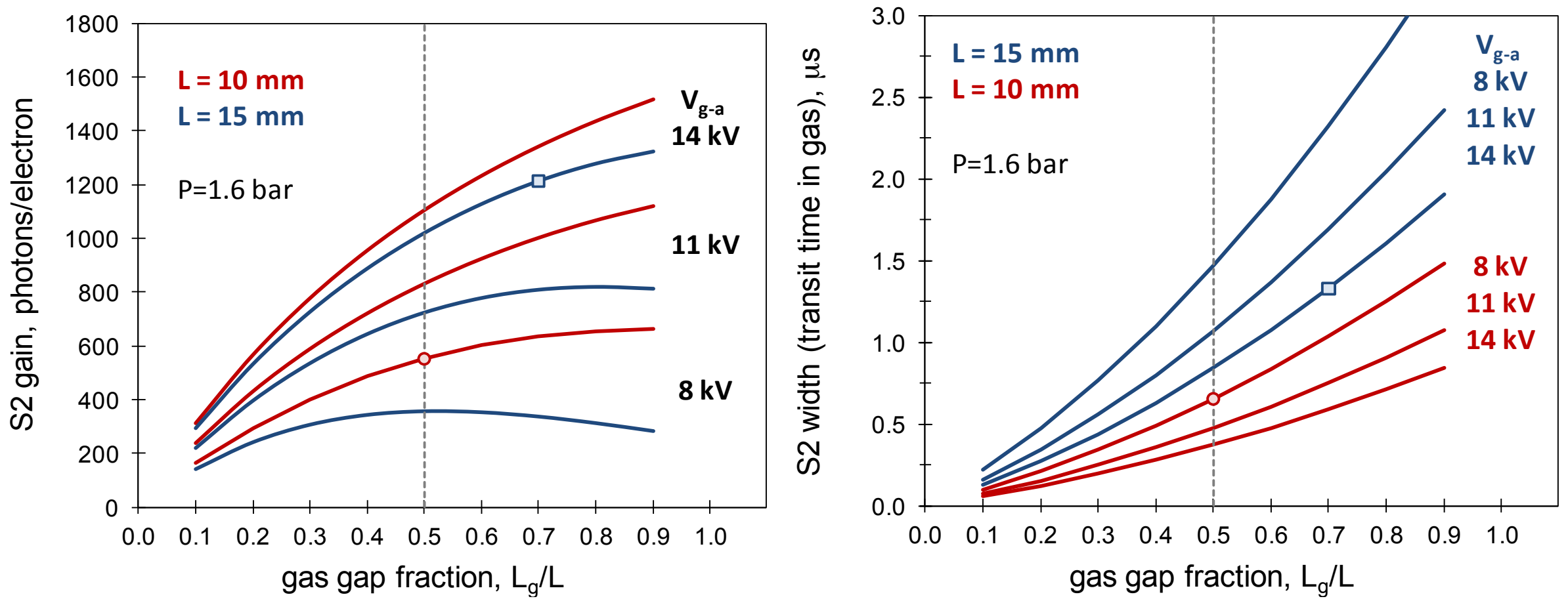

**Figure 6.6.2. Variation of S2 gain (left) and S2 width with gas gap for anode-gate distance of 10 mm and 15 mm, and for several gate-anode voltages. The S2 gain includes electroluminescence photon yield and emission probability. The S2 width applies only to emitted electrons since it ignores diffusion in the liquid. Baseline and design maximum scenarios are indicated by the round and square markers, respectively.**

### Electrode Configuration and S2 Energy Resolution

In addition to appropriate S2 gain and pulse width, we must ensure that the S2 resolution is as good as it can be, primarily so that that ER/NR discrimination at low energies is not compromised by the adopted S2 design and, more generally, that the S2 channel has high resolution, especially at higher energies. This is intimately related to the quality of our calibration (e.g., $^{85}$Kr signals with ~1,000 ionization electrons) and the characterization of detector backgrounds (e.g., radioactivity gamma rays with up to ~$10^5$ electrons). Our goal is for the ER bandwidth (a key parameter for ER/NR discrimination) to be dominated by recombination fluctuations in the liquid — which affect the number of electrons extracted from particle tracks — and S1 light-collection and photoelectron statistics, since these parameters cannot be improved easily. Therefore, fluctuations related to S2 photon production and measurement must remain small, at the level of a few percent. This motivated the detailed study of electroluminescence and electrode grid configuration, which is summarized below. Other factors contributing to the S2 resolution are the uniformity of response in the horizontal plane over the scale of the whole TPC diameter (e.g., wire sagging and electrostatic deflection), though those in principle can be calibrated and hence removed.

Aside from diffusion (in both gas and liquid phases), drifting electrons follow electric field lines and therefore their length and the field strength close to the wires must be carefully controlled to avoid substantial dispersion or even the possibility of significant charge multiplication (the first Townsend coefficient for cold Xe vapor at our operating pressure reaches ~1 e/mm at 35 kV/cm [20,21]). Two additional concerns, which are intimately related, are the HV resilience of the electrodes and their VUV reflectivity, which depend strongly on the wire surface quality and material or coating; we are investigating these issues through the dedicated R&D activities described in Section 6.10.

The optimization of the anode geometry involves a compromise between optical, electrostatic, mechanical, and electroluminescence properties. The latter were assessed through full electron transport modeling, in particular examining the S2 photon production statistics from single electron drifts in the gas phase of the various candidate geometries. A combination of software was used for this purpose. Garfield++ is a Monte Carlo simulator for electrons in drift chambers [22]. Electrons are microscopically tracked as they drift, and the locations of any excitations or ionizations are recorded. An excitation is assumed to produce one photon, and ionizations give extra electrons, which are also tracked. It can calculate electric field maps for simple configurations where an analytical solution exists. This limits it to 2-D geometries consisting of planes and wires. To calculate the electron transport properties of the gas, Garfield++ is interfaced to Magboltz [21], which relies on elastic and inelastic cross sections for gases to calculate the relevant transport parameters (drift, diffusion, and gain). To simulate electric fields from 3-D geometries, field maps were created using the Elmer solver [23] and the meshing tool Gmsh [24]. The field map was then read by Garfield++ for tracking. In Figure 6.6.3, we show equi-field contours for unit cells of three of the wire-grid geometries considered: a simple crossed-wire arrangement, which allows low mechanical deformation relative to a parallel wire plane; a fine-woven mesh such as that used in LUX; and an etched hexagonal grid-set similar to that used in XENON100 [25]. The latter geometry required implementation of the gate electrode to simulate electron focusing; these hexagons were created with three different wire-rounding values of 0%, 10%, and 40%.

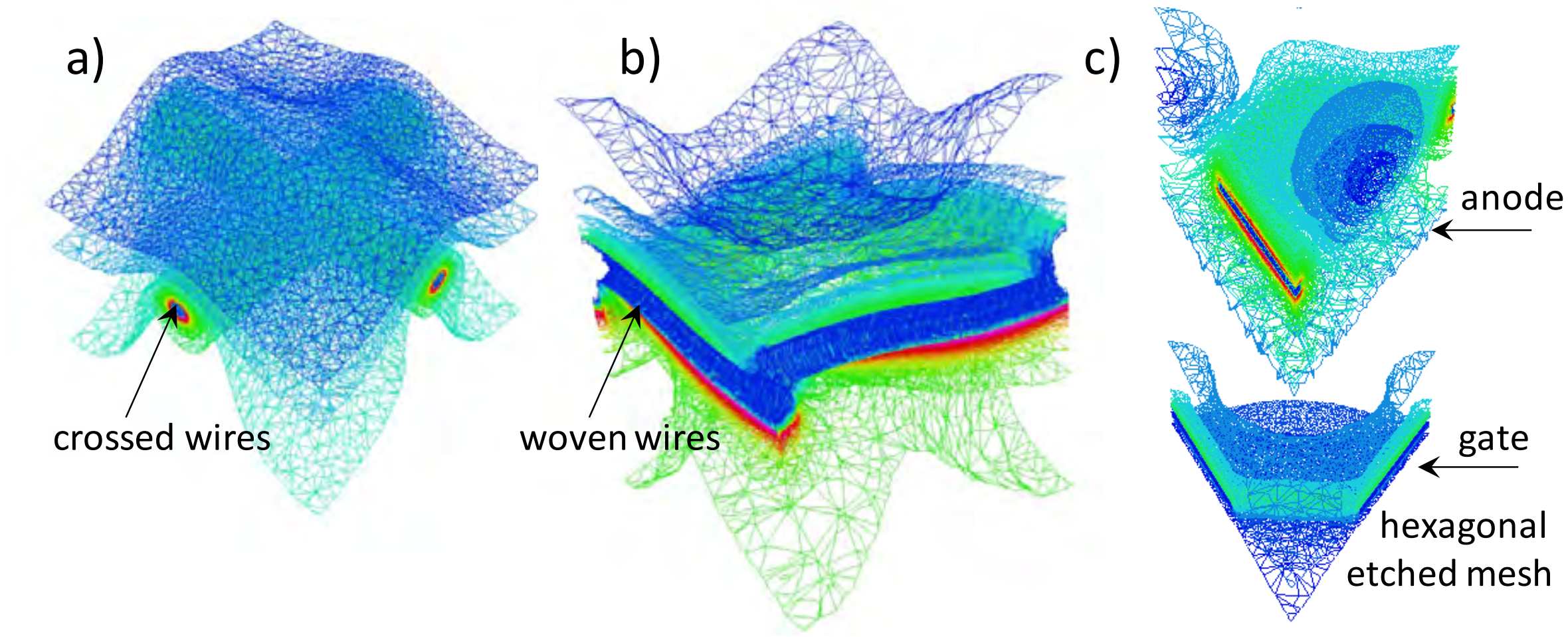

**Figure 6.6.3. Electrostatic models for several anode configurations calculated using Elmer [23] and meshed with Gmsh [24]. The crossed-wire mesh (a) and the LUX-style woven mesh (b) are both candidate configurations for the LZ anode; in (c) we show the unit cell for the XENON100-style hexagonal etched meshes (gate also shown).**

In our simulation study, we confirmed that the woven mesh used in LUX produces very small dispersions of photon production (typically <1%) for a range of operating conditions, never far from the values obtained for the parallel configuration in ZEPLIN-III, where the anode was a solid plate (no top PMT array and no gate grid) [26]. We compare these values with the fluctuations expected from photoelectron statistics for single electron signals, which is of order 15% and, more importantly, with the best S2 resolution achievable at MeV energies by EXO, which is 1–2% [27]. It is therefore desirable to keep photon-production fluctuations to a maximum of ~2% so that this does not compromise the performance of the instrument for background characterization, calibration, and non-WIMP signals.

The fine LUX woven mesh has 30-μm wire diameter and 250-μm pitch and is our baseline for LZ if a suitable manufacturer can be identified that can accommodate the large diameter required. The optical transparency of this mesh is acceptable (80%) and the maximum electric field at the wire surface is modest. The properties for this choice are listed in Table 6.6.2.

**Table 6.6.2. Main electroluminescence properties calculated for a LUX-like woven-mesh anode (LZ baseline); a gate-anode voltage difference of 8 kV is assumed, along with a nominal gas gap of 5 mm. The S2 photon yield is calculated as described above and also from the simple parallel-plate approximation and experimental electroluminescence yields from [18], for direct comparison with Figure 6.6.1.**

| | Value | Notes |
|---|---|---|
| **Wire diameter** | 30 μm | |
| **Wire pitch** | 250 μm | |
| **Optical transparency** | 79.7% | Normal incidence |
| **Maximum wire field** | 33 kV/cm | Elmer |
| **Photon yield** | 583 ph/e | Garfield++ |
| **(Parallel plate)** | 550 ph/e | Parameterized yield |
| **Photon RMS** | 0.26% | Garfield++ |

**Table 6.6.3. Alternative anode configurations explored through Garfield++ simulations. The table lists, for parallel and crossed-wire configurations: the mean S2 photon yield per electron emitted from random locations below the unit cell (ph/e) and relative width of the photon distribution (percent *rms*), both given to last (statistical) significant digit; the maximum electric field at the wire surface, E*, in kV/cm; the optical transmission at normal incidence, T. A gate-anode voltage of 7.5 kV was assumed in this case, which is lower than the adopted value of 8 kV; a nominal gas gap of 5 mm is considered in all cases.**

| Wire pitch → | 1 mm | | | | 2 mm | | | | 3 mm | | | |
|---|---|---|---|---|---|---|---|---|---|---|---|---|
| | ph/e | *rms* | E* | T | ph/e | *rms* | E* | T | ph/e | *rms* | E* | T |
| **Parallel wires** | | | | | | | | | | | | |
| **50.8 μm** | 530.5 | 1.23% | 60 | 94.9% | 572 | 8.9% | 105 | 97.5% | 1066 | 47% | 140 | 98.3% |
| **101.6 μm** | 529.9 | 0.63% | 36 | 89.8% | 529.1 | 2.53% | 62 | 94.9% | 560 | 9.7% | 82 | 96.6% |
| **203.2 μm** | 530.3 | 0.53% | 22 | 79.7% | 527.1 | 1.23% | 36 | 89.8% | 523.2 | 3.1% | 48 | 93.2% |
| **Crossed wires** | | | | | | | | | | | | |
| **50.8 μm** | 530.7 | 0.69% | 30 | 90.6% | 527.2 | 1.91% | 56 | 95.1% | 528.8 | 4.3% | 76 | 96.7 |
| **101.6 μm** | 531.1 | 0.68% | 19 | 82.4% | 528.4 | 1.36% | 51 | 90.6% | 526.3 | 2.49% | 47 | 93.6 |
| **203.2 μm** | 531.5 | 0.72% | 15 | 69.1% | 528.3 | 1.41% | 21 | 82.4% | 525.2 | 2.19% | 25 | 87.7 |

We studied also three other anode configurations that could be suitable for LZ: grids made from parallel wires or from crossed wires, as well as the hexagonal anodes of the type used in XENON100, made from chemically etched SS plate. The latter offers high optical transparency and acceptable photon production dispersion (≈1.2%), but it leads to significant fields at the metal surface (~100 kV/cm) and the potential for some charge multiplication, which indeed we recorded in some of our simulations.

We focused instead on the parallel- and crossed-wire configurations listed in Table 6.6.3. Clearly, the parallel-wire grid is less attractive from a mechanical point of view, since it lends itself to significant deformation — and it leads to higher fields at the wire surface (up to a factor of 2 relative to the crossed-wire version) and higher S2 dispersion in general. A good alternative to the very fine-woven mesh used in LUX is the crossed-wire grid with 100-μm wire at 2-mm pitch. The remaining configurations are mostly acceptable, except for the 50-μm wire at 3-mm pitch (*rms*>4%) and 200-μm wire at 1-mm pitch (T=69%).

## 6.7 Design and Optical Performance of the Skin Detector

The design of the TPC located inside the LXe inner vessel requires that a physical buffer region be between them. This region provides necessary mechanical clearance to allow detector assembly, houses detector instrumentation including the PMTs, and, most importantly, is used to limit the maximum electric field gradients by providing a standoff between biased TPC components and the electrically grounded inner vessel. We instrument this buffer region, or "Xe skin," as part of our anticoincidence strategy to identify backgrounds that scatter within these regions with high efficiency. The Xe skin is divided into two primary functional regions: a cylindrical (side) skin region outside of the main TPC field rings, and a dome skin region underneath the TPC, below the bottom PMT array.

The side and dome skin regions contain a total of more than 2 tonnes of LXe and are viewed by 180 dedicated 1-inch R8520 PMTs. This Xe skin detector performs a similar and complementary function to the outer LS detector. If the skin regions were filled with passive filler material, the efficiency of the background rejection by the veto would be substantially degraded due to the absorption of secondary scatters that would otherwise have been tagged in the outer detector. The goal for the Xe skin detector design is to achieve a clear anticoincidence detection threshold for ER scatters directly from gamma-ray backgrounds, or gamma rays from thermal neutron capture, at a deposited energy of 100 keVee in 95% of the volume of the Xe skin.

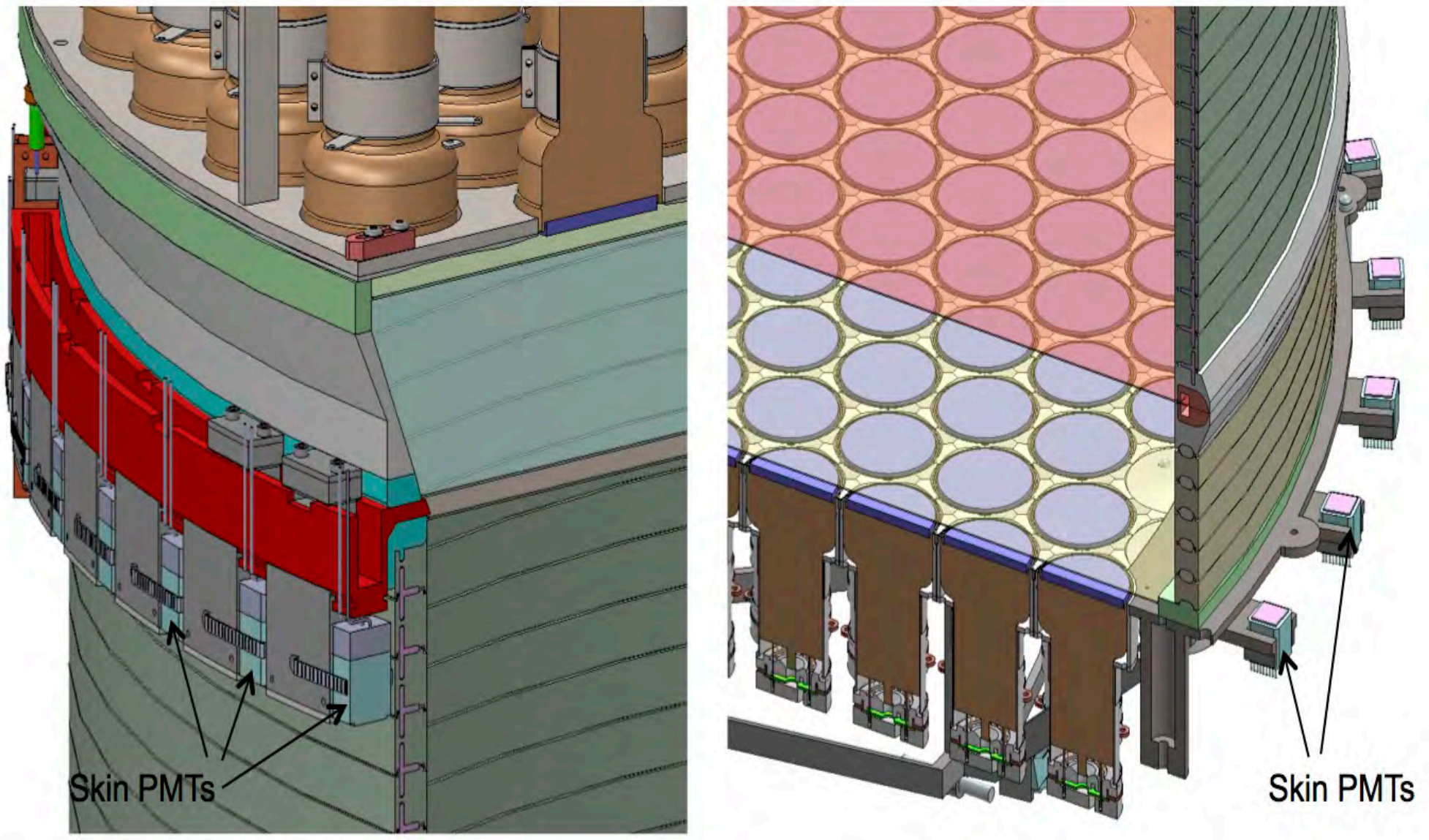

**Figure 6.7.1. Arrangement of skin photomultipliers. Left: Side PMTs near top TPC array. Right: Side and dome PMTs below bottom TPC array. Extensive PTFE lining is required to minimize photon extinction in the skin region.**

It should also be noted that for events that deposit energy in both the main TPC and the LXe skin region, leakage of light from the skin to the TPC can compromise the primary S1/S2 background rejection if the (non-leaking) signal in the skin is below threshold. This is because the leakage light from skin will add to the S1 signal from the TPC and lower the S2/S1 ratio used for particle discrimination. Random coincidences between the two regions can also lead to false vetoing of fiducial interactions. For these reasons, it is important to directly instrument the outer LXe to clearly identify events with a scattering vertex in the outer region and also to minimize light leaks between inner and outer regions with good design of the intermediate wall.

The side skin region will be 4 cm wide near the top of the TPC, increasing to 8 cm in the lower half due to the tapered vessel shape. In the baseline design, this region is instrumented with 60 1-inch R8520 PMTs viewing down located just below the LXe surface, and a further 60 looking up located at the same level as the lower PMT array. The inside surface of the inner cryostat vessel is lined with thin PTFE sheets for improved light collection. The outer surface of the TPC also provides a PTFE reflector.

The dome skin region below the lower PMT array will be instrumented with another 60 1-inch R8520 PMTs. PTFE will be used to cover the components in this region to improve light collection with an overall goal of 95% coverage. The R11410 and R8520 PMTs will have reflective PTFE sleeves to reduce photon absorption on their side and rear walls. The placement of the skin photomultipliers is depicted in Figure 6.7.1.

The skin region uses the 1-inch PMTs, rather than the larger model in the TPC due to mechanical constraints. The Hamamatsu R8520 is specifically designed for LXe operation. It was the primary PMT used, for example, in the XENON10 and XENON100 detectors [12,25]. This is a very compact 1-inch-square PMT with quartz window and bialkali photocathode with a typical QE of 30% at 175 nm. A gain of $10^6$ is provided by an 11-stage metal channel dynode chain. These can be operated with passive voltage divider bases with either negative or positive bias.

The design studies for the LXe skin used a detection threshold goal of 100 keVee for at least 95% of the volume of both the side and dome regions. Threshold detection requires a 95% efficiency for observing at least 3 phe coincident (in a 100-ns window) in the R8520s from the primary scintillation light event. The R8520 PMTs are assumed to have typical 30% QE, and the light yields in the LXe are assumed to be suppressed (by as much as 65% from zero field values) by the electric fields between the TPC and the inner vessel wall, which will be present during the operation of the TPC. A PMT coverage of 60 upper + 60 lower 1-inch PMTs was able to achieve this veto threshold performance goal in the dome skin region. As shown in Figure 6.7.2, the lowest efficiency for light collection in the skin region occurs in the region equidistant between the upper and lower PMTs, where it falls by over a factor of 5 compared with the regions closer to the skin PMTs. Given the simulated light collection, the energy threshold target of 100 keV in the side skin can be achieved assuming a conservative value of 95% reflectivity for the walls of the lateral skin region. The effective energy threshold scales close to linearly with the number of PMTs used in this region.

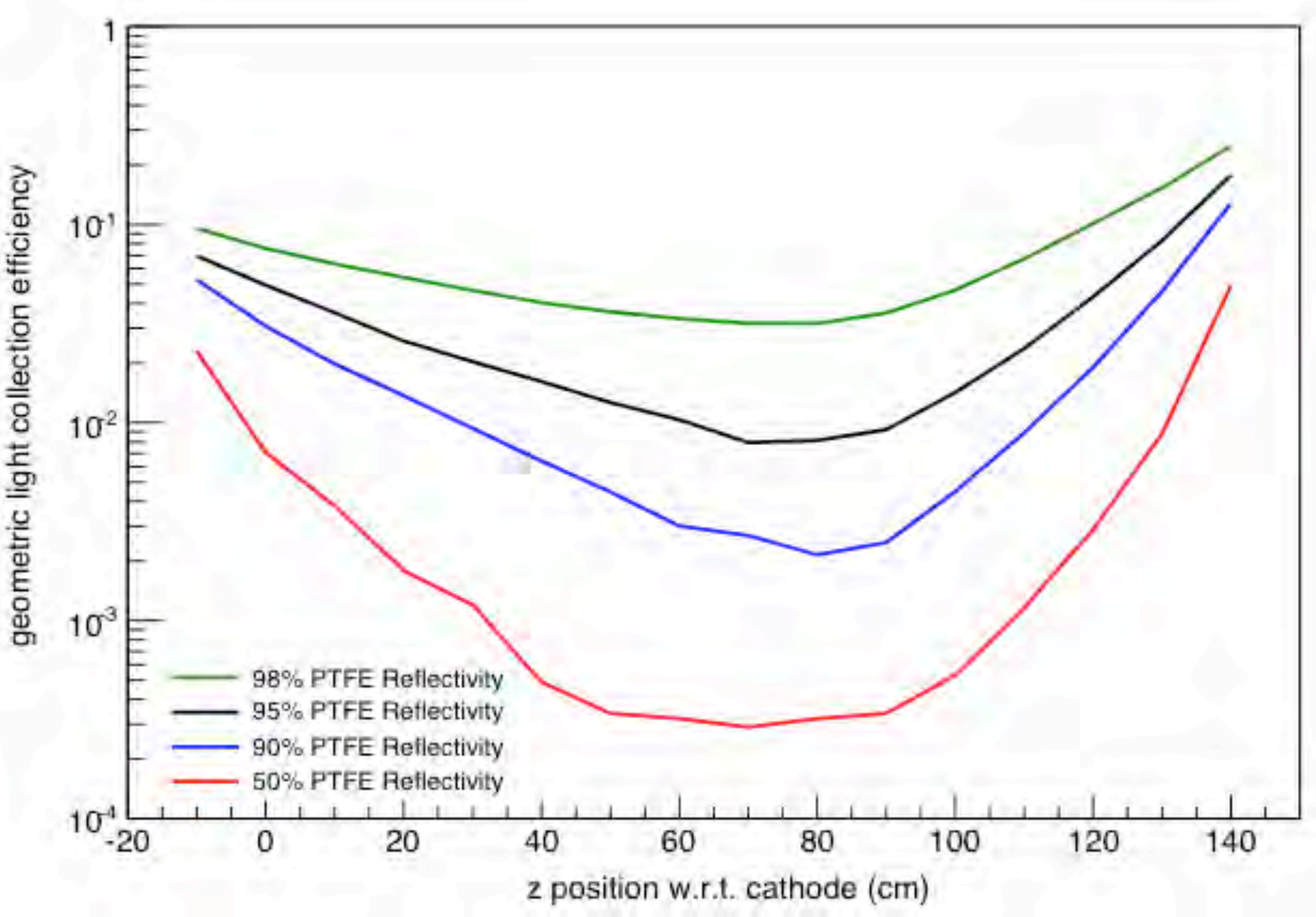

**Figure 6.7.2. Effective light-collection efficiency calculated from light simulations, for a vertical profile in the LXe side skin region, using a range of assumptions for the reflectivity of the PTFE lining the region. The position of the interaction is measured relative to the cathode in the TPC. The vertical scale is light-collection efficiency, taking into account absorption in the walls and liquid (but not the QE of the tubes). A 100 keVee event in a high field region yields 2,150 VUV photons. A value of just under 1% LCE (bottom of 95% reflectivity curve) corresponds to a 95% detection efficiency for 3 phe when using 60 top and 60 bottom R8520 PMTs.**

The dome skin region is a less regular shape than the side region, and houses a number of components, including the bottom TPC PMTs, the Ti plate and trusses to support the array, bases, cabling, and LXe fluid plumbing. More conservative assumptions were made in the simulations for the net reflectivity of the surfaces in this space, using only 90% average reflectivity to account for breaks in the PTFE reflectors. In simulations, in order to achieve a 100 keVee identification threshold (with 95% likelihood of ≥3 phe detected) for interactions in 95% of the LXe volume (i.e., voiding the requirement for the 5% of LXe skin that is most difficult to collect light from), it is necessary to use 60 x 1-inch PMTs. The baseline design calls for thirty PMTs to be housed in a ring at the periphery of the region, interspersed with the lower PMTs of the LXe skin, but pointing downward, rather than up. The other 30 will be distributed on the truss structure. To maximize the light signals, the rear and sides of the R11410 and R8520 PMTs are sleeved in PTFE (see Figure 6.4.3.2) and their bases covered with a PTFE cap. In addition, the internal fluid piping and cabling trusses and the surface of the inner vessel will be covered with, or made from, that material. Again, in the light propagation models tested, increasing PTFE coverage to include all components significantly reduces the number of PMTs required to cover this region. Economizing on the number of PMTs comes at the expense of more PTFE reflectors.

## 6.8 Internal Fluid System

Efficient purification of LXe is a significant challenge, and is especially important given the large size of LZ and the resultant long electron drift lengths and long photon path lengths. Purification is discussed in detail in Chapter 9, and the overall internal flow diagram is shown schematically in Figure 6.8.1. Liquid in the detector is continuously circulated to a purification tower located outside the water tank, where it is evaporated in a two-phase heat exchanger and passed to a gas system. There, it is purified by a commercial heated getter. While the getter is highly efficient in a single pass, continuous purification has proved necessary in most previous such detectors, primarily because of the large amount of PTFE and other plastics (e.g., cables) in the TPC that serve as a long-term source of outgassing. After passing through the getter, the Xe returns to the liquid tower, where it is recondensed in the two-phase heat exchanger, degassed and subcooled, and then passed back to the detector. In addition, separate gas flow through the external purification system purges the spaces above the liquid. The overall flow rate is 500 slpm of gas, or roughly 1 liter/min liquid flow.

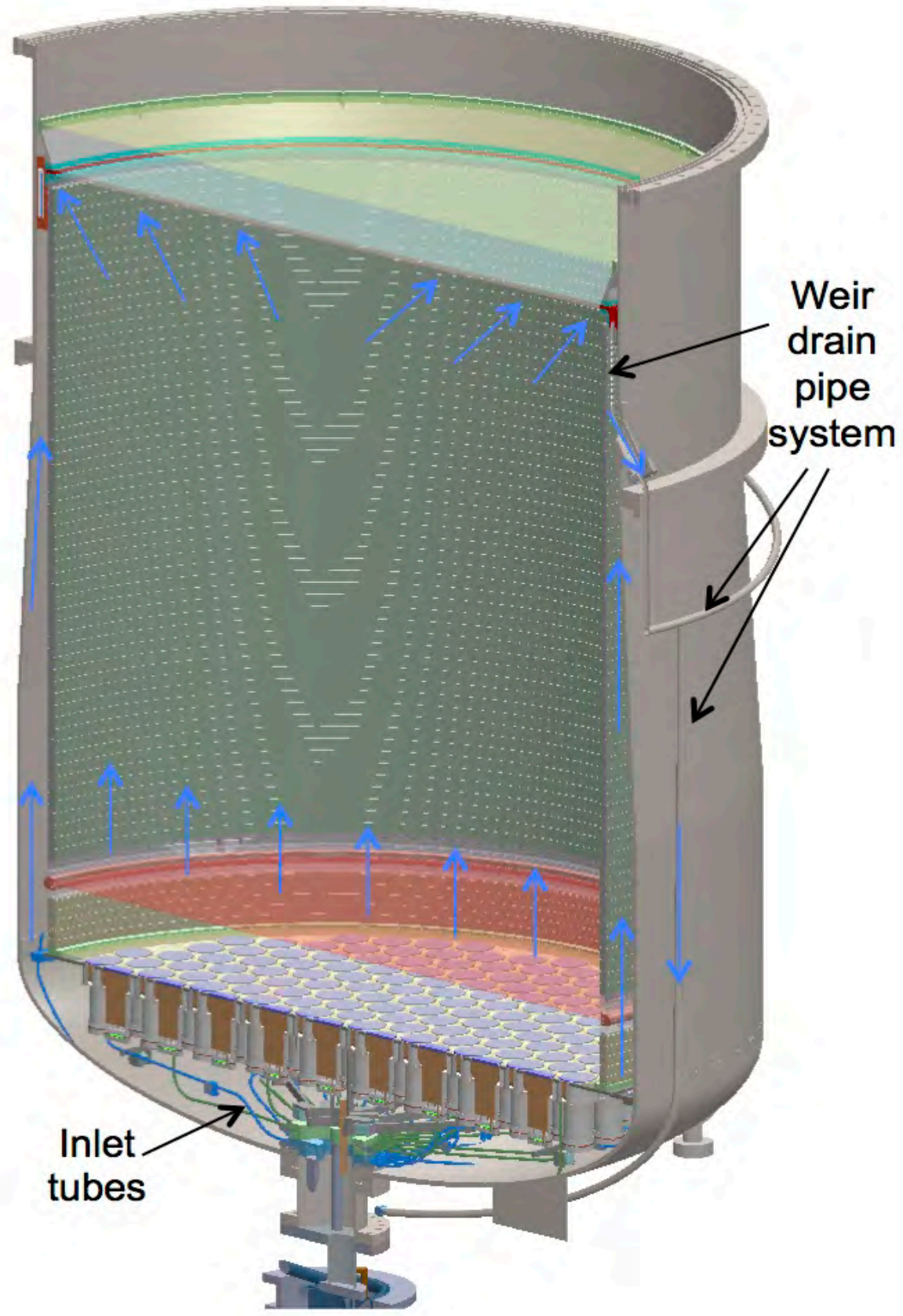

**Figure 6.8.1. Schematic of the internal plumbing system, showing flows into the TPC and skin regions, and weir and drain system for LXe circulation in the TPC region, with flow direction sketched in blue.**

While most of the functionality and complexity of the system is external to the Xe detector system and is described in Chapter 9, several important elements of the design are in the Xe detector system and are described here. This “internal circulation system” has several goals: effective circulation of both liquid and gas; establishing and controlling the thermal environment of the detector, including suppressing bubble formation and providing convective mixing to disperse internal radioactive sources; and maintaining a stable and quiet liquid level at the surface of the detector.

The liquid circulation paths are designed to efficiently sweep all of the liquid regions. Separate tubing sets are used to direct individually controlled flows into the bottom of the TPC, the lower part of skin, and the two liquid-filled conduits (HV and bottom cabling/fluid). The plumbing paths in the central region are shown in Figure 6.8.1, and the distribution of tubes in the bottom dome is shown in Figure 6.8.2. The flow through the conduits proceeds from the furthest points from the detector into the skin. The flow in both the skin and TPC is upward, with the liquid collected in a set of equal-height weirs with a common drain. A central goal in all of these flows is to eliminate as far as practical any stagnant “dead” regions — the prime example of which would be the conduits if they were not purged. Such dead spaces, once impure, serve as a slow source of diffusively driven impurities that can greatly complicate purification. This is an issue not only for purity that affects charge and light collection, but also following the use of radioactive tritium introduced as a calibration source (see Chapter 10), which must be subsequently removed.

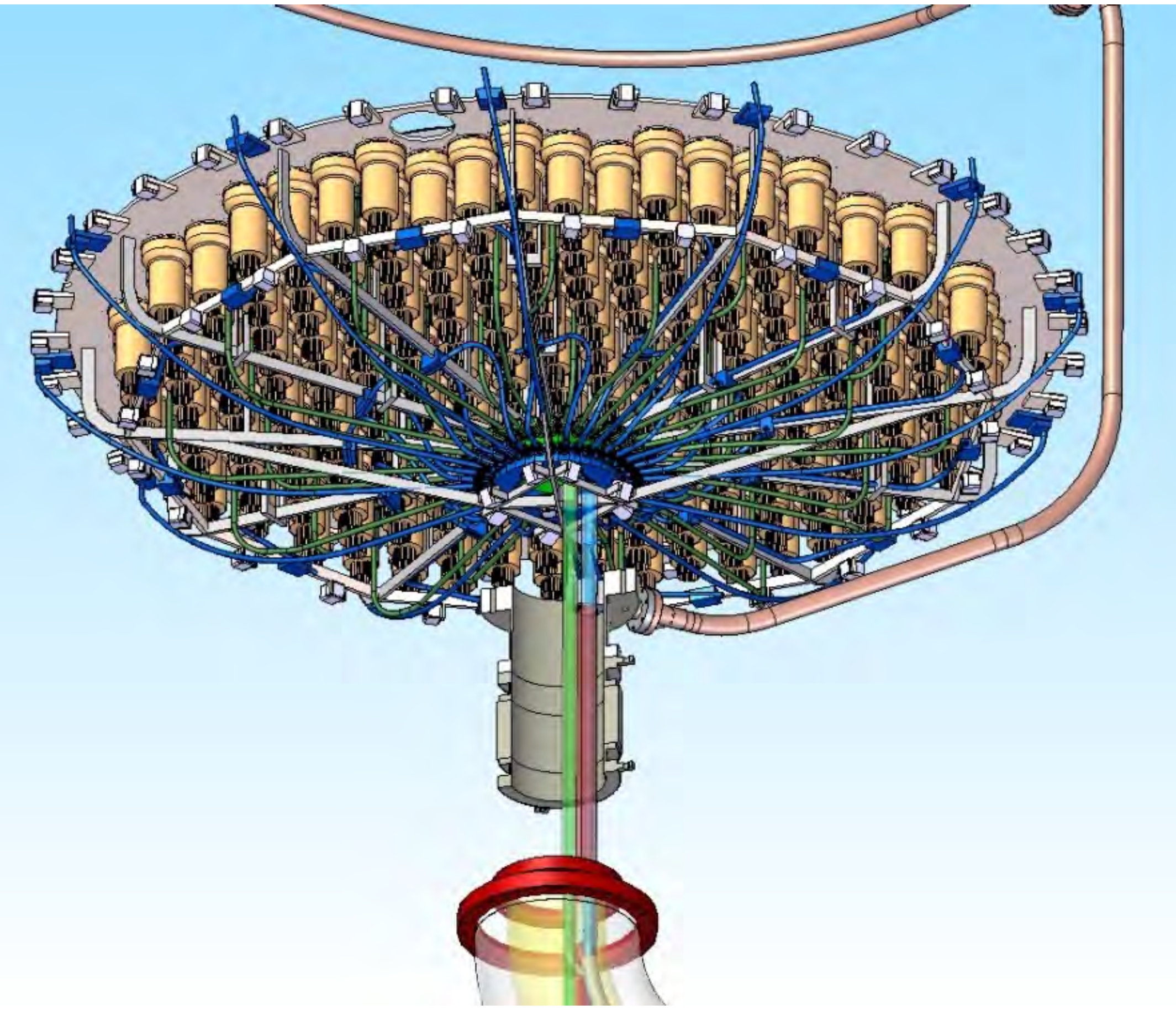

**Figure 6.8.2. Fluid distribution in the bottom dome region. Fluid tubes in blue distribute flow into the dome and wall skin regions, while tubes in green distribute fluid into the TPC. The drain from the weirs is also shown — it transitions from inside the Xe vessel to the vacuum space and back into the Xe space below the dome, thus avoiding the highest field region of the skin.**

We have chosen not to have any plumbing in the challenging high-field regions of the TPC. This limits locations where fluid lines can access the central TPC volume to the bottom PMT array, and to the perimeter of the TPC near the liquid surface, both of which regions have voltages near ground. We thus distribute the inlet tubes across the bottom PMT array holder, and line the circumference of the liquid surface with a set of weirs embedded into the wall of the TPC between the gate and anode grids. The placement of the inlet and outlets will be designed using computational fluid dynamics (CFD) to enable a uniform flow pattern. Similarly, the entry of fluid into the skin will be via a set of tubes that are distributed in the dome, with their placement again guided by CFD calculations.

These flows will also be used to control the operating temperature of the detector and, as much as possible, the thermal profile and behavior of the fluids throughout the system. The purification tower separately controls both the temperature and flow rate of the liquid in each flow path. The primary heat loads are on the wall of the vessel, the PMT bases, and the divider chains in the reverse and forward field sections of the TPC. The incoming fluid to the skin especially will provide the cooling to counter these heat loads and, guided by CFD.

An important goal is to obtain sufficient convective flow inside the detector to fully mix internal radioactive sources sufficiently quickly. An important source is $^{83m}$Kr, which we wish to have mixed on the time scale of its 1.83 hour half life. Convection will likely be driven by some combination of the PMT base heaters attached to the bottom PMT array plate, and possibly by slightly warming the fluid returning to the TPC center. It is also possible that the flow of fluid can sufficiently drive convection. These issues will be studied in part using a full 3-D CFD simulation.

Initial cooling of the detector must be done with care in order to avoid large thermal gradients in the TPC structure, which has no obvious thermal anchor point. We thus plan to circulate Xe gas, cooled in the purification tower, to slowly cool the entire system at a controlled rate. We will make use of the separate flow streams in the TPC and skin, as well as the gas purge in the top dome area, and are using CFD to plan the rate of cooling. Sensors described in the next section will be used to monitor the cooling.

We also will attach two thermosyphon cooling (evaporator) heads directly to the inner vessel, one near the top and one near the bottom of the vessel. These will maintain detector temperature during any periods when we are not circulating fluid, but we do not anticipate them to be as effective in controlling the detector temperature as controlled fluid circulation.

The liquid level in the TPC and, secondarily, the skin is set by having both regions drain over a set of weirs. These weirs are integrated into a mechanical assembly that is integrated with the rings that hold the gate and anode grids, and which contains a common drain trough located at the top of the skin region: See Figure 6.2.1.6. A set of weirs drains Xe through slots in the plastic between gate and anode. The design of the weirs is simple, but their length must be selected carefully to avoid instability in flow over the lip, and to minimize the relationship between lip height and flow. This design will also be studied by CFD calculation, and will be tested in the system test (Section 6.10). The collection trough spans the circumference of the detector and is composed of three separate units, each of which has a separate drain line. These lines are routed outside the Xe vessel near the top of the liquid because we do not allow any mechanical elements, either conducting or nonconducting, in the skin in the vicinity of the HV cathode in order to maximize the HV standoff ability of the skin.

In addition to liquid circulation, we will have flows that purge all the spaces above the liquid: in the detector above the main surface both below and above the top PMT array; in the conduits above the dome; and above the bottom PMT conduit and HV conduit. The conduits must be purged because of the outgassing from the plastics in the cables in them, especially at their warm ends for which outgassing is orders of magnitude higher than from cold plastics. The gas above the detector will be circulated in a loop with controlled input and output flows, which we will operate in a balanced mode so that we neither evaporate nor condense liquid in the detector region. To ensure effective purge of the space above the liquid of the main TPC, this gas-purge system will use a set of tubes distributed in the top PMT array, in a manner similar to what is done on bottom array.

All tubing and associated fittings and weir structures in the skin of the vessel will be made from high-reflectivity PTFE so as to minimally interfere with light collection. All tubing and weir structures in the skin space will have as little optical footprint as possible.

## 6.9 Xenon System Monitoring

Several aspects of the detector require monitoring beyond that provided by PMT signals. Good resolution of the S2 signal relies on achieving a calm liquid surface at the right level, which we will monitor through precision level sensors, acoustic bubble sensors, and an optical inspection system. The thermal profile of the detector is an important aspect of liquid circulation and the stability of the liquid surface, and is measured by an array of thermometers. The ability of the system to sustain high voltages is very important and the optical system will help locate any sources of discharge, while a set of loop antennae will not only measure discharges but may also detect precursor signals to full discharge. Bubbles encountering a high-field surface can also lead to discharge, and so detection of bubbles is an important aspect of achieving high voltages. Finally, it is important to confirm that the significant thermal contraction of the plastic field cage system behaves as expected, and so this motion will be monitored by a set of position sensors. This section discusses these monitoring systems in detail.

### 6.9.1 Thermometers

Temperatures need to be monitored at approximately 80 positions throughout the Xe detector and nearby. The chosen temperature sensor is a platinum (PT100-type) resistor, the precise make and shape of which is to be determined, taking into account installation and radioactivity level restrictions and constraints. The readout method is 4-wire throughout and the cabling to be used as much as possible is a semi-rigid polyimide-SS layered composite structure with parallel strip pairs inside and shielding/ground planes on the outside. The design of the semi-rigid cabling is individual to each sensor, or group of sensors. Semi-rigid cabling will be used inside the Xe and vacuum spaces to a maximum length of 1.5 meters, transitioning at connector blocks to conventional shielded 4-core wiring to cover the long stretches toward the breakout boxes, where mechanical robustness and threading capability is important. At the vacuum barriers of the breakout boxes, standard DB25 connectors are used, with 80 thermometers (320 wires) requiring at least 13 such connectors. The provision of DB25 connectors marks the interface to the slow-control work package (Chapter 11). Alternatively, higher-density connectors could reduce this number. More economic readout schemes, for example one in which some sensors are grouped and connected in series so that a common I+/I– wire pair can be used together with individual V+/V– taps on the thermometers, have been considered but dismissed as adding too much risk (a single wire failure could result in losing a whole group of thermometers). At positions where the potential effects of intrinsic component radioactivity are minimized, commercial pin headers and sockets can be used (black plastic is often a source of radioactivity). Nearer to the detectors, a combination of pins and clean PTFE, PEEK, or Delrin connector bodies should be used. The readout method implemented by the controls group will be based on modulating the sense current to avoid effects of thermo-power and other D.C. offsets. Calibration of the platinum resistors can be either via a generic table available for these, or individually tested together with the slow-control DAQ. Checks of intrinsic radioactivity levels are necessary, particularly for the PT100 and semi-rigid cabling near the main volume of the detector.

Cryogenic, laminated, layered semi-rigid cabling will likely be used for reading out at least the bulk of the thermometers. This cabling and the low-radioactivity connectors (if needed) can be made in-house with individual design for easy installation on site. The cable is based on polyimide/Kapton, to which stainless steel is laminated. These raw materials are etched to produce individual cabling and are laminated to receive shielding and ground layers on either side. A minimum track width of 150 μm and pitch of 300 μm can be achieved. The electronic capacitance between wire pairs is ~80 pF/m, depending on detailed geometry and operating temperature (via the temperature-dependent permittivity of Kapton).

### 6.9.2 Level Sensors

For level sensors, two main designs are needed: a parallel plate type for precision surface sensing, and long coaxial types. The precision surface sensor has the plates installed horizontally, straddling the boundary between the LXe and the electroluminescence region to measure the liquid level with high precision, allowing any tilt of the detector to be readily measured, and seeing variations in the liquid surface, for example from bubbles. Coaxial sensors monitor primarily the liquid levels during filling and emptying of the TPC and in the various elements of the purification tower. These sensors have been shown to work in principle, but detailed studies are still required to explore linearity, fringe field effects, systematics, capillary and meniscus effects, mechanical reproducibility, etc. The current design foresees three coaxial sensors that will span the height of the weir trough and the full 10-mm gate-anode distance, and three parallel plate sensors at the weir overflow openings. The sensors will be read out at high frequency with the aim of monitoring the condition of the Xe surface (level, ripples, waves) and, as such, will require a precision of ~10 μm. Further sensors will monitor the bottom skin region during filling and emptying, and a long level sensor will be used inside the PMT cabling standpipe to monitor the filling process. A pressure sensor at the bottom of the cryostat vessel will measure the head of the liquid and provide further information about the Xe level during filling and emptying. The readout method is via determination of capacitance with respect to a reference capacitance (all sensors employ three electrodes and a feedback readout circuit). This arrangement greatly reduces systematic effects arising from the long cabling in LZ and its variable capacitance (mechanical and thermal effects). The feedback readout circuit is the same as used for the position sensors and is based on modulated readout with a minimum number of analogue components and the bulk of front-end complexity absorbed into the firmware of a field-programmable gate array (FPGA). For feedthroughs at the vacuum barrier, a standard flange with sufficient numbers of coaxial connectors is foreseen. Each capacitive level or precision surface sensor requires three coaxial cables/feedthroughs. A drawing of a precision surface sensor and data from a prototype are shown in Figure 6.9.2.1.

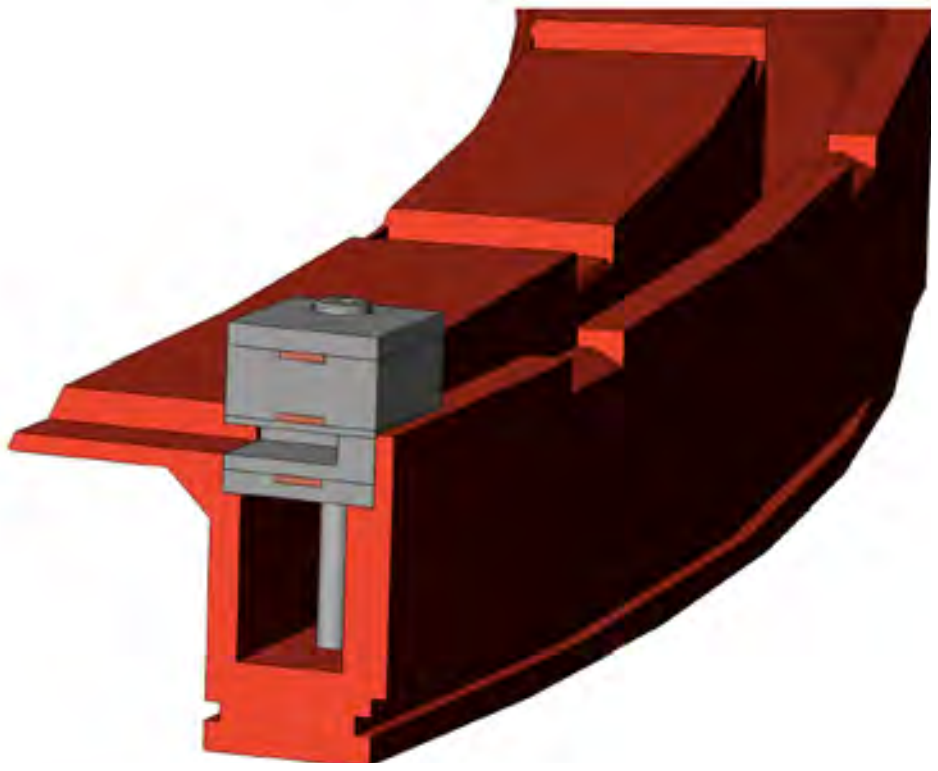

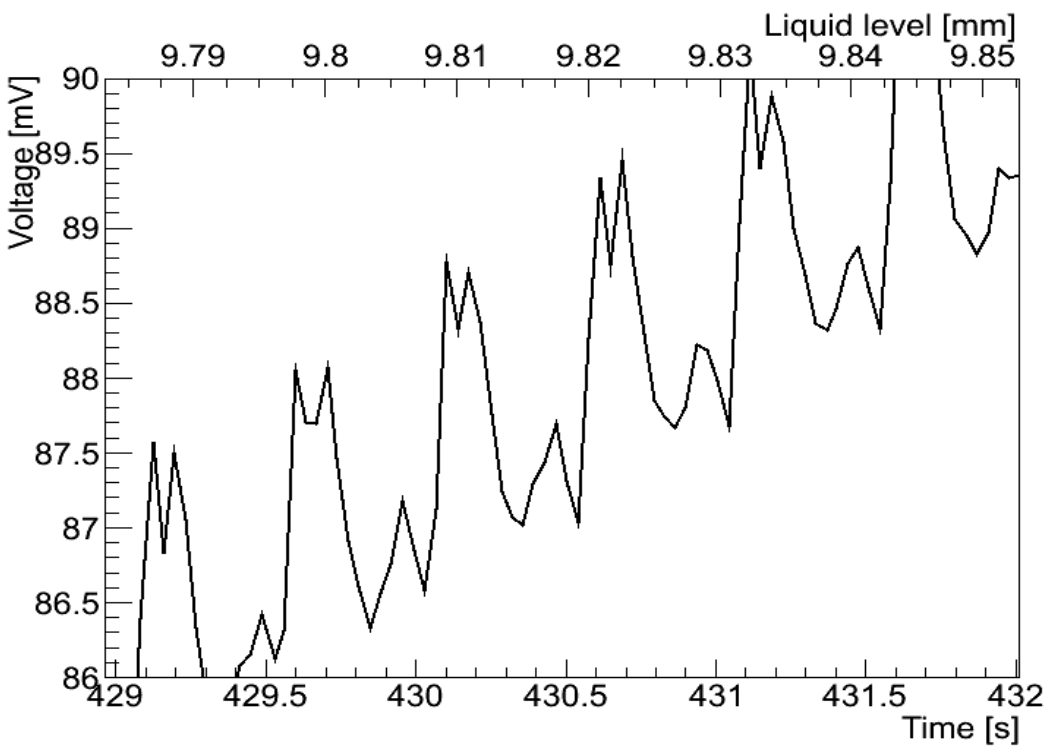

**Figure 6.9.2.1. Left: Drawing of an example of a precision level sensor (gray with copper electrodes) situated in the weir region (red) of the LZ TPC. Any variation in the liquid level between the probe (middle) and excitation (lower) electrodes corresponds to a change in capacitance relative to the reference capacitance between the probe and inverse excitation (upper) electrode. Right: Response of a prototype sensor to an increasing liquid level created by accumulating drops of oil (with ~8 µm level change per drop). Shown is an overall increase in the liquid level punctuated by excursions of measured voltage caused by the drops entering the container.**

### 6.9.3 Optical System

A system to carry out internal visual inspection at critical points is a powerful tool for monitoring the TPC. Small cameras exist and are most likely a cost-effective solution, but there is concern regarding radioactivity levels introduced by the presence of such cameras and their ability to operate at LXe temperatures. Both concerns require installation of cameras away from the TPC and at a higher temperature with optical systems (fibers, lenses, etc.) installed to transport the optical image from the internal region of the detector to the camera on the exterior of the cryostat. The cameras must be equipped to allow for temperature control as well as having an illumination system, to provide light inside the closed detector volume. This illumination can easily be supplied by LEDs. Ideally, such cameras are installed very close to or on the connector flange to the outside, with light guides and fibers installed internally. We plan to test this configuration on the system test stand (Section 6.10) before finalizing the design.

### 6.9.4 Acoustic Bubble Detection

A large range of acoustic sensors exists, based on different materials and physics effects. The best known are piezoelectric materials, but the majority of these are probably too radioactive to be acceptable. Another possibility is polymer film such as PVDT. Alternatively, sensors can be fabricated in-house from fully characterized materials. The sensors should be installed in contact with the outside of the Xe cryostat to pick up internal sound. The exact positioning of the sensors is to be determined through simulations of excitation modes of the vessel, supported by trials on existing systems, to determine the optimum locations of the sensors. Eight sensors are foreseen, with three at 120° spacing near the lower cylindrical section of the cryostat, three near the top of that section, and one each on the top and bottom dome. The vacuum-barrier electrical connection is via a standard flange with a DB25 connector or individual coaxial connectors, to interface to dedicated readout electronics. Connection to slow control or faster DAQ is via optical fiber/USB standard solution. The readout will be continuous at 200 kS/s, or faster, to allow for sufficient bandwidth, properly anti-aliased at the analogue input, with further filtering and decimation in FPGA firmware. Also, triggers are necessary so that only data that are a clear departure from baseline noise are identified and transmitted downstream, i.e., allowing "significant events" to be recorded by slow control without necessarily recording the full data stream.

### 6.9.5 Loop Antennae for Discharge Detection

To monitor for the absence or occurrence of HV breakdown events, loop antennae capable of picking these up should be installed in critical positions. Care must be taken so that these metal aerials do not interfere with the presence of HV; therefore, eight such antennae will be installed on the top and bottom PMT trusses. A more detailed analysis on types and locations of possible HV breakdowns will further inform the exact positioning of these sensors. The antennae should be suitably decoupled via HV blocking capacitors or transformers before interfacing to fast-readout electronics. Cabling should be routed so as to not interfere with the HV present in the TPC. A standard flange with a set of coaxial or other high-speed signal feedthrough is required. The readout electronics will use fast sampling (200 MS/s), based on existing readout/optical fiber/USB interface system, with data reduction in an FPGA. Triggers for significant events will have to be developed.

### 6.9.6 TPC Alignment Sensors

Position sensors will be fitted on and around the top PMT array to monitor the thermal contraction and expansion of the TPC during cooling and warming, which is particularly large (~2 cm) in the vertical direction, as discussed in Section 6.2.3. Fitting capacitive parallel plate position sensors allows the monitoring of vertical, horizontal, and helical motion of the TPC, giving vital information for determining whether and what countermeasures should be applied to ensure uniform cooling or warming of the TPC. These sensors are necessary because the alternative — equipping the length of the TPC with temperature sensors — is not feasible, as this would require readout wires to cross high-field regions. The position sensors are of simple design and are made from radiopure materials. The electronic readout is based on the same feedback circuit used in the level sensors, which acts to minimize the effect of cable capacitance on the sensor output. Eight sensors (three for vertical movement, three for horizontal movement, and two for helical movement) are planned. These sensors will also give important information on any lateral displacements that would alter the skin region gap, and the alignment of the top PMT array with respect to the TPC anchor points. The design of such a sensor and prototype test data are shown in Figure 6.9.6.1.

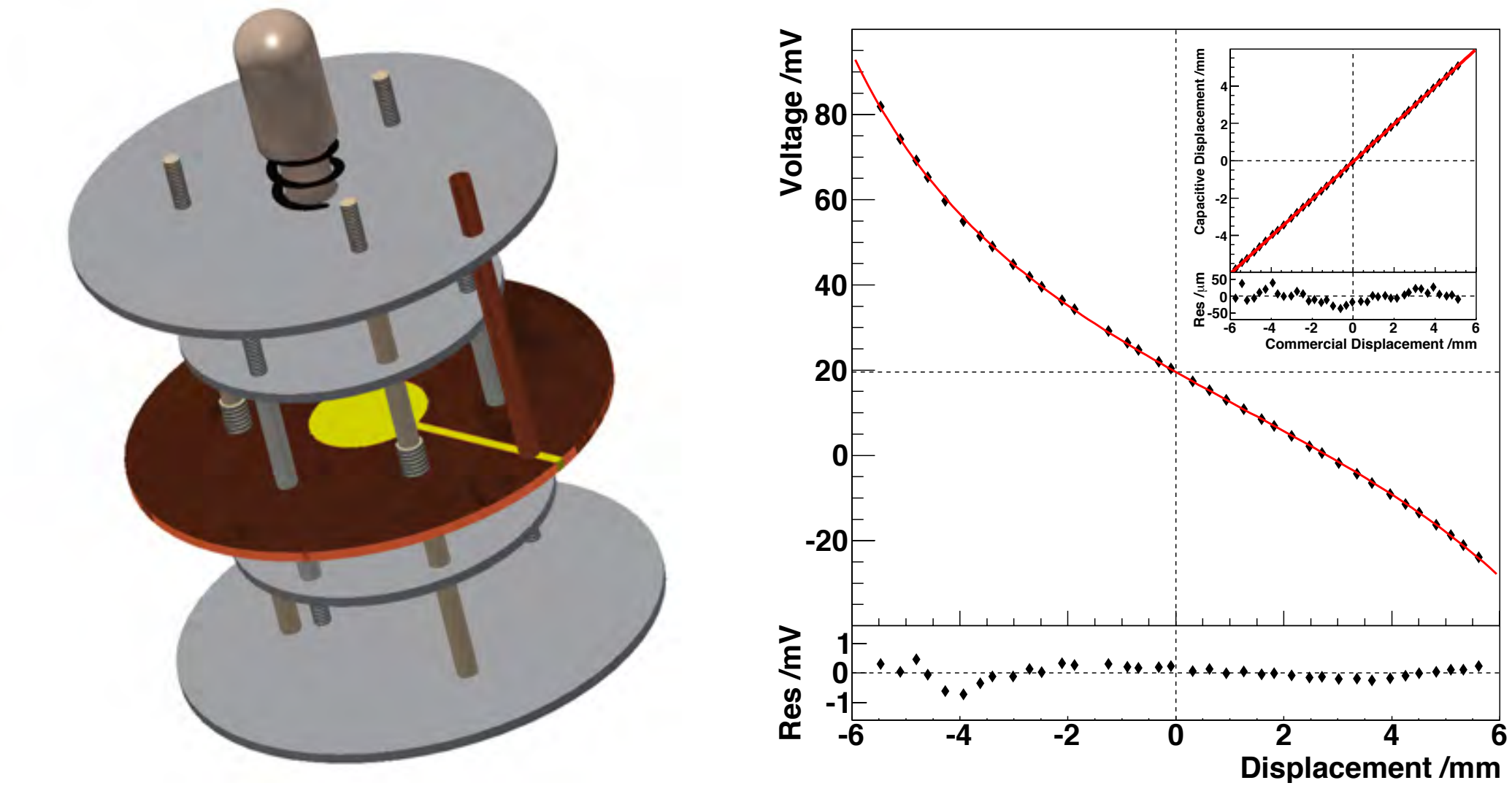

**Figure 6.9.6.1. Left: Drawing of the position sensor. The top and bottom plates are guard plates and held at ground potential. Within these are the plates that carry the excitation signals. The central plate contains the sensing electrode (yellow) surrounded by a guard electrode that protects against the effects of fringe fields. The central plunger moves in and out, which varies the relative positions of the excitation electrodes to the sensing electrode. A central spring is used to allow a single fixed point. Right: (Main plot) Response of the sensor as a function of displacement fitted using the expected response function. The residuals of this fit are also shown. (Inset) The variables extracted from the fit to the response are used to calculate the displacement of the sensor vs. a commercial alternative, with residuals shown.**

## 6.10 Integrated System Testing

A critical part of our planning is integrated or system testing of combined elements of key aspects of the Xe TPC and associated systems. The sections below summarize the smaller and then larger test systems available to the collaboration to evaluate critical aspects of the design of WBS 1.5 components and to test large-scale prototypes. Our testing plan is summarized in Section 6.10.6.

### 6.10.1 Study of Single Cathode Wires at High Field in Liquid Xenon

To ensure the successful delivery of HV to the LZ TPC, we take a comprehensive approach, beginning with an experimental study of the physics processes involved in the electric breakdown of individual cathode wires at a microscopic (quantum) level. A small double-phase Xe chamber was built and is now being operated at Imperial College London for this purpose. Instead of a cathode grid at the bottom of the liquid region, a single metal wire is used as a test sample. This configuration ensures that very high electric fields can be achieved at the wire surface by applying modest voltages of ~10 kV, approaching 300 kV/cm for a 100-μm sample, and ~1 MV/cm for 20-μm wire. The chamber has a single internal photomultiplier viewing down from the gas phase to detect both photon and charge emission from the upper surface of the wire sample. The electroluminescence response has single-electron sensitivity, allowing us to measure minute electron currents preceding macroscopic breakdown.

Most practical cathode electrodes in double-phase Xe detectors have been limited to surface fields of 40–65 kV/cm [17,28-32] — although the Xed chamber at Case Western Reserve University operated with substantially higher values of up to 220 kV/cm [33]. These fields are much lower than the published onset of electroluminescence or charge multiplication in the liquid, the former being 400–700 kV/cm for LXe [34,35]. The standard electrostatic design methodology adopted in previous experiments is therefore unsuitable and the adoption of a new “maximum allowable field” that can be sustained at the surface of metal surfaces must be conservative until new data can illuminate how to improve this. This justifies the currently adopted target value of 50 kV/cm.

Electron emission from metal surfaces can be caused by local enhancement of the electric field, the presence of thin insulating layers, or other effects that result in a lower effective work function. This can be accompanied by simultaneous photon emission. Our study focuses on the phenomenology associated with the onset of electrical breakdown. In particular, we are exploring its dependence on electric-field magnitude and direction, wire material, diameter, surface quality, history, etc. We will also investigate possible mitigation steps such as electropolishing, chemical etching, and conditioning in gas, to inform the production of the LZ wire grids.

This R&D activity employs the small chamber shown in Figure 6.10.1.1 (left) to test cathodes made from a single wire. The Xe vessel contains 4 kg of liquid in equilibrium with gas at 1.6 bar. Gate and anode grids 14 mm apart straddle the liquid surface, ensuring S2 yields that are mostly independent of the cathode voltage and sufficiently high for efficient cross-phase extraction and detection of single electrons. The 130-mm-long cathode wire is mounted 25 mm below the gate electrode, stretched between two feedthroughs that can deliver up to −10 kV to the liquid. The electric field is highest on the upper surface of the sample, so that any electron emission is likely to lead to electroluminescence signals in the gas.

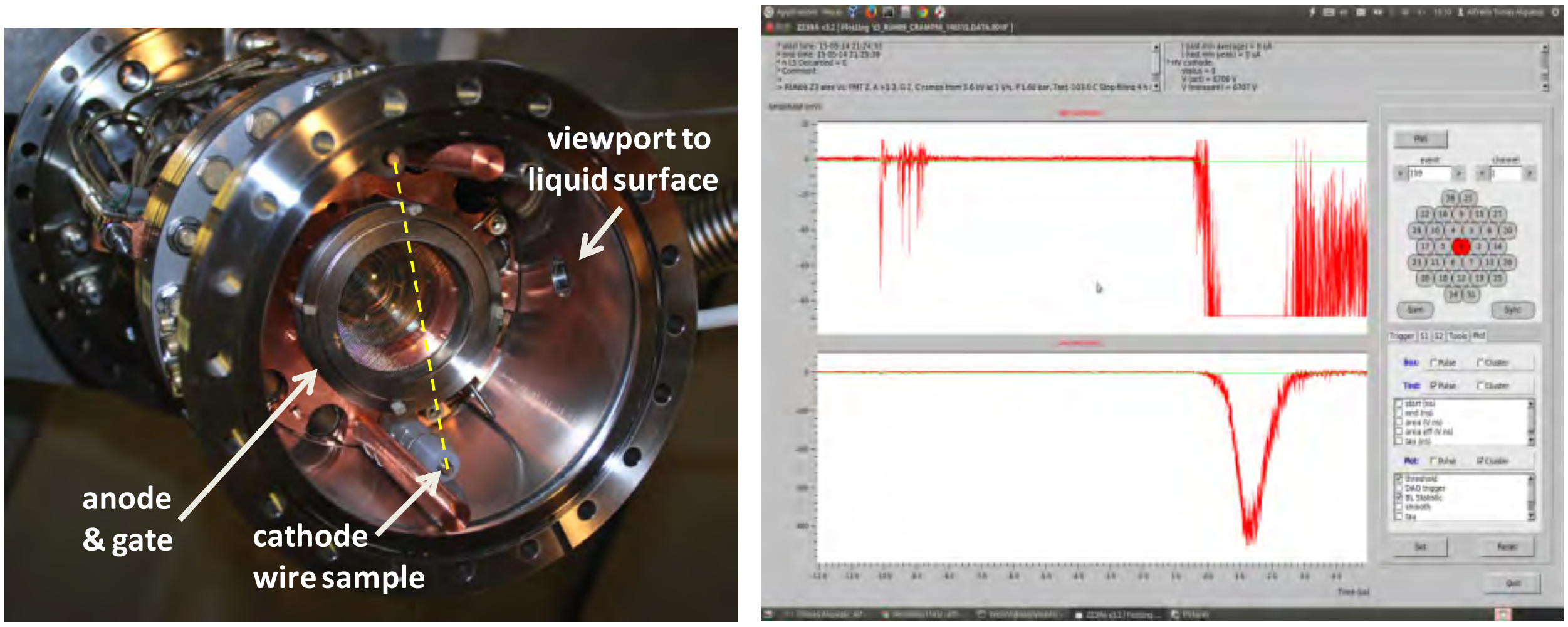

**Figure 6.10.1.1. Left: Internal view of Imperial LXe chamber looking up from below. A photomultiplier is just visible through the gate and anode grids, which establish electroluminescence above the liquid surface (nominally located between them). The cathode wire sample (highlighted) is stretched between the two feedthroughs shown. Right: A typical waveform, probably unrelated to discharge, recorded with a sample of 100 μm ZEPLIN-III wire with ≈145 kV/cm on the wire surface. A large S2-like signal (seen clearly in the lower, low-sensitivity channel) is preceded by an S1-like optical pulse and a single electron cluster.**

Most subsystems required to operate the chamber were inherited from the ZEPLIN-III experiment (gas handling and purification, slow controls, data acquisition). The ZE3RA data-reduction software allows full exploitation of the 2-ns-sampled waveforms [36], which are recorded in high- and low-sensitivity channels to cover both very small signals and larger S1 and S2 pulses.

The chamber is cooled by means of a 1-inch cold finger immersed in liquid nitrogen, providing 10–15 W of cooling power and autonomy of 7–9 hrs between LN fills, and very simple and reliable operation. Prior to condensing, the Xe gas is purified with a heated SAES getter for approximately one week to ensure sufficient electron lifetime (≳20 μs) during the test. Cooldown is achieved overnight, and the chamber is filled, operated, and emptied in a single day.

Prior to the cathode test, the gate-anode system is biased to establish two-phase operation. Then the voltage applied to the wire sample is ramped up slowly (~1 V/s) to several kV until the power supply trips, with PMT data being digitized simultaneously. This voltage and other slow-control data are embedded with the main data set for analysis. Electron emission from the cathode can, if accompanied by prompt light, be reconstructed to the cathode depth by electron drift time. A sample of the 100-μm wire used in ZEPLIN-III has already been tested and observed to reach 145 kV/cm before tripping. Figure 6.10.1.1 (right) shows an event acquired close to the trip voltage, showcasing the types of pulses we can measure: from left to right, an S1-like pulse from prompt light, then a small cluster of photoelectrons corresponding to a single emitted electron, and finally a large S2 pulse delayed by ≈10 μs. Most other events close to breakdown feature more complicated topologies that we are now analyzing. This breakdown field is clearly much higher than achieved in ZEPLIN-III with the same type of wire — which sustained stably 62 and 40 kV/cm in the first and second runs, respectively [29,31]. However, we point out that the total length of the ZEPLIN-III cathode wire was 117 meters, which is probably a very relevant parameter.

We will continue to test wires before and after treatment, in particular to try to understand the origin of photon and electron emission around the early onset of instability. We plan to examine wire samples from Xed [33], LUX gate and cathode grids [17], and candidate wires for LZ. Apart from SS, we will consider beryllium copper and tungsten, as well as coated samples (gold, silicon nitride, etc.). Small- and medium-scale wire grids can then be built and tested as described below.

### 6.10.2 Study of Bare Cathode Frames and Gridded Cathodes at High Field in Liquid Xenon

In concert with the efforts described in Section 6.10.1, we plan additional studies of the onset and origin of photon emission in high electric fields in LXe. These studies will utilize small dual-phase LXe emission detectors at institutions within the LZ collaboration. An example at LBNL is shown in Figure 6.10.2.1. This test bed is functionally similar to that described in Section 6.10.1, but with the ability to evaluate small cathode wire planes consisting of an SS frame and stretched wires.

Initial tests will study the maximum surface electric field that can be applied to the cathode frame alone, in the absence of wires. This is clearly an important test, as the cathode frame is a requisite component in deploying a gridded TPC. Subsequent tests will study the maximum surface electric field with a single grid wire stretched across the diameter of the frame. As long as effects from the frame are subdominant, these studies should obtain results that can be cross-checked with those from Section 6.10.1. Separate samples of wire have been obtained for these studies, with a focus on SS, beryllium copper, and gold-coated tungsten. The effect of surface treatments such as electropolishing will also be explored.

Given the importance of a robust expectation for electric-field performance in LZ, we intend full duplication of the study of all wire samples and surface preparations. We also will search for effects due to LXe liquid purity, temperature, and thermodynamic history.

A final step in this program will be to string complete small cathodes and verify their performance in terms of the maximum surface electric field that can be sustained prior to the onset of photon emission. A key question is whether this performance can be simply inferred from the single-wire studies, or if (as stated in Section 6.10.1) the length of wire is in fact a critical parameter. A complete cathode grid frame will allow an approximate factor-of-10 increase in wire length compared with a single wire. A dependence on wire length would point clearly to the importance of a more aggressive surface finish and treatment program, with continued testing as already described. A series of other tests using small test chambers available to the collaboration are also planned over the next year. These will include component, PMT base and temperature, and other sensor testing in LXe.

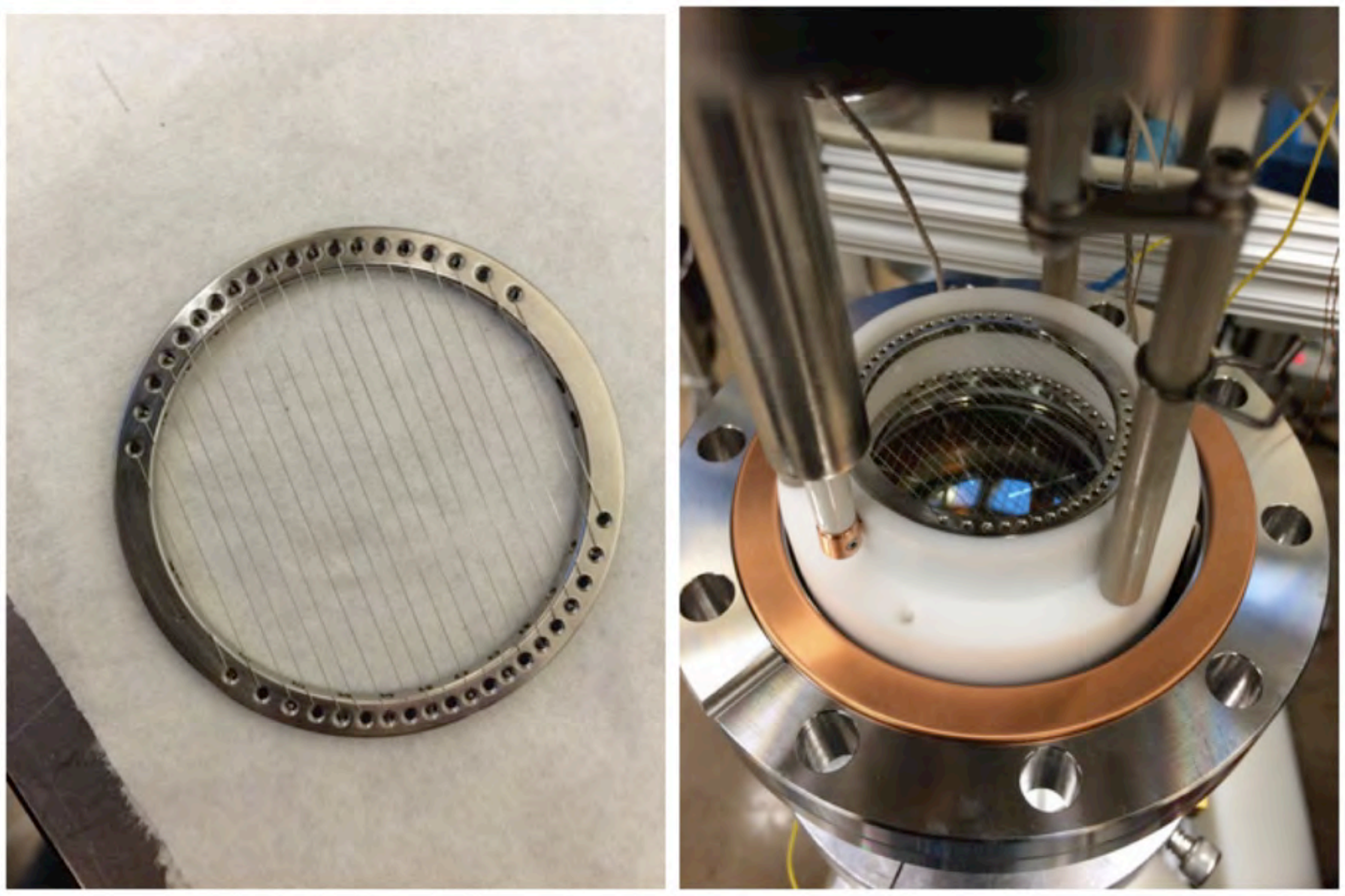

**Figure 6.10.2.1. Left: Example of a wire grid that will be tested in the small LBNL test chamber or other test chambers. Right: View of the LBNL test chamber.**

### 6.10.3 Large-scale High-voltage Testing of Critical Assemblies in Liquid Argon

Three assemblies in the LZ detector must accommodate the large negative-cathode voltage. These are the forward-grading structure (connecting the cathode grid to the gate grid), the reverse-grading structure (connecting the cathode grid to the bottom grid), and the cable-grading structure (connecting the cathode grid to the grounded shield of the cathode HV cable). Each structure contains a series of conductive rings connected by resistors to produce a controlled grading of the voltage. It is critical that these assemblies sustain the applied cathode voltage without producing light from electrical discharges across their components or to the inner wall of the cryostat. We have developed a large-scale system for testing these critical assemblies at HV in liquid argon, which acts as a cost-effective proxy for LXe. A schematic view of the setup is shown in Figure 6.10.3.1 (left) and dewar and HV connection (right).

The HV tests are performed within a 240-liter cryogenic dewar of 16-inch bore. The assembly under test is supported from below by a platform that hangs from the top flange of the dewar. The bottom of the tested assembly is grounded by the platform; the top is connected to HV that can be ramped to –200 kV to simulate the LZ cathode. The HV is delivered through a polyethylene cable that originates at a feedthrough located 8 feet above the top of the dewar. The feedthrough is connected to a DC power supply made by Glassman High Voltage. A controlled electrostatic environment is maintained around the test assembly by surrounding it with a highly transparent grounded metal mesh that shields any nearby structures. Seven lenses at various angles view the assembly from just outside the mesh. These are connected to fiber bundles that route the images to a charge-coupled device (CCD) camera located just above the top flange of the dewar, providing a real-time view of any electrical discharges that occur during testing. Below the tested assembly is a quartz window coated with fluorescent tetraphenyl butadiene (TPB) wavelength shifter. This window is viewed by an 8-inch-diameter PMT, giving efficient detection of ultraviolet light with single-photon sensitivity.

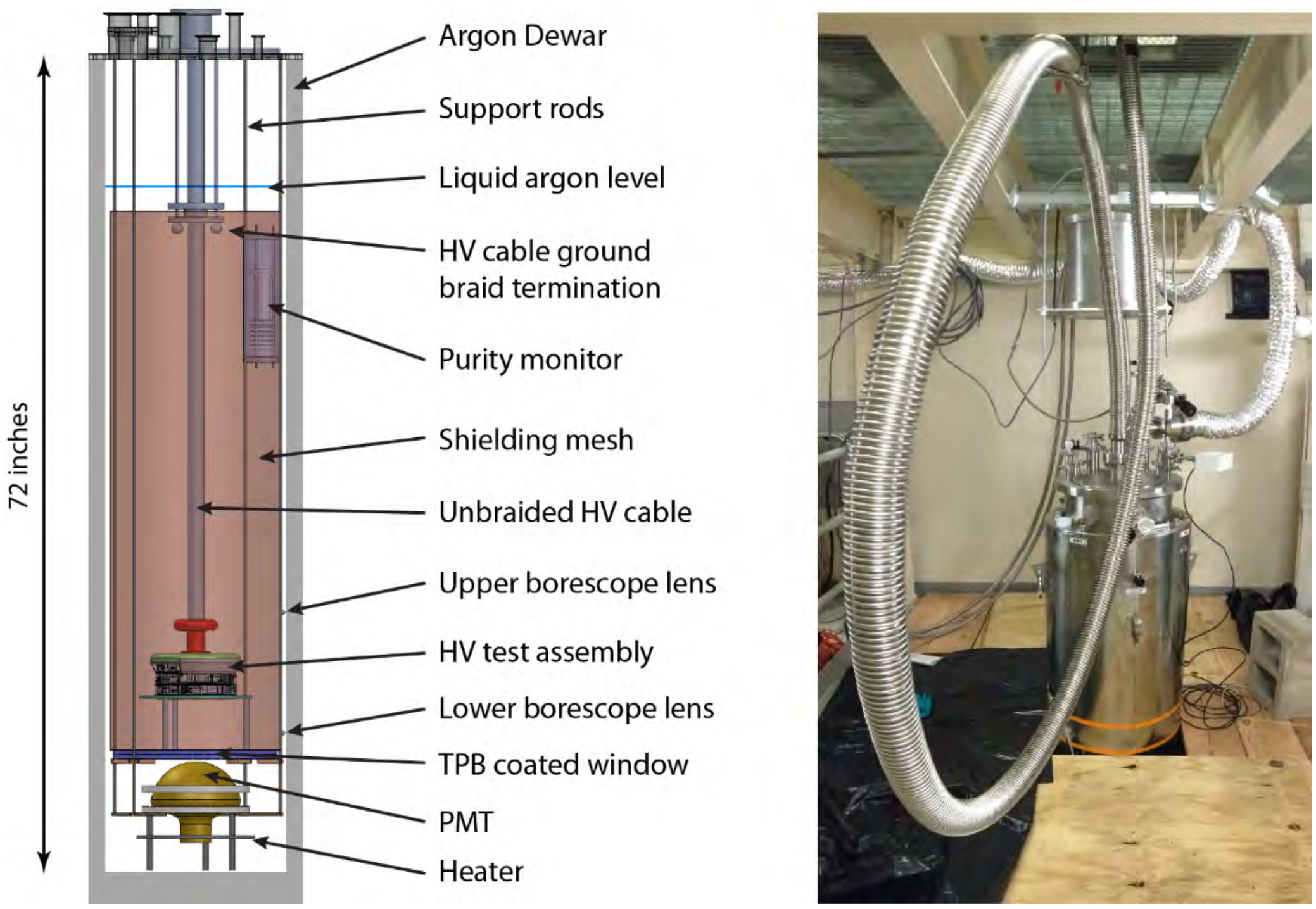

**Figure 6.10.3.1. Left: Side view showing the internal components of the liquid argon test system. Note that the vertical support rods (shown out-of-plane in this view) are positioned outside the shielding mesh. For clarity, only two of the seven borescope lenses are shown. The HV test assembly shown here is a portion of the grading structure of the reverse-field region of the detector. Right: The liquid argon dewar and HV cable conduit at Yale University. The HV feedthrough sits above the square hole in the steel grating at the top of the image.**

Commercially available liquid argon typically contains impurities of 10 ppm. These are known to substantially enhance the dielectric strength of liquid argon, and must be removed before making meaningful tests [37]. The dewar is filled with commercially available liquid argon that is passed through a liquid phase purification system based on molecular sieve and activated copper filters [38]. The purity of the argon is measured at the top of the dewar by a compact monitor based on a device developed by the ICARUS collaboration [39]. Conversion electrons from a $^{207}$Bi source ionize liquid argon in an applied electric field. The resulting electrons are drifted through two charge-sensing regions separated by a 6-cm distance. A charge-sensitive amplifier measures the charge induced by the electrons in the two regions to infer the fraction of electrons lost while drifting between them. The electron lifetime is then computed from the lost fraction and the time required to traverse the 6-cm distance. The monitor is sensitive to lifetimes less than 100 μs, corresponding to 2.5 ppb of oxygen.

In a typical testing cycle, the tested HV assembly is mounted to the platform that hangs below the top flange of the dewar. The top flange is then lowered by crane into the argon dewar. The HV cable is then lowered through a port on the top flange until it engages a socket at the top of the test structure. The dewar is flushed with argon gas, evacuated, and filled over several hours with purified liquid argon. High voltage can be applied once the liquid argon has submerged the termination of the HV cable braid, with ramping to full voltage requiring a few hours. A 1-kW heater is used to boil the argon at the bottom of the dewar after HV testing. The HV cable and top flange are removed once the dewar is warm. This arrangement allows for testing of a unique assembly every four days, allowing for rapid design iteration of the critical HV assemblies.

### 6.10.4 Case Western and SLAC Test Systems

The large-scale components whose test in liquid argon is discussed in the previous section must ultimately be tested in LXe. A challenge for such tests is the cost of and complexity of handling large amounts of LXe systems. Test systems must therefore accommodate TPC test structures that are as narrow as possible while still maintaining the high electric fields in the critical region around the cathode that will be encountered in LZ, and also allow a full test of the reverse-field region. Construction of the LXe test system is planned in two phases.

In Phase I, a 10-inch-diameter vessel is used to test a version of TPC structure discussed in Section 6.10.6 with an overall height up to 70 cm, and requiring roughly 100 kg of LXe. We anticipate operating this system at voltages up to 100 kV. The Phase I testing platform, shown in Figure 6.10.4.1, has been developed at Case Western, and is being commissioned at SLAC, where the Case group has relocated. The system has two basic architectural features in common with LZ: a separate purification tower housing a heat-exchanger system connected to the main vessel via a vacuum-insulated plumbing run below both vessel sets; and a side-entry HV feedthrough at the cathode level. The HV feedthrough system uses a commercial 100-kV ceramic feedthrough installed on a side port directly on the cold vessel. One part is immersed in LXe, while the nominal vacuum side is on the exterior and immersed in Fluorinert FC-770. This is an insulating fluorocarbon with good dielectric properties, and is liquid at both room temperature and at 165 K. A commercial >100-kV-rated HV cable is immersed in this fluid for the run between cryogenic and room temperatures, and continues uninterrupted to a commercial power supply.

The implementation of the TPC prototype in this vessel is driven by many of the same constraints as LZ, and shares several features. It has an HV standoff skin layer of LXe whose thickness is the minimum needed to achieve sufficiently low electric fields on all HV surfaces. Cabling and plumbing feedthroughs are located on both the top and bottom of the vessel (visible in Figure 6.10.4.1) so that the skin has no cables or fluid lines in the HV region around the cathode. In addition, the fluid circulation will use a weir system based on the LZ design. The test platform has ample breakout hardware for multiple PMTs, level sensors, thermometers, and other instrumentation such as loop antenna discharge sensors. We plan to outfit the TPC with a PMT at the top and bottom with conical reflectors. This will allow a two-phase S1/S2 readout that will provide a highly sensitive measure of Xe purity. We will also deploy a version of the camera system being developed at Texas A&M (Section 6.10.5) and also being deployed in the Yale system to image discharge phenomena.

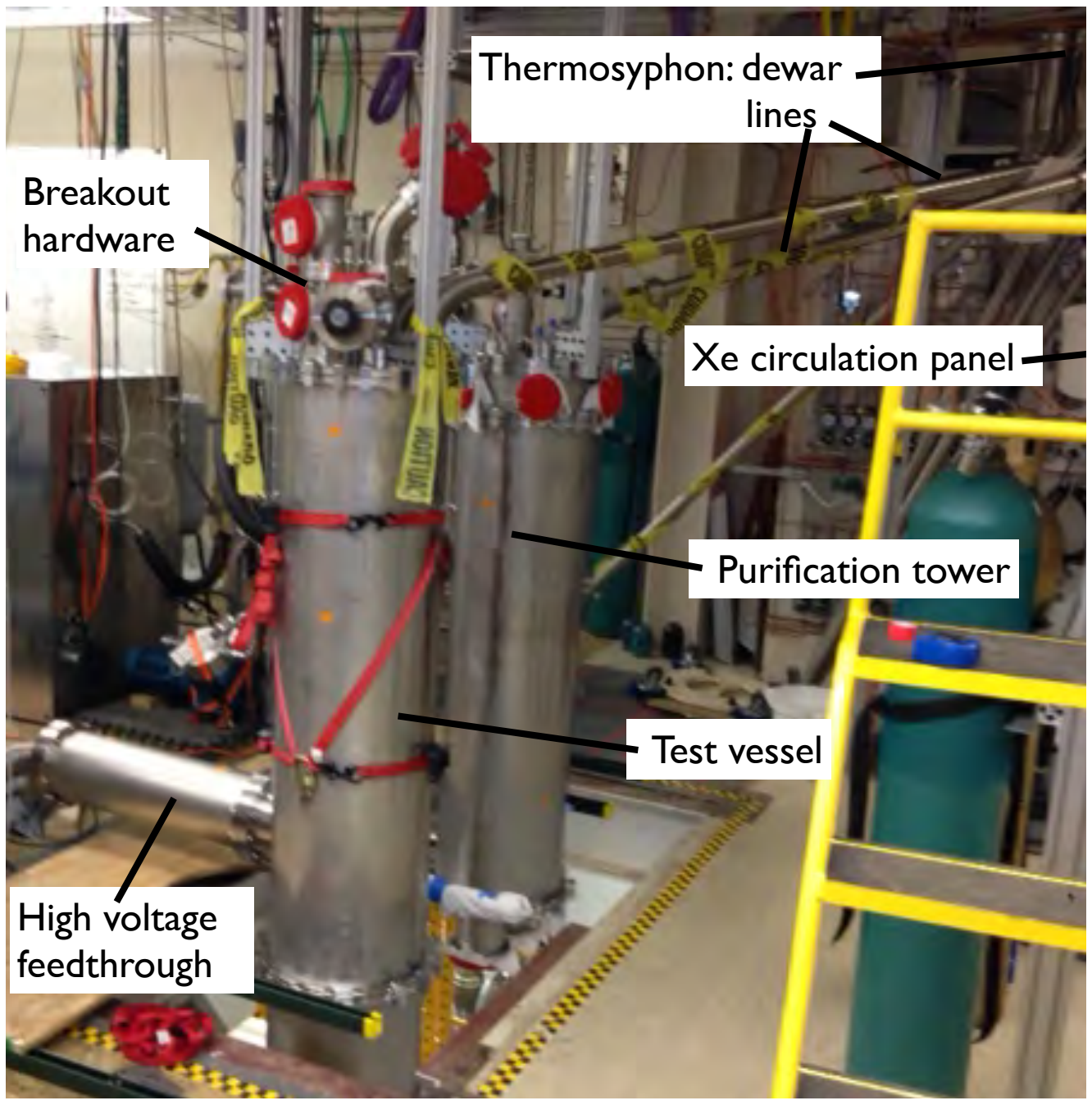

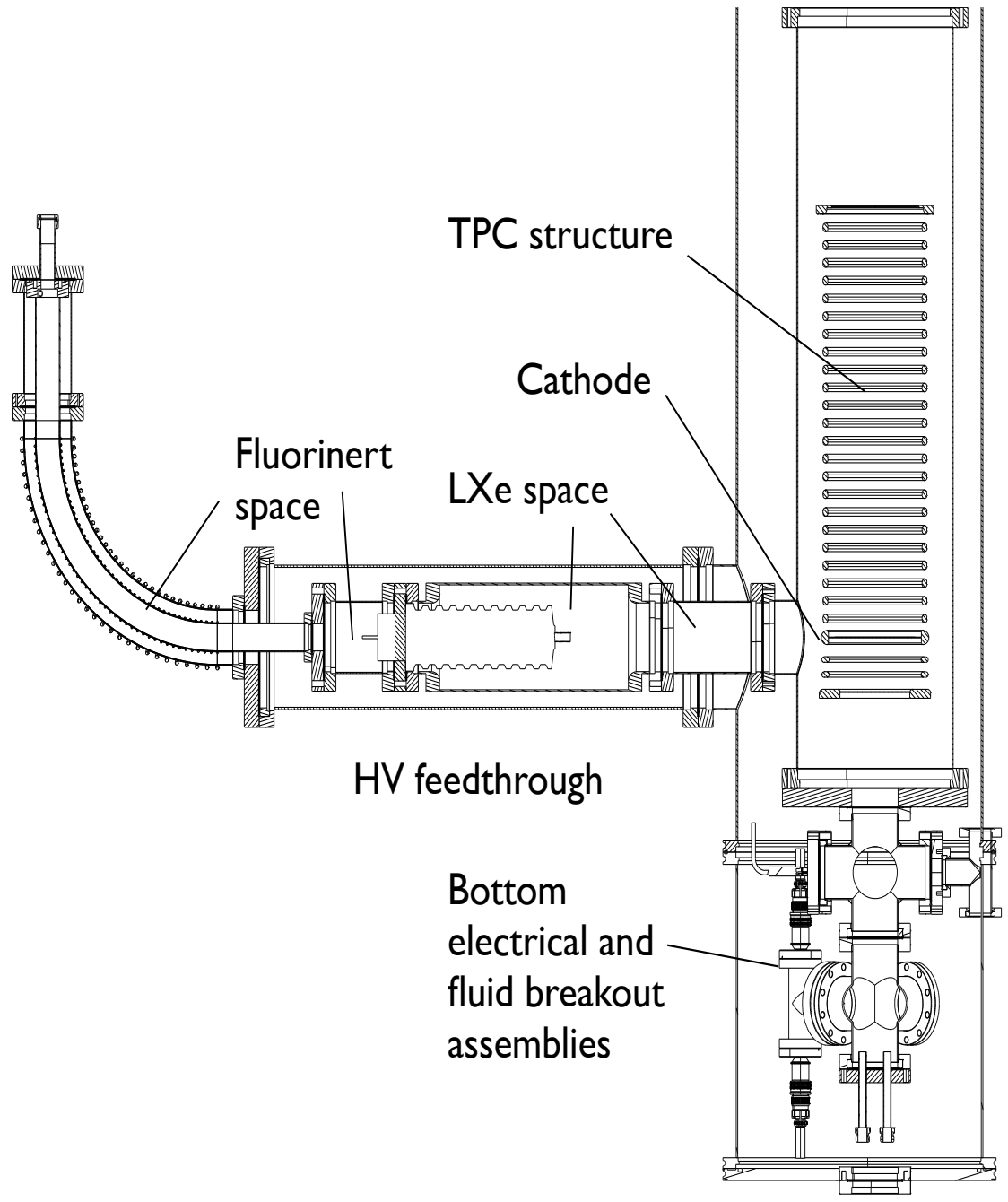

**Figure 6.10.4.1. Left: Phase I system. Right: Cross-section view of the Phase I test vessel showing the HV feedthrough and TPC structure installed.**

In Phase II, a significantly larger vessel with a nominal 18-inch diameter that requires some 500 kg of LXe will be used to test a larger TPC structure. The cathode of this TPC can be operated at the full 200-kV design voltage of LZ. Once again, the radius of the vessels and the thickness of the Xe skin are sufficient to reproduce the final fields that LZ will have in the cathode region, and to test the reverse-field region at full voltage. Critically, the cryostat will be sized to accommodate a copy of the full-size LZ HV feedthrough system.

The staging of both phases at SLAC is shown in Figure 6.10.4.2. The former BaBar counting room and surrounding space in the IR2 experimental hall is being renovated for the purpose of hosting this system, providing ample room for operations and later expansion. The vessels will be deployed under HEPA units so we can establish a soft-wall clean area. The support systems for these test vessels, partially visible in Figure 6.10.4.1 (left) and Figure 6.10.4.2, are extensive. They build on developments from LUX and ZEPLIN and serve as prototypes of what will be used on LZ. Cryogenics for both phases are supplied by a thermosyphon "backbone," which consists of a multiport thermosyphon dewar capable of providing more than 12 separate PID (proportional-integral-derivative controller)-controlled cooling heads. Two will be used for each vessel set, and one or more in the purification tower, with several more used for automated cold traps in the gas-handling and -sampling systems. The purification tower will include prototypes of the elements of the circulation system planned for LZ: a weir reservoir, two-phase heat exchanger, gas-phase heat exchanger, and a "subcooling" thermosyphon head on the condensing stream, along with an extensive set of fluid level sensors and thermometers.

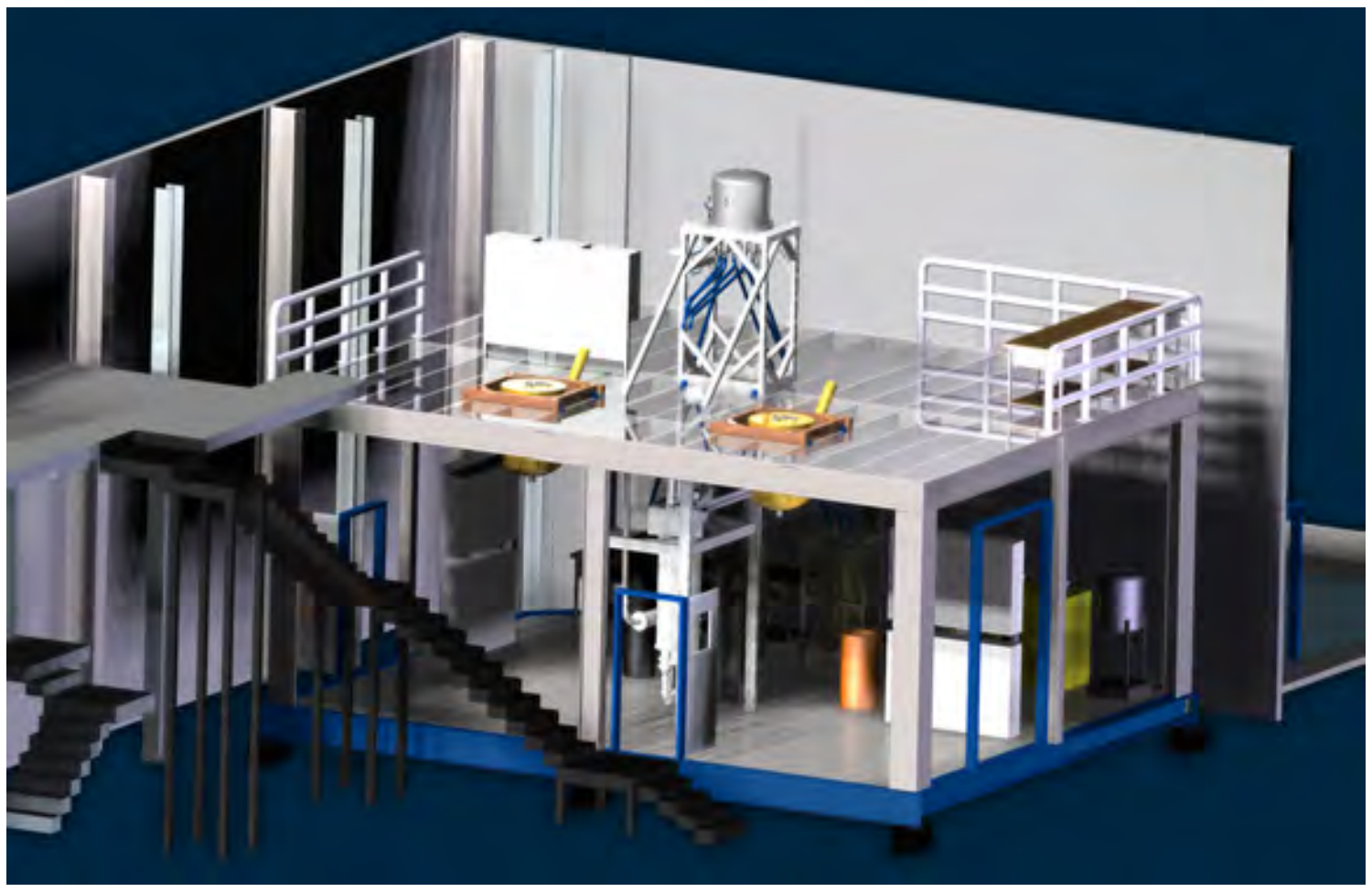

**Figure 6.10.4.2. The System Test Platform housed at the former counting room for BaBar at SLAC.**

The system for online purification through a hot getter uses highly automated gas-handling panels based on ½-inch-diameter tubing that will accommodate flow rates well in excess of 100 slpm. This will (1) allow the large test platforms to achieve higher purity quickly (by contrast, LUX, with 300 kg, circulates Xe at roughly 25 slpm), allowing faster testing cycles; and (2) allow tests of the heat-exchange system at high flow rates. This system also features a high-flow capacity metal diaphragm compressor as the circulation pump. This technology allows very high flow rates and has been identified (WBS 1.4 and Chapter 9) as the technology for LZ, but to our knowledge has not been used in any previous similar Xe experiment. Thus, the system will provide an important test of these pumps.

Critical elements of control and fail-safe recovery of the Xe will also be developed as part of the system test platform, including integration of process loop controllers (PLCs) for critical systems and integration with larger slow-control system development. Phase I Xe recovery uses a thermosyphon-driven storage and recovery vessel patterned on a similar device used for LUX. For Phase II, we will use a compressor-based recovery into standard storage cylinders, as is planned for LZ. This requires a highly reliable system with generator-based backup power. For Phase II, we will also deploy a passive recovery "balloon" for final fail-safe recovery and containment of the 500 kg of Xe. Elements of the planned LZ online and slow-control systems will be developed for the system test to allow a high degree of test automation. The gas system is also designed to be closely integrated with an automated, high-sensitivity purity-monitoring system initially developed by the Maryland group. This will be important in order to achieve purity for successful HV testing; it will also allow us to check our understanding of various factors that will be important for LZ purification. Finally, the system is designed to accommodate the range of gaseous radioactive calibration sources deployed in LUX and planned for LZ.

### 6.10.5 Camera Systems

The camera system discussed in Section 6.9.3 is first being deployed as an essential part of the System Test, and so is further described here. Two-phase Xe operation requires a series of grids and field-shaping rings at different voltages that provide a uniform drift region and electron extraction from a nonturbulent LXe surface. While the design goes to great lengths to minimize problems, sparking or turbulence (perhaps bubbling) could occur, and it is important to understand the cause and location. While some problems can be detected by the photomultipliers and capacitive level sensors, visual inspection of the location of possible sparks and observation of bubbles or floating contaminants has clear advantages. The Texas A&M group has designed, built, and tested at LN temperatures a prototype system that includes a CCD camera to record images from a coherent fiber-optic bundle, which can view the entire active region of the system test TPC (see Figure 6.10.5.1). The plan for the system test is to build a multifiber bundle to attach to this camera to enable inspection of up to seven regions inside the cryostat. This multibundle fiberscope will allow inspection of the TPC internal region, HV feedthrough region, reverse and forward field regions during the commissioning, and operation of the system test stands.

A sketch of the proposed system is shown on the right in Figure 6.10.5.1, indicating the location of the CCD camera and the fiber bundles that will be used to observe the space inside the liquid argon system test cryostat. A similar camera and fiber configuration is being planned for the SLAC System tests using LXe. Both systems will have seven 0.72-mm silica fiber bundles containing 30,000 fibers to view the internals of the detector. These fibers will be arrayed in a 2-3-2 close-packed arrangement and imaged on the CCD plane of the camera.

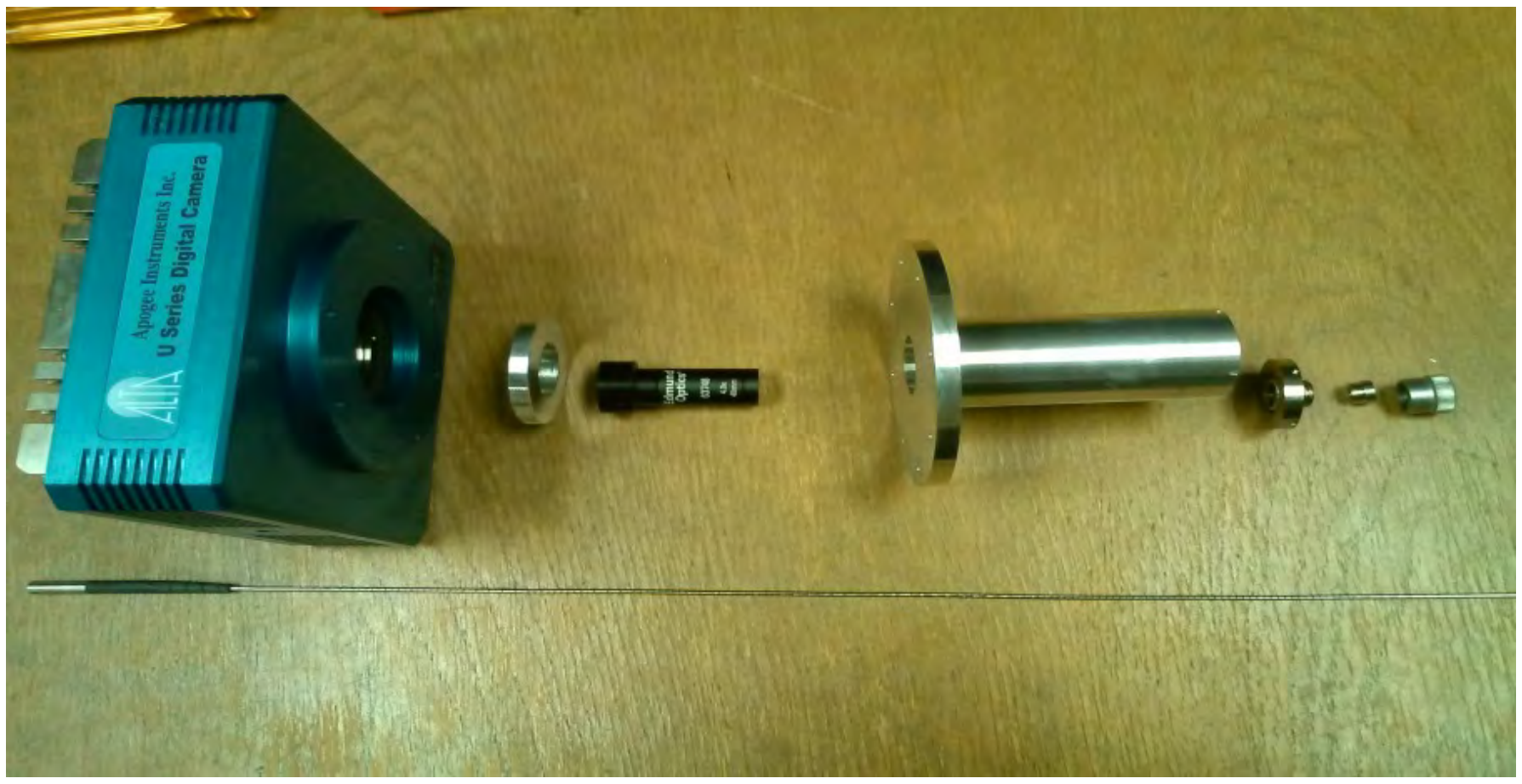

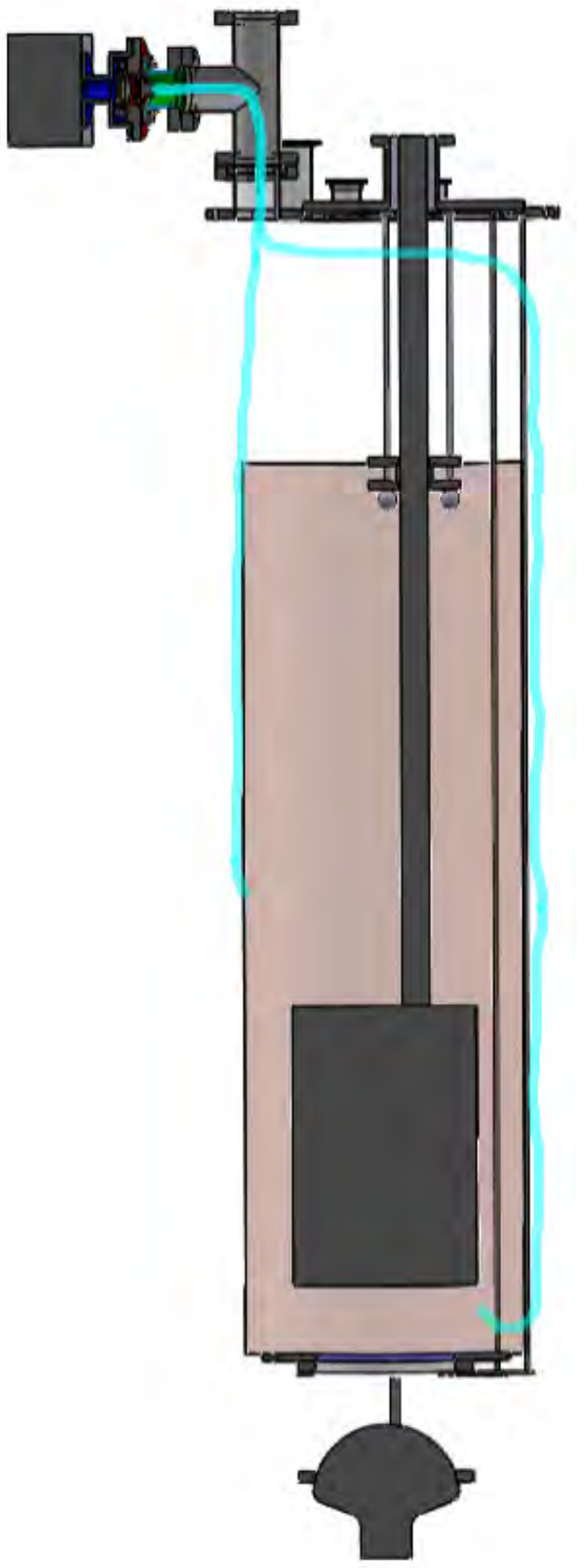

**Figure 6.10.5.1. Above: Disassembled camera setup. From left, they are: camera, F-C mount converter, telecentric lens, camera/optical fiber adapter (disassembled). Beneath: Optical fiber.**

**Right: Survey camera arrangement proposed for the Yale system test stand. The camera is at the top left of the figure and the blue lines indicate the locations of the fibers inside the cryostat.**

## 6.10.6 Summary of Integrated Testing Plans

The testing capabilities and facilities described above will be used to test critical prototype elements of the TPC and HV systems. The first goal of these integrated tests is to demonstrate that the TPC field cage structure and grids can reach the electric fields required. Below is a brief description of these phased tests.

Prototype TPC structures are under construction, to be tested both in liquid argon at Yale and in LXe at SLAC. Models of these prototypes are shown in Figure 6.10.6.1. The model shown on the left is a prototype of the reverse-field region of the TPC as it would be tested initially (Section 6.10.3) and uses the camera system described in Section 6.10.5. A prototype of the HV cable would first be connected to a simple plate at the top of the structure and operated up to 100 kV. This prototype is designed so that 100 kV simulates the electric fields seen in LZ at 200 kV and thus would test the TPC design under these conditions. A second version of this prototype would subsequently and possibly concurrently be operated in the Phase I LXe test system at SLAC (Section 6.10.4), albeit in a different HV configuration and taking into account lessons from the tests in liquid argon. Parts for two of these TPC prototypes have been fabricated primarily by LBNL. Prototype grid structures may also be tested in this prototype and these would be informed by the single-wire and small-grid testing described previously.

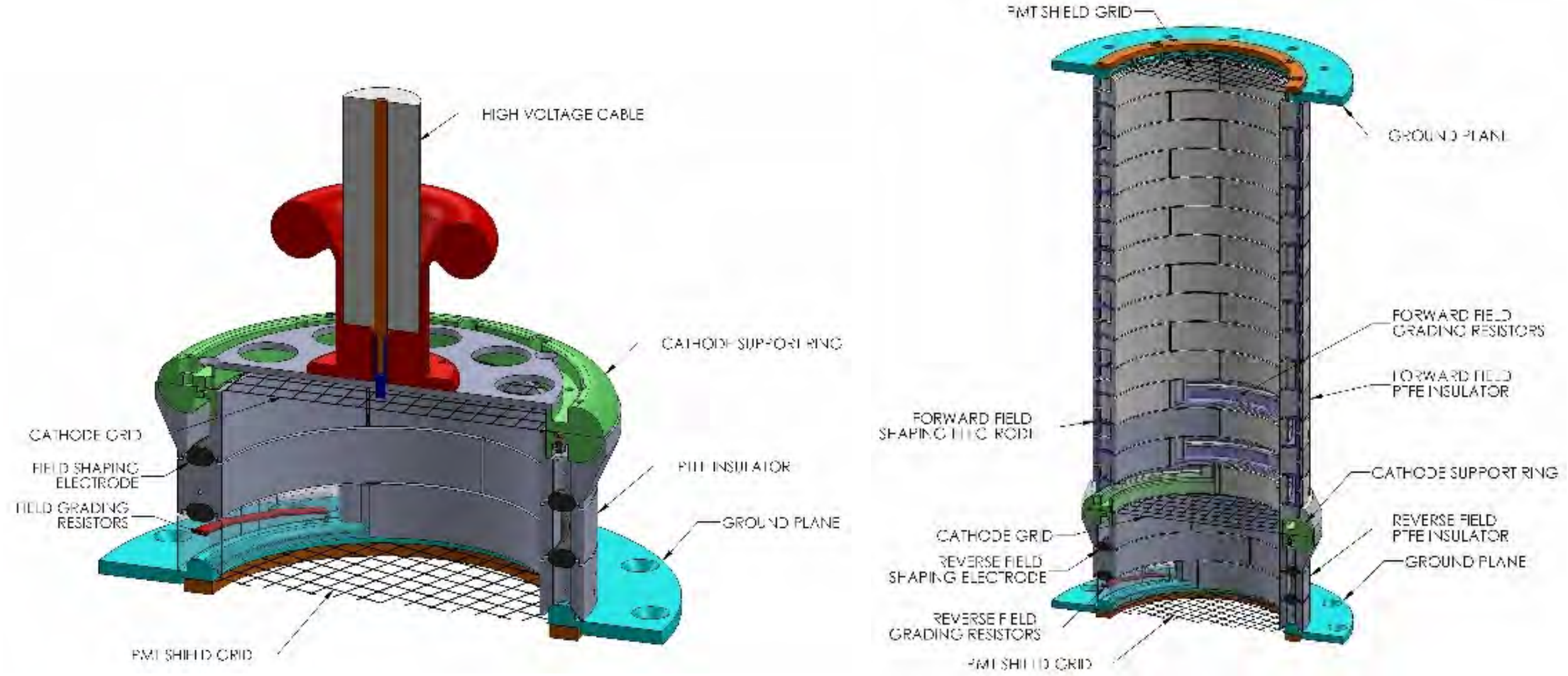

**Figure 6.10.6.1. Left: Phase I, reverse-field region prototype described in the text. Right: Schematic of larger-scale TPC and HV grid prototypes that would be tested in a phased approach described in the text.**

Ultimately, a larger-scale TPC prototype would be tested in the Phase II LXe test system at SLAC and possibly a version tested in liquid argon at Yale, depending on what is measured in Phase I prototype testing. This Phase II prototype is shown in Figure 6.10.6.1 (right) and would incorporate realistic grid structures, updated designs for the TPC structure, and other features. Prototype internal temperature, level, and other sensors would be included. The primary purpose of this prototype program would be to verify most aspects of HV operation in a realistic structure. However, operation in S2/S1 mode with purification and PMT readout would be the long-term goal and thus this would become a test bed for additional studies during the fabrication of the LZ components.

## Chapter 6 References

# 7 Outer Detector System

## 7.1 Introduction

In this chapter, we describe the performance and design of the outer detector system for LZ. The principal signal we seek, that of a WIMP scatter depositing 5-50 keV of energy in the central volume of LXe, will never be accompanied by deposited energy in the surrounding detector components. In contrast, the dominant backgrounds that might fake a WIMP signal will deposit energy not only in the central Xe detector but also in the material surrounding it. If we are able to detect these secondary interactions, we can veto the background event. Table 3.8.1.1 shows the major backgrounds in LZ, which include signals from gamma rays with energies in the few-MeV range and neutrons from ($\alpha$,n) reactions or created by cosmic-ray interactions.

To reduce these backgrounds to the level required, we surround the large active Xe volume with an integrated detector capable of tagging gamma rays and neutrons with high efficiency. Three detector elements are used to achieve this performance:

- The instrumented "skin" of the Xe, the region outside the LXe TPC (see Chapter 6),
- The gadolinium-loaded liquid scintillator (Gd-LS), and
- The portion of the surrounding water that is instrumented as a muon veto.

The outer detector system comprises the scintillator and water systems. In addition to the performance of the integrated veto system, this chapter describes the design of the outer detector system and modifications needed in the water tank to accommodate the LZ experiment.

## 7.2 Function and Performance of the Outer Detector

The outer detector serves two critical functions:

1. **To veto neutron and gamma backgrounds with high efficiency.** Although the outer region of the Xe shields the inner region very efficiently, the outer half of the Xe could not be used as part of the fiducial mass without an external veto. By instrumenting the outer skin of the Xe and adding the scintillator veto, we are able to double the fraction of the Xe in which very-low backgrounds are achievable. The outer detector is particularly important for vetoing neutrons, the background that most closely mimics dark-matter scattering.

   One risk to the performance of the LZ detector is that some material very close to the Xe could have a concentration of radioactive impurities higher than expected. The combination of the Xe skin and the outer detector serves to mitigate this risk. The integrated veto can suppress most backgrounds even if they are significantly higher than the design goals, with only a slight reduction in fiducial volume.

2. **To help characterize and measure the background.** A claim of a WIMP signal would require extraordinary supporting evidence. The outer detector will provide crucial supporting evidence necessary to establish a discovery. In particular, the only way to measure the neutron background reliably is by measuring the number of low-energy deposit scatters in the TPC that are followed by neutron captures in the outer detector.

The major non-neutrino background sources in LZ are neutrons and gammas from components within the cryostat and beta decays from radon and krypton distributed throughout the Xe. The levels for radon and krypton are designed to be low enough that the activity from both sources combined will be only 20% of that expected from astrophysical neutrinos (see Table 3.8.1.1). The principal goal of the integrated veto system is to reduce the effect of neutron and gamma backgrounds to a level smaller than that caused by radon and krypton over a very large fraction of the active Xe.

Neutrons represent a particularly troublesome background in the absence of an external veto. A neutron scatter produces a nuclear recoil (NR), as does a WIMP scatter, and after scattering they can escape the TPC and skin more easily than a gamma. The neutron background, which is principally produced in (α, n) reactions in materials near the Xe, is more difficult to predict than the gamma background. If a possible WIMP signal is seen, the outer detector will be needed to identify and measure the neutron background with good systematic error. The design requirement for NR background is to limit the background to less than 0.1 count in the 5,600 tonne-day exposure. Without the outer detector, the neutron background in a 5.6-tonne fiducial volume is a few events, so meeting the target requires a veto efficiency of greater than 90% for neutrons escaping the TPC.

The principal sources of gamma background are components in direct contact with the Xe volume, such as the PMTs, the PTFE reflectors, and the titanium inner vessel. Gammas in the few-MeV range can scatter at small angles in the outer region of the TPC, depositing 0.5-10 keV of energy, and then exit the TPC without a second scatter. The primary LZ requirement is that the number of electron recoil (ER) events in the 5,600 tonne-day exposure is less than 27. Meeting this requirement requires a veto efficiency of >70% for such gammas.

We have carried out simulations to characterize the impact of the outer detector on the background characteristics of the detector. The simulation package used for this work is the same as the one used in LUX, one that is known to reproduce the measured background in LUX very accurately — see also Chapter 4. The results of these simulations are captured in Figure 7.2.1, which shows the spatial distribution of all major radioactive backgrounds that scatter once in the Xe volume.

The left panel of Figure 7.2.1 shows the distribution of single scatters from backgrounds in the Xe TPC. The background in this plot is the sum of the NR background from neutrons, the ER background from gammas, and a uniform source of ER scatters from pp solar neutrinos. The plot assumes S2/S1 cuts that produce a 99.5% rejection of ER backgrounds with acceptance of 50% for NR backgrounds. The figures plot depth (Z) versus radius-squared, so that the area on the plot is proportional to the volume of Xe. The central region is, as expected, extraordinarily free of background. But the background is much higher, within 20 cm of the outer structures, and this is from neutrons and gammas. The white line indicates the fiducial volume, defined as the region in which the background satisfies the design goals for ER and NR background limits defined above. Without using information from the Xe skin or the outer detector, the fiducial mass is 3.3 tonnes, or about 45% of the active Xe. Most of the active Xe in this case is used as a veto rather than as target material for WIMPs. If one of the component materials in the cryostat were to have a larger amount of radioactivity than the design target, even less of the active Xe would be in the fiducial volume.

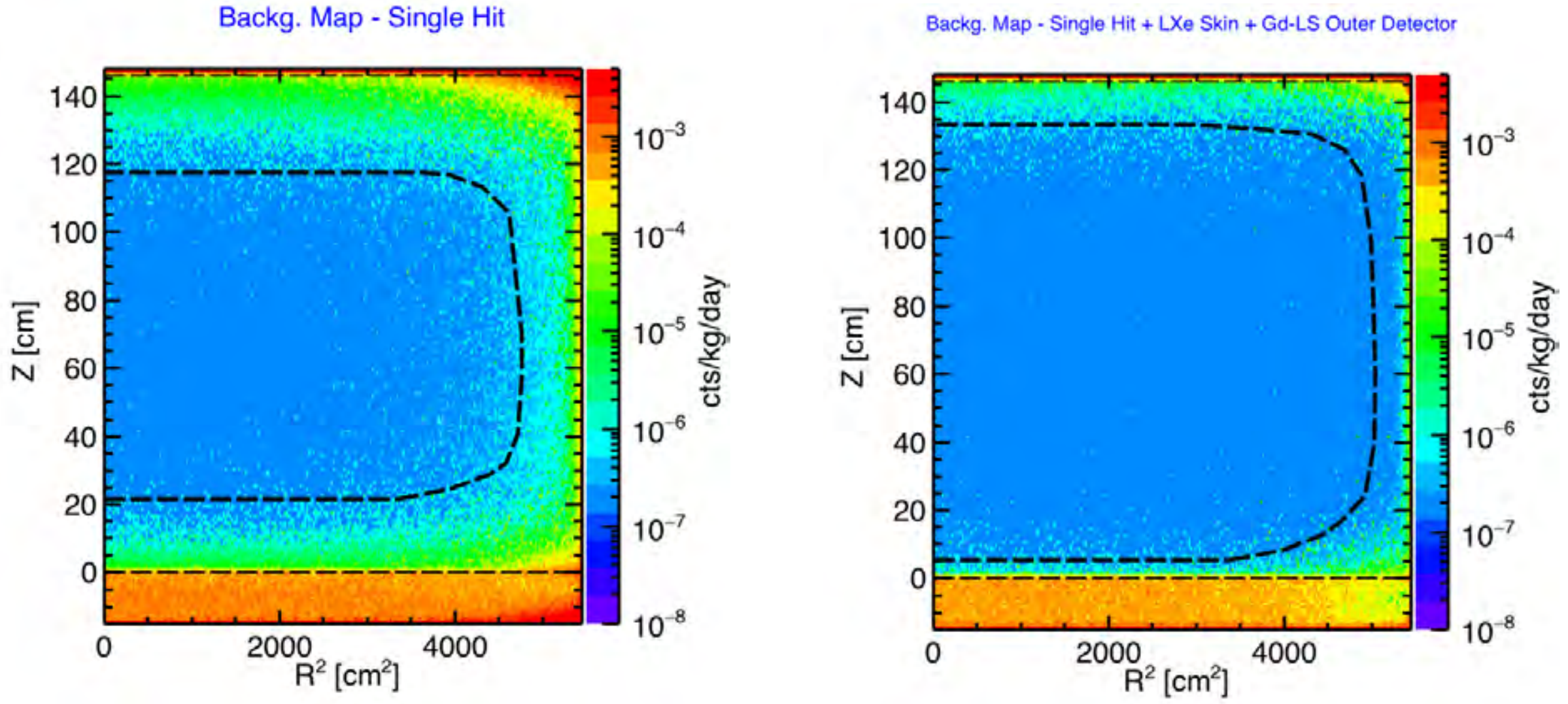

**Figure 7.2.1. Total NR background plus ER leakage from sources external to the LXe in the TPC. A discrimination efficiency of 99.5% is applied to ERs from gamma rays and solar pp neutrinos. Left: All single scatters in the TPC. Right: Single scatters in the TPC, vetoing on signals in the instrumented Xe skin and LS detector. Approximate fiducial masses, denoted by the black boundary line, are 3.8 and 5.6 tonnes for the two cases. These plots are taken from Figure 3.8.5.1, which also contains the cases with only LXe skin veto and only Gd-LS veto.**

The right panel of Figure 7.2.1 shows the performance when vetoing events that also deposit energy in either the instrumented Xe skin or the outer detector. These two systems operate as an integrated high-efficiency veto for neutrons and gammas. The white line shows that the fiducial volume can be extended to within a few centimeters of the edge of the active Xe. The fiducial volume with the integrated veto system is 5.6 tonnes, 1.7 times as large as for a stand-alone Xe TPC. Even if the neutron and gamma backgrounds were significantly higher than assumed in this study, the very-low background needed for effective operation of LZ could be maintained by reducing this fiducial volume by only a small amount.

Because of the large surface area of the LXe vessel, the fiducial volume increases by about 270 kg for every additional centimeter thickness of Xe at the boundary. To meet the LZ background requirements over a 5.6-tonne fiducial volume without using an external veto would require a TPC containing 11 tonnes, 4 tonnes more than the LZ design value.

## 7.3 Overview of the Outer Detector System

The proposed layout of the LZ outer detector is shown in Figure 7.3.1. A hermetic detector is built from nine vessels fabricated from UVT acrylic. The use of segmented vessels allows fabrication to take place at the manufacturer's facility at considerable cost savings. The sizes of the vessels are chosen to allow straightforward insertion into the water tank and assembly of the full detector inside the water tank. Structural finite element analyses (FEAs) of the vessels have been performed to validate the design without introducing more inert material than is needed for safe operation. The acrylic for the side vessels is 1 inch thick; for the top and bottom vessels, the acrylic is 0.5 inch thick except for the top wall of the top vessel and the bottom wall of the bottom vessel.

The vessels will be viewed by 120 8-inch Hamamatsu R5912 PMTs. The PMTs are mounted on stainless steel frames in the water tank, separated from the Gd-LS vessels by 80 cm. This arrangement gives a light-collection efficiency of about 7% averaged over the volume of the outer detector, corresponding to a light yield of about 130 photoelectrons for a 1-MeV energy deposit. The water shields the Gd-LS from gammas that originate in the R5912 tubes. A low-density water displacer will be used to fill in the gaps between the cryostat and the acrylic vessels and the gaps around the penetrations, to reduce the probability of absorption in inert material. A white diffuse reflector will be placed inside the outer detector vessels to improve collection of the scintillation light.

Simulation of the veto performance showed that the veto efficiency varies slowly with the thickness of the scintillator in the outer detector over the range 50 to 80 cm. This thickness is therefore optimized to reduce the risk of problems during fabrication and assembly. To insert the side vessels into the water tank easily, the scintillator thickness needs to be significantly less than 70 cm. Cleaning the vessels during fabrication requires that the thickness be no less than 61 cm, however, so we chose 61 cm for that thickness.

The liquid scintillator is based upon linear-alkylbenzene (LAB), a hydrocarbon chain with one benzene ring attached. LAB has a flashpoint that exceeds that of diesel fuel, and the safety aspects of diesel fuel in an underground facility have been explored and defined. The LAB is loaded with Gd, 0.1% by mass, via an organic chelating agent, trimethyl hexanoic acid (TMHA). This scintillator mix with 0.1% Gd doping was used by Chooz [1], Palo Verde [2], and Daya Bay [3]. The specific approach adopted by LZ is very similar to that used in the Daya Bay neutrino experiment, but with additional purification to achieve a lower uranium/thorium (U/Th) background.

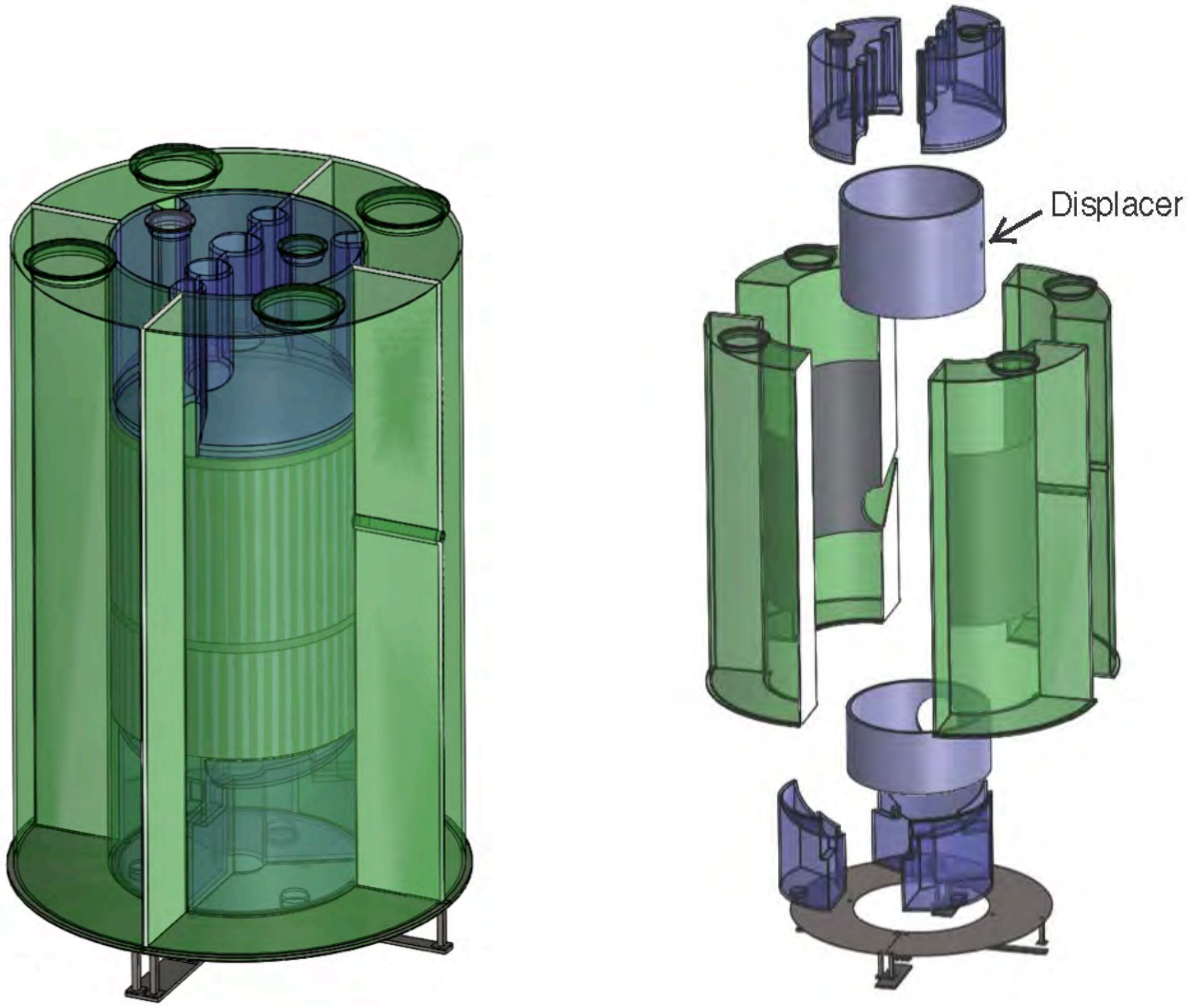

**Figure 7.3.1. Layout of the LZ outer detector system, which consists of nine acrylic tanks. The largest are the four quarter-tanks on the sides. Two tanks cover the top, and three the bottom. The exploded view on the right shows the displacer cylinders placed between the acrylic vessels and the cryostat.**

Gadolinium is added to the scintillator to increase the efficiency for tagging neutrons while maintaining low veto deadtime. The benefit of using Gd and scintillator for this purpose was demonstrated in the ZEPLIN series of experiments. Neutrons moderated to thermal energies in the scintillator are captured 90% on $^{157}$Gd or $^{155}$Gd, releasing 3-4 gammas with total energy of 7.9 MeV ($^{157}$Gd) or 8.5 MeV ($^{159}$Gd); the remaining 10% of the neutrons are captured on hydrogen, producing a single 2.2-MeV gamma. The Gd captures are tagged with higher efficiency because of the multiple gammas produced and the high energy of those gammas. The Gd reduces the neutron capture time to about 30 μs, compared with about 200 μs in scintillator without Gd. Figure 7.3.2 shows the simulated capture time for low-energy neutrons entering the outer detector. To maintain low deadtime for the veto system requires maintaining excellent radiopurity of the liquid scintillator, even with a veto window of 125 μs that is matched to the capture time with Gd. Without Gd, the veto window would be close to 1 ms, and the radiopurity requirements for the scintillator would be very difficult to meet.

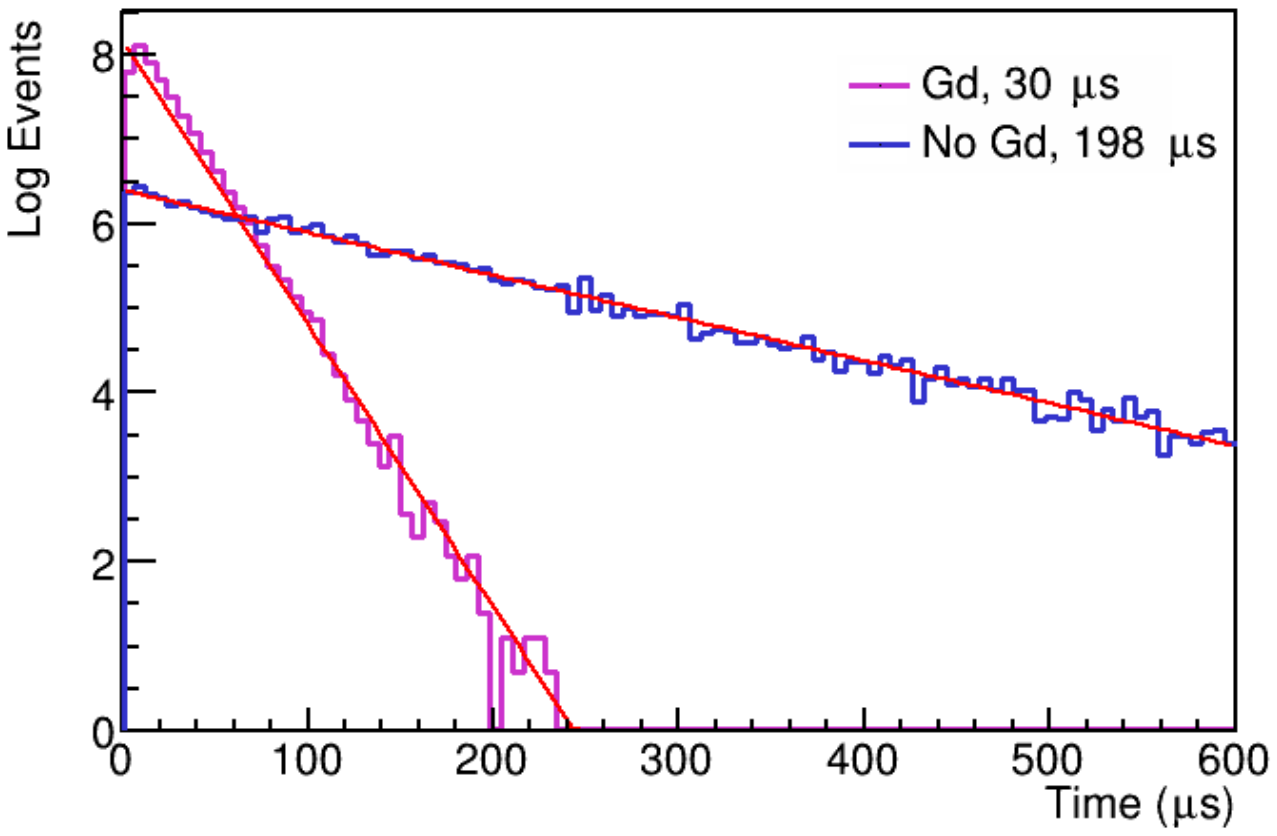

**Figure 7.3.2. The simulated distribution of capture times for thermal neutrons in the outer detector for LS with and without Gd.**

## 7.4 Mechanical Design and Systems

The 20.8 tonnes of scintillator liquid are contained in nine acrylic vessels, as shown in Figure 7.3.1: four tall vessels on the sides, two vessels that form a plug on the top, and three vessels that form a plug at the bottom. Taken together, the LS system forms a 61-cm-thick detector surrounding the Xe vessel, with several penetrations for connections to the Xe detector and for calibration systems. Similar acrylic vessels were used for the Daya Bay Antineutrino Detectors [3].

The masses and volumes of the nine vessels are shown in Table 7.4.1. The side vessels represent the largest part of the veto mass, holding about 88% of the scintillator. Each of the four side vessels is 427 cm high, extends in radius from 100.3 to 161.3 cm, and covers one-quarter of the full azimuth. A vessel is supported and anchored to a stainless steel base frame, which is in turn anchored to a base plate installed on the floor of the water tank. The net upward force on each side vessel when filled is 5826 N.

The two top vessels form a 61-cm-thick plug that fits inside the side vessels. They will be anchored from the top of the outer cryostat vessel. The three bottom vessels form a plug of the same thickness at the bottom. The penetrations for services and calibrations are positioned within the gaps between the acrylic vessel and in cutouts in the acrylic vessels.

The scintillation light is viewed by PMTs in the surrounding water, so the acrylic used to construct the vessels is chosen to be transparent to photons with wavelengths greater than about 300 nm. The vessels will be filled with LS at the same time that the water tank is filled with water, minimizing the differential pressure on the vessel walls and the stresses on them. This makes it possible to engineer the vessels with acrylic 2.54 cm thick.

The flanges on the detector cryostat protrude about 2 inches outside the cylindrical surface. To avoid building recesses in the acrylic vessels to accommodate these protrusions, a low-density foam will be installed as a water displacer around the outer vessel of the cryostat. This maintains low absorption of gammas between the scintillator and the Xe skin detector.

The vessels will be cleaned inside and leak-checked at the fabrication vendor. They will be wrapped in protective sheets at that time, and placed in double bags before being crated for shipping. The protective sheets will be removed after they are installed in the water tank. The final cleaning of the outside of the vessel will be done at that time.

As a feasibility study, a mock side vessel was slung under the Yates cage, taken down the shaft, and transported to the cart-wash area just outside the LUX experimental hall. We have studied the process of installing the acrylic vessels into the LUX/LZ water tank using a detailed computer model. The acrylic vessel will be transported in a horizontal position to the deck immediately above the water tank. The vessel will then be rotated using lifting eyes at the top and bottom. Figure 7.4.1 demonstrates one step of this process, near the point that requires maximum clearance above the deck. The vessel is lowered in vertical position into the water tank and then transported radially outward to near the wall of the tank. Figure 7.4.2 shows the assembly step at which all of the quadrant vessels are in the tank, and the first one is being brought into place around the cryostat. A white diffuse reflector, Tyvek, is placed at the inner surface of the scintillator vessels, the surface facing the cryostat.

**Table 7.4.1. The mechanical characteristics of the nine scintillator vessels.**

| | Acrylic Volume ($m^3$) | Acrylic Mass (kg) | LAB Volume ($m^3$) | LAB Mass (kg) |
|---|---|---|---|---|
| **Side Tank (each** | 0.643 | 759 | 5.172 | 4462 |
| **Top Tank (each)** | 0.214 | 253 | 0.862 | 744 |
| **Bottom Tank (each)** | 0.132 | 156 | 0.561 | 484 |
| **Total** | 3.396 | 4007 | 24.095 | 20789 |

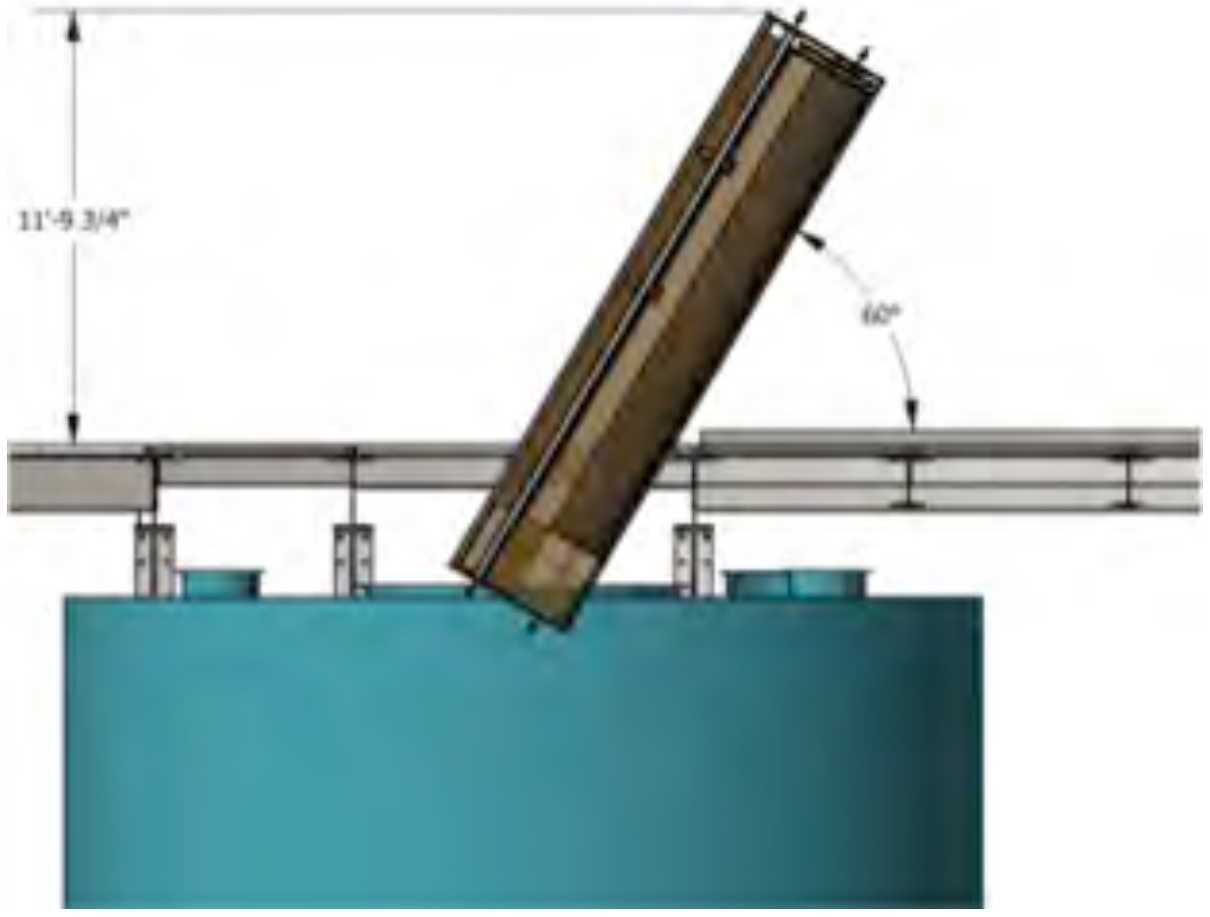

**Figure 7.4.1. A step in the assembly sequence for the outer detector system. This shows one of the quadrant vessels at 60°, the point of maximum height above the water tank.**

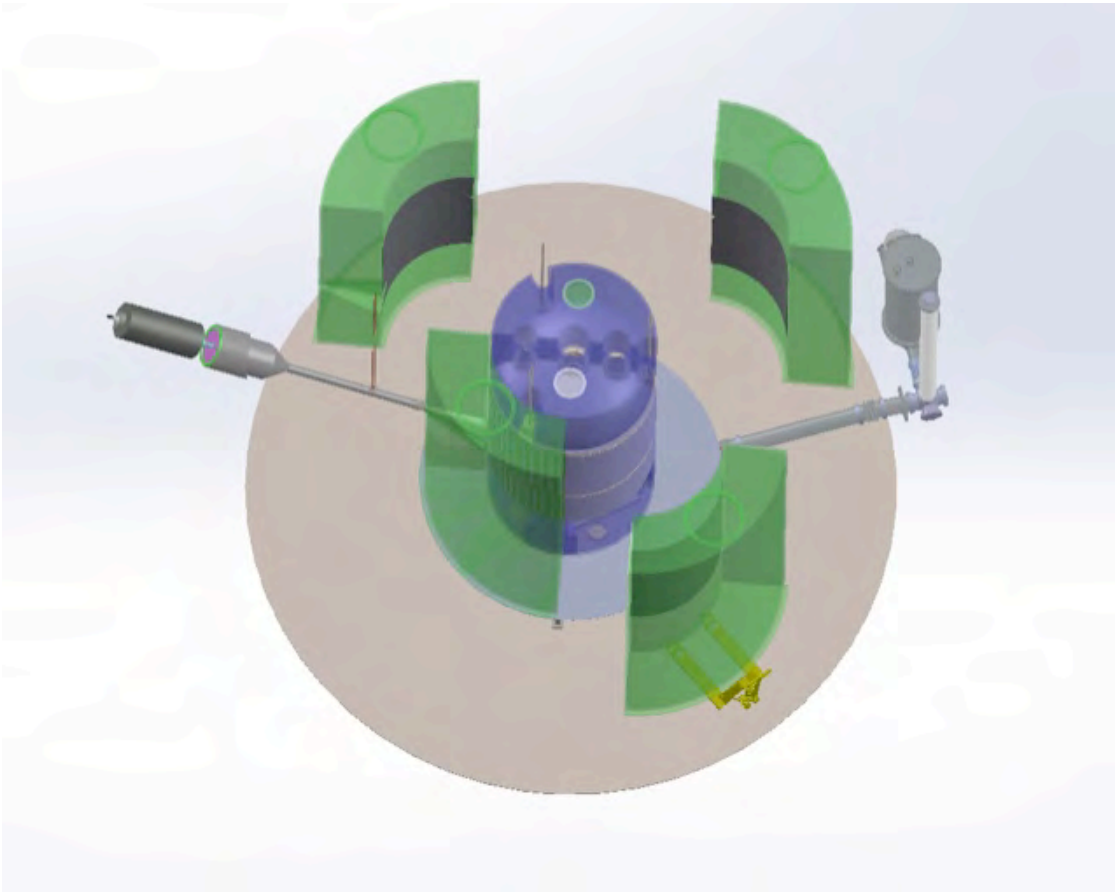

**Figure 7.4.2. Another step in the assembly sequence for the outer detector system. This shows the four quadrant vessels in the tank, with one already moved into place around the cryostat.**

Preliminary quotations for fabrication of the vessels have been obtained from vendors experienced with previous acrylic vessel fabrication for high-energy physics experiments. A site visit to one of the vendors to discuss design and fabrication has been completed.

## 7.5 Technical Description of the Liquid Scintillator

We chose Gd-LS for the detection medium to achieve excellent efficiency for neutrons and gammas. Various organic liquid scintillators have been used for neutron tagging due to their production of relatively large numbers of photons at low energies of a few MeV. The neutron-capture reaction occurs on the hydrogen in the organic scintillator, $n + p \rightarrow d + \gamma$, but the cross section is small at 0.332 barns, with a neutron-capture time of about 200 μs. The 2.2 MeV γ–ray is in the energy range of natural radioactivity, which extends to 2.6 MeV.

Gadolinium-loaded scintillators have been used in several experiments designed to measure inverse beta decay from reactor antineutrinos, including Palo Verde, RENO, and Daya Bay. There are several compelling advantages of adding Gd to the scintillator:

- The (n,γ) cross section for natural Gd is very high, 49 kilobarns, with major contributions from $^{155}$Gd and $^{157}$Gd isotopes. Because of this high cross section, only a small concentration of Gd, 0.1% by mass, is needed in the LS.
- The neutron-capture reaction on Gd releases 8 MeV of energy in a cascade of 3-4 γ rays. The efficiency for detecting at least one of these gammas is very high.
- The time delay for the neutron-capture is also significantly shortened to 30 μs in 0.1% Gd as compared with 200 μs in undoped scintillator. This shortened delay time reduces the accidental background rate by a factor of 7.

To detect the low-neutron/gamma backgrounds with high efficiency, the Gd-LS must have the following key properties:

1. Long optical attenuation length, >10 m at 430 nm (the emission spectrum is shown in Figure 7.5.1);
2. High light yield, 9,000 photons/MeV;
3. Ultralow impurity content, mainly of the natural radioactive contaminants, such as U and Th; and
4. Long-term chemical stability, over the lifetime of the experiment.

All of these properties have been achieved on a large scale for the Daya Bay experiment. We are adopting the same formulation and fabrication techniques to take advantage of this proven performance.

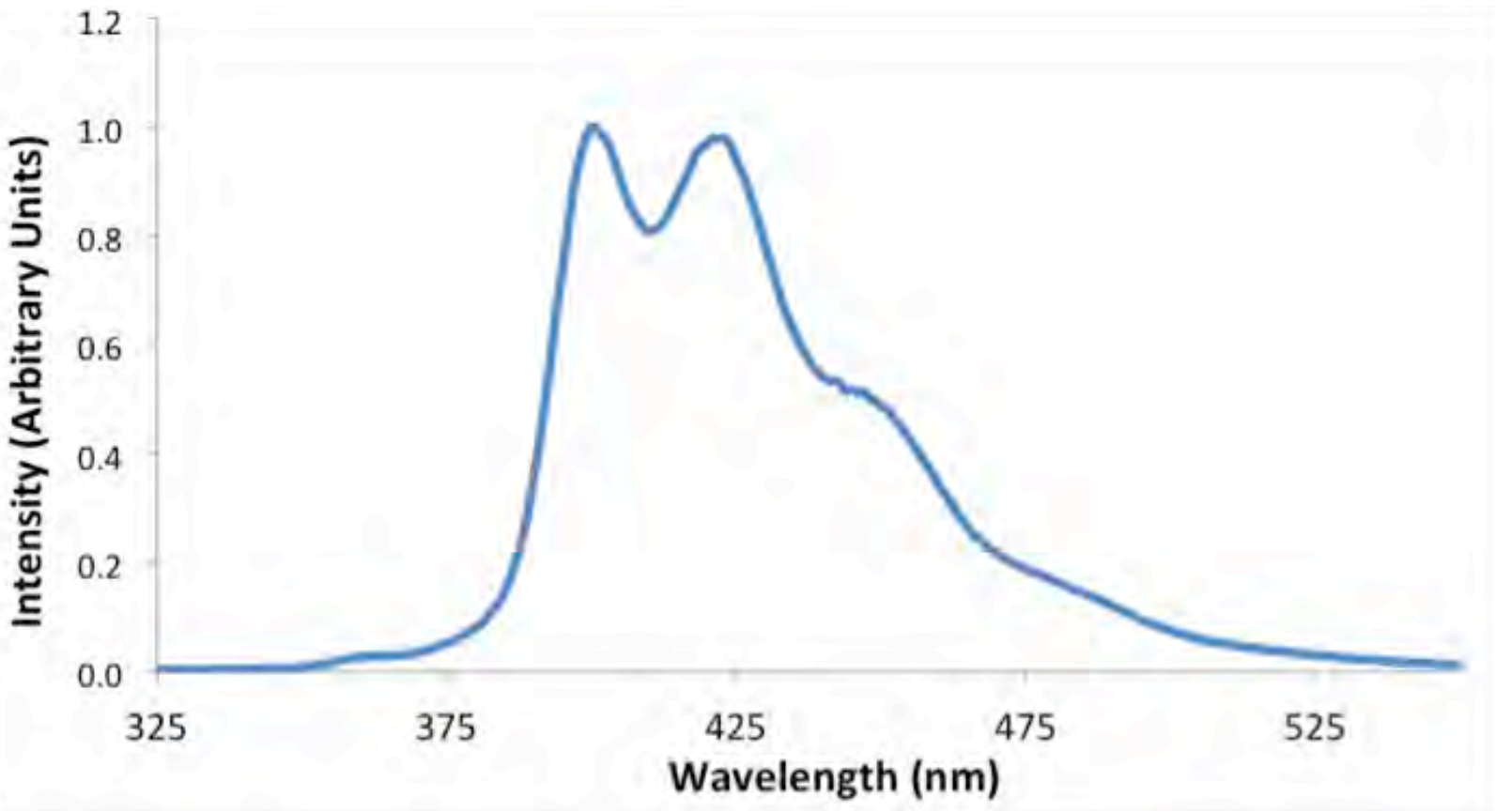

**Figure 7.5.1. The emission spectrum of Gd-LS with Gd concentration of 0.2%, based on measurements made of the Daya Bay scintillator.**

It is necessary to avoid any chemical decomposition, hydrolysis, formation of colloids, or polymerization, which over time can lead to development of color, cloudy suspensions, or formation of gels or precipitates in the scintillator, all of which can degrade the scintillator. Recent successful demonstration of the above-mentioned key items has been done by reactor electron antineutrino experiments using LAB-based, 0.1% Gd-loaded scintillator.

Assessments of the U/Th contaminations in the Daya Bay scintillator show that Daya Bay has already achieved levels nearly acceptable for the LZ experiment. The LZ requirement is based upon limiting gamma rays in the LXe fiducial volume to a rate lower by a factor of 4 than that expected from the PMTs, cryostat, and PTFE. We estimate the U/Th contamination requirements to be <1.7 and <3.2 ppt by mass, respectively. The three most important ingredients — $GdCl_3$, LAB, and PPO — of the scintillator mixture have already been analyzed for U/Th contamination by the Daya Bay group. The required levels of contamination for each component are summarized in Table 7.5.1. The required level of $^{238}$U is a factor of 12 below that of Daya Bay. In addition, a level of 0.6 ppt of $^{40}$K is the required, which is a factor of 12 lower than Daya Bay. KamLAND and Borexino reached contamination levels orders of magnitude lower by filtering and stripping the scintillator solvent. We will be able to meet the targets by inserting a second pass of purification into the production process, one step beyond that applied for Daya Bay. Underground sources of carbon must be used for all organics to meet the $^{14}$C, and the scintillator must be kept out of contact with the atmosphere to avoid $^{85}$Kr.

The principal development of LZ scintillator will be led by the Brookhaven National Laboratory (BNL) group. The level 3 manager there has considerable experience, including supervision of the development and production of the Gd-loaded LS for the Daya Bay experiment. The development of LZ scintillator will be undertaken in two phases: a demonstration phase (to reach ppt levels), followed by a production phase for deployment. The BNL group has a state-of-the-art Liquid Scintillator Development Facility equipped with a variety of instruments (e.g., UV, IR, XRF, LC-MS, medium-scale mixing reactor, thin-film distillatory, 2-m attenuation length system, etc.) for quality assurance that are essential to quality control of the scintillator. The cost-effective plan for the 20.8 tonnes of LZ low-background Gd-LS production will be to carry out the production at the BNL facility and ship the synthesized scintillator to SURF for filling. The scintillator will be produced at a rate of 0.5 tonne per week and will be stored in 55-gallon PTFE-lined drums, which are then shipped to SURF for storage. The purification methods for all components of the Gd-doped scintillator are developed and will be applied to each component before synthesis.

**Table 7.5.1. U/Th impurities in LAB-based Gd-LS. The first column shows required contamination levels for various components of the Gd-LS. Propagation of these contributions to the proposed Gd-LS is summarized in the last two columns.**

| Part | Raw Values (ppt) | | | | Gram per liter Gd-LS | In 0.1% Gd-LS (ppt) | | | |
|---|---|---|---|---|---|---|---|---|---|
| | $^{238}$U | $^{232}$Th | $^{40}$K | $^{14}$C | | $^{238}$U | $^{232}$Th | $^{40}$K | $^{14}$C |
| **LAB** | 1 | 0.5 | 0.4 | $2.3\times10^{-6}$ | 860 | 1.0 | 0.5 | 0.5 | $2.0\times10^{-6}$ |
| **$GdCl_3$** | 300 | 1200 | 20 | | 0.86 | 0.5 | 2 | 0.04 | |
| **PPO** | 20 | 70 | 10 | $50\times10^{-6}$ | 3 | 0.07 | 0.24 | 0.04 | $0.2\times10^{-6}$ |
| **TMHA** | 20 | 70 | 10 | $50\times10^{-6}$ | 3 | 0.07 | 0.24 | 0.04 | $0.2\times10^{-6}$ |
| **bis-MSB** | 4000 | 14000 | 2000 | $7\times10^{-3}$ | 0.015 | 0.07 | 0.24 | 0.04 | $0.2\times10^{-6}$ |
| **Total** | | | | | | 1.8 | 3.4 | 0.6 | $3\times10^{-6}$ |
| **Daya Bay** | | | | | | 20 | 4 | 7 | |

We are building an LS screener, consisting of an acrylic vessel that can hold 30 kg of Gd-LS viewed by three PMTs. We will fill this with scintillator liquid samples from the Gd-LS production line and place it in the LUX water tank, or similar, underground at SURF. By operating in this environment, we will be able to check that the scintillator meets the radiopurity standards.

## 7.6 Photomultiplier System

Building on the successful use of the Hamamatsu R5912 PMTs in experiments such as MILAGRO, AMANDA, and most recently Daya Bay, the LZ outer detector will use 120 of these same PMTs.

The two existing models for this PMT are R5912 and R5912-02. The latter possesses four more dynode stages and provides higher gain, at the cost of higher dark current and slightly degraded timing characteristics. As LZ does not need the additional gain, the less-expensive model R5912 was chosen. Even for this model, several subcategories exist that represent various quality levels of glass window and photocathode materials. Our assessment of radioactivity levels concluded that the basic model was sufficient for LZ needs and requirements.

The R5912 PMTs are suitable for the outer detector for the following reasons:

1. The spectral response ranges from 300 nm to 650 nm, with a peak wavelength at 420 nm. This matches well the scintillation light from the LAB mix between 390 and 440 nm. The quantum efficiency also covers the relevant range, with an average expected value of ~25% at 430 nm.
2. The PMTs will be submerged in up to 6 m water and must be able to operate in this environment. Daya Bay has been able to demonstrate successful operation of the R5912 assembly at higher pressures than those required for LZ. This experiment used the same assembly that LZ will use.
3. The radioactivity levels of the PMTs and assembly are a fairly weak constraint, thanks to the minimum 80 cm of water that separate them from any active detector volume. Eighty cm of water typically reduces the integrated flux of incoming gammas by more than 2 orders of magnitude, before taking into account geometric effects. For the event rate in the Xe target, the 80 cm of water plus the thickness of the scintillator make the R5912 contribution largely subdominant to internal sources, for both gammas and neutrons. In the scintillator itself, the simulated event rate from PMT radioactivity is 20 Hz (1% deadtime would be caused by 125 Hz).

The R5912 PMTs and waterproof assemblies will undergo rigorous individual testing to fully characterize the response and to validate uniform operation and long-term stability for the lifetime of the detector. These tests will include individual electrical behavior, gain measurements, linearity, after-pulsing, dark current, and dark count. In addition, the PMTs will be radioactively screened to make sure the activities are consistent with the values listed in Table 7.6.1.

Table 7.6.1. Additional characteristics of the R5912 PMTs, provided by Hamamatsu.

| Characteristic | Value |
|---|---|
| Number of dynode stages | 10 |
| Window material | Borosilicate glass |
| Photocathode material | Bialkali |
| Minimum photocathode effective area | 284 $cm^2$ |
| Typical bias voltage for 1e7 gain | 1500 V |
| Maximum voltage | 2000 V |
| Single phe rise time | 3.8 ns |
| Single phe FWHM | 2.4 ns |
| Single phe fall time | 55 ns |
| Typical single phe spectrum peak-to-valley ratio | 2.5 |
| Mean QE at 390 nm | 25% |
| Anode linearity at ±2% deviation | 20 mA |
| Pressure rating | 0.7 MPa |
| Radioactivity levels per PMT | $^{238}$U: 900 mBq, $^{232}$Th: 470 mBq, $^{40}$K: 3 Bq |

## 7.7 Water PMT Support System and Optical Calibration System

The LAB scintillation light is viewed by 120 8-inch PMTs in a cylindrical array of 20 ladders with six PMTs each. Figure 7.7.1 shows the plan view of the water-support system. The PMT faces are positioned 70 cm from the outer-detector tank wall. The water between the PMTs and the scintillator vessel shields the active detector elements from radioactivity in the PMT assemblies. In this location, the PMTs also see the Cherenkov light from cosmic-ray muons passing through the water.

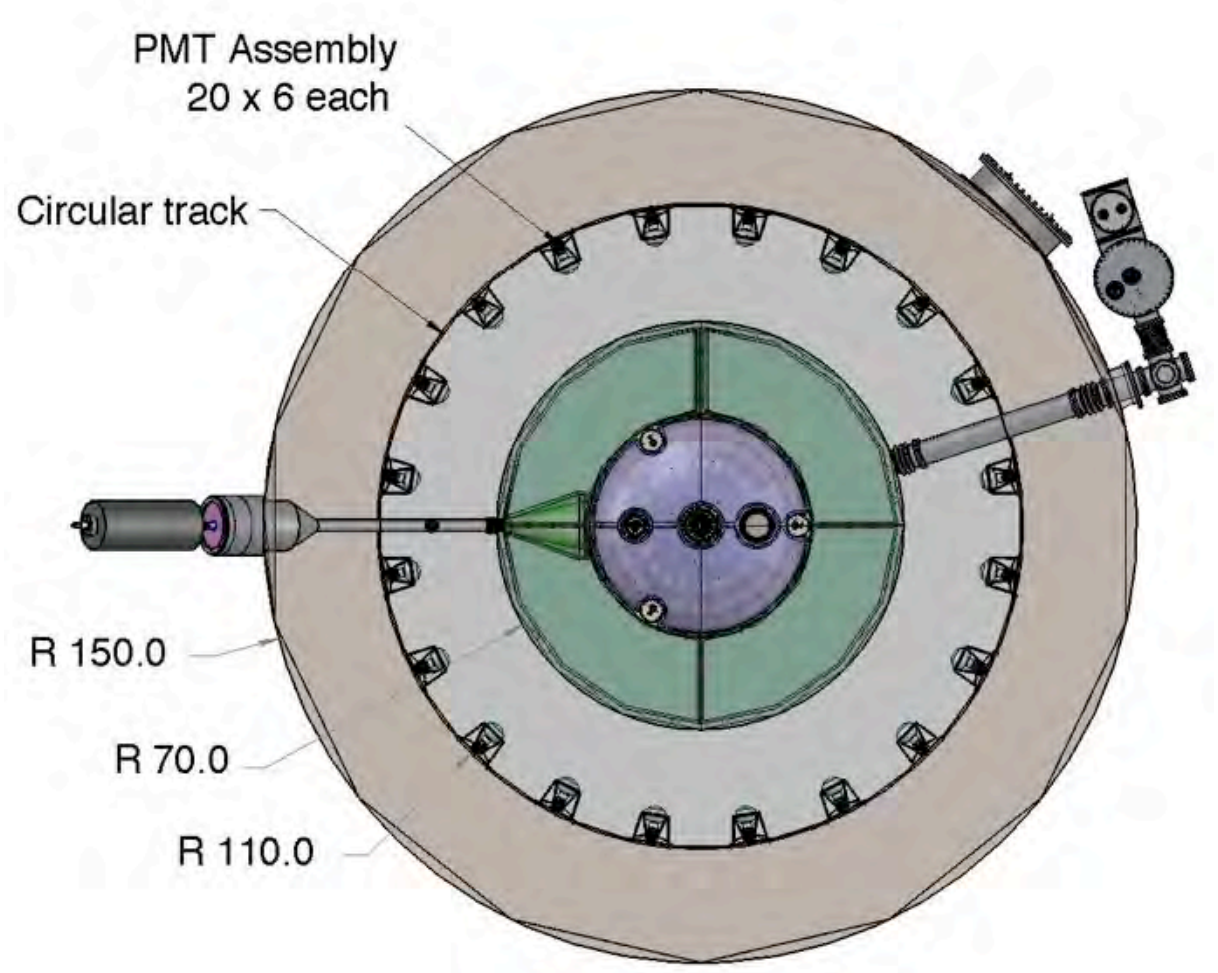

**Figure 7.7.1. Plan view of the water PMT support system. The 20 PMT ladders are mounted to a circular track on the roof of the water tank.**

The PMT ladders are supported from a circular track mounted on the top of the water tank, with a diameter of 18.5 ft. Attachment points on the tank floor keep them located properly.

The PMT frame is modified slightly from the Daya Bay version. Each PMT has a Finemet magnetic shield to isolate the performance from the local magnetic field. The PMT cables run directly from the PMTs through one of the ports in the top of the water tank to an electronics rack outside. Strain relief is applied at the port and on the ladders.

An optical calibration system will be used to monitor the performance of the outer detector and to maintain calibration. The light will be emitted from 30 LED-driven fibers mounted on the PMT support system, one placed at the center of four PMTs. Small diffusive reflectors will reflect some of the light back to the PMTs. In addition, the Tyvek reflectors inside the side vessels will also reflect light to the PMTs. This system will make it possible to cross-calibrate the responses of the water PMTs and to maintain that calibration over the life of the experiment. In addition, the calibration system will be used to check the optical properties of the water and of the scintillator. A prototype of the LED driver for the optical calibration system has already been built and tested. The pulse is less than 2 ns wide, so the width of the observed pulse will be limited by the PMT characteristics.

Fibers will also be installed at the bottom of the side tanks. These will be used to measure the light-attenuation properties of the LS over time. According to simulation, the number of photoelectrons observed per thousand photons produced in the scintillator varies from 4.2 with an absorption length of 9 m to 5.6 with an absorption length of 15 m.

## 7.8 Scintillator Distribution System and the Filling Process

Each of the nine scintillator vessels has one input and one output line. The side vessels have an input port in the top and an output port in the bottom. The bottom vessels have both ports in the bottom. The top vessels have both ports in the top, with an internal line to the bottom for emptying. The lines are Teflon® tubing, with strain relief to the PMT ladders. All 18 lines terminate at a feedthrough panel in the 2-foot flange on the north side of the lid to the water tank. To reduce pressure on the acrylic tanks, a 100-gallon reservoir will be suspended from the floor beams above and next to the north flange of the water tank. The LS will be taken underground in 55-gallon drums. Secondary containment will be provided for all of the lines carrying scintillator. The gas volume above the reservoir will be kept filled with dry nitrogen to minimize the amount of radon entering the liquid. In addition, the vessels will be purged with nitrogen before filling.

To reduce the differential pressure inside and outside the vessels, they will be co-filled with the water tank that encloses them. The hydrostatic pressure on the outer surface of the side vessels varies during the filling process from 1 to 7.8 psi, while the pressure on the inside from the scintillator varies from 1.7 to 7.6 psi. The 100-gallon reservoir is connected to the side tank and filled to match the internal pressure to the outside pressure. Table 7.4.1 shows the net buoyant force on the various tanks when they are filled with scintillator and the surrounding water is in place.

We carried out an FEA of the side vessels during filling to model the stresses on the acrylic during the fill process. The optimal process is to maintain the level of the water outside the vessels to between 10 and 20 inches below the level of the scintillator inside. The FEA shows the maximum tensile stress is 420 psi, occurring when the vessel is about half-full. The design requirement is to keep that stress below 750 psi to avoid crazing. The maximum stress in the long term after filling is 250 psi.

## 7.9 Threshold, Background Rate, and Deadtime

As described above, a neutron capture in the Gd-LS releases either a few gammas of total energy 8 MeV or a single gamma of energy 2.2 MeV. The neutron-detection efficiency is therefore quite high with the 100 keV threshold planned for the outer detector, which corresponds to about 13 photoelectrons observed. Figure 7.9.1 shows the simulated inefficiency for vetoing 1-MeV background neutrons as a function of threshold in keV. These simulations indicate that the efficiency of the combined veto for neutrons escaping the Xe TPC is about 96%, and the veto efficiency for 1-MeV gammas is about 73%. The inefficiency for gammas is dominated by Compton scattering in the inert material, especially the inner wall of the acrylic, and is fairly insensitive to the threshold.

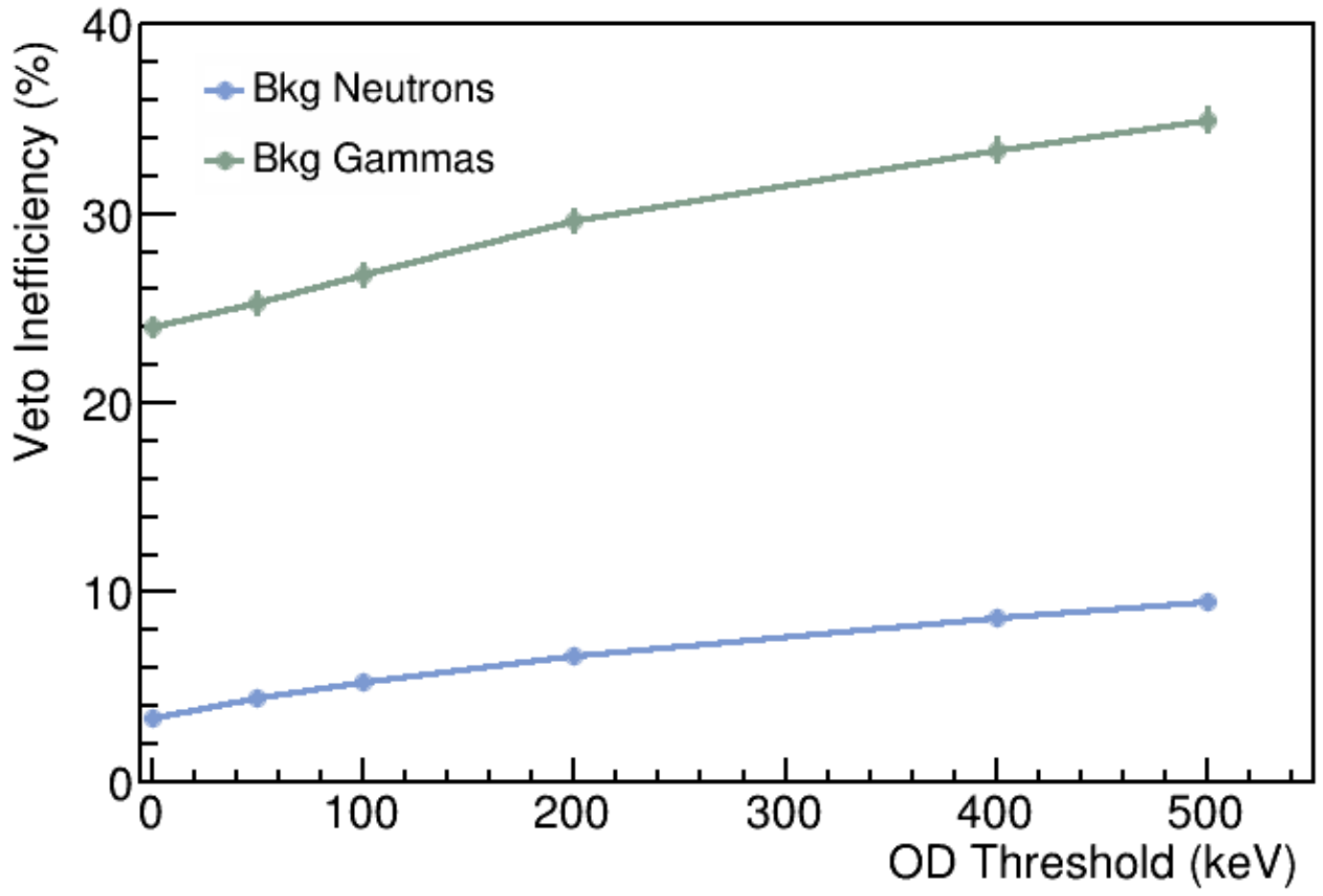

**Figure 7.9.1. The simulated inefficiency for vetoing background 1-MeV neutrons and 1-MeV gammas as a function of threshold for the outer detector.**

We will read out the outer-detector PMTs each time the Xe TPC produces a trigger, without need of a hardware threshold. We will therefore be able to apply the veto offline, with parameters carefully selected to optimize the veto efficiency. We will apply a neutron veto over a time window of 125 μs with a threshold of no more than 200 keV. In addition, we will apply a gamma veto over a much tighter time window of 1 μs with a threshold of 100 keV. We will set up an outer-detector hardware trigger to calibrate and monitor the background environment.

The sensitivity of the outer detector is estimated to be about 130 photoelectrons/MeV from simulation of LZ and this agrees with benchmarks against similar detectors. Thus, a 100-keV threshold corresponds to about 13 photoelectrons, and the trigger rate will be dominated by alphas, gammas, and betas from the U and Th chains.

Because the time of the low-energy scatter in Xe is determined well by measuring the prompt S1 light, the neutron veto window will be determined by the capture time for neutrons. We presently assume a 125-μs veto window, which is >4 capture lifetimes. To be conservative, we set a goal of keeping the deadtime for this time window at around 2%. This requires keeping the background rate in the outer detector below 160 Hz, which can be met by the radiopurity targets in Table 7.5.1. The width of the veto window needed for high-efficiency neutron vetoing will be determined by studying it with neutron-calibration data.

## 7.10 Environment, Safety, and Health Issues

The most important safety issue to consider underground is flammability. The flashpoint for LAB is 120-140 °C, which makes it a Class IIIB liquid, the lowest hazard category. A Class IIIB liquid is a combustible liquid with a flashpoint at or above 200 °F (93.4 °C). OSHA flammable and combustible liquid regulations do not apply to IIIB liquids. The boiling point is greater than 250 °C, and the melting point is below -70 °C, so it is very stable as a liquid. The density of LAB is 0.86 gm/cm$^3$, significantly less than water. In addition, it has very low solubility in water.

The primary risk to consider is a crack in one of the acrylic vessels. The first preventive step is to thoroughly check the integrity of the vessel before introducing the scintillator liquid into it. The second is to monitor the vessel carefully during the filling process. The water tank serves as a secondary containment system for the scintillator. We will install a sensor system into the water-circulation loop that can detect small amounts of LAB. We will also be able to separate LAB from the water in the water treatment. If a significant leak is observed, we will skim the surface of the water to recover most of the scintillator, and then remove the residual scintillator by distillation.

## Chapter 7 References

# 8 Cryostat

This chapter discusses the design of the LZ cryostat, the extensive program to acquire suitable materials for it, and the simulations of associated backgrounds. The baseline design for the cryostat assumes the use of titanium, although stainless steel is being evaluated as an option. A critical requirement for the cryostat is to limit the contributions to radioactive backgrounds in the sensitive region of the Xe detector. We summarize our simulations of the expected backgrounds from the cryostat in Section 8.1. The results of the R&D program to assay potential cryostat materials are given in Section 8.2. Finally, we described the cryostat design and key interfaces in Section 8.3.

## 8.1 Background Simulations

The material baseline is commercially pure titanium, Grade 1, due to its low radioactivity, high strength-to-weight ratio, and low density. As a result of these properties, the neutron and gamma absorption in the cryostat is minimized, making it possible to veto backgrounds more efficiently. Titanium of this grade was used successfully for the LUX cryostat [1]. In case we are unable to obtain titanium with the required radiopurity, we are also considering stainless steel (SS) such as SS316L/Ti and SS304L.

The backgrounds have been assessed using the LUXSim [2] code modified to reproduce in detail the design of the LZ detector. Simulation is based on the well-established and validated GEANT4 package [3]. The maximum allowed background for each relevant LZ component is set to be less than 10% of the rate from astrophysical neutrino sources before any S2/S1 rejection. Therefore, the main limitation in terms of background for LZ will be given only by the astrophysical sources (see Chapter 12).

For neutrons, ($\alpha$,n) reactions and spontaneous fission neutron energy spectra were generated using the SOURCES software [4]. These spectra were then embedded into the LUXSim framework, which propagates neutrons isotropically emitted from the cryostat. For the $^{238}$U and $^{232}$Th decay chains, we used the approach described in [5], while for the $^{40}$K and $^{60}$Co decays, we used the standard GEANT4 process [3]. Neutron and gamma yields for Ti and SS are presented in Tables 8.1.1, obtained from [4] and [6], respectively.

Nuclear and electron recoil backgrounds from the cryostat, within the corresponding energy ranges of 1.5-6.5 $keV_{ee}$ and 6-30 $keV_{nr}$, have been estimated selecting single scatter events in the sensitive LXe volume, assuming radial and horizontal position resolution of 3 and 0.2 cm, respectively. Signals from the LS veto and the LXe TPC skin layer have been used to veto events with energy depositions above 100 keV and within an 800 μs time window. No additional efficiency or S2/S1 cuts have been applied. In the simulation, 5.6 tonnes fiducial mass and 1,000 days' exposure time have been considered, as shown in Figure 3.8.5.1, evaluated after all the veto systems are applied.

The "Maximum Allowed Radioactivity (mBq/kg)," shown in Table 8.1.2, indicates the radioactivity from each isotope — $^{238}$U, $^{232}$Th, $^{40}$K, and $^{60}$Co — that corresponds in total to one-third of 10% of the background from the astrophysical pp neutrinos (ER) and nuclear scattering (NR).

**Table 8.1.1. Neutron and gamma rates in the background simulation described in the text.**

| | Titanium | Stainless Steel |
|---|---|---|
| | Neutron Yield ($10^{-6}$ n/s at 1 Bq/kg) | |
| **U** | 3.1 | 1.5 |
| **Th** | 5.3 | 1.4 |
| | Gamma Yield (gamma/decay) | |
| **U** | 2.23 | |
| **Th** | 2.74 | |
| **K** | 0.1 | |
| **Co** | 2 | |

**Table 8.1.2. Results from cryostat gamma background simulations for titanium and stainless steel.**

| Maximum Allowed Radioactivity [mBq/kg] | | |
|---|---|---|
| | **Titanium** | **Stainless Steel** |
| $^{238}$U | 0.75 | 0.67 |
| $^{232}$Th | 0.51 | 0.74 |
| $^{40}$K | 16.81 | 8.50 |
| $^{60}$Co | - | 2.28 |

We have considered the Ti assay results by LZ, detailed in Section 8.2, and the SS considered for the Neutrino Experiment with a Xenon TPC (NEXT) collaboration [7] and our own results, also described in Section 8.2. Results for Ti from TIMET [8] and potentially also SS show that these materials would satisfy the allowed activity for the fiducial mass of 5.6 tonnes. Titanium has lower backgrounds, as described below, and is our preferred choice.

## 8.2 Cryostat Material Searches

The baseline design is to use CP-1-grade Ti for the inner and outer vessels of the cryostat for LZ, due to its low radiological background content. The Ti used in LUX contains <0.25 mBq/kg of $^{238}$U, <0.2 mBq/kg of $^{232}$Th, and <1.2 mBq/kg of $^{40}$K; however, such low values appear to be rare, and readily procuring Ti with similar levels of contaminants has proved difficult for several other experiments. LZ groups at Rutherford Appleton Laboratory (RAL) and University College London (UCL) have embarked on a program of R&D, working with partners from the Ti production industry, to identify the points of inclusion of contamination during the Ti roll stock manufacturing processes, to control their effect, and to procure sufficiently clean material — with contamination levels comparable to the LUX Ti.

The production of Ti metal is a complex procedure that involves a number of stages in which additives and inclusions are deliberately introduced. Several such points in the production cycle may contribute to contamination of the final metal with elements containing radioactive isotopes; $^{238}$U, $^{235}$U, and $^{232}$Th are of particular concern. The refinement of the mineral concentrates, particularly for ilmenite, involves the addition of or exposure to coke, coal, oil, and tar prior to the chlorination process. It is not uncommon for such materials to contain relatively high levels of U and Th. However, the $TiCl_4$ produced at this stage undergoes chemical treatment and filtering to remove chlorides and sludge before pure liquid $TiCl_4$ is created, carrying away most impurities, including U and Th. Ultrapure $TiCl_4$ is commercially available, as are titanium hydride and titanium nitride powders that are produced through plasmochemical processing from the $TiCl_4$. Beyond this stage, lack of sufficient contact controls with surfaces during the Kroll process and metallothermy present other potential sources of U and Th. However, the elements introduced — largely Mg and Ar — will probably not be problematic. The Ti sponge post-Kroll processing is exposed to several stages in which U and Th can enter the chain. Ti ingots and slabs are produced by pressing and melting the Ti sponge, yet often Ti alloy and Ti scrap is added at this stage. Other alloys such as aluminum and vanadium may also be included. Radioactive contamination contained within the scrap and alloys is then carried through to the Ti ingot and into the roll stock. The major stages of this production process, indicating inclusion points, are depicted in Figure 8.2.1 [9].

During this R&D period we have engaged several titanium providers including VSMPO [10], TIMET Supra Alloy [11], Honeywell [12], and PTG [13] to provide sample material, taken from various stages along the production process, in order to determine where radioactivity, particularly U and Th, enters the chain. In our campaign we have received 22 samples including: eight sponges, one sample of very-high-purity Ti (Honeywell), one Grade-1 with 10% scrap (VSMPO), two Grade-2 sheets (Supra Alloy and PTG), eight Grade-1 sheets (Supra Alloy, PTG and TIMET) — and also Ti bolts and nuts.

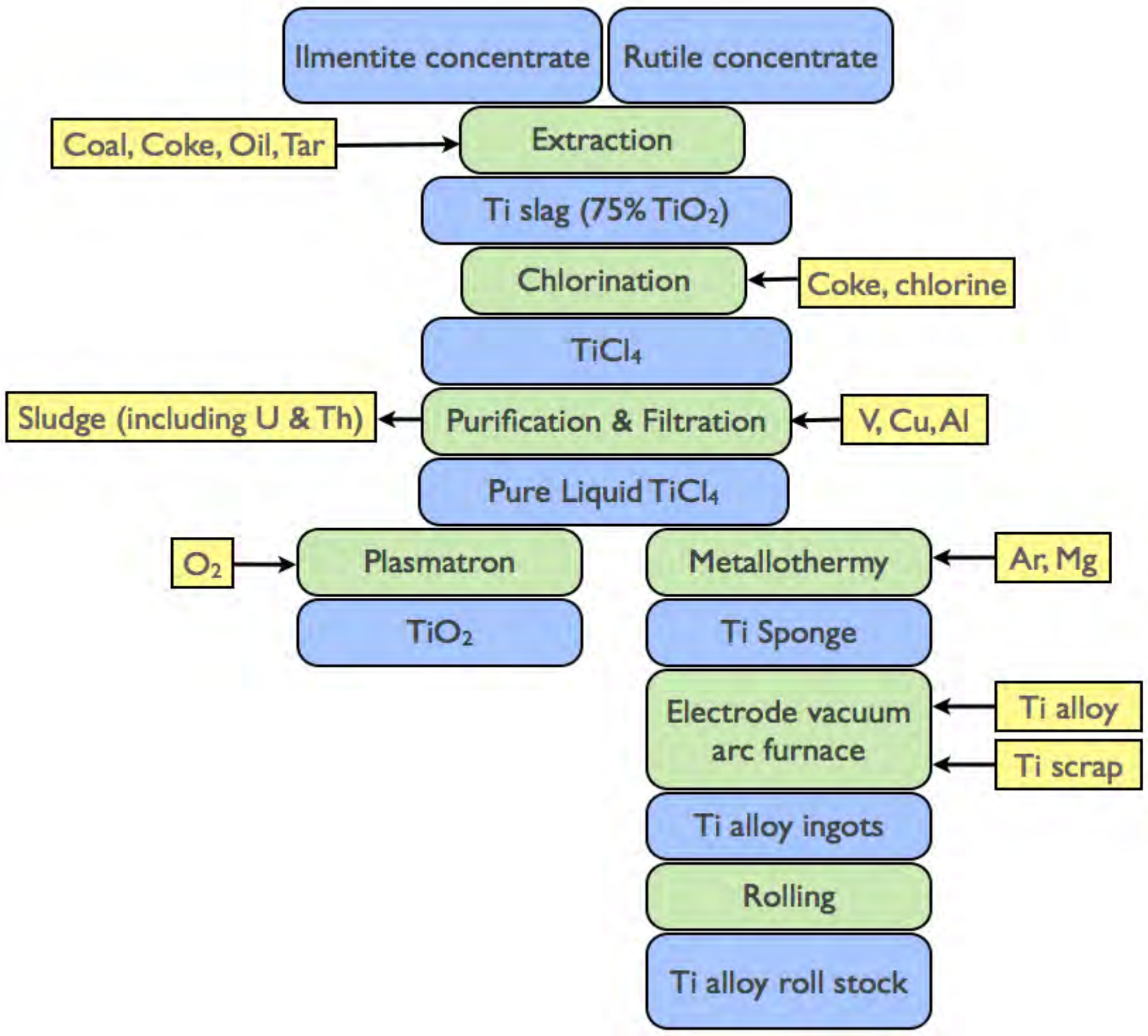

**Figure 8.2.1. The commercial production of Ti metal, indicating the major stages (green boxes), the post-processing products (blue boxes), and the additives, as well as reductions during the procedure (yellow boxes). Figure adapted from [9].**

Table 8.2.1 summarizes all the Ti samples that have been radioassayed in this campaign. Results from the LUX material campaign are included for comparison. We note that sample #22 in Table 8.2.1 represents the lowest radio-impurity contamination ever reported for a titanium sample, with activities of <1.6/<0.09 mBq/kg for $U_e$ /$U_l$ and also 0.28/0.23 mBq/kg for $Th_e$/$Th_l$, respectively.

Low-background experiments searching for dark matter or neutrinoless double-beta decay, such as XENON-1T [14] and PANDA-X [15], or GERDA [16] and NEXT, respectively, use stainless steel for their cryostats, with the majority of materials coming from German stockholder NIRONIT [17]. We procured 13 samples from NIRONIT for radioassay. We also conducted independent assays of samples received directly from the GERDA and NEXT experiments, to cross-check published results [7-18] and to measure radioisotopes not reported, as well as early U and Th activity.

The 13 SS samples were received in November 2014 (a total of 152 kg). They originate from different heats and were made by different mills: seven samples at TyssenKrup Nirosta (Germany) and six samples at Aperam (Belgium). All samples were electropolished at LBNL and prescreened with a surface HPGe counter (MERLIN), particularly for excessive $^{60}$Co. Samples with <20 mBq/kg of $^{60}$Co were forwarded for more sensitive tests underground at SURF and at the University of Alabama. Stainless steel radioassays are summarized in Table 8.2.2.

An impact of the titanium from TIMET and stainless steel from NIRONIT on LZ background is presented in Figure 12.3.3.1.

**Table 8.2.1. Summary of the 22 samples assayed for the LZ cryostat, including various grades and types from multiple suppliers.**

| # | Supplier | Sample name | 238U (mBq/kg) | | 232Th (mBq/kg) | | 40K (mBq/kg) | Ti grade/type | Report Date |
|---|---|---|---|---|---|---|---|---|---|
| | | | early | late | early | late | | | |
| 1 | Supra Alloy | Carlson 8J102 | 31.00 | 4.10 | 0 | 2.80 | *1.80* | CP-1 Sheet | Jan 2013 |
| 2 | TIMET | Osaka 26-29461 | *2.50* | 248 | 0 | *4.10* | *12.00* | Sponge | Aug 2013 |
| 3 | TIMET | Tangshan TX027594 | *2.50* | 6200 | 0 | *2.50* | *15.00* | Sponge | Aug 2013 |
| 4 | TIMET | Toho C 12009C | *2.50* | *62* | 0 | *1.60* | *12.10* | Sponge | Aug 2013 |
| 5 | TIMET | TOHO W 11266W | *2.50* | 124 | 0 | *1.60* | *12.00* | Sponge | Aug 2013 |
| 6 | TIMET | Zaporozhye 6680-12 | *2.50* | 744 | 0 | *1.60* | *12.00* | Sponge | Aug 2013 |
| 7 | TIMET | ZUNY TX027641 | 25.00 | 2480 | 0 | 4.1 | *12.0* | Sponge | Aug 2013 |
| 8 | TIMET | HN0021-B sample1 | 11.00 | *0.60* | 0 | *0.60* | *2.50* | CP-1 Sheet | Aug 2013 |
| 9 | TIMET | HN0021-B sample 2 | 4.90 | 3.33 | 2.85 | 0.80 | *1.50* | CP-1 Sheet | Sept 2013 |
| 10 | PTG | ATI W74M | 46.00 | 2.80 | 0 | 2.80 | *1.80* | CP-1 Sheet | Jan 2013 |
| 11 | Supra Alloy | Timet BN3672(2) RMI 404666 (9) | 110.00 | *2.40* | 0 | 170.00 | 2.40 | CP-2 | Jan 2013 |
| 12 | PTG | Thyssen Krupp 611292 | *9.60* | 3.60 | 0 | 2.40 | *2.10* | CP-2 | Jan 2013 |
| 13 | TIMET | Henderson 22-49312 | *3.70* | 2480 | 0 | 12.30 | *18.00* | Sponge | Aug 2013 |
| 14 | S6MB annulus | Bolts | 1300.00 | *6.00* | 0 | 160.00 | *60.00* | Bolts | Jan 2013 |
| 15 | EE-33 full | Nuts | 500.00 | *8.40* | 0 | 80.00 | *60.00* | Nuts | Jan 2013 |
| 16 | Honeywell | T149858991 | 3.70 | 4.69 | 0 | 1.63 | *1.50* | CP-1 Sheet | Sept 2013 |
| 17 | VSMPO | 528 g | *61.70* | *6.20* | 0 | *4.10* | *31.00* | CP-1 Metal (10% scrap) | Jan 14 |
| 18 | VSMPO | 996 g | 17.28 | 12.35 | 0 | *4.10* | *6.20* | CP-1 Sponge | Jan14 |
| 20 | TIMET | HN2470 | 8.51 | 0.37 | 0 | 0.61 | 0.52 | CP-1 Sheet | Nov 14 |
| 21 | TIMET | Master ID #46 | 8.00 | *0.124* | 0 | *0.12* | *0.62* | CP-1 Sheet | Sept 14 |
| 22 | TIMET | HN3469 | *1.60* | *0.09* | 0.28 | 0.23 | *0.54* | CP-1 slab | Jun-16 |

**Table 8.2.2. Summary of the 13 stainless steel samples radioassayed for LZ. Due to a high $^{60}$Co content detected in samples 10-13 during prescreening, these were not assayed further and as such early-chain contents were not measured (the prescreen detector is not sensitive to early-chain decays).**

| # | Sample name | 238U (mBq/kg) | | 232Th (mBq/kg) | | 60Co (mBq/kg) | 40K (mBq/kg) |
|---|---|---|---|---|---|---|---|
| | | early | late | early | late | | |
| 1 | NIRONIT 311113 | 7.3 | 0.35 | 1.1 | 4 | 14.5 | *0.53* |
| 2 | NIRONIT 511803 | 1.2 | 0.27 | 0.33 | 0.49 | 1.6 | *0.4* |
| 3 | NIRONIT 512006 | *1* | 0.54 | 0.49 | 1.1 | 1.7 | *0.59* |
| 4 | NIRONIT 512844 | 1.4 | 0.5 | 0.5 | 0.32 | 2.6 | *0.5* |
| 5 | NIRONIT 521663 | 1.9 | 0.38 | 0.81 | 0.73 | 5.6 | *0.46* |
| 6 | NIRONIT 521994 | 0.5 | 1.9 | 1.7 | 1.5 | 4.5 | *0.5* |
| 7 | NIRONIT 124113 | 0 | 1.1 | 0 | 4.1 | 8.2 | *3* |
| | NIRONIT (Alab) 124113 | 0+/-22 | 4.89 | 0 | 5.37 | 14.6 | 1.7 |
| 8 | NIRONIT 211093 | 0 | *0.6* | 0 | *0.8* | 7.4 | *3* |
| | NIRONIT (Alab) 211093 | 0+/-11 | 2.46 | 0 | 0.37 | 14 | 0 |
| 9 | NIRONIT 528292 | 0 | *0.6* | 0 | *0.9* | 6.5 | *3* |
| | NIRONIT (Alab) 528292 | 0+/-22 | 2.22 | 0 | 0.67 | 9.69 | 0 |
| 10 | NIRONIT 832090 | 0 | 4 | 0 | 2.2 | 26 | *4* |
| 11 | NIRONIT 407156 | 0 | *0.6* | 0 | 4.8 | 32 | *2* |
| 12 | NIRONIT 528194 | 0 | *0.8* | 0 | 2.1 | 32 | 5 |
| 13 | NIRONIT 828660 | 0 | *1.4* | 0 | *1.5* | 335 | *4* |

We have determined that the highest radiopurity is achieved using a combination of commercially pure titanium (0.03% N, 0.1% C, 0.015% H, 0.2% Fe, 0.18% O, 99.475% Ti as per ASTM B265 required by the ASME BPVC) without added scrap material and cold hearth electron beam (EB) refining technology ("EB melting"). Cold hearth melting provides an important mechanism by which high-density contaminants are removed by gravity separation and settle in the cold hearth. By contrast, vacuum arc remelting (VAR) technology provides virtually no refining of the raw material. These two factors — no scrap and EB melting — are the key elements in the titanium production that assure a low level of radioactive contamination.

## 8.3 Cryostat Design

### 8.3.1 Material, Working, and Transportation Conditions

The baseline material for the LZ cryostat is commercially pure Ti, Grade 1 per ASME SB-265, with additional low-radioactivity background requirements as presented above. The design of the vessels complies with the following codes: ASME BPVC [19], 2012 Int. Building Code, and ASCE 7, with site soil classification Class B (Rock) for seismic conditions. The 2008 U.S. Geological Survey hazard data [20] for this location are: $S_S$ = 0.121 g, $S_{MS}$ = 0.121 g, and $S_{DS}$ = 0.081 g. The Seismic Design Force for LZ is 0.054g, which imposes a force of 5297 N at the center of mass of the cryostat during an event. The outer-vessel support and base fasteners are adequate to withstand this force; the inner-vessel tie rod supports need careful consideration when taking this force into account. A combination of reaction points will be incorporated into the vessel design to react to the vertical and horizontal forces.

The assembly of the cryostat underground assumes that the inner vessel along with its contents will be moved as an assembly down the Yates shaft at SURF. The width of the shaft is nominally 1.85 m, and the maximum payload width is 1.70 m to clear features in the shaft cross section and provide some margin of safety. The outer vessel may be moved down the Yates shaft in pieces.

The LZ cryostat operational temperatures (°C) and pressures (bar absolute) are summarized in Table 8.3.1.1.

**Table 8.3.1.1. Pressures and temperatures for cryostat operational conditions: normal, bakeout, and most severe failures.**

| Vessel | Pressure (bar absolute) | | Temperature (°C) | Condition |
|---|---|---|---|---|
| | Internal | External | | |
| Inner | ≤ 4.0 | Vacuum | -112 to 37 | Normal |
| | Vacuum | 1.01 | ≤ 100 | Bakeout (dry, no water in water tank) |
| | Vacuum | 1.48 | -112 to 37 | Failure Mode (water flooded between inner and outer vessels) |
| Outer | Vacuum | 1.48 | 0 to 37 | Normal |
| | Vacuum | 1.01 | ≤ 100 | Bakeout (dry, no water in water tank) |
| | 1.48 | 1.01 | 0 to 37 | Failure Mode (Xe gas leak between inner and outer vessels and no water in water tank) |

### 8.3.2 Baseline Vessel Design

The baseline vessel design is a conventional cylindrical geometry with ellipsoidal heads, as shown in Figure 8.3.2.1. The inner vessel is split once near the top head with a flange pair. To minimize the passive volume filled with LXe, the diameter of the inner vessel is tapered at its half-height and the bottom head has an ellipsoidal shape with a 3:1 aspect ratio (the other three heads have 2:1 aspect ratios).

To comply with ASME code, a stiffening ring has been located at the smaller aperture of the conical section of the inner vessel. This ring acts as a support line and therefore allows a thinner wall of the top section of the vessel, which can be reduced, i.e., from 8 to 6 mm. The outer vessel is split twice, once near the top head and again in midsection. Because the inner vessel is making full use of the Yates shaft cross section, it is impossible for the outer vessel to be conveyed together. With a three-piece design, the outer vessel can be transported down the Yates shaft in pieces and reassembled once in the Davis Cavern. Considerable thought was put into finishing fabrication of the outer vessel underground, but this idea was dropped in favor of adding a second flange pair. Welding and pressure-testing underground would add unnecessary schedule risk to the project.

For internal pressure, the thickness of the vessel walls in the cylindrical section is governed by a straightforward formula in the pressure vessel code. The minimum wall thickness is a function of the material, vessel diameter, pressure, and degree of inspection. Three materials were considered for this experiment. At the required internal pressure, the minimum wall thicknesses in the cylindrical section are listed in Table 8.3.2.1.

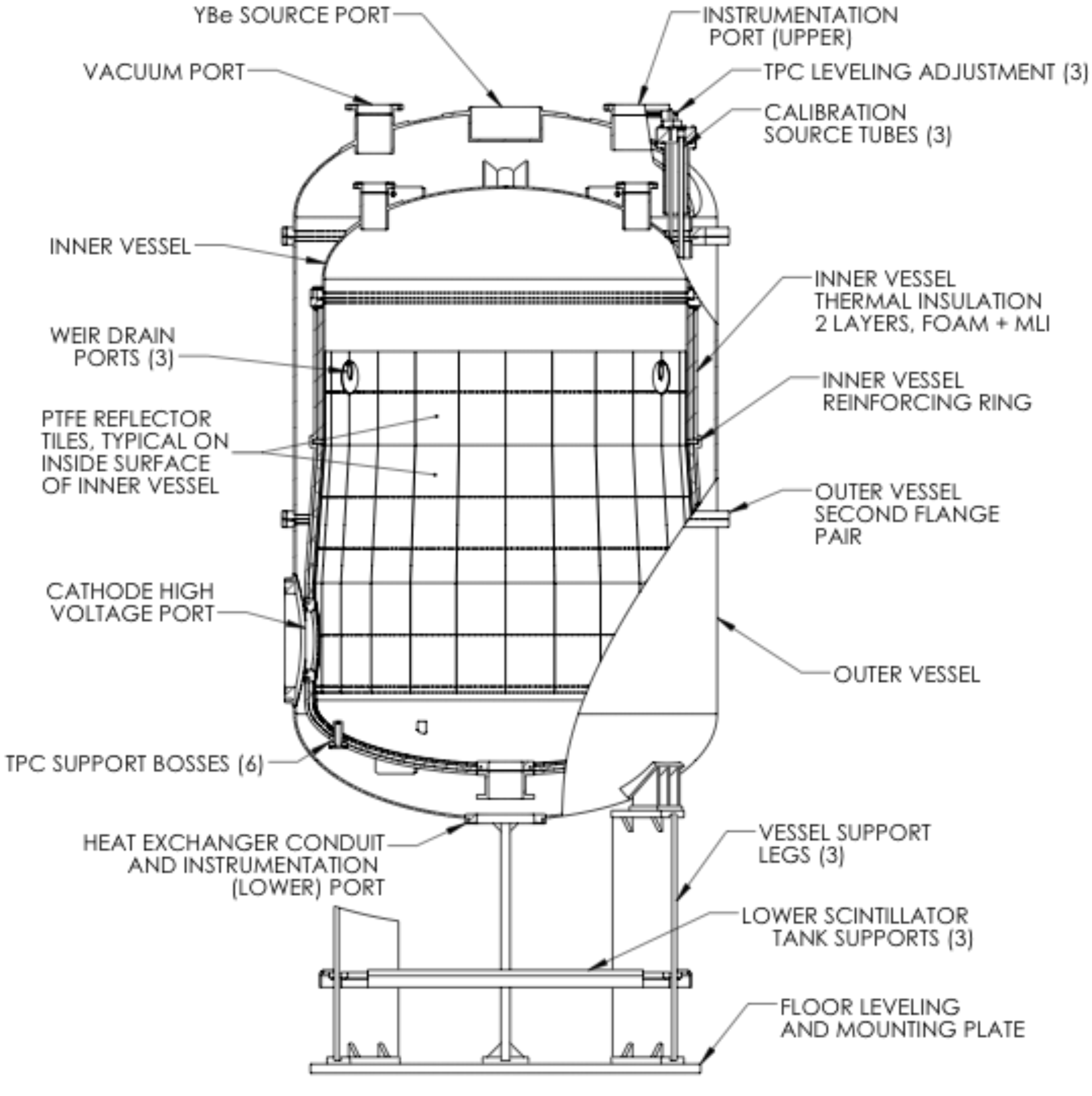

**Figure 8.3.2.1. View of the LZ cryostat. Main parts of the inner and outer vessel assembly are highlighted.**

**Table 8.3.2.1. Minimum wall thickness based on internal pressure and materials considered for the LZ cryostat.**

| Material | Inner Vessel Wall Thickness (mm) |
|---|---|
| **Ti, CP, Grade 1 (baseline)** | 5.5 |
| **Cu103** | 8.2 |
| **SS316L** | 3.3 |

A number of ports are necessary to carry fluids and electrical signals to and from the inner detector. These are added to the top and bottom heads, as well as a side penetration for cathode HV. The latter is discussed in a separate section of Chapter 6.

Buckling is an important failure mode to consider for vessels that see external pressure (vacuum in this case). The ASME BPVC specifies safe external working pressures based on material, temperature, wall thickness, diameter, and length. If the allowable external pressure is insufficient, a vessel designer has a couple of options. The first is to increase the wall thickness. In the case of LZ, this is undesirable for a number of reasons: Most notably, it creates more background radiation and reduces veto efficiency. The other option is to add reinforcing rings. Reinforcing rings essentially shorten the length of the vessel from a buckling perspective. Because the outer vessel is already segmented for transportation reasons, this makes a natural reinforcement against buckling.

Comparing the values for internal pressure 4 bar versus external pressure 1.48 bar, it is evident that the vessel design is driven by external pressure. It should also be noted that the minimum wall thickness is the minimum as-built, not the nominal. During the head-forming process, for instance, flat material is drawn or spun into shape, and in that process thinned from its original thickness. It should also be remembered that material is commercially available in discrete increments as opposed to infinitely variable thickness. Total mass and minimum thicknesses required by the ASME code for each segment of the inner and outer vessel made of Ti and SS are summarized in Table 8.3.2.2

To minimize the amount of LXe between the TPC and the inner cryostat, the shape of the inner vessel is tapered at its half-height. Studies of the electric-field distribution show that the electric field is below the maximum allowed value of 50 kV/cm.

The outer vessel will be sealed (twice, since there are two flange pairs) with differentially pumped double Viton O-rings. This is the most reliable and cost-effective sealing solution for this large room-temperature application. The inner vessel will be sealed with a sprung metal C-seal because at the experimental temperature, O-ring seals with typical elastomeric materials are not suitable. Additionally, the large diameter prohibits the use of a knife-edge flange. Smaller ports on the vessels will be sealed with either a sprung metal C-seal or traditional knife-edge flange. Commercially pure Ti itself is too soft to support a knife-edge feature, but the smaller ports can make use of readily available explosion bonded flanges that feature an SS face and Ti back.

**Table 8.3.2.2. Vessel-wall thicknesses (mm) and total mass (kg) imposed by the external pressure at the normal, bakeout, and failure conditions.**

| | Inner Vessel | | | | | | Outer Vessel | | | |
|---|---|---|---|---|---|---|---|---|---|---|
| Material | Top head (mm) | Upper wall (mm) | Conical section (mm) | Lower wall (mm) | Dished end (mm) | Total mass (kg) | Top head (mm) | Wall (mm) | Dished end (mm) | Total mass (kg) |
| **Titanium** | 8 | 6 | 8 | 8 | 12 | 736 | 8 | 8 | 8 | 1091 |
| **SS316L** | 6 | 5 | 6 | 6 | 9 | 1033 | 7 | 7 | 7 | 1844 |

Inner-vessel flanges with knife-edge or C-seal gaskets for cryogenic service will also feature a secondary O-ring seal to facilitate room-temperature leak detection.

### 8.3.3 Leveling System

Leveling will be done in two stages, with a different degree of precision required for each. The first stage entails surveying and shimming of the outer vessel to level it to on the order of 1:1000. This will be done during installation of the outer vessel and its support legs or base. At this time, it will be convenient to add shims between the legs and the bottom of the water tank, or between the legs and the bottom of the outer vessel. All of the parts above will be manufactured nominally parallel, but obviously there will be some manufacturing variation to mitigate in order to get the gate and anode grids parallel to the liquid/gas interface. Additionally, the result of hydrostatic forces and thermal contraction on the internal parts means that some fine-tuning of the level will be required after the vessel is cooled and filled with LXe, despite a best effort on the assembly when warm and dry. For this second leveling stage, an adjustable three-point suspension system will be used to fine-tune the level of the inner cryostat en masse, much like the method used successfully in LUX.

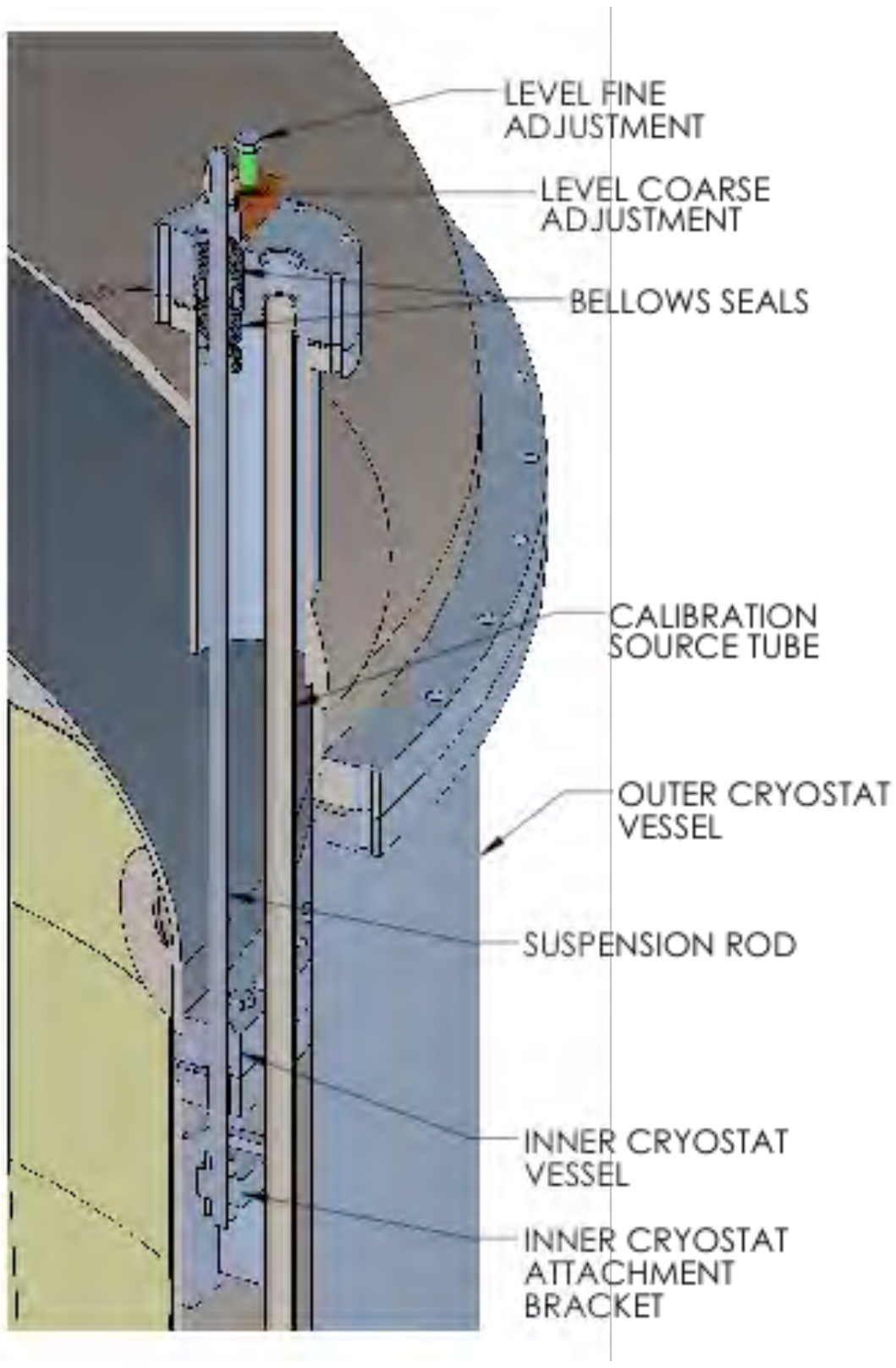

**Figure 8.3.3.1. Model of the suspension system for the inner vessel leveling.**

The LZ suspension rods feature a double bellows seal for both reliability and compliance with the as-built geometry of the vessels. The suspension system is accessible through the water and outer-detector veto tanks. The suspension rods are long enough, and small enough in diameter, to prevent significant heat transfer via conduction. A conceptual model of the suspension is shown in Figure 8.3.3.1.

### 8.3.4 Calibration Tubes

Calibration tubes surround the detector in three vertical locations, allowing suitable scans of the TPC with external sealed sources. Titanium (CP-1) or oxygen-free-copper fabricated calibration tubes are equally spaced and straddled across the HV port, and reside in the vacuum space between the inner and outer vessels of the cryostat  (see Figures 8.3.2.1 and 8.3.3.1). The calibration source tube enters the vacuum space via a sealing boss directly above the tubes, on the outer vessel dished head, ensuring source delivery simplicity. The source tubes run vertically from the top of the outer vessel dished head to the bottom of the TPC. The size of the source tubes has been optimized to allow use of many types of calibration sources, i.e., AmBe, small neutron generator, or gamma sources. The source tubes are evacuated to avoid further radon contamination.

### 8.3.5 Cathode High-voltage Port for the Inner and Outer Vessel

The HV port provides an interface for the HV connection and the TPC. The inner-vessel port design must be able to sustain a local field strength of up to 50 kV/cm. Both inner- and outer-vessel ports are manufactured from the vessel parent material and will require soft-metal sealing technology in the form of a sprung C-seal operating at better than $10^{-9}$ mbar·l/s. The detector vessel supports one 254-mm-inner-diameter HV port with a suitable flange arrangement and a large aperture for the outer vessel, enabling installation of the vacuum jacket (VJ), promoting ease of assembly, and ensuring integrity of seals, as shown in Figure 8.3.2.1. Both ports are designed to ASME BVPC, Division 1, and will be made from a single metal block using a computer-controlled milling machine.

### 8.3.6 Cryostat Base

The cryostat base supports the mass of the cryostat and the TPC and also provides a point of attachment to the water-tank floor, thus overcoming any buoyancy effects. The base is attached to the pre-aligned mounting plate, which is anchored to the floor, according to the local civil engineering requirements. This mounting plate defines the apparatus datum; all subsequent surveys will refer to the center point, X:0,Y:0,Z:0. As shown in Figure 8.3.2.1, the support base comprises a three-fin leg design with cross braces, minimizing the number of lower veto tanks required, and providing greater veto efficiency and reduced cost. Veto tanks rest and clamp to a structural support shelf. The base is leveled with respect to the ground mounting plate, and fine adjustments are made by shim plates at the base vessel interface; hence, the cryostat is aligned correctly.

### 8.3.7 TPC Support Bosses

The interface between the TPC and inner vessel will be six bosses on the bottom head of the inner vessel, as shown in Figure 8.3.2.1. The lower PMT array mounting plate will have mating features that will include a lead-in feature and nut plates so that when the TPC assembly is lowered into the inner vessel, screws can be inserted through the outside bottom head of the inner vessel to secure the TPC assembly. After securing the TPC, the bosses, which from the outside are essentially small ports, will be blanked off in the same manner as all round ports on the vessels. The mating surface of the bosses and the screw holes will be machined after welding to the bottom head to provide the most accurate interface, and to keep them concentric with the inner vessel, mutually co-planar, and perpendicular to the inner vessel axis.

The load on these bosses varies at different stages of assembly, transportation, and operation. During assembly, the load is vertical and in compression; during underground transportation with a horizontal vessel axis, the bosses are designed to handle the shear loads and movement of a horizontal TPC (supports at the top of the TPC ensure that the TPC is not entirely cantilevered off these bottom bosses). In operation, the load is again vertical but this time in tension, as the TPC is buoyant in the very dense LXe. The load and the resulting stress in the bottom head during operation are dominated by pressure, not the TPC bosses. The bosses are far enough away from the highest stress area of the bottom head "knuckle" to be of little consequence.

### 8.3.8 PTFE Reflector Coating

Photomultiplier tubes (PMTs) in the skin region (outside the TPC but inside the inner vessel) are part of the LZ veto strategy. Polytetrafluoroethylene (PTFE) is required on the inside surfaces of the inner vessel to efficiently reflect light generated by the scintillation of LXe in this skin region. Without PTFE, light that strikes the walls of the inner vessel would be absorbed and never reach the PMTs to be counted.

PTFE is used very successfully as a reflective material in LUX, and no other dielectric material is known to be as efficient in the VUV. Several other polymers in the Teflon "family," ETFE (ethylene tetrafluoroethylene) and PFA (perfluoroalkoxy), for instance, would be desirable because of their manufacturing properties but these materials are not yet proven for this application. It would be very convenient, for instance, to coat the inner vessel in the manner of typical household cookware, but this method is not possible with PTFE, and the resulting coating is too thin. Crucially, small dielectric particles must be avoided at all cost near HV areas of the TPC, and therefore a thicker PTFE material is preferred. It has been shown that thicker films, of order of millimeters rather than micrometers, are required for maximum VUV reflectivity [21]. PTFE has a coefficient of thermal expansion that is an order of magnitude greater than typical metals. The material shrinks approximately 1.45%, going from room temperature to the working temperature in the detector. A 1.5-m part would therefore shrink about 2 cm on cooldown, so if the PTFE lining in the vessel was a continuous part, it would pull away substantially (about 1 cm radially) from the walls of the vessel when cooling. With this in mind, the only way to mitigate the difference in thermal expansion is to tile the inside surfaces of the vessel with discrete parts that overlap. Individual tiles would be attached with a Teflon central screw, and in cases where the tile is substantially rectangular as opposed to square, there would be a central screw with a round hole and a second screw with a slotted feature in the tile. The Teflon screws would thread into nut plates welded to the inner surface of the vessel wall. An example of this concept is shown in Figure 8.3.2.1.

### 8.3.9 Thermal Insulation

It is important to limit the heat transfer between the inner vessel and the environment to minimize the amount of refrigeration needed underground. In addition, reducing heat transfer helps to prevent unwanted convection currents in the LXe fluid. The vacuum between nested inner and outer vessels essentially eliminates thermal conduction and convection, in typical Dewar fashion. The dominant mode of heat transfer is therefore radiation, and with a large surface area (~16 square meters) and a large temperature difference, it is potentially substantial. Multilayer insulation (MLI or superinsulation) is proposed as the baseline solution to reduce this thermal load during normal operation. MLI is a well-known insulating material for in-vacuum cryogenic service. The thermal load with and without this material varies by approximately an order of magnitude. In this case, the expected heat load with bare vessels would be several hundred watts, and with MLI several tens of watts. In this experiment, the proposed amount of total refrigeration is about a kilowatt, so MLI is the clear choice.

MLI works well during normal operation, but is ineffective in the event of a failure in which a gross amount of liquid water enters the volume normally occupied by vacuum between the vessels. To mitigate this failure mode, a closed-cell polyurethane foam will be applied to the outer surface of the inner vessel anywhere it is in contact with LXe (basically everywhere below the main seal flange). The proposed foam thickness is 2 cm over the bottom head, and 1 cm over the remainder. MLI will be wrapped over the foam, and covers the entire inner vessel and its cold appendages. With the foam in place, the maximum heat transfer in this failure mode is expected to be 3600 W, which corresponds to a Xe boil-off of 450 standard liters per minute (slpm). This rate is within the Xe-recovery capacity of the system.

# Chapter 8 References

# 9 Xenon Handling and Cryogenics

## 9.1 Overview

The Xe handling and cryogenics strategy of LZ derives strongly from the successful LUX program and takes maximum advantage of key technologies demonstrated there: chromatographic krypton removal, high-flow online purification, high-efficiency two-phase heat exchange, sensitive in situ monitoring of Xe purity, and thermosyphon cryogenics. In many cases, the primary technical challenge of the LZ implementation is how to best scale up from LUX.

Two classes of impurities are of concern: electronegatives such as $O_2$ and $H_2O$ that suppress charge and light transport in the TPC, and radioactive noble gases such as $^{85}Kr$, $^{39}Ar$, and $^{222}Rn$ that create backgrounds that are not amenable to self-shielding. Both types of impurities may be found in the vendor-supplied Xe, and both may also be introduced into the detector during operations via outgassing and emanation. The purity goals are summarized in Table 9.1.1.

We have three mitigation techniques to control these impurities. First, electronegatives are suppressed during operations by continuous circulation of the Xe in gaseous form through a hot (450 $^{o}$C) zirconium getter. Gettering in the gas phase is made practical by a two-phase heat exchanger. A metal-diaphragm gas pump drives the circulation and the system is capable of purifying the entire Xe stockpile in 2.3 days.

Second, the problematic radioactive species, $^{85}Kr$ and $^{39}Ar$, which have relatively long half-lives (10.8 years and 269 years, respectively) and no supporting parents, can be removed from the vendor-supplied Xe prior to the start of operations. LZ will perform this removal with a high-throughput chromatography system closely modeled on the Case Western system that was successfully employed by LUX. Due to underground space considerations and other resource constraints, the online Xe purification system will not have noble-gas removal capability, so it is necessary to exercise careful control of the Xe after the chromatographic processing so that $^{85}Kr$ and $^{39}Ar$ are not subsequently re-introduced.

Third, a careful program of screening and monitoring is integrated into the LZ design. We rely primarily on an improved version of the coldtrap/mass-spectrometry monitoring system that has been successfully applied to LUX [1,2]. Radon and krypton emanation screening is also employed, as described in Sections 12.4.1 and 9.7, respectively.

The outline of this chapter is as follows: We describe the chromatographic krypton removal system in Section 9.2. The online purification system is described in Section 9.3. The recovery system, including emergency recovery, is described in Section 9.4. Xenon storage and transportation is described in Section 9.5. Xenon sampling and assay is described in Section 9.6. Section 9.7 describes the outgassing model that informs our purification strategy. Cryogenics infrastructure is described in Section 9.8, and Xe procurement is discussed in Section 9.9.

**Table 9.1.1. LZ Xe purity goals.**

| Quantity | Specification |
|---|---|
| Natural krypton concentration | < 0.015 ppt (g/g) (live-time averaged) |
| Air content (krypton equivalent) | < 40 standard cc |
| $^{222}Rn$ decay rate in 7 tonnes active volume | < 0.67 mBq |
| Charge attenuation length | > 1.5 meters |
| Equivalent $e^-$ lifetime at 2 mm/μs | > 750 μs |
| Equivalent $O_2$ concentration | < 0.4 ppb (g/g) |

## 9.2 Krypton Removal via Chromatography

Commercially available research-grade Xe typically contains trace krypton at a concentration of ~(10–100) ppb[1] (parts per billion). $^{85}$Kr is a $\beta$ emitter with an isotopic abundance of ~2 × 10$^{-11}$, an endpoint energy of 687 keV, and a half-life of 10.8 years. To reduce the rate of $^{85}$Kr ER backgrounds to 10% of the *pp* solar neutrino ER rate, we require that the total concentration of natural krypton be no more than 0.015 ppt (parts per trillion), a factor of ~(0.7-7) × 10$^{6}$ below that of research-grade Xe. Since the krypton is dissolved throughout the Xe, $^{85}$Kr decays cannot be rejected via self-shielding, nor does the getter remove krypton during in situ purification due to its inert nature. Gas vendors have indicated that they could supply Xe with a krypton concentration as small as 1 ppb, however this would still far exceed the LZ background goal.

Trace argon is also a concern due to the presence of the $\beta$ emitter $^{39}$Ar (endpoint energy of 565 keV and half-life of 269 years), and we require that the background rate due to $^{39}$Ar be no more than 10% of that of $^{85}$Kr. Due to the low isotopic abundance of $^{39}$Ar (8 × 10$^{-16}$) [3], this implies an argon concentration requirement of 4.5 × 10$^{-10}$ (g/g), substantially less demanding than the krypton requirement. Furthermore, because argon is amenable to the same removal techniques as krypton, no special measures are required to satisfy the argon goal.

Two methods exist to separate krypton and xenon: distillation [4-6] and chromatography [7]. Both are technically challenging owing to the similar physical and chemical properties of these atomic species. During the XENON10 and LUX programs, the Case Western group developed a gas chromatographic method based on a charcoal column and a helium carrier gas [7]. Using this method, the Case group processed the 400 kg of LUX Xe at a rate of 50 kg/week down to a krypton concentration of 4 ppt, exceeding the LUX goal of 5 ppt and rendering this background subdominant to the leading background by a factor of 10 during the 2013 LUX WIMP search [8]. This is the lowest krypton concentration achieved by any LXe TPC to date. In addition, one 50 kg batch was processed twice for LUX, resulting in an upper limit on the concentration of <0.2 ppt, merely a factor of 13 short of the ultimate LZ goal. As described in Section 9.6, greater screening sensitivity is being developed to determine the ultimate krypton-removal limit; however, we note that a concentration as high as 0.2 ppt would result in only a modest relaxation of the LZ WIMP sensitivity (equaling the rate of *pp* solar neutrino ER events).

Chromatographic separation relies on the weaker van der Waals binding of krypton to carbon relative to Xe, which causes krypton atoms to flow through the charcoal column more rapidly than Xe. This property allows for a separation cycle by feeding Xe (with trace krypton) into the column at a fixed rate and duration under the influence of a fixed flow of helium carrier gas. The rates and durations are tuned so that the last of the krypton exits the column prior to the earliest Xe. This permits the krypton to be purged by directing the krypton-helium stream to a coldtrap that captures the krypton. This feed-purge cycle is followed by the Xe recovery cycle, in which pumping parameters are altered to accelerate the rate at which the Xe exits the column. Figure 9.2.1 illustrates this process by showing the time profiles for the two species as they exit the column. As the purified xenon-helium stream exits the column, the Xe is separated from the helium using a cryogenic condenser. After a number of feed-purge-recovery cycles, the helium is pumped away, the condenser is warmed, and the Xe is transferred to a storage cylinder. The purity of the Xe is measured using the coldtrap/mass-spectrometry technique developed by the Maryland group [2] and described in Section 9.6.

**Table 9.2.I. LZ krypton removal goals. The $^{85}$Kr isotopic abundance is assumed to be 2 × 10$^{-11}$.**

| | LZ Requirement | Unit |
|---|---|---|
| Nominal Kr concentration of vendor-supplied Xe | ~(0.1-1) × 10$^{-7}$ | g/g |
| Allowed Kr concentration | 1.5 × 10$^{-14}$ | g/g |
| Required Kr rejection factor of removal system | ~(0.7-7) × 10$^{6}$ | |
| Processing rate of removal system | 200 | kg/day |
| Allowed Kr mass in 10 tonnes Xe | 1.5 × 10$^{-7}$ | g |
| Air allowed in 10 tonnes of Xe (from Kr spec.) | 40 | std. cc |
| Allowed $^{85}$Kr decay rate in 6 tonnes fiducial Xe mass | 0.026 | mBq |
| $^{85}$Kr decays in 1,000 days in 6 tonnes fiducial Xe mass | 2283 | events |
| After energy cut (ΔE = 5 keVee) | 26 | events |
| After ER discrimination (99.5%) | 0.13 | events |

[1] All concentrations are quoted in grams of natural Kr per gram of Xe.

In addition to purity requirements, the processing rate and the Xe acquisition schedule influence the krypton-removal system. Owing to the forecast for the market price of Xe (See Section 9.9), we expect to purchase the bulk of the Xe at the latest possible date, requiring a rapid removal of the krypton to begin experimental operations on time. Therefore in Figure 9.2.2, we show an expanded version of the LUX system capable of processing 200 kg of Xe per day, or 10 tonnes in 60 days at 85% uptime. This design has better isolation between the chromatography loop and the recovery loop than did the LUX design, which achieved $3 \times 10^4$ reduction per pass and was likely limited by cross-contamination. While we may expect some improvement of the krypton rejection factor as the system is tuned during initial operations, we conservatively plan for two complete passes of the Xe (120 days), which is more than sufficient to reduce the krypton concentration in the raw stock by a factor of $10^7$ to ~0.01 ppt. Fast processing also provides schedule insurance in the event of an unintentional contamination event. We note that the introduction of just 40 $cm^3$ of air at standard temperature and pressure (STP) corresponds to 0.015 ppt of krypton in 10 tonnes of Xe. A single pass through the chromatographic system could easily correct an error as large as hundreds of liters. (The corresponding requirement for argon is less demanding [270 $cm^3$ of air at STP].)

As illustrated in Figure 9.2.2, the three principal processing stages are: (1) chromatography to separate Xe and krypton, and capture the krypton in a cooled charcoal trap; (2) recovery of the Xe from the column into a cooled condenser to strip the Xe from the carrier gas; and (3) storage and analytical sampling of the purified Xe into DOT-rated cylinders suitable for transport to the experiment. The system architecture follows two general principles. First, we minimize the components exposed to both the raw and purified Xe streams to minimize cross-contamination, which leads to the two mostly separate loops in the figure (green and blue). Second, the arrangement matches the feed-purge cycle time to the Xe recovery time, so that by alternating between the two columns, we achieve twice the duty cycle of single-column operation. Two condensers are also used so that one is always available in the recovery path when the other is in the storage phase. The double-swing system with two columns and two condensers gives a factor-of-4 increase in production rate relative to a single-column-single-condenser system at modest additional cost and complexity.

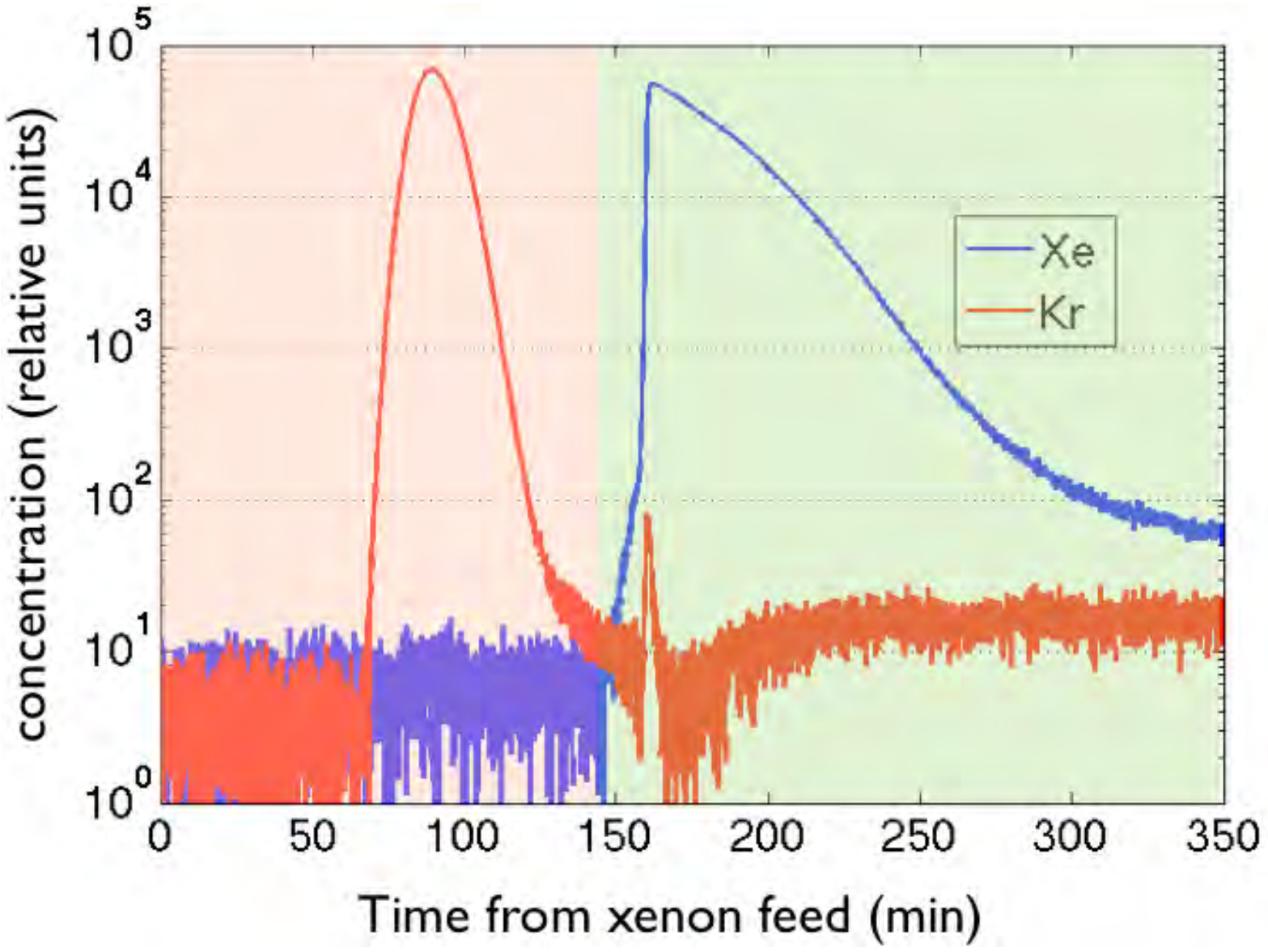

**Figure 9.2.I. Test data taken with the LUX Kr removal system illustrating the time profiles for krypton and xenon as they exit the column for a sample of Xe spiked with approximately 1% krypton. Data are acquired with a sampling RGA. During production, the ~100 ppb krypton in the raw Xe is below the RGA sensitivity, so spiked samples are used to tune the chromatography parameters.**

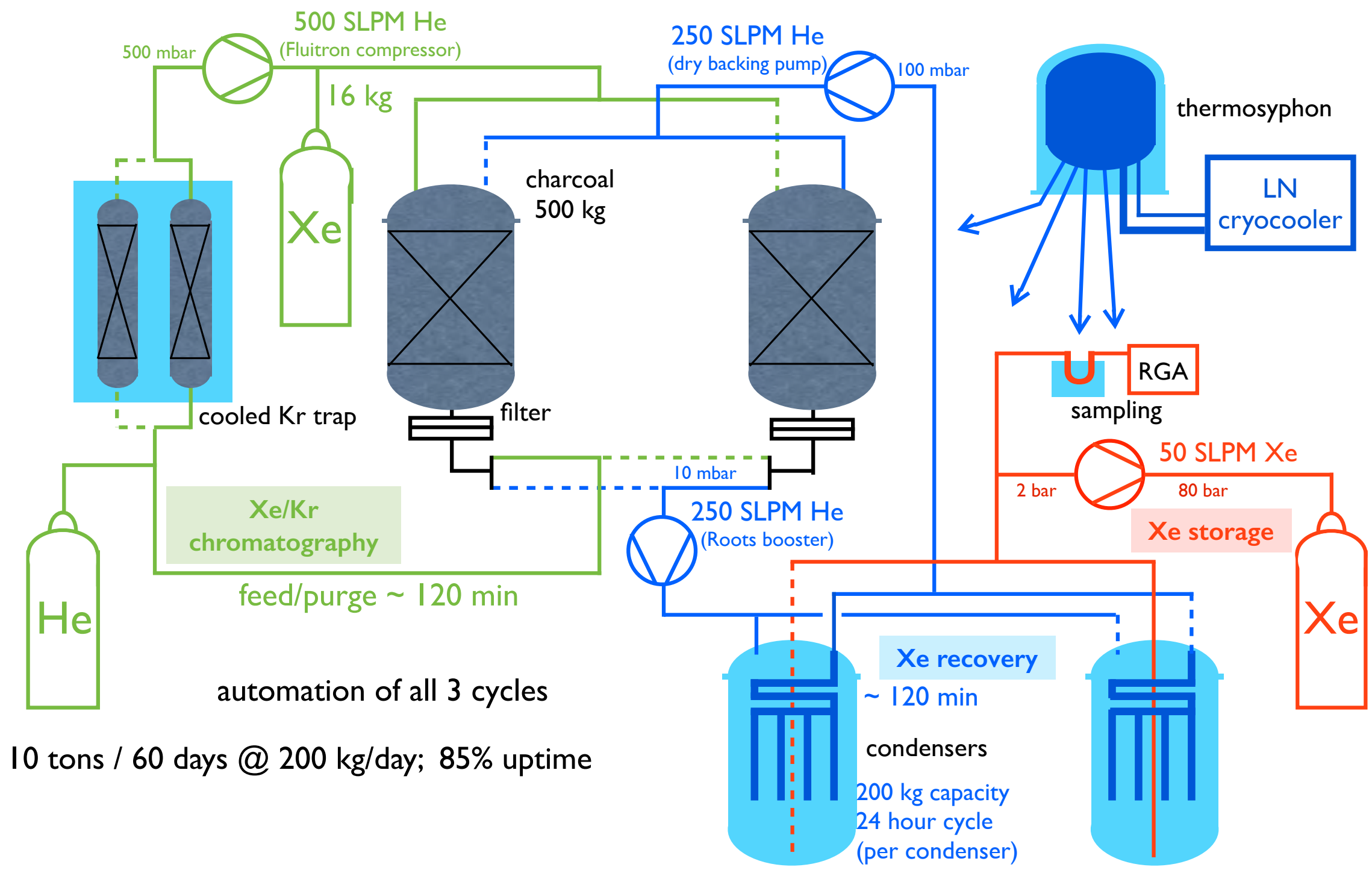

**Figure 9.2.2. The main components and flow paths of the LZ krypton-removal system. Colored lines indicate the path for a given process. The green line shows the left-hand charcoal column in the chromatography feed-purge separation cycle with krypton being collected in the right trap; a second trap is valved off for cleaning (dashed green line). The blue line shows the subsequent part of the cycle in which Xe is being recovered from the column and collected in the condenser. The red line shows Xe from the condenser being stored and analyzed. Key pressures and flow rates are indicated. The light-blue background indicates components that are cooled by the LN thermosyphon system.**

The overall processing rate is governed by the differential transit time of the krypton and Xe in the column, whereas all other components are sized to match this rate. In the LUX system, each feed cycle consisted of a 2 kg slug of Xe into a column of 60 kg of charcoal at 50 slpm (standard liters per minute) helium flow rate. The krypton purge was completed 120 minutes after the start of the feed, at which time recovery commenced. The ultimate purification rate is limited either by cross-contamination or by the exponential tail of residual krypton remaining in the column. Straightforward scaling to reach 200 kg per day calls for a 16-kg slug every two hours, which in turn requires an approximately eightfold scaling in charcoal mass and flow rate. Some further tuning will be required based on further studies of chromatography and cross-contamination using the LUX system. The remaining parameters, as shown in Figure 9.2.2, are appropriately scaled from the existing system.

The system architecture uses computer-controlled pneumatic valves for changing all of the process states between the various cycles. Pumps, mass flow controllers, heaters, and diagnostic sensors (pressure, temperature, etc.) are controlled by this system as well. We are investigating the use of programmable logic controllers (PLCs) for process control, as they would reduce latency and improve reliability compared with the proprietary LUX software used previously.

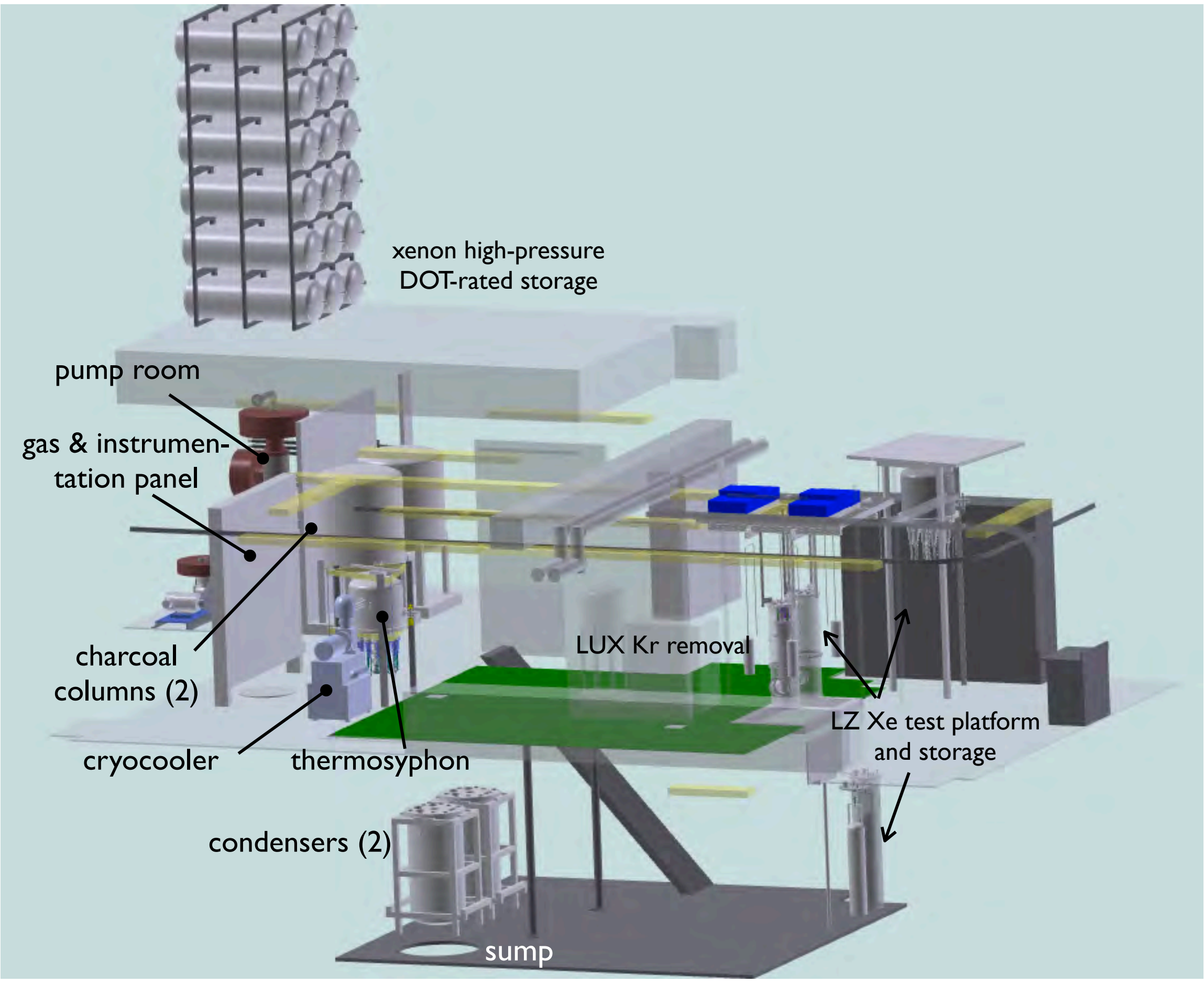

**Figure 9.2.3. A preliminary layout of the LZ krypton-removal system. The main floor, which is the middle level in the figure, is approximately 1,000 square feet and is adjoined below by a 300-square-foot "pit" and above by a high-bay area at grade level with external doors to the building's driveway. The main floor holds the primary control panels, pumps and compressors, the thermosyphon cooling system, traps, and auxiliary equipment. The pit contains the condensers, since they must be gravity-fed by the thermosyphon cooling loops.**

The krypton-removal system has three cryogenic elements that are cooled by a liquid nitrogen (LN) thermosyphon system, which is now a well-established technology that allows for automated temperature control. These components are highlighted by a blue background in Figure 9.2.2 and consist of the condensers, the krypton traps, and the sampling trap. All three cycles are fully automated, as well as krypton trap cleaning and analytic sampling, minimizing the chance of operator error.

During processing, several hundred kilograms of Xe will be in various parts of the system. In the event of an unexpected loss of cooling power, we rely upon a gas compressor backed up by a diesel generator to ensure that no Xe is lost due to warming. The control systems also have redundant power.

High-flow-rate pumps for circulating the carrier gas during chromatography are being investigated. The most likely solution is an all-metal diaphragm compressor manufactured by Fluitron. This technology is of interest because models at the lower flow rate of the series (100 slpm) are well suited for testing in the LUX system and/or LZ test platform, and models at higher flow rates are well suited to the production system. The same technology could also be utilized for the LZ online purification system as described in Section 9.3, with accordant benefits of working with a single vendor that can also meet our purity requirements. A 100-slpm unit has been received and is being prepared for use in the LZ test platform as well as for use in krypton removal to ensure that purity is not compromised.

In Figure 9.2.3, we show a physical layout of the major system components, including 10 tonnes of Xe storage. The system will be installed in Building IR2 at SLAC, where sufficient overhead space is available. We are investigating the possibility of storing the Xe stockpile at modest overburden during Kr removal to limit the cosmogenic production of $^{127}Xe$, which decays via electron capture with a half-life of 36 days and is expected to be present at a level of 0.25 mBq/kg in secular equilibrium at sea level.

## 9.3 Online Xenon Purification System

This section describes the LZ online Xe purification system, which is intended to control the concentration of electronegative impurities such as $O_2$ and $H_2O$ through continuous purification. Such species must be suppressed below 0.4 ppb $O_2$ equivalent to allow for adequate charge and scintillation transport in the TPC (see Table 9.1.1).

The centerpiece of the LZ online purification system is a hot zirconium getter, a purifier technology capable of achieving >99% one-pass removal efficiency for virtually all non-inert impurities of interest, including hard-to-remove species such as $N_2$ and $CH_4$ [9]. Impurities chemically bond to the surface of zirconium pellets, irreversibly removing them from the Xe. The getter operates at elevated temperature (450 $^{o}$C) to allow the captured impurities to diffuse into the bulk of the getter pellets, leaving the surface free for additional gettering. The getter material must be replaced when it becomes saturated with impurities. Besides being highly effective purifiers, zirconium getters are also notable for not emanating substantial quantities of radon, a basic requirement for any detector component that is continuously exposed to the LZ Xe.

The Xe circulates through the purification system at a rate of 500 slpm (3 kg/min), a value that was chosen based upon previous experience with the LUX system and based on economic and space constraints. At this flow rate, the 10 tonnes of Xe can be purified in 2.3 days, comparable to the 1.7-day turnover time of the LUX recirculation system. As described in Section 9.7, the recirculation rate and the charge attenuation goal constrains the allowed outgassing rate of the detector, and this drives the outgassing plan of the detector. Prior experience with LUX and other liquid noble detectors indicates that the scintillation absorption length goal (>15 meters) is less demanding than the charge attenuation length goal and will be satisfied by the same requirements.

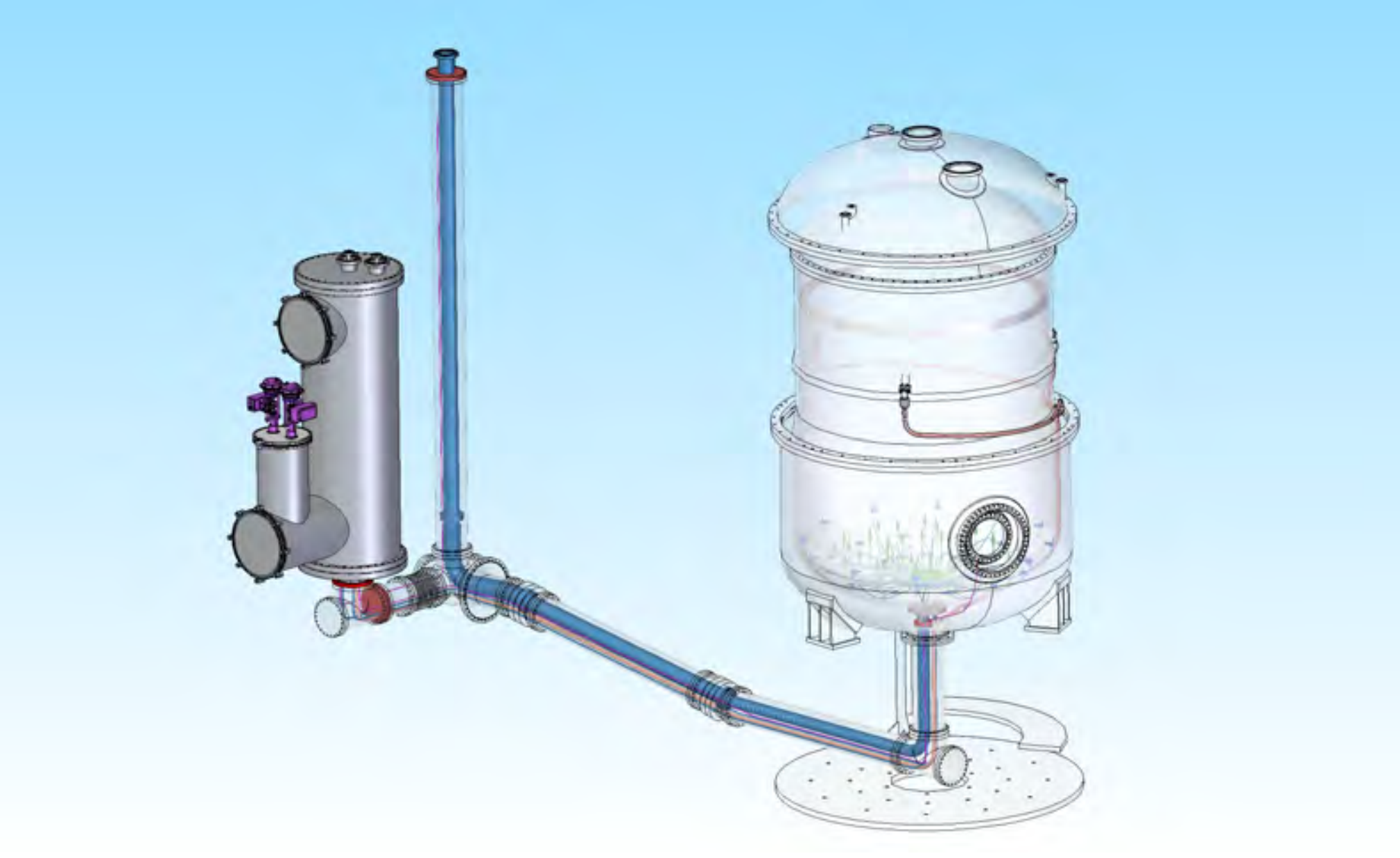

**Figure 9.3.1. The LXe tower (left), PMT standpipe (middle), and the cryogenic transfer lines (lower center) and their relationship to the detector (right).**

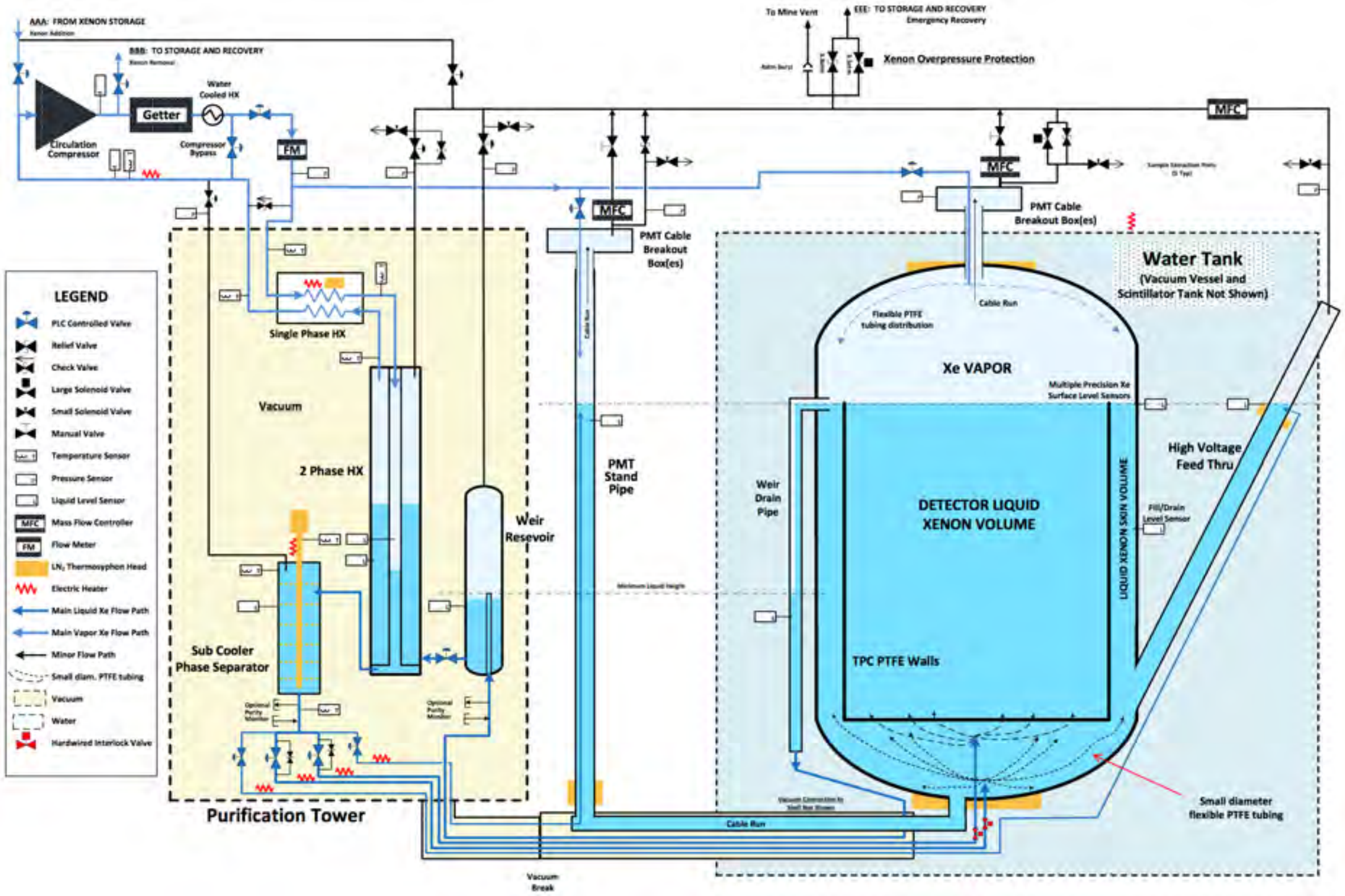

**Figure 9.3.2. LXe flow block diagram. At right is the LZ TPC, with its angled HV conduit and its weir drainpipe controlling the liquid level in the TPC. The dashed box surrounding the TPC indicates the water shield. Liquid xenon is supplied to the TPC and collected from the weir drainpipe through the horizontal transfer lines at the bottom of the detector. Cables from the lower array of the TPC and other instrumentation are delivered to a room-temperature feedthrough at the top of the PMT standpipe in the middle of the figure. At left is the LXe tower, containing the weir reservoir, the two-phase heat exchanger, the gas-gas heat exchanger, and the subcooler / two-phase separator. Liquid xenon returns to the detector through four circuits, each of which is regulated by a remotely operated valve. Above the liquid system is the gas pump and zirconium getter. Purge-gas lines also feed into the gas pump from each of the conduits.**

Because the hot getter operates on gaseous Xe only, the purification system must vaporize and re-condense the liquid to create a continuous purification circuit. This process is made thermally efficient by a two-phase heat exchanger developed for LUX [10]. In LZ, the two-phase heat exchanger is located outside the TPC in a separate cryostat known as the LXe tower, as shown in Figure 9.3.1. This choice simplifies the TPC design and relaxes the space constraints and radioactivity requirements on the heat exchanger and its associated support hardware. It also permits separate checkout and commissioning prior to its installation underground. The Xe is transferred in liquid form between the tower and the detector through vacuum-insulated transfer lines that penetrate through the wall of the water shield and connect to the bottom of the TPC vessel.

The primary purification loop is shown schematically in Figure 9.3.2 and operates as follows: Xenon is removed from the detector at 500 slpm (3 kg/min) primarily as a -95 $^{o}$C saturated liquid at about 1.0 l/min from the weir trough near the top of the TPC (see Chapter 6). The collected liquid exits the cryostat, flows to the LXe tower through the transfer lines, and is evaporated in the two-phase heat exchanger by absorbing heat from the warm Xe gas returning from the getter. An additional gas-gas heat exchanger warms the gas to room temperature. These two heat exchangers substantially reduce the large thermal load that would otherwise be required to vaporize and recondense the Xe (2 × 4.7 kW watts at 500 slpm for the phase change alone).

The Xe gas exiting the tower enters an all-metal triple-diaphragm compressor that drives the flow through the system. This compressor uses an electric motor to drive a single-piston oil pump that pressurizes the bottom diaphragm, transmitting force to the top diaphragm to compress the gas. A middle diaphragm with grooves creates a space that is monitored for leakage of either oil or Xe. The all-metal design (including a metal seal) minimizes the ingress of krypton and radon into the Xe. The compressed gas is preheated to ~300 $^{o}$C before passing into the heated getter (model PS5-MGT150 from SAES). The preheating of the gas ensures that it does not substantially cool the getter. The Xe gas entering and exiting the getter passes through an additional gas-gas heat exchanger for thermal efficiency. The purified gas returns to the LXe tower, where it recondenses in the two-phase heat exchanger. The gas plumbing for the circulation is pre-cleaned stainless steel (SS) tubing connected with orbital welds wherever possible. VCR fittings with metal seals will be used where necessary to open a connection. Valves will be of the bellows-seal type to keep the system hermetic.

The temperature difference between the counterflowing vaporizing Xe and condensing Xe streams drives the heat transfer. $LN_2$ thermosyphon cooling makes up the difference in heat-exchanger efficiency and the various internal and external heat loads. By returning the LXe in a subcooled state 1 $^{o}$C above the freezing point to ensure no disruptive bubbles or boiling, we need 270 W of net cooling to the cryostat at 500 slpm.

In addition to the LXe circulation loops described above, two secondary gas-only recirculation loops act on the conduits that house the cables for the PMTs and instrumentation cables (from Xe vessel top and the PMT standpipe). These cables are immersed in LXe on one end, but they penetrate the liquid surface and terminate at room-temperature feedthroughs filled with Xe gas at the other end. The nonmetal components of these cable bundles are a potential source of problematic outgassing, particularly the warm ends where the diffusion constants of the insulating plastics are the largest. We manage this outgassing by purging these conduits with a continuous flow of Xe gas away from the TPC at a modest flow rate. This purge gas flow merges at the gas pump inlet with the primary circulation flow exiting the LXe tower and returns back to these conduits, bypassing the LXe tower. A gas-gas heat exchanger minimizes the required heat input and output for this gas circuit.

The purge-gas flow rate required to ensure that back-diffusion is negligible is characterized by the unitless Peclet number, $P = VL/D$, where $V$ is the linear velocity of the gas in the conduit, $L$ is the diffusion distance of interest, and $D$ is the diffusion constant of the impurity species. For a 4-inch-diameter conduit, a gas flow rate of 0.5 slpm per conduit, a diffusion distance of 10 cm, and a diffusion constant of 0.086 cm$^2$/sec (valid for $O_2$ diffusion in Xe gas at room temperature), the Peclet number is 8.2, indicating that back diffusion is negligible. We plan for a total of 10 slpm of flow across all conduits to provide for additional safety margin.

The circulation system must also accommodate detector calibration and purity assays. As discussed in Chapter 10, radioactive sources such as $^{83m}$Kr and tritiated methane will be introduced into the Xe to calibrate the central regions of the TPC, and the online purification system has a valve and port system to allow for this. Due to its inert nature, $^{83m}$Kr may be injected into the circulation stream upstream of the getter, while tritiated methane must be injected downstream. There are also valves to allow for the sampling of the LXe in the tower and gaseous Xe before and after the getter, as described in Section 9.6.

The underground installation of the purification system is shown in Figure 9.3.3.

The LXe tower is shown in detail in Figure 9.3.4 and contains the following major components: a weir reservoir for collecting the liquid returning from the detector and preparing it for evaporation; a two-phase tube-in-shell heat exchanger for evaporating the liquid and liquefying the gas; a single-phase gas-gas heat exchanger to bring the exiting Xe gas to room temperature and pre-cool the returning gas; a subcooler/phase separator for cooling the return liquid and removing gas bubbles; two thermosyphon heads; liquid sampling ports; and five cold valves with actuators. Components are hung with high-strength metal rods, stainless or Inconel 718.

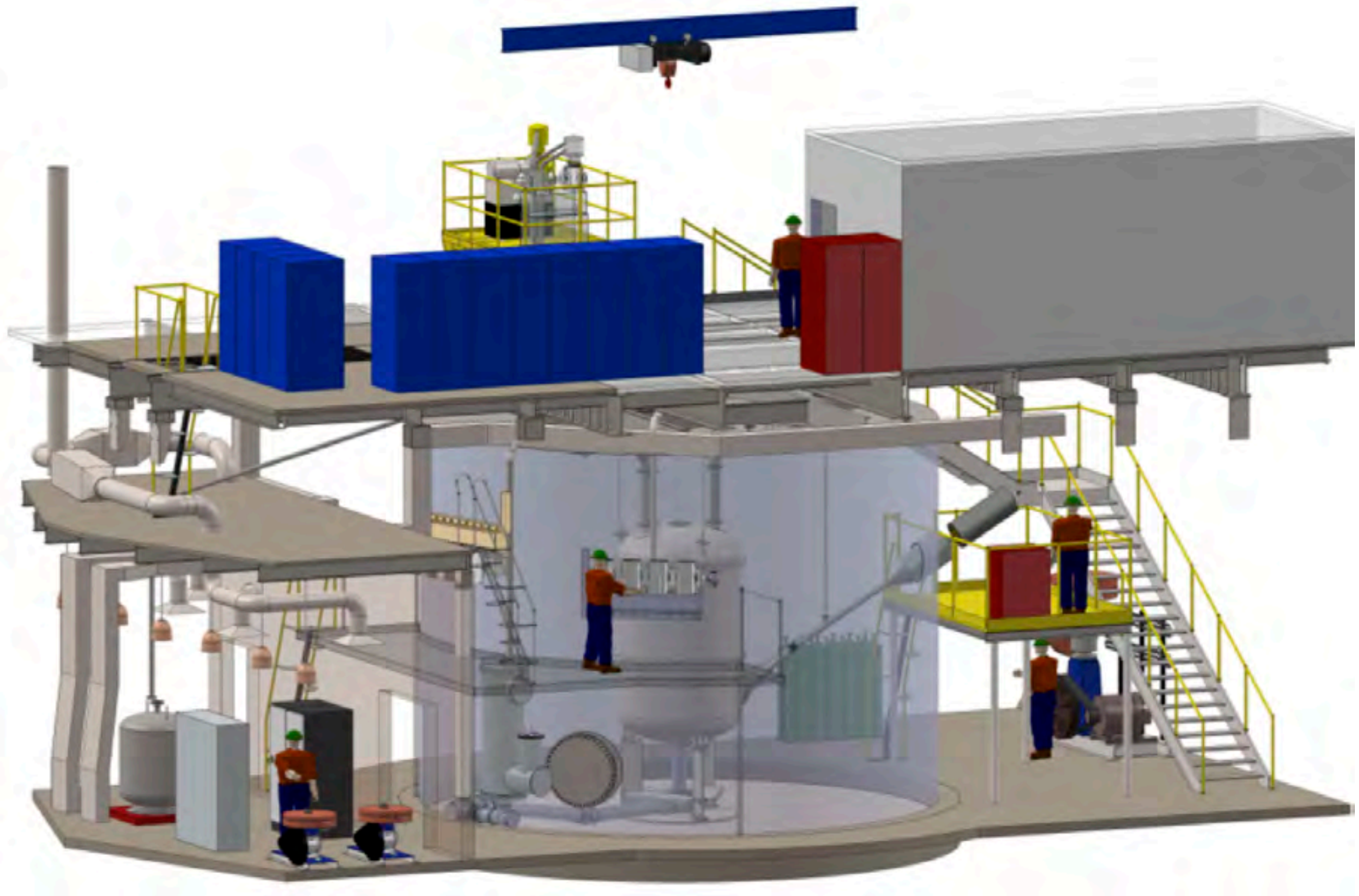

**Figure 9.3.3. Underground installation of the online purification system, showing the TPC, water shield, and LXe tower (center); getter and gas pump (left); and recovery compressor (right, underneath stairs).**

Liquid flow from the subcooler passes through four cold control valves. Two of these are ½-inch pipe size and can partition the flow between the TPC volume and skin volume. This is very useful for normal operation and also vessel cooldown. The other two ¼-inch pipe size valves route flow to the PMT standpipe and to the HV feedthrough. Return flow from these two volumes is back into the bottom of the Xe vessel, thus eliminating contamination buildup problems. Approximately 90% of the LXe goes directly to the Xe vessel from the large cold valves. Each of these four flow streams has an electric heater and an associated thermometer for cooldown and warmup control. A fifth cold control valve is placed between the weir reservoir and the shell side of the two-phase heat exchanger. This valve works in conjunction with the compressor suction control to establish the height of the boiling LXe. It will also provide a small amount of Joule-Thomson cooling by using the energy supplied from the compressor. All five of these cold control valves are bellows sealed and designed for use in helium liquefiers.

The weir reservoir is a simple vessel that acts as a volume buffer between the weir drainpipe and the two-phase heat exchanger. Besides establishing the liquid level in the weir drainpipe in the detector, it also provides a location to collect nonvolatile impurity species that may be present in the LXe. Such species, if they exist, would become harmlessly and irreversibly trapped in the weir reservoir, since the only mechanism to leave is through a phase change.

Liquid flows from the bottom of the weir reservoir into the bottom (shell) of the two-phase heat exchanger where it is vaporized, while Xe from the getter is condensed. This is a delicate balancing act. The bottom of the weir reservoir is placed at the same height as the bottom of the two-phase heat exchanger. Liquid is drawn up into the shell by the lower pressure of the suction side of the gas pump. Xe boils from the condensing heat load and, to a lesser extent, from cavitation caused by compressor suction pressure lower than saturation pressure. The two-phase heat exchanger condensing tubes are internally finned copper. Scaling from the LUX experience, approximately seventy ½-inch-diameter × 0.75-meter-long tubes are needed for the LZ heat exchanger. A careful study of condensation in the two-phase heat exchanger is under way using ANSYS CFX and will be benchmarked by the experience with LUX, as well as LZ test platform data.

Xenon is transferred in liquid form between the tower and the detector through a vacuum-insulated transfer line that penetrates the wall of the water shield and connects to the bottom of the TPC vessel. The PMT standpipe that contains cabling from the bottom of the detector is part of this transfer line. All bottom cabling is routed from the vessel in a 4-inch-diameter LXe-filled tube up into the PMT standpipe, where a free surface height is established by equilibrating pressure with the Xe vessel vapor phase pressure. Two ⅝-inch tubes transport subcooled LXe to the skin and TPC volumes of the vessel. One 1-inch tube returns the weir overflow. Two ⅜-inch tubes carry LXe to the top of the PMT standpipe and the HV feedthrough. There is a vacuum break in this transfer line at the PMT standpipe junction tee. Figure 9.3.4 shows the LXe tower and transfer line details.

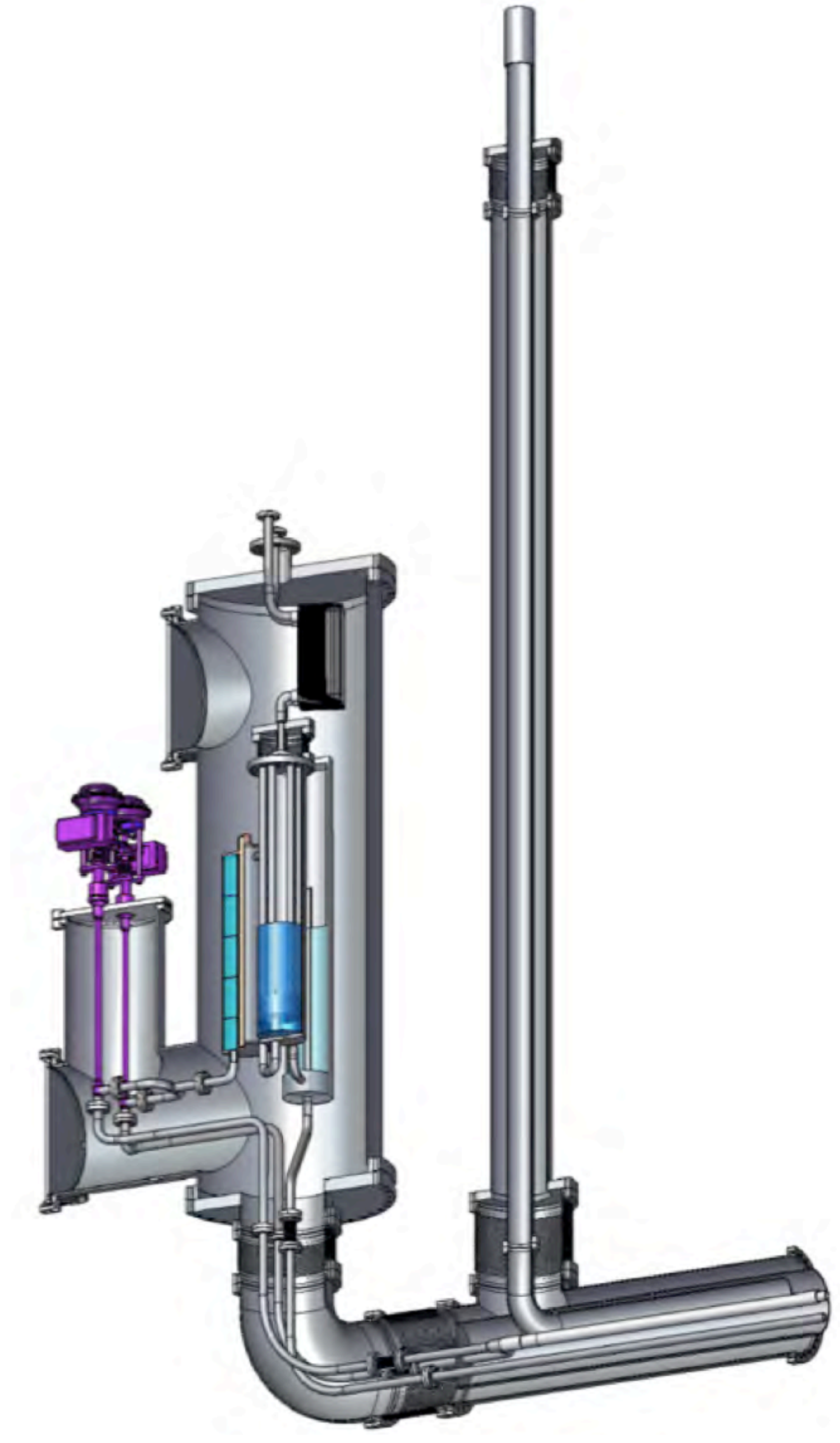

**Figure 9.3.4. LXe tower and PMT cable standpipe. The two-phase heat exchanger is shown infilled with liquid (dark blue), while the weir reservoir is shown as light blue. The purple valves control the return of the LXe to the detector through four control loops.**

The subcooler is the last major component in the LXe tower. It is made with perforated copper plates attached to a central copper rod that forms a $LN_2$ thermosyphon head at the top. Copper mesh or similar will be placed between the copper plates (depending upon analysis) to ensure a more than adequate heat transfer area. The thermosyphon will be able to take the liquid down to close to the freezing point 161.5 K. With a detector operating point of 175 K and 1.73 bara, this will give 13 K of subcooling. Returning the LXe in a subcooled state 1 K above the freezing point, we provide 270 W of net cryocooler cooling at 500 slpm. The subcooler also has a vapor vent line at the top in case it is ever needed.

The LZ circulation and purification system described here is similar to the LUX system currently in operation. A thermodynamic model of the LUX system has been built and verified against data taken with LUX to understand and measure heat-exchanger performance. The thermodynamic model was then altered to reflect the larger LZ system and to predict performance. The compressor for LZ is a different technology than LUX, using hydraulic oil to drive the diaphragm instead of a direct coupling between a piston and the diaphragm. A Fluitron compressor has been purchased for the LZ system test and will be used to verify the choice of compressor technology. Operation of the compressor for the system test (see

Section 6.7) will exercise it over a large range of operating conditions and test the control strategy. The getter system planned for LZ is also a larger version of the turnkey SAES getter used for LUX.

The LUX purification program provides relevant experience and some unresolved questions. For example, in considering the performance of the purification system, we usually assume that the getter removes impurities with an exponential time constant that is characteristic of the volume-exchange time of the TPC. This assumption is reasonable if the Xe is well-mixed in the detector. In the case of LUX, the nominal flow rate is 25 slpm, and with a total Xe mass of 370 kg, the LUX volume-exchange time is 42 hours. On multiple occasions, a small quantity of methane was injected into LUX, in both natural and tritiated form, and these injections gave the opportunity to measure the behavior of the purification system. Exponential removal was indeed observed, both with the Xe gas-sampling system (natural methane injections) and via the activity in the detector (tritiated methane injection). Surprisingly, the removal time constant was found to be a mere six hours — seven times faster than expected. Although this result is not fully understood, it seems likely that the LUX purge-gas purification loop may be playing a strong role in increasing the removal speed beyond naïve expectation. Nevertheless if this phenomenon could be fully understood and exploited, it would lead to better-than-baseline performance of the LZ purification system.

## 9.4 Xenon Recovery

The Xe recovery system removes the Xe from the detector and returns it to the storage cylinder packs. Recovery may be initiated during normal operations, but it may also be manually or automatically triggered in the event of an emergency, such as a lack of cooling power, where Xe pressure threatens to approach the cryostat pressure rating. Xenon may also be removed from the detector in a controlled way to optimize operating conditions or to allow repair and maintenance of the detector.

The online purification system (Section 9.3) has redundant circulation compressors that each receive Xe gas at 1.6 atma (atmospheres absolute) and compress it to 4 atma at a rate of 500 slpm (3 kg/min). The same compressors can output gas at 5 atma (74 psia [pounds per square inch absolute]) at a flow rate of 400 slpm with an input pressure of 1.3 atma. These compressors could be used to pressurize the empty storage packs. This could store about 132 kg of Xe without operation of the recovery compressor. This could be used for operational adjustment of Xe levels. A vacuum pump is incorporated into the system to allow emptying the packs fully back into the detector.

A high-purity 270 slpm (1.5 kg/min) recovery compressor can compress Xe to 1200 psia and restore all the Xe from the detector into the packs. This compressor will most likely be a two-stage triple-diaphragm compressor with an interstage pressure of about 230 psia. During normal recovery at 1.5 kg/min, it would take five days to transfer the Xe from the detector to the packs. Once the Xe level in the detector drops below the weir, liquid can only be extracted by drawing liquid from a dedicated return line in the PMT cable standpipe. To vaporize Xe at this rate, 2.5 kW of heat is applied to the LXe in this line with electric heating coils. Xenon can be removed at a slower rate by bypassing part of the output from the recovery compressor back to its input and reducing the heat input to the LXe proportionally. The circulation compressors would continue to operate normally to keep the input to the recovery compressor at 3 atma. The circulation-control system coordinates the operation of valves, the circulation compressors, the recovery compressor, and the heat input. Pressure, temperature, and flow sensors in the detector and Xe circulation system as well as pressure sensors and scales on the packs provide additional monitoring information.

We consider an emergency to be any unplanned event in which the pressure may rise in the cryostat. For example, if there is no cooling of the experiment and the vacuum insulation is good, Xe gas production would be less than 100 slpm. The cryostat is rated for a pressure of 4 atma at the bottom of the LXe, so during a cooling emergency the Xe must be vented before the gas pressure exceeds 3.4 atma. The recovery compressor receives gas from the circulation system after the circulation-system compressor (with an intake pressure of 3 atma) and compresses the Xe sufficiently to fit back into the packs. During emergency recovery, the circulation compressors can be on or off, as gas will flow through the valves if the pressure on the suction side is greater than that on the output side. For operational simplicity and power conservation, the circulation compressors would be turned off if the recovery compressor were operating in emergency mode. The recovery compressor is activated if the gas pressure in the detector reaches 2.4 atma.

A more serious emergency would be an air leak to the vacuum space between the inner and outer cryostat. Air in this space would increase the heat transfer to the inner detector sufficiently to vaporize Xe at 270 slpm (1.6 kg/min). This is a conservative number that was calculated assuming the air in the former vacuum space stays at 25 $^{o}$C and ignores frost buildup. If the cryo cooler is still working, it partially compensates for the increased heat transfer, but the recovery compressor is sized to handle the flow rate if there is an air leak and cooling failure at the same time.

The worst-case accident is a breach of the outer vacuum jacket (VJ), resulting in water from the water shield spilling into the cryostat-insulating vacuum and causing a large heat load directly into the inner titanium vessel. If the water wets the entire outer surface of the inner titanium vessel, the instantaneous heat transfer is 3 MW, but then drops rapidly as ice forms next to the vessel and provides some insulation. A foam layer 10 mm thick on the vertical walls and 20 mm thick on the bottom head of the inner titanium vessel is incorporated into its design to limit heat transfer in this scenario. With this foam in place, the maximum heat-transfer rate drops to 3.5 kW, equivalent to vaporizing 450 slpm of Xe. The foam also provides a mechanical cushion if ice does form, to reduce compressive loads on the cryostat. There is no point in the system where Xe can leak directly into water with one containment failure. This flow rate is above the flow rate of the recovery compressor. A backup piston-recovery compressor can accommodate this higher flow rate, but it is not as clean. The Xe recovered with the backup recovery compressor will probably need to be repurified (via krypton removal) before it can be used again in LZ. Of course, in the case of a water breach in the VJ, significant downtime will be needed to disassemble, repair, and reassemble the experiment.

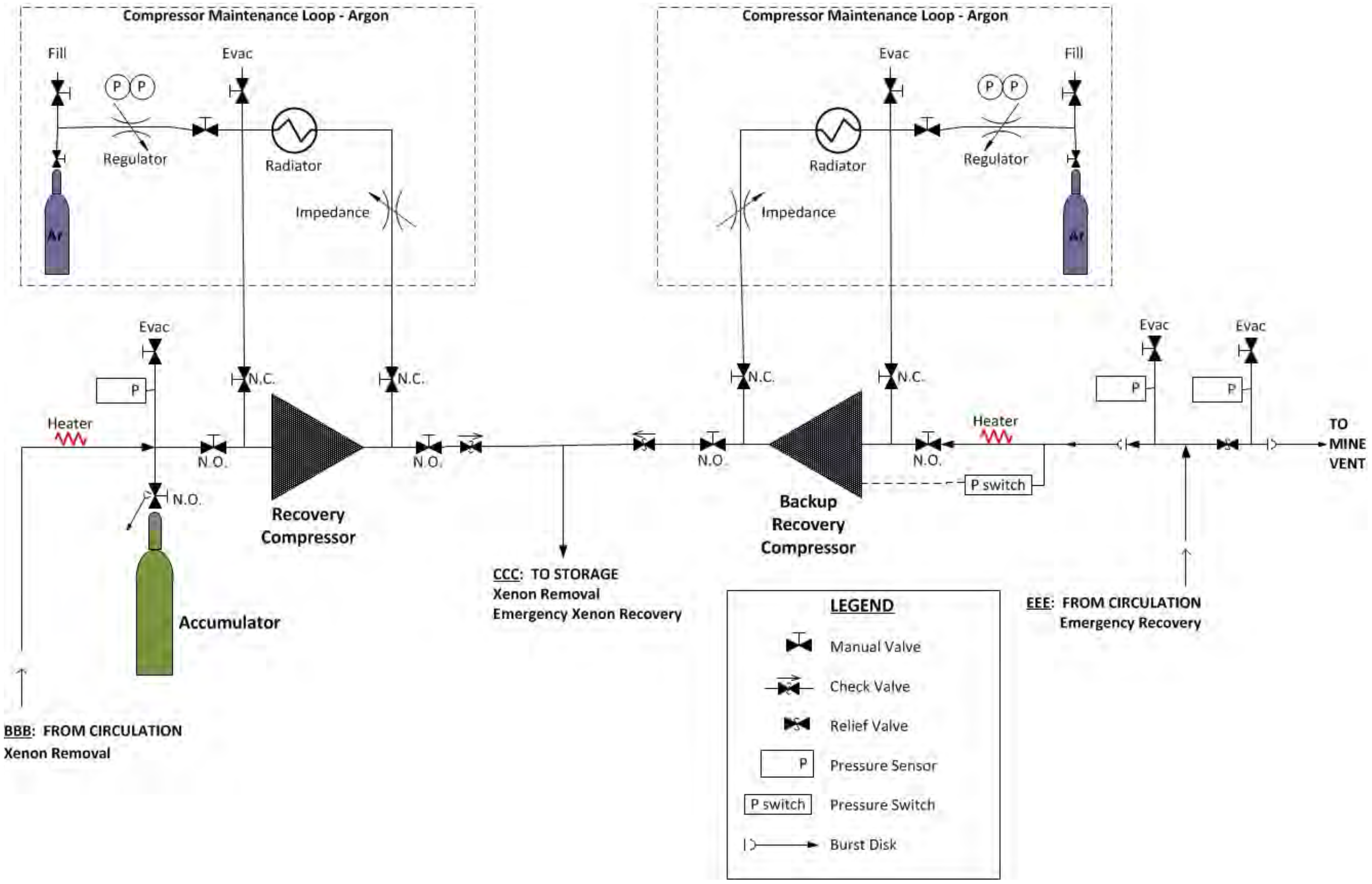

**Figure 9.4.1. Schematic diagram of Xe recovery.**

SLAC possesses an underutilized helium compressor that may be used in LZ as the backup recovery compressor. The backup recovery compressor will be isolated from the system by a burst disk. If at any time the pressure in the detector exceeds 2.9 atma, this burst disk will rupture and a pressure switch will turn on the backup recovery compressor. There is also a burst disk to the mine exhaust at 3.4 atma if in an emergency all compressors should fail.

In the event of an emergency, the Emergency Safety System controls the safe recovery of the Xe to the storage facility. This includes cases in which neither underground access nor remote human operation is possible. We are considering several architectures to implement this system. In the first model, it is implemented as part of the normal circulation-control system, which is based upon a programmable logic controller (PLC). PLCs are typically used in industrial control applications and are renowned for their reliability. The PLC software will monitor the detector as well as AC power and LN storage to determine the most appropriate mode of operation. The emergency system will be integrated with the slow control so that in the standby mode, it will be possible to operate the recovery hardware remotely, using standard slow-control tools. In recovery modes, the emergency system will operate autonomously, though it still will be possible to return it to standby mode (for a limited period of time) by a command from slow control. A redundant pressure switch hard-wired to the recovery compressor and activated at a pressure of 2.5 atma could provide an additional layer of backup capability if the PLC is inoperable. We are also considering implementing the Emergency Safety System on a second standalone PLC independent of the circulation-control system, or as a hardware-only system composed of relay logic.

The recovery compressor requires up to 5 kW and the backup recovery compressor requires up to 7.5 kW. These compressors must be capable of running at all times. A 30 kW generator provides backup power to the experiment. The primary load for this generator is the 11 kW needed by the detector cryocooler. If emergency recovery is initiated when the experiment is running on backup power, the compressors will have priority and the cryocooler may be disabled.

Normal detector operating pressure is about 1.6 atma. If the pressure exceeds 2.2 atma, an alarm sounds and the operator is notified. The operator could manually initiate Xe recovery to keep pressure under control. If the pressure exceeds 2.4 atma, the emergency safety system will automatically initiate emergency recovery with the recovery compressor. If this fails, the redundant pressure switch will open the valve and start the recovery compressor at 2.5 atma. If the pressure continues to rise, the burst disk to the backup recovery compressor will burst at 2.9 atma and a pressure switch will start the backup recovery compressor. Finally, if the pressure is still rising, Xe will be vented to mine exhaust at 3.4 atma by a burst disk.

## 9.5 Xenon Transportation, Storage, and Transfer

The LZ Xe will be sourced from multiple gas suppliers (Section 9.9), shipped to SLAC for krypton removal (Section 9.2), shipped to SURF to be moved underground, and finally liquefied into the LZ detector. The Xe storage, transport, and transfer must be done without losses due to its high cost, and also without introducing more than 40 standard cc of air once the krypton has been removed. The storage system must be compatible with filling systems of the gas suppliers and safe to transport on U.S. roads. DOT-rated high-pressure cylinders are a standard solution for transporting gas.

The critical point for Xe is near-room temperature (289.73 K), so the density as a function of pressure is very nonlinear, as seen in Figure 9.5.1. As a consequence, the storage capacity of a cylinder of fixed volume increases by almost a factor of 3 between 800 to 1000 psig and then increases slowly above that pressure. Limiting the storage pressure to approximately 1000 psig helps control compressor cost without significantly increasing storage capacity requirements.

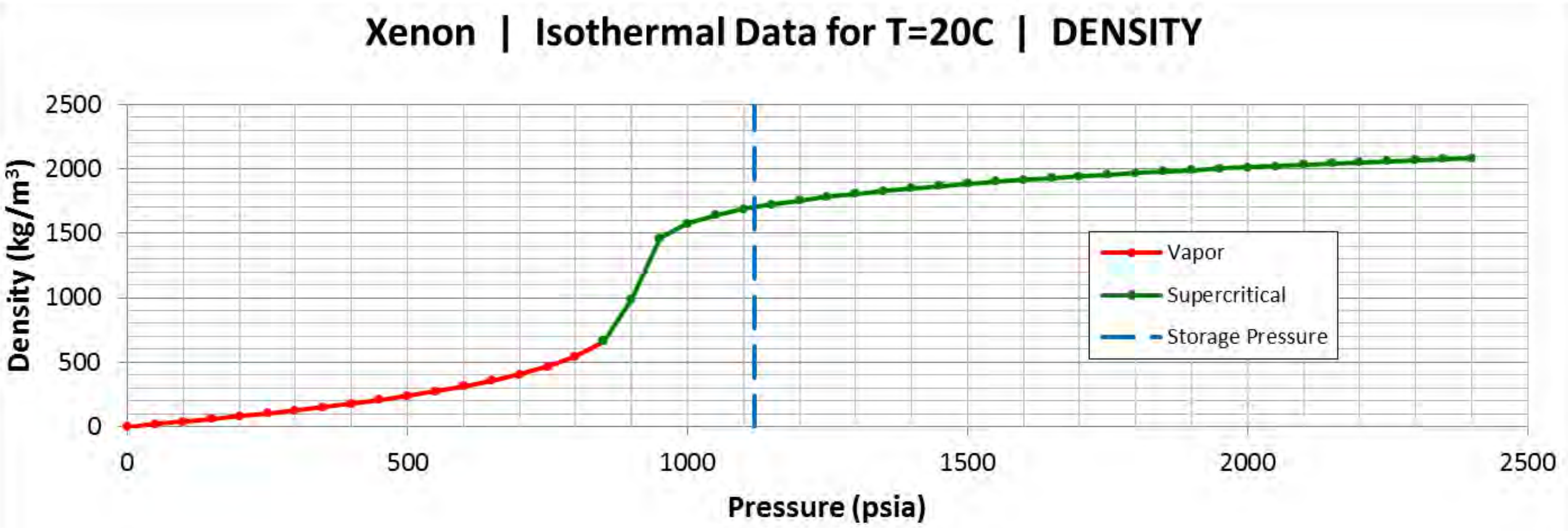

**Figure 9.5.1. Xenon gas density as a function of pressure at 20 °C.**

A variety of standard DOT-rated compressed gas cylinders have working pressures of 2400 psig. A common and economical size is 9.28-inch-diameter × 55.5-inch-tall with a volume of 49.1 liters, such as a Norris 8BC300 cylinder. One of these cylinders filled to 1120 psig at room temperature holds 83 kg of Xe and has a total weight of 144 kg. For LZ, groups of 12 cylinders are housed in a pre-engineered freestanding pack that can be handled with a pallet jack, forklift, crane, or rolled by hand. The packs are rated for transportation with valves and manifolds installed. One such 12-pack holds 996 kg of Xe at 1120 psig and has a total mass of about 2000 kg. Ten packs are required to store the full quantity of Xe for LZ.

Each cylinder is sealed with a Ceodeux D304 tied diaphragm valve with combined soft and metal seal, marketed to the semiconductor industry for ultra-high-purity gas applications. The valves have a burst disk as an overpressure safety device. The rated helium leak rate for the valve is $10^{-8}$ mbar-l/s, equivalent to $1.5 \times 10^{-10}$ grams of krypton per year from the air, or 0.002 ppt krypton contamination per year for a 49-liter cylinder. The allowable time-averaged contamination rate for the experiment is 0.004 ppt krypton per year.

The connection between the valve and cylinder is a high-quality ¾-inch 14 NGT. We have explored the possibility of making a seal weld between the SS valve and the steel cylinder after valve installation, but this would invalidate the DOT rating of the cylinder and risk damage to the valve. Valve and cylinder vendors indicate that a properly installed valve will have sufficient leak tightness to meet our requirements. We are also exploring the possibility of coating the threads of the valve with indium before installation to act as a lubricant and sealant. Indium can be applied to SS with an ultrasonic soldering process or electroplating. The ZEPLIN-III experiment used indium on cylinder valve threads successfully but did not quantify the leak rate achieved.

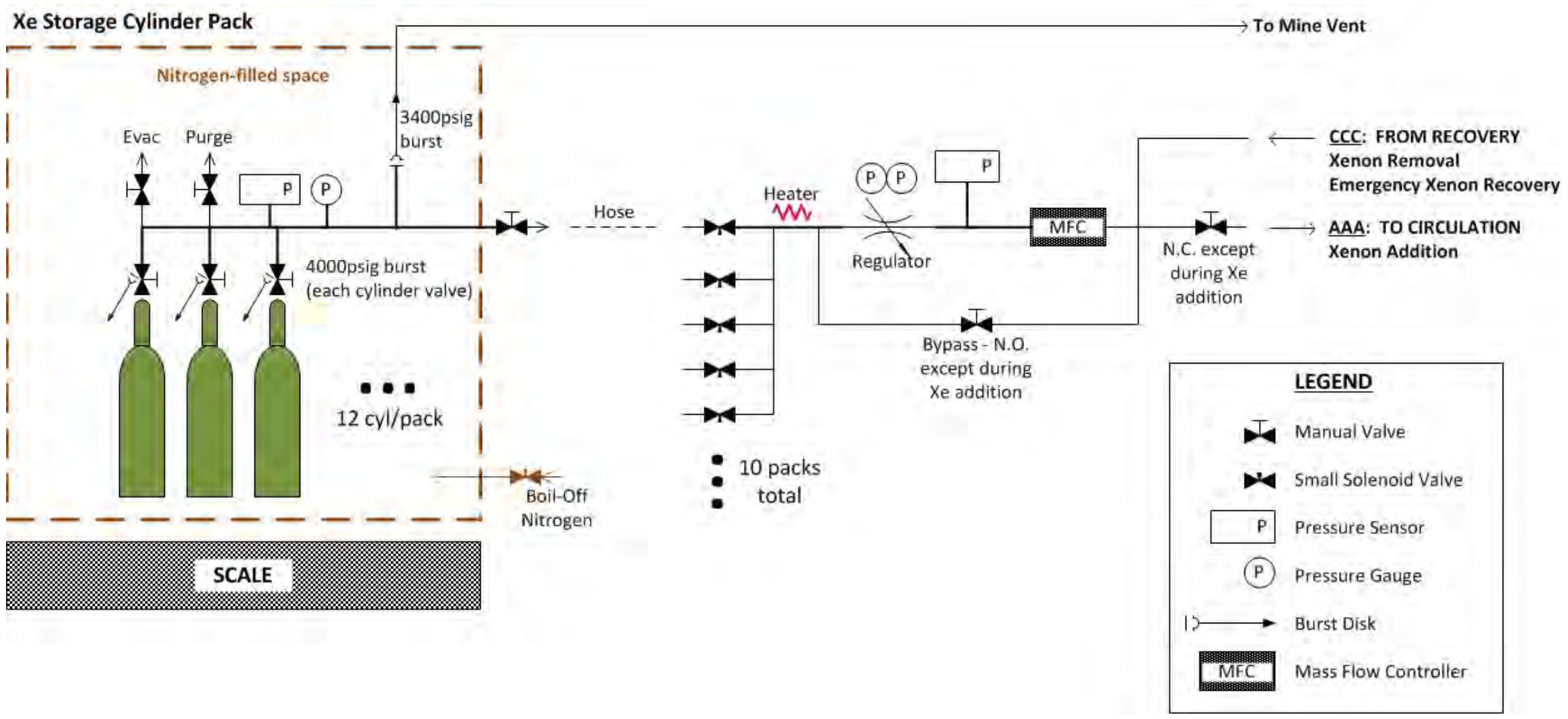

**Figure 9.5.2. Xenon storage and supply system.**

Figure 9.5.2 shows a schematic of the Xe storage and supply system plumbing. The cylinder outlet valve connection is DISS 718, a VCR-like fitting that makes an all-metal seal, including a metal gasket. The cylinders in the pack are connected with an all-welded SS manifold that mates to the DISS fittings. The manifold has a supply-side pressure transducer, evacuation port, purge port, and a manifold valve. As a secondary precaution against air contamination, the cylinder packs are housed in a sealed sheet-metal box that can be purged with boil-off nitrogen. Figure 9.5.3 shows a model of the pack with the purge covering. This purge system reduces the krypton contamination rate of the Xe from diffusion and permeation through the seals and from small leaks in the plumbing. The purge system is intended for use during long-term storage, but not during transport.

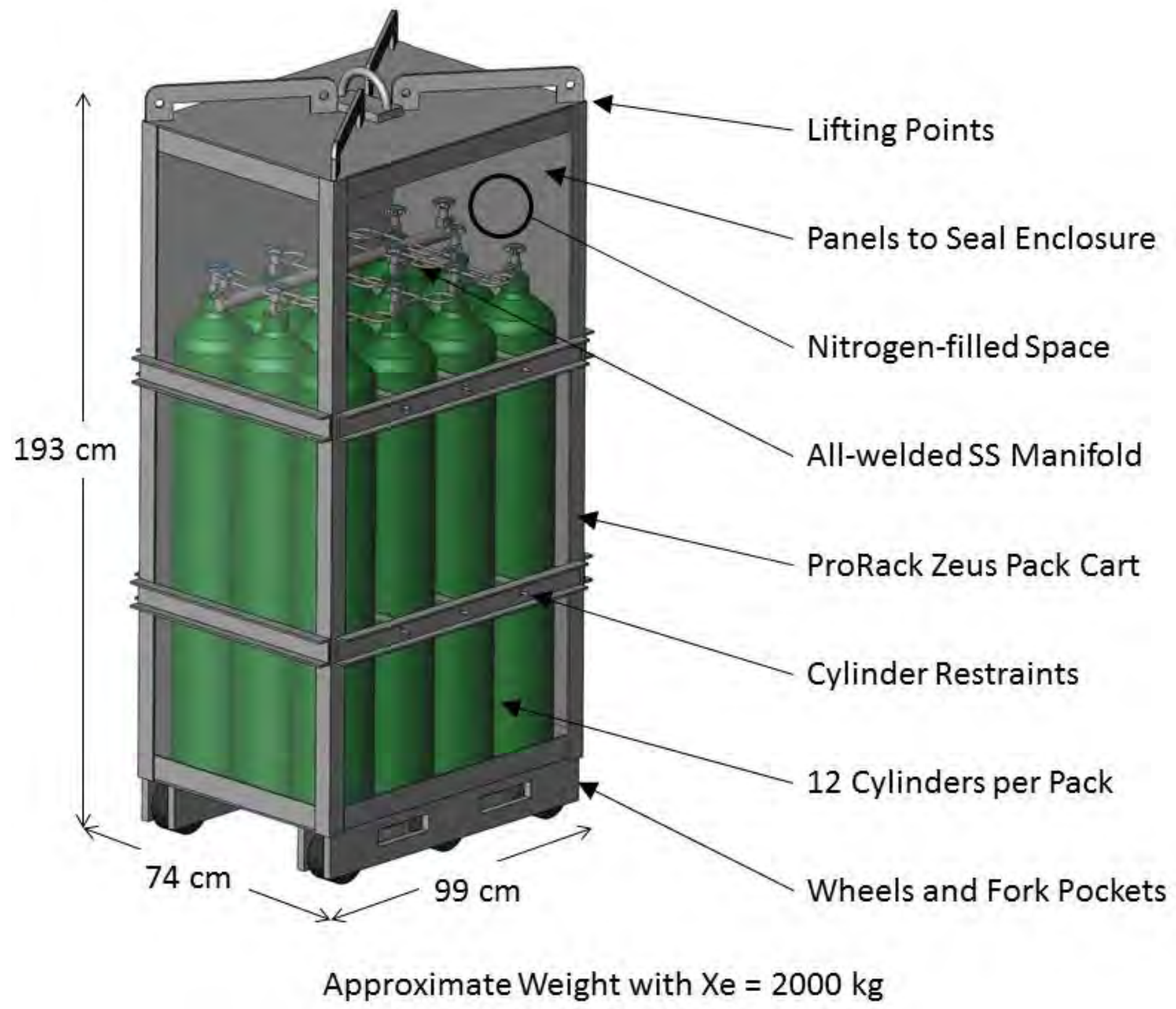

**Figure 9.5.3. A single cylinder pack containing 12 Xe cylinders.**

The manifold valve is connected to a pressure regulator with an output pressure transducer and a mass flow controller when Xe is delivered from the pack. This delivery system has controlled electrical heating to prevent Xe from liquefying in the regulator outlet due to Joule-Thomson cooling during expansion. The pack rests on a scale to monitor the quantity of Xe in the pack and provide an accurate measure of Xe delivered.

The cylinders are tested to 4000 psig, and they also have a 4000 psig burst disk. This pressure would be reached if a cylinder filled with 83 kg of Xe were heated to 96 $^{o}$C. If the pressure relief failed and remained closed, the cylinder minimum burst pressure of 5760 psig would be reached at 142 $^{o}$C. The temperature of the cylinders must be monitored and controlled during storage and transport. A sprinkler system in the storage location at SURF and any other long-term storage location protects the Xe from damage in the event of a fire. A temperature-controlled truck could be used for transport.

The cylinders, valves, manifolds, and pack are purchased separately from the manufacturers. We intend to have the cylinder manufacturer clean the cylinders, install the valves, and perform a preliminary leak check of the assembly. The pack, cylinders, and manifolds are integrated at the University of Wisconsin's Physical Sciences Laboratory (UW-PSL). Depending on the results of the indium coating experiments the valves may also be installed at UW-PSL. A final helium leak test is done on each pack before shipping to the Xe supplier for filling. The pack is shipped evacuated and ready to receive Xe.

The filled pack will be shipped to the krypton-removal facility at SLAC. During krypton removal, a single pack delivers the vendor-supplied research-grade Xe to the chromatography loop at a controlled rate and a second pack receives dekryptonated Xe from a compressor connected to the Xe condenser. The facility will have secure storage with adequate emergency ventilation, temperature monitoring, and fire protection for Xe packs.

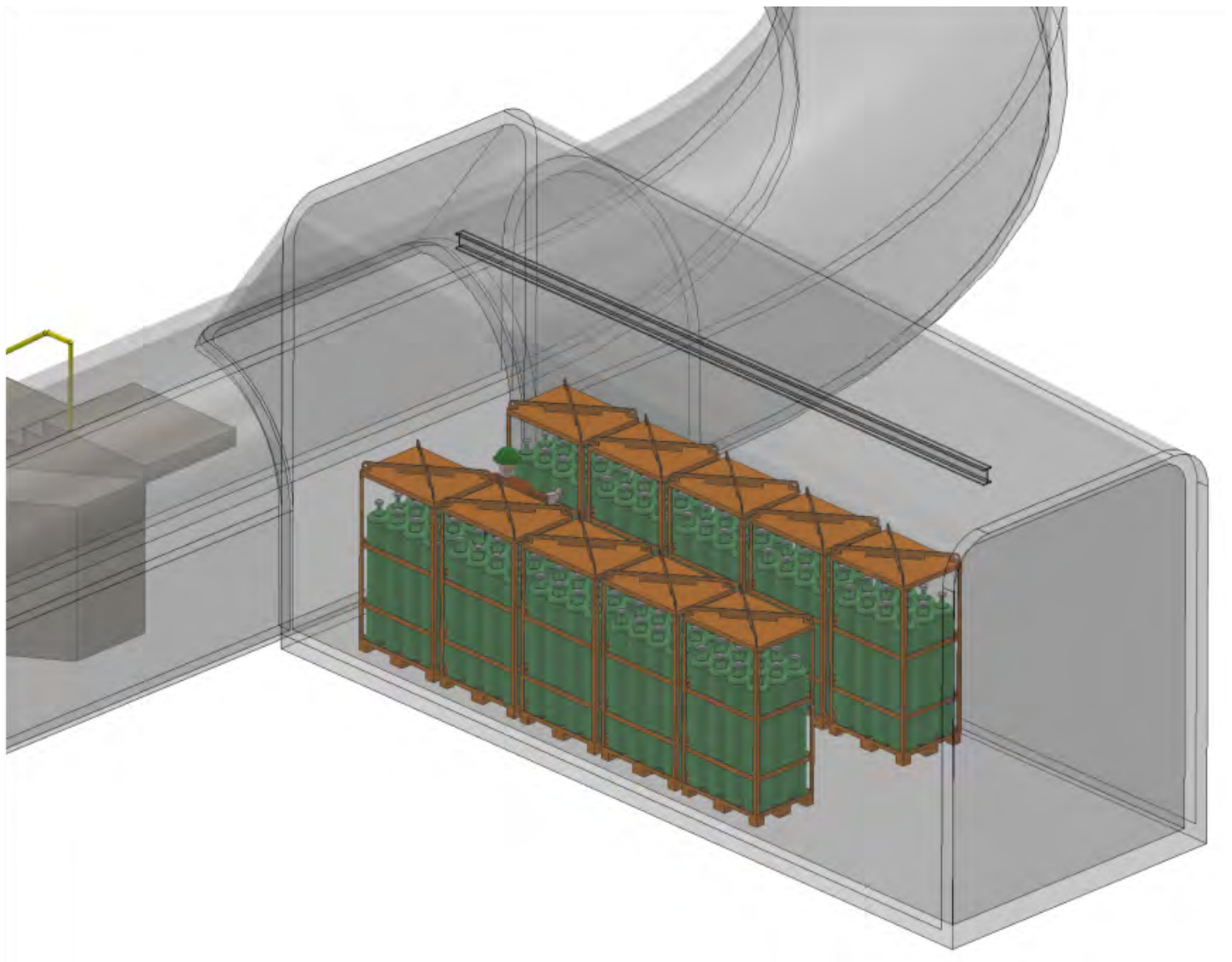

**Figure 9.5.4. Underground storage of cylinder 12-packs at SURF.**

Packs filled with processed Xe are shipped by truck to SURF and delivered to underground storage. The packs will be loaded on a small rail car in the Yates headhouse, loaded into the cage, lowered to the Davis Campus level, and rolled into the Davis Campus on the small rail car. Part of the Davis Campus infrastructure work (see Chapter 13) is to prepare a storage space for the Xe using an existing excavation in the LN storage room access drift, as shown in Figure 9.5.4. A monorail will be installed to allow lifting the pack off the small rail cart and into place. The Xe storage space will have a boil-off nitrogen supply to purge the enclosed packs, a sprinkler system to control temperature during a fire, and an emergency ventilation and oxygen monitoring system in case of an accident. Supply and return plumbing from the Xe storage area will connect the packs to the LZ detector plumbing.

After the LZ detector is assembled and plumbed to the Xe purification system and vacuum system, it is evacuated and held at vacuum for two months to remove residual gases in the plastic components (see Section 9.7). Once the residual outgassing has reached an acceptable level as monitored by the Xe sampling system (see Section 9.6), the cryostat is cooled by the thermosyphons (Section 9.8) until it is just above the Xe condensation temperature. The outgassing rate at the cold temperature is also monitored and confirmed to be acceptable. Then Xe is delivered from the packs to an input port in the online purification circuit (described in Section 9.3). Xenon passes through the getter system on the way into the detector. With the circulation loop operating, the temperature is lowered until the Xe starts to condense. The mass of delivered Xe is monitored with the pack scales and the liquid level in the detector is monitored with internal liquid level sensors. At the end of the fill, all the residual Xe is transferred into one pack and the other packs are stored under full vacuum.

## 9.6 Xenon Sampling and Assay

Sensitive purity monitoring is integrated into the LZ Xe handling plan. The basic monitoring technology that we employ is the coldtrap/mass-spectrometry method developed at Maryland for LUX and EXO-200 [1,2,11]. Here we describe our experience with the method as applied to LUX and our plans for its implementation in LZ.

The simplified schematic shown in Figure 9.6.1 (left) illustrates the basic concept of the sampling technique. A gaseous Xe sample of interest flows through a precision vacuum leak valve to an RGA, where the partial pressures of its impurity species are measured. The observed partial pressures are proportional to the flow rate, so good sensitivity is achieved by flowing the gas as fast as possible. In practice, the flow rate is limited by the large pressure of the bulk Xe gas, which causes the RGA to trip at pressures much greater than ~$10^{-5}$ torr. Much larger flow rates can be achieved by selectively removing most of the Xe from the sample between the leak valve and the RGA. This is accomplished by means of an LN coldtrap that holds the pressure of the Xe at its output at a constant value of 1.8 x $10^{-3}$ torr (the vapor pressure of Xe ice at 77 K). A section of low-impedance plumbing further reduces this pressure to <$10^{-5}$ torr at the location of the RGA. The virtue of the coldtrap is that it allows most impurity species of interest, including $N_2$, $O_2$, Kr, Ar, He, and $CH_4$ to pass through in proportion to the flow rate, while holding the Xe pressure at a constant and modest level independent of the flow. The method has a demonstrated sensitivity of ~0.1 ppb for $N_2$ and $O_2$. For krypton, which has a unique high-mass signature at 84, 86, and 82 atomic mass units, the sensitivity has been shown to be 0.2 ppt [2]. Further sensitivity improvements are discussed below.

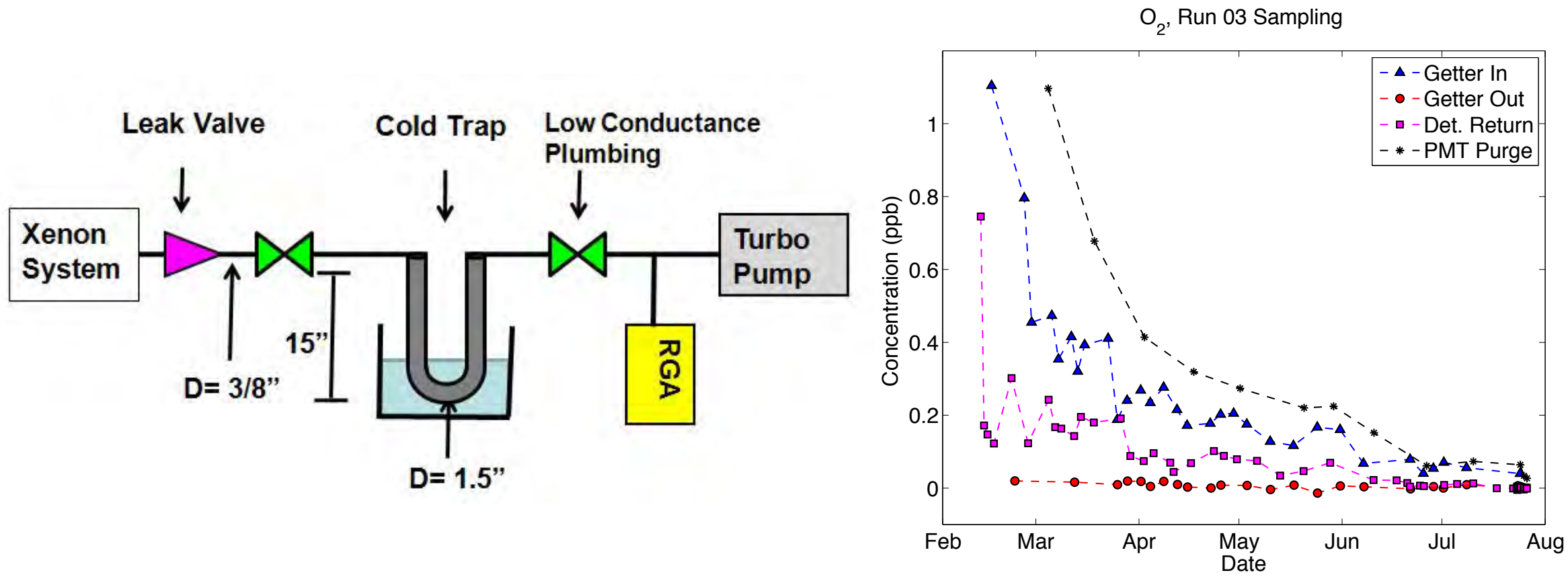

**Figure 9.6.1. Left: Schematic diagram of the Xe sampling system. Right: Trending plot of the $O_2$ concentration in LUX during the 2013 physics run. Four sampling locations are shown. The decreasing concentration over time is due to the action of the online purification system and indicates that the sources of the $O_2$ outgassing were mostly exhausted over this period.**

The usefulness of the technique is illustrated Figure 9.6.1 (right), where LUX $O_2$ concentrations measured during the 2013 physics run are shown. Besides confirming the effectiveness of the LUX purification system, this monitoring data also confirms that its Xe plumbing remains leak-tight. Furthermore, certain classes of purity problems, if they occur, can be identified by their unique characteristics. For example, an air leak gives a characteristic ratio of $O_2$ and $N_2$, and a constant rate of increase of argon. Residual outgassing of detector components, on the other hand, slowly decreases over time and favors the species of purge gas that was used during the detector checkouts (usually $N_2$). Also, the presence of a very large and out-of-balance $N_2$ concentration could indicate that the getter is beginning to saturate. (The getter has a certain capacity for each impurity species, and $N_2$ is often the first impurity for which the getter fails.)

The LUX system is almost fully automated, requiring human intervention only to fill the coldtrap LN dewar. For LZ, the system will be completely automated, allowing programmatic and semicontinuous monitoring data to be collected. The LN coldtrap will be replaced with a thermosyphon head to facilitate this automation.

The LZ sampling program begins when the first empty Xe storage pack is assembled. To confirm that the pack is suitable for use in LZ, we will perform a long-term test of the storage pack in which it will be partially charged with a quantity of Xe (~100 kg), the purity of which will be monitored periodically over the course of a year or more. The construction, commissioning, and operation of the krypton-removal system will also be guided by an integrated and highly sensitive analytical system. For example, the chromatographic processing parameters may be tuned to maximize the krypton rejection factor, and problematic elements of the system can be located, modified, or removed. During bulk processing of the Xe, frequent sampling allows the purified Xe to be certified for use in LZ.

Ultimately, the analytical system will be installed underground in the Davis Cavern and integrated into the online purification system. There it monitors the outgassing of the TPC component parts, measuring the rate at which krypton and other impurities accumulate while clean Xe gas circulates at room temperature or elevated temperature. This monitoring program will continue as the detector is cooled for the first time. Finally, the Xe can be transferred into the detector and liquefied once the outgassing rate observed at the cold temperature is acceptable. Automated and frequent monitoring continue throughout the physics run to confirm that no problematic leaks or outgassing have occurred. Sampling locations within the Xe handling system include the liquid return stream, the purge-gas flow, and before and after the getter.

Compared with LUX, LZ requires that the sensitivity of the cold/mass-spectrometry technique be improved from the current-best ~0.3 ppt to ~0.015 ppt. An additional factor of 10 in sensitivity beyond this would be ideal, allowing krypton impurities to be identified well before they reach the problematic level. We are currently pursuing multiple routes to gain this additional sensitivity, including higher flow rates, lower electronic noise, and more sophisticated statistical analysis of the data. We are also pursuing trapping methods based upon chromatography and other techniques to integrate the small krypton signal over a longer time period. The development of the screening technique proceeds in concert with the krypton removal, since highly purified Xe is required to demonstrate good sensitivity, while good sensitivity is required to develop the krypton removal capability.

## 9.7 TPC Impurity Burden and Outgassing Model

LZ detector materials are chosen primarily based upon the radioactivity and light-collection goals of the experiment as well as the engineering constraints. Here we consider the effect of those detector materials on the purity of the Xe with respect to electronegatives and krypton. Ingress of krypton into the Xe during operations is a particular concern because it cannot be removed by the online gas-purification system.

All nonmetal detector components are laden with trace quantities of atmospheric gases, including krypton and oxygen. Outgassing begins when LZ is first evacuated and will continue as the detector is filled with gaseous Xe and then LXe. The initial impurity burden of each detector component is determined by the solubility of each impurity species in the material, by the component volume, and by the impurity's abundance in air. The outgassing rate is determined by the diffusion constant, by the geometry and dimensions of the part, and by the pumping impedance. As shown in Table 9.7.1, we have measured the solubility and diffusion constant for important impurities species in key LZ detector materials.

Using the solubilities shown in Table 9.7.1, we can approximately account for the total impurity load of the detector materials prior to outgassing by summing the burdens from the TPC materials (the PTFE and polyethylene parts) and the cathode HV cable. The result is shown in Table 9.7.2, for both LZ and LUX (for comparison).

We find that prior to outgassing, the $O_2$ and $N_2$ burdens of the LZ detector materials compared with the mass of Xe are on the order of 1 ppm (part per million) and are smaller than that of LUX. This reflects the fact that the plastics scale as the surface area of the detector, whereas the Xe scales as the volume. Similarly, the $O_2$ and $N_2$ burdens are small compared with the capacity of the getter, implying that the getter will not need to be replaced during operations as long as the system remains leak-tight.

The krypton burden of the LZ detector materials, also shown in Table 9.7.2, is a larger concern because the absolute quantity of krypton exceeds the LZ goal by a factor of ~2,050 and because krypton cannot be removed during operations. We have two strategies to mitigate the risk that this krypton will contaminate the Xe during operations: (1) reduce the krypton burden by outgassing the detector components prior to operations, and (2) freeze the krypton in place by cooling the detector components to LXe temperature, thereby dramatically reducing the diffusion constant.

**Table 9.7.1. Diffusion constants (D) and solubilities (K) of several impurity species and materials of interest for LZ. The measurements have been done as part of the LZ R&D program.**

| | Teflon (LUX sample) | | Polyethylene (LUX sample) | | Viton | | LZ HV cable (Polyethylene) | |
|---|---|---|---|---|---|---|---|---|
| | K (%) | D ($10^{-8}$ $cm^2/s$) | K(%) | D ($10^{-8}$ $cm^2/s$) | K(%) | D ($10^{-8}$ $cm^2/s$) | K(%) | D ($10^{-8}$ $cm^2/s$) |
| $N_2$ | 5.2±0.3 | 15.1±0.2 | 1.96±0.25 | 16.2±0.8 | 48±27 | 2.2±1.0 | | |
| $O_2$ | 10.9±0.7 | 31.4±0.4 | 1.82±0.56 | 39.0±1.9 | 8.3±0.5 | 6.776±0.061 | | |
| Kr | 29.1±2.5 | 5.56±0.08 | 10.7±1.1 | 6.44±0.31 | 12.5±1.1 | 1.248±0.015 | 7±1 | 11.2 |
| Xe | 78.3±6.9 | 0.80±0.01 | 56.8±5.8 | 2.03±0.31 | 15.0±1.6 | 1.70±0.10 | | |
| Ar | 3.2±0.3 | 16.85±0.24 | 3.42±1.07 | 20.1±1.0 | 6.0±1.9 | 4.00±0.11 | | |
| He | 2.8±0.2 | 1,268±18 | 0.35±0.04 | 435±21 | 5.0±0.4 | 435.89±3.88 | | |

**Table 9.7.2. An approximate accounting of the impurity burdens of the LZ (and LUX) detector components prior to outgassing. The burdens are calculated from the volume of the parts, the concentrations of the species in air, and the solubilities shown in Table 9.7.1. Three basic components are included: the PTFE and PE components of the TPC structure and PMT arrays, and the PE HV cable. Component volumes for LZ are from the engineering model; component volumes for LUX are estimated based upon approximate dimensions.**

| TPC plastics | LZ | LUX | Units |
|---|---|---|---|
| PTFE volume | 156 | 11 | liters |
| PE volume | 4 | 48 | liters |
| Kr | $3.1 \times 10^{-4}$ | $3.8 \times 10^{-5}$ | g |
| $O_2$ | 9.4 | 0.92 | g |
| $N_2$ | 15 | 2.0 | g |
| **Cathode HV cable** | | | |
| PE volume | 3.5 | 1.4 | liters |
| Kr | $1.3 \times 10^{-6}$ | $5.2 \times 10^{-7}$ | g |
| $O_2$ | 0.017 | 0.007 | g |
| $N_2$ | 0.066 | 0.027 | g |
| **Total** | | | |
| Kr | $3.1 \times 10^{-4}$ | $3.8 \times 10^{-5}$ | g |
| $O_2$ | 9.4 | 0.93 | g |
| $N_2$ | 15 | 2.0 | g |
| Kr/Xe | 31 | 103 | ppt (g/g) |
| Kr/(Kr goal) | 2,048 | 21 | |
| $O_2$/Xe | 943 | 2,512 | ppb (g/g) |
| $N_2$/Xe | 1,491 | 5,443 | ppb (g/g) |
| $O_2$/getter capacity | 0.17% | 0.77% | |
| $N_2$/getter capacity | 2.40% | 8.06% | |

First, we may outgas the plastic components prior to cooldown using a vacuum pump or by circulating and discarding a small quantity of purified Xe gas. This may be done at room temperature or at a slightly elevated temperature (consistent with the temperature specifications of the TPC and cryostat) to promote krypton diffusion. This reduces the krypton content of the detector components prior to their exposure to the LZ Xe.

Applying the diffusion constant measurements shown in Table 9.7.1, we calculate the remaining impurity load for $O_2$, $N_2$, and krypton as a function of outgassing time by solving the diffusion equation. For this calculation, a 2-cm-thick cylindrical shell with a diameter and height of 1.5 meters represents the PTFE reflectors, while a 4-meter-long solid polyethylene cylinder with a diameter of 3.4 cm represents the HV cable. Figure 9.7.1 shows the impurity burden remaining in the detector components as a function of outgassing time. The exponential time constants associated with the outgassing are listed in Table 9.7.3.

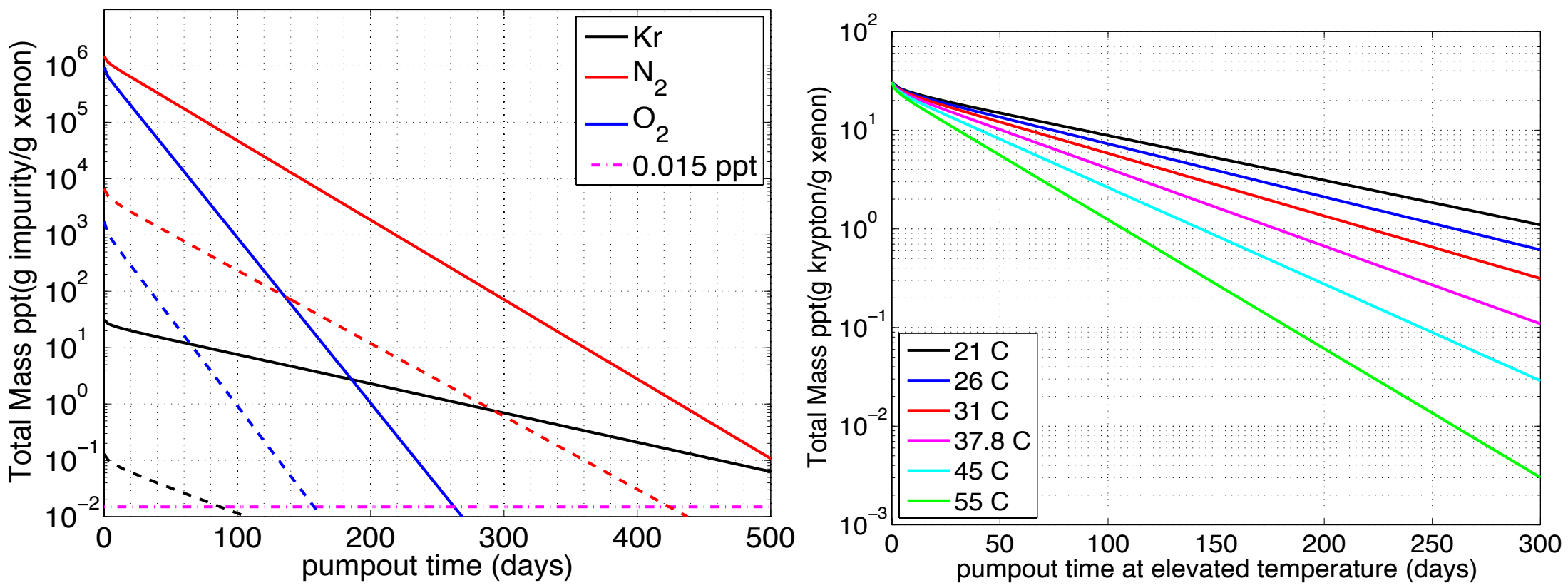

**Figure 9.7.1. Left: Impurity burden of the LZ detector components as a function of outgassing time while pumping at room temperature. The quantity of impurities remaining in the plastic components is compared to the total LZ Xe mass stockpile (10 tonnes). The 2-cm-thick PTFE reflectors and the polyethylene HV cable are included in this calculation. Solid lines: summed impurity burden of both components. Dashed lines: impurities in the HV cable alone. Right: Kr burden of the detector materials as a function of outgassing time at several elevated temperatures.**

**Table 9.7.3. Outgassing time constants of two important plastic components of the LZ TPC.**

| | PTFE reflectors (2 cm thick) | Polyethylene HV cable (1.3-in diameter) |
|---|---|---|
| | τ (days) | τ (days) |
| $N_2$ | 31.1 | 33.5 |
| $O_2$ | 15.0 | 14.0 |
| Kr | 84.4 | 48.6 |

This calculation indicates that even at room temperature, the $O_2$ and $N_2$ burdens of the materials will drop rapidly due to their substantial diffusion constants. The total amount of $O_2$ present in the detector material is expected to approach a mere 1 ppb compared with the total Xe mass (10 tonnes) in 100 days of pumping, indicating that the recirculation system should have no difficulty reducing the equilibrium value in the Xe far below the goal of 0.4 ppb. These expectations will be tested during the initial pump-out of the detector by monitoring the $O_2$ level with the Xe gas-sampling system.

The krypton burden, however, is difficult to reduce significantly by outgassing alone due to its low diffusion constant in PTFE at room temperature ($5.6 \times 10^{-8}$ cm$^2$/sec, compared with $31 \times 10^{-8}$ cm$^2$/sec for $O_2$) and because of the substantial thickness of the PTFE components (2 cm for the TPC reflector rings). Figure 9.7.1 (left) indicates that 100 days of room-temperature pumping reduces the dissolved krypton to ~9 ppt when compared to the Xe mass, still ~600 times higher than the krypton concentration goal of LZ. At this point, the krypton outgassing rate of the entire detector is expected to be about 35 ppt/year, dominated by the 2-cm-thick detector PTFE parts. We neglect the thin PTFE liner on the inner surface of the cryostat vessel because the time constant is expected to be much shorter. Also, the krypton content of the polyethylene HV cable will be reduced in this time to ~0.01 ppt compared to the total Xe mass, effectively eliminating it as a long-term concern. Elevating the outgassing temperature to 45 $^{o}$C is expected to reduce the PTFE content to ~2.6 ppt in 100 days, a modest improvement, as shown in Figure 9.7.1 (right).

Fortunately, for those components such as the PTFE reflectors that will be immersed in the cold LXe, the krypton diffusion constant will be dramatically lower during physics running because it obeys an Arrhenius equation: $D(T) = D_0 \exp(-E_a/k_BT)$, where $k_B$ is the Boltzmann constant, $E_a$ is a characteristic activation energy, and $D_0$ is the diffusion constant at infinite temperature. This will largely immobilize the krypton inside the plastic components at LXe temperature.

Specifically, to maintain our krypton goal throughout the physics run we require the total krypton outgassing rate to be < ~0.005 ppt/year, implying a PTFE diffusion constant suppression factor of about $10^4$ compared to room temperature. With a $\Delta T$ of ~120 K between 298 K and ~178 K, this implies $E_a$ > 0.17 eV. (Typical values for $E_a$ are in the few eV range). We are working to measure $E_a$ in a representative sample of PTFE so that this effect can be fully quantified. Other components of the detector that remain partially at room temperature, such as the HV cable and the PMT and instrumentation, must be fully outgassed prior to filling the detector with Xe so that their krypton burden does not impact the experiment.

We note that the toy impurity model discussed here is useful for guiding the design and implementation of the purification system, however it is not comprehensive. It assumes that the pumping geometry is ideal, that the component shapes are simple, and that the only relevant impurities are krypton, $O_2$, and $N_2$. In fact, there is evidence from the field dependence of the electron lifetime in LUX that the dominant impurity species is not $O_2$. We are working to benchmark the model to the LUX experience to gain further insight into the reliability of the model. Ultimately, the outgassing of the LZ TPC will be carefully monitored via the Xe gas-sampling system beginning when the cryostat is first sealed and evacuated. The bulk of the LZ Xe stockpile will not be introduced into the detector until the krypton outgassing is properly characterized *in situ* and deemed acceptable both at room temperature and at LXe temperature.

## 9.8 Cryogenics

Cooling power to maintain Xe in the liquid phase is provided by a cryogenic system that distributes nitrogen in gas and liquid phases. By utilizing the approximate 100 K temperature difference between the nitrogen and Xe evaporation temperatures, sufficient gradient exists to provide for thermal control and temperature modulation. Existing infrastructure from LUX, including 450-liter LN storage tanks; VJ pipe; and miscellaneous valves, sensors, and fittings are modified and re-utilized to provide a front end to the primary cooling equipment for the experiment while providing supplemental LN storage. During operations, distribution of LN is from a VJ central 1,000-liter distribution tank that is co-located with a Stirling cycle cryocooler. The cryocooler liquefies cold evaporated nitrogen gas in a closed-loop cycle operating at atmospheric pressure. Multiple thermosyphon heat pipes are connected to the LN in the storage tank that is used as a heat sink to provide for heat removal at four locations within the Davis Cavern. Those four locations are the detector, HV feedthrough, LXe tower, and Xe sampling system. The underground installation is shown in Figure 9.8.1.

Exterior to the Davis Cavern is an LN storage room that contains four 450-liter LN storage tanks mounted on mass scales for monitoring LN consumption. A commercially purchased VJ piping system that has a complete implementation of control valves, relief valves, and pressure-monitoring equipment connects the tanks to equipment in the Davis Cavern. A separate small-diameter tube furnishes boil-off nitrogen that is utilized to purge the freeboard of the water tank and the water-purification system vacuum pump. The current four storage tanks are required to provide an initial liquid volume for startup of the cryocooler and to provide backup to the cryocooler until a second cryocooler is procured at a future date. Alterations to the VJ piping system are required to allow for rerouting of LN to new destinations in the Cavern that will differ from the LUX LN use locations. The rail-mounted 1,100-liter LN storage tank utilized by LUX will provide for the resupply of LN from the surface during brief interruptions in the operation of the underground cryocoolers.

A cryocooler based on the Stirling thermal cycle is selected for cooling of LZ. It is depicted schematically in Figure 9.8.2. Included in the cryocooler design is a cryogenerator, a work platform sized for two cryogenerators, a 1,000-liter LN storage tank to provide a distribution reservoir, and a complete monitoring and control system. The design of the cryocooler is closed cycle. Nitrogen inside the storage tank is reliquefied rather than being vented to the cavern as was done during LUX operations. The cryocooler is capable of 1000 W of cooling power at the 77 K boiling temperature of LN at atmospheric pressure. Preliminary estimates of total heat load on the cryogenic system are between 700 and 900 W. The 1000 W configuration provides a margin against unanticipated heat load.

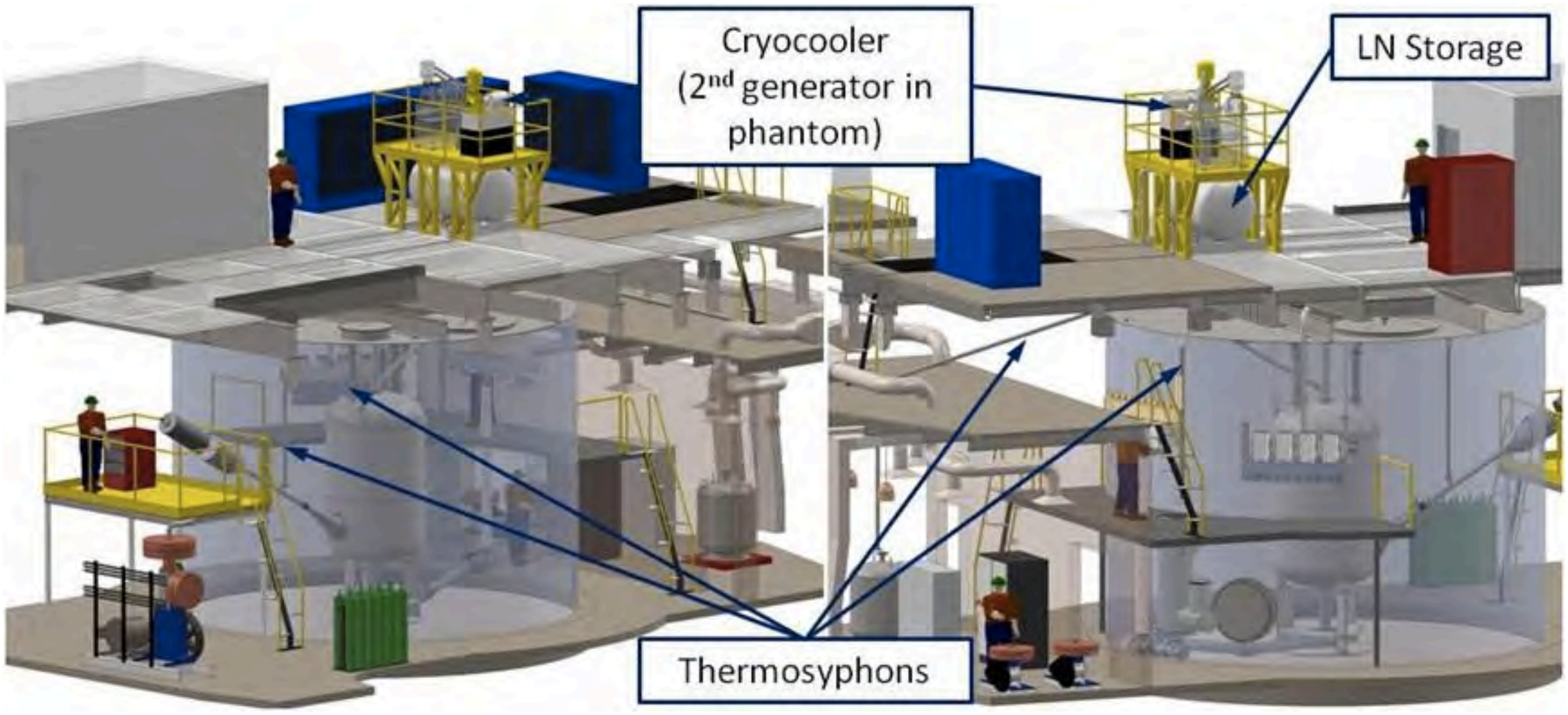

**Figure 9.8.1. Cryogen system cavern layout — two views.**

Cryocoolers based on the Stirling cycle have additional advantages: quick startup, low energy consumption (each unit draws 11 kW of electricity), and variable-drive motors that allow adjustable cooling levels below 1000 W. These units have been deployed at many institutions worldwide, including installations at SNOLAB; Gran Sasso (Icarus); and multiple university laboratories in the United States,

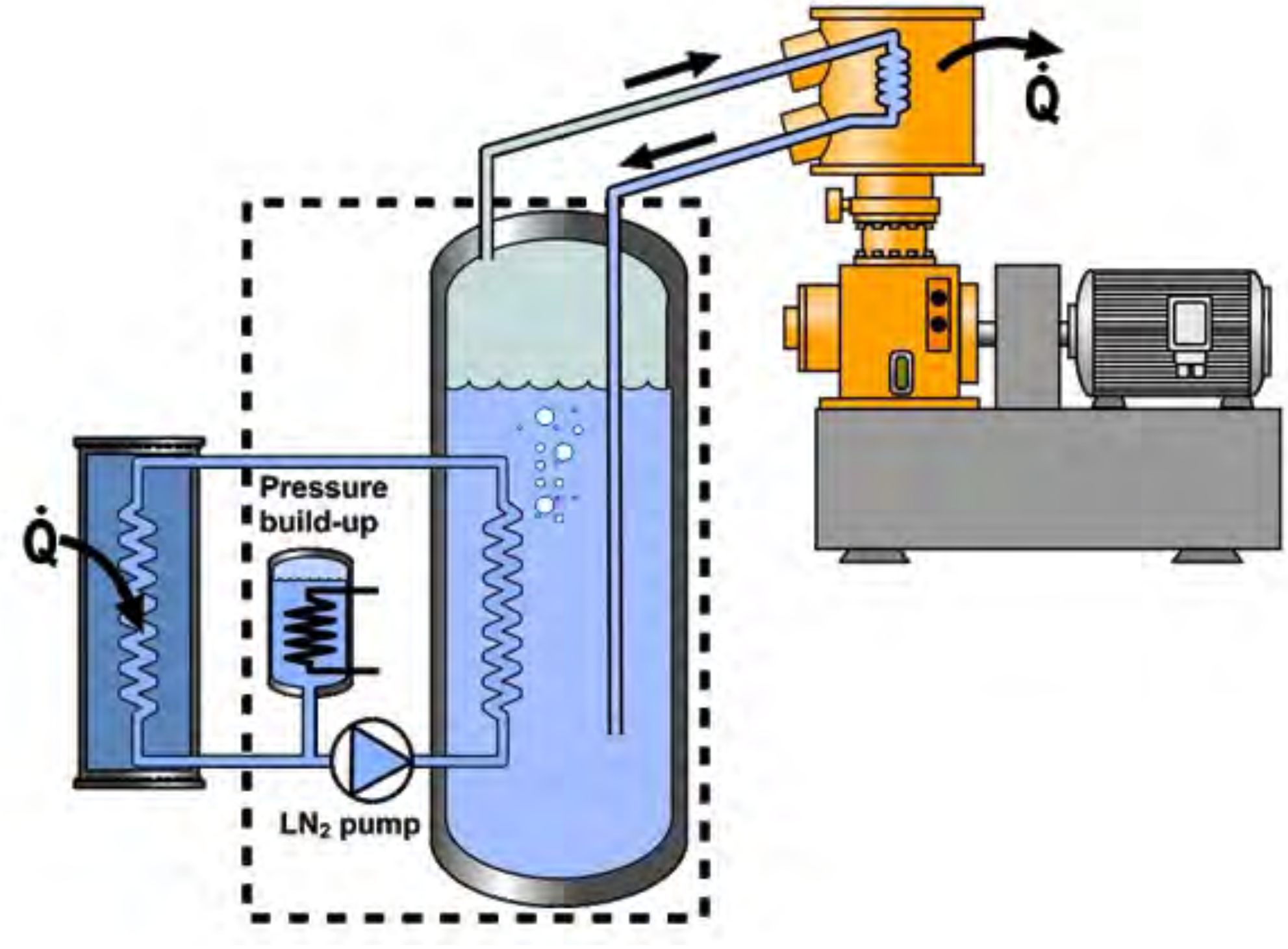

**Figure 9.8.2. Cryocooler schematic. A cryogenerator removes heat, Q, from a closed LN storage reservoir. Other devices requiring heat removal are thermally connected to that reservoir. All of the systems are closed-loop.**

Russia, and Asia. Maintenance of the cryocooler is required after 6,000 hours of continuous use, so the system is designed for the future addition of a second cryocooler (likely purchased as a component of operations). Transport of LN from the surface to the Davis Campus can sustain operations during short-term maintenance.

LUX receives deliveries of LN two to three times per week. Even though LZ is better insulated, scaling indicates that LZ requires one LN delivery every 24 to 36 hours to sustain operation. For long-term operations, that would increase the inherent transport risk and manpower requirements related to LN transfer. Therefore, the addition of a second cryogenerator at some point would significantly reduce the risk to long-term operations.

Distribution of cooling power from the cryocooler to the Xe-containing experimental systems is accomplished by use of thermosyphons. A thermosyphon is a type of heat pipe. Thermosyphon technology was successfully applied to LUX following development at Case Western Reserve University [12]. Thermosyphons comprise condenser and evaporator heads connected by small-diameter tubing wrapped with multilayer insulation (MLI) and charged with nitrogen gas at modest pressure. In this application, the condenser is immersed in an LN bath, causing the nitrogen gas in the thermosyphon to liquefy. The liquid flows in the small-diameter tubing by gravity to the evaporator that is physically attached to the device that requires cooling. Heat from the evaporator causes the LN pooled in the evaporator to change phase, removing heat via the nitrogen latent heat of vaporization. The warmed nitrogen gas then returns to the condenser by buoyancy effect.

Cooling of the Xe relies upon the temperature difference between the LN and Xe vaporization temperatures as well as the adjustable cooling power of the thermosyphon. Changing the mass of circulating nitrogen can modulate the thermosyphon cooling power. As the mass is increased, pressure in the condenser rises, effectively increasing the boiling temperature in the thermosyphon and inducing an increased heat flux to the LN bath that remains at atmospheric pressure and 77 K.

Thermosyphon cooling is completely passive, with a fixed amount of nitrogen in the secondary cooling loop. Therefore no pumps are required and there is no direct path for large quantities of LN (typically stored in dewars) to get into the Xe via a leak. As the thermosyphon transport tubing may be 9.5 mm or 12.7 mm outer diameter (OD), a minimal amount of nitrogen can be placed into the tubes, on the order of 15 grams, to reach 10 atma. This mass gradually increases as the nitrogen condenses but the mass remains modest. Overpressure protection is by relief valves and burst disks. The thermosyphons are charged with nitrogen via high-pressure cylinders. Control of the gas mass (and cooling power) is accomplished with pneumatic valves, mass flow controllers, sensors, and processing devices connected to slow control via Ethernet. Thermosyphons deployed in LUX were easily capable of delivering more than 200 W cooling for each deployed evaporator head. The single largest device needing cooling is the LXe tower, in which two-phase and gas-gas heat exchangers are anticipated to require 300-400 W of cooling to recondense Xe returning to the detector. Two thermosyphon evaporators are able to provide that energy removal.

Vacuum-pumping systems take advantage of tubing in place to route thermosyphons and Xe transfer lines. Packages of combined scroll pumps and turbomolecular pumps are deployed on the decking of upper Davis Cavern and near the Xe heat exchanger tower in lower Davis Cavern. These pump combinations evacuate: the annulus between the inner and outer vessels, detector internals, Xe heat-exchanger tower, Xe gas-sampling system, and the entire Xe gas system. Vacuum pumps remaining from LUX operations are utilized. Two vacuum leak-check carts are provided so that there is a minimum of one cart available at any time in the Surface Assembly Laboratory and the underground Davis Campus. A variety of gauges are deployed, including manual gauges (that can show vacuum in power outage), thermocouple gauges, and ion gauges.

PMT and signal cables from both the inner and outer cryostats must be routed to locations where they can transition from either a pure Xe or high-vacuum environment to atmospheric conditions in the Davis Cavern. PMT and signal cables from the inner cryostat are routed through VJ conduits. Conduits exit the cryostat from both the top and bottom and are routed to a SURF-installed mezzanine level on the north side of the water tank. The inner tube of the VJ is sized to allow pulling of all cables during installation. The outer jacket has "tee" connections to the vacuum-pumping system. Breakout boxes are installed on the conduits at the mezzanine for installation of hermetic feedthroughs at the point where cables must be terminated in order to exit the Xe space. Breakout boxes and inner conduits are compatible with the 3.4-atma maximum Xe pressure. Signal cables from the vacuum space are similarly routed through a vacuum conduit to another breakout box, with hermetic feedthroughs to keep the vacuum space sealed.

Control of temperature, pressure, and liquid levels within the cryogen system is needed both for the correct operation of detector systems and for safe operation. Control of these parameters is via connection to the slow control system described in Chapter 11. Sensor deployment is estimated largely based on a scale-up of LUX use. Each thermosyphon utilizes multiple thermometers, pressure transducers, a vacuum gauge, a mass flow controller, and controller boxes to digitize signals and transmit via Ethernet. Similarly, the LN storage room, VJ distribution piping, cryocooler, and vacuum pumping system need active monitoring and control of temperature and pressure.

## 9.9 Xenon Procurement

Approximately 10 tonnes ($1.8 \times 10^6$ gas liters) of Xe must be procured for the LZ detector. This assumes that the current Xe inventory in LUX (about 370 kg) and from other sources (primarily at Case Western Reserve and SLAC, about 150 kg) will be used. In our baseline plan, the South Dakota Science and Technology Authority (SDSTA) will procure most of the Xe. About 30% of the Xe would be an in-kind contribution from collaborators in China and Russia. Significant funding from the U.S. funding agencies would be reserved as contingency to augment the funding from SDSTA (see Chapter 17). We note that there is substantial flexibility with regard to timing and procurement arrangements through SDSTA.

The historical price of Xe has fluctuated by roughly a factor of 5. The SDSTA commissioned a private study of the Xe market and pricing by an experienced consultant. This 50-page report was delivered in October 2013 and documented the sources of Xe, the current usage, and predicted future usage. An updated was issued in October 2014. The current production of Xe is about $10 \times 10^6$ liters per. The demand from the lighting industry is anticipated to decline (as LED lighting becomes more prevalent, supplanting Xe lighting) and a modest growth is foreseen in the supply side as new capabilities come online.

The procurement in the United States would be dictated by availability and pricing, since the time needed for Kr removal for the full 10 tonnes is less than six months. We have engaged the services of the experienced consultant referenced above to facilitate interactions with the Xe vendors, including sources in Russia and Ukraine. Russian collaborators own a modest stock of Xe for current experiments. This, together with some additional purchases, makes up the roughly 1 tonne of the Russian in-kind contribution, which includes the 0.3 tonnes already in hand. Similarly we are in discussions with potential collaborators in China to join LZ. They currently have about 1.2 tonnes of Xe and there is significant availability of Xe in China.

We have reached out to some other potential scientific users of large amounts of Xe. This includes a NASA proposal for an asteroid mission that would use Xe as a propellant and the nEXO collaboration. A joint statement of need has been created and is being used in discussions with potential vendors.

## Chapter 9 References

# 10 Calibration Systems

A rigorous calibration strategy is required for the unambiguous direct detection of dark-matter particle interactions in the LZ detector. The basic questions about any event in the detector are: (1) How did the particle interact, and (2) how much energy did it deposit? LZ has adopted a comprehensive calibration strategy in order to accurately answer these questions, achieve its science goals, and be ready to address the widest possible range of predicted dark-matter signatures. The principal calibration techniques planned for LZ have been used successfully in previous experiments, especially LUX.

## 10.1 Overview

This section is an overview of the essential calibrations LZ will require. Subsequent sections contain details of how each calibration method will be implemented.

### 3-D (x,y,z) Position Reconstruction

As described in Chapter 3, signals in LZ consist of scintillation photons (S1) and ionized electrons (S2). As digitized, these signals will exhibit spatial (x,y,z) variations in both S1 and S2, specifically:

1. Variation in the (x,y) S2 response due to position-dependent light collection, as well as small non-uniformities in the electroluminescence region. This is expected to be an O(10%) effect.
2. A z-dependent decrease in the S2 response, due to the capture of drifting electrons by residual electronegative impurities. This is generally characterized by the free-electron lifetime.
3. Variation in (x,y,z) S1 response due to the optical properties and photon collection of LZ. This is expected to be as large as a factor of 2.

Item 1 is a detector property and not expected to exhibit significant time dependence. The electron lifetime depends strongly on Xe purity, and the light collection depends weakly on Xe purity. Xenon purity is expected to be time-dependent during operation, which means that periodic calibrations of the detector response are essential.

Calibration and correction for variations of the photomultiplier response are discussed in Chapter 6. Once this PMT calibration is done, the next step in calibrating the LZ detector is to measure the normalization of the (x,y,z) response using an internal $^{83m}$Kr gamma source, as used successfully in LUX. The primary result of this calibration will be a high-statistics map of the absolute S1 and S2 photon collection. The $^{83m}$Kr source will be quasi-homogeneously distributed within the TPC volume due to convective mixing. The LUX experience shows that sufficient statistics are obtained from all volume elements of the active TPC with a single injection prior to decay below quiescent background.

The free-electron lifetime, $\tau$, is most directly measured from the z-dependence of a mono-energetic gamma source. The internal $^{83m}$Kr decay, though not strictly mono-energetic in S2, satisfies the requirements of this important calibration. This is because the two-step decay is localized to less than a micron spatially, and has a half-life between decays of 150 ns. In the S2 channel, these two steps appear as a single energy deposition.

### Definition of Background and Expected Signal Bands

The dominant background in the WIMP search energy range will be low-angle Compton scatters. An internal tritium source will be used to provide a high-statistics definition of the distribution of background events. This has already been successfully demonstrated in the LUX detector, and LZ will use the same technique. A suite of externally deployed neutron sources will be used to provide a high-statistics calibration of the response to WIMP scatters, since in both cases the interaction with a Xe nucleus results in a recoiling Xe atom.

An example of the response from a tritium source (proxy for ER background) and neutron sources (proxy for signal) from the first run of LUX is shown in Figure 10.1.1.

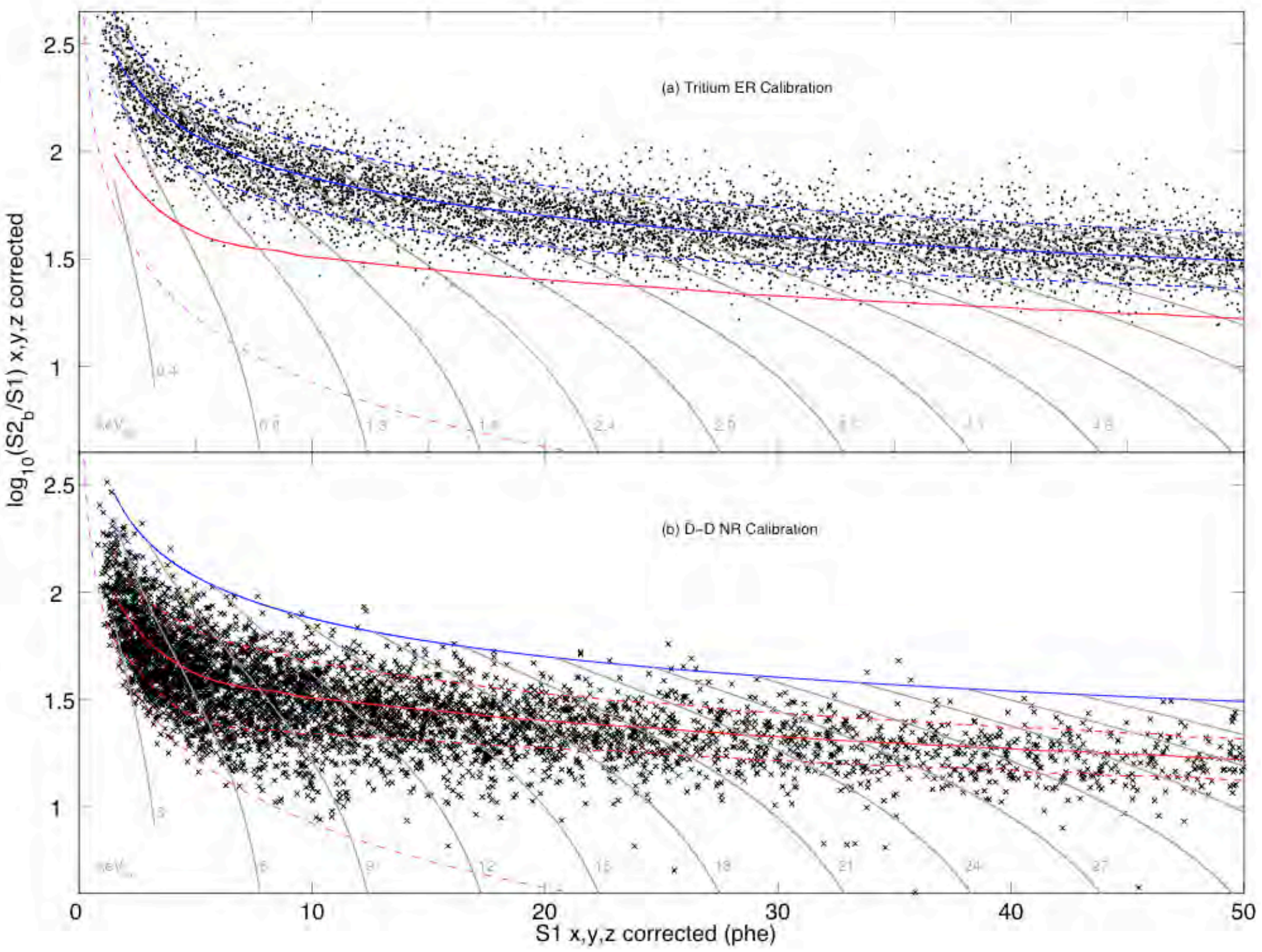

**Figure 10.1.1. Calibration of the background (top panel) and expected signal (bottom panel) obtained by the LUX detector [1]. Inset keV contours show the energy scale reconstructed from the formula given in the text.**

### Measurement of Background Discrimination

An in situ measurement of the discrimination against ER background will be obtained from the observed response to calibration sources, as shown in Figure 10.1.1. It has been conventional to define this metric for 50% acceptance of broad-spectrum nuclear recoils (NRs). We will follow this convention to ensure we have met our design goal, but will additionally employ more sophisticated methods to optimize the discrimination as a function of energy deposit.

### Energy Reconstruction of Expected Signal

The reconstructed energy scale depends on the incident particle type. The energy scale for electromagnetic interactions is linear and independent of the applied electric field, when reconstructed from the sum of photons ($n_{ph}$) and electrons ($n_e$): $E = \varepsilon(n_e + n_{ph})$. The average energy to produce a single quanta is $\varepsilon = 13.9$ eV, with an uncertainty of a few percent [2]. A number of Xe activation lines, in conjunction with the metastable internal $^{83m}$Kr source, will be used to calibrate the absolute detector response for a given energy deposit.

WIMPs are generally expected to interact with the atomic nucleus. Nuclear-recoil energy reconstruction is best obtained from $E = \varepsilon(n_e + n_{ph})/f_n$, where $f_n$ is the quenching factor for NRs. This is the fraction of the NR energy that is ultimately transferred to electrons, and is thus measurable. It is mildly energy-dependent, and appears to be consistent with the expectations of Lindhard theory [3], namely, $f_n \approx 1/5$ [4].

As described in Chapter 3, numerous measurements of the NR quenching factor in LXe exist. LZ will not need to simply rely on these measurements, but instead will be able to make various in situ measurements.

### In situ Nuclear Recoil Calibration

The in situ NR calibration involves three components:

1. Measurement of the S2/S1 response of LZ for NRs from threshold to several hundred keV, as demonstrated in Figure 10.1.1.
2. Measurement of NR endpoints at 4.5 keV (YBe), 40 keV (AmLi), and 74 keV (DD generator).
3. Measurement of low-energy NRs via tagged low-angle scattering of DD neutrons. This type of in situ measurement was pioneered by LUX, and will be improved upon by LZ.

The 74 keV endpoint in particular (from 2450 keV DD neutrons) will also be used as a "standard candle" for translating between LZ response and ex situ calibrations. The source offers a robust, precise feature far from detector threshold (cf. Figure 10.3.1). Recent advances in the understanding of energy reconstruction in LXe have rendered the standard candle less of a requirement and more a provision for redundancy. For example, LUX was able to rely on absolute measurements of $n_e$ and $n_{ph}$, along with the simple model described in [4], as implemented by the Noble Element Simulation Technique (NEST) [5].

Previous experiments [6] have used an electromagnetic source as standard candle. Drawbacks to that approach are a strong dependence on the applied electric field, and the difficulty of introducing an electromagnetic signal into a detector of the size of LZ. LZ will bypass these systematic uncertainties by using an NR energy deposition as the standard candle. Previous work has shown that NRs of this energy are only very weakly affected by the applied electric field [7].

### Ex situ Nuclear Recoil Calibration

To fully exploit the discovery potential of LZ, a detailed understanding the NR response of LXe at the smallest energies is needed. LZ collaborators have proposed a series of dedicated measurements of the NR signal from very-low-energy NRs, to be performed in an external test bed. An ideal source of the requisite low-energy neutrons is the $^{7}$Li(p,n) reaction. As described in [8,9], this reaction provides highly mono-energetic, collimated neutrons at either 24 keV or 73 keV. The neutrons will produce Xe recoils with endpoint energies of 0.73 and 2.2 keV. For direct comparison, the test bed will also measure the DD endpoint at 74 keV.

### Time Stability

Stability of the detector response in time will be monitored by regular calibrations with an internal metastable krypton source, as described previously and in Section 10.2.

### Xenon Skin and Liquid Scintillator Veto

The LZ design goal conservatively assumes a 100-keV energy threshold in the Xe skin region. Attainment of this goal will be demonstrated using a suite of gamma sources: $^{133}$Ba (356 keV), $^{57}$Co (122 keV), $^{207}$Bi (83 keV), and $^{241}$Am (60 keV).

Calibration of the liquid scintillator (LS) will be demonstrated using the same radioactive sources used for the Xe skin. This is facilitated by the fact that the source tubes are positioned between the Xe skin and the scintillator veto in the vacuum space between the inner and outer cryostat vessels.

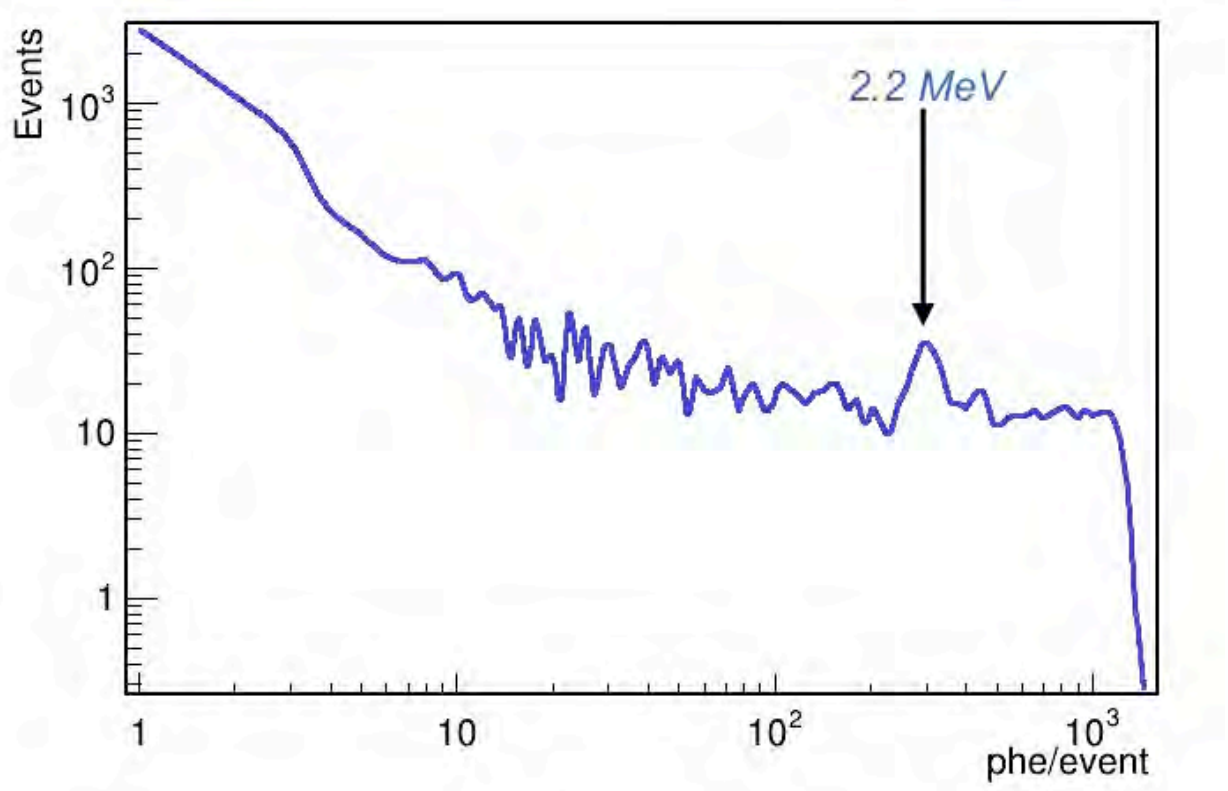

**Figure 10.1.2. Preliminary simulation of the LS outer detector, showing response to neutrons. Capture on hydrogen results in 2.2 MeV, indicated by the arrow.**

The LZ design goal also assumes a 100-keV energy threshold in the LS veto component of the outer detector. This is quite conservative, considering that detailed Monte Carlo studies indicate an average signal response of >10 photoelectrons per 100 keV of deposited energy (see Chapter 7). Additional higher-energy calibration of the light yield will be obtained from neutron capture on hydrogen (2225 keV), $^{60}$Co(2506 keV from two distinct gammas), and neutron capture on gadolinium (approximately 8 MeV gamma cascade). These calibrations have previously been used successfully by Daya Bay. We note that the LZ outer detector does not capture all the energy from high-energy gamma cascades as frequently as does the thicker Daya Bay Antineutrino Detectors. Nevertheless, such neutron-capture events will still provide important veto capabilities.

## 10.2 Internal Calibration Sources

As discussed in the next section, the self-shielding ability of LXe is a double-edged sword. External calibration sources face severe difficulty in probing the central detector volume. Low-energy calibrations in the detector's fiducial volume can only be accomplished through long (>>1 day) exposure to high-energy (>1 MeV) sources. Internal calibration sources are thus essential to accomplishing the science goals of LZ.

### Metastable Krypton ($^{83m}$Kr)

Metastable $^{83m}$Kr has served as the calibration workhorse of LUX. Its homogenously distributed low-energy decays enable the construction of high-resolution 3-D maps of S1 and S2 detector response. An illustration of this mapping in LUX is shown in Figure 10.2.1 for the z dependence of the signal, and Figure 10.2.2 for the (x,y) dependence of the S1 signal.

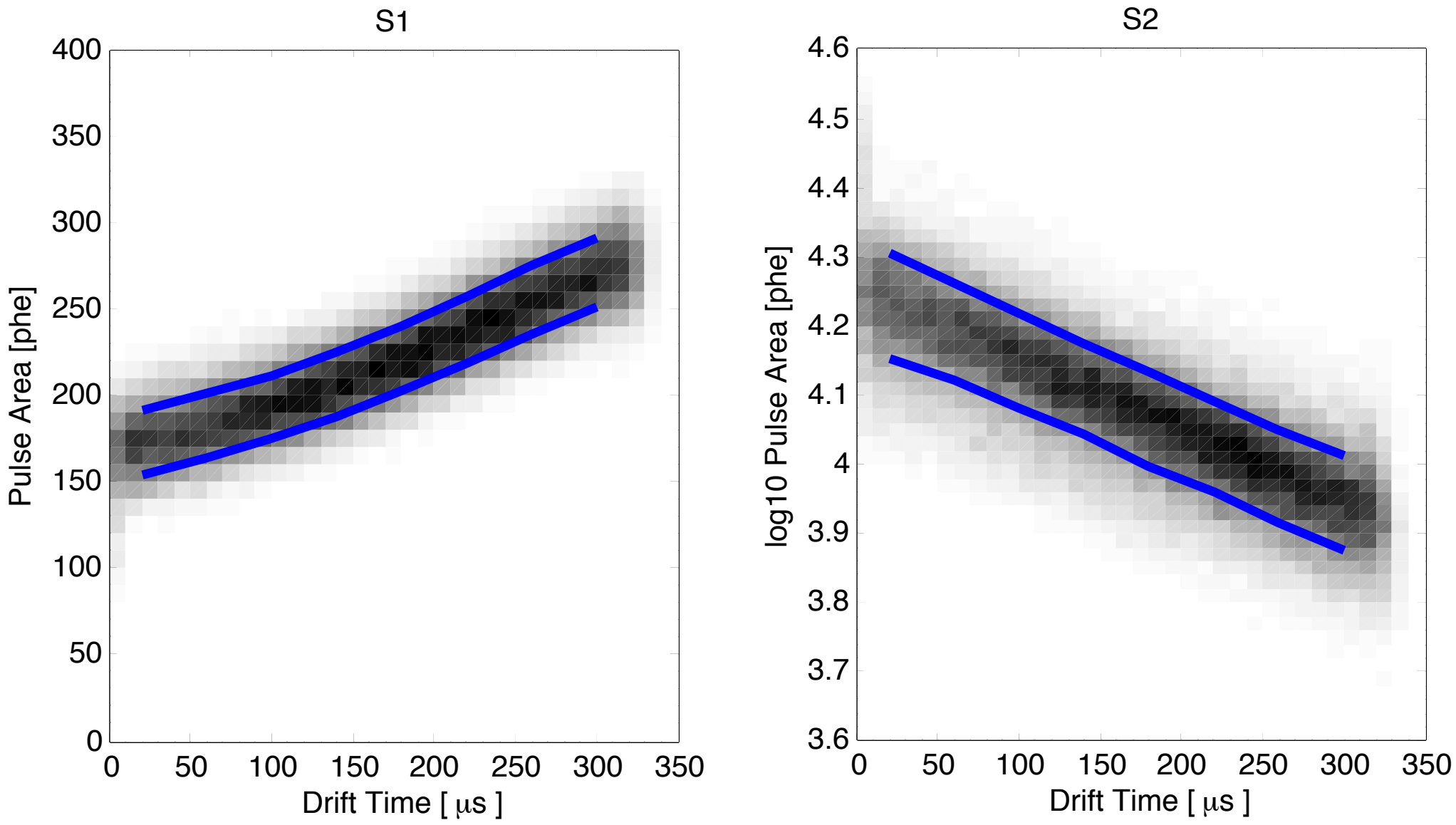

**Figure 10.2.1. Data from the LUX detector, showing the S1 (scintillation) and S2 (electron) response to $^{83}$Kr, as a function of the event drift time. The source gives a direct calibration of the detector response, which varies significantly as a function of the z coordinate. Blue lines indicate the 68% contours.**

The $^{83m}$Kr will pass through the getter in the Xe handling system, to remove any impurities. Its 1.83-hour half-life is short enough to allow an injected dose to decay quickly below quiescent background, but is long enough to be easily injected and observed. $^{83m}$Kr is continuously produced within a simple generator consisting of $^{83}$Rb-infused charcoal. $^{83}$Rb exhibits a conveniently long half-life of 86.2 days. $^{83m}$Kr decays consist of a distinctive and easy-to-tag two-step process: A 32.1-keV de-excitation is followed by a 9.4-keV transition to ground, with half-life between the two of 154 ns. In LZ, as in LUX, the two S1 signals will be separately resolvable, but the two S2 signals will appear as one combined pulse with the total energy of 41.6 keV.

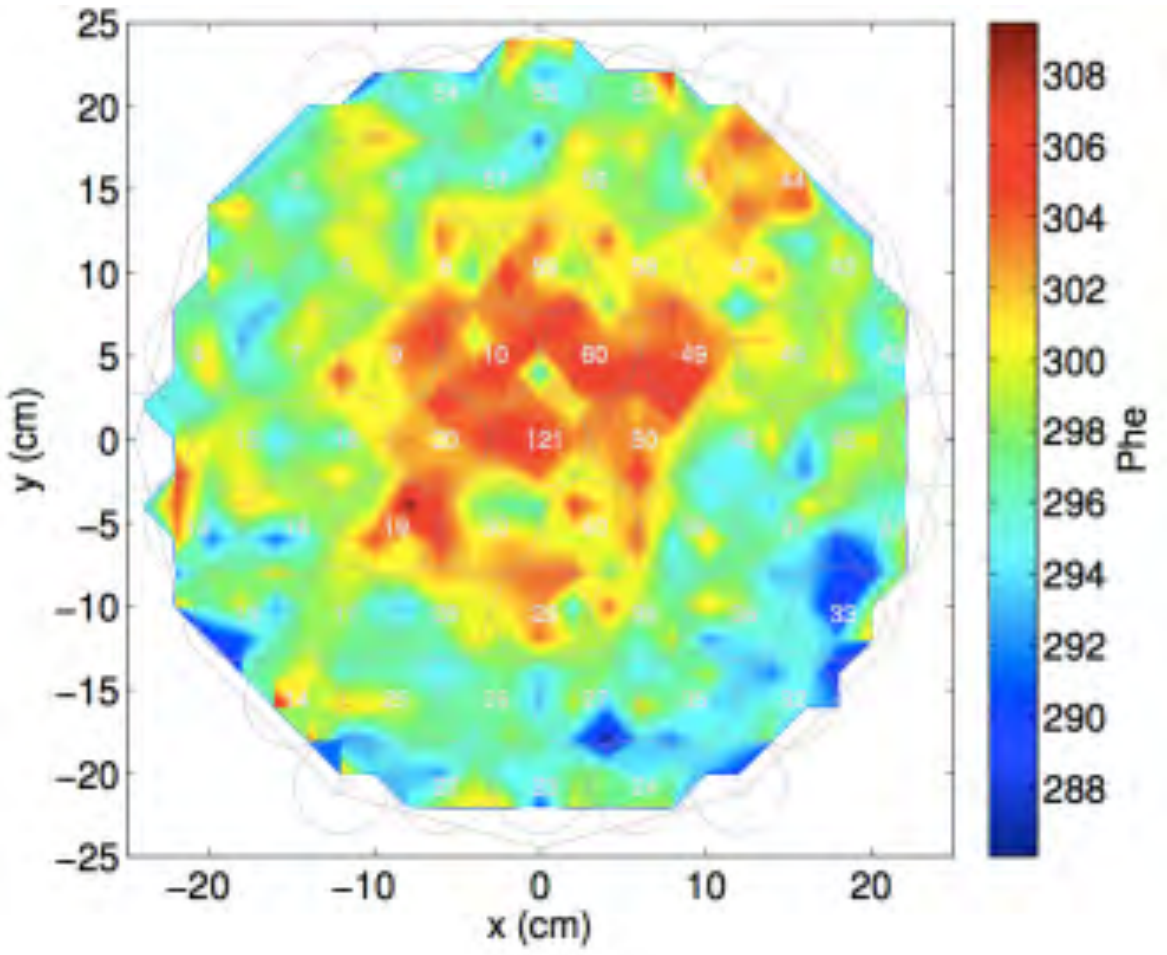

**Figure 10.2.2. Data from the LUX detector, showing the S1 (scintillation) response to $^{83m}$Kr as a function of (x,y).**

Regular twice-weekly injections have been used successfully in LUX to monitor the response of the detector. The primary utility of this is to measure the free-electron lifetime, which depends on Xe purity. Similarly, this temporal calibration can detect any variation in the photomultiplier response. Such variations would likely indicate photocathode degradation, which has been observed in, e.g., the XENON10 experiment (though not in LUX).

#### Tritium ($^{3}$H)

As shown in Figure 10.1.1, LUX has recently demonstrated the first successful calibration of the low-energy ER band using $^{3}$H in the form of tritiated methane. The same technique will be employed for LZ. This calibration depends on the ability to completely remove the $^{3}$H after the calibration, since the 12.3-year half-life and 18.6-keV endpoint would otherwise be catastrophic for a successful dark-matter search. LZ will have more stringent background requirements than LUX, and careful measurements have been made to ensure efficient methane-removal capabilities are implemented, consistent with these requirements.

Injecting tritium in the form of tritiated methane ($CH_3\,^{3}H$) enables the rapid removal of the tritium, both because getters are extremely effective at adsorbing methane, and because methane has a low absorptivity into PTFE detector components relative to bare tritium. Dedicated tests of methane-removal capabilities have shown efficient and complete removal from a small LXe detector test bed. The LUX experiment observed complete removal with a 6.7-hour half-life [10], while LZ is planning for a somewhat longer purification timescale of ~2 days.

#### Activated Xenon

The detector medium itself has been used as an internal calibration source in Xe-based detectors [11], taking advantage primarily of the metastable isotopes $^{129m}$Xe (236 keV, 8.9-day half-life) and $^{131m}$Xe (164 keV, 11.8-day half-life). These states result from fast neutron exposure, and are populated by cosmogenic (muon-induced) neutron activation. These activated Xe states are useful, but not ideal, as their energies are much higher than the WIMP-search region of interest. Recently, LUX has observed the electron capture lines from $^{127}$Xe [12]. These are mono-energetic at the binding energy of the Xe atom, and thus in principle very useful for calibrating the low-energy ER response. However, the isotope decays with a 35-day half-life, and so is also a potential background. LZ is taking precautions to minimize cosmogenic activation of the Xe (see Chapter 9). If $^{129m}$Xe and $^{131m}$Xe could be produced through specialized neutron exposures that avoid $^{127}$Xe production, these 236-keV and 164-keV decays could be useful measures of the Xe skin response. The practicalities of this specific activation are currently under study.

#### $^{220}$Rn

Some regions of the Xe skin may have poor light-collection efficiency, requiring an even higher-energy internal source for successful calibration. We plan to use $^{220}$Rn (6.4 MeV alpha emitter, 56-s half-life) to study skin regions. $^{220}$Rn can be generated using a thorium oxide powder generator, and results in no long-lived daughters (stable $^{208}$Pb is reached in 11 hours). The practicalities of $^{220}$Rn are currently under study, particularly the question of whether this source is sufficiently dispersible given its short half-life.

#### Plumbing, Instrumentation, and Sources

The internal calibration system consists of both the sources themselves and the requisite injection plumbing, shown schematically in Figure 10.2.3. The $^{83m}$Kr generator is constructed by placing $^{83}$Rb (in aqueous solution) on clean charcoal, and then placing that charcoal inside a section of tube sealed at both ends by particulate filters (to prevent the loss of charcoal). The $^{83}$Rb activity is in the μCi range, practically obtained by placing a discrete number of drops of solution onto the charcoal. $^{83}$Rb emits gammas at 520.4, 529.6, and 552.6 keV, which are convenient for measuring the activity of the $^{83m}$Kr generator. Because of the anticipated high frequency of $^{83m}$Kr injections, the LZ $^{83m}$Kr injection system will be fully automated: Pneumatic valves on either side of the generator will be remotely controllable, and the dose activity will be controlled through this valve timing in conjunction with an output-side low-flow mass flow controller. Additional R&D will precisely calibrate the delivered dose activity.

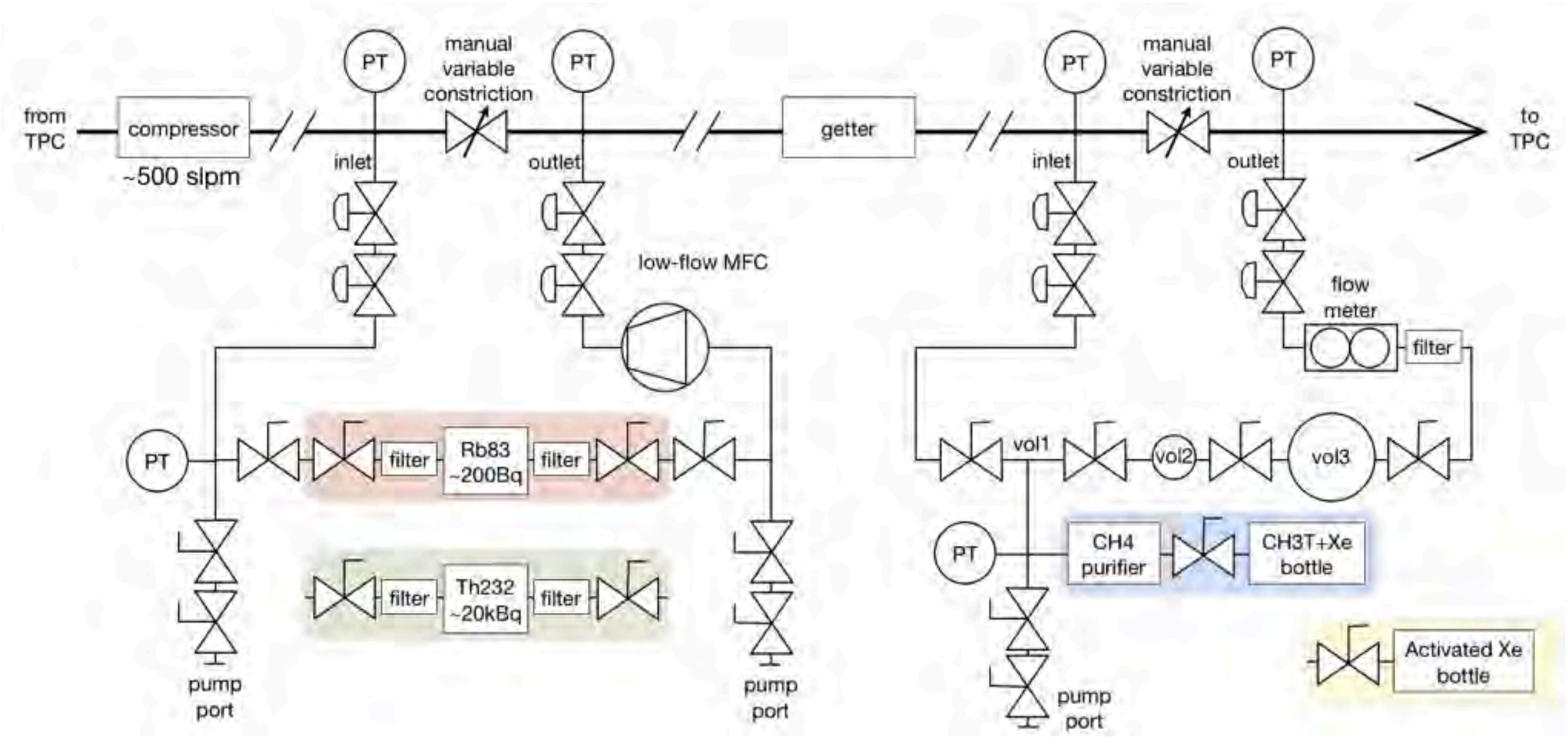

**Figure 10.2.3. Schematic of the internal source handling and deployment.**

The $CH_3{}^3H$ source system consists of a storage bottle containing a mixture of purified Xe and a tiny quantity (about one pCi) of tritiated methane. A series of dose volumes are defined, such that the storage-bottle activity can be carefully injected in increasingly active steps (to avoid accidentally overdosing the detector). Because methane cannot pass through the getter, this injection must take place downstream of the getter. Studies at the University of Maryland have shown that non-methane hydrocarbons can be present in any methane sample, which is remedied by employing an inline methane purifier.

$^{220}$Rn and activated Xe will only be injected very occasionally (if at all) and can be injected using the $^{83m}$Kr and $CH_3{}^3H$ injection plumbing, respectively.

The dispersal of injected sources within the LXe of the TPC is an object of study and design within WBS 1.5. We require mixing times similar to the decay half-lives of the injected sources.

## 10.3 External Calibration Sources

LZ will have three vertical source tubes in the vacuum space between the inner and outer titanium cryostats, as discussed in Chapter 8. The source tubes are constrained to have an inner diameter of 30 mm, large enough to accommodate deployment of commercial sources in nearly all cases. The necessary source strengths are well within the available range. Sources that cannot be obtained commercially in our requisite form factor and rate will be fabricated by LZ. The source tubes will be sealed at both ends and pumped out when not in use, to mitigate the plating of radon daughters. Additionally, two of the neutron sources (YBe and DD) require their own dedicated conduits, as described below.

In this section, we first describe the physics requirements met by each source. We then describe the physical deployment of the sources.

### Neutron Sources

A suite of four neutron sources will provide a broad-spectrum NR calibration, with the additional benefit of four distinct kinematic endpoints, as shown in Figure 10.3.1.

### Americium Beryllium (AmBe) and Americium Lithium (AmLi) Neutron Sources

AmBe neutron sources have typically been the broad-spectrum neutron source of choice, as they cover the range, from threshold to in excess of 300 keV recoil energy. The motivation to also use an AmLi source is the lower maximum neutron energy of about 1.5 MeV, which results in a fairly distinct endpoint at about 40 keV.

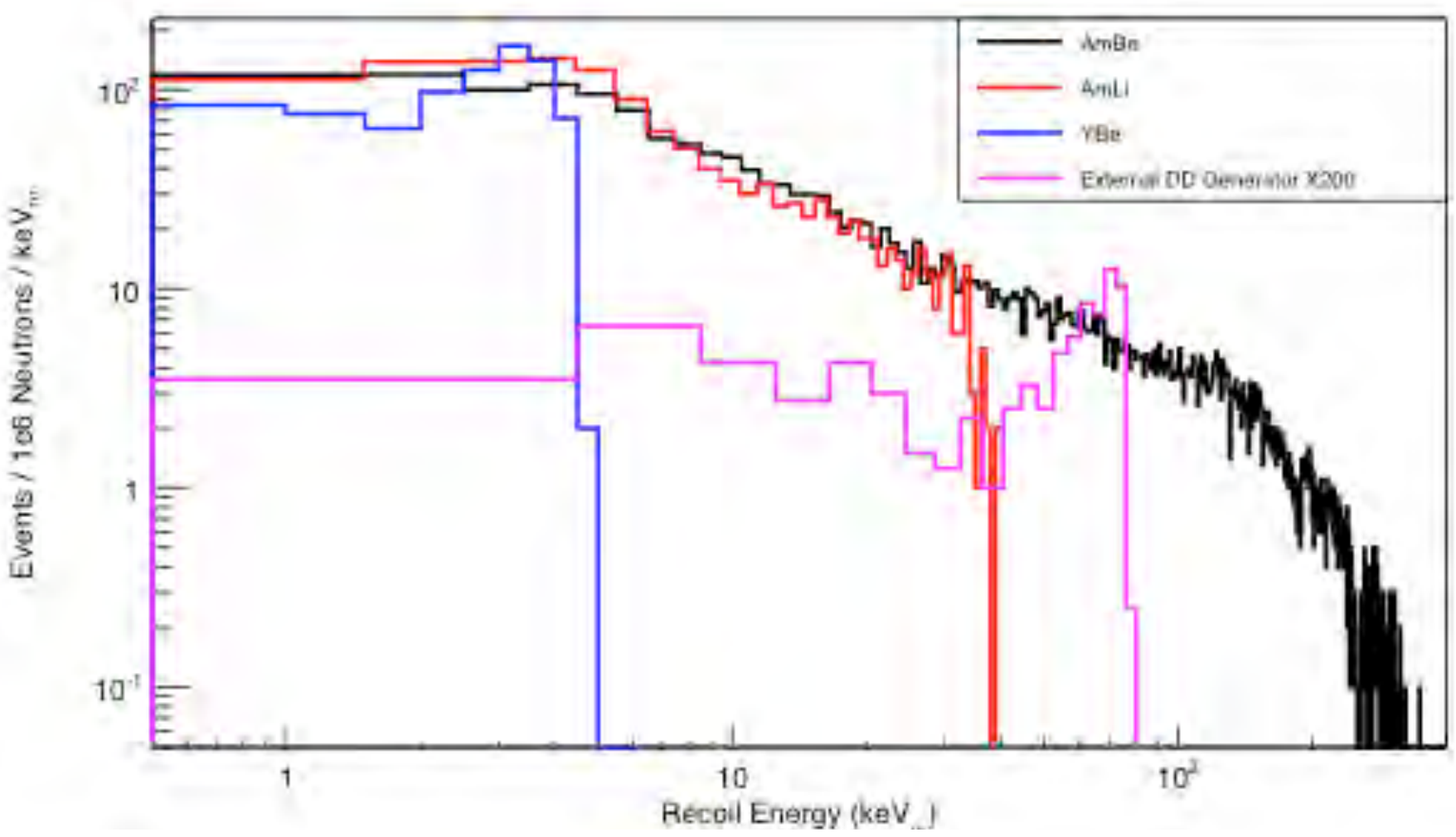

**Figure 10.3.1. Energy spectrum obtained from each of the four primary neutron-calibration sources, showing broad-spectrum coverage and kinematic endpoints.**

Simulations show that the yield of single-scatter NR candidates with energy less than 25 keV is comparable to AmBe but with an enhanced fraction of events at low recoil energy (less than 10 keV). The rates shown in Figure 10.3.1 would be obtained in about three hours of live time, assuming 100 neutron/s source strength. It is notable that the (α,n) yield is lower for AmLi than for AmBe, so that a higher americium activity of about 2.5 mCi is required to obtain a source strength of 100 neutron/s. The sources will be encapsulated and there are no additional safety concerns.

### Yttrium Beryllium Neutron Source (YBe)

A YBe neutron source [8] produces mono-energetic 152-keV neutrons from a (γ,n) reaction on the Be nucleus; $^{88}$Y provides predominantly 1.8-MeV gammas leading to a 4.5-keV NR endpoint. This will provide an anchor for the NR signal response near detector threshold. It will also be extremely useful for understanding the background and signal from $^{8}$B coherent elastic neutrino nucleus scattering. This source will also produce $10^4$ more 1.8-MeV gammas than neutrons. It therefore requires significant gamma shielding, and cannot be deployed in the source tubes. An independent, dedicated conduit at the top of the cryostat will allow a requisite 28 cm of tungsten shielding between the source and the detector. This is indicated by the arrow in Figure 10.3.2. The tungsten shielding will have a total mass of about 100 kg.

If the YBe source were to become stuck in the conduit, the low-energy detector response would be overwhelmed, which would be unacceptable. Two different techniques are being considered, and a conceptual design will be generated for each to assess their risks. In the first design, the source is counterweighted from three steel cables, so as to be neutrally buoyant. In the second design, the source is suspended using the existing overhead crane at SURF. In both cases, the YBe will be situated in a removable insert within the larger shield. This affords the possibility of removing it from the tungsten shielding even in the unlikely event that the tungsten mass were to become lodged in place. This insert will also provide shielding to make the YBe source safer to handle.

### Deuterium (DD) Neutron Source

A deuterium-deuterium neutron generator provides monoenergetic 2.45-MeV neutrons, emitted isotropically in bunches. The generator will be deployed outside the water shield. This will allow significant collimation of the neutrons via a fixed, air-filled conduit through the water and scintillator veto, one of which is shown in Figure 10.3.2. It will be 6 cm in diameter and will be positioned on a radial axis, with its symmetry axis 10 cm below the gate grid. This will ensure that NRs from the source experience minimal attenuation of the S2 signal. It will also ensure that the calibration is minimally limited by pileup (by limiting the event drift time). Monte Carlo studies indicate that the chosen size of the conduit maximizes the usable neutron flux while minimizing the impact on the shielding efficacy. A second conduit for DD neutrons will be located at an upward angle (not shown in Figure 10.3.2).

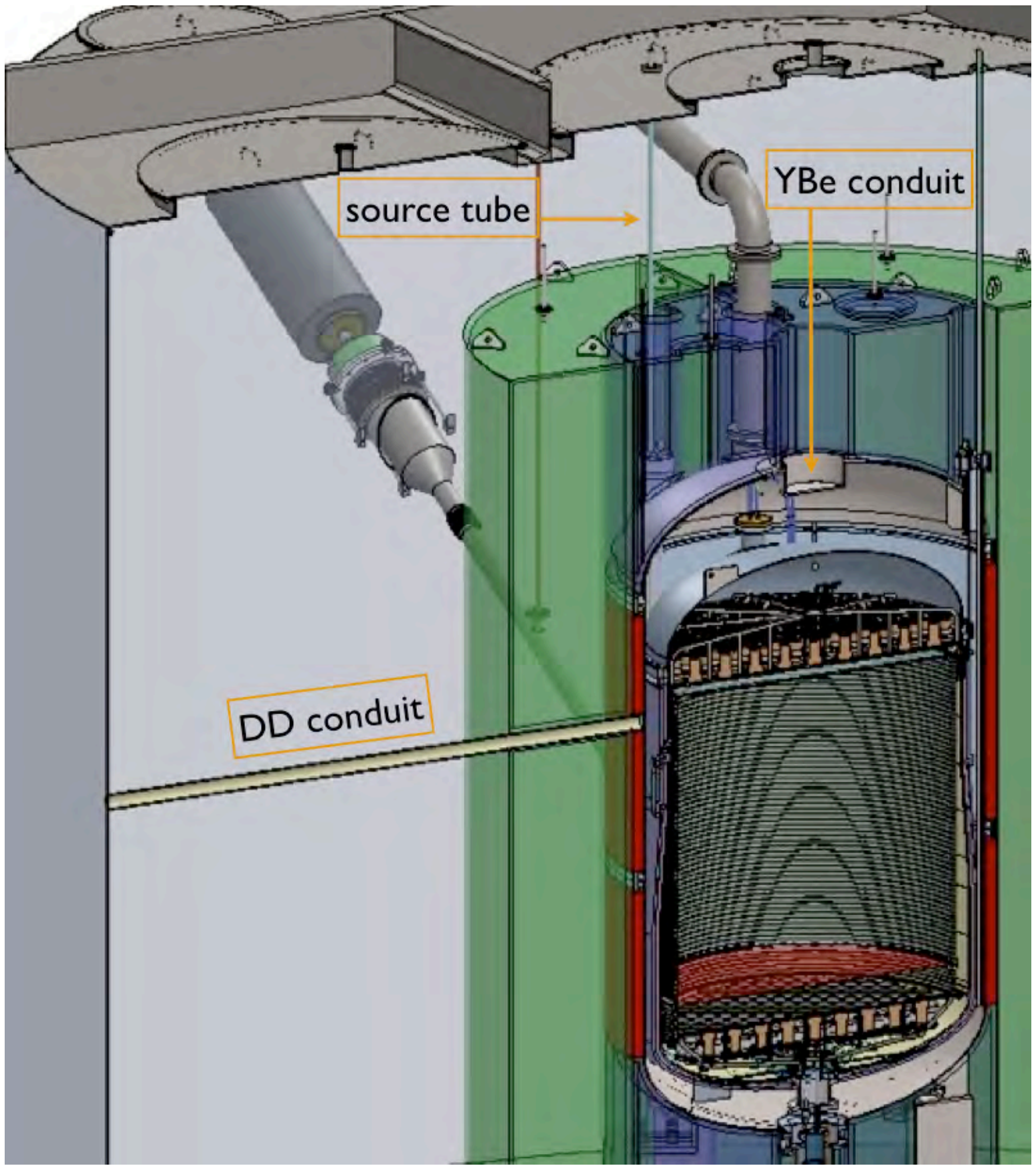

**Figure 10.3.2. Partial cutaway view of the LZ detector, showing the location of the YBe and DD neutron calibration conduits, as well as one (of three) source tubes. The YBe tungsten shielding block is not shown. A second (angled) DD conduit is also not shown.**

Preliminary single-scatter NR data from a DD calibration of LUX [13] is shown in Figure 10.3.3. Simulations indicate that a similar spectrum will be obtained in LZ, with a significant number of low-energy events and a robust endpoint.

The DD generator used by LUX will continue its service for LZ. An important planned upgrade of this generator will reduce the bunch width of emitted neutrons from several tens of microseconds down to about 100 ns. This will allow improved time-of-flight tagging.

This source provides a highly monoenergetic sample of neutrons arriving at the active region of the detector. This permits the possibility of tagging a second scatter to infer the recoil energy of the first scatter. Preliminary LUX analysis has demonstrated that very-low recoil energies, as low as about 1 keV, can be calibrated successfully with this technique. Dual-scatter events are used to calibrate the ionization yield (S2), since the two scatters are observed at separate times. The photon yield (S1) can then be inferred by using the calibrated S2 signal to assign an energy deposit to single-scatter events.

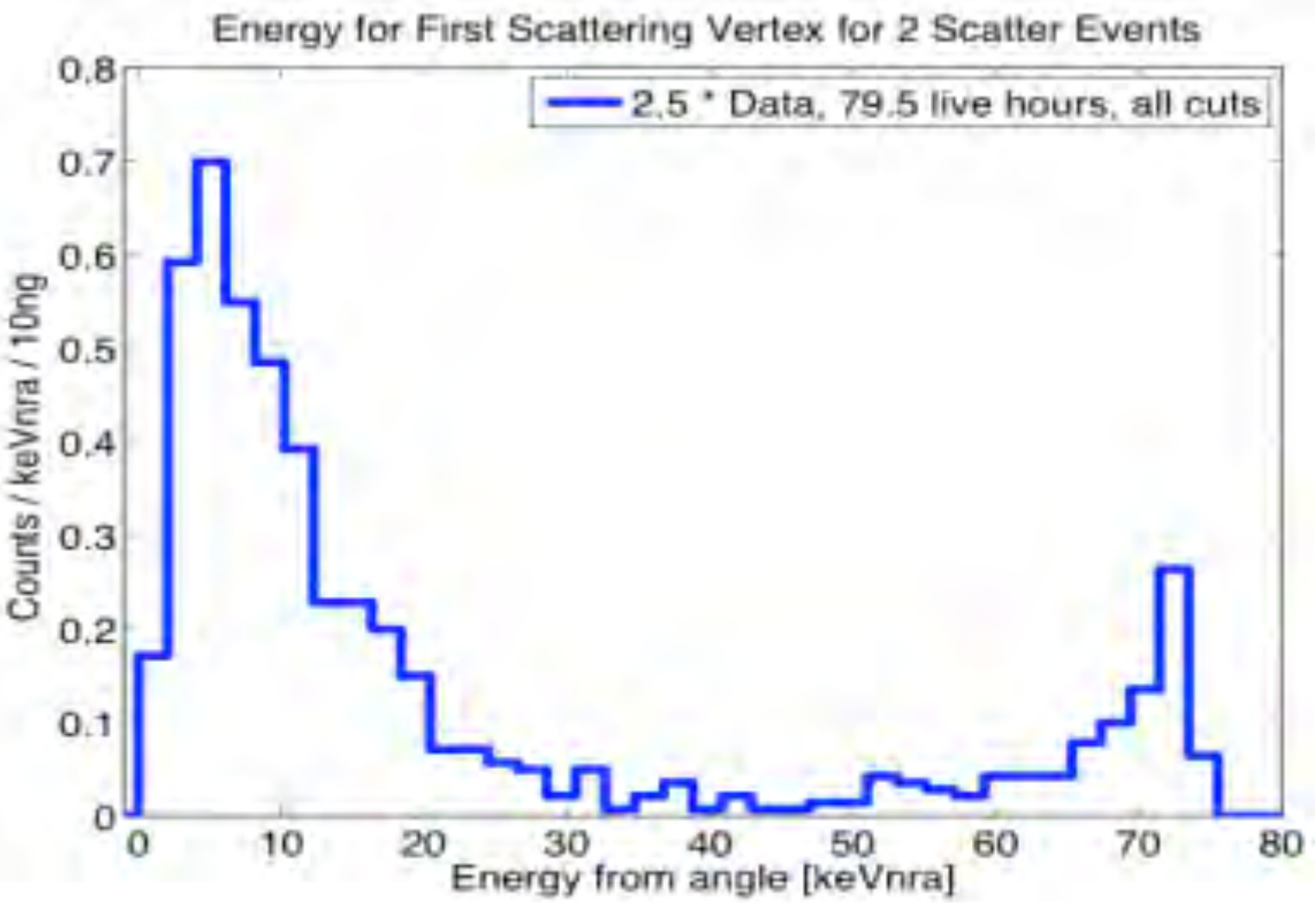

**Figure 10.3.3. Preliminary single-scatter NR S2 (electron) energy spectrum obtained with the LUX detector. Expectations for LZ are very similar, as shown in Figure 10.3.1.**

The larger size of LZ suggests it may be possible to use time-of-flight of the 2.45-MeV neutrons between scatters to separately resolve the S1 signals. This could lead to increased accuracy of the measurement of primary scintillation signal yields. For example, a 100-cm path length gives rise to a 50-ns transit time. Studies are under way to explore the feasibility of this approach.

#### Effect of Gd Doping in the Outer Detector

As described in Chapter 7, the outer LS detector is doped with Gd to improve the neutron-capture efficiency and signal generation. Just as the self-shielding capability of the LXe benefits dark-matter search operation at the expense of the ease of calibration, so too does Gd doping. Specifically, the approximately 8-MeV gamma cascade following Gd neutron capture has the potential to introduce gamma pileup during neutron calibrations. A detailed study of this situation is under way, and preliminary results indicate that LZ will be able to comfortably calibrate at a rate of 100 neutrons per second, resulting in reasonable calibration times (measured in hours). Because the Gd capture time is about 30 µs, our baseline design provides neutrons from both the DD and YBe sources close to the top of the active LXe target. This minimizes the event drift time, and thus the effects of gamma pileup.

#### Gamma Sources for Calibration of the Active Xenon TPC

External gamma sources are not required for any of the primary calibrations of the active Xe TPC. This is by design, as the active region is intended to be self-shielded against external gammas. Nevertheless, several calibrations of secondary importance may be obtained from external gammas. These include studies of higher-energy backgrounds and signal fidelity near the edge of the TPC. A full suite of high-energy gamma sources may be used for this purpose, including $^{137}$Cs (662 keV), $^{60}$Co (1173 keV and 1332 keV), and $^{208}$Tl (2614 keV).

#### Gamma Sources for Calibration of Xenon Skin and the Scintillator Veto

As discussed above, both the Xe skin veto and the organic scintillator veto are designed to have fully efficient scintillation detection capabilities for electromagnetic energy depositions $E > 100$ keV. A suite of low-energy sources will be deployed in the source tubes to verify this performance. Several higher-energy sources will be used for higher statistics characterization of the signal yields. Additional details about the veto signal readout are given in Chapter 7.

#### Source Tubes Deployment

The source tubes are a fixed part of the cryostat system and are described in Chapter 8. Source tubes will be maintained under vacuum when not in use, to prevent Rn plate-out. The source deployment system will be a simple mechanical winch with welded steel cables attached to a small metal canister. The canister will be guided through the source tubes by PTFE fins, to prevent it from becoming lodged in the tube.

An unacceptable failure mode of this system would be a source stuck in a tube. As an additional engineering control against this eventuality, the sources are designed to be separately removable from the canister using a redundant, dedicated steel cable. Because the source diameter is more than a factor of 2 smaller than the source tubes, it is always possible to raise them back to the upper deck without any friction. A full-scale mockup of the design of this system is being constructed and will be exhaustively tested prior to deployment.

## 10.4 Ex situ Energy Calibration

An external calibration program offers the flexibility to probe a wider variety of energies and detector conditions without sacrificing valuable dark-matter search time. Although the funding for such measurements is not included in the LZ proposed baseline cost, we describe here a suite of ex situ “benchtop” measurements whose aim is to robustly characterize the low-energy NR response of LXe, as a function of both recoil energy and applied electric field.

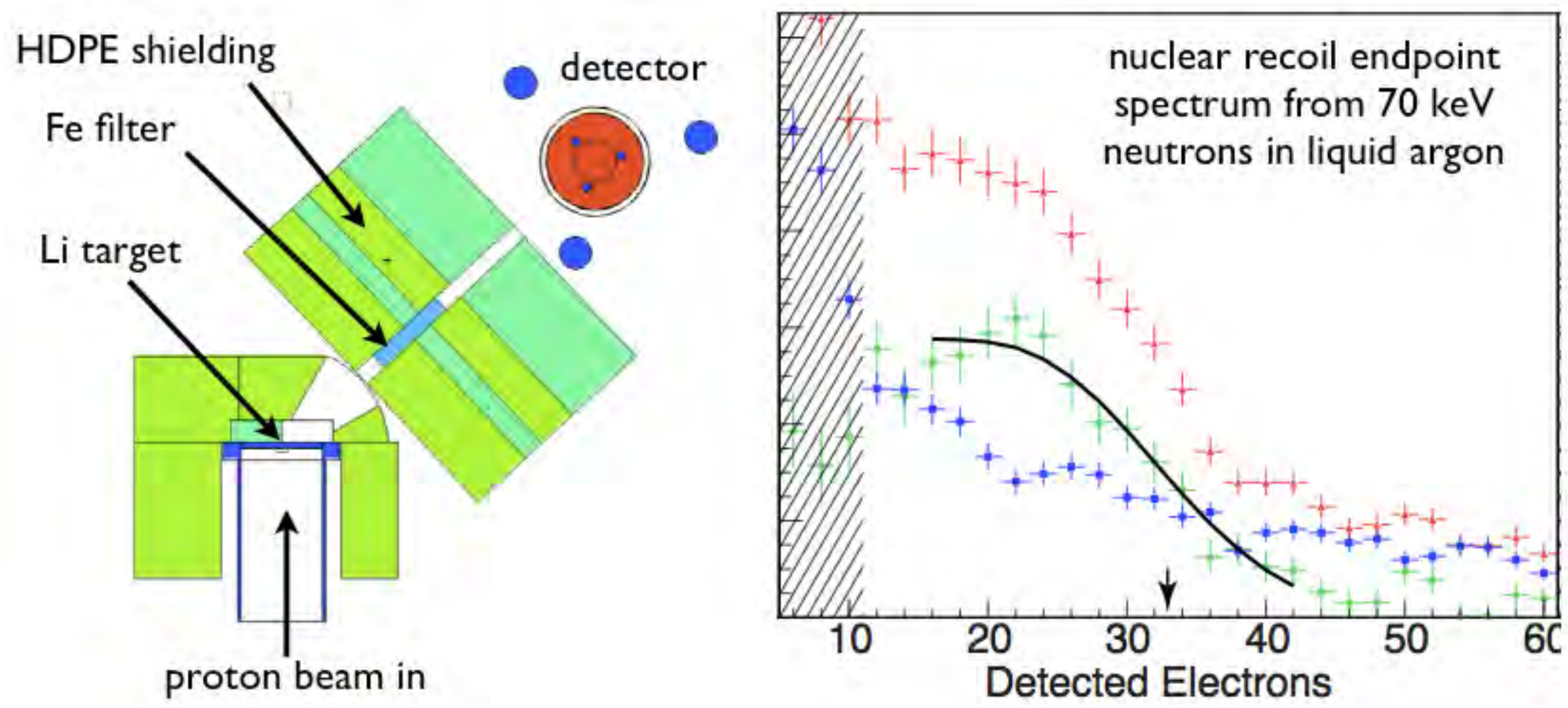

**Figure 10.4.1. Left: Schematic of the neutron source, shielding, and detector as deployed at LLNL CAMS for low-energy NR measurement. Right: Data obtained at CAMS using a small dual-phase liquid argon TPC.**

**Table 10.4.1. Primary low-energy NR calibration energies accessible at the LLNL CAMS facility in its present configuration. Other neutron energies may be obtained by simply changing the isotopic filter.**

| Fast Neutron Energy | Nuclear Recoil Endpoint Energy | ne |
|---|---|---|
| **24 keV** | 0.73 keV | 3-4 |
| **73 keV** | 2.2 keV | 10-12 |
| **2450 keV** | 74 keV | 210 |

One of the primary neutron sources that will be used in this campaign has already been developed at the LLNL Center for Accelerator Mass Spectrometry (CAMS) Facility. The accelerator produces a tunable beam of approximately 2 MeV protons, which then impinge on a lithium target. A low-energy (approximately 10-100 keV) neutron beam results, via the $^{7}Li(p,n)^{7}Be$ reaction. This reaction is kinematically constrained, with the outgoing neutron energy depending on both the outgoing neutron angle as well as the proton energy. This results in some width to the neutron-energy distribution, so the monochromaticity of the outgoing neutrons is further ensured using an isotopic notch filter. Figure 10.4.1 is a schematic of the neutron source as built at CAMS. The neutron beam is described in detail in [9], and a first series of measurements in a dual-phase liquid argon detector [14]. That work used the 70-keV notch in natural Fe to produce a kinematic endpoint at 6.7 keV in liquid argon.

The same source would result in 2.3 keV NRs in LXe. A dual-phase LXe detector with target mass of about 1 kg has recently been built. Initial measurements will focus on the S2 (electron) response. Later optimization of the photon-detection efficiency of the test-bed detector will permit a detailed study of the S1 (photon) response.

The present configuration provides either 24-keV or 70-keV neutrons, using natural iron as a filter. Other energies are accessible by changing to materials such as vanadium. The kinematic endpoint energies of these calibrations are summarized in Table 10.4.1, along with the expectations for the 2.45-MeV DD endpoint.

The 74-keV endpoint from DD neutrons will be used to ensure compatibility of response of the test-bed detector with respect to the LZ detector. Historically, this translation was performed using 122-keV gammas from $^{57}Co$. The disadvantages of that method are that (1) the gammas are strongly affected by small variations in applied electric field, and (2) the gammas are not sufficiently penetrating to be useful for large LXe detectors.

As an additional cross-check, an ex situ measurement of the 4.5-keV kinematic endpoint from YBe could be made.

## Chapter 10 References

# 11 Electronics, DAQ, Controls, and Online Computing

This chapter describes the LZ signal processing electronics, data acquisition and trigger systems, detector-control system, and online data processing.

## 11.1 Signal Processing

The processing of the signals generated by the Xe PMTs is schematically shown in Figure 11.1.1. The Xe PMTs operate at a negative HV supplied by the LZ HV system, described in more detail in Section 11.7. HV filters are installed at the HV flange on the breakout box. The PMT signals leave the breakout box via a different flange and are processed by the analog front-end electronics, described in more detail in Section 11.3. The amplified and shaped signals are connected to the data acquisition (DAQ) and sparsification systems, described in more detail in Sections 11.5 and 11.6. The digitized data are sent to Data Collectors and stored on local disks.

The PMTs of the outer-detector system operate at positive voltage. The processing of the signals from these PMTs, shown schematically in Figure 11.1.2, is slightly different from the signal processing of the Xe PMTs shown in Figure 11.1.1. The HV filters of the outer-detector PMTs extract the PMT signals directly from the HV line; no separate signal flanges are required. The same type of amplifier used for the Xe PMTs is used for the outer-detector PMTs, except that only a single gain channel will be used to amplify and shape the outer-detector PMT signals.

The data flow is schematically shown in Figure 11.1.3. The event builder assembles the events by extracting the relevant information from the Data Collector disks, DAQ1–14. This step is discussed in more detail in Section 11.10. The event files are stored on local RAID arrays, RAID 1 and RAID 2, before being distributed to the data-processing centers for offline data processing and analysis.

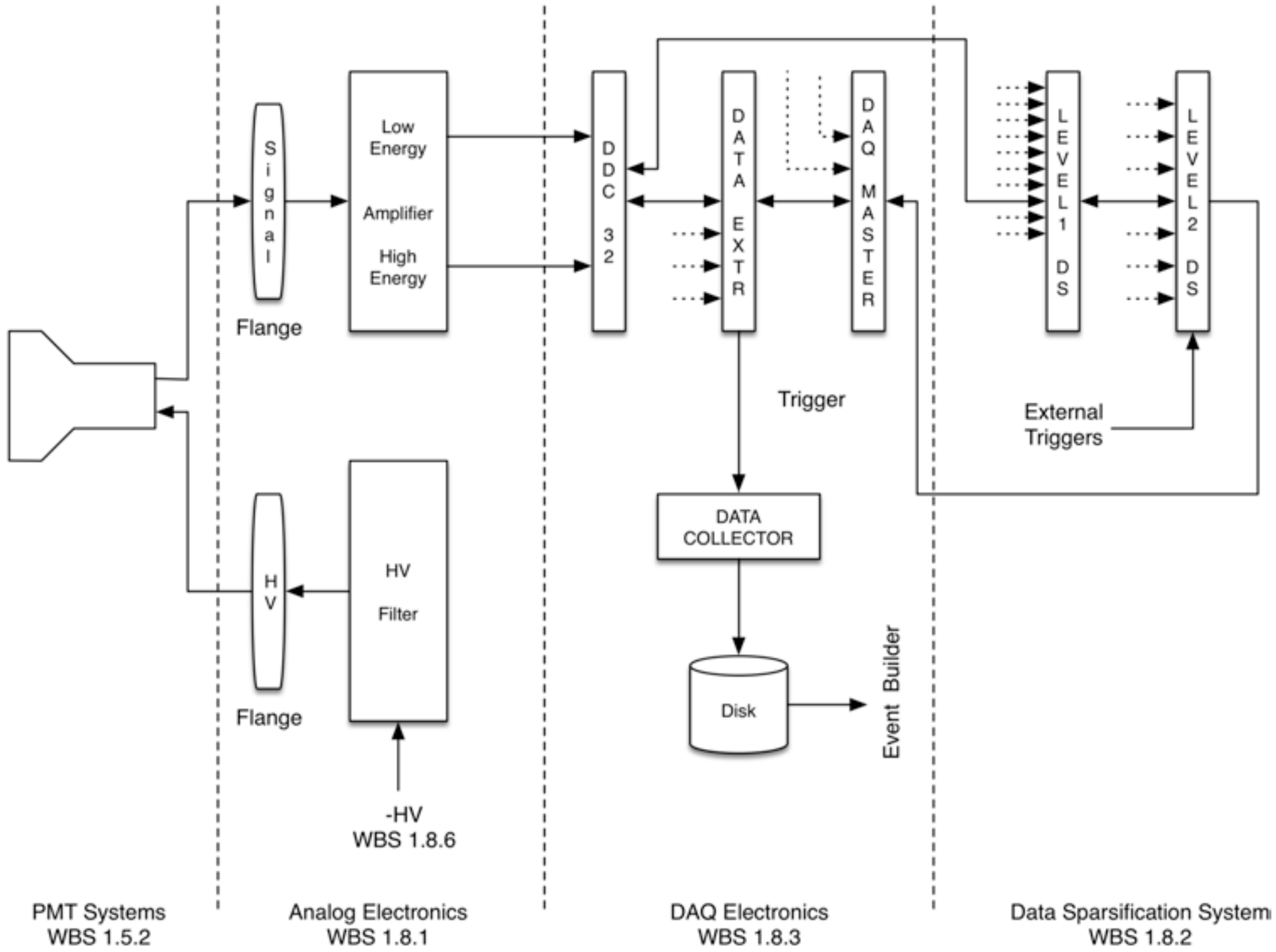

**Figure 11.1.1. A schematic of the signal processing of the Xe PMTs. The PMTs of the TPC array use dual-gain signal processing. The PMTs of the skin PMTs only utilize the low-energy section of the amplifiers.**

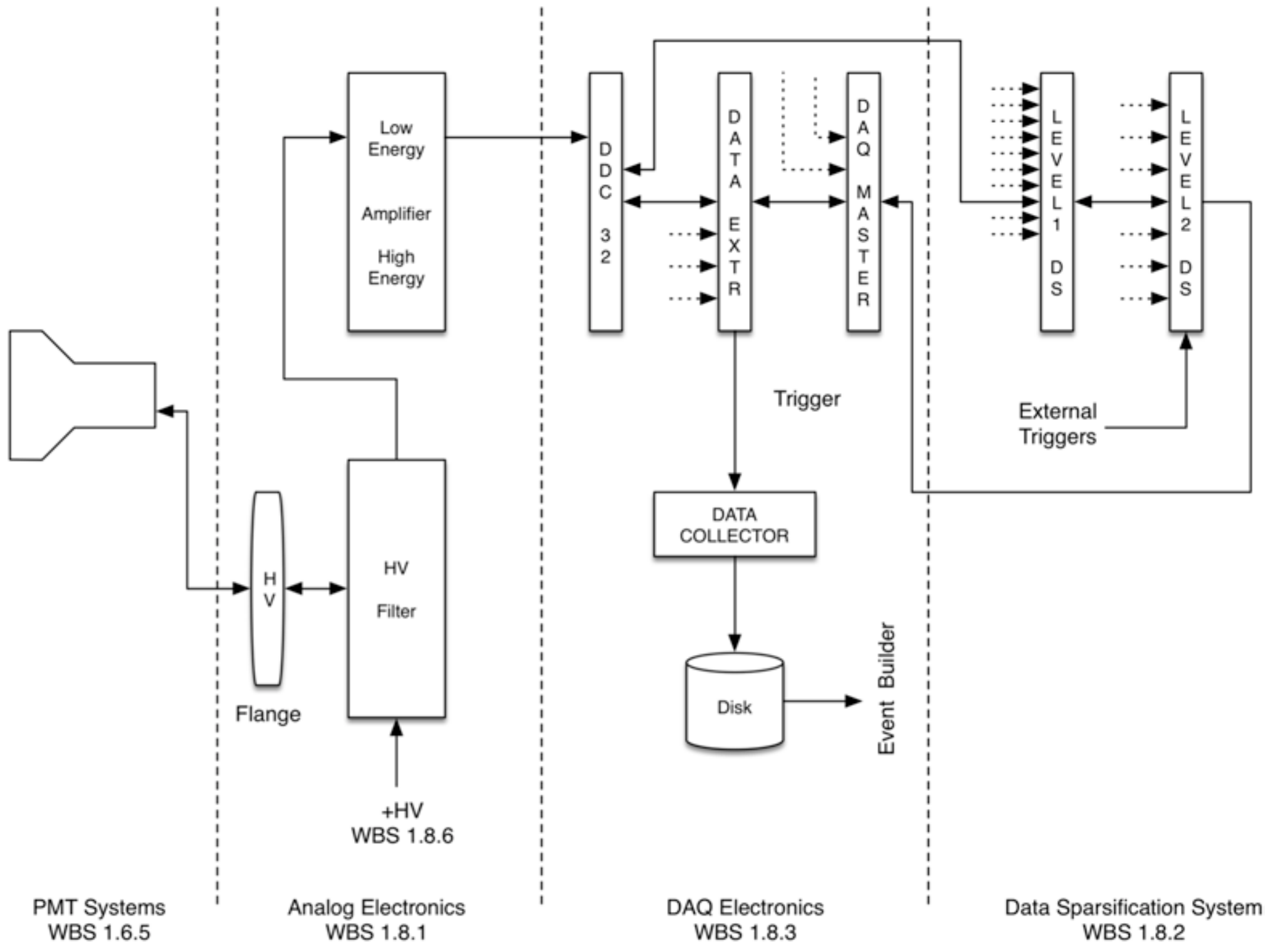

**Figure 11.1.2. A schematic of the signal processing of the outer-detector PMTs.**

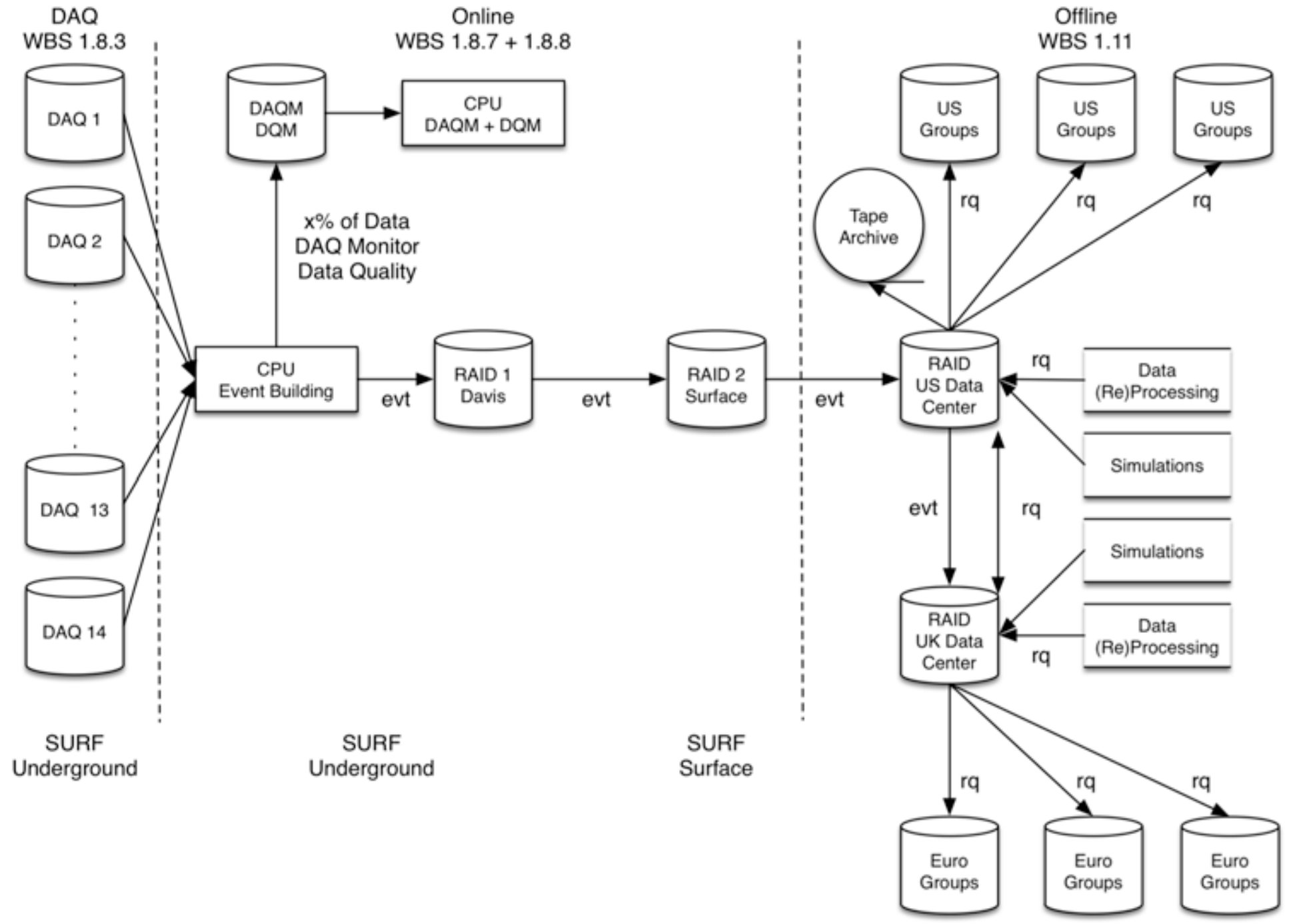

**Figure 11.1.3. A schematic of the data flow.**

## 11.2 Requirements

The parameters of the analog and digital electronics are defined based on the properties of the PMT signals, the required dynamic range of the different PMTs, and the expected calibration rates.

Three different PMTs are used in LZ. The TPC PMTs will see S1- and S2-type signals. The skin and outer-detector PMTs will only see S1-type signals. The relevant properties are listed in Table 11.2.1.

The cables that connect the Xe PMTs to the analog electronics are 40–60 ft long; the actual length depends on the final design of the conduits and the location of the area in which the analog electronics will be installed. These internal cables will be of the same type as those used for LUX. The manufacturer of the LUX cables (Gore) specifies an attenuation of 29 dB/100 ft at 400 MHz, similar to that of an RG178 cable. Measurements with the LUX cables have shown an amplitude reduction of S1 signals by approximately 30% for the 42-ft-long cables.

The design of the analog electronics is constrained by the required dynamic range of the LZ signals. Our design relies on the assumption that a single photoelectron (SPHE) detected in a TPC PMT generates a 13.3 mVns pulse at the input of the amplifiers. The required dynamic range is defined by the sources used to calibrate LZ and the desire to detect high-energy events for background studies. Chapter 10 provides details on the LZ calibrations.

The DAQ system design is based on our experience with LUX and the required LZ calibration rates. Typical calibration rates are listed in Table 11.2.2. The TPC source calibration rates are limited by the maximum drift time of 700 µs in LZ. During source calibrations of the TPC, a 150 Hz calibration rate results in a 10% probability of detecting a second calibration event within the drift time of the previous calibration event. Neutron calibrations are carried out to define the NR band in the TPC. These calibrations utilize external neutron sources and a neutron generator. External sources, inserted into the source tubes around the central cryostat, are used to calibrate the skin and the outer-detector PMTs. The count rates for these calibrations are not limited by the drift time in the TPC and they can be carried out at substantially higher rates. Weekly LED calibrations are done to examine the SPHE response of the PMTs and to monitor the PMT response. These calibrations can be carried out with rates as high as 4 kHz.

The data volume to be handled by the DAQ system can be estimated on the basis of our experience with LUX. In WIMP search mode, we will focus on events with energy depositions below 40 keV. During krypton calibrations, in which the total energy deposition is 41.6 keV, the average LUX event size was 203 kB (or 1.7 kB/channel). The size of each event was dominated by the width of the S2 signals. The event size of LZ can be estimated by scaling the LUX event size by the ratio of the number of PMT channels. Because LZ has four times as many TPC PMTs as LUX, and taking into account the dual-gains, we expect that the event size in LZ will be roughly 8 times as large as the LUX event size, or 1.6 MB. This estimate does not include the data volume associated with the outer-detector and the skin PMTs. If each outer-detector and skin PMT detects a single S1-like pulse, that increases the event size by about 45 kB. With compression, the typical event size for LZ is estimated to be 0.53 MB. Monte Carlo simulations show that the total background rate in LZ will be about 40 Hz. The background rate in the WIMP search region (0–40 keV) will be about 0.4 Hz. At 40 Hz, the data rate is 21 MB/s. In 1,000 days, LZ will thus collect 1.9 PB of WIMP-search data. Other estimates, including energy depositions above 41.6 keV, result in an estimated total data volume of 2.8 PB. By optimizing the event selection, we expect to be able to reduce the total volume of WIMP-search data by a factor of 2 (see Section 11.6).

**Table 11.2.1. Properties of the PMTs.**

| System | Type PMT | # | Gain | HV |
|---|---|---|---|---|
| **TPC** | R11410-20 | 488 | $<5 \times 10^6$ | <1500 V |
| **Skin** | R8520 | 180 | $<1 \times 10^6$ | <800 V |
| **Outer detector** | R5912 | 120 | $<1 \times 10^7$ | <1500 V |

**Table 11.2.2. Calibrations and expected count rates.**

| Calibration Type | System | Typical Count Rate | Frequency |
|---|---|---|---|
| **Internal sources** | TPC | <150 Hz | Twice a week (Kr) |
| **External gamma sources** | Skin and outer detector | TBD | TBD |
| **Neutrons** | TPC | <150 Hz | TBD |
| **LEDs** | TPC and outer detector | 4 kHz | Weekly |

## 11.3 Analog Electronics

### 11.3.1 Design Criteria

The analog front-end design for the LZ experiment has benefited immensely from the experience with the LUX detector. In what follows, we have retained all the features of LUX electronics that performed well, improving some areas. The most important figure of merit of the LUX analog front end was its low noise characteristics, which allowed us to set our thresholds well below the SPHE level. Assuming the LZ noise characteristics are the same as those for LUX, we expect to be able to run with a threshold of 0.25 PHE (photoelectron) for each PMT. Figure 11.3.1.1 shows a simulated distribution of total S1 pulse area for events in which two PMTs detected one PHE each. A threshold of >0.25 PHE is set for each PMT. The distribution shown in Figure 11.3.1.1 is governed primarily by the ~35% rms width of the PMT signal, and receives almost no contribution from electronic noise. Setting a threshold at 1 PHE for the total S1 signal yields an efficiency of ~95% for events that produce at least two PHEs in the PMTs. It corresponds approximately to the *lowest* energy threshold achievable in such a detector, and hence drives the noise specifications for the front end.

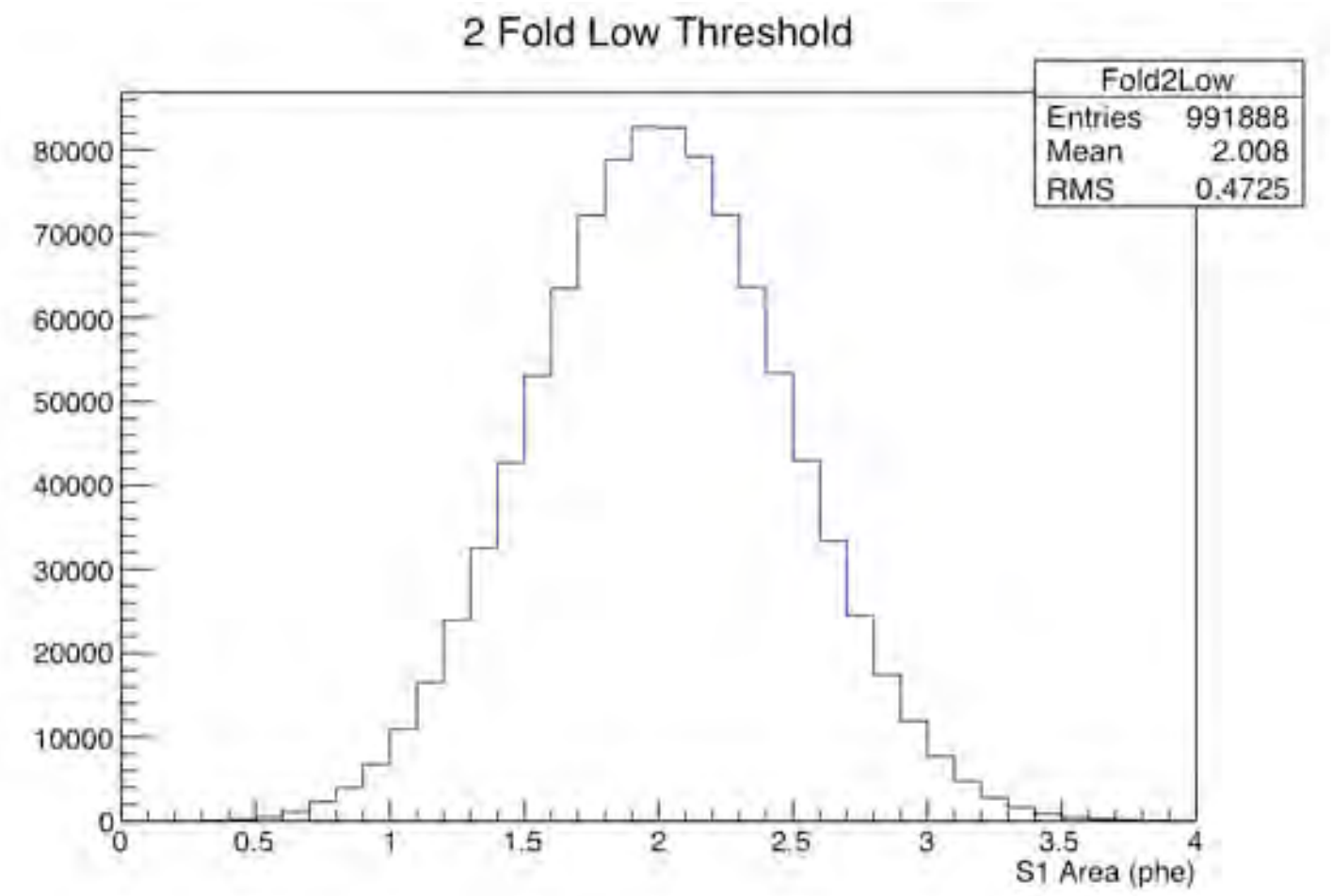

**Figure 11.3.1.1. A simulated distribution of total pulse area for events in which two photoelectrons produce a signal >0.25 PHE in two PMTs.**

The other key design parameter for the LZ front end is the dynamic range, defined by the sources used to calibrate LZ. Isotopes such as Kr-83m (32.1 and 9.4 keV transitions), activated Xe (236 keV and 164 keV transitions), and tritium (endpoint at 18.2 keV) will be present or injected directly into the LXe volume and will be used to calibrate the detector periodically. LZ will retain the ability to detect high-energy events that saturate the PMTs of the top array by using only the light collected by the bottom array to determine the total S2 area.

The LZ electronics will provide excellent resolution for single liquid electrons, which are expected to yield at most 50 PHEs, depending on the strength of the electric field in the gas region of the TPC. The typical duration of such pulses will be about 0.5–1 μs. At the same time, we will need to provide extremely clean measurements of SPHEs in order to have a sharp turn-on of the S1 efficiency. SPHE spectra also help with maintaining an in situ calibration of the PMT gains. To meet these requirements, the analog electronics provides one low-energy (high-gain) and one high-energy (low-gain) output for each PMT.

Figure 11.3.1.2 shows this concept. The high-energy channel has low gain and a 30-ns full width at tenth maximum (FWTM) shaping-time constant. Its dynamic range is defined by the 236-keV Xe activation line. The low-energy channel has a 10× higher gain and wider shaping. It is optimized for excellent SPHE response and dynamic range. The shaping times and gains are derived from one assumption: the DAQ will have a usable dynamic range of 1.8 V at the input and will sample the pulses at 100 MHz with 14-bit accuracy. A 0.2 V offset is applied to the digitizer channels in order to measure signal undershoots of up to 0.2 V. Other relevant parameters are:

- SPHE response of the PMTs at the amplifier input: 13.3 mV-ns pulse.
- S1 light yield at 236 keV: between 1400 PHEs (for zero field) and 660 PHEs (for 1 kV/cm).
- S1 light distribution: evenly distributed over all PMTs. The bottom array receives 80% of the total S1 light.
- S2 light yield for ERs: 27 liquid electrons/keV and 50 PHEs/liquid electron (conservative maximum, assuming perfect liquid electron extraction).
- S2 light distribution: 22% of the S2 light is in a single top PMT. The S2 light is distributed evenly over all bottom PMTs; the bottom array receives 45% of the total S2 light.

The op-amps indicated in Figure 11.3.1.2 are very similar to those used in LUX. The gain and shaping-time constants of the

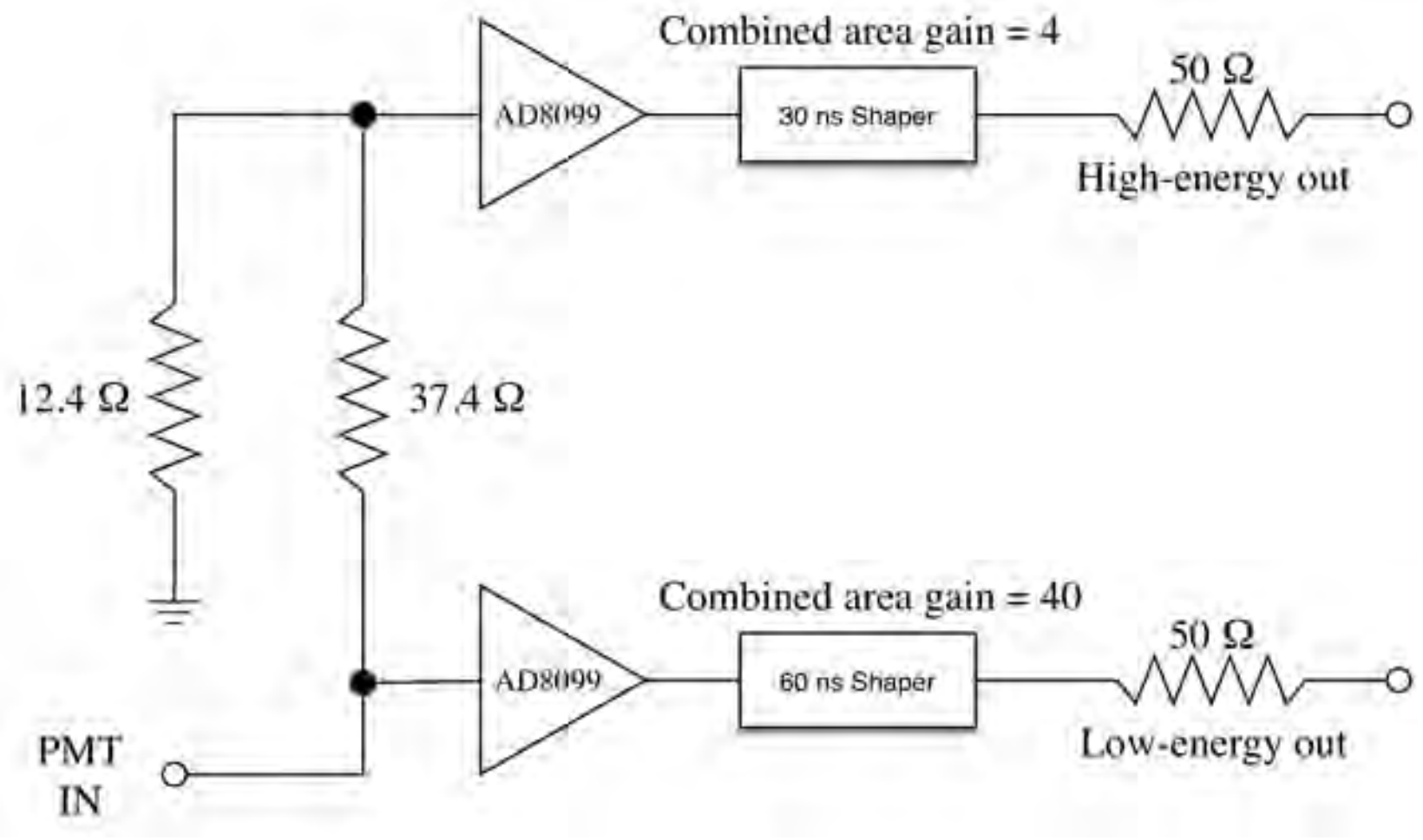

**Figure 11.3.1.2. A schematic diagram of the LZ amplifier.**

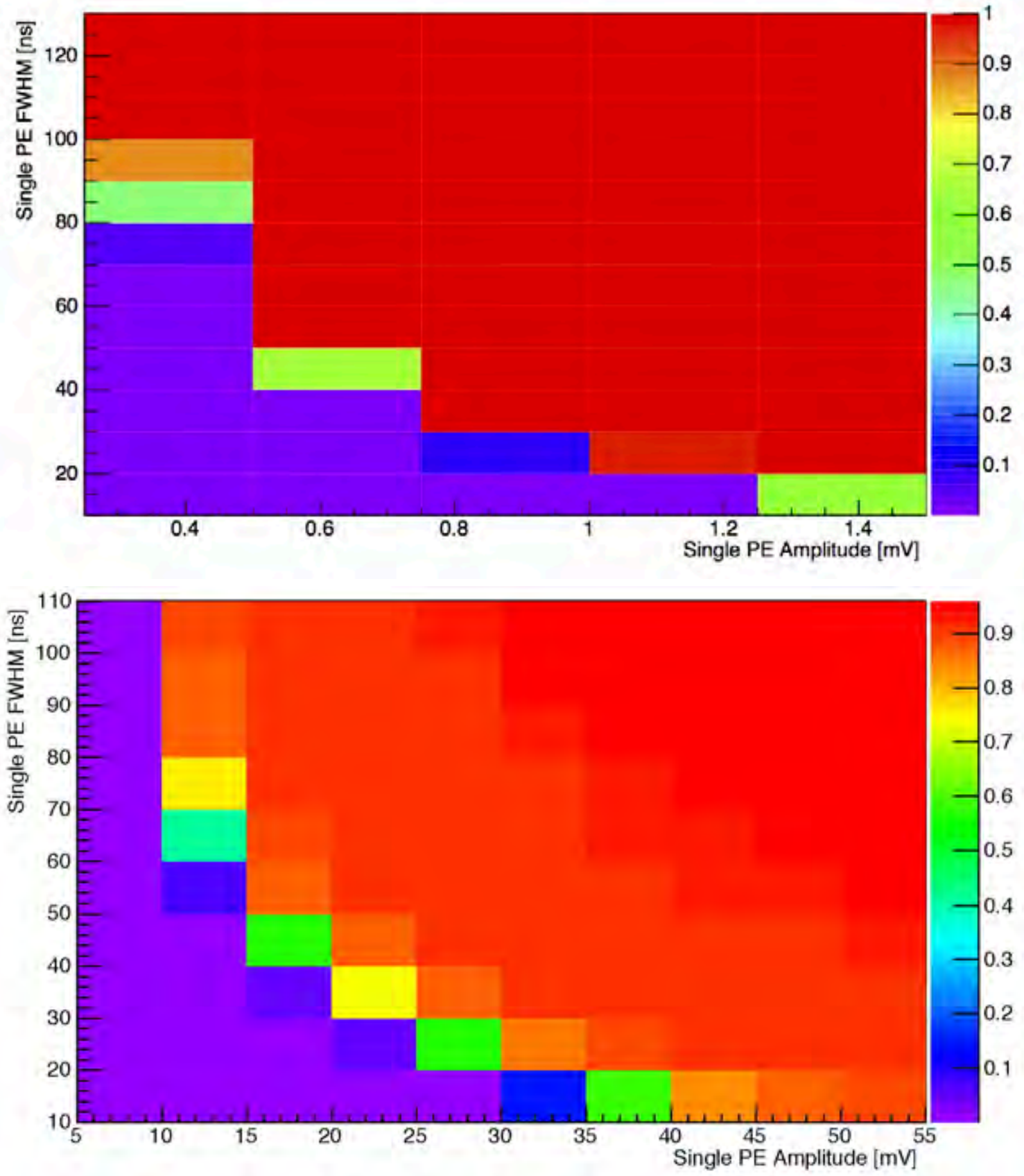

**Figure 11.3.1.3. Top: A simulation study of the S2 response for a 236-keV Xe transition in the center of the detector, as seen by the top PMTs. The gain and the shaping width of the SPHE response are varied. The color code shows the fraction of events in which the peak PMT saturates. Bottom: A simulation study of the S2 response for a 3-MeV energy deposition in the center of the detector, as seen by the bottom PMTs. The color code shows the fraction of PMTs of the bottom array that saturate.**

amplifiers were optimized using simulations. Figure 11.3.1.3 shows the results of simulations of S2 pulses associated with the 236-keV transition in activated Xe (top) and a 3-MeV energy deposition (bottom). Figure 11.3.1.3 (top) shows that S2 saturation in the top PMT array for the 236-keV transition is not a problem if the amplitude of a single PHE is less than 1.5 mV (12 ADCC). If we allow one PMT to saturate, single PHEs of up to 3.0 mV can be accommodated. Figure 11.3.1.3 (bottom) shows that the S2 associated with larger energy depositions will not start to saturate the PMTs of the bottom array if the amplitude of a single PHE is less than 15 mV (120 ADCC). The two outputs of the amplifiers have the following properties:

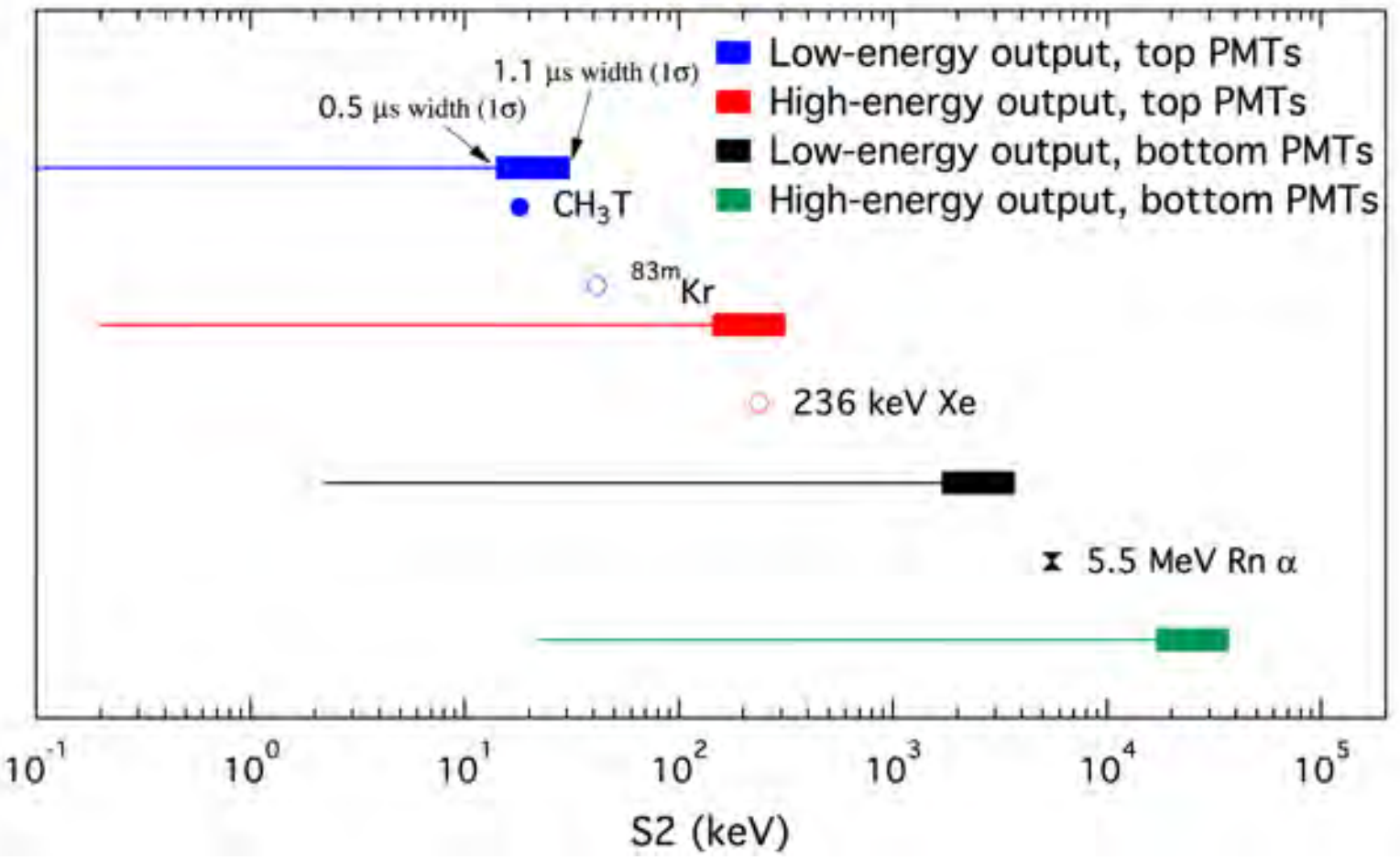

**Figure 11.3.1.4. Dynamic range for S2 signals for the top and bottom PMTs. The bars indicate variations in the upper level of the dynamic range due to the variations in the width of the S2. The lower and upper ends of these bars show the dynamic range for 0.5-µs-wide and 1.1-µs-wide pulses, respectively.**

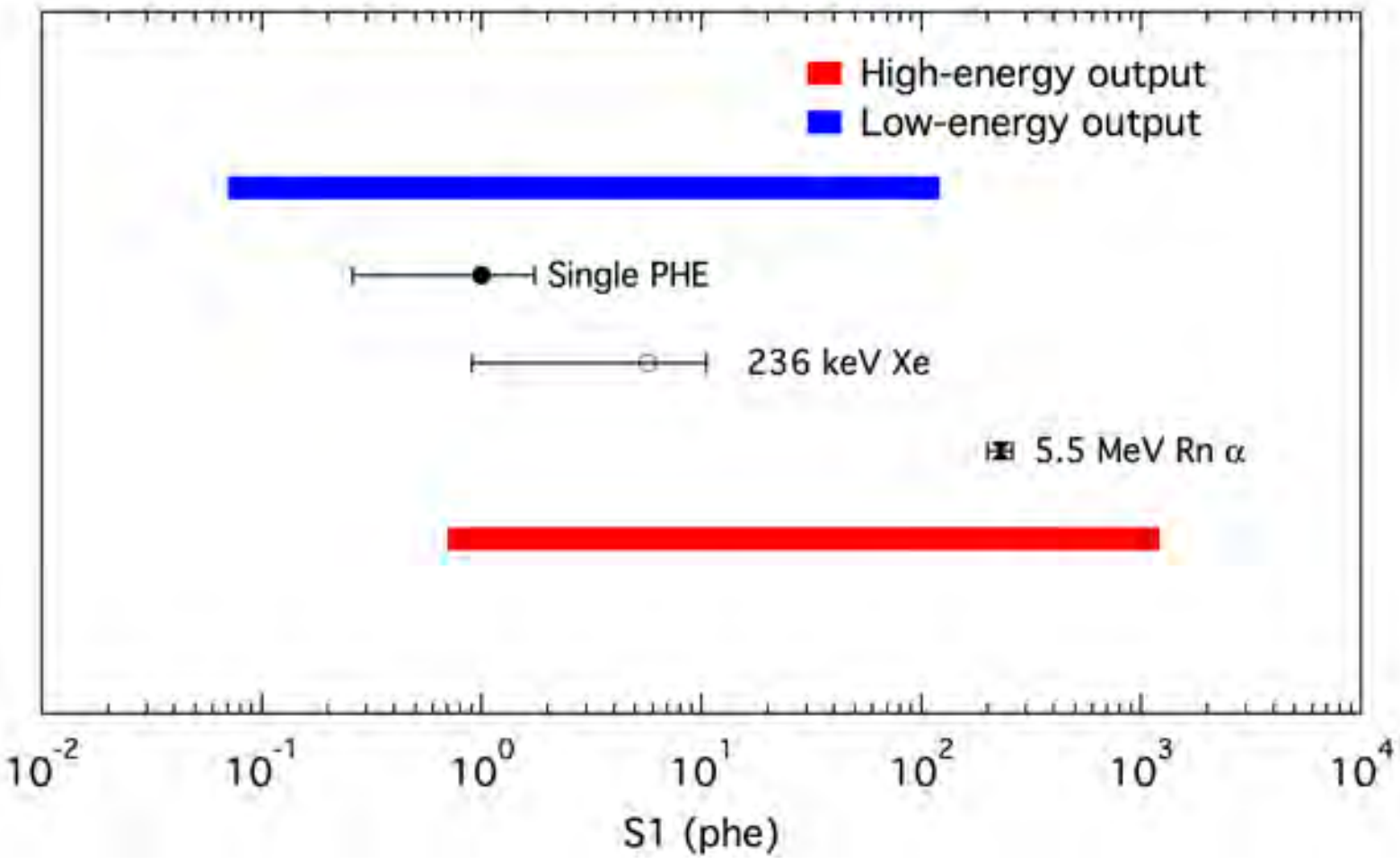

**Figure 11.3.1.5. Dynamic range for S1 signals detected in the bottom PMTs. The range required for the Xe activation lines and alpha particles from radon decay are also shown.**

- **Low-energy output**: Gain 40, 1 PHE = 122 ADCC (amplitude), S1 dynamic range: 121 PHEs (4,600 keV), S2 dynamic range 14–31 keV (top) and 1,700–3,700 keV (bottom) for 0.5–1.1 µs wide pulses (1σ) with no saturation.
- **High-energy output**: Gain 4, 1 PHE = 24 ADCC (amplitude), S1 dynamic range: 600 PHEs, S2 dynamic range 140–310 keV (top) and 17,000–37,000 keV (bottom) for 0.5–1.1 µs wide pulses (1σ) with no saturation.

The dynamic range for S2 signals is shown in Figure 11.3.1.4. The low-energy channel of the amplifier provides the dynamic range required for the tritium and krypton calibrations. The high-energy channel is required to provide the dynamic range required to measure the activated Xe lines. S2 signals induced by alpha particles from radon decay will saturate one or more channels of the top array, but the S2 pulse area can still be reconstructed using the low-energy channels of the bottom PMTs.

The dynamic range for S1 signals is shown in Figure 11.3.1.5. The figure shows the dynamic range of a bottom PMT. Also shown are the number of PHEs associated with the full-energy deposition of the 236-keV Xe activation line and a 5.5-MeV alpha particle. The dynamic range provided by the dual-gain channels is sufficient for all LZ calibrations.

The final gain and shaping parameters of the amplifiers will be fixed after more detailed simulations, including the electronics response, have been carried out and the noise of all components of the electronics chain has been measured during a chain test.

**Table 11.3.1.1. Summary of the number and type of the 1276 analog signals.**

| | High-gain Signals | | | Low-gain Signals | | |
|---|---|---|---|---|---|---|
| **PMT Type** | **#** | **Gain** | **Shaping (FWTM)** | **#** | **Gain** | **Shaping (FWTM)** |
| **Top TPC** | 247 | 40 | 60 | 247 | 4 | 30 |
| **Bottom TPC** | 241 | 40 | 60 | 241 | 4 | 30 |
| **Skin** | 180 | TBD | TBD | 0 | NA | NA |
| **Outer detector** | 120 | TBD | TBD | 0 | NA | NA |
| **Total** | 788 | | | 488 | | |

The same amplifier design will be used for the skin and outer-detector PMTs although the gain and shaping may be adjusted. For these PMTs, only the low-energy channel will be instrumented. A summary of the number and type of analog signals is shown in Table 11.3.1.1.

## 11.3.2 LZ Amplifier Prototype

Figure 11.3.2.1 shows the first amplifier prototype with four input channels. The final amplifiers will have eight input channels. The amplifier will be installed on a signal flange and connected to the PMT signal lines using the DB-25 connector visible on the left side of the figure. The DB-25 connector allows the signal lines to be interleaved with two ground lines.

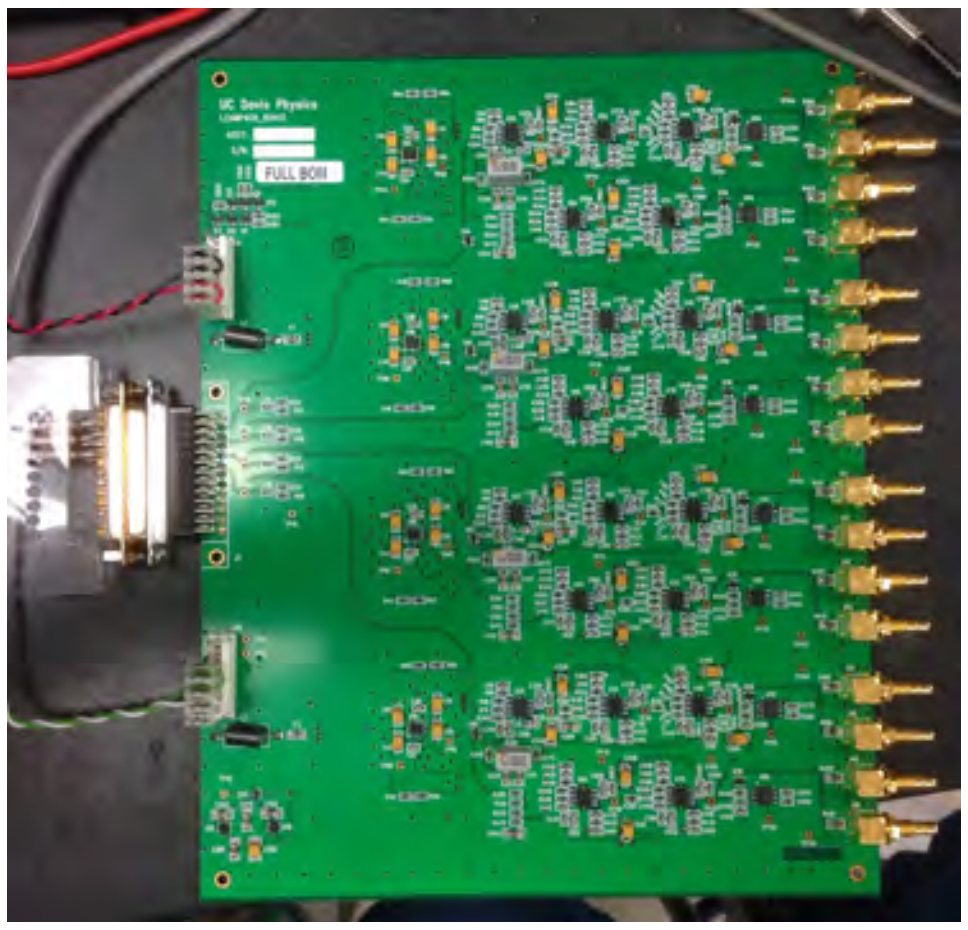

**Figure 11.3.2.1. Photograph of the four-channel amplifier prototype. The input signals are connected to the DB-25 connector at the left side of the board. Each channel has four outputs, visible on the right.**

A waveform captured with the pre-prototype of the digitizer is shown in Figure 11.3. 2.2. The digitizer captured the output of the low-gain channel of the amplifier when a 50 mV S1-like pulse was presented to the input of the amplifier.

The RMS ADC noise of the free-running digitizer channels was measured to be 1.19 ± 0.01 ADCC. When one of the low-gain channels of the amplifier, with the corresponding amplifier input terminated with 50 Ω, was connected to the digitizer, the measured noise remained virtually unchanged (1.20 ± 0.01 ADCC). When one of the high-gain channels of the amplifier was connected, the measured noise increased to 1.58 ± 0.02 ADCC. The noise added by the high-gain channel was estimated to be 0.38 ± 0.02 ADCC (46 ± 2 μV).

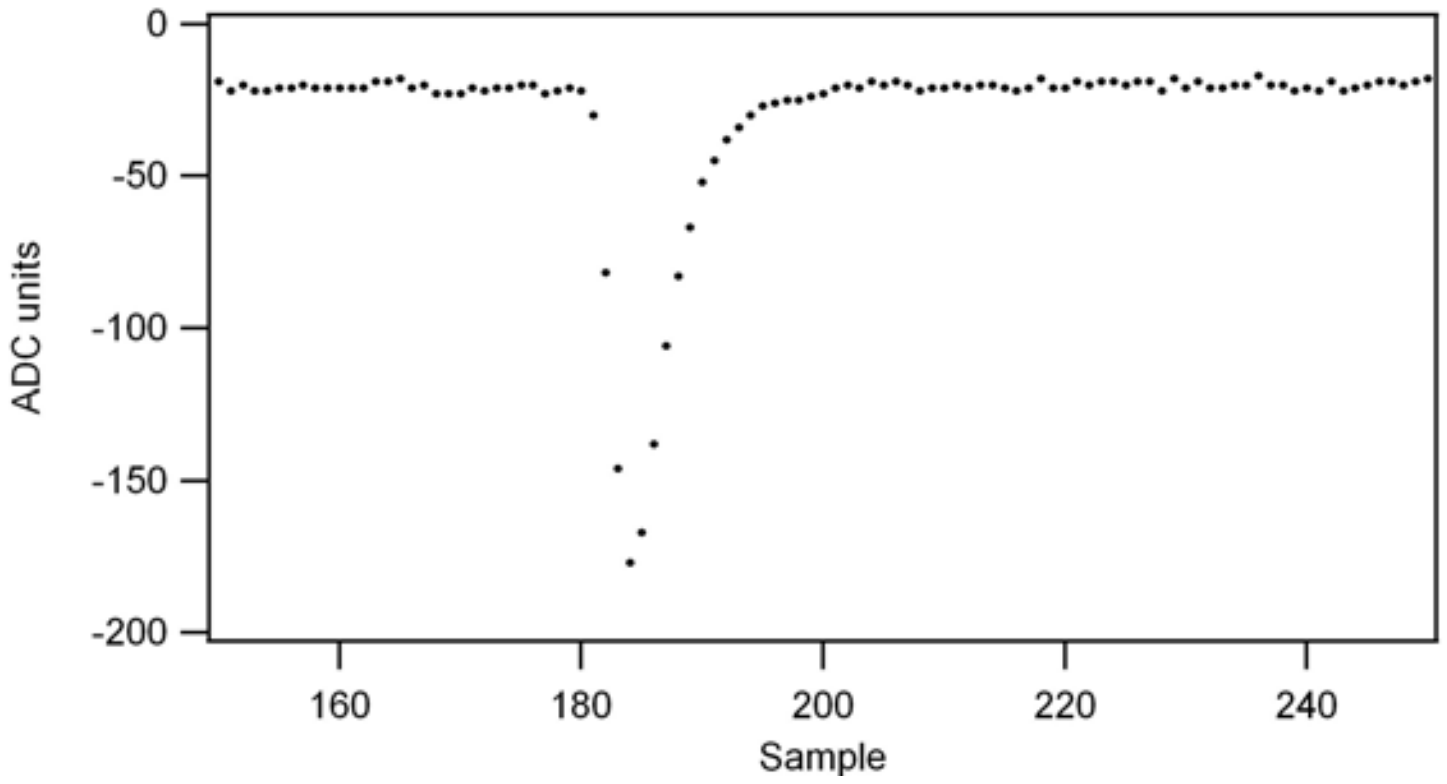

**Figure 11.3.2.2. A waveform from the low-gain channel of the prototype LZ amplifier, captured with the pre-prototype of the LZ digitizer, sampling at 100 MHz. This waveform was generated when a 50 mV S1-like pulse was delivered to the input of the amplifier.**

The crosstalk between the individual amplifier channels is shown in Figure 11.3.2.3. A pulse applied to the input produced a 1.2 V pulse at the high-gain output (Ch 1). The associated low-gain signal is visible on Ch 2. Other channels show no evidence of crosstalk.

The linearity of the amplifier was studied using S1-like test pulses. The signals from the high-energy output saturate the dynamic range of the digitizers before nonlinear effects in the amplifier become important. Examples of the results of these linearity studies with S1-like pulses are shown in Figure 11.3.2.4 The input pulses had a rise time of 18 ns and a fall time of 78 ns. The low-gain output shows a linear behavior across the entire range of input signals used. The high-gain output becomes nonlinear when the area of the input signal exceeds 10 Vns. Although the response is nonlinear in this region, there is still a one-to-one correlation between the area of the digitized output pulse and the area of the input pulse.

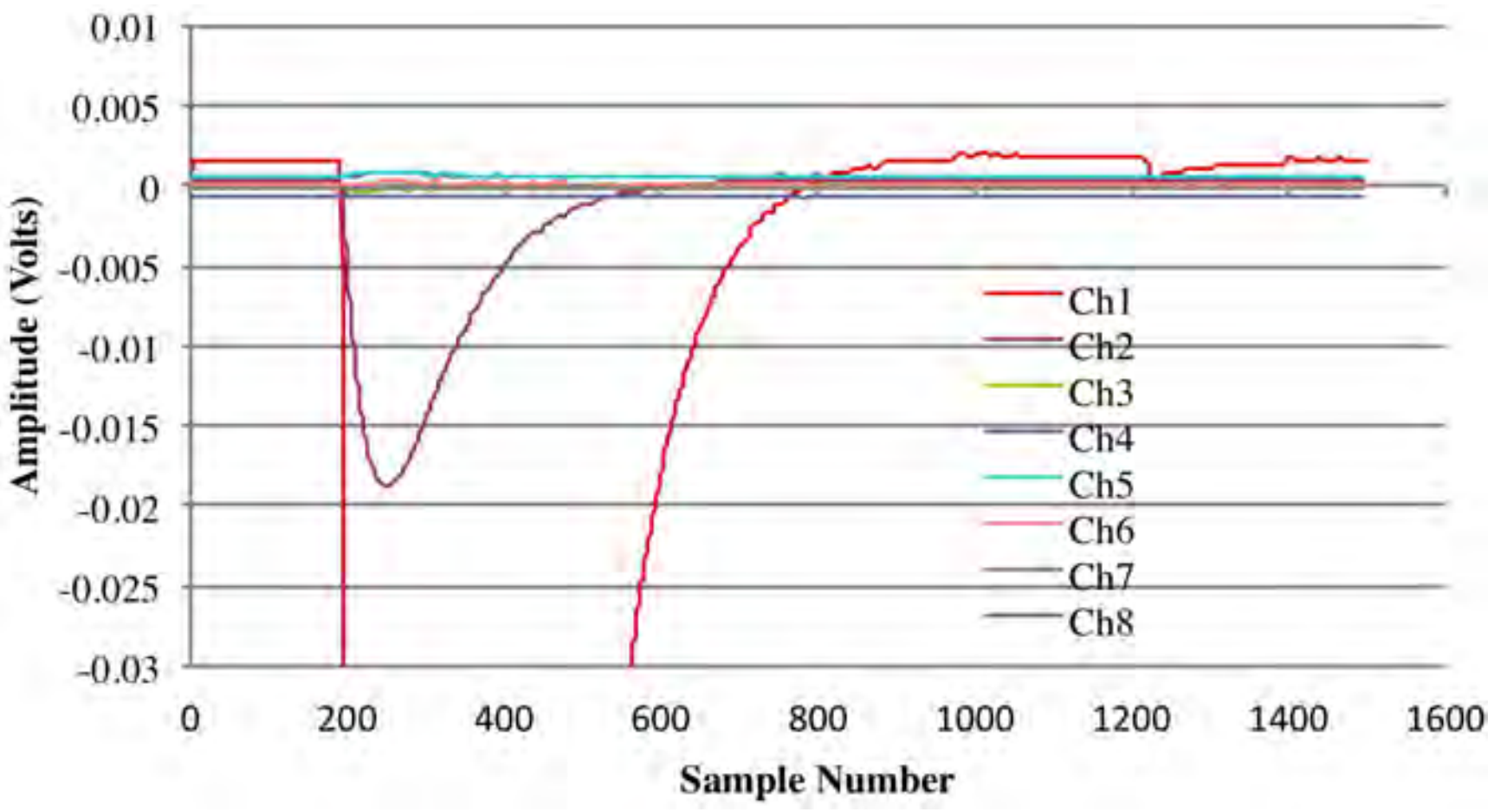

**Figure 11.3.2.3. Output waveforms when a pulse on input 1 generates a 1.2 V output pulse on output 1.**

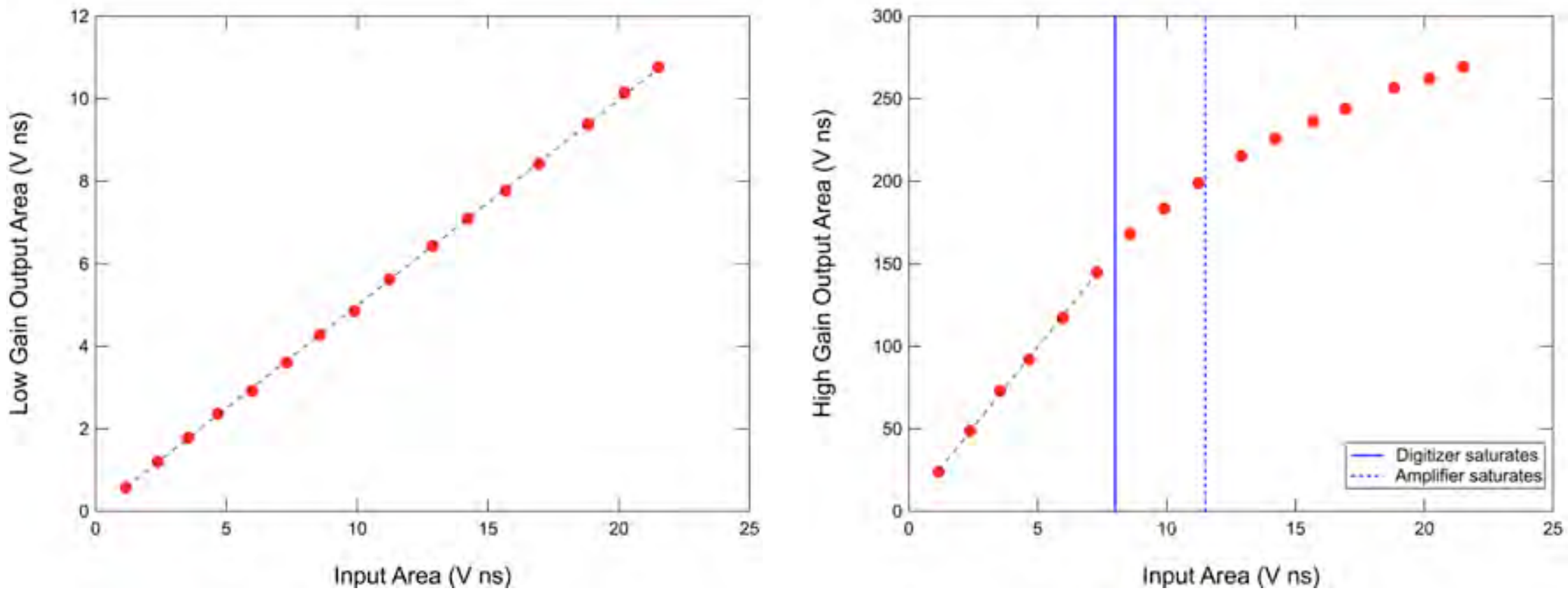

**Figure 11.3.2.4. Results of linearity measurements. The area of the output pulse, captured with our digitizer, is plotted as function of the area on the input pulse. The results obtained for the low-gain (high-energy) channel are shown on the left while the data collected for the high-gain (low-energy) channel are shown on the right.**

## 11.4 Digital Electronics

The LZ digital electronics is based on a digital platform (a motherboard); a prototype of this platform is shown in Figure 11.4.1. The final LZ motherboard will be based on this design, but will operate with a more powerful Kintex field-programmable gate array (FPGA) from Xilinx, either the new Series-7 or the newest UltraScale [1]. The final motherboard will provide gigabit Ethernet, RS-232, and low-voltage differential signaling (LVDS) interfaces, and four logic outputs, either TTL or NIM. Waveform memory (3,578 kB) will be provided by the FPGA. A large event-buffer memory of up to 128 MB will be provided by the dual-core processor. The onboard clock can be driven externally in order to synchronize multiple boards to the same clock source. Very high processing power, nominally 52 giga-operations per second, will be provided by the onboard FPGA. Two daughter card connectors can host two separate daughter cards, or one daughter card of twice the size. The I/O pins of these connectors are arranged as differential pairs, supporting either the differential or single-ended signals. A dual-core processor will be connected to the FPGA with the 32-bit memory bus, as well as two dedicated 16-bit-wide FIFOs. Readout of the FPGA data can be performed either via the memory bus or via the FIFOs, depending on the application. The board can be hosted in a 6U VME crate, or it can be powered with a tabletop power supply. Power consumption is minimized by using low-voltage chips.

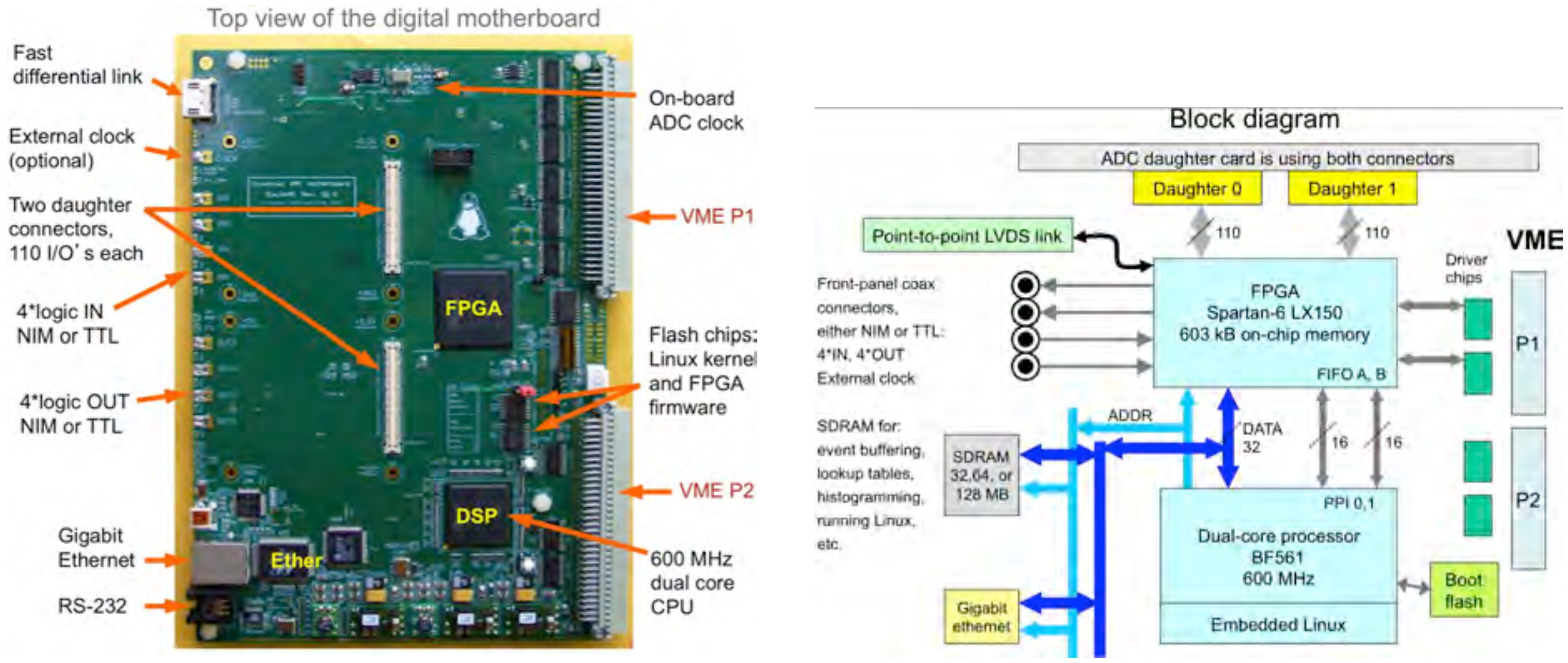

**Figure 11.4.1. The digital motherboard used to develop the LZ digitizers. It provides gigabit Ethernet, RS-232, USB-2, VME, and LVDS interfaces. The FPGA and the dual-core processor are rated at 52 and 2.4 giga operations per second, respectively.**

The onboard processing power and multiple interfaces provide flexibility that can be applied to almost any project. The LVDS links enable custom communication architectures. The Ethernet provides support for distributed experiments and/or standalone remote applications. The processor is running Linux, which is popular, free, and fully customizable, allowing each board to perform on-the-fly data processing and online diagnostics.

The 32-channel ADC card shown in Figure 11.4.2 implements the digitizer front end. It provides 32 channels of digitization and two waveform reconstruction outputs for diagnostic. The ADC channels feature remote DC offset control. The card is connected to the two daughter connectors of the digital baseboard that provide the control signals and power. The printed circuit-board layout can accommodate quad A/D chips with sampling frequency up to 125 MHz. This card, installed on the digital motherboard, is referred to as the DDC-32 in the remainder of this chapter.

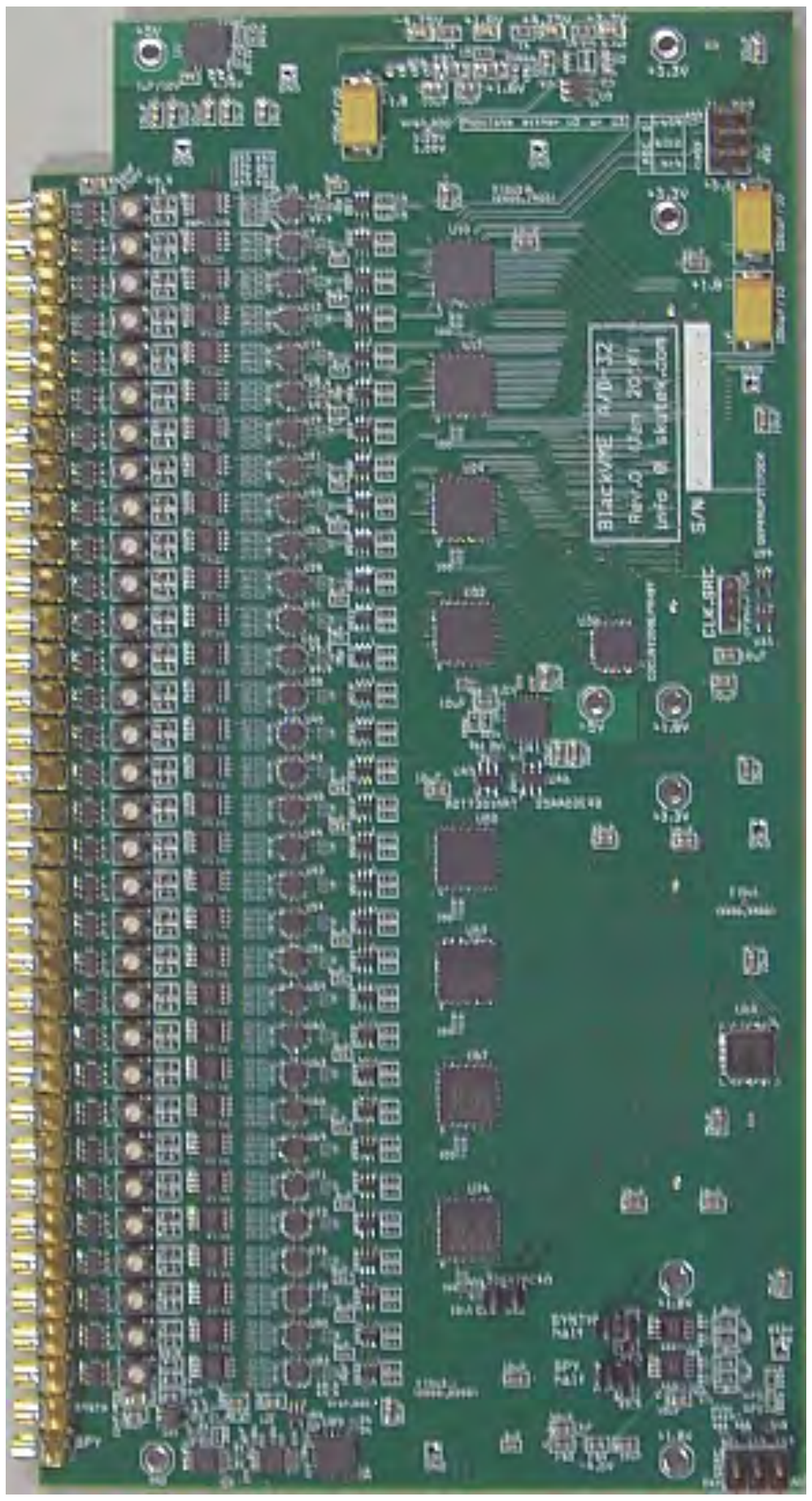

**Figure 11.4.2. Prototype ADC daughter card with 32 channels.**

## 11.5 DAQ

The top-level architecture of the LZ DAQ system is shown schematically in Figure 11.5.1. The DDC-32s continuously digitize the incoming PMT signals and store them in circular buffers. When an interesting event is detected, the Data Extractor (DE) collects the information of interest from the DDC-32s. The DEs compress and stack the extracted data using their FPGAs and send the data to Data Collectors (DCs) for temporary storage. The Event Builder (EB) takes the data organized by channels and assembles the buffers into full event structures for online and offline analysis. The DAQ operation is controlled by the DAQ Master (DM) for high-speed operations such as system synchronization and waveform selection, and by the DAQ Expert Control/Monitoring (DECM) system for slow operations such as running setup/control and operator diagnostics. The entire system runs synchronously with one global clock.

Figure 11.5.2 shows a more detailed overview of the different key elements of the DAQ system. The digitizers are sampling at 100 MHz with 14-bit resolution over a 2-V range. During normal operation, the boards will collect waveforms in a Pulse Only Digitization (POD) mode, which is expected to effectively reduce the raw waveform volume by a factor of 50 [2]. The amount of memory assigned to each channel is set so that no data truncation is expected even if the POD mode reduces the data volume by only a factor of 20. The POD waveforms are stored in dual-buffered memory that is divided into sections that hold the header information and the actual POD samples, as shown in Figure 11.5.3. Separate POD header and payload memories will improve the performance of extracting waveform data when the Data Sparsification Master (DSM) detects an event of interest [3].

The extreme flexibility that comes with using FPGAs and their internal memories allows us to assign the entire on-chip memory to just one specific channel when needed. This feature will be used for system diagnostics and noise measurements where capturing long, continuous (non-POD) waveforms is important. In such a mode, the DDC-32 will be able to capture 10-ms-long waveforms, suitable for power-spectral-density analysis.

The DEs and the DM use the same hardware but different firmware. They use the same motherboard as the digitizer boards, with different daughter cards that enable communication with multiple DDC modules and the DCs. Each daughter card can serve up to 12 DDC-32s over HDMI links and one DC over a dedicated gigabit Ethernet connection.

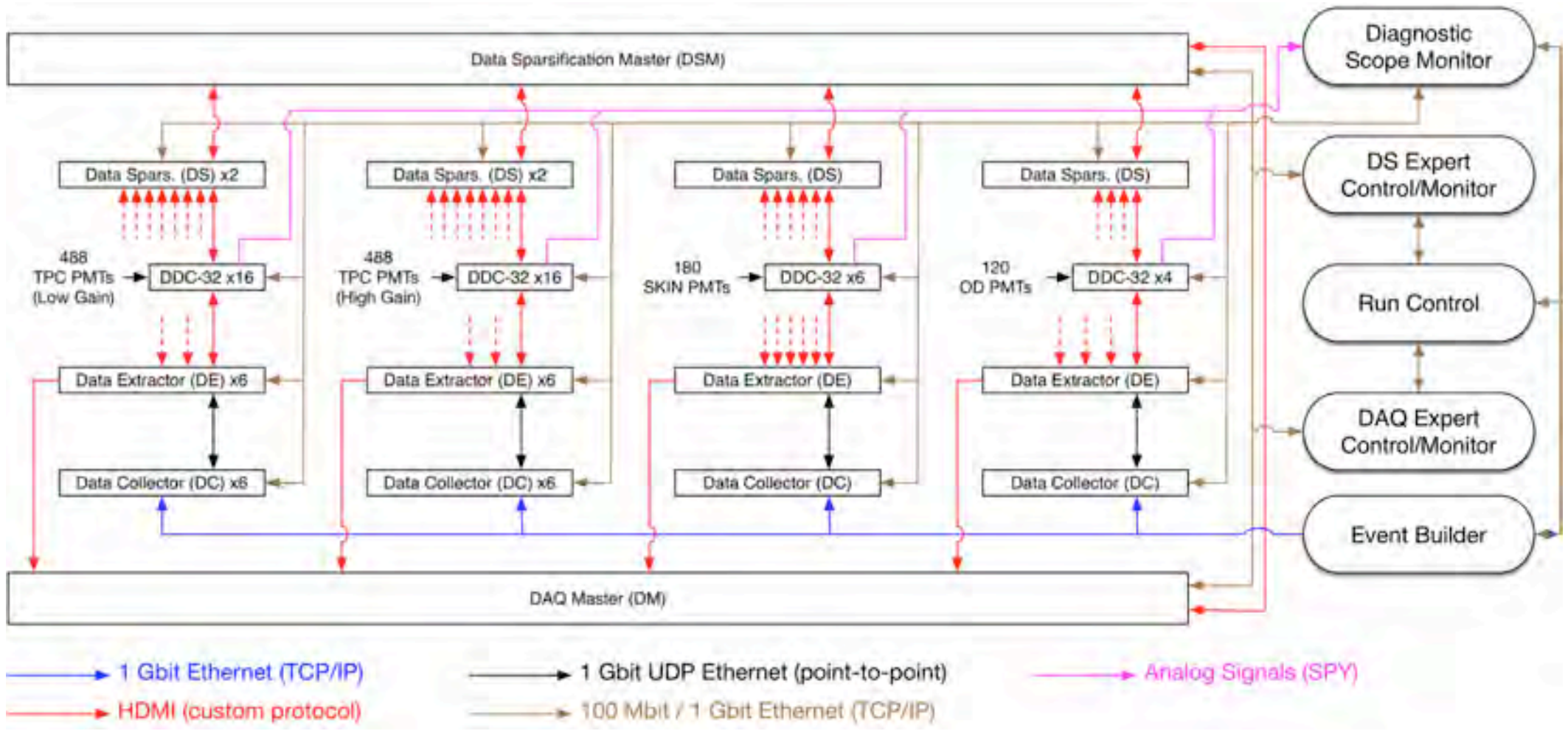

**Figure 11.5.1. Diagram of the DAQ architecture. Groups of digitizers (DDC-32) capture the amplified and shaped signals from the Xe, skin, and outer-detector PMTs. The waveforms of interest are extracted from the DDC-32s and compressed by the Data Extractors (DEs) before they are passed to Data Collectors (DCs) for temporary storage. The DAQ Master Board (DM) coordinates the high-speed operation of the entire DAQ system when the Data Sparsification Master (DSM) signals the detection of waveforms to be preserved. The global clock distribution system is not shown in this diagram.**

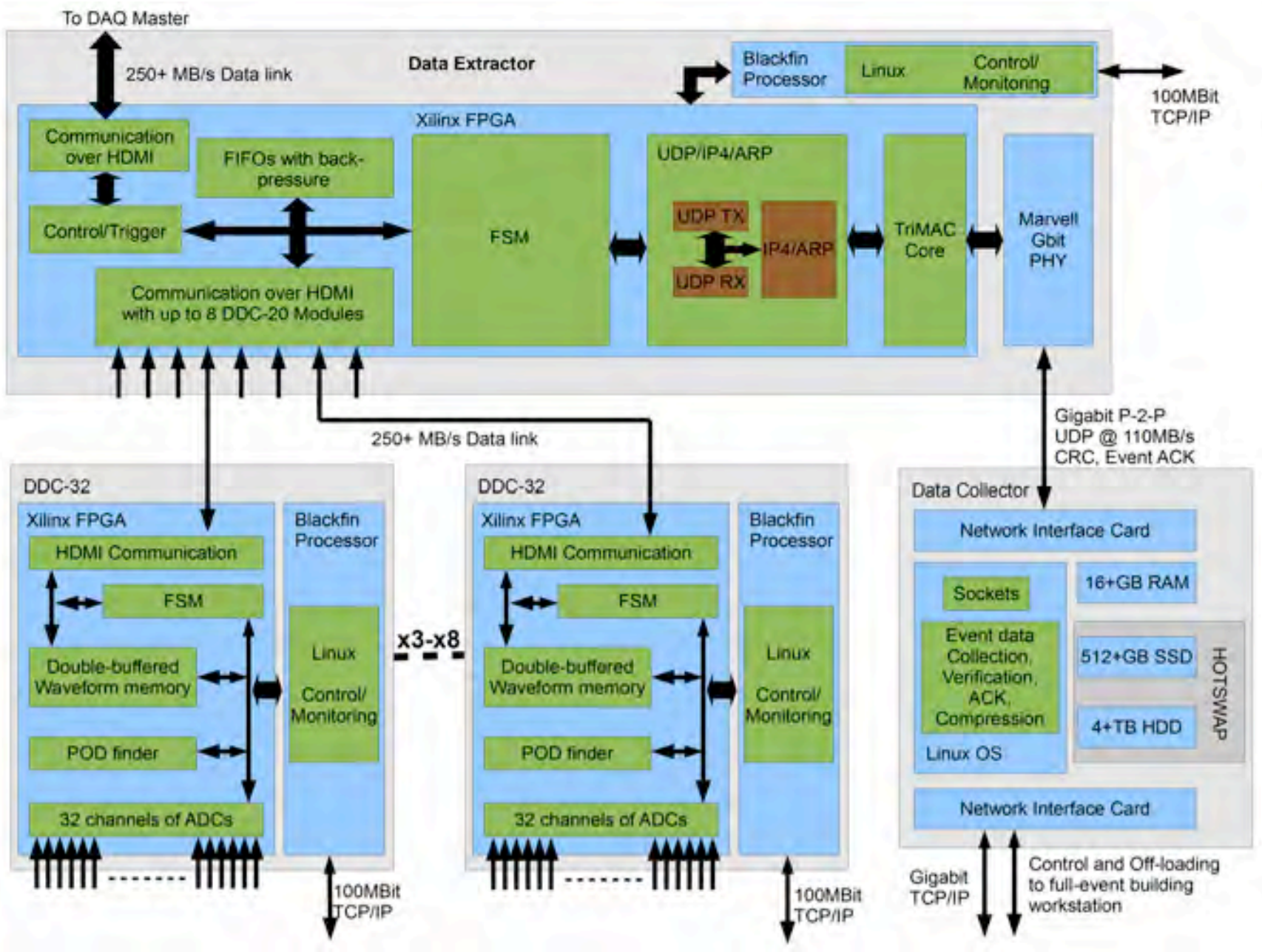

**Figure 11.5.2. Detailed depiction of the inside of and interaction between the key elements of the proposed DAQ system.**

The HDMI link has seven single-ended lanes used for communicating states of the finite state machines (FSMs) between boards and four LVDS lanes used for fast offloading of the waveforms from the digitizers. On the current motherboard, we have used input/output serial/deserializer (IOSERDES) elements, offered in the Spartan-6 FPGA series, and have confirmed the advertised 1-Gbit throughput per LVDS lane.

The gigabit Ethernet link between the DE and the DC utilizes the User Datagram Protocol (UDP). To mitigate the limitations of UDP, we will add for each event a cyclic redundancy check (CRC) and acknowledgment, and also use the dedicated Ethernet link in a point-to-point topology to eliminate congestion and packet loss. All UDP packets formed by the DE's FPGA will be sent over a single gigabit Ethernet cable to a dedicated network port on the DC.

The DCs will be implemented using server-grade, rack-mountable (2U) workstations. We have tested a prototype DC with the parameters shown in Table 11.5.1. We were able to confirm reliable data transfer from the DE to the solid-state drive (SSD) in the DC. With appropriate modifications to the network interface card (NIC) drivers, we were able to transfer data at a constant rate of 109.8 MB/s, with no data loss or corruption. Table 11.5.2 summarizes the tested link performances and their expected utilization.

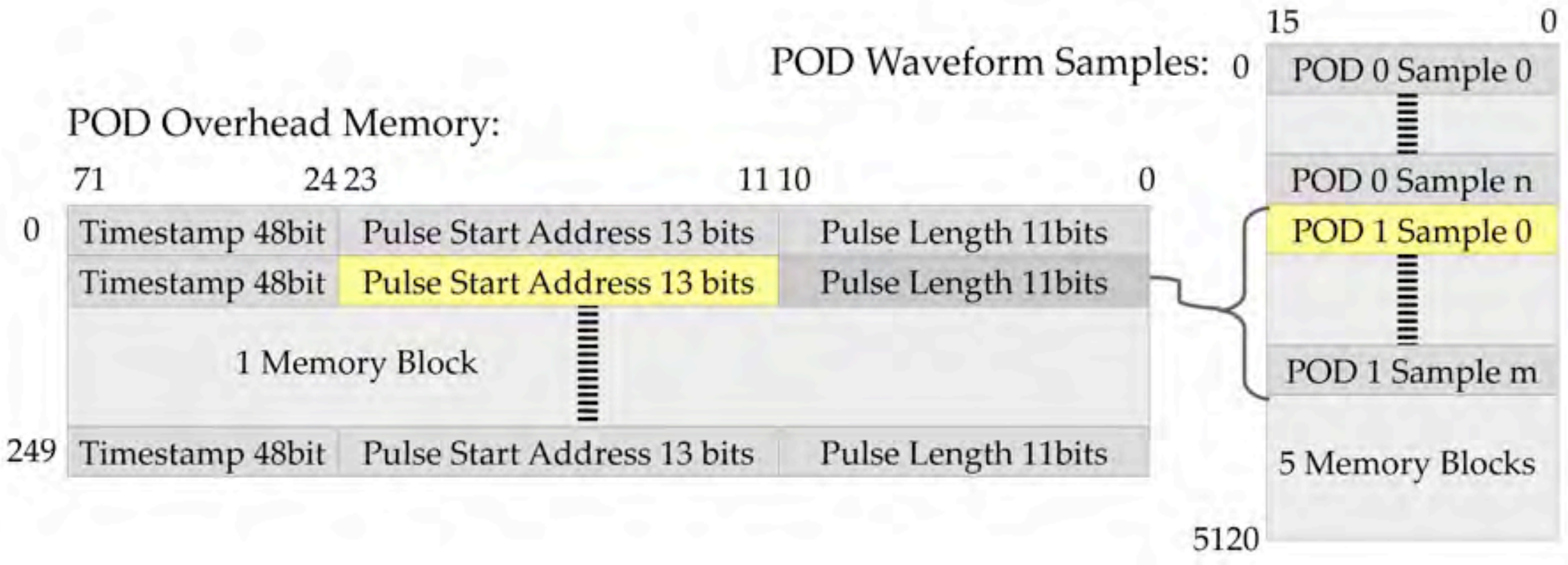

**Figure 11.5.3. Depiction of the proposed memory organization of POD waveform storage in the FPGA for a single buffer (out of two) of a single channel. Separation of POD overhead and POD samples will improve the performance and ease of extracting information from memory.**

**Table 11.5.1. Key parameters of the prototype Data Collector.**

| | | | |
|---|---|---|---|
| **Processor:** | Intel Xeon E3-1270V3 3.5GHz Quad-Core | **HDD:** | SAMSUNG 840 Pro Series 256GB SSD |
| **Motherboard:** | ASUS P9D-V ATX | | Western Digital RE 4TB 7200 RPM |
| **Memory:** | 16GB Kingston DDR3 SDRAM ECC | **Case:** | NORCO RPC-270 2U Server Case |
| **NIC:** | Intel Ethernet Server Adapter I350-T2 | **Hot Swap:** | ICY DOCK 3.5" and 2.5" SATAIII 6Gps HDD Rack Tray |

**Table 11.5.2. Summary of the performances of the DAQ links and their expected utilization levels.**

| Link | Expected Performance | Maximum Expected Usage | Usage |
|---|---|---|---|
| LVDS over HDMI | 250+ MB/s | 8.6 MB/s | < 3.5% |
| DE -> Gigabit UDP -> DC | 109 MB/s | 34.4 MB/s | < 33% |

**Table 11.5.3. Summary of the expected storage and buffering capabilities on the Data Collectors during calibrations.**

| Source | Data Rate per Data Collector (uncompressed) | 512 GB Solid-state Drive | | 4 TB Hard-disk Drive | |
|---|---|---|---|---|---|
| | | Raw | 7z | Raw | 7z |
| Kr-83 | ~5.2 MB/s | ~1.1 days | ~4.4 days | ~9 days | ~35 days |
| LED | ~34.4 MB/s | ~4 hours | ~16 hours | ~1.4 days | ~5.3 days |

The data stored on the DCs are processed by the Event Builder (EB). The EB builds events by sampling the different DC disks and collecting all information associated with a given event. The resulting event (EVT) files are compressed in order to achieve the smallest possible file size for storage and transmission. The EVT files are written to RAID array #1, located in the Davis Cavern, before being transferred to the RAID array #2, located on the surface at SURF. The RAID arrays have sufficient storage capacity for a full month of data taking.

The DAQ system is designed to be able to allow LED calibrations of the TPC PMTs in about 10 minutes. This requires an event rate of 4 kHz, resulting in a ~340 MB/s total waveform data rate. The fast (400+ MB/s for SSDs) storage drives in the individual DCs allow for sustaining such collection rates, and the amount of space offered by each DC permits significant buffering in case of connection problems to the off-site permanent storage, as shown in Table 11.5.3.

We are planning to use 7z compression. Nominally, 7z uses the LZMA algorithm, but we have found that the PPMd algorithm [4] is better suited for the waveforms we are going to collect. Based on comparisons made using LUX data, the 7z PPMd compression offered an additional 33% data reduction for waveforms and a 15% data reduction for reduced quantities over gzip, which was used in LUX.

Because the POD mode relies on time stamping, the entire DAQ system will run with a global 100 MHz clock. The DDC-32s, DEs, and DM will run with the external clock source. We are considering two ways of distributing the clock, either by using a proven method of dedicated NIM logic clock FAN-IN/OUTs or using clock recovery from the serial links of the HDMI cables; the latter method is currently being evaluated. We are also looking at the possibility of using one of the single-ended channels of the HDMI cable for a dedicated clock signal.

The configuration, acquisition control, and monitoring will be done using the DECM workstation. For improved robustness, the DECM will communicate with all DAQ elements using an isolated 100/1000 LAN; the run-control (RC) system, described in Section 11.10, will not have direct access to the front-end digitizers. We have successfully tested and are planning to use ICE middleware as our communication framework [5].

## 11.6 Data Sparsification

The design of the LZ data sparsification system is based on our experience with event selection in the LUX trigger system. The top architecture of the LZ data sparsification system is shown schematically in Figure 11.5.1. The DAQ and data sparsification firmware operates in parallel on the FPGAs of the DDC-32 digitizers.

The DDC-32s continually process the incoming pulses and extract specific required quantities such as pulse area, pulse height, and time of occurrence. By using digital filters with various filter lengths, the system can distinguish S1 and S2 pulses. The parameters extracted from the waveforms by the DDC-32s are sent to the Data Sparsifiers (DSs) for further processing and generation of secondary quantities such as the multiplicity vectors of groups of PMTs and their total energy. These secondary quantities are sent to the Data Sparsification Master (DSM), where the final waveform selection decision is made. The decision to preserve the current waveforms is sent to the DAQ Master.

The LZ DAQ/Sparsification system is capable of handling event rates up to ~250 kHz. This is much higher than the highest event rates expected in the LZ detector, which will occur during LED calibrations (~4 kHz).

As in LUX, the PMT signals will be processed by parallel digital integrating filters, allowing for discrimination against area of the incoming pulses. The filter with an integration width of about 60-100 ns is tailored to S1-like pulses; the second filter with a few-microsecond width is optimal for wider S2-like pulses. The filters are designed to also perform automatic baseline subtraction.

As shown in Figure 11.5.1, the TPC, skin, and outer-detector PMTs will use separate DDC-32s and DSs because these PMTs have different purposes and the firmware will be tailored to their specific needs. Such separation also makes it easier to apply different scaling factors for the digitization of signals from different PMT groups. The three major waveform-selection modes for the central TPC PMTs are summarized in Table 11.6.1.

If a waveform selection condition is met, the DSM sends a signal to the DM. At the same time, a packet containing the selection parameters used to make the decision is merged with the captured waveform data. This allows for offline evaluation of the selection decision for every individual event. This feature has proved extremely valuable in monitoring LUX data quality.

**Table 11.6.1. Summary of three major waveform-selection modes for the central TPC PMTs.**

| Trigger Mode | Summary |
|---|---|
| **S1 Mode** | Detection of coincident S1-like signals across selected channels. No fiducialization. |
| **S2 Mode** | Detection of coincident S2-like signals across selected channels. Fiducialization in the x,y plane. |
| **S1 & S2 Mode** | Detection of S1-like signals followed by S2-like signals within a selected drift time range. Fiducialization in x, y and z planes. |

Similarly to the DAQ system, the DS system will be controlled and monitored by a dedicated DS expert control/monitoring (DSECM) workstation over an isolated 100/1000 LAN network. The DS system will operate on the same 100 MHz global clock as the DAQ.

The LZ DS system will adopt and expand on the monitoring capabilities of the LUX system. Capabilities such as performing continuous noise sweeps (monitoring S1 and S2 filter crossing rates as a function of threshold) or monitoring channel hit distributions of the selected events, all in parallel to regular uninterrupted data-sparsification operation, have been invaluable in LUX and surely will be in LZ.

Each of the DDC-32s has a dual 14-bit analog reconstruction output (SPY) allowing for diagnostics and scope monitoring. The SPYs can be sourced with individual incoming channels or their digital sum. They can also be sourced with S1 and S2 filter outputs, individual or summed, or any other signal internal to the FPGA. This feature will be very useful, especially in development and deployment stages, as proven in the LUX project.

## 11.7 PMT High-voltage Supplies

LZ will use the Wiener Mpod LX HV system to bias the Xe and the outer-detector PMTs. This system is currently being used for LUX. The system will provide negative HV to the Xe PMTs and positive HV to the outer-detector PMTs. The HV modules are part of the EDS 201-30x 504 series. Each module provides 32 HV channels and uses a common floating ground. The voltage ripple is less than 5 mV. The HV can be set with a resolution of 10 mV, and the current on individual channels can be measured with a resolution of 50 nA.

HV connections to the HV filters are made using Kerpen cable with Redel connectors on both ends. Each Kerpen cable carries HV for 32 channels. Important properties of the HV system are listed in Table 11.7.1.

**Table 11.7.1. Details of the PMT HV system.**

| HV Module | Maximum HV | Max Current per Channel | PMTs | Channels | Required Number of Modules |
|---|---|---|---|---|---|
| **EDS 201 30n** | -3000 V | 500 μA | TPC/Skin | 662 | 21 |
| **EDS 201 30p** | +3000 V | 500 μA | Outer-detector | 120 | 4 |

## 11.8 Cables

All network, signal, and HV cables are low smoke zero halogen (LSZH) cables. LZ uses the same cables that have been approved for LUX use in the Davis Laboratory. The exact lengths of the HV and the signal cables will be fixed once the location of the analog amplifiers is fixed and the route of the cable trays has been finalized.

The types of cable and their lengths are listed in Table 11.8.1. The networking cables will have various lengths; individual lengths will vary based on the location of the network switched and the location of the devices to be connected. About half of the network cables are used by the slow-control system; the other half will be used for the networks associated with the DAQ, trigger, and online systems.

We have not been able to identify a suitable HDMI cable that provides good performance and uses LSZH materials. The total length of these cables is small and the electronics racks are enclosed and vented directly into the exhaust system of the Davis Campus. This arrangement will be assessed by the SURF safety team. Note that rack enclosures and connections to the exhaust system are also used in LUX for this same reason.

**Table 11.8.1. Information on LZ signal, logic, HV, power, and network cables.**

| Cable Type | Type | Length (ft) | Number of Cables | Notes |
|---|---|---|---|---|
| **Network** | Belden 7936A | 10,000 (total) | 750 | Cat 6, LSZH, various lengths |
| **Signal** | LMR-100A-FR | 56 | 128 bundles (8 cables/bundle) | LSZH |
| **Logic** | LMR-100A-FR | 2 | 500 | LSZH |
| **HV** | Kerpen | 56 | 25 | LSZH, 32 channels per cable |
| **Power cords** | CordMaster | 6 | 100 | LSZH |
| **HDMI** | TBD | TBD | 99 | Lengths TBD. |

## 11.9 Slow Control

### 11.9.1 Requirements

Based on experience with LUX, we expect modest requirements for the bulk of the slow-control channels. For most channels, the rates will be less than 1 Hz and a latency of 1 s is acceptable. Typical data rates will be no more than a few kB/min/channel. The interfaces that will be used for most channels include the four-wire interface for the resistance temperature detectors (RTDs), analog inputs and outputs, relays, serial links (RS232 or RS485), and Ethernet links.

The hardware for these "standard" interfaces is included in WBS 1.8.5. Other interface types, not listed above, will be provided by the corresponding subsystem. The estimated requirements for the "standard" channels are summarized in Table 11.9.1.1.

**Table 11.9.1.1. Slow-control interface needs.**

| WBS | | RTD | Analog | Digital | Relay | Serial | Ethernet |
|---|---|---|---|---|---|---|---|
| 1.3 | **LN system** | 20 | 14 | 10 | 10 | 21 | 14 |
| 1.4 | **Gas system** | | | | | 12 | |
| 1.4.1 | **Sampling** | 12 | 8 | 24 | 24 | 14 | 4 |
| 1.4.2 | **Kr removal** | 25 | 38 | 75 | 75 | 17 | 12 |
| 1.4.4 | **Delivery and recovery** | 30 | 23 | 12 | 12 | 31 | 4 |
| 1.4.5 | **Recirculation and purification** | 18 | 25 | 28 | 28 | 12 | 4 |
| 1.4.6 | **Liquid purification tower** | 20 | | | | 10 | |
| 1.5 | **Detector** | | 5 | | | 5 | 6 |
| 1.5.5 | **Xe system monitoring** | 100 | | | | 10 | |
| 1.6 | **Veto** | 20 | 20 | | | | |
| 1.7 | **Calibration** | | 4 | 10 | 10 | 8 | |
| 1.8 | **DAQ and electronics** | 36 | 182 | | | 10 | 40 |
| | **Environment** | 40 | 10 | 10 | | 5 | 5 |
| | **Total** | 321 | 331 | 169 | 159 | 155 | 89 |

### 11.9.2 Platform Choice

Not all the sensors/equipment monitored and controlled by the slow-control system are of the same importance. They can be broadly classified as “critical” sensors/equipment for which a failure or human error can damage equipment or result in loss of Xe, and “standard” sensors/equipment for which readout or control errors will not pose any immediate danger. We decided to use a hybrid platform, schematically shown in Figure 11.9.2.1, with an Allen-Bradley programmable logic controller (PLC) based system servicing the critical sensors/equipment. This system will be programmed using graphical ladder diagram language [6] — commonly used for implementing process control in industrial applications — to guarantee fast and reliable response in critical situations. The noncritical inputs and outputs will be serviced by a database-centered system. The interface between the PLC and the database will be implemented using SQL bridge [7]. Such an approach allows us to minimize the slow-control system cost without compromising its reliability. As an additional safety feature, a standalone emergency system will provide redundancy, taking control over critical hardware in the unlikely case of a PLC failure.

The noncritical system is built around the main slow-control database that serves as a hub between the slow-control core, device drivers, and the user interface. The core is very compact and lightweight. The hardware-specific device drivers transfer the information between the hardware devices and the database. It is important to note that the drivers run as separate processes, and the system as a whole remains stable if one of the drivers fails. Hot-swapping the drivers is not a problem either. The components of the system can be distributed across different computers or even different hardware platforms; e.g., part of the drivers can run on an embedded system if necessary. The safety of operation is ensured by two constantly running pieces of software: the main watchdog that is responsible for starting, stopping, and monitoring hardware drivers, and the alarm system, which constantly monitors the critical parameters and alerts all interested parties if they go outside a predefined range.

The database is constantly mirrored to a separate computer in the Davis Laboratory and to several computers on the surface. Underground access to the database is configured for high availability [8], with the primary mirror seamlessly taking over from the master database in the case of problems or required maintenance. The slow-control software also uses an ICE server for direct communication with run control (RC) (see Section 11.10.1). Through this server it is possible to monitor and control a limited number of important parameters, bypassing the database. This additional redundancy increases the overall

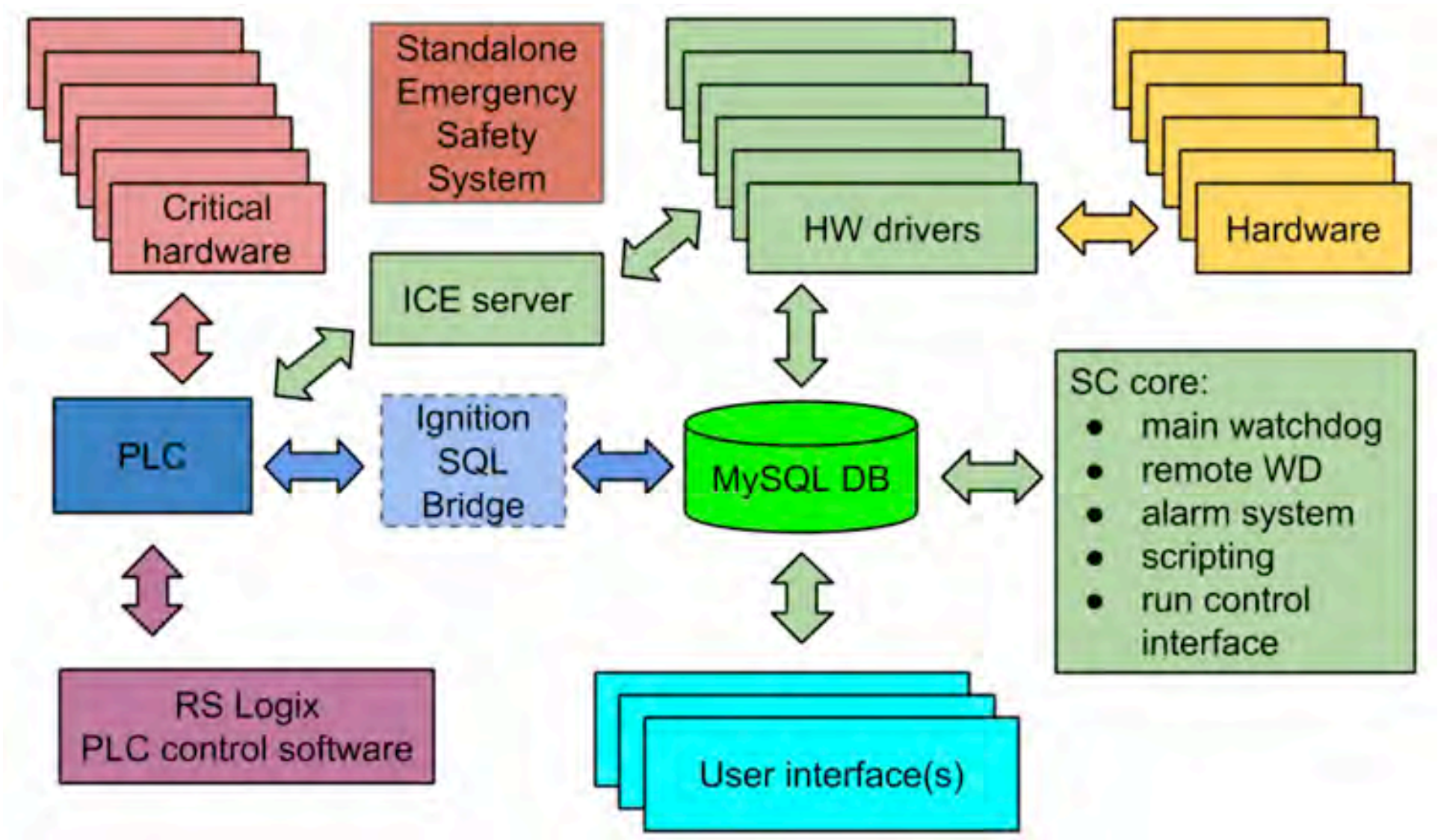

**Figure 11.9.2.1. Slow-control functional diagram.**

robustness of the system in critical situations.

The noncritical control and monitoring software will be based on the open-source slow-control system developed for and currently used by LUX. This choice is based on several factors. The LUX system can be easily scaled up to the expected number of channels and data rate for LZ. From a safety point of view, it is very robust due to its modular architecture and is designed to provide quick response in emergency situations. An important additional factor is the existing level of experience in the collaboration.

### 11.9.3 Special Requirements

According to our estimates, the platform described above will be able to handle up to a few thousand standard channels (<1Hz, >1s latency) without overload. These rate and latency limits are not strict, but are guidelines to guarantee that all the channels play "nicely" together in the large system.

A small fraction of channels are expected to go beyond the standard requirements. For example:

- The temperature and pressure sensors in the circulation and Kr removal subsystems
- The valves for Kr injection
- The nonstandard sensors related to the Xe vessel and TPC: acoustic, radio, and camera

Faster readout and control (~10 Hz, ~0.1 s latency) for a limited number of channels is not a problem even for the base system. If faster acquisition (up to few kHz) is required, the solution would be a dedicated acquisition board with an embedded Linux system running a slow-control hardware driver. The data will be stored in a dedicated database to reduce the load on the main system. Such an add-on can be seamlessly integrated into the main slow-control system due to the modular design of the latter.

### 11.9.4 Network

The Ethernet network (Figure 11.9.4.1) is the backbone of the slow-control data flow. For this reason, a high-performance local network is built on site. Each subsystem is equipped with a dedicated managed switch and a number of unmanaged switches if necessary. All the "standard" interfaces are ultimately routed through an Ethernet link: via ADAM modules for RTDs, analog input/output, and relay controls; and via serial-to-Ethernet adapters for RS232 and RS485. The subsystem switches are linked to the main 48-channel managed switch connected to the main slow-control computer via a gigabit uplink. The cabling will be made with LSZH cables, ensuring triple redundancy for the links to the main subsystems.

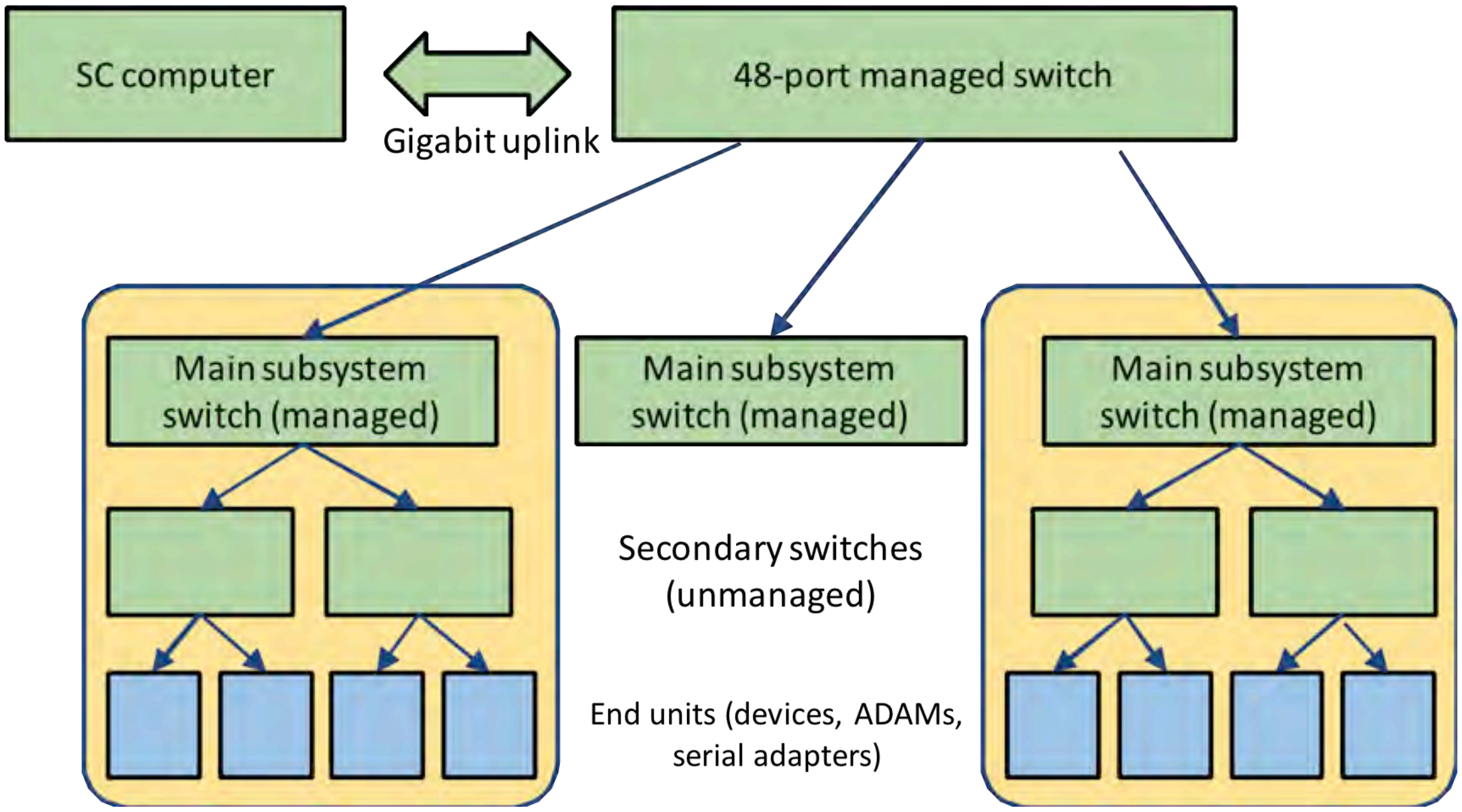

**Figure 11.9.4.1. Slow-control system network.**

## 11.10 Online System

The core of the online system is the run-control system (RC), schematically shown in Figure 11.10.1. It comprises the software and hardware required to allow the operator to define and initiate data collections of different types, control and monitor subsystems (DAQ, slow control, EB, and offline software), and log key information to the database. The physical system is hosted on a single rack-mount computer, identical to the hardware used for the event builder. The use of identical hardware with an on-site spare minimizes any downtime of this critical system. RC software is maintained on a mirror system throughout the project and operation periods to allow off-line debugging and code development.

The work on the RC system started with defining the architecture of the system and the development of detailed specifications and interface documents with the various subsystems. An important decision was the selection of the middleware to be used for communications between RC and the subsystems. We have past experience with a number of middleware packages, in particular the industry-standard CORBA (Common Object Request Broker Architecture) system. Given the limitations of CORBA, e.g., scalability, throughput, reliability, and thread-safety, a new solution was considered with similar functionality but improved performance. A recent study [9] to identify middleware to run CERN accelerators ranked ICE (Internet Communications Engine) and ZeroMQ as the top two evaluated systems. The criteria for this evaluation included reliability and speed, and the ability of the system to handle a large number of messages per second (both small and large messages) and publish to a large number of clients in minimal time. Based on preliminary comparisons ICE has been identified as the best choice for LZ middleware. ICE is an object-oriented system similar to CORBA, but improves on many of CORBA's limitations. ICE provides both a request-reply and publish-subscribe service. ICE supports C++ calls and runs on Linux. The RC software will be developed using a combination of C++ and Python, and run on a Linux-based server. We have not yet selected the final graphics/GUI (graphical user interface) library, but PyQt [10] appears to be the most promising and will be the basis of our initial development. To make our code somewhat independent of the specific choices for the system architecture, a library of wrappers will be developed for various low-level system calls, including

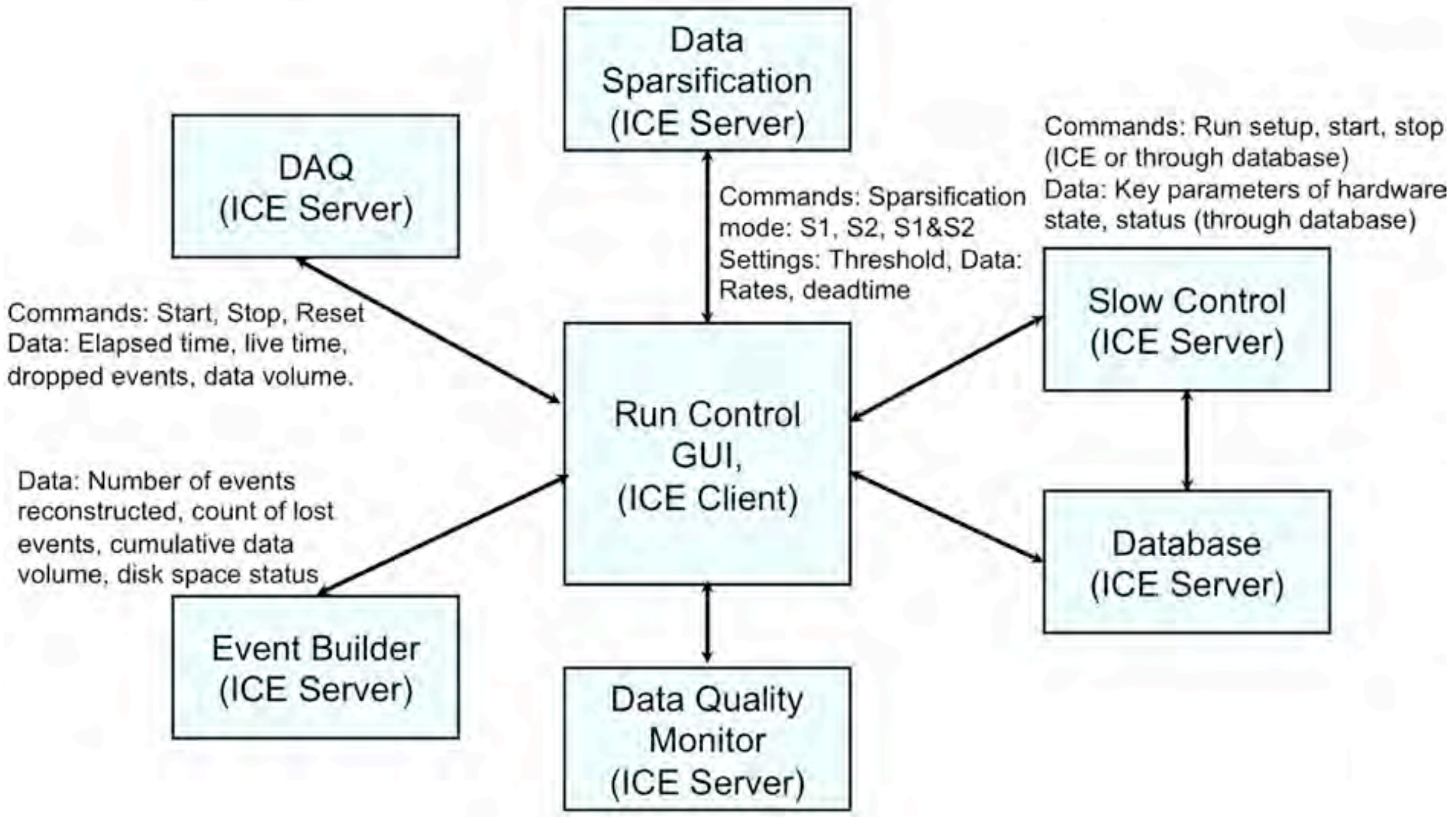

**Figure 11.10.1. Block diagram showing the primary interfaces between the RC system and other LZ subsystems. The users use a GUI to interact with the RC system.**

common command/control interfaces, database queries, and graphical display/plotting commands.

To provide a seamless interface to the DAQ system, each data collection mode (e.g., calibration runs, data runs) is associated with a corresponding set of command sequences to the various subsystems. The list of data collection modes includes normal data-taking modes (S1 only, S2 only, S1 and S2; see Table 11.6.1 for more details) and a number of calibration modes, including the LED, krypton, tritium, and neutron calibration (see Chapter 10 for more details). The RC group will interact closely with the groups responsible for the various subsystems to determine the required interfaces, including defining what status information is shown, what commands are provided, and how status information should be displayed. Based on this information, the RC group will develop an application programming interface (API) and skeleton (server/daemon) program for each subsystem, including the appropriate handles for communication with RC, covering commands and sharing of status information. These skeleton server programs will be further developed by each group for their specific subsystem, working in close consultation with the RC programmer. This model has worked very well in the past, expediting code development compared with the model in which each subproject is expected to develop its own interface code from scratch.

Throughout the course of the development, the run-control and slow-control groups will work together closely and try to adhere to common standards for coding, code-management, and common libraries for communication and interfaces. Most data sharing between run-control and slow-control will occur through the database, but some of the most important commands will occur through a direct ICE connection between slow-control server computers and the run-control system.

The final RC system provides the user with a simple GUI to the DAQ, data quality monitoring, and the slow-control systems. The GUI will guide the user through setting up and starting/stopping various data-collection modes and starting runs, making sure all settings, data, and operator comments are logged into the database. The GUI includes key alerts for out-of-bound values, as determined by the slow-control system, and provides a small number of user-selected plots that allow the user to quickly monitor the health of the LZ system.

A skeleton version of the RC software will be developed for the electronic system test. During this development period, the options for the basic system framework will be determined and the system hardware will be purchased. The RC group will work with software developers for the other subsystems to develop interface control documents (ICDs) specifying command sequences, settings, and critical data to be exchanged directly through an ICE link or indirectly through the database. Lessons learned from the electronic system test and the feedback from early users will provide input for the final design phase, during which ICDs will be finalized and the full set of interfaces between RC and the subsystem implemented.

Because the code will be developed and maintained by an experienced professional programmer, it is expected that the RC system will be well maintained throughout the LZ project and operation periods, with minimum downtime.

## 11.11 Installation

The electronics will be installed in two locations. The analog electronics for the Xe PMTs will be installed on the mezzanine level, as shown schematically in Figure 11.11.1. The amplifiers will be arranged in eight to 10 racks, with 10 amplifier cards installed in each rack. The racks will be located horizontally on the wall of the water tank in order to improve cooling and provide better access to each amplifier crate. Exhaust ducts may must be installed at this location to facilitate cooling of the electronics. Because it will be difficult to enclose the electronics crates at this location, it is important the air flowing over the electronics is directed into the exhaust system. This will prevent any smoke generation as a result of an electrical problem, e.g., overheating of components, from spreading into the Davis Campus. The outputs of the amplifiers are routed in cable trays to the DAQ and trigger systems installed in the electronics racks installed at deck level (see Figures 11.11.1 and 11.11.2).

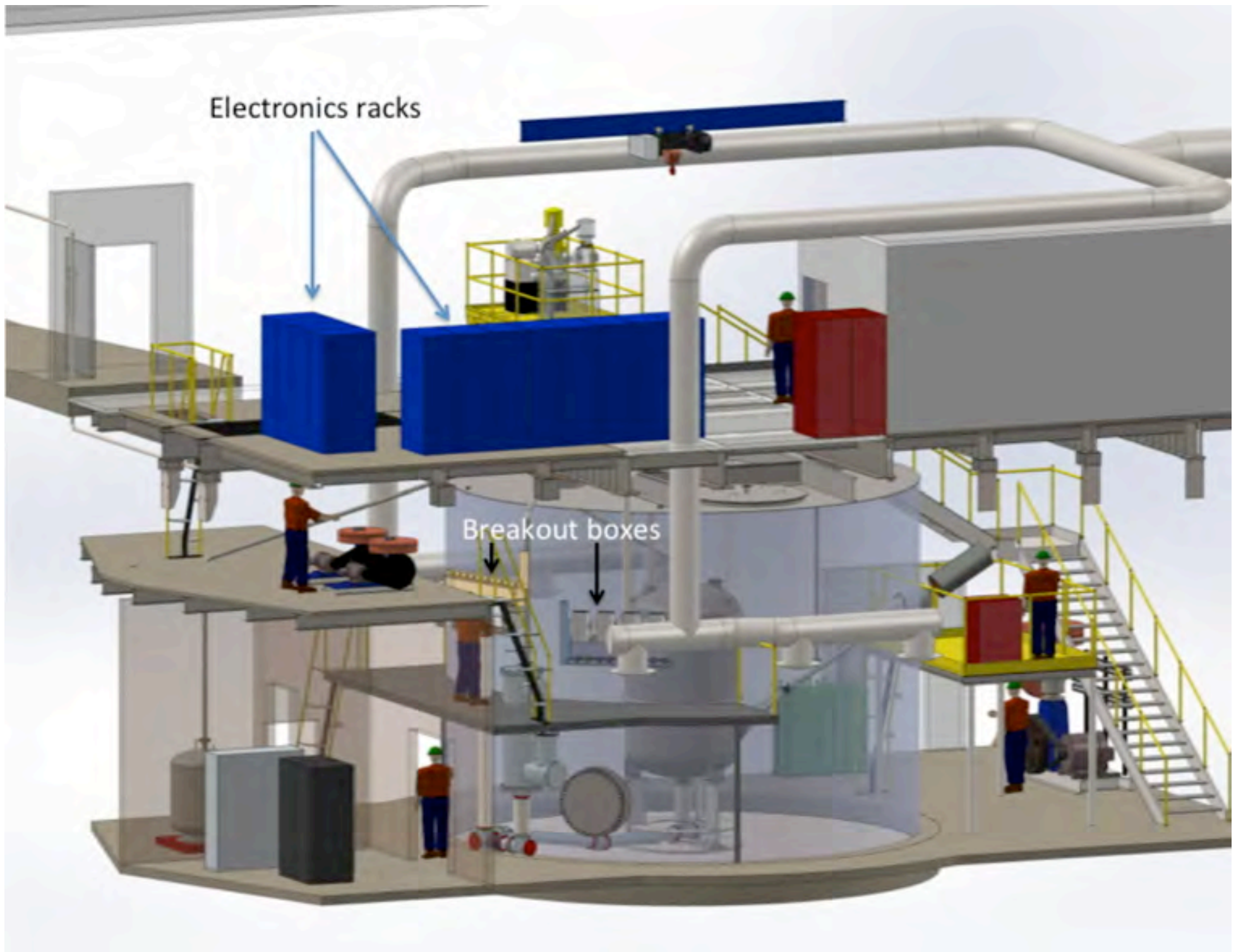

**Figure 11.11.1. Installation of the analog electronics for the Xe PMTs at the mezzanine level and the electronics racks at the deck level.**

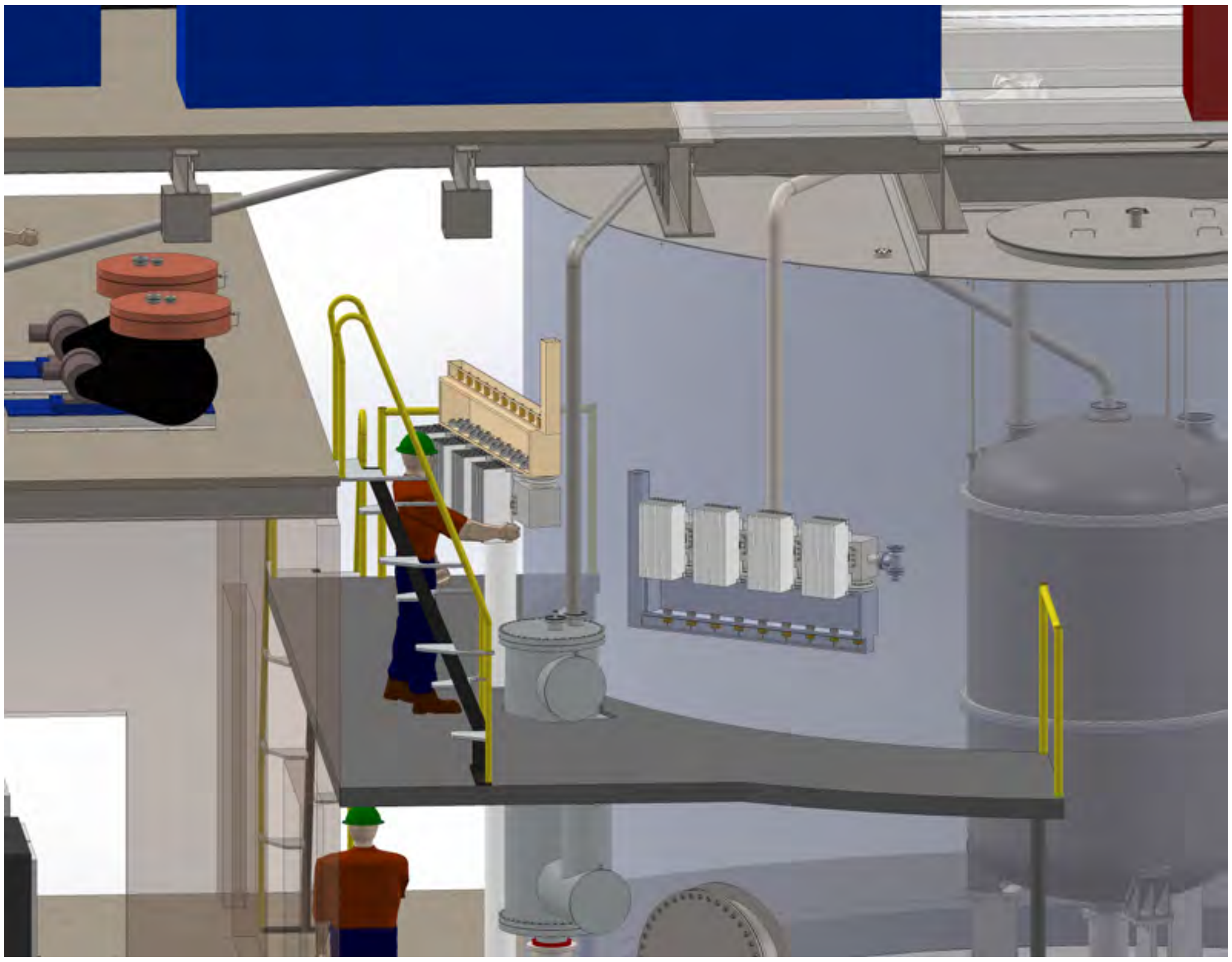

**Figure 11.11.2. Close-up of the mezzanine level showing the amplifier racks installed on the breakout boxes.**

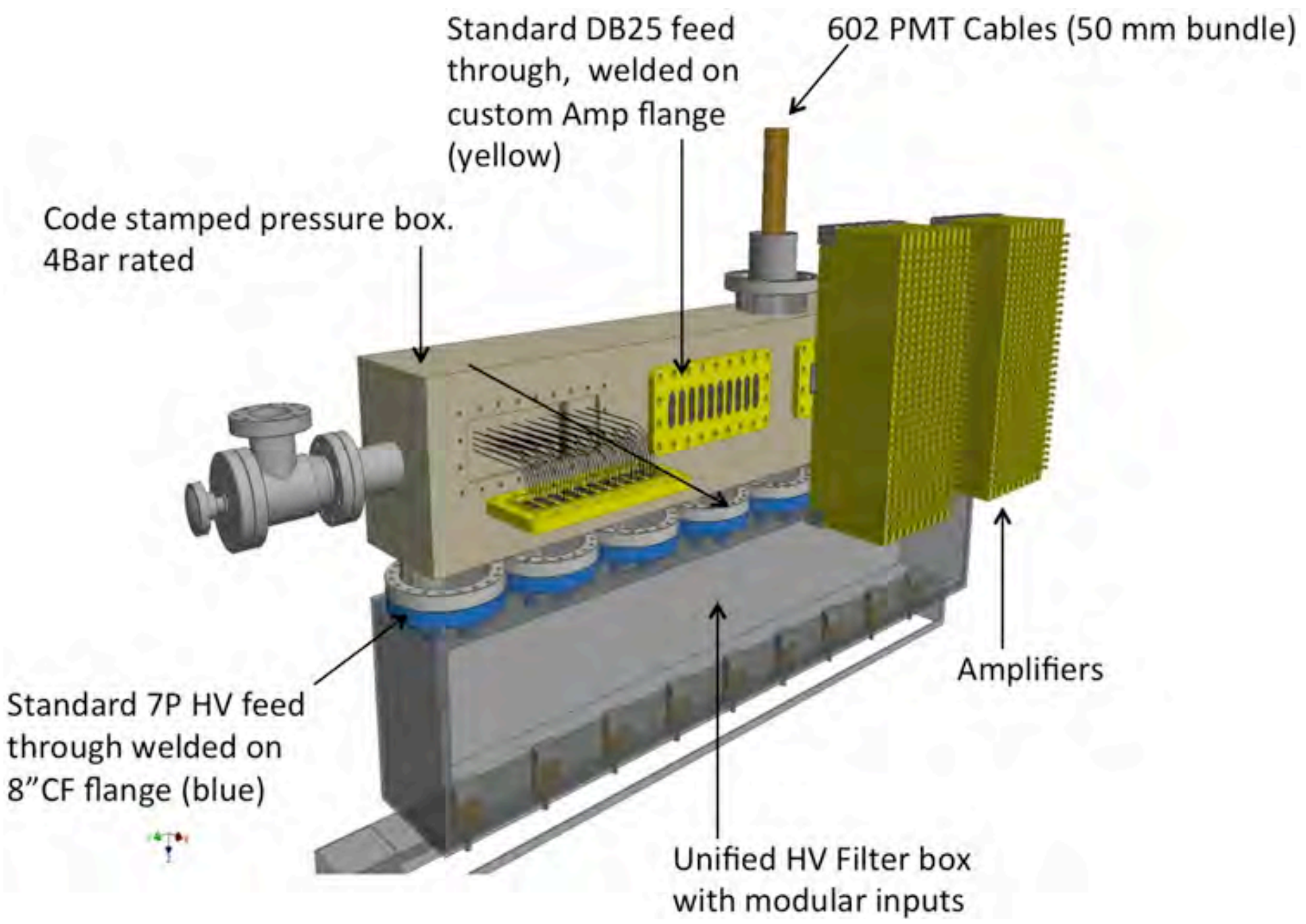

**Figure 11.11.3. The breakout box for the PMTs located in the top of the cryostat. Two electronics crates with 10 amplifiers each are attached to two flanges on the breakout box. The grey box, mounted below the pressure box, contains the HV filters. The location of the breakout boxes is shown in Figure 11.11.1.**

A schematic of the breakout box, with the amplifier racks disconnected from the box, is shown in Figure 11.11.3. The internal signal and HV cables enter the breakout box through the conduit at the bottom of the box. The signal cables are routed to the rectangular DB-25 connectors on the rectangular flanges with the amplifier crates. The HV cables are routed to the round HV flanges, visible on the top of the breakout box.

**Table 11.11.1. LZ power requirements in the Davis Laboratory**

| System | Predicted load (kW) | Maximum load (kW) |
|---|---|---|
| Thermosyphon tower | 4.2 | 4.2 |
| Circulation panel | 7.9 | 7.9 |
| Storage panel | 0.1 | 0.1 |
| Storage system | 10.9 | 10.9 |
| Emergency control | 0.5 | 1.5 |
| $LN_2$ storage | 22.6 | 22.6 |
| Gas control | 1 | 1 |
| Breakout carts | 4.6 | 4.6 |
| Amplifiers | 4.8 | 12 |
| DAQ | 30.5 | 42 |
| Data Sparsification | 20 | 24 |
| Monitoring computers | 2.1 | 6 |
| HV and slow control | 12.5 | 23 |
| Ultrasonic cleaning | 2.3 | 2.3 |
| Water system | 13.1 | 13.1 |
| Control room | 5.8 | 5.8 |
| Tools | 1.3 | 1.3 |
| **Total** | **146** | **188** |

The amplifiers and the HV pickoffs for the outer-detector system will be installed in the electronics racks shown in Figure 11.11.1. The HV cables from the outer-detector PMTs are routed via one of the flanges on top of the water tank to this location. The design of the flange is such that no connectors are required to ensure a light-tight and radon-tight connection. Not using connectors on the flange improves signal quality.

A total of 12 electronics racks are installed at the deck level, as shown in Figure 11.11.1, to provide space for all electronics systems, except the Xe amplifiers, and the computer and disk systems. The racks are fully enclosed and the forced airflow across the electronics is directed into the exhaust system of the Davis Laboratory.

The power requirements of the electronics system in the Davis Laboratory have been estimated and are summarized in Table 11.11.1. These estimates are based on the measured power consumptions of the various components of the system. The maximum load is the load of the equipment at startup. The power required by the electronics systems is provided by uninterruptible power supplies installed in the electronics racks. The total maximum load of the electronics is 109 kW.

Table 11.11.1 also shows the power consumption by other LZ systems.

## Chapter 11 References

# 12 Assay, Screening, and Cleanliness

## 12.1 Introduction

For LZ to realize a sensitivity to a WIMP-nucleon cross section above $2 \times 10^{-48}$ cm$^2$ at a 50 GeV/c$^2$ WIMP mass from an approximate 5.6-tonne fiducial mass within three years of WIMP data taking, the maximum tolerable ER background in the WIMP search energy region (1.5 – 6.5 keV$_{ee}$) from non-astrophysical sources should be approximately $3 \times 10^{-6}$ events/day/kg/keV (dru) before discrimination through S2/S1 and vetoes from the LXe skin and OD are applied. Although this represents an unprecedented low background rate for dark-matter detectors, it can be achieved by employing a number of proven low-background and assaying techniques that have been successfully employed in recent rare-event searches for dark matter, as well as in neutrinoless double-beta decay and neutrino experiments [1-9]. These techniques include:

- A comprehensive material-screening campaign to select components that satisfy stringent radioactivity constraints such that the single scatter background particle interaction rate within the fiducial volume and WIMP search energy range is reduced below 1 $\mu$ dru
- Removal of radioactive elements such as Kr, Ar, or Rn from the LXe to limit their single scatter background in the WIMP search energy range to below 2 $\mu$ dru
- Adherence to cleanliness protocols for control of airborne radioactivity and particulates, both during parts manufacture and during final assembly and integration

The material-screening campaign is the primary route to controlling the ER and NR backgrounds resulting from radioactivity in and on detector materials in the experiment. These are primarily the gamma-ray-emitting isotopes $^{40}$K, $^{137}$Cs, and $^{60}$Co, as well as $^{238}$U, $^{235}$U, $^{232}$Th, and their progeny. The U and Th chains are also responsible for neutron production following spontaneous fission and ($\alpha$,n) reactions, and therefore represent the most serious source of background. Kr, Ar, and Rn outgassing from materials into the Xe also results in ER backgrounds, and $\alpha$-emitting Rn daughters can contribute to neutron backgrounds. Generally, contaminants in massive components or those closest to the central active volume of Xe present more stringent cleanliness and radiopurity requirements for the experiment. These are the PMTs; PMT bases and cables; the TPC components, including the PTFE sheets and support structures; the Xe target itself; and the cryostat. Our screening campaign includes several mature techniques for the identification and characterization of radioactive species within these bulk detector materials, namely gamma-ray spectroscopy, mass spectrometry, neutron activation, and measurement of Rn emanation from components before their integration into LZ. These complementary techniques collectively produce a complete picture of the radiological contaminants. Table 12.1.1 provides a summary of the techniques. A detailed description and list of facilities available to LZ appears in Section 12.3.

**Table 12.1.1. Primary material radioassay techniques, indicating isotopic sensitivity and detection limits, as well as typical throughput or single-sample measurement duration.**

| Technique | Isotopic Sensitivity | Typical Sensitivity Limits | Sample Mass | Destructive/ Nondestructive | Sampling Duration | Notes |
|---|---|---|---|---|---|---|
| **HPGe** | $^{238}$U, $^{235}$U, $^{232}$Th chains, $^{40}$K, $^{60}$Co, $^{137}$Cs (any gamma emitter) | 50 ppt U, 100 ppt Th | kg | Nondestructive | Up to 2 weeks | Very versatile, not as sensitive as other techniques, large samples |
| **NAA** | $^{238}$U, $^{235}$U, and $^{232}$Th, K (top of chain) | $10^{-12}$ to $10^{-14}$ g/g | g | Destructive | Days to weeks | Sensitive to some contaminations |
| **ICP-MS** | $^{238}$U, $^{235}$U, and $^{232}$Th (top of chain) | $10^{-12}$ g/g | mg to g | Destructive | Days | Requires sample digestion; preparation critical |
| **GD-MS** | $^{238}$U, $^{235}$U, and $^{232}$Th (top of chain) | $10^{-10}$ g/g | mg to g | Destructive | Days | Minimal matrix effects, can analyze ceramics and other insulators |
| **ICP-OES** | $^{238}$U, $^{235}$U, and $^{232}$Th (top of chain) | $10^{-9}$ g/g | g | Destructive | Days | Requires sample digestion; preparation critical |
| **Rn emanation** | $^{222}$Rn, $^{220}$Rn | 0.1 mBq | kg | Nondestructive | Days to weeks | Large samples, limited by size of emanation chamber |

Gamma-ray spectroscopy with ultralow-background high-purity germanium (HPGe) detectors can typically measure U and Th decay chain species with sensitivities down to ~20 ppt (g/g). HPGe can assay $^{60}$Co, $^{40}$K, and other radioactive species emitting gamma rays. This technique is nondestructive and, in addition to sample screening, finished components can be assayed prior to installation. Under the assumption of secular equilibrium, with all isotopic decays remaining within the same volume, the activity and concentration for any particular isotope in the chain may be inferred from the measured U and Th content, assuming natural terrestrial abundance ratios. Unfortunately, it is relatively simple to break secular equilibrium through removal of radioactive daughter isotopes during chemical processing or through emanation and outgassing. HPGe is less sensitive to measures of the progenitor isotopes or the low-energy or low-probability gamma-ray emission from the early-chain decays of $^{238}$U and $^{232}$Th, but readily identifies the concentrations of isotopes from mid- to late-chain isotopes, particularly those with energies in excess of several hundred keV [10]. The early chain of $^{238}$U refers to measurements of isotopes down to $^{226}$Ra, and late chain beyond this point. The early chain of $^{232}$Th refers to measures of $^{228}$Ra and $^{228}$Ac, and late chain to $^{228}$Th and beyond. Since radioactivity from different parts of the $^{238}$U chain can vary considerably, and since the ratio of early- to late-chain $^{232}$Th content may change with time, it is important to measure full chain activity, and do so periodically.

Neutron activation analysis (NAA) increases the sensitivity to U and Th by neutron-activating samples of material and, with the shortened induced half-lives, achieves sensitivities up to ~1,000 times better than direct counting. Samples must be specially prepared and compatible with neutron irradiation in a reactor. As with inductively coupled mass spectrometry (ICP-MS), this is a destructive technique, requiring small sample masses; additionally, assumptions of secular equilibrium need to be made since this technique measures the top of the U and Th chains.

ICP-MS assays U and Th contamination in small samples. Sensitivities for U and Th can be a factor of ~100 more sensitive than HPGe gamma-ray counting. The samples are atomized and measured with ultrasensitive mass spectroscopic techniques. Care must be taken to avoid contamination of solvents and reactants. ICP-MS directly assays the $^{238}$U, $^{235}$U, and $^{232}$Th content for progenitor activity informing the contribution to neutron flux from ($\alpha$,n) in low-Z materials, but also the contribution from spontaneous fission, which in specific materials can dominate. However, it is limited in identifying particular daughter isotopes that are better probed by HPGe and typically contribute the bulk of the alpha and gamma-ray emission through the U and Th decay series.

Most of the facilities for the assaying measurements are operated directly by LZ groups or exist at LZ institutes, allowing us to maintain full control of sample preparation, measurements, analysis, and interpretation of data necessary to ensure sufficient sensitivity with reliable reproducibility and control of systematics. Commercial facilities that can provide ICP-MS, glow-discharge mass spectrometry (GD-MS), and inductively coupled plasma optical emission spectroscopy (ICP-OES) are available to the collaboration, and may be exploited for additional throughput. However commercial service providers are typically limited in sensitivity due to regular exposure of their instruments and sample preparation infrastructure to materials with high concentrations of contaminants.

Detector components will be matched with the appropriate assay technique depending on the material and requisite sensitivity, defined by Monte Carlo simulations described in Section 12.2. In some cases, the final detector material or components are assayed. When manufacturing or processes are complex, raw components as well as final components will be assayed to assist in maintaining purity through the manufacturing process and to assist in selecting those processes that do not introduce additional contamination.

Stringent constraints are also applied to "intrinsic" contamination of the Xe by radon and krypton. The LZ Xe purification program will remove $^{85}$Kr from the Xe down to the level of 0.015 ppt (g/g) using chromatographic techniques (see Chapter 9). All components that come into contact with any Xe in the experiment, whether within the primary instrument or in the gas storage, circulation, or recovery pipework, are screened for fixed contaminants, Rn emanation, and Kr outgassing to ensure that the intrinsic background remains within defined limits. These emanation and outgassing measurements are performed in dedicated chambers built and operated by LZ institutions. Similarly, techniques to measure bulk contamination of materials with radon progeny that emit alpha and beta particles that are not readily identified using HPGe, ICP-MS, or NAA are being developed by the collaboration. This is particularly important for large amounts of plastic such as the PTFE reflector panels within the TPC. Prolonged exposure to radon-contaminated air during manufacture will result in the presence of alpha-emitting $^{210}$Pb. The high cross section for ($\alpha$,n) reactions on the fluorine in the PTFE will result in neutron emission. A program to measure $^{210}$Pb forms part of the ongoing R&D efforts (described in Section 12.6).

The results of screening and analytics measurements are entered into an LZ materials database, building on the existing LUX screening campaign, described in Ref [11]. The database collates assay results from materials selected for use and identifies the components that contain them. These are referenced to results from Monte Carlo simulations that detail the background from the components in LZ. The contributions from several other sources in addition to bulk contamination of the materials and components will be included in the database. The first is the contribution from radon-daughter plate-out on components, especially during component transport, storage, and assembly. This is controlled through use of dedicated clean rooms available to the project at SURF (Chapter 13); active monitoring of the environment for radon; and following established cleanliness, handling, and storage procedures. Selected lightweight plastics and rubbers with low radon-diffusion coefficients are used to enclose materials in transit and during temporary storage. The second contribution comes from cosmogenic activation of components before they are moved underground, such as $^{46}$Sc production from Ti activation emitting 889 and 1120 keV gamma rays with a half-life of 84 days. While contribution to the background from potential activation products are assessed with a number of simulation toolkits, data from the LUX experiment in particular is able to provide considerable input to reduce the systematic uncertainty for such calculations and accurately assess the time-varying impact to the background radiation budget of the LZ experiment.

While LZ's sensitivity goals require unparalleled low background contamination control for dark-matter experiments and consequent severe constraints on material contamination, the screening campaign outlined here builds on the demonstrated substantial experience of the collaboration and established procedures or techniques employed by rare-event searches for background mitigation to meet these challenges with confidence. In the following subsections we detail the screening program, beginning with expected background rates derived from Monte Carlo simulations and requirements on material-screening sensitivity.

## 12.2 Monte Carlo Simulations and Background Rates

The LZSim Monte Carlo simulation package has been constructed to model the experiment and inform the design, determine optimal performance parameters, and define tolerable rates from background sources. Developed using the GEANT4 toolkit [12], the framework inherits from, and closely follows, the successful implementation of the LUX model [13], with evolving design of all parts of the experiment, including the inner detector, the cryostat, and the veto outer detector (OD), reflected in appropriate changes to the model geometry.

Simulations are performed to assess the background contribution from all expected background sources, including astrophysical neutrinos, intrinsic radioactivity in the Xe, and emission from every component in the experiment. This extends to subcomponents, with the model accurately representing the physical distribution of contaminants, particularly since (α,n) neutron yields can vary by many orders of magnitude, depending on the primary constituents of the materials containing the alpha-emitting uranium and thorium and decay products. Similarly, the physical distribution of gamma-ray, alpha, and beta particle emitters are modeled, as electrons created by these may produce detectable photons through Cherenkov and Bremsstrahlung processes, particularly in quartz or plastics close to or in contact with Xe.

Energy depositions from interactions in the Xe and outer detector are recorded in LZSim. Where necessary, optical tracking is performed following scintillation and ionization generation implemented using NEST [14]. LZSim models photon hit patterns and timing to mimic S1 and S2 signal generation in LZ, and allows for accurate studies of rare mechanisms that might produce backgrounds such as MSSI (multiple-scintillation single-ionization) events or background pile-up. Such detailed characterization and quantification of all background sources and their impacts are necessary to assign confidence to expected background event rates, their spectra, and their physical distribution in the detector. As a discovery instrument, the expected background in LZ must be well understood and quantified before any significance can be ascribed to observation of any potential signal and WIMP discovery.

With the exception of astrophysical neutrinos, the major sources of background in LZ will be radioactivity from construction materials surrounding the central fiducial volume, and radon and krypton distributed throughout the xenon. The goal for the maximum unvetoed differential single-scatter ER rate from each of these non-astrophysical sources after cuts in the WIMP search energy range has been set to 1 μdru. This is approximately 10% of that expected from irreducible pp solar neutrinos deduced from our Monte Carlo simulations of the detector (see Table 12.2.2). Similarly, an upper limit of 0.4 unvetoed WIMP-like single-scatter NRs due to neutron emission from material radioactivity is the goal for a 1,000-day exposure, reduced to 0.2 after a 50% NR efficiency is applied. This level of background allows LZ to achieve its sensitivity goal within 3 years of live-time (Chapter 4).

Acceptance of screened materials for use in LZ depends on the Monte Carlo simulations and the overall radioactive background budget. When a component is identified as required in LZ, it is incorporated into the LZSim model and a preliminary estimate of maximum tolerable activity from that component is calculated. This requirement necessarily depends on activity from other components and the overall budget, and initial inputs to LZSim for detector-related backgrounds are based on measured values, or from screening results from previous experiments. The maximum tolerable activity for the new component, including contingency for dominant materials such as the PMTs, is then translated to a required screening sensitivity for radioassaying a material sample. This in turn informs the screening technique and facility that will be employed for the assay. Screening results are then fed back into LZSim to produce an accurate assessment of electron and NR background and overall impact. The acceptance of the component depends on whether it meets requirements, or if it can be accommodated given related constraints and achieved radiopurity in other components and materials. In some cases, and as is justified by our assay experience, we may employ sampling of complete components. As the materials are assayed, this screening provides "as-built" input for the LZ background model.

Figure 12.2.1 shows the flow diagram, from identification of a component requirement, to determining assay requirements through Monte Carlo simulations, to performing screening and iterating on the simulation results, determining impact on background, and finally deciding whether the component is acceptable for use in the experiment. Where materials or components exceed tolerable limits, alternatives are sourced.

In building the background model for LZ and determining maximum tolerable rates, the initial inputs to LZSim for detector-related backgrounds come from our screening results, results from previous experiments, or literature values. Given the experience with the ZEPLIN detectors, LUX, XENON10, and other similar LXe TPCs such as EXO, as well as the materials and components defined in the design of LZ at present, contributors to background are identified and listed in Table 12.2.2, along with their calculated impact on background rates. These values are updated as the LZSim detector model is modified and as screening results become available.

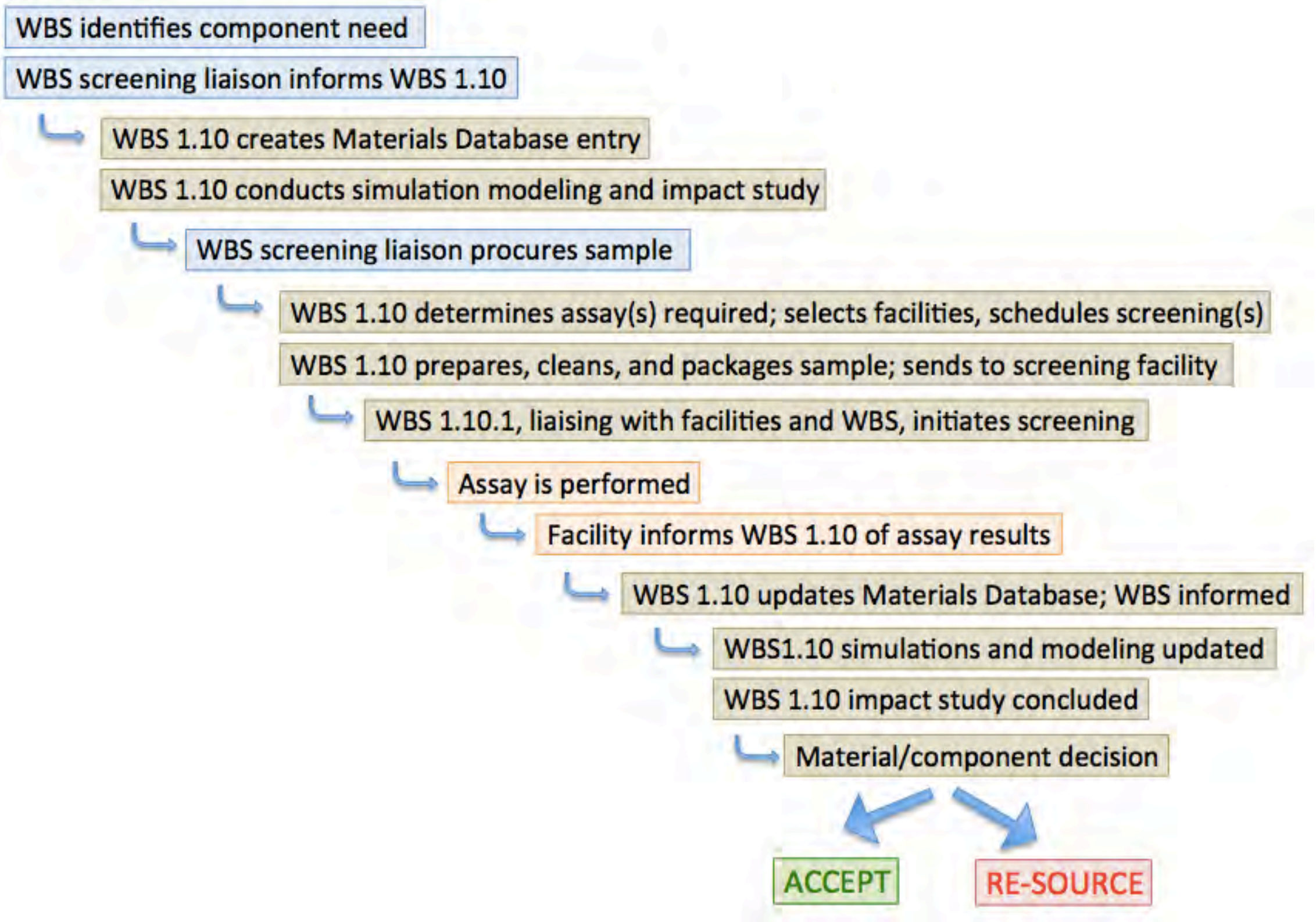

**Figure 12.2.1. Flow diagram depicting the process of component identification, screening, iterative Monte Carlo simulations and impact studies, determination of impact, and finally decision on accepting or re-sourcing material.**

The total activity as determined through Monte Carlo simulations with LZSim for the dominant material components and non-astrophysical sources satisfies the LZ sensitivity requirements, with 0.33 NR events in 1,000 live-days with a fiducial volume of 5.6 tonnes of Xe, and a WIMP search energy regime defined by 1.5 – 6.5 $keV_{ee}$ from U and Th contamination, and 67 ER events (approximately 2.4 $\mu$ dru) before discrimination or NR efficiency is applied. The following list presents the justification for the assumed values of $^{238}U$, $^{232}Th$, $^{60}Co$, and $^{40}K$ content used as initial input to LZSim for the major components. In Table 12.2.1, we present the full list of materials and references for the assayed values of radioactivity. LZ's assay campaign has made good progress in measuring the main contributors to LZ's ER and NR backgrounds. We anticipate completing assays for all these items with LZ-specified materials and assembling a detailed background model prior to the start of integration and assembly. The impact of these backgrounds is presented in Table 12.2.2.

- The activity for early production models of the 3-inch R11410 PMTs has been measured by the LZ collaboration [15]. Multiple batches of PMTs (25 total) were fully counted. In addition, all individual component materials used in PMT fabrication were counted. Measurements of components and finished assemblies will continue as PMT fabrication takes place, as described in subsequent subsections. The XENON1T collaboration [16] recently published extensive assays of the same 3-inch PMT, including assays of the components used in its fabrication as well as a large sample of the finished product. The LZ assays are consistent with the XENON1T assays. Recent LZ assays of the assembled PMT with the Chaloner and Maeve detectors produced limits on the early-U content, useful in understanding the spontaneous fission neutron backgrounds, with some modeling of the PMT required. Assays of the production materials for LZ's PMTs are currently under way, with all components to be assayed and accepted prior to the production of the PMTs.
- The two cryostat vessels will include flanges and CP-1 grade titanium with activities recently measured by LZ. The assayed U and Th levels are substantially lower than those recorded by the LUX collaboration. These assays are being confirmed by ICP-MS and with additional direct counting assays with other LZ detectors at Boulby.
- The contamination assumed in the PTFE arises from measurements by the EXO collaboration. The EXO-200 double-beta decay experiment used ultrapure Teflon, manufactured by DuPont; employed sintering techniques developed with Applied Plastics Technology Inc. (APT) in Bristol, Rhode Island; and assayed with NAA. Similar DuPont TE-6472 or equivalent raw material, sintered at APT, may be used in LZ. LZ is advancing its NAA program with the PTFE as well as assays of the $^{210}$Pb content of the bulk materials used to fabricate the PTFE.
- The field-shaping rings are constructed from 260 kg of Ti whose activity is the same as the cryostat vessel.
- The LXe skin PMTs are 1-inch-square Hamamatsu R8520 devices. These are the same type of PMTs used in the TPC of the XENON100 experiment. Several groups have measured their activity, and we use the results reported by the XENON collaboration as input for our Monte Carlo studies [4].
- The internal supports include the cathode, gate grid, anode, and PMT support structures. These components are constructed from titanium. The wires for the grids themselves are also made from stainless steel and their background contributions are included in our model.
- The HV umbilical contains stainless steel, copper cable, and insulating material; the utility and HV conduits will be titanium and stainless steel. As with the internal supports, stainless steel values are as measured by LZ, as is the case for titanium, and copper as measured by LUX and LZ.
- The liquid scintillator, formed of LAB and additives, has been assayed by LZ. These results improve on the limits set by Daya Bay for a similar composition.
- The activity for the acrylic vessels for the scintillator is informed by measurements from the SNO and Daya Bay collaborations. The other components have been assayed by LZ or taken from EXO or SuperNEMO.
- R5912 Hamamatsu PMTs are used to collect light from both the Gd-LS veto and the Cherenkov water shield. These PMTs are the same as those used in the LUX water shield at present, and their activity has been measured by LZ.
- The base of the OD support stand (stainless steel), and water PMT stands total some 620 kg of material and their impact has also been assessed. Stainless steel values that are input to the Monte Carlo simulations are as measured by LZ.

Table 12.2.1 presents the assay values and citations for the comprehensive list of materials in LZ design. LZ and LUX assays are maintained on the collaborations database. XENON100 assays are taken from [17]; EXO-200 from [18]; XENON1T [16]; MAJORANA, GERDA, and SuperNEMO by private communication or conference presentations; SNO from [19]. LZ has assayed two critical components: the Ti and prototype PMTs. The collaboration will assay all materials used in the detector as they are designed and procured. The contamination values are used to generate mass-weighted average activities for Monte Carlo simulations for ER backgrounds, and in critical components the individual components are used to generate NR rates from ($\alpha$,n) and spontaneous fission.

Table 12.2.2 presents the impact of the background sources assembled in Table 12.2.1, as these materials dominate given their mass and proximity to the LXe target. Most of the entries in Table 12.2.2 are formed from composite materials, where some 130+ components or subcomponents contribute to make up the different elements. The detector CAD model is used to establish the detailed geometry of the components. The NR and ER background calculations take into account radioactivity from all the components. LZ's screening and cleanliness-maintenance procedures will be applied to all materials to guarantee adequate background control and accurate modeling. Requirements on different materials and components vary, with their impact depending on material and position. Nonetheless, the simulations inform the necessity for screening U and Th in materials at the order of tens of ppt levels, tens to hundreds of ppb for $^{40}$K, and 5 fCi/kg for $^{60}$Co. Materials of sufficient radiopurity have been successfully deployed in rare-event search experiments and will be procured and incorporated in the LZ project following sample/component measurements with available technology and facilities that incorporate screening, cleanliness maintenance, and outputs from the R&D program. These measurements and procedures aim to reliably identify clean materials and maintain their purity throughout the chain, from fabrication to installation and operation. The assaying program to achieve these minimum limits is detailed in the following section.

**Table 12.2.1. Materials in the LZ design listed with radioactivities (mBq/kg) as determined by direct assay data from the LZ screening program, and from other published experimental results. The light-brown numbers represent 90% CL upper limits on isotopes, and the black numbers are based on error weighted averaged values for gamma line detection.**

| Material | U-early | U-late | Th-early | Th-late | $^{60}$Co | $^{40}$K | Reference |
|---|---|---|---|---|---|---|---|
| mBq/kg | | | | | | | |
| General Materials | | | | | | | |
| Ti | <1.60 | <0.09 | 0.28 | 0.23 | 0.00 | <0.54 | LZ |
| PTFE | <0.02 | <0.02 | <0.01 | <0.01 | 0.00 | <0.10 | EXO-200 |
| PEEK | 8.50 | 8.50 | <2.40 | <2.40 | 0.00 | 0.00 | LUX |
| LEDs | 2.00E3 | <100 | <200 | <100 | 0.00 | <1.00E3 | LZ |
| Cu | <0.04 | <0.04 | <0.01 | <0.01 | 0.00 | <1.55 | EXO-200 |
| Cable RG174 | <29.8 | <1.47 | 3.31 | <3.15 | <0.65 | 33.14 | XENON100 |
| Stainless steel | 1.20 | 0.27 | 0.33 | 0.49 | 1.60 | <0.40 | LZ |
| Epoxy | <0.55 | <0.55 | <0.10 | <0.10 | <0.00 | <0.63 | EXO-200 |
| Tyvek | <6.00 | <6.00 | <2.20 | <2.20 | 0.00 | 5.10E3 | SuperNEMO |
| HDPE | 5.96 | <0.37 | 0.63 | 0.62 | 0.00 | 3.40 | SNO |
| Rubber | <124. | <124. | <41.0 | <41.0 | 0.00 | 24.5 | EXO-200 |
| Viton | 2.63E3 | 2.49E3 | 220 | 220 | <10.0 | 2.15E3 | LZ |
| Aluminum | 1.13 | 1.13 | 0.37 | 0.37 | 0.00 | 25.5 | GERDA |
| Polyurethane | 57.0 | 57.0 | 9.00 | 9.00 | <6.00 | <80.0 | LZ |
| Ceramic – TPC resistors | 617 | 247 | 122 | 122 | 0.00 | <186 | LZ |
| UHMW-PE | <6.20 | 22.2 | <1.22 | <1.22 | 0.00 | <9.30 | LZ |
| Delrin | <4.00 | <0.70 | <0.18 | <0.18 | <0.30 | 18.0 | SuperNEMO |

| Material | U-early | U-late | Th-early | Th-late | $^{60}$Co | $^{40}$K | Reference |
|---|---|---|---|---|---|---|---|
| **Liquid Scintillator** | | | | | | | |
| LAB | <0.00 | <0.00 | <0.00 | <0.00 | 0.00 | <0.00 | LZ |
| $GdCl_3.6H_2O$ | <1.24 | <1.24 | <0.41 | <0.41 | 0.00 | <0.00 | LZ |
| PPO | <1.85 | <1.85 | <2.60 | <2.60 | 0.00 | <0.00 | LZ |
| TMHA | <0.25 | <0.25 | <0.29 | <0.29 | 0.00 | <0.00 | LZ |
| bis-MSB | <2.60 | <2.60 | <0.78 | <0.78 | 0.00 | <0.00 | LZ |
| **Outer Detector Components** | | | | | | | |
| PMT glass | 1.51E3 | 1.51E3 | 1.07E3 | 1.07E3 | 0.00 | 3.90E3 | LZ |
| Acrylic | <0.01 | <0.01 | <0.01 | <0.01 | 0.00 | <0.07 | SNO EXO |
| Polyurethane foam | 20.0 | 57.00 | <2.60 | <9.00 | <6.00 | <80.0 | LZ |
| **PMT Bases** | | | | | | | |
| Resistors | 1.17E3 | 369 | 227 | 227 | <9.27 | 4.36E3 | LZ |
| Capacitors | 4.05E3 | 1.13E4 | 3.89E3 | 3.89E3 | <10. | 300 | LZ |
| Cirlex board | 23.9 | 19.1 | 3.19 | 3.19 | <0.63 | <15.1 | LZ |
| Solder; Elsold | <58.2 | <11.8 | <10.7 | <10.7 | <2.24 | <31.8 | LZ |
| Connector | 4.60 | 4.60 | 5.80 | 5.80 | 0.00 | 0.00 | MAJORANA |
| CuBe spring | 795 | 795 | 41.0 | 41.0 | 0.00 | 0.00 | MAJORANA |
| **R11410 PMTs** | | | | | | | |
| Faceplate | <11.0 | 1.20 | <0.40 | <0.37 | <0.15 | <2.70 | LZ, XENON1T |
| Pure Al seal | <46.7 | <1.20 | <1.10 | 1.10 | <0.20 | <9.50 | LZ, XENON1T |
| Co-free body | <118 | 3.33 | <4.62 | <4.36 | 0.90 | <12.7 | LZ, XENON1T |
| Electrode disk | <110 | <3.05 | <5.24 | <4.02 | 8.78 | <7.80 | LZ, XENON1T |
| Dynodes | <73.6 | 3.75 | <1.31 | <1.31 | 0.83 | 8.33 | LZ, XENON1T |
| Shield | <92.5 | 3.25 | 2.00 | 2.00 | 0.50 | <8.00 | LZ, XENON1T |
| L-shaped insulation | <13.9 | 2.01 | <0.76 | <0.49 | <0.16 | 4.86 | LZ, XENON1T |
| Faceplate flange | <43.9 | 2.06 | <1.00 | <1.00 | 14.44 | 3.89 | LZ, XENON1T |
| Stem | 150 | 16.3 | 14.4 | 6.88 | <1.25 | 68.8 | LZ, XENON1T |
| Stem flange | <46.4 | <5.93 | <2.14 | <0.54 | 15.7 | 5.00 | LZ, XENON1T |
| Getter | <1.21E4 | 603 | <345 | <500 | <72.4 | 1.38E3 | LZ, XENON1T |
| Full PMT assay | <120 | <4.50 | <10.5 | 12.5 | 0.00 | 45.5 | LZ, XENON1T |
| **R8520 1" PMTs** | | | | | | | |
| Main metal package | 19.0 | 19.0 | <13.0 | <13.0 | 40.0 | 90.0 | XENON100 |
| Glass in stem | 970 | 970 | 340 | 340 | <10.0 | 2.3E3 | XENON100 |
| Spacer betw electrodes | 780 | 780 | 260 | 260 | <12.0 | 800 | XENON100 |
| Seal betw window metal | 17.0 | 17.0 | 370 | 370 | <27.0 | 5.00 | XENON100 |
| Electrodes | 19.0 | 19.0 | 18.0 | 18.0 | 12.00 | 0.15 | XENON100 |
| Window | <0.50 | <0.50 | <1.80 | <1.80 | <0.10 | 18.0 | XENON100 |
| **TPC Components** | | | | | | | |
| Reverse field-shaping resist. | 617 | 247 | 122 | 122 | 0.00 | <186 | LZ |
| Loop antenna | 0.20 | 0.20 | 0.12 | 0.12 | 0.00 | <1.86 | EXO-200 |
| Internal thermometer | <1.00E3 | <1.00E3 | <424 | <424 | <14.2 | <2.05E3 | LUX |
| Acoustic sensor PVDF | <0.09 | <0.09 | <0.02 | <0.02 | 0.00 | <0.37 | EXO-200 |

**Table 12.2.2. The estimated intrinsic contamination and physics-generated background signals in LZ for the 1,000-day-long, 5.6-tonne fiducial volume, and energy window of 1.5 – 6.5 $keV_{ee}$ exposure. Mass-weighted average activities obtained from Table 12.2.1 are shown for composite materials. All significant elements of the LZ detector are represented. The estimated ER and NR events are modeled using the LZSim package that includes the physics and detector characteristics. With respect to neutron emission backgrounds, the modeling for these results includes all neutrons emitted by (alpha,n) and spontaneous fission (dominant in the U-early chain neutron emission numbers). However, the modeling does not currently include the additional vetoing effect arising from the emission of multiple high-energy gammas, and simultaneous emission of multiple neutrons in spontaneous fission. This is expected to very effectively provide an additional veto, with an efficiency >90%, for the dominant spontaneous fission component of NR events. This process will be more fully integrated into the LZ Monte Carlo codes in the future.**

| Intrinsic Contamination Backgrounds | Mass | U-early | U-late | Th-early | Th-late | $^{60}$Co | $^{40}$K | n/yr | ER | NR |
|---|---|---|---|---|---|---|---|---|---|---|
| | (kg) | (mBq/kg) | | | | | | | (cts) | (cts) |
| Upper PMT structure | 40.2 | 1.45 | 0.10 | 0.25 | 0.21 | 0.00 | 0.50 | 3.96 | 0.01 | 0.002 |
| Lower PMT structure | 64.1 | 0.85 | 0.06 | 0.15 | 0.12 | 0.00 | 0.33 | 5.49 | 0.01 | 0.003 |
| R11410 3" PMTs | 93.7 | 67.1 | 2.68 | 2.01 | 2.01 | 3.86 | 62.1 | 372.5 | 1.24 | 0.203 |
| R11410 PMT bases | 2.7 | 525. | 74.6 | 29.1 | 29.1 | 3.60 | 109. | 76.7 | 0.17 | 0.033 |
| R8520 Skin 1" PMTs | 4.2 | 60.5 | 5.19 | 4.75 | 4.75 | 24.2 | 333. | 11.4 | 0.09 | 0.002 |
| R8520 Skin PMT bases | 0.9 | 513. | 58.3 | 24.2 | 24.2 | 3.91 | 108. | 23.3 | 0.06 | 0.003 |
| PMT cabling | 85.5 | 29.8 | 1.47 | 3.31 | 3.15 | 0.65 | 33.14 | 89.5 | 0.92 | 0.008 |
| TPC PTFE | 343. | 0.02 | 0.02 | 0.01 | 0.01 | 0.00 | 0.10 | 24.1 | 0.17 | 0.007 |
| Grid wires | 0.33 | 1.20 | 0.27 | 0.33 | 0.49 | 1.60 | 0.40 | 0.02 | 0.01 | 0.000 |
| Grid holders | 69.6 | 1.60 | 0.09 | 0.28 | 0.23 | 0.00 | 0.54 | 6.92 | 0.02 | 0.003 |
| Field-shaping rings | 262. | 5.89 | 1.81 | 1.13 | 1.08 | 0.00 | 1.83 | 32.2 | 1.22 | 0.004 |
| TPC sensors | 0.90 | 8.76 | 7.28 | 1.37 | 1.37 | 0.20 | 5.39 | 0.72 | 0.08 | 0.000 |
| TPC thermometers | 0.70 | 332. | 329. | 136. | 136. | 4.90 | 658. | 85.2 | 3.67 | 0.010 |
| Xe recirc. tubing | 5.2 | 0.02 | 0.02 | 0.01 | 0.007 | 0.00 | 0.10 | 0.37 | 0.00 | 0.000 |
| HV conduits – cables | 138. | 1.80 | 2.00 | 0.40 | 0.60 | 1.40 | 1.20 | 15.6 | 0.72 | 0.001 |
| HX and PMT conduits | 200. | 1.05 | 0.21 | 0.27 | 0.38 | 1.18 | 0.60 | 11.9 | 0.41 | 0.000 |
| Cryostat vessel | 2.14E3 | 1.60 | 0.09 | 0.28 | 0.23 | 0.00 | 0.54 | 213. | 0.86 | 0.019 |
| Cryostat seals | 4.5 | 102. | 102. | 34.0 | 34.0 | 7.27 | 22.6 | 40.3 | 0.79 | 0.001 |
| Cryostat insulation | 23.8 | 18.9 | 18.9 | 3.45 | 3.45 | 1.97 | 51.7 | 85.2 | 0.92 | 0.003 |
| Cryostat Teflon liner | 70.7 | 0.02 | 0.02 | 0.01 | 0.01 | 0.00 | 0.10 | 4.97 | 0.00 | 0.000 |
| Outer detector tanks | 4.00E3 | 0.15 | 0.37 | 0.02 | 0.06 | 0.04 | 4.32 | 101. | 0.14 | 0.0002 |
| Liquid scintillator | 2.08E4 | 0.01 | 0.01 | 0.01 | 0.01 | 0.00 | 0.00 | 22.9 | 0.00 | 0.00 |
| Outer detector PMTs | 122. | 1.50E3 | 1.50E3 | 1.07E3 | 1.07E3 | 0.00 | 3.90E3 | 2.09E4 | 0.08 | 0.022 |
| OD PMT supports | 620. | 1.20 | 0.27 | 0.33 | 0.49 | 1.60 | 0.40 | 37.0 | 0.25 | 0.00 |
| $^{222}$Rn (0.67 mBq) | | | | | | | | | 23.2 | - |
| $^{220}$Rn (0.07 mBq) | | | | | | | | | 4.68 | - |
| $^{nat}$Kr (0.015 ppt g/g) | | | | | | | | | 24.5 | - |
| $^{nat}$Ar (0.45 ppb g/g) | | | | | | | | | 2.47 | - |
| **Subtotal (Non-ν counts)** | | | | | | | | | **66.7** | **0.33** |
| **Physics Backgrounds** | | | | | | | | | | |
| $^{136}$Xe 2νββF | | | | | | | | | 53.8 | 0 |
| Astrophysical ν counts (pp+$^{7}$Be) | | | | | | | | | 271 | 0 |
| Astrophysical ν counts ($^{8}$B) | | | | | | | | | 0 | 0 |
| Astrophysical ν counts (Hep) | | | | | | | | | 0 | 0.002 |
| Astrophysical ν counts (diffuse supernova) | | | | | | | | | 0 | 0.113 |
| Astrophysical ν counts (atmospheric) | | | | | | | | | 0 | 0.385 |
| **Total** | | | | | | | | | **392** | **0.83** |
| **Total (with 99.5% ER discrimination, 50% NR efficiency)** | | | | | | | | | **1.96** | **0.41** |
| **Sum of ER and NR in LZ for 1,000 days, 5.6-tonne FV, with all analysis cuts** | | | | | | | | | **2.37** | |

## 12.3 Radioassay and Screening Campaign

Experience from LUX, EXO, and similar low-background experiments suggests that several hundred samples will require screening. ICP-MS, HPGe, and NAA will all be utilized, as no single technique has sensitivity to all radioactive isotopes within all materials, nor can any single technique at present provide sensitivity to the full $^{238}$U and $^{232}$Th decay chains. Use of all three techniques will provide the required accurate model of a material's full gamma-ray and neutron emission, and the subsequent impact on the radiation budget and sensitivity of the experiment. Any rare-event search would benefit from the availability of all three techniques, which vary in their sample throughput and screening durations, requirements for sample size, access to instruments, ability to do bulk screening of complete components, and preservation or destruction of samples in the assaying process.

ICP-MS, operated in surface laboratories without the need for shielding, requires only grams of sample material, is generally less expensive than NAA or HPGe, and is considerably faster, allowing rapid throughput of samples, of ~day, prior to or even during component or detector construction. For the EXO experiment, of the ~500 samples screened, almost 50% were screened with ICP-MS only. Additionally, where samples must be assayed before manufacturers are permitted to use materials from particular batches, such as for the many components within the R11410 PMTs, rapid turnaround and feedback on material suitability is crucial. In principle, GD-MS also exhibits the benefits of fast throughput and is available to the collaboration commercially. However it has poorer sensitivity (~0.1 ppb U/Th) and can only be used with conductive or semiconductive solids. The limitation of ICP-MS is that the sample must be soluble — typically in mixtures of HF and $HNO_3$ — and that several samples from materials must be screened to probe contamination distribution and homogeneity. NAA probes the bulk contamination simultaneously, can also be performed on small samples, and is not limited by the composition of the material since no sample digestions or ablations are required. Indeed, of all known techniques, NAA can provide the best sensitivity to U and Th concentration. However, concurrent activation of trace contaminants of little interest or from the primary constituents of the sample can produce high gamma-ray fluxes that present a background to the U and Th measurements, severely compromising sensitivity. It is also a relatively slow and costly process. Finally, HPGe also probes the bulk contamination within almost any material; is nondestructive; and, as well as having sensitivity to most of the U and Th chains, can measure the most problematic gamma-ray-emitting isotopes $^{40}$K and $^{60}$Co that are inaccessible or difficult to measure with ICP-MS. However, HPGe measurements do necessarily require large sample masses (~1kg) and long measuring times (weeks per sample).

### 12.3.1 Inductively Coupled Plasma Mass Spectrometry

Achieving instrument sensitivity in practice depends critically on sample preparation and its introduction to the ICP-MS system, as well as on subsequent analysis. Extreme care must be taken to ensure samples are not contaminated; systematics are reduced to an absolute minimum; and calibrations, quality control, and consistency checks are performed throughout the measurements.

All sample preparation, measurements, and analysis will be conducted with the instrument never exposed to high concentrations of contaminants that would be present in materials routinely screened by commercial systems, compromising the sensitivity of the devices. Sample material preparation for input to the detector will occur in clean rooms with dedicated hoods for sample dissolution and digestion to limit contamination, and will follow the procedures presented in [1] and [20]. Typically, samples are dissolved in $HNO_3$/HF acid and facilitated where necessary by a microwave digestion system, raising the temperature and pressure to increase the rate of dissolution of heavy metals. The digested materials may then be separated with chromatography resin, with the Th eluted from the resin with 0.5 M oxalic acid and the U with 0.02 M HCl acid. Both U and Th fractions are then evaporated to dryness and reconstituted with nitric acid before measuring them with the mass spectrometer. The newly released Agilent 7900 is capable of measuring samples containing up to 25% total dissolved solids, a factor of 100 greater than the traditional matrix limit for ICP-MS, allowing flexibility for novel radiochemical techniques and sample dissolutions. This is achieved with little compromise to the resolution of interferences that can contribute to the masses of interest due to matrix elements introduced into the sample following digestion; therefore, minimum detectable activity (MDA) can be maintained at the ppt level for U/Th in most materials to be used in LZ.

The EXO experiment has also demonstrated sensitivity to $^{40}K$, to which ICP-MS is typically insensitive due to interferences with the $^{40}Ar$ gas in the instrument. Sensitivity down to ppm levels was achieved using chemical resolution to remove interfering polyatomic or isobaric species from the ICP-MS ion beam with controlled ion-molecule chemistry.

An Agilent 7900 ICP-MS, with sensitivity to U and Th in principle below $10^{-12}$ g/g, has been procured at University College London (UCL) for dedicated use in the LZ screening program and will come online for LZ assays in 2015. The UCL system should exhibit sensitivity similar to that achieved by EXO for $^{40}K$, with species differentiation capability and both kinetic energy discrimination and chemical discrimination within its octopole reaction cell. Microwave ashing and digestion ovens for sample preparation — in order to realize ppt-level sensitivity following sample digests and to enable required rapid throughput and systematics control — will be installed in fall 2015. Ultrapure acids are produced in-house with acid distillation and reflux instrumentation that will be commissioned at the same time.

ICP-MS systems are available to the collaboration at the University of California at Davis and the University of Alabama, and are already being employed for LZ material screening. In addition, limited access may become available to an ICP-MS at the University of Edinburgh, which includes a laser ablation head that can be used to rapidly ablate material from a sample; the ablated material is introduced as plasma directly into the ICP-MS without the necessity of digesting the sample. This allows the probing of the physical distribution of U and Th within materials, especially on surfaces; however, it can suffer from difficulty with quantitative analysis. Nonetheless, the complete sample screening period is reduced to hours or less.

### 12.3.2 Neutron Activation Analysis

The LZ collaboration has access to two facilities for neutron activation of samples: one at UC Davis and the other at MIT. Samples are irradiated with neutrons from the reactor to activate some of the stable isotopes, which subsequently emit gamma rays of well-known energy that are detected through gamma-ray spectroscopy. Elemental concentrations are then inferred, using tabulated neutron-capture cross sections convoluted with the reactor neutron spectra.

UC Davis oversees a TRIGA Mark II reactor with typical output of 1.5 MW (2 MW maximum), or approximately 1 GW per 20 ms when pulsed [21]. The reactor provides easy access for LZ material screening. Once activated, samples are screened using one of four Canberra HPGe detectors of 8%, 25%, 50%, and 99% relative efficiency. A fifth 85% relative efficiency HPGe detector manufactured by ORTEC is presently being installed with a hermetic NaI Compton veto for background suppression. This reactor was used to screen Ti during the LUX screening campaign [22], providing excellent sensitivity to K, but limited in sensitivity to U and Th due to background from Sc activation within the Ti. No difficulty was presented for screening of plastic samples, with a sensitivity of <0.7 ppb achieved for U. Studies are ongoing to determine Th sensitivity. The UC Davis reactor and HPGe suite will again be used for the LZ screening campaign, with the addition of a $D_2O$ module developed for increasing the fast/thermal neutron flux. This module is presently being calibrated at a 0.25 MW TRIGA reactor at UC Irvine. The LBNL Berkeley Low Background Facility (BLBF) also has access, experience, and established procedures for NAA with the UC Davis.

The 6 $MW_{th}$ MIT Reactor II (MITR-II) is a double-tank reactor with an inner tank for light-water coolant moderator and an outer one serving as heavy-water reflector [23]. Two pneumatic sample insertion facilities are available. Steady-state thermal neutron fluxes of up to $5 \times 10^{13}$ n/(s·cm$^2$) can be achieved.

The sample insertion facilities can accommodate multiple samples that range in size but are typically a few mm in diameter and several cm in length. The two sample insertion facilities offer differing thermal over fast neutron flux ratios. Sample irradiations ranging from minutes to days can be performed, allowing accumulation of very large neutron fluences, which is key for reaching high analysis sensitivity. LZ samples will be prepared, surface cleaned (when needed), and hermetically sealed at the University of Alabama prior to activation. The counting of the activated samples will also be performed at Alabama, utilizing three shielded HPGe detectors and a double differential time-energy analysis. The typical shipping delay of 24 hours is acceptable compared with the half-lives of the typical activation products of interest ($^{42}$K, $^{233}$Pa, $^{239}$Np). The Alabama group routinely achieved $10^{-12}$ g/g sensitivity for Th and U using these techniques, as reported in [1], appropriate for the LZ material screening campaign. Indeed, NAA results obtained by the same group for the KamLAND experiment reached sensitivity to U and Th at the $10^{-14}$ – $10^{-15}$ g/g level [8] for liquid scintillator.

### 12.3.3 Gamma-ray Spectroscopy with High Purity Germanium Detectors

Several HPGe detectors located in facilities both above- and underground are available to the LZ collaboration, with differences in detector types and shielding configuration providing useful dynamic range both in terms of sensitivity to particular isotopes and physical sample geometries. The majority of these facilities, previously used for LUX or ZEPLIN, are managed and operated by LZ collaborating institutes and are already in use for the LZ material screening campaign. The detectors are typically several hundreds of grams to several kilograms in mass, with a mixture of N-type, P-type, and broad energy Ge crystals, providing relative efficiencies at the tens of percent through to in excess of 100%. While P-type crystals can be grown to larger sizes and hence require less counting time due to their high efficiency, the low energy performance of the N-type and broad energy crystals is superior due to less intervening material between source and active Ge. Specifically, an N-type Ge permits assaying down to very low gamma-ray energies, ~keV, and in principle detection of many atomic X-rays that otherwise would be attenuated by the lithium diffusion layer in a P-type detector.

Cleaned samples are placed close to the Ge crystal and sealed for several days to weeks in order to accrue sufficient statistics, depending on the MDA. The detectors are generally shielded with ancient Pb and Cu, flushed with dry nitrogen to displace the Rn-carrying air, and often are surrounded by veto detectors to suppress background from Compton scattering that dominates the MDA for low-energy gamma rays. To reduce backgrounds further, the detectors are operated in underground sites, reducing the muon flux by several orders of magnitude. Background-subtracted gamma-ray counting is performed around specific energy ranges to identify radioactive isotopes. Taking into account the Monte Carlo-generated detector efficiency at that energy for the specific sample geometry allows calculation of isotopic concentrations.

The LZ experiment requires multiple HPGe detectors due to the long counting times required to achieve sensitivities at the tens to hundreds of ppt level, coupled with the large number of samples that require screening. The three most extensively used HPGe detector facilities — for LZ at present, or previously by the LUX and ZEPLIN experiments — are Maeve, Morgan, and CUBED (SURF); SOLO (Soudan Mine); and Boulby (UK). These are being supplemented with further capability at SURF and Boulby. Late in 2015, the Black Hills State University Underground Campus (BHUC) will be completed and Morgan, Maeve, CUBED, and SOLO will relocate to this dedicated low-background facility. We anticipate that additional counters will be added to the array from the South Dakota School of Mining and Technology (SDSMT), UC Berkeley, and the University of South Dakota (USD). A new well-type detector and pre-screening instrument will be installed summer 2015 at Boulby to complement existing counters. The HPGe detectors available to LZ are shown in Table 12.3.3.1.

LBNL operates a two-site facility with both surface and underground detectors. The surface BLBF is within a 4π shielded room with 1.5-m-thick low-activity serpentine rock concrete walls surrounding a 115% relative efficiency N-type low-background HPGe (Merlin). The HPGe detector head is mounted on a J-hook to reduce line-of-sight for background from electronics and the cryostat, and is housed in a Pb and OFHC Cu castle. It has an MDA of approximately 0.5 ppb (6 mBq/kg) to $^{238}$U for O(kg) samples from 1 day of counting, and 2 ppb (8 mBq/kg) for $^{232}$Th. Sensitivity to $^{40}$K and $^{60}$Co is at the level of 1 ppm and 0.04 pCi/kg, respectively. The underground detector (Maeve), formerly situated at Oroville, is now operational at SURF. This 85% P-type HPGe detector is housed at the 4850L in the Davis Campus in a low-activity Pb- and Cu-shielded and Rn-flushed chamber. For ~kg samples, the MDA after approximately a week of counting is over an order of magnitude lower than from the surface detector. $^{238}$U and $^{232}$Th sensitivity to 10 ppt (~0.1 mBq/kg) and 25 ppt (~0.1 mBq/kg), respectively, is achieved as is 100 ppb for $^{40}$K and 4 fCi/kg for $^{60}$Co. These MDAs are sufficient for the LZ experiment. Figure 12.3.3.1 shows the assays of LZ's Ti and stainless steel samples measured by the Maeve detector. The most recent assay is the lowest-activity Ti found to date. LZ's Ti assays and the impact on the ER and NR backgrounds are presented in Figure 12.3.3.1.

**Table 12.3.3.1. Gamma-counting facilities available for LZ material radioassays. Sensitivities shown are approximate detectable activities after 2 weeks of counting and samples of order-kg mass. Typical cavity size within the shielding of these detectors within which samples may be placed is 0.03 m$^3$. With the exception of SOLO, all are managed by LZ institutes.**

| Detector | Site | Site Depth (mwe) | Crystal Type | Crystal Mass & Relative Efficiency (kg) | Sensitivity, U (mBq/kg) | Sensitivity, Th (mBq/kg) | Detector Status |
|---|---|---|---|---|---|---|---|
| **Chaloner** | Boulby | 2805 | BEGe | 0.8 (48%) | 0.6 | 0.2 | Online |
| **CUBED** | SURF | 4300 | N-type | 1.2 (60%) | 0.7 | 0.7 | Fall 2015 |
| **GeII** | Alabama | 0 | P-type | 1.4 (60%) | 4.0 | 1.2 | Online |
| **GeIII** | Alabama | 0 | P-type | 2.2 (100%) | 4.0 | 1.2 | Online |
| **Lunehead** | Boulby | 2805 | P-type | 2.0 (92%) | 0.7 | 0.2 | Online |
| **Lumpsey** | Boulby | 2805 | Well-type | 1.5 (80%) | 0.4 | 0.3 | Fall 2015 |
| **Wilton** | Boulby | 2805 | BEGe | 0.4 (18%) | 7.0 | 4.0 | Fall 2015 |
| **Merlin** | LBNL | 180 | N-type | 2.3 (115%) | 6.0 | 8.0 | Online |
| **Maeve** | SURF | 4300 | P-type | 2.1 (85%) | 0.1 | 0.1 | Online |
| **Morgan** | SURF | 4300 | P-type | 2.1 (85%) | 0.2 | 0.2 | Online |
| **SOLO** | Soudan | 2200 | P-type | 0.6 (30%) | 0.1 | 0.1 | Online |

The SOLO Low Background Counting Facility, Soudan Mine, at a depth of 2200 mwe, houses a nitrogen-flushed Pb shield that has a minimum thickness of 30 cm (50 Bq/kg $^{210}$Pb activity) with a 5-cm inner liner of 150-year-old low-activity Pb (50 mBq/kg $^{210}$Pb activity). A counting chamber of 8000 cm$^3$ contains the 0.6 kg "Diode M" HPGe detector. This detector achieves sensitivities to the 10 ppt level for $^{238}$U and $^{232}$Th and 25 ppb for $^{40}$K for multi-kg samples. The SOLO facility is scheduled for exclusive use and 100% live-time screening of LZ PMT components. The R11410 3-inch PMT was developed by the LZ PMT R&D program at Brown in conjunction with Hamamatsu [15,24] and has evolved to become the lowest-background photosensor per unit area photocathode coverage suitable for large-scale liquid Xe operation. This followed extensive and iterative screening by LZ collaborators and others to identify component sources of background and to screen suitable alternatives. The PMT arrays remain a significant contribution to the backgrounds in LZ, excluding the irreducible neutrino background. Hamamatsu has agreed to continuous monitoring with SOLO of the batch materials to be used in the construction of LZ PMTs. The PMTs will also be screened as complete units following their delivery at SOLO, Morgan, CUBED, and detectors at Boulby.

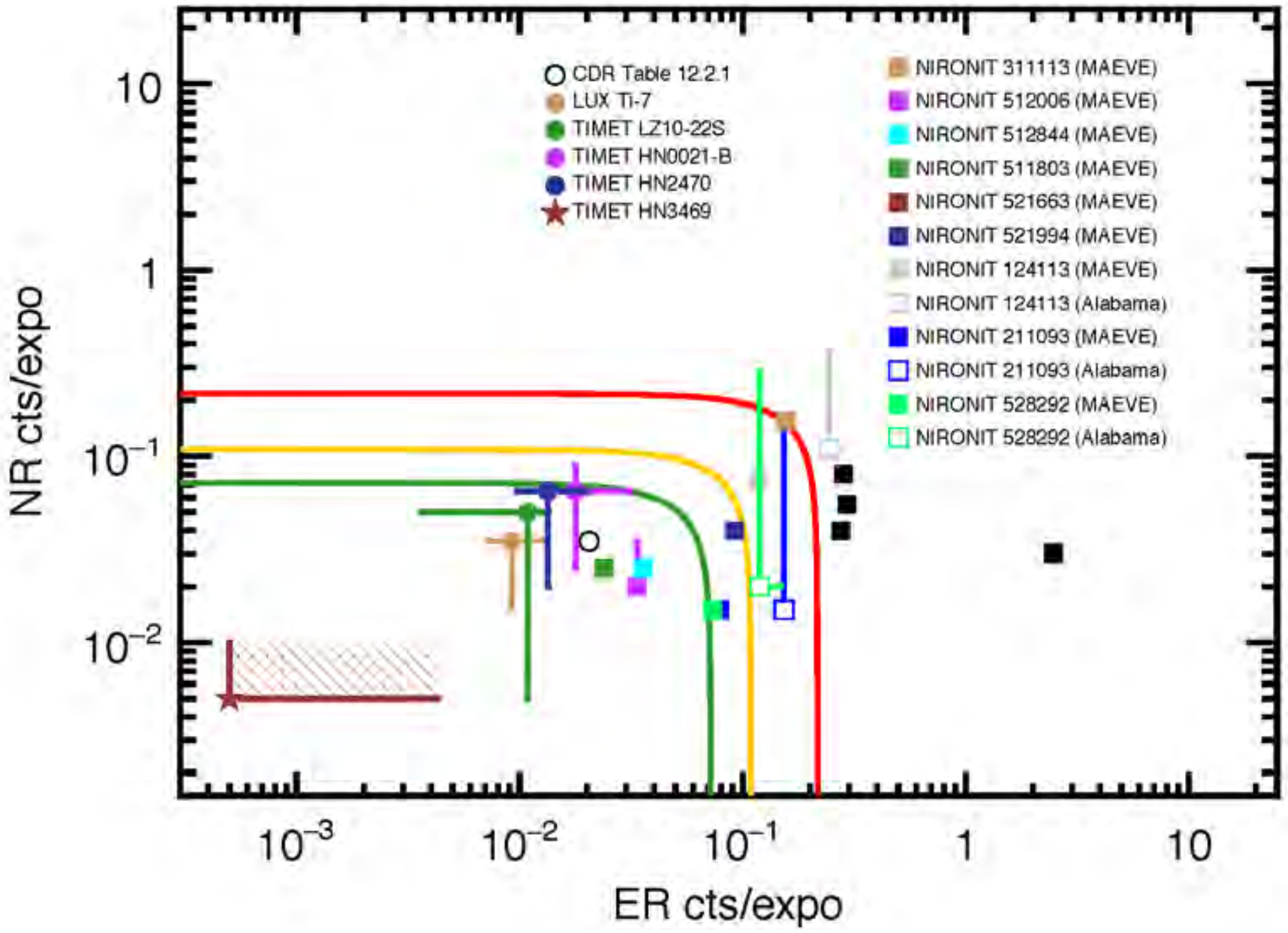

**Figure 12.3.3.1. The plot shows the background counts resulting from LZ's Ti and stainless steel samples from the LXe cryostat in the full exposure of a 5.6-tonne fiducial volume, 1,000 day exposure after all the veto systems are applied, for ER events within [1.5 – 6.5] keVee, with 99.5% rejection, and within [6 –30] keVnr, and 50% acceptance, for NR events. The red curve corresponds to the sum of 10% of the pp solar neutrinos background for ERs and of 0.1 NR events. The yellow curve is for the sum of 5% of the pp neutrino background and 0.05 NR events. The green curve corresponds to the sum of 3.3% of the pp neutron background and 0.03 NRs, the requirement for the LZ cryostat. The markers indicate actual (positive) radioactivity determinations, the upper bars correspond to screening upper limits, and the lower bars show the effect of excluding spontaneous fission contributions, which we expect to self-veto efficiently due to the high gamma multiplicity. The Ti identified by the assay program is well below LZ's requirements.**

The Boulby underground facility at 2805 mwe previously housed a 20-cm-thick Pb and Cu castle for an ORTEC GEM-XX240-S P-type HPGe of 92% relative efficiency. This detector has been used extensively for the screening campaign of the ZEPLIN-III experiment, particularly for its second science run and screening of low-background PMTs and the veto detector components [25]. In response to the needs of improved sensitivity at Boulby, the detector has undergone refurbishment at ORTEC, where the entire detector (except the Ge crystal) has been overhauled and retrofitted with ultralow-background components to become a GEMXX-95-LB-C-HJ. This detector, with sensitivity to about 50 ppt of $^{238}$U and $^{232}$Th, came online in August 2014, and features a J-type neck with remote pre-amplifier through a new custom-designed Rn-proof Pb and Cu shield, and a 95-mm-diameter carbon-fiber window.

Taking advantage of technological advancements in development of low-background Ge detectors in recent years, a second and more sensitive detector has been procured for Boulby to significantly enhance the UK's screening underground capability for LZ. A Canberra broad-energy Ge (BEGe) BE5030 (0.8 kg Ge, 48% relative efficiency), also housed within a custom-built Rn-proof ($N_2$-flushed) shield with Pb and Cu from the existing underground stock at Boulby, came online alongside the ORTEC detector. Both detectors are housed in a clean room in a dedicated low-background counting facility area of the laboratory, which is undergoing a complete upgrade through 2015; this has no impact on the live-time of the Ge detectors. The crystal in the BEGe detector is configured in a unique planar geometry, yielding greater peak-to-Compton ratios at the gamma-ray energies of interest, and improved energy resolution (by

~30% at 122 keV over typical P-type detectors). BEGe detectors also have considerably lower energy thresholds due to a factor-70 reduction in dead layers around the Ge crystal, providing useful efficiency to 10 keV (as opposed to ~80 keV for the existing Boulby detector and similar P-type HPGe) and consequently can directly measure isotopes such as $^{210}$Pb — a problematic source of background that is particularly difficult to quantify with other techniques. The BE5030 achieves <50 ppt sensitivity to $^{238}$U and $^{232}$Th for typical samples, thanks to its ultralow-background material construction and carbon-fiber entrance windows (since the MDA for any isotope is proportional to the resolution and the efficiency of the detector at the energy corresponding to gamma-ray emission from the isotope or its decay, and inversely proportional to the background rate in the detector at that energy).

Further HPGe facilitation is in progress by the USD group, which is deploying ultralow-background detectors underground in the BHUC at SURF. CUBED presently hosts a single 1.2-kg ORTEC N-type coaxial HPGe detector with a 254 cm$^3$ active volume and a relative efficiency of 60%. The sample chamber, with a dimension of 8000 cm$^3$, is surrounded by a 10-cm-thick 99.9% OFHC copper shield, enclosed in a stainless steel box that is itself sealed by 10 cm of lead. The first detector will begin screening samples in fall 2015. The sensitivity of this detector is anticipated to be better than 100 ppt for $^{238}$U and $^{232}$Th.

Finally, in addition to the surface detectors at UC Davis for NAA, surface screening capability for LZ is also available at the University of Alabama with the GeII and GeIII instruments. Two lead-copper-shielded low-background HPGe detectors equipped with active cosmic-ray veto systems are operated on the surface, with shield cavities for large samples. These devices can reach 0.3 ppb sensitivity for $^{238}$U and $^{232}$Th with two weeks of counting and provide useful prescreening for LZ components.

## 12.4 Radon

Particular attention must be paid to radon, as it is a noble gas consisting solely of radioactive isotopes; is produced in the decay chains of uranium and thorium; and due to its chemical inertness and subsequent long diffusion lengths through solids, has the ability to enter Xe volumes. Outgassing of radon from a material in which it has been produced is commonly termed “radon emanation.” Especially for materials in contact with or in close proximity to Xe, radon emanation must be taken into account in setting the levels of U/Th that can be tolerated (with stringent limits particularly on U) due to the presence of $^{222}$Rn in the $^{238}$U decay chain, as well as $^{220}$Rn from $^{232}$Th decay. Unlike radioactivity from fixed contaminants, LXe cannot provide self-shielding against the dispersed Rn. Given its radiogenic origin, radon emanation from a material may be estimated by simulation once its U/Th decay chain content (in particular its $^{226}$Ra content) has been assayed during the screening process using HPGe detectors. However, assay and simulation must be supplemented by direct screening for radon emanation for critical materials due to limited sensitivity in HPGe, systematic error from assumptions on equilibrium chain states, and uncertainties in describing radon transport in materials.

A related phenomenon is radon plate-out, discussed in Section 3.9.2, in which charged radon progeny are deposited onto the surfaces of materials exposed to air that typically contains concentrations of $^{222}$Rn ($T_{1/2}$=3.82 days) ranging from tens to hundreds of Bq/m$^3$ [26,27]. The decay daughters can be embedded into material as they recoil due to subsequent decays. Beyond radon concentration and surface area, the susceptibility to plate-out depends on the material and factors such as air-flow rates, which are difficult to predict and therefore must be measured wherever possible. In the cases where measurements are not available, conservative estimates must be used to predict contamination risk from plate-out. Plate-out may be further enhanced in the presence of an electric field, since positively charged radon daughters are deposited on negatively charged surfaces such as electrodes within the TPC. The background due to radon daughters on the surfaces arises predominantly from neutron production. In particular the long-lived $^{210}$Pb ($T_{1/2}$=22.3 years) in the decay series decays to $^{210}$Po (via $^{210}$Bi), which emits an $\alpha$ that feeds ($\alpha$,n) reactions. In addition to neutron background, progeny from Rn plated onto the inner surfaces of the TPC, particularly $^{206}$Pb, can lead to spatial leakage of mis-reconstructed events at the TPC walls, rapidly reducing the fiducial mass. Furthermore, incomplete charge collection of these recoils at the edges of the TPC can cause them to overlap with the low-energy NR band. We are conducting detailed simulations, utilizing the position reconstruction algorithm successfully deployed in both LUX and ZEPLIN-III and adapted for the TPC, extraction electrodes, and top PMT array configurations of LZ, to study position reconstruction of such edge events. We further discuss control of radon emanation and radon plate-out below.

### 12.4.1 Radon Emanation

To achieve the desired LZ sensitivity, the experiment can tolerate a maximum $^{222}$Rn activity within the LXe volume of only 0.67 mBq, which corresponds to a steady-state population of approximately 300 atoms. This rate is dominated by the "naked" beta decay of $^{214}$Pb to $^{214}$Bi, whereas the $^{214}$Bi beta decay itself is readily identified by the subsequent $^{214}$Po alpha decay that would be observed within an LZ event timeline ($T_{1/2}$=160 μs). Similar coincidence rejection also occurs where beta decay is accompanied by a high-energy gamma ray, which may still be tagged by the LXe skin or external Gd-LS vetoes even if it leaves the active Xe volume. Radon-220 generates $^{212}$Pb, which decays with a short-timescale Bi-Po (beta-alpha delayed coincidence) scheme similar to $^{214}$Pb. Radon daughters are readily identified through their alpha decay signatures, as demonstrated in LUX, and can be used to characterize the $^{222}$Rn and $^{220}$Rn decay chain rates and distributions in the active region, providing a useful complement to estimating radon concentration from the beta decay contribution to the ER background. Indeed, these isotopes were the only sources of alpha decay in LUX [11].

There are multiple potential sources of radon emanation (e.g., PTFE reflectors, PTFE skin, PMT glass, PMT and HV cables, grid resistors, components in the circulation system), and radon emanation screening must be sensitive to sources that individually sustain smaller populations. We use 0.67 mBq in the 7-tonne target within the TPC as a hard upper limit for $^{222}$Rn. For $^{220}$Rn, we set a target of 0.07 mBq, based on the ratio of species observed in LUX.

Critical materials will be screened for radon emanation, defined as all that are within the inner cryostat or come into direct contact with Xe during experimental operation. Several methods and technologies exist for Rn-emanation screening as adopted by rare-event search experiments. We expect at least three dedicated LZ radon-emanation screening stations to be required, building on prototypes that are under construction and evaluation. In the first, developed at Case Western Reserve University and being commissioned at the University of Maryland, radon atoms and daughters are collected electrostatically onto silicon PIN diode detectors to detect alpha decays. In the second prototype, at Alabama, the radon atoms are collected by passing the radon-bearing gas through liquid scintillator, with the decay detected through coincidence counting between two PMTs viewing the scintillator. LZ as some access to the well-understood systems employed by the SuperNEMO group at UCL that already achieve 0.09 mBq sensitivity with a PIN detector and Rn concentration line [28]. An emanation chamber for LZ is being assessed with this existing infrastructure to establish MDAs for a similar system that would be constructed and dedicated to LZ screening. A new faculty member joining SDSMT has extensive experience with radon reduction and monitoring efforts, including the use of Rn-emanation chambers. SDSMT has developed a system very similar to the Case Western Reserve University/University of Maryland system, which is anticipated to bring a fourth system to assist with screening.

The prototypes will be evaluated on the basis of background rates and efficiency measured using calibrated sources of radon. Screening a single sample for LZ is expected to take about two weeks, including emanation and collection/detection times and repeated measurements to check reproducibility, as well as minimum sensitivity requirements and typical radon emanation MDAs. The radon-emanation screening campaign, coordinated through dedicated management in the screening working group, extends beyond initial material selection. As pieces or sections are completed during installation of gas pipework for the LZ experiment, they will be isolated and assessed for Rn emanation and outgassing for early identification of problematic seals or components that require replacement or correction. Based on the sensitivity and operational experience of the screening systems developed at the individual institutions described above, we will construct a screening program underground by relocating one of the university systems in order to screen large-scale assembled detector elements and plumbing lines.

### 12.4.2 Radon Plate-out

The first and most effective defense against radon plate-out is limiting exposure of detector parts to air. Using conservative estimates of deposition rates based on measurements [26], exposure time limits have been calculated for LZ components, with particular attention to materials such as Teflon and titanium, as they are large-surface-area components in direct contact with the LXe target. For Teflon, which has a large ($\alpha$,n) yield of $6.8 \times 10^{-6}$ neutrons/$\alpha$ due to its high fluorine content, the exposure limits are 490 days and 28 days for surface air (20 Bq/$m^3$) and mine air (350 Bq/$m^3$), respectively. For Ti, whose ($\alpha$,n) yield is much lower, the tolerable exposure times are considerably longer, at the level of years. Tolerances on exposure to surface air are expected to be satisfied without difficulty; however, the possibility of radon plate-out begins at the time of manufacture, which can be many months before final assembly and integration. For this reason, we are working with vendors to minimize exposure to air and, where appropriate, we will install radon monitors to quantify exposure for input to the background model. Furthermore, while in storage, detector components will be covered in a material such as plastic or rubber, which strongly inhibits radon diffusion, and if necessary, the storage space will be flushed with gas. R&D is being carried out to identify and test promising candidate materials for covering against radon. One such material recently identified and successfully tested by the SuperNEMO collaboration [28] is styrene-butadiene rubber (SBR), which is also inexpensive. Since the inner cryostat will be assembled on the surface before moving underground, plate-out risk will be limited if not mitigated. Cleaning techniques to remove surface contamination from most materials employed in LZ are well established. Further details on radon plate-out mitigation during storage, handling, and transport are described in Section 12.8.

For materials such as Teflon, which are produced in granular form before being sintered in molds, plate-out comprises an additional dimension of risk because surface contamination of the granular form becomes contamination in bulk when the granules are poured into molds. Detecting $^{210}$Pb in bulk plastics is an active area of investigation within the collaboration and is further described in Section 12.6 (R&D).

An XIA-Ultralow 1800 surface alpha detector system has been procured for operation initially at Brown, and then underground at Sanford Lab, to assess the levels of surface contamination on the TPC inner components, most notably the PTFE liners. The instrument has been shown to achieve sensitivities of 0.003 alphas/$cm^2$/day in samples of areas of 700 $cm^2$. This survey work builds on work from the Southern Methodist University group, the XMASS collaboration, and others who have successfully demonstrated low-background alpha screening using this instrument [29].

## 12.5 Internal Backgrounds

Similar dispersed backgrounds throughout the volume can be generated by krypton contamination, and as a result of cosmogenic activation of Xe.

Krypton-85 is a beta-emitter with a half-life of 10.8 years and a dominant (99.6% branching ratio) bare beta decay mode of endpoint energy 687 keV. Its presence in the atmosphere comes from cosmogenic causes, production in nuclear power plants, and past production and testing of nuclear weapons. Coupled with a long half-life and diffusion properties as a noble gas, it can become a significant contaminant in the course of production and storage of Xe. The research-grade Xe procured for LUX contained an average 130 ppb (g/g) $^{nat}$Kr/Xe upon procurement, with an estimated $^{85}$Kr concentration of $2 \times 10^{-11}$ g/g [30]. This was reduced to 3.5 ± 1.0 ppt g/g in LUX, resulting in a measured event rate of 0.17 ± 0.1 mdru [11]. To control its contribution to the background budget in LZ, its concentration in the LXe must be less than 0.015 ppt (g/g). The primary approach to achieve this level is Xe purification and the corollary requirement of 0.01 ppt sampling sensitivity. Levels of 0.2 ppt (g/g) have been demonstrated already during the LUX production run by double-processing a 50 kg LXe batch. The Xe purification program is described in Chapter 9. Measurements of Kr outgassing from materials are also under way at the University of Maryland.

Trace quantities of argon are also a concern due to beta-emitting $^{39}$Ar, with a 269-year half-life and 565-keV endpoint energy. This background is constrained to be less than 10% of $^{85}$Kr, resulting in a specification of $4.5 \times 10^{-10}$ (g/g) or 2.6 μBq. The Kr removal system, which also removes Ar, should easily achieve this goal.

The other source of radioisotopes intrinsic to the LXe is cosmogenic activation. While cosmogenic production of radioisotopes underground in LXe may be neglected due to the low cosmic-ray flux, Xe production and storage takes place at surface facilities, where there is no shielding from cosmic rays. Simulation packages have been developed and validated against data to estimate cosmogenic production [31,32]. For natural Xe, radionuclides produced include tritium, tellurium, and cadmium [33]. Of the isotopes produced, most have short half-lives and/or low production rates. The exception is tritium, with a 12.3-year half-life and production rate of ~15/day/kg at the Earth's surface, which will require reduction to negligible levels through Xe purification during experiment commissioning.

Radioisotopes of Xe that can be produced through cosmogenic or neutron activation — such as $^{127}$Xe ($T_{1/2}$=36.4 days), $^{129m}$Xe ($T_{1/2}$=8.9 days), $^{131m}$Xe ($T_{1/2}$=11.9 days), and $^{133}$Xe ($T_{1/2}$=5.3 days) — however, cannot be removed with purification nor self-shielded. The $^{127}$Xe, $^{131m}$Xe, and $^{129m}$Xe radioisotopes can provide useful energy calibration points, whereas the production rate and half-life of $^{133}$Xe renders it unmeasurable following transport of the Xe underground in LUX [11]. The $^{131m}$Xe and $^{129m}$Xe do not generate any significant WIMP search background. However, $^{127}$Xe produces energy depositions within the WIMP search region of interest and also poses a background for axion searches. $^{127}$Xe undergoes electron capture that results in an orbital vacancy that is filled by electron transitions from higher orbitals, resulting in an X-ray or Auger electron cascade. An 85% probability for the capture electron coming from the K shell results in a cascade with a total energy of 33 keV. A further 12% of captures from the L shell generate cascades of 5.2 keV total energy deposition, and the remaining 3% of decays come from higher shells (M and N) to deposit up to 1.2 keV. For a WIMP search energy window of 1.5 – 6.5 keV, significant numbers of L, M, and N shell decays will generate background. However, the daughter $^{127}$I nucleus is left in either a 619, 375, or 203 keV excited state (with no direct population of the ground state as part of the electron capture). The subsequent decay of the $^{127}$I to the ground state emits internal conversion electrons or gamma rays that permit almost all of the $^{127}$Xe background to be rejected by coincidence tagging — only those X-ray/Auger events for which the associated gamma rays are not detected contribute to the low-energy ER background in the Xe active region. This effect is expected predominantly at the edge of the Xe target. For example, with a mean free path of 2.6 cm in LXe, the 375-keV gamma ray can potentially escape the active region, reducing the efficiency of coincidence rejection further for events at the edges of the LXe, as seen in LUX [11]. In LZ, the skin and external veto systems significantly aid rejection and characterization of this background. The Xe may be stored underground for >>1 month prior to commencement of the WIMP exposure to further mitigate $^{127}$Xe background even in the early stages of a WIMP search exposure. Since $^{127}$Xe is also produced efficiently through neutron capture, neutron activation of the Xe has been assessed and may be controlled through shielding during storage, as described in Section 12.8.

## 12.6 R&D

The collaboration has broad capabilities to control backgrounds through materials screening and following procedures to maintain cleanliness, from manufacture and fabrication through integration and operation. Nonetheless, as mentioned in previous sections, some areas require R&D to improve capability. Those efforts, together with their major milestones, are described below.

**Detection and mitigation of $^{210}$Pb in bulk plastics.** Initial estimates indicate that $^{210}$Pb concentrations in bulk Teflon exceeding a few hundred mBq/kg pose a dangerous source of background for the experiment through (α,n) production on fluorine by $^{210}$Po, a $^{210}$Pb daughter nucleus. The mechanism by which $^{210}$Pb could appear in bulk Teflon was described in Section 12.4. The techniques of direct gamma counting, neutron activation analysis, and mass spectrometry cannot readily be applied to detect $^{210}$Pb or its daughters at the mBq/kg level. R&D is under way at the University of Alabama to explore alternate techniques for detecting $^{210}$Pb in bulk at low levels and to avoid contamination during the manufacturing process.

The R&D program has two objectives. The first is to determine the levels of $^{210}$Pb in bulk Teflon that can be detected by direct gamma and beta spectroscopy. Plastic samples having the same geometry as that used for direct gamma and beta gamma counting will be spiked in bulk with $^{210}$Pb. These samples will be sent for counting at low-background HPGe detectors within the collaboration with detection capability for very-low-energy gammas in order to measure the $^{210}$Pb detection efficiency in these samples by counting the 46.5 keV gamma rays. In parallel, the potential for detecting $^{210}$Pb decays near the surface by beta spectroscopy using large-area silicon detectors will be explored. The second objective is to work with Teflon fabricators to establish procedures, ensuring that the risk of radon plate-out during the fabrication and storage of granules is adequately controlled so that little $^{210}$Pb ends up in the bulk during the molding process. The aim of the program is to measure the levels of $^{210}$Pb that can be detected by gamma and beta spectroscopy, develop the protocols for producing Teflon with low Rn exposure, and implement the production protocol by summer 2015. If $^{210}$Pb detection by direct counting proves to have adequate sensitivity, the $^{210}$Pb content of Teflon components produced for the experiment will be screened using this method.

As already discussed, it should be noted that particular configurations of ultralow-background Ge detectors may be sensitive to $^{210}$Pb with high efficiency and this will provide confidence with useful upper limits to complement the R&D described above. The CUBED detector at SURF will have a low threshold of about 10 keV and useful sensitivity to the gamma-ray peak from $^{210}$Pb at 46.5 keV. Similar counting is available with the BEGe detector from Canberra at Boulby Mine. Configured with a flat-disc geometry to suppress Compton background specifically at these low energies, the BE5030 has >30% efficiency at 46.5 keV. Finally, development of well-type detectors by the USD group, described later in this section, and the Canberra well-type detector coming online at Boulby will vastly increase this sensitivity.

**In-house ICP-MS screening.** Mass spectrometry is a sensitive and widely applicable technique for screening for radio-contaminants in materials and has the advantage of fast turnaround (several days compared with two to four weeks by other methods of comparable sensitivity).

Furthermore, compared with other methods, it has high sensitivity to early chain U/Th content. Consequently, in order to carry out screening for the experiment in a timely manner, we expect to screen a large fraction of material samples by mass spectroscopy. Mass spectroscopy services are commercially available but at significant cost (approximately $500/sample); moreover, considerable time and effort are required to find a service that maintains the integrity of the samples and produces reliable results.

R&D efforts to establish throughput at ppt levels of several samples per week is being conducted at UCL using a newly procured Agilent 7900 ICPM-MS system. Off-the-shelf equipment must first be installed and commissioned in a clean environment, and then systems developed for sample handling, preparation, and measurements, including training for chemistry techniques that do not cross-contaminate samples or otherwise compromise measurements with interferences. This iterative developmental stage for faster and more sensitive turnaround is required for rapid analysis of material samples for LZ with high reproducibility and reliability and at the ppt level for U and Th identification. Microwave ashing and digestion methods are also being developed in partnership with Analytix/Milestone to provide requisite sample throughput, ppt-level sensitivity, and reproducibility.

An ICP-MS facility is also available at the University of Alabama, with demonstrated capability to detect U/Th down to the level of tens of ppt and availability to screen tens of samples on the timescale of a few months. This facility does not, however, provide services for sample digestion and separation/concentration of U/Th content. The University of Alabama group is carrying out a systematic program to certify the capability of the facility, develop protocols to prepare samples for analysis without risk of cross-contamination, and set up a laboratory where common digestion and separation/concentration procedures can be done. In parallel, a certification program is being carried out for a commercial ICP-MS laboratory. Certification of capability and development of protocols for producing samples without cross-contamination is expected to be completed in summer 2015 and a laboratory set up by the end of 2015.

The UC Davis group is supporting an effort by the campus ICP-MS laboratory to assay U/Th content in Ti at the ppt level. This effort is targeting separation of U/Th from Ti while maintaining high efficiency for retaining any U/Th content. A method based on using TRU resin to separate Ti from Th has been attempted but with only very limited success. An alternative anion exchange resin method is now under investigation.

**Maintaining cleanliness during storage, transport, and processing.** Identifying materials that have sufficiently low radioactive content is just the first step in delivering a detector that operates with low backgrounds. Materials and components must be processed and assembled into subassemblies and integrated with other subassemblies and installed. Transport must be handled carefully and many components will need to be stored for various periods of time, awaiting assembly and integration.

All LZ groups will follow required protocols that include airborne-particulate and Rn control in order to maintain cleanliness throughout fabrication, transport, storage, assembly, and integration. To establish and validate the protocols, the USD group is building a portable clean room that will initially be set up in their laboratories but may eventually be moved to SURF for further development and testing of cleanliness protocols on site. Dust and radon monitoring under different protocols will be carried out and the effectiveness of different storage methods will be evaluated. Further on-site cleanliness and storage development is being conducted at SDSMT.

**Detector development.** Several groups are engaged in R&D of ultralow-background detector technologies to significantly enhance the capability and capacity of the LZ screening and cleanliness campaign as well as background model development. Outlined below, these include the production of Ge well detectors and large-area Si detectors.

The USD group is developing an in-house program of large HPGe-well-detector construction for low-background counting at SURF. The well detector will be built using a large-size detector-grade germanium crystal grown at USD. The proposed well will be fabricated with a blind hole (80 mm diameter and 50 mm depth), leaving at least 20 mm and 30 mm of active detector thickness at the side and the bottom of the well, allowing the counting geometry to approach $4\pi$. The detector will combine excellent energy resolution at low and high energies with maximum efficiency for low background. This is particularly valuable where small, low-mass components and materials need to be selected but where only upper limits are recorded prior to construction as a result of detector sensitivity or throughput constraints. The sensitivity of the detector is expected to approach approximately 50 μBq/kg for U and Th as a result of the $\sim 4\pi$ coverage and construction from ultralow-background components.

The LZ group at LBNL is developing large-area Si detectors to directly assay surface contamination of critical detector materials. LBNL has a well-documented history of creating novel solid-state detectors and proposes to develop 88,000 $mm^2$ detectors to detect the alphas principally originating from radon-decay daughters. Commercial detectors are typically limited to ~500 $mm^2$ active area and modest internal contamination, coming primarily from detector mounting components. Large-area Si detectors have already been manufactured at LBNL with material screened and verified as sufficiently pure in terms of U/Th content. R&D is being conducted to assess a variety of mounting designs to reduce the mass of the mounts as well as pursue ultralow-background materials to construct these mounts. The R&D program will make use of the LBNL BLBF as well as direct-assay spectrometers at SURF. In addition to developing the detectors, the group will engineer housings to facilitate the assay of critical detector materials and maintain detector cleanliness. Surface alpha counting with Si detectors is also being developed at SDSMT.

## 12.7 Laboratory Backgrounds

Radioassays of detector materials and controls to limit further contamination do not mitigate external radioactivity arising from the laboratory environment or from cosmic-ray muons. However, these external backgrounds, specifically gamma rays and neutrons from the rock and muon-induced neutrons, are rendered negligible due to use of the water shield surrounding the detector and the Gd-LS. The cylindrical water shield is 7.6 m in diameter and 6.1 m in height. The LUX experiment is demonstrating the efficacy of the same active water Cherenkov shield that will surround LZ, and in the same location. Monte Carlo simulations of the LZ experiment as described in Section 12.2 account for the reduction in background attenuation due to detector dimensions, as well as for the impact of the detector not being centered on the same point as LUX in the Z coordinate.

The Davis Cavern at SURF's 4850L is surrounded by Yates formation (HST-06) with a modest rhyolite intrusion (less than 1/10$^{th}$ of the surface of the cavern), and 5-20 cm of concrete covering the rock. The rhyolite has $^{238}$U, $^{232}$Th, and $^{40}$K activities of approximately 100 Bq/kg, 45 Bq/kg, and 900 Bq/kg, respectively. The concrete is considerably lower in activity, with $^{238}$U, $^{232}$Th, and $^{40}$K activities of 25 Bq/kg, 5.5 Bq/kg, and 640 Bq/kg, respectively, based on typical mixes [34,35]. The majority of the Yates formation is very-low-radioactivity rock at $^{238}$U, $^{232}$Th, and $^{40}$K activities of 2 Bq/kg, 0.8 Bq/kg, and 50 Bq/kg, respectively [36,11]. Gamma-ray fluxes have been measured at SURF [37] as 2.16 ± 0.06 cm$^{-2}$s$^{-1}$ above 0.1 MeV and 0.632 ± 0.019 cm$^{-2}$s$^{-1}$ above 1 MeV. The water tank, the outer detector scintillator, steel shielding, and LXe skin of the LZ detector are expected to suppress this flux by about 5 orders of magnitude [38,39]. Further reduction will be achieved through event selection in the energy range of interest, single scatter selection, and application of anti-coincidence cuts with the veto systems, leading to an event rate comparable to that from the cryostat. This is approximately 10 events in the 5.6-tonne fiducial volume in 1,000 days.

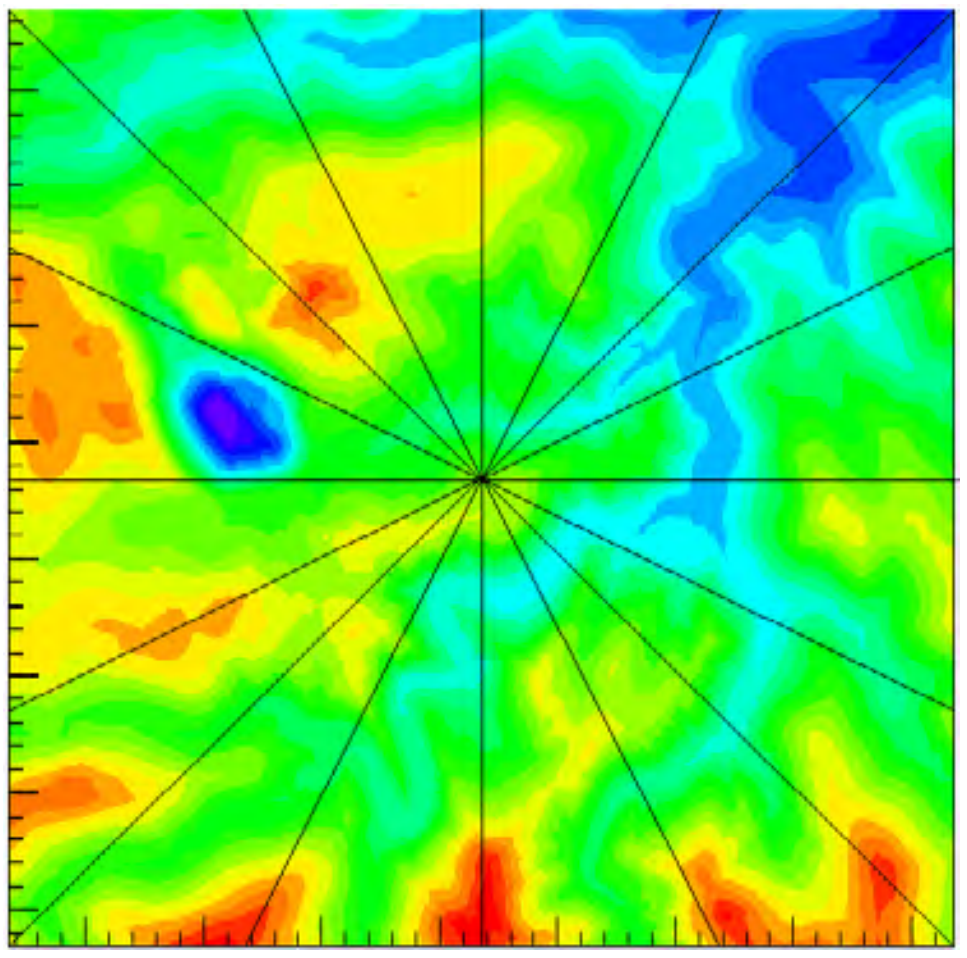

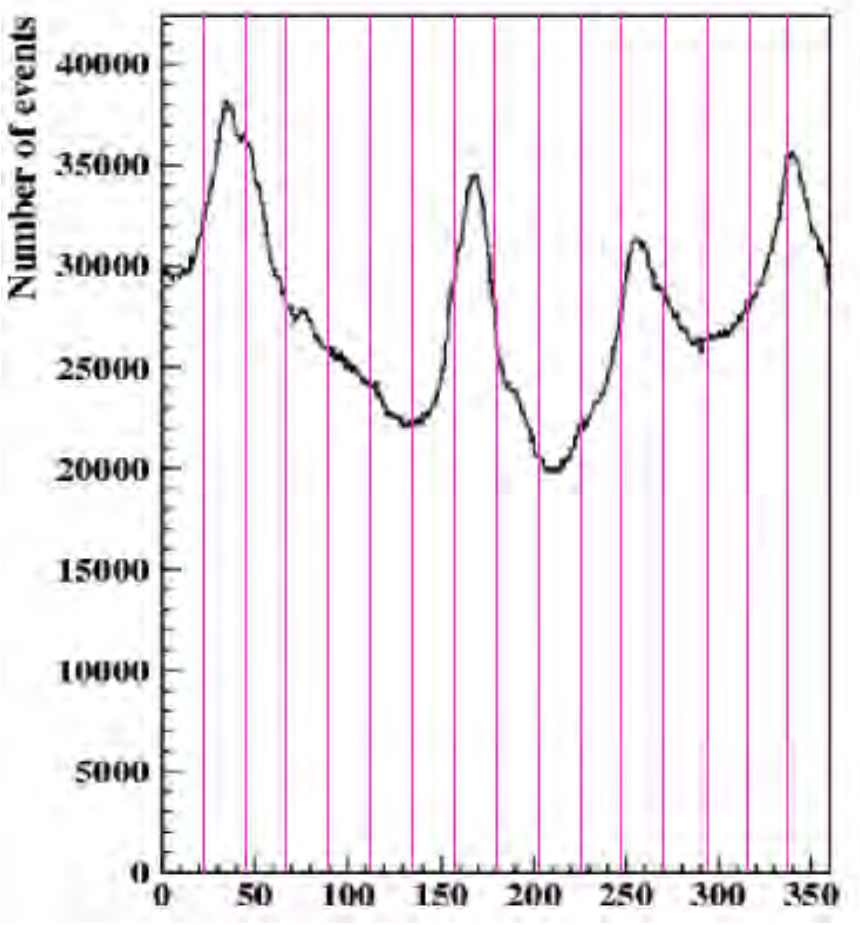

**Figure 12.7.1. Left: Surface profile above the Davis Campus (center of map). The lines drawn from the center divide it into sectors of similar open angle (20° — 25°) to guide the eye. Right: Muon azimuth angle distribution. Vertical lines show approximately the division of the sectors on the left figure, where the azimuth angle is calculated from east (pointing to the right on the left figure). Moving from east to north and further on counterclockwise on the map on the left shows how the peaks and valleys on the surface profile correspond to variations in the number of muons through the laboratory.**

The neutron flux through the cavern is primarily due to spontaneous fission of $^{238}$U and ($\alpha$,n) reactions from alpha particles emitted in the decay series of U and Th. This fast neutron flux is estimated to be about $10^{-6}$ neutrons/(cm$^2$ s) for neutron energies over 1 MeV [37]. This background is moderated very efficiently by the water and scintillator shield [38]. The minimum attenuation is expected below the cryostat where the minimum thickness of hydrogenous material (water or scintillator) is about 70 cm. This is enough to suppress the neutron flux by more than 6 orders of magnitude. The whole shielding will reduce the NR rate from rock neutrons to a level below the neutron rate from the cryostat. This is less than 0.01 count in 5.6 tonnes in 1,000 days.

At the 4850L of SURF, shielding from muons is provided by (4300 ± 200) mwe rock overburden. The measurement of the muon flux at this depth at SURF performed by the active veto system of the Davis experiment found $(5.38 \pm 0.07) \times 10^{-9}$ cm$^{-2}$ s$^{-1}$sr$^{-1}$ for the vertical flux [40]. Recent detailed assessment with the MUSIC and MUSUN simulation packages [41,42] developed by the Sheffield group, which incorporate the profile of the surface above the Davis Campus (shown in Figure 12.7.1), agrees very well with this vertical flux: $5.18 \times 10^{-9}$ cm$^{-2}$ s$^{-1}$sr$^{-1}$.

Muons crossing the LZ water tank are readily detected via Cherenkov emission, and any that deposit energy in LZ are similarly easily detected. However, muon-induced neutron production through spallation, secondary spallation, or photonuclear interactions by photons from muon-induced EM showers in high-Z materials can generate background for the experiment [43-46]. Furthermore, the neutron flux decreases more slowly than the muon flux because higher-energy muons yield more neutrons. The total muon-induced neutron flux at SURF is $(0.54 \pm 0.15) \times 10^{-9}$ n cm$^{-2}$ s$^{-1}$, where approximately half of this flux produces neutrons of energies greater than 10 MeV, and approximately 10% of the flux is from neutrons with energies in excess of 100 MeV. The high-energy component (E >10 MeV) is reduced by approximately a factor of 3 by the water shield for the LUX experiment, with an integrated rate of $10^{-7}$ neutrons s$^{-1}$ impinging on the LUX cryostat resulting in only 60 ndru of single-scatter events in the WIMP search energy range. In LZ, an active neutron veto supplements the shielding such that the muon-induced neutron rate from the rock is expected to be much less than the neutron emission from material radioactivity.

Muon-induced neutrons generated in water are also considered, since although water is low-Z, several hundred tons provides a substantial target. With a neutron yield similar to polyethylene at approximately $2.5 \times 10^{-4}$ neutrons muon$^{-1}$g$^{-1}$cm$^{-2}$, the production rate is of order $10^{-9}$ neutrons kg$^{-1}$s$^{-1}$. However, the water also self-shields very effectively such that the rate on the LUX cryostat is reduced to $6 \times 10^{-7}$ neutrons s$^{-1}$ resulting in 120 ndru of single scatters in the WIMP search energy range, and is expected to give an insignificant contribution to the LZ background given further veto suppression. Other high-Z materials in and around the cryostat may contribute to the muon-induced neutron rate but the efficient muon and neutron veto system will suppress this background by a large factor. Full simulation of muon-induced neutron background is in progress.

Background may also be generated due to trace U, Th, and $^{40}$K within the water itself. With activities for these below 2 ppt, 3 ppt, and 4 ppb, respectively, the gamma-ray and neutron flux is extremely low. Radon in the water tank is reduced with an $N_2$ purge blanket. Further mitigation is achieved by establishing a vertical temperature gradient and limiting convection such that the Rn is transported to the edge of the tank, far from the detector. These techniques have already been successfully applied by LUX inside the water shield.

## 12.8 Cleanliness

Once materials and components have met material screening requirements and components have been procured, they must be kept clean during fabrication, storage, transport, and final assembly and integration into the experiment. We refer to this task as cleanliness. The three major sources of contamination that must be addressed by the LZ cleanliness program are radon diffusion and daughter-nuclei plate-out, dust and debris, and cosmogenic activation. Other sources of contamination must also be addressed, for example residual chemicals from fabrication processes, removal of which should be effected at the same time as cleaning to remove dust.

We require that the effects of radon plate-out, dust deposition, and cosmogenic activation do not significantly increase the expected backgrounds after materials procurement, for which rigorous materials screening is carried out as described in earlier sections. Quantitatively, we require that cleanliness controls allow no more than a 10% increase in non-astrophysical backgrounds. For Rn-daughter plate-out, the dominant source of background is production of NR candidates via the $(\alpha,n)$ process on PTFE (fluorine) from $^{210}$Po decay. The corresponding requirement on cleanliness control for radon plate-out is then that any increase in NR counts is less than 0.02, which is 10% of the total radioactivity NR background budget of 0.2 events in 5.6 tonnes over 1,000 days. This corresponds to an activity of 20 mBq/m$^2$ from $^{210}$Pb deposited on PTFE surfaces. We define the requirement for maximum $^{210}$Pb activity on the PTFE at half of this value, i.e., 10 mBq/m$^2$. Further possible constraints from $^{210}$Pb ion recoils being mis-reconstructed into the fiducial volume are under evaluation.

In the case of dust deposition, there are two considerations: The first is evaluating the contribution to the ER and NR backgrounds from the gamma rays and neutrons emitted by the dust due to its intrinsic radioactivity. The second consideration is the contribution due to radon emanation from the dust. Radon produced in small dust particles is expected to have a significant probability of escaping the dust by diffusion or recoil, making this mechanism particularly dangerous for materials in close proximity to the LXe. Of these two considerations, radon emanation leads to the more stringent limit on dust deposition. Since radon emanation into the LXe volume due to dust deposition must be limited to 10% of 0.67 mBq, and typical $^{238}$U activity in dust is on the level of 10 mBq/g, the dust mass allowed on the PMTs, PTFE reflectors, TPC components, and mechanical supports must be less than about 5 mg altogether. It should be noted that the value of 10 mBq/g, while typical for U activity measured in dust at various locations, is being checked by radioassay of dust samples collected at representative locations at SURF. Measurements of the dust-particle count density, size, and deposition rate have been carried out [47]. The presence of dust may also have seriously adverse effects on other subsystems such as HV delivery. Particulates settling on electrodes or otherwise distorting the electric field can lead to electrical discharges and breakdown. Therefore, cleanliness goals cannot be derived on the basis of controlling backgrounds alone. Once the tolerances on dust deposition from considerations of other factors such as electrical stability and optical transparency are known, the cleanliness program will be augmented as needed to accommodate them.

The LZ cleanliness program will be coordinated by the LZ Cleanliness Committee, comprising collaborators with extensive experience dealing with cleanliness issues on LUX, EXO, KamLAND, MAJORANA Demonstrator, and other low-background experiments. The committee will initiate development of cleanliness protocols, review protocols prior to implementation, and address cleanliness issues as they develop. Each subsystem will have a cleanliness liaison who will be involved in the development and promulgation of cleanliness protocols and is responsible for ensuring that the appropriate protocols are followed and documented. All protocols and documentation will be stored in the LZ information repository.

As described in Section 12.4, procedures will be implemented to control radon plate-out. Radon-reduced air will be used where necessary and exposure to air will be limited, the tolerable exposure time depending on the radon content of the air and the sensitivity of the experiment to contamination of the component. Clean rooms with radon scrubbers and radon-free glove boxes will be used for assembly of radon-sensitive items. Large surface components such as the inner cryostat will be overpressured with $N_2$ dewar boil-off to prevent back diffusion of radon. Radon monitors will be operated at fabrication sites, both on site at SURF and off site. When not being processed, components will be covered in radon-inhibiting materials and storage volumes will be purged with nitrogen gas if necessary. An important component of the program is to implement procedures to verify that adequate measures have been taken to control radon plate-out. For critical items, coupons, handled in the same way as the detector material, will be collected and assayed upon arrival on site before installation and integration. In general, facilities used to screen candidate materials in the first phase of construction will be used during the integration and installation phase for cleanliness checks. Efforts to develop large-area low-background silicon detectors and high-efficiency Ge detectors for this task are described in the R&D section. No material or component will be accepted for final integration without adequate documentation in the information repository demonstrating handling following these protocols and successfully passing assay tests.

Control of dust will be achieved by application of certified cleaning protocols, transport and storage under clean conditions, and assembly in glove boxes and clean rooms under full access control and with personnel wearing overgarments, masks, and hair coverings. Clean-room environments and areas used for storage and work prior to final cleaning and assembly will be monitored with particle counters. Large-capacity ultrasonic baths and deionized water supplies, along with stocks of cleaning supplies and fresh chemicals, will be used for final cleaning during the installation and integration phase. For transfer and storage of components, double-bagging will be employed so that as components move from an uncontrolled environment to a controlled environment, the outermost layer can be removed just before entering the controlled environment while keeping the component itself continuously covered. Custom-built cases, some air-sealed, and specialized transport services will be employed to assure cleanliness of detector components during shipping. As for radon plate-out monitoring, coupons will be analyzed for dust control. Protocols for some items will require swipes that will be collected and counted before acceptance for integration. No component will be accepted for integration without supporting documentation that the appropriate cleanliness protocols have been followed.

A third major source of contamination risk addressed by the cleanliness program is cosmogenic activation. Given the strong attenuation of the muon flux by the rock overburden at the Davis Cavern, together with further attenuation of spallation neutrons by the water shield, cosmogenics are not expected to be a significant contributor to backgrounds after the detector is installed underground. However, cosmogenic activation of $^{46}$Sc in the Ti cryostat, $^{60}$Co in any copper components, and various radioisotopes of Xe are expected to be significant at the surface. While the $^{46}$Sc and $^{60}$Co backgrounds will be efficiently self-shielded in LZ, since the gammas are produced at the edge of the LXe volume, no such self-shielding occurs for Xe radioisotopes, some of which have atomic de-excitation energies of a few keV. The longest half-life is about 36 days, such that sufficient cool-down during commissioning is expected. However, the possibility of storing the dekryptonized LXe underground prior to commissioning is also being considered. The addition of sheets of neutron absorber around the gas tanks would effectively suppress $^{127}$Xe production by thermal neutrons. In all scenarios, exposures of major components and materials will be documented throughout the entire period from production to detector assembly and cool-down underground for implementation into the LZ cleanliness database and background model. A thorough evaluation and assessment of activation products that may be produced within each material proposed for use in LZ is being performed using toolkits developed by LZ collaborators, such as the muon-generator software by the Sheffield group, and common packages such as ACTIVIA [32]. Measurements of cosmogenic activation in LUX materials provide valuable data against which to validate our simulations. Transport and storage planning for the detector-construction phase will take the results of this assessment into account.

## Chapter 12 References

# 13 Infrastructure at the Sanford Underground Research Facility

This section describes the SURF surface and underground infrastructure improvements and additions needed in order to facilitate the assembly and installation of LZ. A detailed assembly and installation sequence for LZ is presented in Chapter 14. Infrastructure at SURF includes the following items: (a) surface laboratory space for assembly of the Xe detector (WBS 1.5) into the inner cryostat vessel (WBS 1.2); (b) staging space for scintillator veto tanks and scintillator (WBS 1.6), for the outer cryostat subcomponents (WBS 1.2) and other detector components; (c) secure storage space for Xe; (d) custom tooling for lowering the fully assembled Xe detector sealed inside the inner cryostat down the Yates shaft; and (e) modifications to the Davis Campus at the 4850L of SURF. These infrastructure elements are described in the subsections below. Surface and infrastructure improvements will be funded by the South Dakota Science and Technology Authority (SDSTA). The design of the infrastructure improvements will be a joint effort of the SDSTA engineering and technical staff and the LZ collaboration.

## 13.1 Surface Infrastructure

The principal components of the surface infrastructure specific to LZ are: (a) the Surface Assembly Laboratory (SAL); (b) the Surface Storage Facility (SSF); and (c) office and meeting space at the SDSTA administration and education buildings. Office and meeting space already exists for ongoing experiments at SDSTA; general improvements to these capabilities will serve the broader experimental community at SURF, and are not described here. The locations of the SAL and the SSF at SURF are shown in Figure 13.1.1.

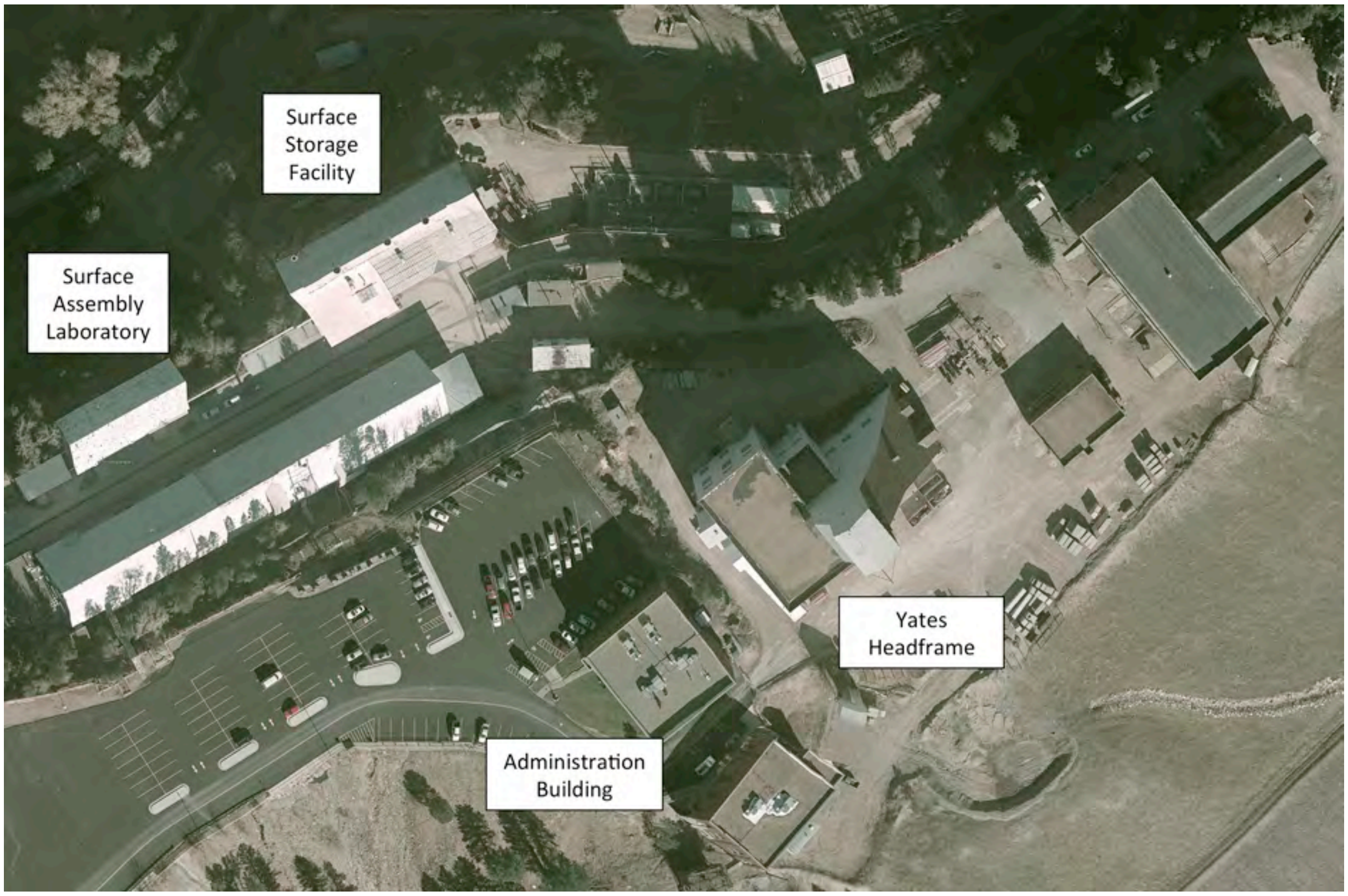

**Figure 13.1.1 Aerial view of the SURF site showing the locations of the Surface Assembly Laboratory and the Surface Storage Facility.**

The SAL was utilized for the assembly of the LUX detector (Chapter 5) and for operation of LUX in a water tank located in the SAL. Significant upgrades to the SAL are planned for assembly of the LZ detector and are described later. The most significant improvement to the SAL is the addition of radon-reduced air-handling capability. The SSF is currently used for general storage. Modest improvements are needed for the SSF.

### 13.1.1 Surface Assembly Laboratory (SAL)

The SAL building is a wood-frame structure comprising four levels: the surface level and levels -1, -2, and -3, which are successively deeper below the surface. This building was renovated for LUX assembly and testing and has a current sprinkler system, new HVAC, and 220 kVA installed electrical capacity. An existing clean room of 2,900 $ft^3$ will be used for cleaning and staging of metal parts and bagged PTFE parts prior to Xe detector assembly in the low-radon clean room. A new low-radon clean room will provide workspace beneath an existing monorail and over a pit in Level -1. New floor supports will support the assembly of the inner detector. Transport of the detector will be facilitated by rolling it to the building shipping dock. The layout of the SAL is shown in Figure 13.1.1.1.

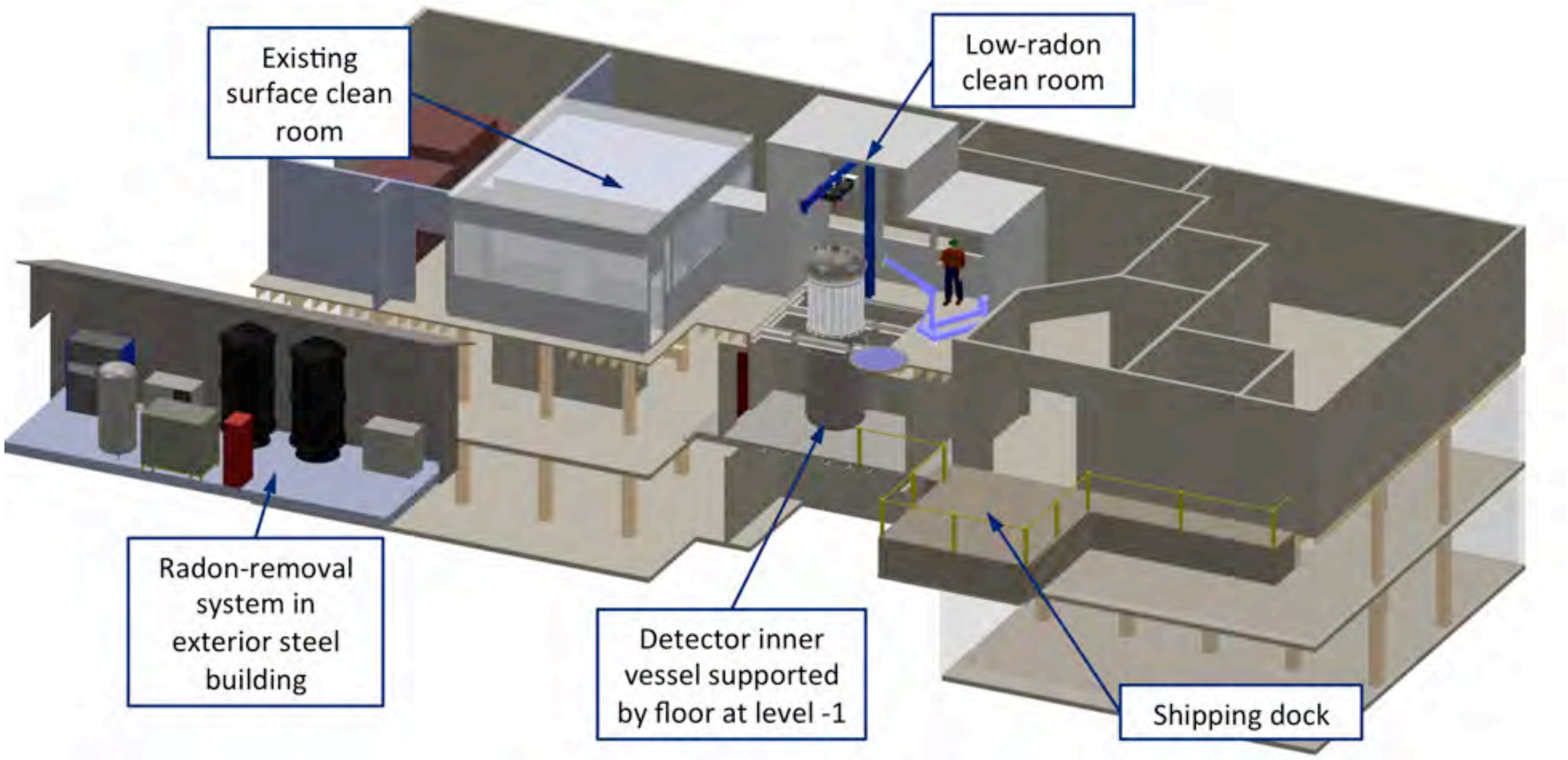

**Figure 13.1.1.1. The Surface Assembly Laboratory.**

### 13.1.2 Radon-reduced Air System

Radon is one the highest-risk contaminants for low-background experiments because it can easily escape from bulk material and quickly diffuse into active parts of the detectors. The innermost materials and parts of ultralow-background experiments must be manufactured or assembled in a radon-suppressed atmosphere. Final assembly of the LZ TPC will occur in a clean room in the SAL, which has a radon-reduced air system.

Carbon adsorption is the preferred technique for removing radon from air and has been effective for varying degrees of radon contamination, from as low as a few $mBq/m^3$ for low-background experiments to levels as high as ~100 $Bq/m^3$. The radon level at the surface at SURF is expected to be in the range of ~20 $Bq/m^3$. The LZ detector assembly clean room will be designed with a radon-reduction factor of greater than 1,000. It will be flushed with radon-reduced air at a flow-rate of 300 $m^3/h$, which is produced by compressing, drying, cooling, and pushing the air through two 1,600 kg activated-carbon towers. Our baseline plan is to use commercial units provided by ATEKO, which has recently developed and built large commercial radon-removal systems for the Cryogenic Underground Observatory for Rare Events (CUORE) and Borexino experiments. These units are capable of reducing radon concentration in the air by a factor of greater than 1,000.

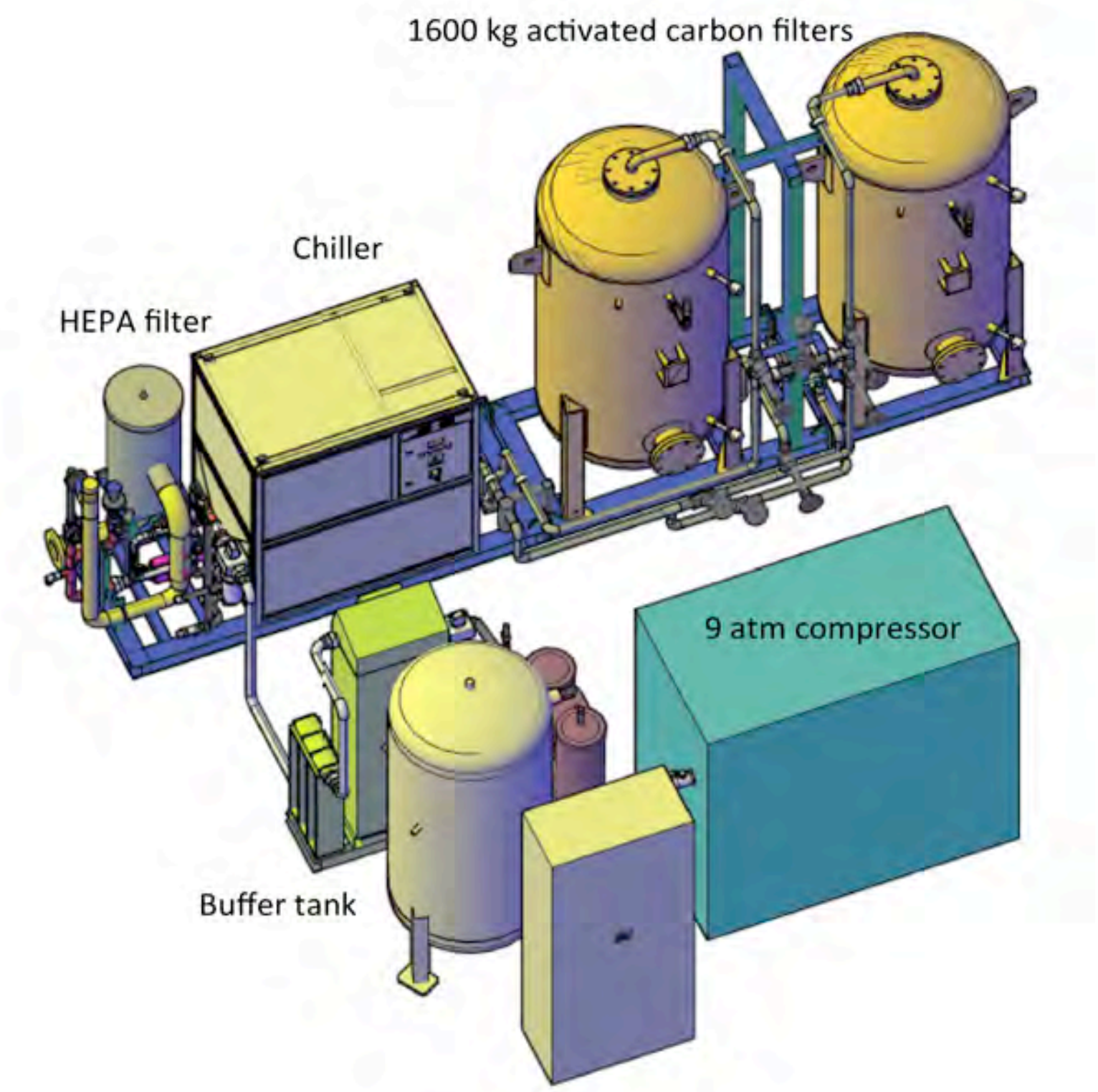

**Figure 13.1.2.1. Typical component layout for a radon-removal system from ATEKO. The total weight of the system is approximately 5,500 kg. Each carbon column weighs 1,600 kg.**

The proposed system is similar to that used by CUORE. A preliminary quotation from ATEKO has been obtained. Discussions to adapt the design to SURF requirements have been held and a site visit made to ATEKO in the Czech Republic . The process is based on compression, cleaning, and drying (dew point –65º C min.) of air, cooling to –55 º C, adsorption of radon from air on activated carbon at –50º C at approximately 8 bars of pressure, followed by heating and pressure reduction of air to ambient pressure and temperature. The system requires a space of about 4.5 m x 7 m with a height of 3.2 m (absorbers with frame). The components of the proposed system from ATEKO are illustrated in Figure 13.1.2.1. Most of this system would be located outside the SAL and will be procured by SDSTA.

### 13.1.3 Surface Storage Facility (SSF)

Short-term storage and staging of equipment needed for experiment assembly will be facilitated by upgrades to an existing building next to the SAL. This includes the large scintillator tanks. The SSF has a large high bay with approximately 445 $m^2$ (4,800 $ft^2$) of lay-down space under a 40-ft-wide 10-ton bridge crane. Hook height on the bridge crane reaches 7.3 m (24 ft). Handling of equipment will also be facilitated by three wall-mounted jib cranes rated at 1 ton, 2 tons, and 3 tons, respectively. Truck entry to the building is via a 4-m (13-ft)-wide x 5-m (16-ft)-tall entry rollup door. Temporary heaters will probably be needed to heat some of the space in this building during the winter months.

## 13.2 Yates Shaft Infrastructure and Custom Transport

The LZ TPC inside the inner cryostat will be transported from the SAL to the Yates headframe in a horizontal orientation. The transport method will be similar to the successful transport of LUX from the same building to the headframe. A large telehandler was used, along with continuous monitor of orientation and G-forces. LUX was transported in the Yates cage but the LZ assembly will be slung under the cage, given its larger size. The LZ detector/inner cryostat will be placed on a custom transport frame.

Once the detector/inner cryostat assembly arrives at the Yates headframe, the assembly on the transport frame will be lowered in the Yates shaft as shown in Figure 13.2.1. Lowering the assembly on its transport cart slung under the Yates cage will take some hours. A crew member will accompany the assembly in the Yates conveyance as was done for LUX. The assembly and transport cart will be extracted from the Yates shaft at the 4850L and rotated back to the horizontal orientation.

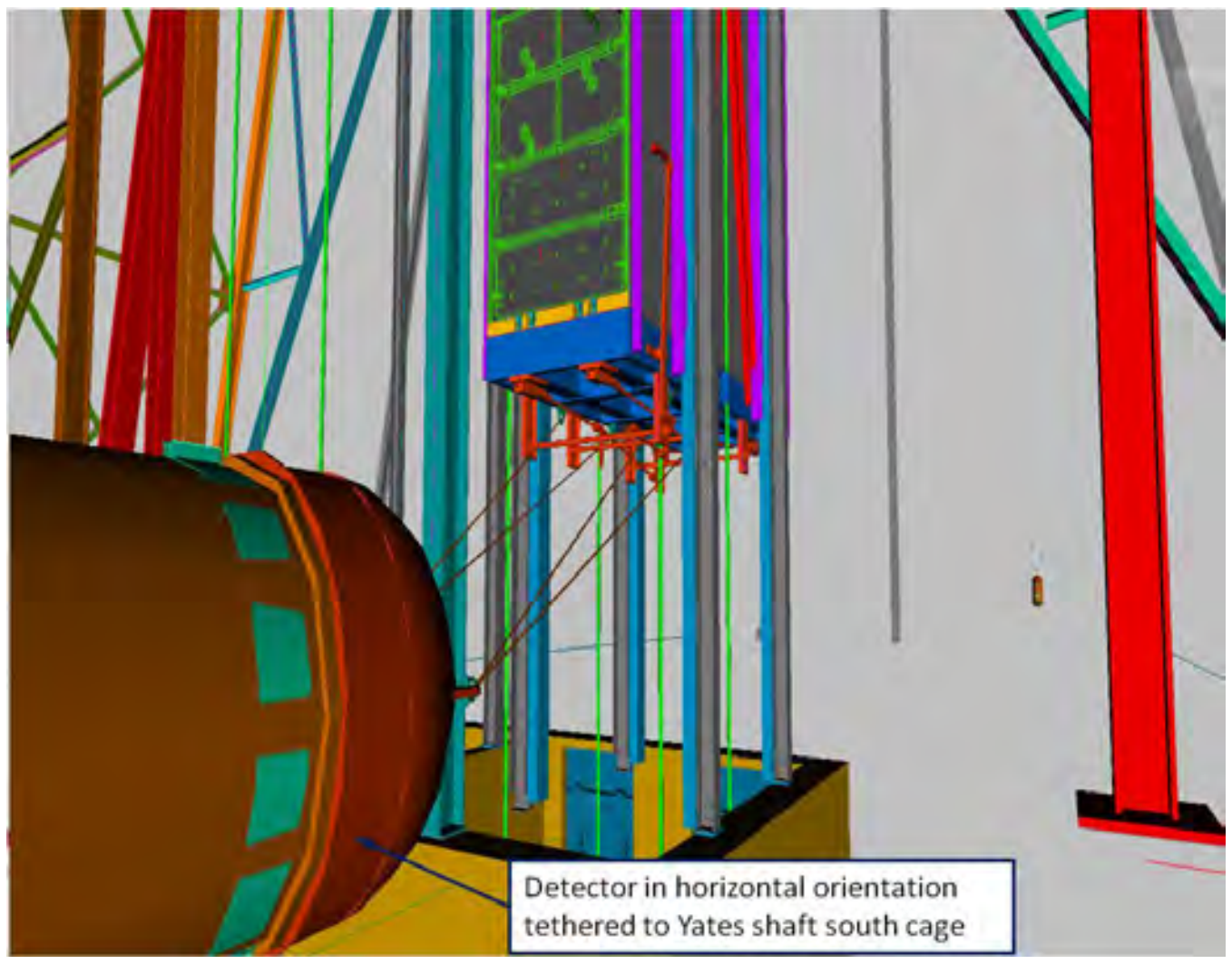

In the first image, the LZ detector lies horizontally at the shaft entrance and the cage bottom is 6 to 7 feet above the floor. Crews will cover the shaft by installing a work platform and attach the slings to the bottom of the cage. These are also attached to the top of the LZ detector. The platform will be removed and the cage hoisted to remove slack. Another wire rope control line from a winch is attached to the bottom of the detector.

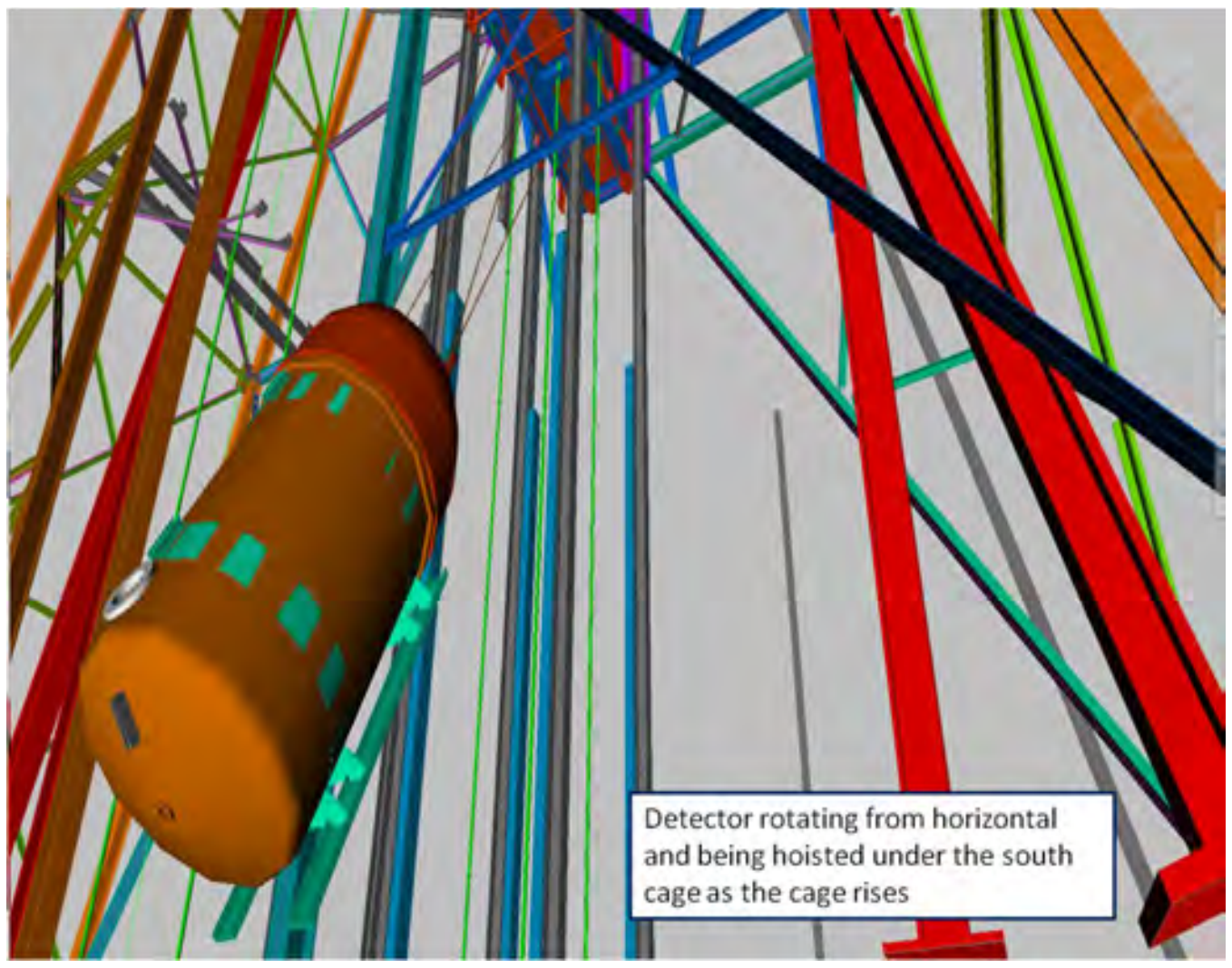

In the second image, the hoist is lifting the LZ detector in the shaft to its vertical position. The control line mounted on the detector bottom will maintain tension in order to control the swing into the shaft. Even though not fully in the shaft, the bottom of the custom transport frame is off the ground. Once hoisting is in process, it will not be stopped unless a difficulty arises. The entire procedure is short-lived.

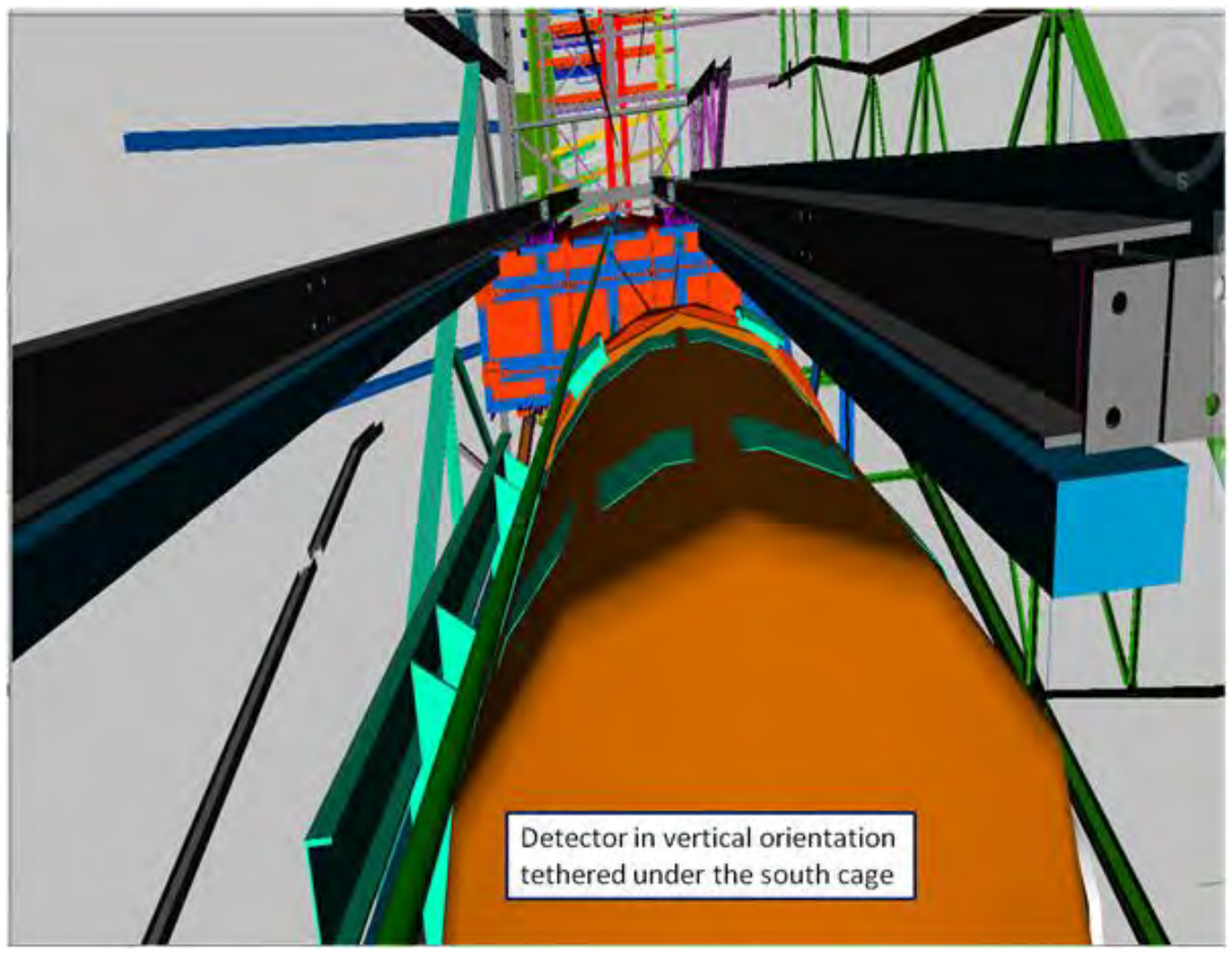

In the third image, looking almost vertically. Here, the detector is hanging vertically. It will be positioned properly in order for the crew to attach the bumpers or guide shoes for clearance and centering during shaft transit. After inspections, a crew will man the cage and watch as it travels underground, being lowered at a rate of 0.7 ft/sec.

**Figure 13.2.1. The transport method of the LZ TPC inside the inner cryostat will be similar to the successful transport of LUX from the same building to the headframe.**

## 13.3 Underground Infrastructure

A plan view of the Davis Campus is shown in Figure 13.3.1. After arriving on the 4850L, the detector will be transported to the Davis Cavern via the Primary Access Drift, passing by the room housing the MAJORANA detector, as depicted in Figure 13.3.1. That is the same access-way utilized for transport of the LUX detector. The larger LZ inner cryostat can be moved through the same space by temporarily removing short segments of the Davis Campus HVAC air-supply ducts to allow for sufficient clearance. All access doorways are sufficiently large to allow for detector passage. Entry to the Davis Cavern via the Primary Access Drift allows the transporter design to take advantage of the 8,100-lb maximum floor loading afforded by design of the Davis Cavern structural steel. It would also be possible to transport the inner detector in the drift that passes by the LN storage room, but this would require the removal (and later replacement) of a wall between the drift and the Davis Cavern.

LZ will take advantage of design features built into the Davis Cavern to allow for deployment of a larger detector than LUX. A 102-inch-diameter flange is built into the 70,000-gallon water tank. Additionally, the redundant 50-ton water chillers and the Davis Campus 1500 kVA substation were implemented with a large future detector in mind.

Figure 13.3.2 shows an isometric view of the LZ deployment in the Davis Campus. Xenon storage cylinders will be securely deployed in a newly upgraded portion of the LN storage room access drift (shown in the lower-right-hand corner). Currently, this is an unfinished area that has been used for rock-moving equipment / storage. SURF will upgrade this area by providing concrete floors and shotcrete-covered walls while also closing off the space to allow for flow-through ventilation. Ventilation will be adequate to mitigate potential oxygen deficiency hazards that could occur in the event of a release from one of the Xe storage cylinders. The Xe cylinders are more thoroughly described in Chapter 9.

The LN storage rooms and control rooms will be largely reused as they were for LUX. LZ will utilize a similar scheme for LN storage tanks until a second cryocooler is secured for long-term operational flexibility. These tanks will supply makeup nitrogen to the cryocooler, assist with transient startup effects, purge gas for the water purification system, and provide some buffer against short-term power interruptions. The control room is expected to remain unchanged, providing minimal office space for underground workers during assembly, commissioning, and data taking.

The size of the ventilation duct that exhausts the LN storage room will be increased to allow for the added flow-through ventilation needed to include the Xe storage.

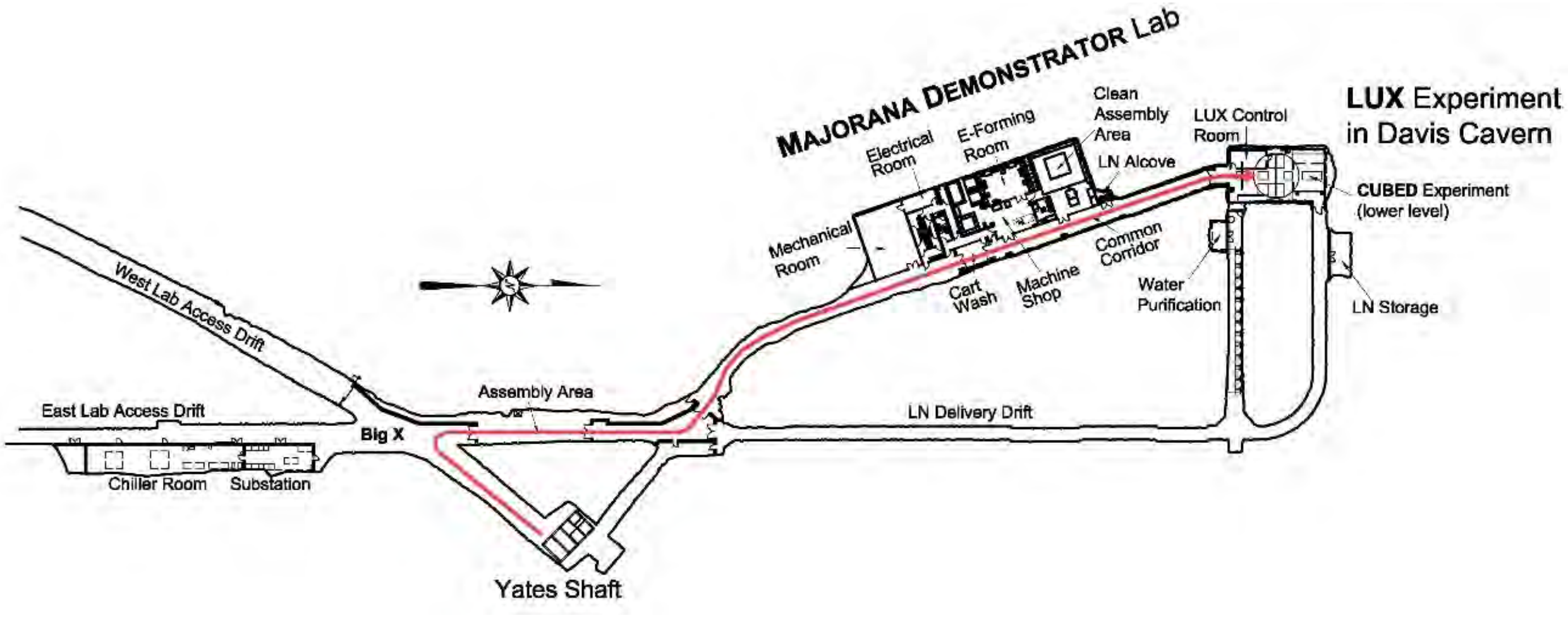

**Figure 13.3.1. Plan view of the Davis Campus at the 4850L.**

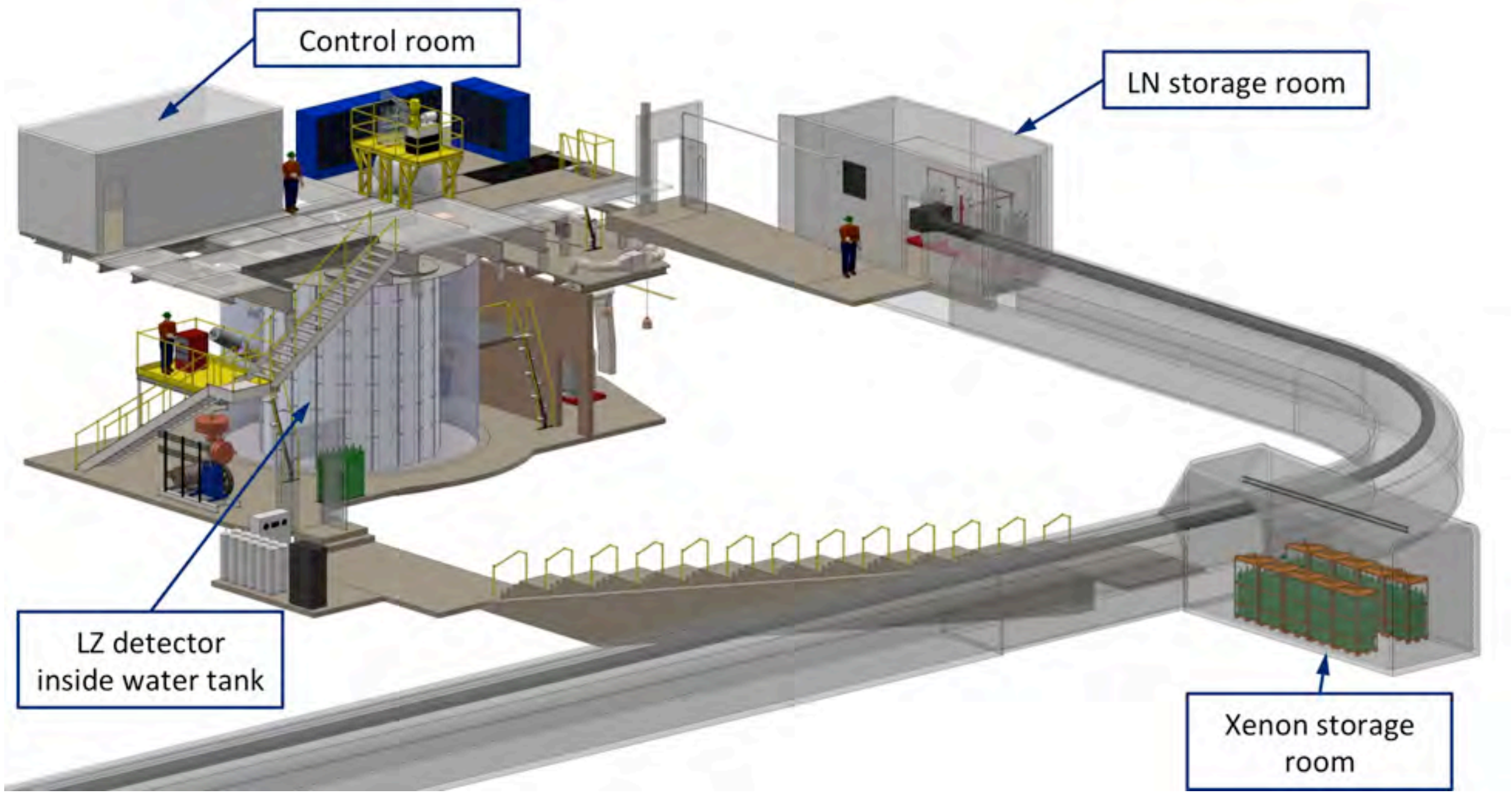

**Figure 13.3.2. Overall layout of LZ in the Davis Campus**

Figure 13.3.3 depicts a close-up view of the Davis Cavern showing supporting infrastructure installed. Key elements include data acquisition (DAQ) cabinets, a Stirling cycle cryocooler to support cooling thermosyphon lines, a platform for a high voltage (HV) power supply, a Xe compressor installed below the steps accessing the lower Davis Cavern, another platform that interfaces to the Xe purification tower, and Xe circulation / storage equipment deployed both in and above the current low-background counting rooms in the lower Davis Cavern.

To facilitate deployment of a larger number of DAQ racks and cryocoolers on the deck of the upper Davis Cavern, SURF will decommission and remove the existing clean room to open up floor space. Platforms deployed near the HV feedthrough and Xe purification tower will also serve to significantly increase the effective floor space in the cavern.

This positioning of the Xe compressor within the cavern places this large piece of 480 V equipment closer to in-cavern existing power panels, while also positioning it to the low-pressure / higher-loss suction side of the Xe gas system. All other Xe circulation and storage equipment take advantage of being co-located in space behind block walls previously utilized for low-background experiments.

The Davis Campus is fed from a 1500 kVA substation. The substation is presently loaded at approximately 60% during worst-case conditions, and has ample reserve capacity to accommodate the projected LZ electrical load of 146 kW (an increase of ~100 kW from LUX) (summarized in Chapter 11). The campus has a 300 kW emergency generator that supplies the air-handling systems, communications, facility control and alarm systems, and egress lighting for up to 48 hours. Floor space is available for an additional 50-100 kW generator to provide backup power for experiments.

To reduce the risk of radon exposure during assembly sequences of the detector inside the water tank, radon-reduced air from the SAL will be plumbed to the assembly area in a pressurized 2-inch pipe installed in the Yates shaft. This takes advantage of the 8-bar discharge pressure of the radon-reduction system to minimize the size of required piping through the shaft and to the Davis Campus. Further study may indicate that the radon content of surface air is sufficient during the short times the sensitive detector components are exposed to air in the water tank. If so, then some small cost reduction will be possible but a dedicated pipe from the surface would still be required. The radon levels in the Davis Campus are 20 times or more larger than at the surface.

The water tank installed for the LUX experiment was designed to be able to accommodate a much larger experiment. The modifications required for LZ are modest.

The pads that support the LUX detector will be removed from the tank floor. Mounting plates will be added to the tank floor for the LZ detector and mounting points for the four side scintillator vessels. At larger radius, mounting points will be installed for the 20 PMT ladders. In addition, a support for the lower conduit that connects the detector to the LXe tower and PMT cable breakout will be installed. Finally, a support for the HV feedthrough will be added.

The detector HV system for LZ requires a feedthrough in the wall of the water tank. In addition, a penetration of the tank wall is needed for routing of the conduit that connects the detector to the LXe tower and the PMT cable breakout. A neutron tube will be installed for the neutron calibration system, and that tube will be filled with water for normal running and nitrogen for calibration. The top of the water tank will have several feedthroughs to accommodate thermosyphons, detector cables, source tubes, scintillator lines, water PMT cables, LED flasher cables, and vacuum pumping.

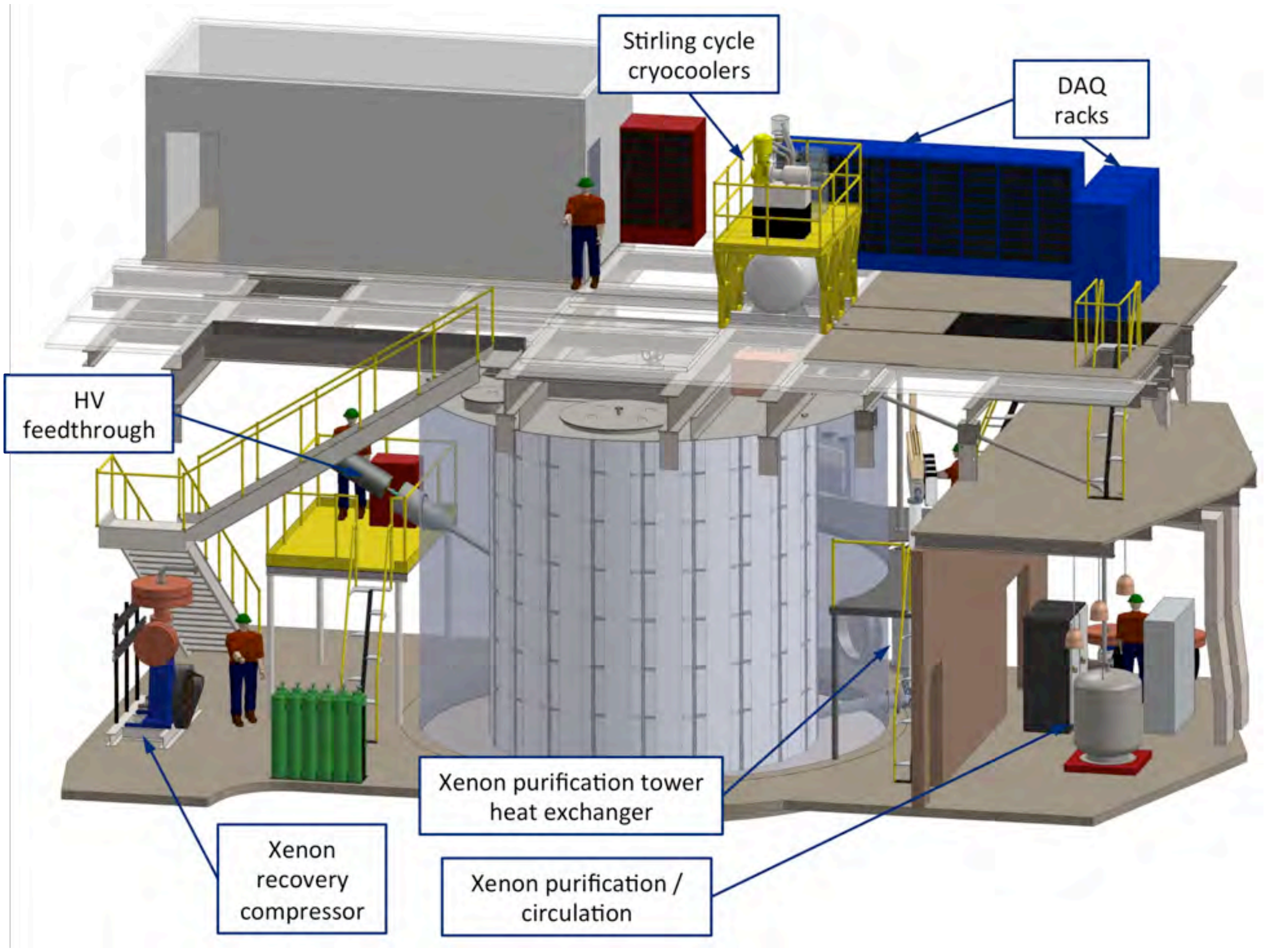

**Figure 13.3.3. LZ and related support systems in the Davis Cavern.**

# 14 Integration and Assembly

This section describes the effort to integrate work at the subsystem level into a coherent design that meets the science requirements and that will result in an operational detector at SURF. A detailed assembly sequence is presented. While the overall scope of integration and assembly is contained in WBS 1.9, significant resources from every subsystem are required. Much of this effort comes from the distributed pool of engineers working for subsystems at many institutions. Coordination of this engineering effort is part of the integration task.

## 14.1 Integration

For LZ to achieve its science goals, it uses primary requirements that define what the subsystems must do; subsystem requirements drive more specific design specifications (see Chapters 3 and 4). A requirements registry has been created to help clearly understand and capture the design drivers. The subsystems must be safe, affordable, timely, compatible with the other subsystems, and possible to assemble and operate with the available infrastructure. The Integration Group provides the necessary management, engineering, design, organizational tools, and administrative effort to assist all subsystems in completing the LZ design. Phone meetings and technical workshops help identify interface issues and hidden constraints created by design choices in other subsystems. The Integration Group maintains a CAD model of the overall LZ detector and a CAD model of the Davis Campus. This helps define physical interferences and design gaps that are not adequately covered. At integration meetings, engineers from each subsystem, and many scientists, share ideas to help with some of the more difficult design challenges.

The Integration Group develops general standards and controls, such as engineering document standards to be used project-wide for CAD file exchange, engineering drawings and design notes, specifications and procedures, engineering change requests and engineering change notices (ECRs/ECNs), and a document-numbering and -organization system. The Integration Group defines component reference names and an overall coordinate system for the experiment. The group maintains a cable and feed-through list with the help of the subsystem management. A key parameters list is maintained as a reference for design, modeling, and science.

As designs mature, they must be documented and reviewed. The Integration Group works with project and subsystem management to arrange and execute design reviews at appropriate times. Conceptual Design Reviews evaluate whether the design meets the requirements, interfaces have been identified and successfully coordinated, and engineering details are sufficiently developed to proceed. Preliminary Design Reviews focus on manufacturability, cost, schedule, risk, and safety. Final Design Reviews ensure that documentation and drawings are complete and the fabrication plan fits within project budget and timeline.

LZ is being assembled and installed in an underground area administered by SURF. Integration includes working with SURF to be sure the infrastructure is adequate to support assembly, installation, and operation of the LZ experiment. SURF engineers are tasked with design and execution of infrastructure projects to support the LZ project. The Integration Group coordinates communication of requirements with SURF and the detector subsystems.

The overall planning of the on-site assembly and installation of the detector at the Davis Campus is part of the integration effort. This work includes defining the sequence of steps to put the detector together, creating a schedule for this work, and developing an understanding of the resources needed to accomplish it. Subsystems support this effort by providing details for handling components and aiding in resource and schedule development.

## 14.2 Assembly

The LZ assembly will happen in three stages — off-site subassembly, surface assembly in the Surface Assembly Lab (SAL), and underground assembly in the Davis Campus — with transportation between stages. The general strategy is to do as much work as practical off site at universities and national laboratories, where highly skilled, specialized labor can easily work with students and scientists. This also reduces travel costs. Parts delivered to the site will be tested, clean, and ready to use, with a few exceptions.

Xe PMTs will be assembled, tested, and characterized prior to delivery to SURF. The PMT vendor(s) will do some QA testing, but burn-in, final electrical characterization, and cold testing of each PMT will be done by LZ. Assembly includes connecting a PMT base to the PMT, attaching fluorocarbon polymer (FCP) reflectors, and final cleaning. Because they have FCP as a reflector, the assemblies must be kept under a nitrogen purge during storage and shipping to prevent radon contamination. Metalized plastic bags can be filled with nitrogen and heat sealed to protect parts during shipment. Internal PMT cables will have a pin-and-socket connection at both ends. They will ship separately and be routed to the PMTs after the PMTs are installed on the support arrays. The cables will be connected to the feed-throughs mounted in flanges after the cables are routed. The cables must be kept clean and under nitrogen purge during storage and shipping because they contain FCP as the primary insulator. Because of concerns about radon contamination, we are still considering an option to do the assembly of all FCP parts to the PMTs at SURF in the reduced-radon clean room of the SAL.

The tested PMTs will be placed into the titanium PMT support structure, including preliminary placement and routing of the cables. The baseline plan is to also do this work at the SAL reduced-radon clean room, but off-site assembly is still under consideration. This work will create an upper and lower populated PMT array. The upper skin PMTs will be attached to the three weir segments. The lower skin PMTs will be directly attached to the lower PMT array.

The wire grids for cathode, gate, anode, and PMT shields will also be manufactured off site. The grids will be cleaned, inspected, packaged, and shipped to SURF. The packaging must keep the wires clean from any debris and protect the fragile wires from shock during shipping.

The field-grading region of the TPC is made from conductive metal rings, insulating FCP spacers, and resistors. These parts will be fabricated by vendors and then inspected and cleaned off site before shipping to SURF. Radon exposure of completed FCP parts will be minimized after final machining. They can be stored and shipped with nitrogen purge or in sealed metal packaging.

The cryostat will be delivered as fully tested and cleaned code-stamped vessels. They will be manufactured in the UK and shipped. A vendor in the United States will perform final cleaning of the vessels just prior to shipment to SURF. The inner and outer vessel will be shipped separately (not nested) in nitrogen-filled sealed bags from the cleaning facility. The reflective liner for the inner vessel will be attached to the wall in the SAL. The plan is to suspend the bottom of the inner cryostat upside down in the vessel-assembly area and to access the inside of the vessel from underneath with a manlift. This will allow a worker to tile the inside while standing on a stable surface.

The HV umbilical will be delivered as a clean, tested, and sealed assembly under vacuum. The heat-exchanger tower subassembly will be built, cleaned, and tested off site, and transported as a sealed assembly under vacuum. The Gd-LS tanks will be manufactured, cleaned internally, tested off site, and shipped in protective packaging. The outside of the Gd-LS tanks will be cleaned on site, underground. The ladders, PMTs, and Tyvek reflectors of the outer detector will be shipped separately.

It is unlikely that the timing of delivery can match the time each piece is needed. A storage location has been identified at SURF (see Chapter 13).

The subassembly work and transportation described above will primarily be the responsibility of the subsystems. Once things arrive at SURF, primary responsibility shifts to the Integration and Installation

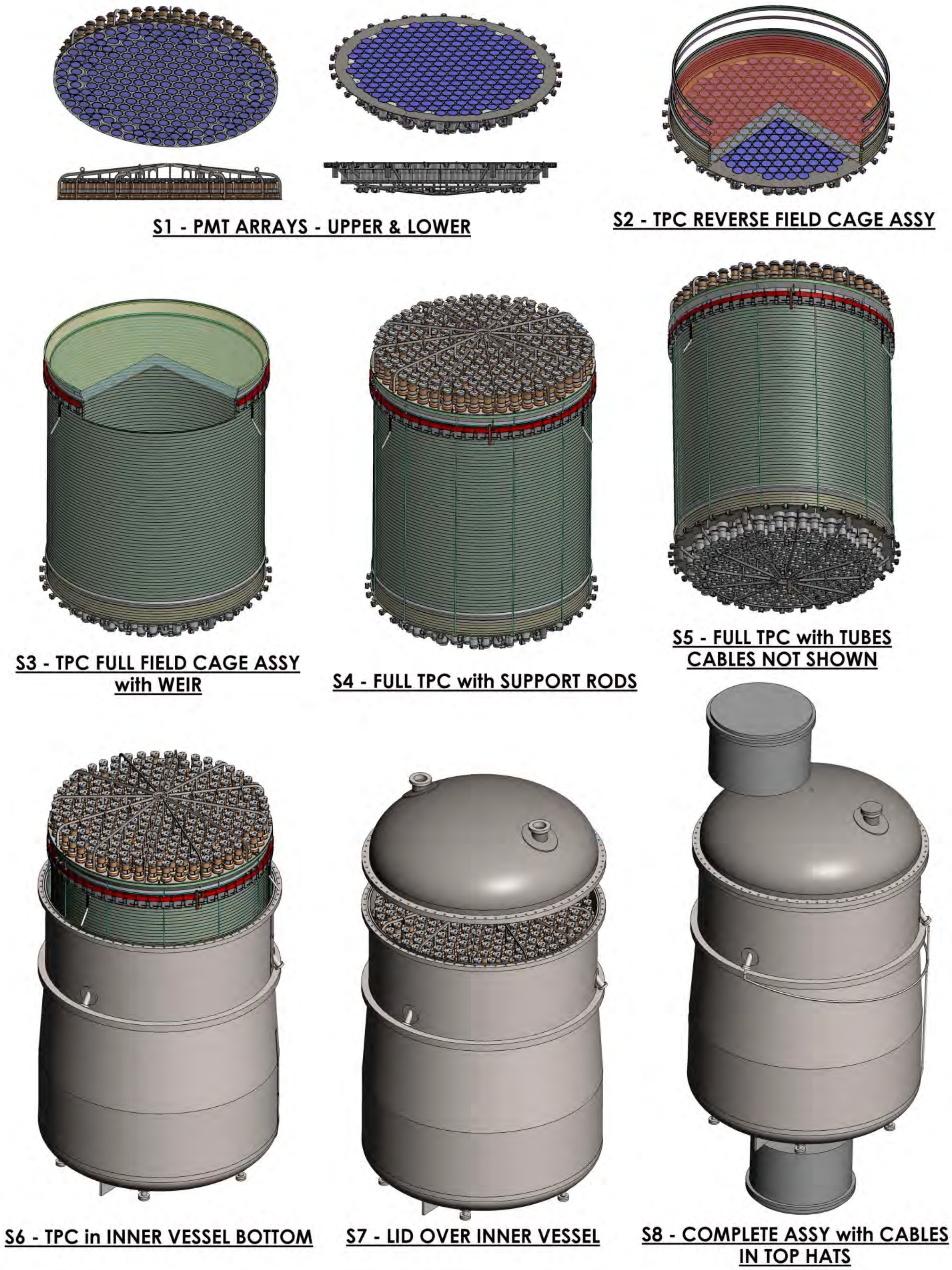

**Figure 14.2.1. LZ surface assembly sequence.**

Team (WBS 1.9). The plan is to work with SURF to hire a pool of local technicians to perform much of the work. This pool will include experts in cleaning, electronics, vacuum, rigging, and mechanical assembly. A lead technician will handle managerial supervision and work direction, but technical supervision will be supplied by engineering and scientific staff. An LZ engineer will be on site to coordinate the work and provide technical oversight. Experts from subsystems will be on site during assembly work focused on those subsystems. Existing SURF engineering and technical personnel will also provide support.

On-site assembly starts aboveground with assembly of the TPC. Cleanliness is critical for this assembly work, both to reduce radioactivity backgrounds in the detector and to control particles that could cause field enhancements and reduce operating voltage. This work will be done in the SAL constructed at SURF with a large reduced-radon clean room (see Chapter 13). Figure 14.2.1 shows the major steps of assembly in the SAL. The first step is receiving and inspection of the parts and subassemblies. The goal of the inspection is to ensure that the parts have not been damaged in shipment and will meet functional requirements. This includes cleanliness.

The next several subsections describe the various stages of assembly. Each subsection concludes with a description of the suite of checkouts that will be performed prior to declaring the stage complete. Final definition of these checkouts is a crucial aspect of the assembly and will be made by WBS 1.9 in close consultation with the relevant subsystem owners.

### 14.2.1 TPC Assembly (Steps S2 – S5)

The assembly will start with the cathode grid supported upside down on a flat surface (see Chapter 6 for descriptions of the components referenced here). The field-grading assembly composed of FCP arcs and complete metal rings will be assembled onto the cathode grid. The arcs and the rings are held together with plastic pins and screws. Resistors that fit into sockets to electrically connect the metal rings will be installed as the rings are stacked. When the field cage has reached the correct height, the PMT shield grid will be installed onto the stack. The lower PMT array will be lifted onto the stack and connected to the PMT shield grid with PEEK screws. The cables for the lower TPC PMTs and the skin PMTs will now be facing up and will get dressed into final positions. Tubes and manifolds for distribution of LXe will be added to the lower PMT array. Temperature sensors and their cables will be installed. Loop antennas for detection of discharge and their cables will be installed. The voltage-control cable for the lower PMT shield grid will also be installed. The lower field grading subassembly (reverse-field region) will be picked up and rotated 180° so the cathode grid is up and lower PMT array is down. This will require a special rigging fixture. The field-grading assembly then proceeds from the cathode grid upward. When the total height is reached, a cylindrical ground plane will be attached to the outside of the field cage to control the field in the region of the skin PMTs. An upper grid subassembly is assembled from the gate grid, three weir sections, the anode grid, and a nonreflective structural spacer. Liquid level sensors and upper skin PMTs will be installed onto the weir segments. The complete upper grid subassembly is then installed onto the top of the field cage. The upper PMT array will then be lifted and set on top of the nonreflective spacer and secured with screws. The cables for the upper TPC PMTs, the upper skin PMTs, and the weir level sensors will get dressed into final position. Temperature sensors and their cables will be installed. The voltage-control cables for the gate, anode, and upper PMT shield grid will then be installed. Tubes and manifold distribution for Xe return gas will be installed on the upper PMT array. The TPC is now assembled and will be run through a series of tests to ensure light-tightness of the field-grading cylinder, function of all the PMTs, function of the HV grids and resistor network, and function of the sensors. Stainless steel threaded rod will be installed to tie the upper PMT array titanium structure to the lower PMT titanium structure temporarily for handling.

### 14.2.2 TPC Insertion into Cryostat with Fluid and Electrical Final Routing (Steps S6 – S8)

The TPC will then be moved over the bottom of the inner cryostat and the bottom inner cryostat raised into position around it. During this process, the PMT and sensor cables coming from the lower PMT array will be routed through the central port in the bottom of the inner cryostat. Xe fluid circulation lines also routed through this port will have been placed into position earlier, but may need adjusting and securing as part of the cable routing. Access for this operation will be through the HV connection port and the bottom port of the vessel. The three Xe tubes from the weir trough must be routed to the ports in the inner cryostat wall. These tubes are FCP bellows that will initially be pointing straight down and then guided into the ports as the inner cryostat is raised. The TPC is supported on six posts projecting upward from the bottom head of the inner cryostat. Tapered guide pins will be installed in the bottom mounting holes of the TPC to engage the holes in the posts and guide the TPC into the correct position. Once the TPC is in the correct place, the guide pins will be removed and bolts will be installed to secure the TPC to the posts. The titanium plate for the upper PMT array will be guided from the inner cryostat near the main flange with tabs that allow vertical motion but constrain radial motion. Access for this work is from the top over the main flange. After these upper guides are secured, the TPC is fixed in the vessel and the temporary threaded rods between the upper and lower PMT arrays will be removed. The location of the weir surfaces that establish the LXe plane will be surveyed relative to known positions on the outside of the inner cryostat. This will allow rough leveling of the weir surface during future assembly steps. The next step is to stage the lid of the inner cryostat over the bottom and install a temporary safety support. The PMT cables, grid cables, sensor cables, and Xe gas lines coming from the upper PMT array need to be routed through a port in the inner cryostat lid. The sensors that monitor the position of the TPC relative to the inner cryostat wall will be installed and cables dressed. The lid can then be lowered onto the bottom of the inner cryostat and the large-diameter seal formed. This seal is designed as two seals (double O-ring) to facilitate a check for leaks by pumping between the seal and looking at the rate of pressure rise. All other ports on the inner cryostat must be sealed for transportation to keep the TPC clean. The ports with cables cannot be sealed with normal blank flanges. Special enclosures — top hats — will be built to house the cables and sealed to the inner cryostat during transport. The sealed inner cryostat will be pumped down to vacuum to ensure the seals are adequate. Other testing will be done at this point. This could include PMT functional tests, sensor tests, light-tightness tests between the skin and central TPC volume, and HV tests. Closed-cell foam insulation and superinsulation blankets will be fit-tested on the outside of the inner cryostat for later underground installation. After these tests, the inner cryostat with TPC assembly will be double-bagged so it is ready to be moved underground.

### 14.2.3 Underground Outer Detector Tank Preparation and Staging (Steps U1 – U2)

The first step of underground assembly is to prepare the water tank (see Chapter 13). All LUX components will have been removed and some infrastructure work on the overhead cranes completed. Some welding is needed in the water tank for attachment points for the cryostat support, the Gd-LS tanks, and the outer detector PMT ladders. New penetrations are needed for the HV umbilical and the HX conduit through the wall of the tank. After these welding operations, the tank will be passivated again to improve corrosion resistance to the pure water. The water tank will then be cleaned and made into a clean room with reduced-radon air delivered from the surface as compressed air through a pipe from the SAL. The tank will be kept with slightly positive pressure to reduce air infiltration. The access door in the side of the tank will be outfitted with a changing room and air lock.

The central top port of the water tank is the only port big enough to allow installation of the Gd-LS tanks. Once the cryostat is installed, this path will be blocked. So the bottom and side Gd-LS tanks must be transported underground and staged in the water tank. Figure 14.2.3.1 shows the sequence of installation in the water tank. The Gd-LS tanks will arrive clean on the inside and covered with protective plastic and rigging frames. The acrylic tanks will be brought into the Yates headframe with a telehandler and set on a cart in front of the Yates cage. The four large side tanks will be moved first. Rigging will be used to attach the steel frame around the tank to the underside of the cage. The cage will be raised to lift the tank until it is vertically under the cage. A drag line will guide the lower end of the tank. The cage will be lowered slowly until the bottom of the tank is at the 4850L. A drag line will be reconnected to the bottom to pull the tank out as it is lowered further. The tank will be placed on a receiving cart. The cart will transport the tank to the entrance to the Davis Campus with a cleaning area (so-called cart wash). The external frame and external packaging around the tank will be cleaned to remove mine dust from transport. The bottom of the tank must enter the Davis area first. One hook from the monorail in the Davis Cavern will connect to two points near the bottom of one side of the rigging frame. A second hook from the same monorail will connect to the top of the tank's rigging frame. The motion of the hooks will be choreographed to lower the acrylic tank into the water tank, keeping the rigging vertical over the lifting points by moving the hooks along the monorail. Once the tank is vertical and set down on the floor of the water tank, it will receive a final inspection and leak test. The external packaging and protective plastic will be removed from the acrylic tank. The outside of the tank will be cleaned with Alconox and water and visually inspected for any cracks. The three bottom tanks and the two top tanks will be transported inside the cage and unpackaged and cleaned in the cart-wash area. The three bottom tanks will be staged inside the water tank. The two top tanks will not be staged, as they will be installed from the top after the cryostat and cable-conduit installation is completed. A custom hoist system will be constructed inside the water tank to move the Gd-LS tanks inside the water tank. Stainless steel support stands that hold the four tall outer Gd-LS tanks will also be staged into the water tank. The HV umbilical and parts for the HX conduit may also be staged in the water tank.

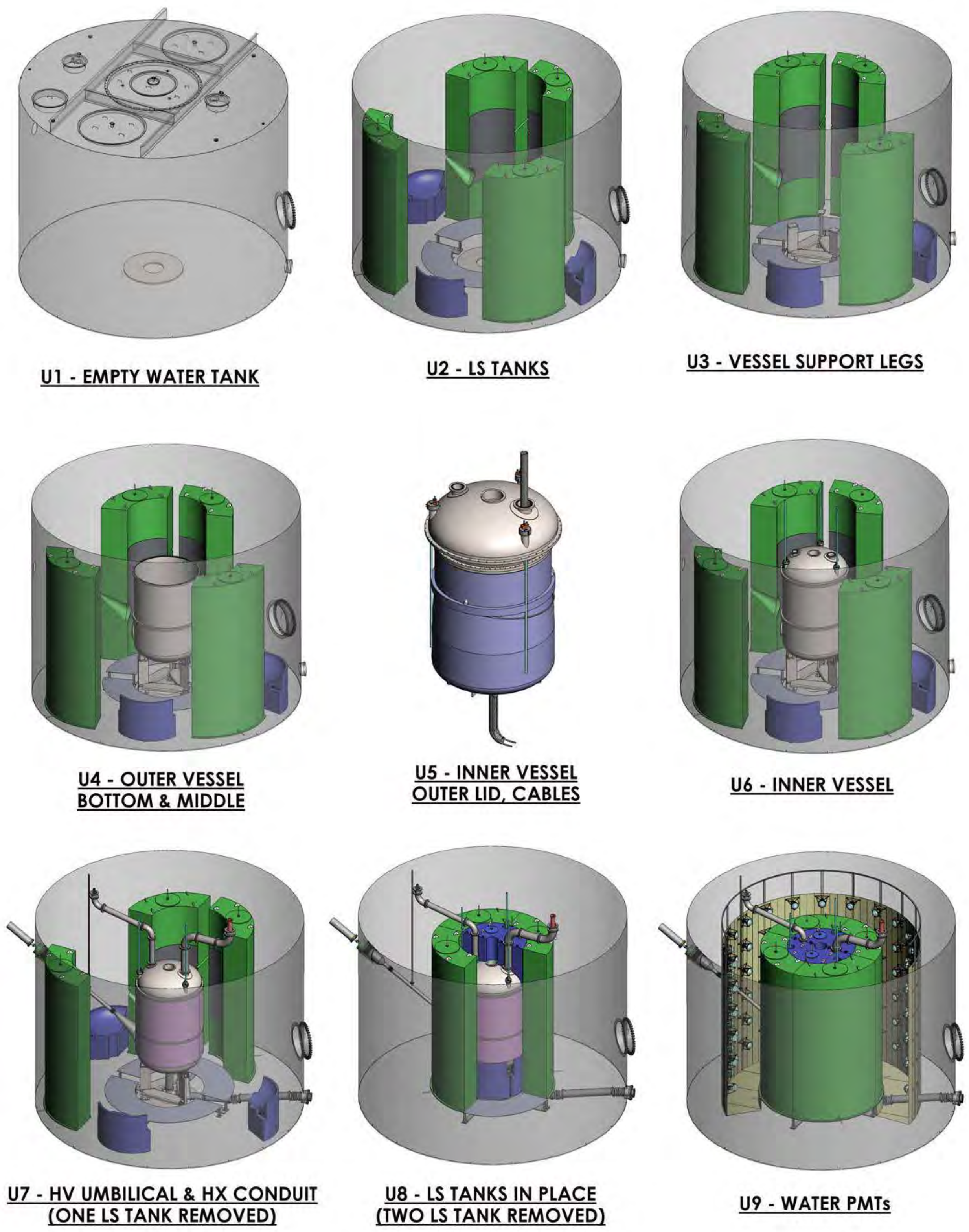

**Figure 14.2.3.1. Underground installation sequence in water tank.**

### 14.2.4 Cryostat Transportation and Underground Assembly (Steps U3 – U6)

The installation of the cryostat starts with the cryostat support. The survey reference system for the detector will be established and a baseplate located on the floor of the water tank. The legs will be brought down in the cage cleaned and double-bagged. The outside bag will be removed at the entrance to Davis. The inside bag will be removed on the top deck near the opening to the water tank. The legs will be lowered into the water tank through the central port with a single hoist. They are then bolted to the baseplate and the supports that contact the outer cryostat will be adjusted to surveyed positions. The outer cryostat has been designed as three pieces so each piece can fit in the Yates cage. Each piece will be brought down in the cage cleaned and double-bagged. The bottom head of the outer cryostat is lowered into the water tank with a single hoist and positioned over the three support legs. The gaps to the legs will be measured and adjusted until an adequate fit is achieved. Once the bottom head is lowered, it will be bolted to the legs. The middle section will be brought in next, rigged into position above the bottom, sealed to the bottom, and leak-tested. The support of the bottom cryostat will be adjusted so the top surface of the outer middle section is level. A displacer around the outer cryostat cylindrical section reduces the amount of water between the Gd-LS tanks and the cryostat. This may be installed around the outer cryostat at this point or after the HV connection is completed. The top lid will be brought into the Davis Campus and staged on the top deck with its inner bag still in place.

The inner cryostat will not fit in the cage, so it must be hung under the cage. It will be horizontal for some of its journey from the SAL to Davis, and vertical for other parts. The TPC support system will be designed to accommodate support in both conditions and the transition between them. An external rigging company will be contracted to rotate the inner cryostat from the vertical assembly position to the horizontal position for removal from the SAL and transport to the Yates headframe. It is anticipated that this move will also be done with a telehandler. There is a detailed plan for transport of the inner cryostat from the surface to the 4850L (see Chapter 13). On the 4850L, a special cart with air skates will be used to bring the inner cryostat from the Yates shaft to the Davis Cavern. The special cart must be cleaned. The outer bag around the cryostat will also be removed in the cleaning area. The cart will move the inner cryostat onto the deck near the entry hole to the water tank. The two hooks on the monorail will be used to lift the inner cryostat off the cart, rotate it back to vertical, and rest it on a temporary support on the top of the deck. A temporary clean room will be built around the inner cryostat. Reduced-radon air from the surface will be used to provide a clean atmosphere and to purge the cryostat once the cable ports are opened. The inner bag around the inner cryostat will be removed. Closed-cell foam insulation will be installed onto the sides and bottom of the cryostat. Then prefabricated superinsulation blankets will be installed. Temporary supports for the outer cryostat lid will be installed onto the inner cryostat. Reduced-radon purge air will be connected to the inner cryostat and the top cable top hat will be opened and removed. The outer cryostat lid that was previously staged will be unbagged, rigged over the inner cryostat, and set on the temporary supports. The upper cables will be routed through the ports on the outer cryostat lid. The permanent support rods that hold the inner cryostat from the outer cryostat will be installed. The supports will be adjusted to position the weir surface of the inner cryostat parallel to the sealing surface of the outer cryostat lid. The lower section of the three calibration ports will also be installed. The lower cable top hat will be opened and removed. The lower cables and LXe lines will be dressed for final installation. The LXe lines from the weir will also be dressed. The outer cryostat lid will then be lifted and the load from the inner cryostat will transfer from the temporary supports on the deck to the permanent supports from the lid. The temporary outer lid supports will be removed and the assembly of the outer cryostat lid and inner cryostat will be lowered into the water tank and into the outer cryostat bottom. As the inner cryostat is lowered, the bottom PMT cables and signal cables will need to be threaded through the central port of the outer cryostat bottom head. The reduced-radon air will continue to purge the inner cryostat during this process. The crane will set the assembly down so the outer cryostat lid rests on the outer cryostat middle section top flange. The inner cryostat will still be hanging from the lid. This flange can then be assembled and leak-checked.

### 14.2.5 Utility Connection (Step U7)

The HX conduit is then installed to the bottom of the cryostat. PMT cables and Xe transport lines share this conduit and connections will be made with an orbital welder whenever possible. Some connections will be VCR fittings. The conduit continues through the outer wall of the water tank to the cryo tower. The PMT cables branch off at a tee to a vertical pipe with an array of flanges and hermetic feed-throughs. This area has many details that must be carefully planned. The upper PMT cables and sensor lines are also installed into a flexible conduit with a vertical pipe and an array of flanges with hermetic feed-throughs.

The HV umbilical attaches to the large side port. To reduce krypton and radon absorption by the internal plastic components, this assembly will also be purged with nitrogen or reduced-radon air whenever it is opened. The umbilical slides as an assembly through the outer port in the wall of the water tank. The mass is supported by cables from the lid of the water tank. The outer vacuum jacket and inner Xe tube are designed so they can slide away from the vessel leaving ~18 inches of space for assembly. The central cable of the umbilical needs to be electrically connected to the cathode. The inner tube of the umbilical then seals against the inner cryostat. This joint is angled, making it difficult to install the bolts at the top of the flange. Studs will be used so the assembly is possible. The flange will have a double O-ring so the seal can be leak-checked at this point. Then the outer vacuum jacket will slide down toward the detector and make a seal to the outer cryostat. This seal will also need a double O-ring so it can be leak-checked. Sealing rings at the water tank wall are installed to seal the inner tube to the outer tube and the outer tube to the vacuum tank. There are no direct water-to-Xe seals.

The final step of cryostat installation is positioning and leveling. The connections for the HX conduit and HV umbilical add load and positional constraints to the hanging inner cryostat. The support rods will be adjusted using feedback from built-in electronic level sensors. We have designed in enough compliance to these connections so the inner vessel can be moved. The cryostat should now be sealed and the reduced-radon air flow can be stopped. The inner cryostat will be pumped down to start long-term outgassing of the internal plastics.

### 14.2.6 Outer Detector Assembly (Steps U8 – U9)

The Gd-LS tanks can now be placed into final position. The three bottom tanks are set on platforms connected to the cryostat legs. The upper Gd-LS tanks are then installed through the top port of the water tank. They are lowered slightly radially outward from their final positions to clear the PMT cable conduit and thermosiphon conduits and, once they are low enough, translated under the conduits to the correct final position. The upper Gd-LS tanks are supported by the top flange of the outer cryostat. The four side tanks will be moved in adjacent to the displacer around the outer cryostat and rest on stainless steel supports. There are notches for the HV umbilical so the tanks have to come in from the proper direction. After positioning, the tops of the tanks are connected for stability. Each tank is secured so it will not float in the water or tip over in an earthquake. Each tank has a fill line and vent line that come to a common overflow reservoir on the top of the water-tank lid. The final system will be visually inspected and leak-checked with a low-pressure gas.

The outer detector PMTs will be installed onto half-ladders and lowered into the water tank through one of the larger off-axis ports in the lid. The half-ladders are assembled together and secured to the roof and floor of the water tank and cables are run up to one of the top water-tank ports. Cables are sealed at these ports. The Tyvek reflectors to direct light lie on the floor, are hung vertically from the top of the ladders, and are stretched across the top of the ladders.

The detector is now ready to be filled with Xe, Gd-LS, and water. Xe filling must wait until the inner cryostat has been at vacuum long enough to get the residual gas content of the plastics to an acceptable level. Warm, low-pressure Xe gas may be circulated to heat the plastic and enhance diffusion. This Xe would be pumped out and repurified or sold. Once plastic outgassing is at an acceptable level, the vessel will be filled with Xe gas and slowly cooled with $LN_2$ until the Xe starts to condense. Filling continues until the Xe liquid is at the desired level. Gd-LS is received on site ready to use in 55-gallon drums. The Gd-LS and water must be filled at the same time to minimize stress on the acrylic walls. The levels do not need to match exactly, so one drum of Gd-LS can be added to the tanks one at a time as the water level rises. It is added by pressurizing the drums with nitrogen gas to force Gd-LS through the filling tubes at the overflow reservoir. Water is purified before it is added and covered with a nitrogen head once the tank is filled. A flowing nitrogen head is maintained over the Gd-LS to protect it from both radon and oxygen.

While the main detector installation and assembly sequence described above are occurring, the support equipment and utilities for the experiment will be installed in Davis. This includes cryogenic cooling equipment, vacuum pumps, $LN_2$ thermosiphons, Xe purification and circulation equipment, TPC HV supplies, PMT readout electronics, PMT HV supplies, calibration source tubes, connections and hardware, the emergency Xe recovery system, DAQ, and control systems. Details of these items are covered in other chapters.

The duration of the work in the SAL from the start of the assembly of TPC to the inner cryostat being sealed and ready to move underground is expected to be seven months. Before underground installation work can begin, LUX will need to be decommissioned and removed and Davis infrastructure work described in Chapter 13 will need to be completed. LZ installation underground has an estimated duration of seven months from staging of the Gd-LS tanks to being ready to fill.

# 15 Offline Computing

## 15.1 Introduction

This section describes the LZ offline computing systems, including offline software for the LZ experiment, the definition of the computing environment, the provision of hardware and manpower resources, and the eventual operation of the offline computing systems. The offline computing organization provides the software framework, computing infrastructure, data-management system, and analysis software as well as the hardware and networking required for offline processing and analysis of LZ data. The system will be designed to handle the data flow starting from the raw event data files (the so-called EVT files) on the SURF surface RAID array, all the way through to the data-analysis framework for physics analyses at collaborating institutions, as illustrated in Figure 15.1.1.

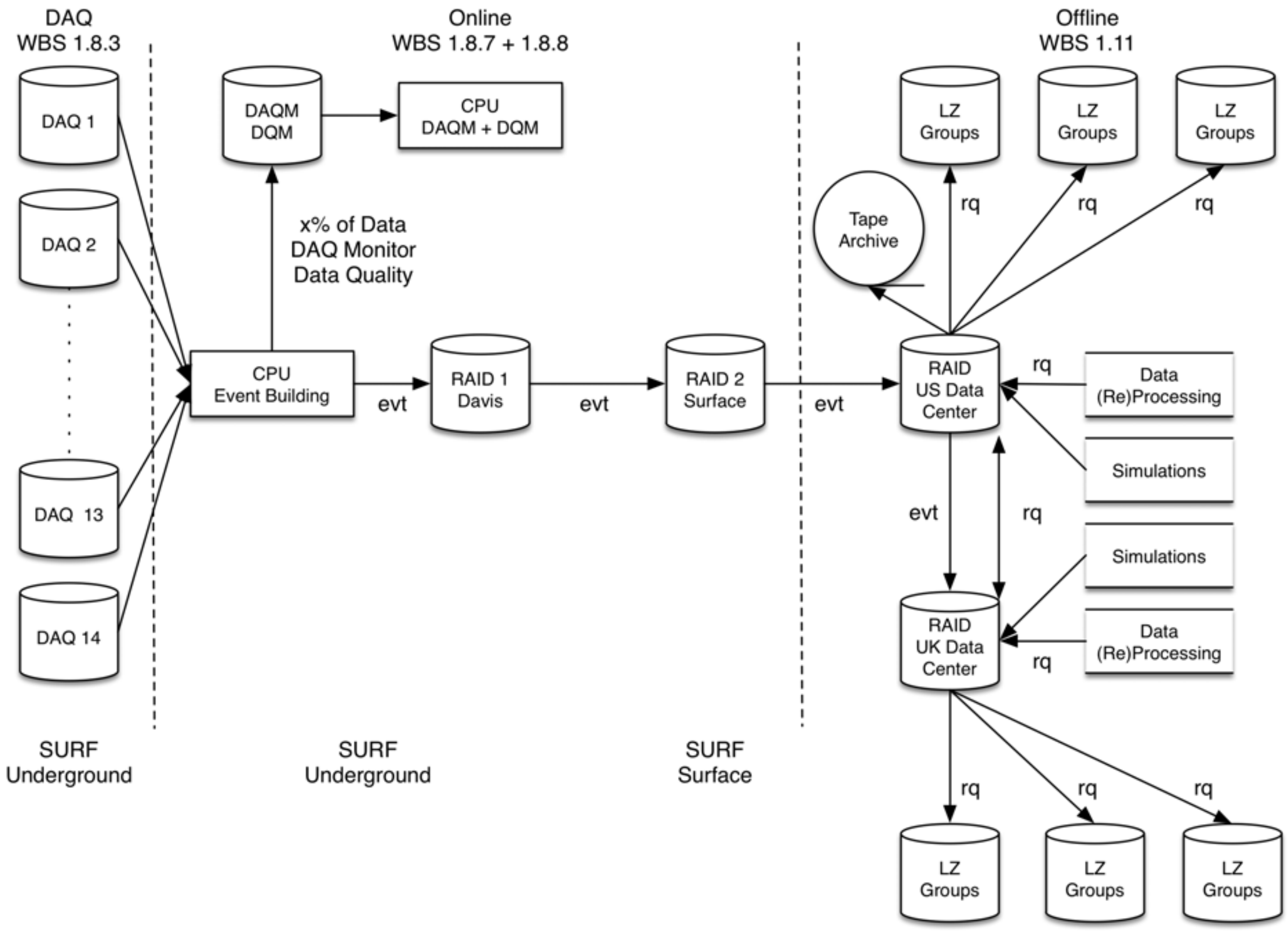

**Figure 15.1.1. Schematic data-flow diagram for LZ.**

## 15.2 Data Volume, Data Processing, and Data Centers

The LZ data will be stored, processed and distributed using two data centers, one in the United States and one in the UK. Both data centers will be capable of storing, processing, simulating and analyzing the LZ data in near real-time. The SURF surface staging computer ships the raw data files (EVT files) to the U.S. data center, which is expected to have sufficient CPU resources for initial processing. The National Energy Research Scientific Computing (NERSC) center at LBNL will contain the resources to act as the LZ U.S. data center. The run processing extracts the PMT charge and time information from the digitized signals, applies the calibrations, looks for S1 and S2 candidate events, performs the event reconstruction, and produces the so-called reduced quantity (RQ) files. These files represent approximately 7% of the size of the original EVT files, based on the LUX experience. The RQ files will be accessible to all groups in the collaboration and represent the primary input for the physics analyses.

**Table 15.2.1. Daily (compressed) data rates in LZ based on scaled LUX data. The scaling factors are as follows: (a) PMT surface area ratio (2) times number of channels ratio — not including the low-gain channels (4); (b) number of channels ratio (8) times rate ratio (13); (c) liquid surface area ratio.**

| Source | LUX (GB/d) | Scaling Factor | LZ (GB/d) | LZ Compressed (GB/d) |
|---|---|---|---|---|
| **Single PE** | 44.00 | 8[(a)] | 352 | 117 |
| **S1** | 0.24 | 104[(b)] | 25 | 8 |
| **S2** | 76.34 | 104[(b)] | 7,939 | 2,646 |
| **Uncorrelated SE** | 20.00 | 9[(c)] | 180 | 60 |
| **Total** | **140.58** | | **8,496** | **2,831** |

The EVT and the RQ files are also mirrored from the U.S. data center to the UK data center (located at Imperial College London) partly as a backup, and partly to share the load of file access/processing, giving better use of resources for all LZ collaborators. The EVT file transfer to the UK data center is done from the U.S. data center as opposed to directly from SURF in order to avoid eating into the bandwidth available to ship the data from the experiment. Subsequent reprocessing of the data (following new calibrations, reconstruction and identification algorithms, etc.) is expected to take place at one or both centers, with the newly generated RQ files copied to the other center and made available to the collaboration.

From the hardware point of view, the system must be able to deal with the LZ data volume in terms of storage capacity and processing. Based on the LUX experience and appropriate scaling for LZ (in terms of number of channels, single/dual gains, rates, etc.), the amount of WIMP search data generated in one year of LZ running is estimated to be 940 TB. Including calibration runs, the total amount of LZ data produced per year is expected to be 1.1-1.2 PB, depending on the amount and type of calibration data collected during yearly operation. This estimate assumes that about three hours of calibration data are collected each week.

The breakdown of the contributions for the different sources of light to the LUX data and their scaling to LZ is given in Table 15.2.1, which clearly shows that the data volume is dominated by the S2 signals. A similar estimate can be obtained by a simple scaling of the average LUX events recorded during krypton calibrations, as described in Chapter 11. These events, with a total energy deposition of 41.6 keV (from 32.2 keV and 9.4 keV conversion electrons), are 203 kB in size, and are dominated by the S2 signal. While these events generate more light than typical events in the WIMP search region, they do provide a useful measure for the data volume. Scaling by the LZ to LUX channel ratio, i.e., by a factor of 8, these events are expected to have 1.6 MB in LZ. With compression, which was shown in LUX to reduce the event size by a factor of 3, this yields 0.53 MB/event. Monte Carlo simulations show that the total background rate in LZ is about 40 Hz, which translates into 21 MB/s, or equivalently 1.83 TB/day. The difference between this rate and the 2.83 TB/day shown in Table 15.2.1 is due to higher-energy background events, some of which can have more than one scatter, i.e., more than one S2 signal. On the other hand, the background rate in the WIMP search region (below 30-50 keV) is expected to be about 0.4 Hz, which means that the total volume of the WIMP-search data can be reduced by optimizing the event selection.

The SURF staging computer will have a disk capacity of 192 TB, enough storage for slightly more than two months of LZ running in WIMP-search mode (at 2.8 TB/day), similar to its underground counterpart. The capacity of the staging arrays was based on the assumption that any network problems between SURF underground and the surface, or the surface to the outside, would take at most several weeks to be fully resolved. The remaining storage capacity can be used to store additional calibration data.

The anticipated data rates imply that the network must be able to sustain a transfer rate of about 0.07 GB/s (which is actually a factor of 2 higher than the nominal rate as a safety margin). Such rates do not represent a particular challenge for the existing networks between SURF and NERSC/LBNL or LBNL and Imperial College.

From the current LUX experience, we expect that processing one LZ event should take no more than six seconds on one core (using a conservative estimate based on an Intel Xeon ES-2670 at 2.6 GHz and 4 GB of RAM per core). Therefore, assuming a data-collection rate of 40 Hz, LZ needs 240 cores to keep up with the incoming data stream. For reprocessing, as software/calibrations are refined, a larger number of cores will be needed to keep the processing time within reasonable limits (e.g., a factor of 10 more CPU cores allows reprocessing of a year's data in approximately one month).

### 15.2.1 The U.S. Data Center

The U.S. data center will be located at NERSC/LBNL. Currently NERSC has four main systems: the Parallel Distributed Systems Facility (PDSF) provides approximately 2,600 cores running Scientific Linux and is the default system for high-energy and nuclear physics projects. The Carver system provides an additional 10,000 cores, while the two CRAY systems, Edison and Hopper, provide 134,000 and 153,000 cores, respectively. All systems can access the Global Parallel File System (GPFS) with a current capacity of about 7.5 PB, which is coupled to the High Performance Storage System (HPSS) with a 240 PB tape robot archive. The LZ resources will be incorporated within the PDSF cluster.

Our planning assumes modest needs for data storage and processing power for simulations, as described in Section 15.4, a rapid growth in preparation for commissioning and first operation, and then a steady growth of resources during LZ operations. The planned evolution of data storage and processing power at the U.S. data center is given in Table 15.2.1.1.

The amounts of raw and calibration data per year are assumed to be 940 TB and 270 TB, as described in the text, while the Monte Carlo data are ramped up to the maximum estimated capacity over the Project period. The processed data are assumed to be 50% of the Monte Carlo simulations and 10% of the data (assuming a slightly higher percentage of 10% in the size of the RQ-files compared to the 7% in LUX). The user data are assumed to be 50% of the Monte Carlo simulations in the years prior to experimental data, and 5% of the total data once LZ is running. The total disk space allocated includes a 20% safety margin with respect to the total amount of calculated data.

The CPU power is slowly ramped up to reach the maximum of 300 cores needed by the simulations one year before LZ operations, after which the yearly CPU capacity is such that it allows continued Monte Carlo production in parallel with real-time data processing, as well as full data reprocessing in a reasonable time.

**Table 15.2.1.1. Planned storage (in TB) and processing power by U.S. fiscal year at the U.S. data center.**

| FY | 2015 | 2016 | 2017 | 2018 | 2019 | 2020 | 2021 | 2022 | 2023 | 2024 |
|---|---|---|---|---|---|---|---|---|---|---|
| Raw data (TB) | - | - | - | - | 470 | 1410 | 2350 | 3290 | 4230 | 5170 |
| Calibration data | - | - | - | - | 135 | 405 | 675 | 945 | 1215 | 1485 |
| Simulation data | 10 | 40 | 60 | 80 | 200 | 200 | 200 | 200 | 200 | 200 |
| Processed data | 5 | 20 | 30 | 40 | 161 | 282 | 403 | 524 | 645 | 766 |
| User data | 5 | 20 | 30 | 40 | 48 | 115 | 181 | 248 | 314 | 381 |
| Total data | 20 | 80 | 120 | 160 | 1014 | 2412 | 3809 | 5207 | 6604 | 8002 |
| Disk space | 24 | 96 | 144 | 192 | 1217 | 2894 | 4571 | 6248 | 7925 | 9602 |
| CPU cores | 75 | 150 | 150 | 300 | 300 | 2700 | 5100 | 7500 | 9900 | 12300 |

We note that the LUX experiment currently sends data from SURF to the primary data mirror at Brown University with an average throughput of 100 MB/s. The primary mirror syncs the data to disk storage at the NERSC center. At NERSC, the LUX data are saved to the RAID 6 disk on the PDSF cluster, and the data are subsequently archived to tape using the HPSS system. Although the U.S. data center will be located at NERSC, other U.S. computing resources are likely to be available to the collaboration. We will utilize resources available to the collaboration in the most effective approach.

### 15.2.2 The UK Data Center

The UK data center will be implemented within the GridPP infrastructure at Imperial College. The UK data center will provide redundancy and parallel capacity for carrying out the first level, near real-time processing of all LZ raw data (when needed), and carrying out reprocessing of the entire data set on timescales of several weeks. Furthermore, the UK data center will contribute to systematic Monte Carlo simulation studies, as well as generating Monte Carlo production runs. In terms of LZ specific software and data, it will be an exact mirror of the U.S. data center and use the same analysis framework, and access the same central database and software repository.

The UK data center at the Imperial College will benefit from a range of local expertise within the HEP group including both technical computing infrastructure and scientific data-processing heritage specific to the LZ requirements. The Imperial College HEP group is a Tier-2 GridPP node and provides the London Group lead and the overall Technical Director for UK GridPP. Local GridPP computing infrastructure includes ~4,000 cores (≡38,890 HEP SPEC), 3.0 PB storage, a 40 Gbit/s network connection (into Janet), and within UK GridPP as a whole there are ~30,000 cores. Within the HEP group are 5 FTE of IT personnel (two SysAdmin, two GridPP, and one other experiment support). Hardware purchased as part of an LZ contribution to the GridPP will be installed into the GridPP and maintained by the local IT personnel and LZ will become an approved project for its entire duration, having both general and dedicated access to GridPP resources. At times of full reprocessing, GridPP will make available on demand at least 1,000 dedicated cores to ensure a several-week turnaround on the full LZ data set. In terms of processing expertise, the HEP group has many people experienced with both CMS and LHCb software coding as well as the lead for the GANGA software used to initiate processing tasks within the GridPP environment. In addition, the Imperial LZ team members who have provided data-center tasks for a number of international projects including ROSAT, ELAIS, ZEPLIN III, and LUX. Working closely with the Imperial College team will be the University of Sheffield and Edinburgh teams, which also bring GridPP expertise and extensive prior experience in direct dark-matter search projects, including LUX and ZEPLIN.

The hardware requirements defined as a contribution to GridPP are 1 PB of storage and 800 processor cores. The hardware will be purchased in two stages, both to defray final costs and to ensure the most up-to-date hardware for the GridPP. At the time of end-use by LZ, GridPP will provide sufficient resources from its available pool and this will be guaranteed for the duration of the LZ Experiment at no further cost to the project. The currently envisioned milestones for the hardware are:

- Early hardware purchase (0.05 PB + 100 cores) by August 2015
- Late hardware purchase (0.95 PB + 700 cores) by April 2017

The early purchase provides sufficient resources to support data-center development and simulation activities at that stage, with the late purchase providing full resources in time to support the experiment commissioning phase.

## 15.3 Software Packages

Among the most important infrastructure offline software packages are the database (DB) and the analysis framework (AF). Figure 15.3.1 shows a schematic flow diagram for the DB. All processes associated with direct control/access to the experiment, i.e., run control (such as run number, time, run configuration, trigger, etc.), slow controls (such as temperature, pressure, HV, etc.), data monitoring, and electronic logs write their data to a secondary DB, located at the experiment (SURF underground). This allows continued data-taking, independent of the connection to the outside world. The secondary DB is mirrored by a tertiary DB, located on the surface at SURF, which in turn is synchronized with the primary DB, located and maintained at the University of Alabama or at the U.S. data center. The data sent from the secondary to the tertiary and subsequently to the primary DB are propagated in near-real time, with latencies of no more than 5-10 minutes. In addition to the information from the underground (secondary) DB, the primary DB records the data-quality information, calibration constants, and run-processing status. Read access from the primary DB is needed for data-quality analysis, calibration, run processing, and ultimately data analysis. The LZ DB will be based on one of the open-source database-management systems — MySQL or PostgreSQL. Although only the latter supports the implementation of bi-temporal data, simpler solutions can be developed to achieve the same functionality.

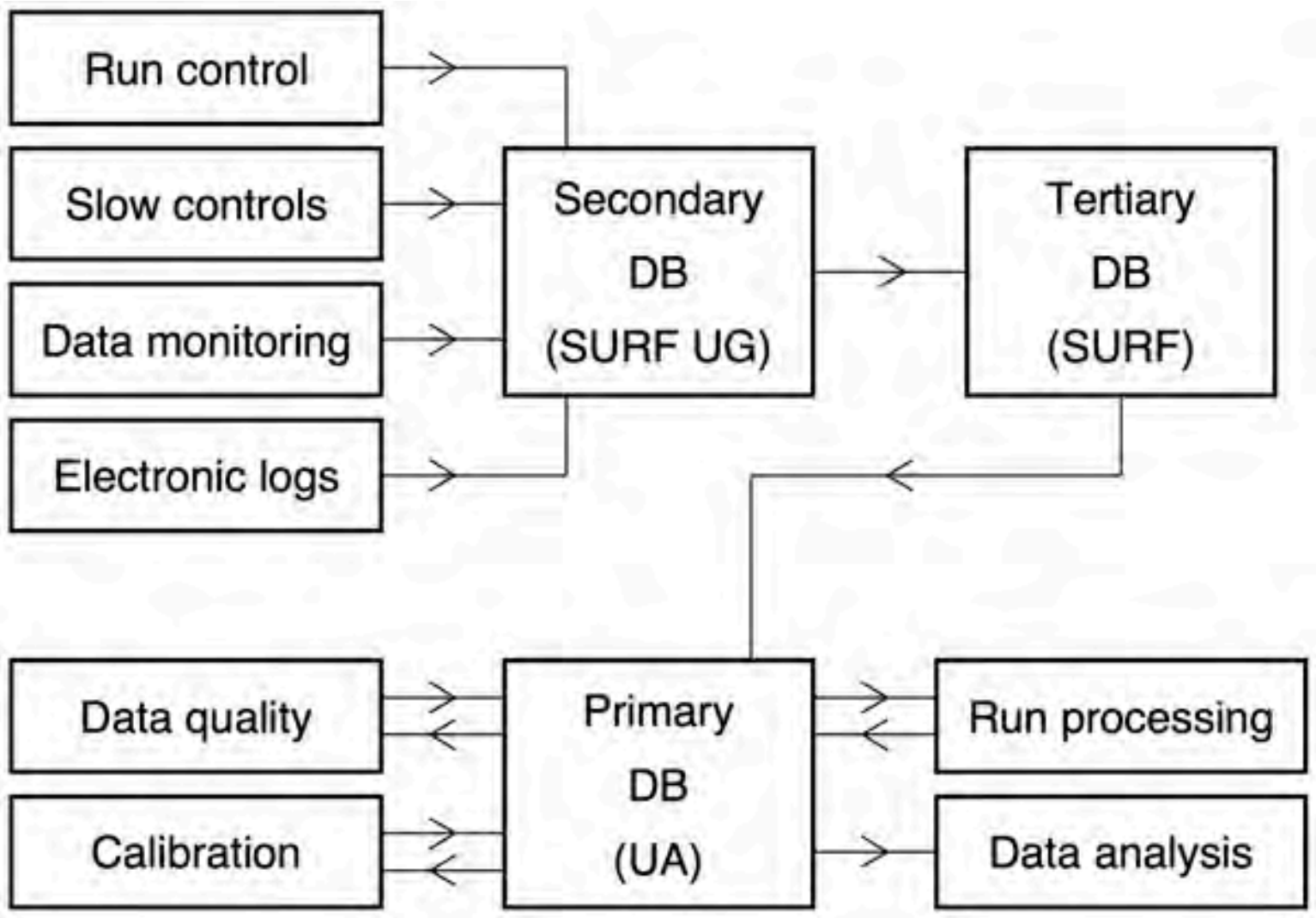

**Figure 15.3.1. Flow diagram for the LZ database system.**

All three DB computers will have identical backup computers ready to take their places should a failure occur. The primary DB itself is backed up by a separate computer, which also ensures the backup of all other offline software components.

The analysis framework will allow users to put together modular code for data analysis to automatically take care of the basic data handling (I/O, event/run selection, etc.). A dedicated task force evaluated various options. In terms of existing frameworks, two ROOT-based frameworks were considered: *Gaudi* (developed at CERN and used by ATLAS, LHCb, MINERvA, Daya Bay, etc.) and *art* (developed at Fermilab and used by MicroBooNE, NOvA, LBNE, DarkSide-50, etc.). In parallel, we evaluated the possibility of evolving the framework developed for LUX, which is based on Python scripts and a MySQL database, and supports modules written in Python, C++/ROOT or MATLAB. For completeness, developing a new framework from scratch was considered as another alternative. However, given the amount of effort this would require (of the order of at least several FTE-years based on estimates from other experiments such as CMS, Double Chooz, MiniBooNE, T2K, etc.), it is an unlikely option. Selection of *Gaudi* as the analysis framework was made in February 2015.

Wherever possible, we anticipate that existing code from the successful LUX and ZEPLIN experiments will be adapted and optimized for LZ, as is, for instance, the case with LUXSim with a geometry option for LZ. In the long run, the LZ simulation is expected to become a stand-alone package, which will be integrated into the data centers frameworks for general LZ community use. The LZ processing and analysis codes will be written to be as portable as possible to ensure straightforward running on both Linux and OSX platforms for those groups who wish to do analysis in-house in addition to (or instead of) running codes on the data centers.

All LZ software (including both online and offline code) will be centrally maintained through a software repository based on *git*, which will include tagged release versions and nightly builds, as well as a suite of well-defined standard performance and integrity tests. A test system based on *gitlab* is currently being evaluated at the University of Alabama.

Cybersecurity risks posed to the offline computing systems relate to the experiment's data and information systems. Much of the LZ computing and data will be housed at major computing facilities in the United States (NERSC/LBNL) and UK (Imperial College), which have excellent cybersecurity experience and records. Specific risks posed to the LZ project relate to data transfer (in terms of data loss or corruption during transfer) and malicious code insertion. File checksums will mitigate the danger of loss or corruption of data during transfer, while copies at both the U.S. and UK data centers provide added redundancy. Malicious code insertion can be mitigated by monitoring each commit to the code repository by the offline group, requiring username/password authentication unique to each contributor to the code repository, eliminating the malicious code from the code repository, and reverting to the previous release.

## 15.4 Simulations

Detailed, accurate simulations of the LZ detector response and backgrounds are necessary, both at the detector design phase and during data analysis. Current LZ simulations use the existing LUXSim software package [1], originally developed for the LUX experiment. This software provides object-oriented coding capability specifically tuned for noble liquid detectors, working on top of the GEANT4 engine. All LZ simulations are expected to be integrated into the broader LZ analysis framework and, as such, this will naturally support ROOT format output at least at the photon level.

The current simulations group is organized into several distinct areas of technical expertise, a structure reflected in the organization of this task:

(a) Definition, maintenance, and implementation of an accurate detector geometry;
(b) Generators for relevant event sources in LZ for both backgrounds and signal;
(c) Maintenance and continued improvement of the micro-physics model of particle interactions in liquid xenon, as captured in the NEST package [2];
(d) Detector response implementation — which transforms the ensemble of individual GEANT4 photon hits at the PMTs to produce an event file of the same format and structure as in the data.

A survey of existing resources within the LZ collaboration shows sufficient CPU power and storage capacity to cover the immediate LZ simulation needs. However, during the LZ project and operation period, the simulations will require an estimated average of 3-5 × $10^4$ CPU hours per week (or equivalently 200-300 cores) and a total of 100-200 TB of disk space (as extrapolated from the current LUX simulation data sets and proper scaling to LZ). These estimates have been fully incorporated in the U.S. and UK data center allocations, both in terms of storage and processing power.

## 15.5 Schedule and Organization

Offline software by its nature is heavily front-loaded in the schedule. To enable the scientists to commission the LZ detector, the software for reading, assembling, transferring, and processing the data must be in place before detector installation. This implies, in particular, that the data transfer, offline framework, and analysis tools themselves will have been developed, tested, debugged, and deployed to the collaboration. We rely on the collaboration's existing experience with the LUX experiment and others (Daya Bay, Double Chooz, Fermi-LAT, DarkSide-50 etc.), which routinely handled similar challenges.

Key offline computing milestones are summarized in Table 15.5.1. After the decision on the choice of the analysis framework for LZ (February 2015), the first framework is expected to be released one year later (February 2016), followed by the first physics integration release another six months later (August 2016). This version includes all necessary modules for real-time processing (i.e., hit-finding algorithms, calibration constants modules, S1/S2 identification, event reconstruction), as well as a fully integrated simulations package (i.e., from event generation through photon hits, digitization, trigger, and data-format output). The first mock data challenge (February 2017) will test both the data flow (transfers, processing, distribution, and logging), as well as the full physics analysis functionality of the framework, separately, while the second data challenge (December 2017) will be dedicated to testing the entire data chain. The third data challenge (June 2018) will also test the entire data chain and is expected to validate the readiness of the offline system just before the LZ cool-down phase.

**Table 15.5.1 Key offline computing milestones.**

| | |
|---|---|
| **Feb. 2015** | Analysis Framework decision |
| **Feb. 2016** | First Analysis Framework release |
| **Aug. 2016** | First physics integration release |
| **Feb. 2017** | First mock data challenge |
| **Dec. 2017** | Second mock data challenge |
| **June 2018** | Third mock data challenge |

Offline computing will be co-led by a physicist experienced in software development and use and a computing professional from LBNL. The software professional will also liaise with NERSC for collaboration on providing LZ compute resources — in particular, provisioning and/or allocating of network, CPU, disk, and tape resources sufficient for LZ collaborators to transfer, manage, archive, and analyze all data for the experiment.

The infrastructure software effort will also involve professional software engineering from LBNL. This person will provide technical leadership, oversight, and coordination of LZ collaboration efforts on infrastructure software as well as the design, implementation, testing, and deployment of critical LZ infrastructure components. LZ infrastructure software includes data management and processing, offline systems and monitoring, offline interfaces to LZ databases, and the analysis framework. The remainder, and bulk, of the software is a collaboration responsibility. Software for simulation, analysis, monitoring, and other tasks will be written and maintained by collaboration scientists.

## Chapter 15 References

# 16 Project Organization, Management, and Operations

The LZ Project (the Project) is international in scope, funding, and organization. This chapter presents an overview of the overall Project organization, safety and risk programs, and the concept for operations. The integrated overall Project Management organization is also described here. This Project organization has the authority and responsibility over all aspects of the Project, including those funded by DOE, NSF, SDSTA, and non-U.S. agencies: the UK's Science & Technology Facilities Council (STFC), Portugal's Fundação para a Ciência e a Tecnologia (FCT), and any Russian- and Chinese-funded scope to be defined. The functions of the Project Advisory Board (PAB) are also described. A detailed discussion of Project Management, management systems, and approaches is described separately in the *Preliminary Project Execution Plan* (P-PEP) document.

## 16.1 LZ Project Organization

The Project's organization from the perspective of DOE is summarized in Figure 16.1.1.

LBNL is the DOE lead laboratory for LZ. As lead laboratory, LBNL will be responsible for Project management and funding from DOE and for ensuring that essential manpower and necessary infrastructure are provided to the Project during the R&D and construction phases. The Project Manager and Project Scientist will be from the lead laboratory and will report to the Physics Division Director. These two Project positions must be jointly approved by LBNL and by the collaboration Executive Board (EB). The LBNL Project Management Office (PMO) will review and provide oversight of the Project and its management systems to ensure that all DOE project guidelines and procedures are followed.

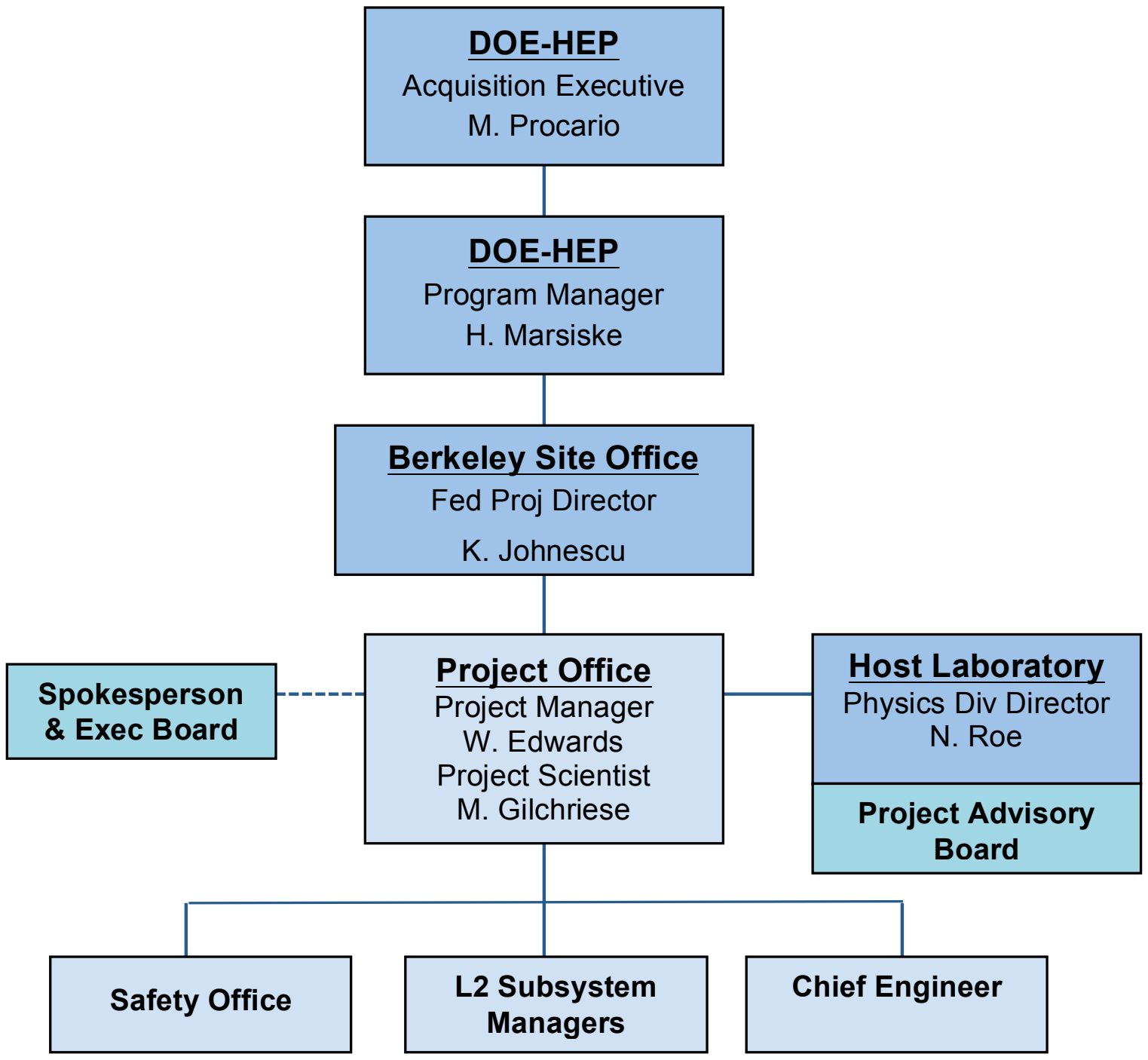

**Figure 16.1.1. LZ Project reporting and responsibility organization chart, with an emphasis on the relationship to DOE.**

Figure 16.1.2 presents the internal organization of the Project. The Spokesperson is elected by the collaborating institutes to represent the scientific interests of the collaboration. The roles and term of the Spokesperson are defined in a governance document. The current Spokesperson is Prof. Harry Nelson (University of California, Santa Barbara). The Spokesperson chairs the EB, which is a representative body of senior collaboration members. The EB will help guide the Project organization in its goal of delivering the experimental apparatus and software that will meet the scientific requirements of the LZ collaboration. An Institutional Board (IB), with representatives from each collaborating institution, meets regularly with the Spokesperson and Project team.

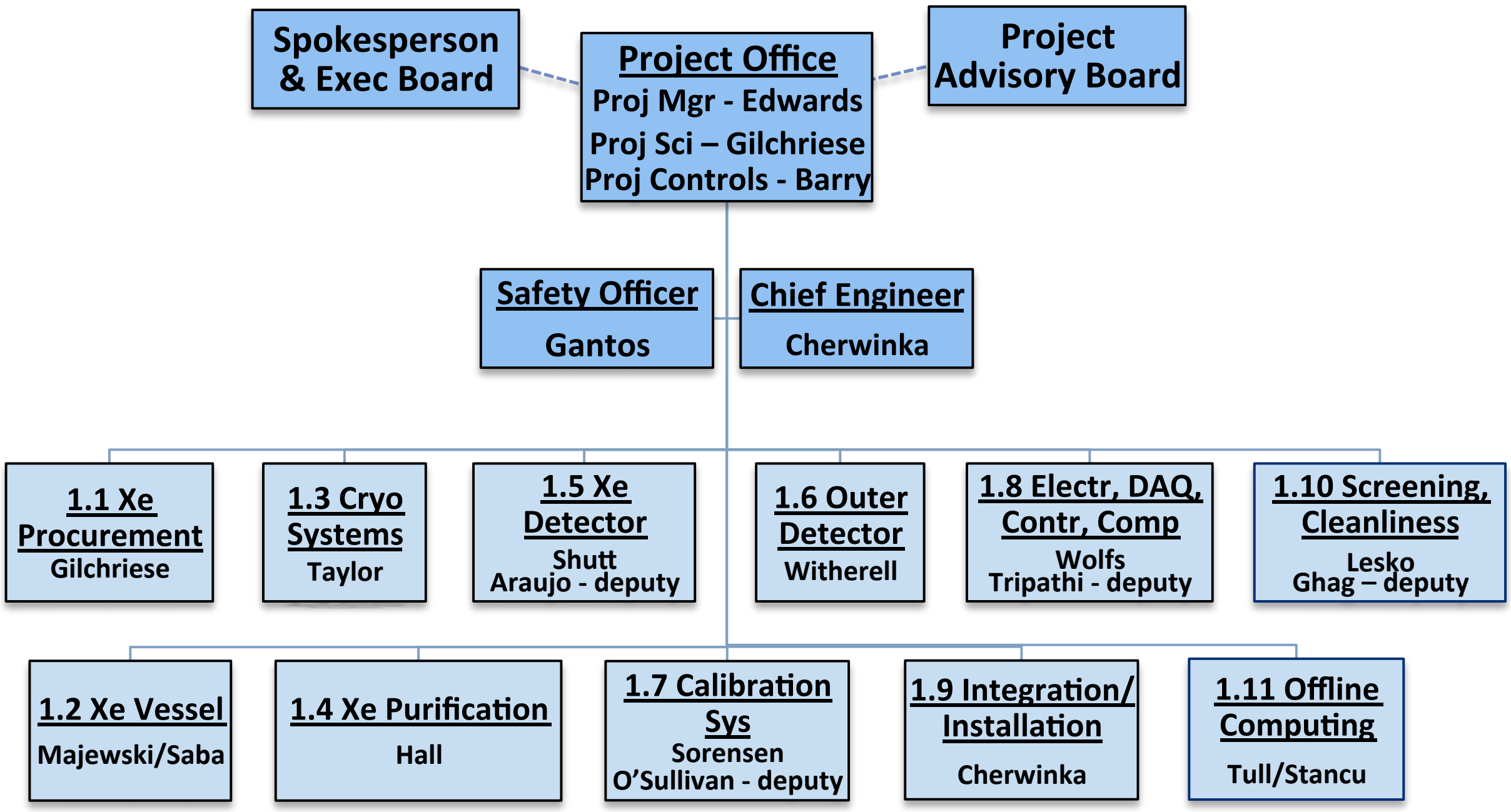

**Figure 16.1.2. LZ project organization.**

### Project Advisory Board

The PAB is an external board, gathered from the U.S. and non-U.S. scientific communities, that has expertise in large scientific projects. This board will provide valuable guidance and advice to the Project over the course of the construction life cycle. The PAB is charged by, and reports to, the Physics Division Director of LBNL. The current members are: David McFarlane (SLAC-chair), Jay Marx (Caltech-retired), Chris Bebek (LBNL), Elaine McCluskey (Fermi National Accelerator Laboratory [FNAL]), Bob McKeown (Jefferson National Accelerator Facility [J-LAB]), Mark Thomson (Cambridge), and Dan Dwyer (LBNL). The PAB may be supplemented as required to provide advice on a specific subject and for specific reviews.

### Project Management Office

The Project Management Office (PMO) personnel include Project Manager William Edwards (LBNL), Project Scientist Murdock Gilchriese (LBNL), Chief Engineer Jeff Cherwinka (Physical Sciences Laboratory, University of Wisconsin), Safety Officer Joseph Gantos (LBNL), and Project Controls Officer Michael Barry (LBNL). Systems engineering and QC/QA functions will also be under the direction of the Project Office.

### Project Work Breakdown Structure

The LZ Work Breakdown Structure (WBS) has 12 major categories, as shown in Table 16.1.1.

### Project Subsystem Organization

The current Subsystem Managers (at Level 2 of the WBS and selected Level 3 managers) and lead engineers are listed in Table 16.1.2. The LZ Technical Board comprises the WBS Level 2 Managers (**Bold**), their deputies, and the Project Office.

The Level 2 or Subsystem Managers, in addition to being members of the LZ Technical Board, are responsible for overseeing the development of the Project baseline with regard to their subsystems. They work with the Project Office to establish a level (L3) organization, helping to ensure that adequate technical resources have been identified, and defining the subsystem-specific requirements as they flow down from the overall Project. The L2 managers oversee the development of the technical design as well as the schedule and cost estimates associated with design, fabrication/execution, assembly, and test of

**Table 16.1.1. LZ Work Breakdown Structure (WBS) shown at L2 and description of what each element covers.**

| WBS | WBS Title | WBS Description |
|---|---|---|
| 1.1 | **Xenon Procurement** | Specification and procurement of the Xe necessary for the LZ experiment. Xe Storage & Transportation Vessels are covered in WBS 1.4, *Xe Purification & Handling*. |
| 1.2 | **Xenon Vessel (Cryostat)** | Labor, materials, and equipment associated with the design, prototyping, materials selection, construction, certification, and delivery, as well as planning and oversight of assembly and testing efforts on site, for the Cryostat Vessel System, its tanks, connecting flanges, insulation, and support structures. |
| 1.3 | **Cryogenic Systems** | Labor, materials, and equipment associated with the design, further prototyping, procurement, construction, assembly, testing, and delivery of the liquid nitrogen Cryogenic System and nitrogen purge system. |
| 1.4 | **Xenon Purification & Handling** | Labor, materials, and equipment associated with the production of high-purity LXe, its storage, delivery to, and recovery from the TPC. This element covers the online purification system, the Xe purity analysis systems, and the automated fail-safe Xe recovery system. A major subcomponent of this element is the stand-alone krypton-removal system, which will be used to purify the Xe prior to experimental operations. |
| 1.5 | **Xenon Detector** | Labor, materials, and equipment associated with the design, prototyping, fabrication, testing, and assembly planning for the central Xe Detector. This element covers the central detector region with its PMTs and the accompanying field-shaping electrodes and reflecting walls. It includes the “skin” veto region outside the main TPC volume and its PMTs. Included are the cathode, anode, and gate HV power supplies and the cathode HV umbilical connection to the TPC cathode and the grid structures, as well as the internal Xe liquid fluid system that brings liquid into the TPC region, providing cooling surfaces for temperature control. Also included is monitoring equipment for temperature, pressure, fluid flow, and other necessary measurements. |
| 1.6 | **Outer Detector System** | Labor, materials, and equipment associated with the design, fabrication, testing, and assembly planning for the Outer Detector system. This includes the acquisition of the acrylic vessels, construction of the scintillator filling system, the acquisition and testing of the Outer Detector PMTs, the mixing and handling of the gadolinium-loaded liquid scintillator, procurement of reflector materials, as well as all the support infrastructure required. It also includes the planning, procedures, and oversight, plus the installation tooling required during the assembly of the system inside the water tank. |
| 1.7 | **Calibration** | Labor, materials, and equipment associated with the design, prototyping, construction, delivery, assembly, and testing, of the Calibration System for the Xe detector and the Outer Detector system, along with the mechanisms, plumbing, valves, and radiation sources required to implement the calibration systems. Included are safety and administrative custodial requirements for source security, handling, and shipping. |
| 1.8 | **Electronics, DAQ, Controls, Computing** | Labor, materials, and equipment associated with the design, prototyping, construction, delivery, assembly, and testing of the analog and digital electronics for the Xe and Outer Detector PMTs, the DAQ and Trigger systems, the PMT HV system, the Detector Control system, and the online and offline hardware and software. This element includes the external signal, PMT HV, and network cables. Not included are the internal HV and signal cables for the PMTs (covered by WBS 1.5) and the detector sensors/instruments required for Detector Control. This element provides standard interfaces for detector sensors/instruments; custom interfaces required to connect custom sensors/instruments to the Detector Control system will not be provided by this element. |
| 1.9 | **Integration & Installation** | Labor and materials necessary to integrate the design effort of the subsystems into an overall detector design, maintain CAD models of the LZ detector and Davis Campus, upgrade the SURF infrastructure to support the detector assembly and operation, and perform on-site surface-level assembly of the detector and installation into the Davis campus underground. Other subsystem elements maintain the responsibility to support integration by communicating design requirements, interface issues, subsystem CAD models, infrastructure needs, and assembly and operation needs. WBS 1.9 supplies planning, management, and skilled labor for assembly and installation, and the subsystems supplies experts on site to support this as needed. |
| 1.10 | **Cleanliness & Screening** | Labor, materials, and equipment associated with specification of radioactive background-level tolerances in the experiment; material radioassaying and control of radioactive background contaminants in the Xe resulting from component outgassing; control of ambient radioactivity; and establishing cleanliness controls, monitoring, and maintenance procedures for manufacture, transport, storage, handling, assembly, and integration of detector components. |
| **1.11** | **Offline Computing** | Software professional labor and computing hardware needed to begin operations of the LZ experiment. Interface to collaboration responsibilities for data processing, analysis, and simulation software. |
| **1.12** | **Project Management** | The cost of labor, travel, and materials necessary to plan, track, organize, manage, maintain communications, conduct reviews, and perform necessary safety, risk, and QA tasks during all phases of the project. Subsystem-related management and support activities for planning, estimating, tracking, and reporting as well as their specific EH&S and QA tasks are included in each of the subsystems. |

**Table 16.1.2. The LZ Project Level 2 and 3 managers and lead engineers.**

| WBS | Description | L2/3 Manager | Deputy or Co-mgr. | Lead Engineer |
|---|---|---|---|---|
| **1.1** | **Xe Procurement** | **M. Gilchriese (LBNL)** | | |
| **1.2** | **Xe Vessel** | **P. Majewski (RAL)** | **J. Saba (LBNL)** | **E. Holtom (RAL)** |
| **1.3** | **Cryogenic System** | **D. Taylor (SDSTA)** | | |
| **1.4** | **Xe Purification** | **C. Hall (UMd)** | | **W. Craddock (SLAC)** |
| 1.4.1 | Xenon Screening | C. Hall (UMd) | | |
| 1.4.2 | Kr Removal | D. Akerib (SLAC) | | |
| 1.4.5 | Xe Gas Recirculation | J. Cherwinka (UW-PSL) | | |
| 1.4.6 | Liquid Xe Tower | W. Craddock (SLAC) | | |
| **1.5** | **Xe Detector** | **T. Shutt (SLAC)** | **H. Araujo (Imperial)** | **J. Saba (LBNL)** |
| 1.5.1 | Cathode High Voltage | D. McKinsey (Yale) | | W. Waldron (LBNL) |
| 1.5.2 | US PMT Systems | R. Gaitskell (Brown) | | |
| 1.5.3 | UK PMT Systems | H. Araujo (Imperial) | | |
| 1.5.4 | TPC | R. Webb (TAMU) | | J. Saba (LBNL) |
| 1.5.5 | Xe Monitoring System | H. Kraus (Oxford) | | |
| 1.5.6 | Internal Fluid System | T. Shutt (SLAC) | | |
| 1.5.7 | System Test | K. Pallidino (SLAC) | | |
| 1.5.8 | Assembly and Installation | J. Saba (LBNL) | | |
| **1.6** | **Outer Detector System** | **M. Witherell (UCSB)** | | **S.Kyre/D.White(UCSB)** |
| 1.6.1 | Scint. Vessels | S. Kyre (UCSB) | | |
| 1.6.4 | Liquid Scint. | M. Yeh (BNL) | | |
| 1.6.5 | Water Tank | D. White (UCSB) | | |
| 1.6.6 | PMT Supports | D. White (UCSB) | | |
| 1.6.7 | PMTs | S. Fiorucci (Brown) | | |
| **1.7** | **Calibration System** | **P. Sorensen (LBNL)** | **K. O'Sullivan(Yale)** | |
| 1.7.1 | Internal Radioact Sources | S. Hertel (Yale) | | |
| 1.7.2 | External Radioact Sources | K. O'Sullivan (Yale) | | |
| 1.7.3 | Mono-energetic N Source | K. O'Sullivan (Yale) | | |
| 1.7.4 | Low-energy nuclear recoil | P. Sorensen (LLNL) | | |
| **1.8** | **Electr., DAQ, Controls, Computing** | **F. Wolfs(URochester)** | **M. Tripathi (UCDavis)** | |
| 1.8.1 | Analog Electronics | M. Tripathi (UCDavis) | | R. Gerhard (UCDavis) |
| 1.8.2 | Trigger Electronics | E. Druskiewics (URochester) | | E. Druskiewics (URochester) |
| 1.8.3 | Data Acquisition System | W. Skulski (URochester) | | W. Skulski (URoch.) |
| 1.8.4 | Eternal PMT HV, Signal | F. Wolfs (URochester) | | |
| 1.8.5 | Slow Control | V. Solovov (Coimbra) | | |
| 1.8.7 | Online HW | J. Buckley (Washington U) | | |
| 1.8.8 | Online SW | J. Buckley (Washington U) | | |
| **1.9** | **Integration & Installation** | **J. Cherwinka (UW-PSL)** | | |
| **1.10** | **Cleanliness and Screening** | **K. Lesko (LBNL)** | **C. Ghag (UCL)** | |
| 1.10.1 | Fixed Contam. Matl Screen | A. Murphy (Edinburgh) | | |
| 1.10.2 | Radon Emanation Screen. | R. Schnee (SDSMT) | | |
| 1.10.3 | Other Screening | C. Hall (UMd) | | |
| 1.10.4 | Cleanliness, Maintenance | J. Busenitz (UAlabama) | | |
| 1.10.5 | Information Repository | J. Busenitz (UAlabama) | | |
| **1.11** | **Offline Computing** | **I. Stancu (UAlabama), C. Tull (LBNL)** | | |
| 1.11.1 | U.S. Data Center | C. Tull (LBNL) | | |
| 1.11.2 | UK Data Center | T. Sumner (Imperial) | | |
| 1.11.3 | Infrastructure Software | S. Patton (LBNL) | C. Faham (LBNL) | |
| 1.11.4 | Simulations | S. Fiorucci (Brown) | P. Beltrame (Edinburgh) | |
| 1.11.5 | Analysis Software | A. Curry (Imperial) | M. Szydagis (SUNY Albany) | |
| 1.11.6 | Subsystem Integration and Validation | S. Patton (LBNL) | M.E. Monzani (SLAC) | |
| 1.11.7 | Subsystem Management | I. Stancu (UAlabama), C. Tull (LBNL) | | |
| **1.12** | **Project Management** | **W. Edwards (LBNL)** | | |

their subsystems. They are also responsible for producing design reports and internal and external reviews. Ultimately, they are responsible for executing the Project Plan with respect to their subsystems.

Design reviews are held for all relevant subsystems and are organized by the Project Office. Each major subsystem and procurement will undergo multiple reviews as the design of the particular subsystem matures and reaches readiness for construction.

## 16.2 Safety

Personnel safety and protecting the environment, as well as equipment safety, are high priorities for the LZ Project. Its scientific goals cannot be achieved without an effective safety and environmental protection program that is integrated into the overall management of the experiment. The details of the ES&H organization are described in an *Integrated Safety Management* document. A separate *Preliminary Hazard Analysis* (PHA) describes the hazards that will be encountered and their associated controls during the execution of the Project. This PHA document received significant input from the L2 subsystem managers who are, and will remain, closely involved in identifying and mitigating these hazards. Many hazards will be similar to those found in past operation of similar experiments (e.g., LUX and Daya Bay). The LZ Project work will take place at multiple institutions in addition to LBNL. Safety of the work at each institution will be the responsibility of the institution and work will be performed in accordance with the requirements and management systems of the home institutions. A sharing of lessons learned for the various locations is expected. Additionally, the LZ Project team will assist collaborating institutions as requested to address any hazard concerns.

Final assembly of the LZ experiment and its operation will take place at SURF, where integrated safety management is well established and will be employed in all phases. SURF ES&H rules and responsibilities will apply to all LZ activity at the SURF site, and SURF will provide safety training for all members of LZ who work on the site.

## 16.3 Risk

The LZ risk program has several key aspects. The first is the early identification of potential risks in each of the detector elements as well as the system as a whole. Second, an early R&D program focuses on understanding, reducing, or eliminating the identified risks. Third is the formal tracking of the remaining risks and mitigation strategies throughout the life of the experiment's construction phase. Last is an accounting for technical, cost, and schedule risk in developing the contingency analysis for the experiment. These first three components (ID, R&D, tracking) will be discussed in this chapter. A Risk Registry for LZ has been assembled and is updated regularly. A Risk Management Plan has also been written.

### Risk Assessment and Tracking

Subsystem Managers have performed a risk assessment of their technical systems. These have been gathered by the Project Office and disseminated back out to the Subsystem Managers, key engineering leads, and the rest of the Project leadership team. Our preliminary *Risk Assessment and Mitigation Strategy* document contains a list of currently identified risks, as well as a further assessment of integrated or system-level risks. A summary list at the end of the document features the highest-priority (a combination of probability and consequence) risks. The *Risk Registry* will be reviewed and discussed regularly in subsystem and overall Project meetings. Updates to the *Risk Registry* will take place several times each year as the Project proceeds, as more information and experience are gathered and risk status changes.

### Risk Mitigation

Several key Project risks can be addressed in the current R&D phase of the project. Therefore, a sizable number of our current and planned R&D efforts are directed toward understanding and mitigating Project risks. Other risk areas require design and the manufacture of prototypes. These are all elements of the risk-mitigation strategy. We will apply the appropriate level of R&D, careful planning, and the judicious assignment of international labor resources within the Project to address all the Project's technical, cost, and schedule risks.

**R&D Plans**

In designing the LZ experiment, R&D efforts focus on developing suitable technologies and helping the collaboration make wise technology and cost decisions. These efforts are also very useful in understanding and reducing risk. The major R&D efforts have been described in previous chapters of this report.

## 16.4 Operations

Experiment Operations will be managed centrally from the PMO in much the same way the construction phase of the Project is managed. Our plan will be based on the successful experience operating the LUX detector at SURF and other projects. The Collaboration will provide much of the necessary resources for shifts and on-call experts. A small engineering and technical group will provide maintenance planning, oversight, and the resources for achieving them. A small computing and software maintenance group will ensure high availability of computing hardware and software and will support the Collaboration's data production and analysis activities.

In the case of LUX, DOE provides LBNL operating funds; NSF provides these funds to two universities. LBNL contracts with SDSTA for on-site support and with university groups to provide travel support to operate the experiment. LBNL also procures materials, supplies, and equipment in support of the experiment. In the case of LZ, the UK will provide substantial aspects of the experiment and may contribute to operations funds for them.

The elements of the LZ operations support are:

- **LBNL operations manager.** Oversight of budget and EH&S matters.
- **EH&S officer.** Provides review of ongoing procedures and EH&S. This is an LBNL position.
- **Engineering Support.** Provides engineering oversight during operations, particularly during the early phases of operations.
- **On-site EH&S officer.** Provides on-site EH&S oversight. In the case of LUX, this is an SDSTA employee under contract to LBNL.
- **On-site operations manager.** Provides on-site maintenance, support for consumables, and interface to SURF staff. The importance of this position in LZ will grow, compared to LUX, given the greater complexity of LZ. This will involve both operations aspects and technical support.
- **On-site procurement.** Materials, supplies, consumables (e.g., liquid nitrogen) under the management of the on-site operations manager.
- **Other procurement.** Materials, supplies, equipment necessary for operations and maintenance. Under the direction of the LBNL operations manager.
- **University travel support.** Support of travel to the site. Under the direction of the LBNL operations manager.
- **Computing support.** Support for professional services for computing hardware and software.

A detailed budget for LZ operations will be developed commensurate with the Critical Decision process. As a point of reference, the LUX operations budget in FY14 (during sustained data taking) is expected to be in the $650K range, but a number of the functions needed for LZ (engineering, computing) are not supported. The LZ operations budget will be higher. We expect that with the added engineering and computing support, the budget will be in the range of $1.5M to $2.5M per year.

# 17 Cost and Schedule Summary

The overall LZ Project plan is summarized in this chapter, including an overview of the Project schedule and the concept for the division of scope and cost among the various funding sources. The planned contributions supported by DOE, South Dakota Science and Technology Authority (SDSTA), the UK's Science & Technology Facilities Council (STFC), China, Portugal, and Russia are outlined. This is a joint project with an international collaboration, and the cost-accounting approaches differ. Therefore, we have attempted to utilize the U.S. DOE cost-accounting approach for the costs presented in this chapter.

## 17.1 Project Schedule

The goal is to begin commissioning and early operations by the start of 2019. The key assumptions behind this date are the timing of the initial CD reviews and the date when we can begin procuring long-lead items such as low-background phototubes for the TPC.

The early critical decision milestones in this plan are a DOE CD-1/3a review in March 2015, followed by a CD-2/3b review in early 2016. Assuming successful reviews and approvals, this enables procurement of long-lead items as early as the late spring of 2015 (by the UK, other) and letting contracts for many other items in early 2016. By late 2017, we should be ready to begin assembling the Xe detector elements into the inner cryostat in the upgraded clean-room assembly space in a surface laboratory at SURF. In parallel with this, decommissioning and removal of the LUX detector will take place. A more complete view of the project schedule is shown in Figure 17.1.1. The dates shown in this schedule correspond to early-finish milestones. Critical decision milestones are given in Table 17.1.1. The future milestones in Table 17.1.1 include schedule float and thus are not indicative of early-finish milestones.

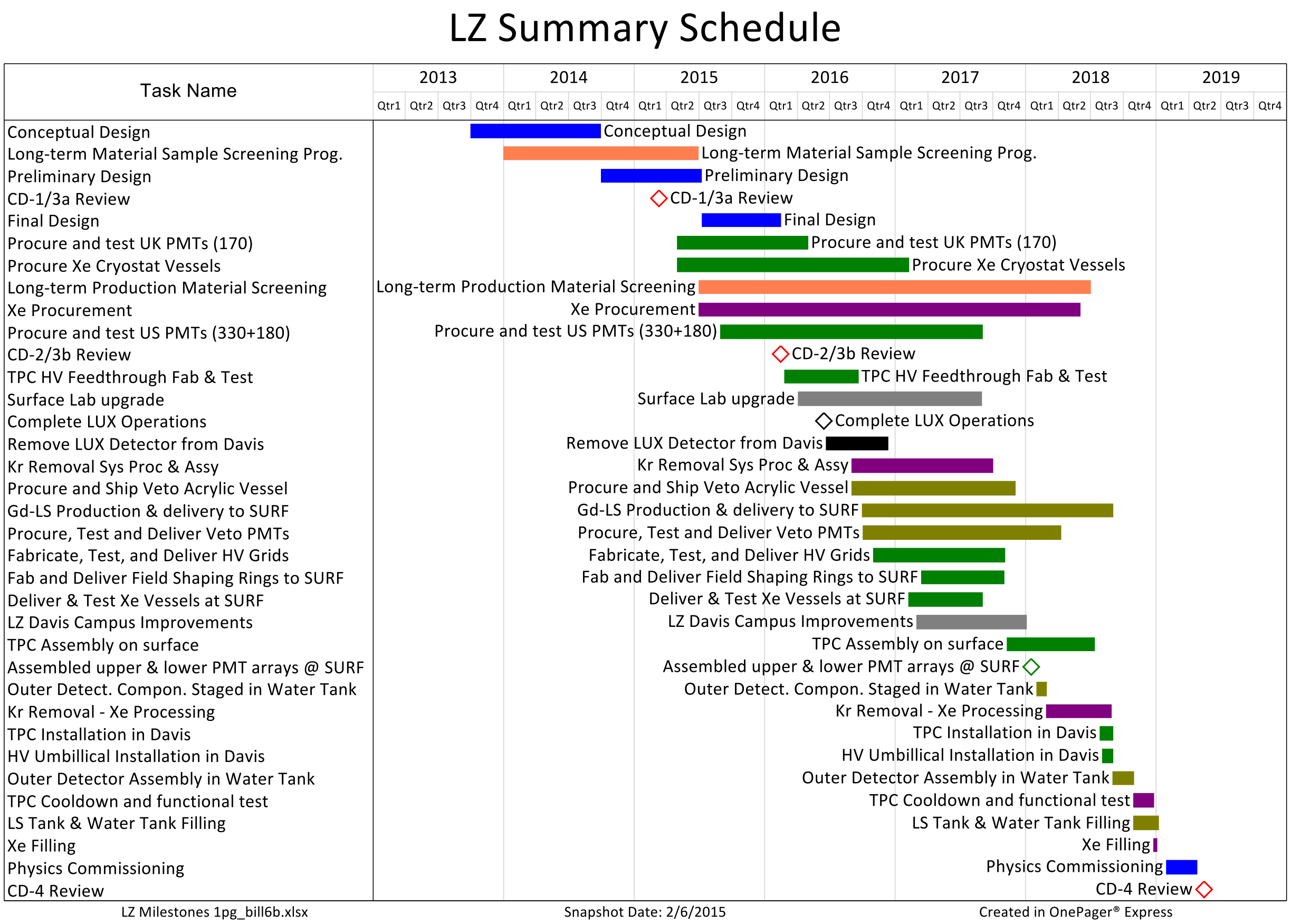

**Figure 17.1.1. Summary of LZ schedule. The dates are in calendar-year quarters, not fiscal-year quarters.**

**Table 17.1.1. Key Project critical decision milestones (U.S. fiscal quarters). Schedule float is included in the future milestones. These are not early-finish dates.**

| Level 1 Milestone | Schedule |
|---|---|
| CD-0, Approve Mission Need | 9/2012 (actual) |
| CD-1/3a, Approve Alternative Selection, Cost Range & Long-lead Procurement | May 2015 |
| CD-2/3b, Approve Performance Baseline & Start of Construction | Q3 FY 2016 |
| CD-4, Approve Project Completion (Ready for Data Taking) | Q1 FY 2021 |

## 17.2 Project Scope

The Project's entire technical scope has been described in the previous chapters. The complete LZ Project includes the detector elements — purified Xe, cryogenic systems, Xe detector, cryostat, veto system, calibration system, electronics, DAQ, trigger, online and offline software, as well as all the integrating activities — system tests, system integration, assembly/installation, on-site infrastructure, and project management.

The planned scope division among the various U.S. and non-U.S. agencies is summarized briefly here. The project scope will be finalized when a formal agreement is completed and approved by the federal, state, and non-U.S. funding agencies.

The major elements of UK scope deliverables include about one-third of the low-background PMTs for the Xe detector, the cryostat set (inner and outer), elements of the low-background counting capability, contributions to integrated system tests, the source calibration delivery mechanism, and extensive contributions to computing and software.

Portugal is expected to contribute to control systems, software, and a measurement system for TPC quality control.

Russia is expected to contribute Xe, and to contribute to integrated system testing and diagnostics.

China may also contribute Xe, possibly including Kr removal, and may contribute to TPC systems, computing, and software. Chinese institutions are not yet members of the LZ collaboration but are starting to be involved in technical and other meetings.

The SDSTA scope includes above- and belowground modifications to required facilities and, as a goal, much of the Xe needed for the experiment.

The NSF is assumed to support scientific efforts for those U.S. collaborating institutions funded by NSF but will not contribute to the Project scope.

The DOE is assumed to fund all remaining Project scope.

## 17.3 Cost Summary

The U.S.-based cost estimate associated with the above scope is shown in Table 17.3.1 in at-year dollars.

**Table 17.3.1. LZ Project U.S.-equivalent cost in at-year dollars. The columns show the DOE base cost (without contingency), planned U.S. equivalent costs of non-U.S. in-kind contributions, the equivalent cost of in-kind contributions planned from SDSTA, DOE contingency, other contingency (from non-U.S. or SDSTA), and the total U.S. equivalent costs.**

| WBS / TASK NAME | DOE Base | Foreign | SDSTA | DOE Cont. | Other Cont. | TOTAL COST |
|---|---|---|---|---|---|---|
| 1.1 Xe Procurement | | 6,060,000 | 14,140,000 | | | 20,200,000 |
| 1.2 Xe Vessel | | 1,937,445 | | | 416,000 | 2,353,445 |
| 1.3 Cryogenic System | 1,483,450 | | | 465,926 | | 1,949,376 |
| 1.4 Xe Purification | 5,782,541 | 247,582 | 279,924 | 2,017,761 | | 8,327,808 |
| 1.5 Xe Detector System | 7,478,267 | 2,706,965 | | 2,279,357 | | 12,464,589 |
| 1.6 Outer Detector | 3,961,426 | 225,865 | | 1,230,873 | | 5,418,164 |
| 1.7 LZ Calibration System | 600,621 | 93,580 | | 180,186 | | 874,387 |
| 1.8 Electronics, DAQ, Controls & Computing | 3,311,935 | 151,065 | | 970,947 | | 4,433,947 |
| 1.9 Integration and Installation | 3,504,935 | | 1,252,800 | 1,057,582 | 300,000 | 6,115,317 |
| 1.10 Cleanliness and Screening | 1,214,661 | 822,926 | | 353,250 | | 2,390,837 |
| 1.11 Offline Computing | 2,127,320 | 399,176 | | 638,195 | | 3,164,691 |
| 1.12 Project Management | 3,222,607 | | | 930,285 | | 4,152,892 |
| Risk Based Contingency | | | | 4,652,000 | | 4,652,000 |
| **TOTAL COST** | **32,687,763** | **12,644,604** | **15,672,724** | **14,776,362** | **716,000** | **76,497,453** |

# LZ Glossary

| | |
|---|---|
| ADCC | analog to digital converter count |
| AF | analysis framework |
| ALP | axion-like particle |
| API | application programming interface |
| APT | Applied Plastics Technology Inc. |
| Ar | argon |
| ASME | American Society of Mechanical Engineers |
| atma | atmospheres absolute |
| BEGe | broad energy germanium |
| BHUC | Black Hills State University Underground Campus |
| BLBF | Berkeley Low Background Facility |
| BNL | Brookhaven National Laboratory |
| BVPC | Boiler and Pressure Vessel Code |
| CAD | computer-aided design |
| CAMS | Center for Accelerator Mass Spectrometry |
| CCD | charge-coupled device |
| CCV | Center for Computation and Visualization |
| CDM | cold dark matter |
| CDMS | Cryogenic Dark Matter Search |
| CFD | computational fluid dynamics |
| CMB | cosmic microwave background |
| CMSSM | constrained minimal supersymmetric standard model |
| CNC | computer numerical control |
| CNS | coherent neutrino scattering |
| CP | charge parity |
| CORBA | Common Object Request Broker Architecture |
| CRC | cyclic redundancy check |
| CTA | Cherenkov Telescope Array |
| CUBED | Center for Ultra-low Background Experiments at Dakota |
| CUORE | Cryogenic Underground Observatory for Rare Events |
| CW | Case Western |
| CWRU | Case Western Reserve University |
| DAMA | DArk MAtter experiment |
| DAQ | data acquisition |
| DB | database |
| DC | Data Collector |
| DCM | DAQ Control and Monitoring |
| DD | direct detection |
| DDC | direct digital controller |
| DE | Data Extractor |
| DEAP | Dark Matter Experiment using Argon Pulse-shape discrimination |
| DECM | DAQ Expert Control/Monitoring |
| DM | DAQ Master |
| DOT | Department of Transportation |
| DRIFT | Directional Recoil Identification From Tracks |
| DS | Data Sparsifier |
| DSCM | DS Control and Monitoring |
| DSECM | DS expert control/monitoring |
| DSM | Data Sparsification Master |
| DSNB | diffuse supernova neutrino background |
| DSP | digital signal processor |
| EB | electron beam |
| EB | Event Builder (Chapter 11) |
| EB | Executive Board (Chapter 15) |
| ECN | engineering change notice |
| ECR | engineering change request |
| ER | electron recoil |
| ETFE | ethylene tetrafluoroethylene |
| EVT | engineering validation test |
| EXO | Enriched Xenon Laboratory |
| FCP | fluorocarbon polymer |
| FEA | finite element analysis |
| FEP | fluorocarbon polymer |
| FIFO | first in, first out |
| FNAL | Fermi National Accelerator Laboratory |
| FPGA | field-programmable gate array |
| FSM | finite state machine |
| FSR | first science run |
| FWHM | full width at half maximum |
| FWTM | full width at tenth maximum |
| G2 | generation-2 |
| Gd | gadolinium |
| Gd-LS | gadolium-liquid scintillator |
| GD-MS | glow-discharge mass spectrometry |
| Ge | germanium |
| GERDA | GERmanium Detector Array |
| GPFS | Global Parallel File System |
| GUI | graphical user interface |
| HDMI | High Definition Multiple Interface |
| HEP | high-energy physics |
| HPGe | high-purity germanium |
| HPSS | High Performance Storage System |
| HV | high voltage |
| HVP | high-voltage connection port |
| HX | heat exchanger |
| I/O | input/output |
| IB | Institutional Board |
| ICD | interface control document |
| ICE | Internet Communications Engine |
| ICP-MS | inductively coupled mass spectrometry |
| ICP-OES | inductively coupled plasma optical emission spectroscopy |
| ID | indirect detection |
| IOSERDES | input/output serial/deserializer |
| IR | infrared |
| J-LAB | Jefferson National Accelerator Laboratory |
| LAB | linear-alkylbenzene |
| LAN | local area network |
| LAr | liquid argon |
| LBF | LBNL Background Facility |
| LBNL | Lawrence Berkeley National Laboratory |
| LCDM | Lambda Cold Dark Matter |
| LCDM | Laboratory for Cosmological Data Mining |
| LC-MS | liquid chromatography-mass spectrometry |
| LD | angular-momentum (L) dependent |
| LHC | Large Hadron Collider |
| LKP | lightest Kaluza-Klein particle |
| LLNL | Lawrence Livermore National Laboratory |
| LN | liquid nitrogen |
| LRF | light response function |
| LS | liquid scintillator |
| LSD | angular-momentum (L) and spin (S) dependent |
| LSZH | low smoke zero halogen |
| LUX | Large Underground Xenon |
| LVDS | low-voltage differential signaling |
| LXe | liquid xenon |
| LZ | LUX-ZEPLIN |
| MC | Monte Carlo |
| MDA | minimum detectable activity |
| Mg | magnesium |
| MIE | Major Item of Equipment |
| MIT | Massachusetts Institute of Technology |
| MITR-II | MIT Reactor II |
| MLI | multilayer insulation |
| MOU | memorandum of understanding |
| MSSI | multiple-scintillation single-ionization |
| mwe | meters water equivalent |
| NAA | neutron activation analysis |
| NAIAD | NaI Advanced Detector |
| NERSC | National Energy Research Scientific Computing Center |
| NEST | Noble Element Simulation Technique |
| NEXT | Neutrino Experiment with a Xenon TPC |
| NIC | network interface card |
| NIM | nuclear instrumentation module |
| NMSSM | next-to-minimal supersymmetric standard model |
| NR | nuclear recoil |
| OSHA | Occupational Safety and Health Administration |
| OVC | outer vacuum cryostat |
| PAB | Project Advisory Board |
| PCB | printed circuit board |
| PDE | photon detection efficiency |
| PDSF | Parallel Distributed Systems Facility |
| PFA | perfluoroalkoxy |
| PHA | Preliminary Hazard Analysis |
| phe | photoelectron |
| PID | proportional integrative derivative controller |
| PLC | process loop controller |
| PLC | programmable logic controller |
| PMO | Project Management Office |
| pMSSM | phenomenological minimal supersymmetric standard model |
| PMT | photomultiplier tube |
| POD | Pulse Only Digitization |
| PPA | particle physics and astrophysics |
| ppb | parts per billion |
| P-PEP | Preliminary Project Execution Plan |
| ppt | parts per trillion |
| PSD | pulse shape discrimination |
| psi | pounds per square inch |
| psia | pounds per square inch absolute |
| psig | pounds per square inch gauge |
| PTFE | polytetrafluoroethylene, or Teflon® |
| QCD | quantum chromodynamics |
| QE | quantum efficiency |
| RAID | redundant array of independent disks |
| RAL | Rutherford Appleton Laboratory |
| RC | run-control |
| RQ | reduced quantity |
| RTD | real-time dispatching or resistance temperature detector |
| S1 | prompt scintillation signal |
| S2 | proportional scintillation signal |
| SAL | Surface Assembly Laboratory |
| SBR | styrene-butadiene rubber |
| SD | spin-dependent |
| SDSMT | South Dakota School of Mining and Technology |
| SDSTA | South Dakota Science and Technology Authority |
| SHM | standard halo model |
| SI | spin-independent |
| slpm | standard liters per minute |
| SMU | Southern Methodist University |
| SPEC | standard performance evaluation corporation |
| SPHE | single photoelectron |
| SRV | storage and recovery vessel |
| SS | stainless steel |
| SSD | solid state disk |
| SSD | solid-state drive |
| SSF | Surface Storage Facility |
| SSR | second science run |
| STFC | Science & Technology Facilities Council |
| STP | standard temperature and pressure |
| SURF | Sanford Underground Research Facility |
| SUSY | supersymmetry |
| TD | technical difficulty |
| TEC | total estimated cost |
| Th | thorium |
| Ti | titanium |
| TMHA | trimethyl hexanoic acid |
| TPB | tetraphenyl butadiene |
| TPC | time projection chamber |
| TTL | transistor – transistor logic |
| U | uranium |
| U/Th | uranium/thorium |
| UC | University of California |
| UCL | University College London |
| UDP | User Datagram Protocol |
| UKDMC | UK Dark Matter Collaboration |
| USD | University of South Dakota |
| UW-PSL | University of Wisconsin Physical Sciences Laboratory |
| VAR | vacuum arc remelting |
| VCR | a type of vacuum fitting |
| VJ | vacuum jacket |
| VUV | vacuum ultraviolet |
| WARP | WIMP Argon Programme |
| WBS | Work Breakdown Structure |
| wDSM | Data Sparsification Master |
| WIMP | weakly interacting massive particle |
| WMAP | Wilkinson Microwave Anisotropy Probe |
| Xe | xenon |
| XRF | X-ray fluorescence |
| ZE3RA | ZEPLIN-III Reduction and Analysis |